\global\def\draftcontrol{0}

   \def\versionno{ On-shell Yang-Mills -- draft   }

\catcode`\@=11

\expandafter\ifx\csname draftcontrol\endcsname\relax\global\def\draftcontrol{0}
\fi

{\count255=\time\divide\count255 by 60
\xdef\hourmin{\number\count255}
\multiply\count255 by-60\advance\count255 by\time
\xdef\hourmin{\hourmin:\ifnum\count255<10 0\fi\the\count255}}
\def\draftdate{\number\month/\number\day/\number\year\ \ \ \hourmin }

\newcommand\makepapertitle{\par
  \begingroup
    \renewcommand\thefootnote{\@fnsymbol\c@footnote}%
    \def\@makefnmark{\rlap{\@textsuperscript{\normalfont\@thefnmark}}}%
    \long\def\@makefntext##1{\parindent 1em\noindent
            \hb@xt@1.8em{%
                \hss\@textsuperscript{\normalfont\@thefnmark}}##1}%
     \newpage
     \global\@topnum\z@   
     \@makepapertitle
     \thispagestyle{empty}\@thanks
  \endgroup
  \setcounter{footnote}{0}%
  \global\let\thanks\relax
  \global\let\makepapertitle\relax
  \global\let\@makepapertitle\relax
  \global\let\@thanks\@empty
  \global\let\@author\@empty
  \global\let\@date\@empty
  \global\let\@title\@empty
  \global\let\title\relax
  \global\let\author\relax
  \global\let\date\relax
  \global\let\and\relax
  \def\version{\let\version\@version\@gobble}
}
\def\@makepapertitle{%
  \newpage
   \ifnum\draftcontrol=1 {}
   \version\versionno
   \vskip 3em%
   \else
   \hfill\hbox to 3cm {\parbox{4cm}{\@pubnum}\hss}%
   \vskip 3em%
   \fi
   \begin{center}%
   \let \footnote \thanks
     {\LARGE {\@title}}%
     \vskip 1.5em%
     {\normalsize
       \lineskip .5em%
       \begin{tabular}[t]{c}%
         \@author
       \end{tabular}\par}%
     \vskip 1.5em%
     {\@bstract}%
     \end{center}%
     \vskip 1.5em
     \@date%
   \par
}

\gdef\@pubnum{}
\def\pubnum#1{%
  \gdef\@pubnum{#1}}

\gdef\@bstract{}
\def\Abstract#1{%
  \gdef\@bstract{%
   \parbox{\textwidth-0pc}{%
   \centerline{\bf Abstract}\penalty1000%
\kern.2cm%
\noindent
\renewcommand\baselinestretch{1.0}%
{#1}}}
}

\def\ps@paper{\let\@mkboth\@gobbletwo%
     \ifnum\draftcontrol=1
    \def\@oddfoot{\hbox to \textwidth{\tiny \versionno \hfil\tiny\draftdate}%
    \hskip -\textwidth \hbox to \textwidth{\hfil\rm\thepage\hfil}}%
     \else\def\@oddfoot{\hbox to \textwidth{\hfil\rm\thepage\hfil}}
     \fi
     \let\@evenfoot\@oddfoot
}

\def\body{\clearpage
          \pagestyle{paper}
    }

\def\@version#1{\ifnum\draftcontrol=1
\typeout{}\typeout{#1}\typeout{}
\vskip3mm\centerline{\hbox{\fbox{\normalsize{\tt DRAFT -- #1 -- }
                   {\draftdate}}}}\vskip3mm
\fi}
\let\version\@version
\long\def\eqlabel#1{\ifnum\draftcontrol=1
                    \tag@false  
                    \tag*{(\theequation) \hbox to -0.2cm{\hspace{0cm}\small{#1}\hss}}
                    \refstepcounter{equation}
                    \edef\@currentlabel{\theequation}
                    \ltx@label{#1}          
                    \else
                    \label{#1}
                    \fi
                    }
\let\st@bibitem\@bibitem
\let\st@lbibitem\@lbibitem
\ifnum\draftcontrol=1
  \def\@bibitem#1{%
    \st@bibitem{#1}\a@@label{#1}\ignorespaces}
  \def\@lbibitem[#1]#2{%
    \st@lbibitem[#1]{#2}\a@@label{#2}\ignorespaces}
  \def\a@@label#1{%
    \gdef\a@lab{\smash{\normalfont\small#1}}
    \ifvmode
      \if@inlabel
        \global\setbox\@labels\hbox{%
          \llap{\a@lab\let\a@lab\relax
                \kern\@totalleftmargin\kern\marginparsep}%
          \box\@labels}%
      \fi
    \fi}
\fi

\documentclass[11pt,A4paper]{article}

\usepackage{amsmath,amssymb,array,calc,epsfig}
\usepackage{cite}
\usepackage{xcolor}
\usepackage[
      colorlinks=true,
      linkcolor=red,  
      urlcolor=blue,    
      filecolor=red,     
      linkcolor=blue,
      citecolor = red,
      pdfstartview=FitV,
      pdftitle={},
        pdfauthor={Paolo Benincasa},
        pdfsubject={},
        pdfkeywords={},
        pdfpagemode=None,
        bookmarksopen=true    
      ]{hyperref}
      
\usepackage{graphicx}
\usepackage{epstopdf}
\usepackage{ccaption}

\ifnum\draftcontrol=1
\fi

\renewcommand\baselinestretch{1.25}
\renewcommand\section{\@startsection {section}{1}{\z@}%
                                   {-3.5ex \@plus -1ex \@minus -.2ex}%
                                   {2.3ex \@plus.2ex}%
                                   {\normalfont\large\bfseries}}
\renewcommand\subsection{\@startsection{subsection}{2}{\z@}%
                                   {-3.25ex\@plus -1ex \@minus -.2ex}%
                                   {1.5ex \@plus .2ex}%
                                   {\normalfont\normalsize\bfseries}}
\renewcommand\subsubsection{\@startsection{subsubsection}{3}{\z@}%
                                   {-3.25ex\@plus -1ex \@minus -.2ex}%
                                   {1.5ex \@plus .2ex}%
                                   {\normalfont\normalsize\it}}
\renewcommand\paragraph{\@startsection{paragraph}{4}{\z@}%
                                   {-3.25ex\@plus -1ex \@minus -.2ex}%
                                   {1.5ex \@plus .2ex}%
                                   {\normalfont\normalsize\bf}}

\numberwithin{equation}{section}

\def\revise#1       {\raisebox{-0em}{\rule{3pt}{1em}}%
                     \marginpar{\raisebox{.5em}{\vrule width3pt\
                     \vrule width0pt height 0pt depth0.5em
                     \hbox to 0cm{\hspace{0cm}{%
                     \parbox[t]{4em}{\raggedright\footnotesize{#1}}}\hss}}}}

\def\sqr#1#2{{\vcenter{\vbox{\hrule height.#2pt
 \hbox{\vrule width.#2pt height#1pt \kern#1pt
 \vrule width.#2pt}\hrule height.#2pt}}}}

\def\aa1{\phi}
\def\cc1{\psi}

\catcode`\@=12

\begin{document}


\title{\bf On-shell diagrammatics and the perturbative structure
of planar gauge theories}

\pubnum{%
arXiv:1510.xxxx}
\date{October 2015}

\author{
\scshape Paolo Benincasa${}^{\dagger}$\\[0.4cm]
\ttfamily ${}^{\dagger}$Instituto de F{\'i}sica Te{\'o}rica, \\
\ttfamily Universidad Aut{\'o}noma de Madrid / CSIC \\
\ttfamily Calle Nicolas Cabrera 13, Cantoblanco 28049, Madrid, Spain\\
\small \ttfamily paolo.benincasa@csic.es \\
}

\Abstract{We discuss the on-shell diagrammatic representation of theories less special than maximally supersymmetric
          Yang-Mills. In particular, we focus on planar $\mathcal{N}\,\le\,2$ gauge theories, including pure 
          Yang-Mills. For such a class of theories, the on-shell diagrammatics is endowed with a decoration 
          which carries the information on the helicity of the coherent states. In the first part of the
          paper we extensively discuss the properties of this decorated diagrammatics. Particular relevance have
          the helicity flows that the decoration induces on the diagrams, which allows to identify the different
          classes of singularities and, consequentely, the singularity structure of the on-shell processes. The
          second part of the paper establishes a link between the decorated on-shell diagrammatics and the
          scattering amplitudes for the theories under examination. We prove that an all-loop recursion relation
          at integrand level holds also for $\mathcal{N}\,=\,1,\,2$, while for $\mathcal{N}\,=\,0$ we are able
          to set up a preliminary analysis at one loop. In both supersymmetric and non-supersymmetric case,
          the treatment of the forward limit is subtle. We provide a fully on-shell analysis of it which is
          crucial for the proof of the all-loop recursion relation and for the analysis of pure Yang-Mills.}

\makepapertitle

\body

\version\versionno

\tableofcontents

\section{Introduction}\label{sec:Intro}

Our understanding of perturbation theory in particle physics is mainly based on its Lagrangian formulation and the 
related Feynman diagrammatics which allows to compute relevant observables such as correlation functions and 
scattering amplitudes. In general, choosing a certain formulation of a theory boils down to establish the basic 
{\it foundational} hypothesis our construction is based on.

The point of view we are most accustomed to is to have unitarity and locality as manifest as possible -- and thus 
they are part of the fundamental set of assumptions -- which typically associates redundancies to the description, 
such as the gauge ones and field redefinitions, while any physical quantity is invariant under gauge and 
field-redefinition choices.

This enhanced freedom in the description, despite of the undeniable success in the exploration of physical
processes, turns out to obscure a great deal of structure of the theory itself. This was already suggested by both 
simple \cite{Parke:1986gb} and recursive formulas 
\cite{Berends:1987me, Mangano:1987xk,Berends:1988zp, Kleiss:1988ne, Berends:1989hf, Kosower:1989xy} 
for scattering amplitudes of gluons,  which are much simpler than what could have been hoped from the Feynman 
expansion.

In the last years, the situation became more and more surprising both for general theories and, in particular,
for the planar sector of $\mathcal{N}\,=\,4$ supersymmetric Yang-Mills theory. In the former case the development of
a very general method to explore the perturbative regime, such as the BCFW-{\it like} deformations 
\cite{Britto:2005aa, Risager:2005vk, Benincasa:2007qj, Cheung:2008dn, Cohen:2010mi, Cheung:2015cba}, revealed
tree-level recursive structures for a quite large class of theories, while in the latter direct loop integral 
analysis have shown that the theory is endowed with a further symmetry, the dual conformal symmetry 
\cite{Drummond:2006rz, Bern:2006ew} (which together with the space-time conformal symmetry forms the
infinite dimensional Yangian \cite{Drummond:2009fd}), as well as the theory turns out to an {\it on-shell} recursion 
relation at  {\it integrand} level \cite{ArkaniHamed:2010kv} at all loops.

The existence of such a type of recursion relation implies that the amplitudes are determined 
in terms of the smallest non-trivial object at all order in perturbation theory and for any number of external 
states.
%
Such a building block is provided by the three-particle amplitudes which are fixed, up to a coupling constant, by
(super)-Poincar{\'e} invariance \cite{Benincasa:2007xk, ArkaniHamed:2008gz}. Therefore one can
turn the table around and start with the three-particle amplitudes, which, as we just saw, are fixed from first
principles, and reconstruct more complicated amplitudes just in terms of these building blocks via 
a prescription which suitably glues them together \cite{ArkaniHamed:2012nw}.
This shows that it is possible to define a physical observable without resorting to the idea of a Lagrangian, and 
the Feynman diagrammatics is replaced by on-shell processes, {\it i.e.} objects whose states are always and all 
on-shell. As a consequence the objects one deals with are always {\it physical} and gauge invariant (contrarily to 
what happens with the Feynman expansion where the individual diagrams break gauge invariance and thus they cannot be
considered as physical in a generic point of momentum space) and there is no need to introduce the idea of
{\it virtual} (off-shell) particles. A further feature is that
locality is generally broken for an individual on-shell process and
it is then restored once all these on-shell processes are summed
up to provide a scattering amplitude. This is not really a 
drawback given that it is exactly the manifest locality and 
unitarity that forces to introduce all the redundancies which
the Lagrangian description is plagued of. However, this is not
the end of the story. The gluing procedure which allows to
generate higher-point/more complex on-shell processes turns out
to preserve Yangian invariance \cite{ArkaniHamed:2012nw} so
that it is no longer hidden. This is particularly evident if
one describes our objects in momentum twistor space or as
an integral over the Grassmannian $G(k,n)$ 
\cite{Mason:2009qx, ArkaniHamed:2009go}. However, the connection
between the on-shell processes and the Grassmannian appears
to be much deeper: on-shell diagrams are related to a
particular stratification of the positive Grassmannian $G(k,n)$ 
whose positivity-preserving diffeomorphisms represent the
Yangian invariance of the amplitudes \cite{ArkaniHamed:2012nw}.
Furthermore, the on-shell diagrams turns out to be intimately
related to permutations, which define equivalence classes
for such objects, and whose adjacent transpositions encode
the BCFW deformation \cite{ArkaniHamed:2012nw}
\footnote{For further discussion on Grassmannian and
combinatorics and their relation to the on-shell diagrams/
bipartite graphs see \cite{Franco:2013nwa}}.

This picture however makes both unitarity and locality somehow 
hidden rather than {\it emergent}. However it can be encoded
in a more general framework where the fundamental object is 
a new geometrical quantity, the amplituhedron 
\cite{Arkani-Hamed:2013jha}, which can be though of as a 
generalisation of the notion of polytopes in momentum twistor space.
The amplitudes are then read off as {\it volumes}, and both
locality and unitarity emerge from the positivity of the geometry
\cite{Arkani-Hamed:2013kca, Bai:2014cna, Franco:2014csa, 
Arkani-Hamed:2014dca}.
Again, the quantities which can be computed at the end of the day
are always the {\it integrands} of the amplitudes themselves.

What we have been describing so far, as already mentioned,
holds just for {\it planar} $\mathcal{N}\,=\,4$ SYM theory, while 
the exploration of the non-planar structure started more recently
\cite{Chen:2014ara, Arkani-Hamed:2014bca,Bern:2014kca, Franco:2015rma}.

Beyond planar $\mathcal{N}\,=\,4$ SYM very little is known, with
the exception of the ABJM theory in three dimensions where
an analogous Grassmannian formulation have been discussed
\cite{ArkaniHamed:2012nw, Huang:2013owa, Kim:2014hva, Huang:2014xza,
Elvang:2014fja}.

A question that is fair to ask is whether a general first principle picture is available for a larger class of
theories, which does not make any reference to a Lagrangian and makes as many structures manifest as possible.

Indeed on-shell diagrams can be defined in general, {\it i.e.}
with no reference to a specific theory: the three-particle
building blocks are fixed by Poincar{\'e} symmetry for general 
helicities \cite{Benincasa:2007xk} and the prescription
for gluing them and generate higher-point/more complex 
diagrams is not really theory dependent given that it boils
down to integrate out the degrees of freedom on
the intermediate lines according to the momentum conservation and on-shell condition
constraints. However, there are important issues with would
need to be addressed. 

First of all, there are some theories for which there is no
BCFW deformation available which returns a recursion relations
at least for some amplitudes already at tree level. Recursive
structures can be found by either using more general deformations
\cite{Cheung:2015cba} or by introducing further data
\cite{Benincasa:2011kn}\footnote{A further approach is provided
by a multi-step BCFW algorithm \cite{Jin:2014qya,Feng:2015qna}.}. In any case,
this suggests that, if also in those cases a(n almost) first
principle on-shell description is possible, there is a modification
needed which takes into account this issue. Secondly, the loop
analysis is even more subtle. It is necessary to unambiguously
define an object representing the {\it integrand}, which in 
$\mathcal{N}\,=\,4$ SYM was possible mainly because of colour 
ordering -- colour ordering allows to unambiguously fix the
loop degrees of freedom among the various possible terms. 
Indeed this still holds for planar theories, but it is not clear how to get rid of such an ambiguity for theories 
whose amplitudes have no ordering whatsoever. This is not however the only issues 
which needs to be solved in order to have a clear identification
between on-shell diagrams and scattering amplitudes. The loop
singularity structure is intimately tied to the single cuts, which
is in general a very ill-defined procedure: even if they vanish
in dimensional regularisation, one is forced to consider also
loops in the external states, which we will refer to as 
{\it external on-shell bubbles}, which makes ill-defined the
single cut by producing a singularity of type $1/(p^2)^2$.
In $\mathcal{N}\,=\,4$ SYM theory, as well as in massless
$\mathcal{N}\,\ge\,1$ and massive $\mathcal{N}\,\ge\,2$, this issue
does not arise because these terms vanish upon summation over
the full super-multiplet \cite{CaronHuot:2010zt}. Thus,
for a general theory, this issue need to be faced. While on one side
one might naively neglect this type of terms because a regularisation
prescription for the {\it integrals} would take care of them,
on the other side, the contributions with external on-shell
bubble at a certain order could contribute as internal loop
at higher order\footnote{We thank Henrik Johansson for discussion
on this point.}. Thus one should either show that somehow
this does not occur, or provide a prescription for the treatment
of these terms.

In this paper we begin the investigation of the on-shell 
diagrammatics and the related mathematical structures for
theories less special then the maximally supersymmetric ones.
In order to reduce the number of ambiguities, for the time being
we focus just on planar $\mathcal{N}\,<\,4$ gauge theories: 
In this way we indeed 
have a well defined object, the {\it integrand}, which can be 
represented via on-shell processes, as well as also the forward
limits are well-defined, except for the non-supersymmetric case,
{\it i.e.} pure Yang-Mills theory.

We focus in particular on the on-shell diagrammatics itself and
its relation to the scattering amplitudes --  a full-fledge 
(algebraic) geometrical discussion will be discussed in a companion 
paper \cite{Benincasa:2015osg}. Differently from the maximally 
supersymmetric case where the asymptotic states are provided
by just single multiplets, the asymptotic states of less 
supersymmetric cases are represented via two multiplets, which
are labelled by the helicity (the multiplets groups states
with the same helicity sign). Therefore, the on-shell
diagrammatics needs to account of these extra data. We discuss
in detail such a {\it decorated} diagrammatics, pointing out
the existence of directed helicity flows with a well-defined
physical meaning. Importantly, the existence of these helicity flows identifies both the presence
of singularities and their class. Furthermore, through them it is possible to define equivalence
classes for the decorated on-shell diagrams.

Once the decorated on-shell diagrammatics has been set up, we establish the link between the decorated
on-shell processes and the scattering amplitudes for $\mathcal{N}\,\le\,2$ supersymmetric gauge theories. We 
investigate the information about the perturbative structure of the theory that the on-shell processes can encode. 
In particular, we provide a fully on-shell proof that for $\mathcal{N}\,=\,1,\,2$ supersymmetric (massless) 
Yang-Mills theories, the all-loop structure of the scattering amplitudes is fixed by the knowledge of the 
factorisation and forward singularities. Even if this was expected, there are many subtleties that need to
be faced. First of all, even for supersymmetric theories, the forward limit is not well-defined and
it is necessary to introduce a {\it suitable} regularisation scheme. Secondly, a single BCFW bridge turns out
to be able to capture the complete cut-constructible information but not all the potentially problematic terms. 
In other words, there is a(n in principle) finite contribution from the boundary term at infinity. As a part
of the regularisation scheme, we provide a prescription to actually include such terms. With such a completion,
we show that those terms which were in principle problematic, upon summation, are of order $\mathcal{O}(\epsilon)$ 
($\epsilon$ being the regularisation parameter) at all loops. For pure Yang-Mills, we discuss in detail the
one-loop structure, for which, upon regularisation of the forward limit, it is possible to identify the on-shell
bubbles as related to poles in the regularisation parameter. In order to obtain the full integrand, one 
has also to introduce a mass-deformation of the forward states. 

The paper is organised as follows in Section \ref{sec:OSdiag} we provide a detailed description of the decorated 
on-shell diagrammatics and the meaning of the helicity flows, putting them in correspondence with certain 
singularities. In Section \ref{sec:DecUndec} we discuss the connection between the decorated on-shell diagrams and 
combinatorics. In particular, as in the maximally supersymmetric case, there exist equivalence relations between 
diagrams. In this section we discuss how also in this case permutations define equivalence classes of
on-shell processes, with the crucial difference that they are {\it selected} by the helicity flows. 
Section \ref{sec:OSampl} is devoted to the connection between {\it decorated} on-shell
processes and scattering amplitude. We discuss the singularity structure both at tree and loop level, as well
as we introduce a regularisation scheme to make sense of the forward limit on general grounds. We provide
the on-shell proof that, for (massless) $\mathcal{N}\,=\,1,\,2$ SYM the all loop structure can be determined
from factorisation and forward singularities upon suitable regularisation, with the potentially problematic terms
which upon summation become of order $\mathcal{O}(\epsilon)$. We also discuss the loop structure for pure
Yang-Mills mainly focusing on the one loop. Finally Section \ref{sec:Concl} contains our conclusion and outlook.


\section{Decorated on-shell diagrammatics}\label{sec:OSdiag}

Let us consider scattering processes in asymptotically Minkowski 
space-times for planar gauge theories with $\mathcal{N}\,\le\,2$ 
supersymmetries. In a regime where asymptotic states can be defined,
the latter are provided by the irreducible representations of the 
Poincar{\'e} group and they are taken to be the direct product of
eigenstates of the momentum operator. Among the unitary 
representations, we will only deal with the ones whose states
are eigenfunctions of the rotation generator of the massless Lorentz
little group $iso(2)$ -- the helicity operator -- and are 
annihilated by the translation generators of $iso(2)$.

Given the isomorphism between the universal covering of the Lorentz 
group and $SL(2\,,\mathbb{C})$, the kinematics can be encoded into 
the spinors $\lambda_a$ and $\tilde{\lambda}_{\dot{a}}$, with the 
first spinor transforming in the fundamental representation of 
$SL(2\,,\mathbb{C})$ and the second one in the anti-fundamental 
representation. In the complexified momentum space, the universal 
covering of the Lorentz group is isomorphic to 
$SL(2\,,\mathbb{C})\,\times\,SL(2\,,\mathbb{C})$
and the two spinors $\lambda_a$ and $\tilde{\lambda}_{\dot{a}}$
transform under a different copy of $SL(2\,,\mathbb{C})$ each.

Taking as a convention that all the external states are incoming, 
the general structure of a scattering amplitude can therefore be 
written as
\begin{equation}\eqlabel{eq:genampl}
 \mathcal{M}_n\:=\:\delta^{\mbox{\tiny $(2\times 2)$}}
  \left(
   \sum_{i=1}^n\lambda^{\mbox{\tiny $(i)$}}
               \tilde{\lambda}^{\mbox{\tiny $(i)$}}
  \right)
  M_{n}
  \left(
   \{\lambda^{\mbox{\tiny $(i)$}}\, ,
   \tilde{\lambda}^{\mbox{\tiny $(i)$}}\, ;
   h_i\}
  \right),
\end{equation}
where the $\delta$-function implements momentum conservation, and
$M_n$ is an analytic function of the Lorentz invariant combination
of the spinors 
$\langle\lambda,\,\lambda'\rangle\,\equiv\,
 \epsilon^{ab}\lambda_a\lambda'_b$  and 
$[\tilde{\lambda},\,\tilde{\lambda}']\,\equiv\,
 \epsilon^{\dot{a}\dot{b}}\tilde{\lambda}_{\dot{a}}
 \tilde{\lambda}'_{\dot{b}}$ 
as well as of the helicities $\{h_i\}$ of the external states
\footnote{For the totally anti-symmetric Levi-Civita symbols 
$\epsilon_{ab}$ and $\epsilon_{\dot{a}\dot{b}}$ we take
$\epsilon_{12}\,=\,1\,=\,\epsilon_{\dot{1}\dot{2}}$}. 
Being the latter eigenfunctions of the helicity operator 
$\hat{\mathcal{H}}^{\mbox{\tiny $(i)$}}$, we take
$\hat{\mathcal{H}}^{\mbox{\tiny $(i)$}}$ to act on an amplitude as it
acts on one-particle states
\begin{equation}\eqlabel{eq:lg}
 \hat{\mathcal{H}}^{\mbox{\tiny $(i)$}}\mathcal{M}_n\:=\:
 -2h_i\,\mathcal{M}_n.
\end{equation}
The action of the Lorentz little group can also be seen as 
a momentum invariant rescaling of the spinors
\begin{equation}\eqlabel{lg2}
 (\lambda^{\mbox{\tiny $(i)$}},\,
  \tilde{\lambda}^{\mbox{\tiny $(i)$}})
 \,\longrightarrow\,
  (t_{i}\lambda^{\mbox{\tiny $(i)$}},\,
  t_{i}^{-1}\tilde{\lambda}^{\mbox{\tiny $(i)$}})
 \:\Longrightarrow\:
 M_n
  \left(
    t_{i}\lambda^{\mbox{\tiny $(i)$}}\, ,
    t_{i}^{-1}\tilde{\lambda}^{\mbox{\tiny $(i)$}}  \, ;h_i
  \right) \:=\:
 t_{i}^{-2h_i}
 M_{n}
  \left(
   \lambda^{\mbox{\tiny $(i)$}}\, ,
   \tilde{\lambda}^{\mbox{\tiny $(i)$}}\, ;
   h_i
  \right).
\end{equation}
In the case one wants to restrict to supersymmetric theories, it
is more convenient to actually use the full Super-Poincar{\'e} group
to define the asymptotic states \cite{ArkaniHamed:2008gz}. If
$Q_{Ia}$ and $\tilde{Q}^{I\dot{a}}$ are the super-charges, coherent
states are defined as
\begin{equation}\eqlabel{eq:cohstat}
 |\lambda,\,\tilde{\lambda};\,\eta\rangle\:=\:
 e^{Q_{aI} w^a \eta^{I}}|\lambda,\,\tilde{\lambda};\,-1\rangle,
 \qquad
 |\lambda,\,\tilde{\lambda};\,\tilde{\eta}\rangle\:=\:
 e^{\tilde{Q}^{\dot{a}I} \tilde{w}_{\dot{a}} \tilde{\eta}_{I}}
 |\lambda,\,\tilde{\lambda};\,+1\rangle,
\end{equation}
with $a,\,\dot{a}\,=\,1,2$ being the usual spinor indices, 
$I\,=\,1,\ldots,\,\mathcal{N}$ the R-symmetry index, while
$w_a$ and $\tilde{w}_{a}$ are two spinors satisfying the conditions
$\langle w,\,\lambda\rangle\,=\,1\,=\,[\tilde{w},\,\tilde{\lambda}]$,
and $\eta^I,\,\tilde{\eta}_I$ are Grassmann variables. Such coherent
states are eigenstates of the super-charges:
\begin{equation}\label{cohstat2}
 Q_{aI}|\lambda,\,\tilde{\lambda};\,\tilde{\eta}\rangle\:=\:
  \lambda_a\tilde{\eta}_{I}
  |\lambda,\,\tilde{\lambda};\,\tilde{\eta}\rangle,\qquad
 \tilde{Q}_{\dot{a}I}|\lambda,\,\tilde{\lambda};\,\eta\rangle\:=\:
  \tilde{\lambda}_{\dot{a}}\eta_{I}
  |\lambda,\,\tilde{\lambda};\,\eta\rangle.
\end{equation}
Except for the maximally supersymmetric case where the helicity 
states get organised into a single multiplet 
\cite{ArkaniHamed:2008gz}, for less supersymmetric theories 
there are two multiplets, which group the states with the same 
helicity sign and thus such a sign can be used to label them. 
Explicitly, for $\mathcal{N}\,\le\,2$ they can be written
as
\begin{equation}\eqlabel{eq:superm}
 \begin{split}
  &|\lambda,\,\tilde{\lambda};\,\eta,\,-\rangle\:=\:
   \sum_{s=0}^{\mathcal{N}}\frac{1}{s!}
   \left(\prod_{r=0}^s\eta_{\mbox{\tiny $I_r$}}\right)
   |\lambda,\,\tilde{\lambda};\,-(1-s/2)\rangle^{I_1\ldots I_s},\\
  &|\lambda,\,\tilde{\lambda};\,\tilde{\eta},\,+\rangle\:=\:
   \sum_{s=0}^{\mathcal{N}}\frac{1}{s!}
   \left(\prod_{r=0}^s\tilde{\eta}^{\mbox{\tiny $I_r$}}\right)
   |\lambda,\,\tilde{\lambda};\,+(1-s/2)\rangle_{I_1\ldots I_s}.
 \end{split}
\end{equation}
In the expression above we chose the $\eta$-representation for the
negative multiplet and the $\tilde{\eta}$-representation for the 
positive one in such a way that the spin-$1$ state appeared as
zero-order term in $\eta/\tilde{\eta}$. It is possible however
to express both the multiplets in the same representation: the 
$\eta$- and $\tilde{\eta}$-representations are equivalent and they
are related to each other by a Grassmann Fourier transform
\begin{equation}\eqlabel{eq:GFt}
 \begin{split}
  &|\lambda,\tilde{\lambda};\,\eta,\,+\rangle\:=\:
   \int d^{\mathcal{N}}\tilde{\eta}\,e^{\tilde{\eta}\eta}
   |\lambda,\tilde{\lambda};\,\tilde{\eta},\,+\rangle\:=\:
   \sum_{s=0}^{\mathcal{N}}\frac{1}{s!}
   \left(\prod_{r=0}^s\eta_{\mbox{\tiny $I_r$}}\right)
   |\lambda,\tilde{\lambda};
   +\left(1-\frac{\mathcal{N}-s}{2}\right)\rangle^{I_1\ldots I_s},\\
  &|\lambda,\tilde{\lambda};\,\tilde{\eta},\,-\rangle\:=\:
   \int d^{\mathcal{N}}\eta\,e^{\eta\tilde{\eta}}
   |\lambda,\tilde{\lambda};\,\eta,\,-\rangle\:=\:
   \sum_{s=0}^{\mathcal{N}}\frac{1}{s!}
   \left(\prod_{r=0}^s\tilde{\eta}^{\mbox{\tiny $I_r$}}\right)
   |\lambda,\tilde{\lambda};\,
   -\left(1-\frac{\mathcal{N}-s}{2}\right)\rangle_{I_1\ldots I_s}.
 \end{split}
\end{equation}

With these coherent states at hand, we can consider them as 
asymptotic states for our scattering processes, and thus an amplitude
can depend on $\eta$ and $\tilde{\eta}$. For each state, either of
the two representations can be chosen. Furthermore, under little
group transformations $\eta$ and $\tilde{\eta}$ behave as $\lambda$ 
and $\tilde{\lambda}$ respectively. In what follows, we
choose the $\tilde{\eta}$-representation for all the states.
Making supersymmetry invariance manifest, the structure of the 
amplitude \eqref{eq:genampl} generalises to
\begin{equation}\eqlabel{eq:genampl2}
 \mathcal{M}_n\:=\:
  \delta^{\mbox{\tiny $(2\times 2)$}}
  \left(
   \sum_{i=1}^n\lambda^{\mbox{\tiny $(i)$}}
   \tilde{\lambda}^{\mbox{\tiny $(i)$}}
  \right)
  \delta^{\mbox{\tiny $(2\times\mathcal{N})$}}
  \left(
   \sum_{i=1}^n\lambda^{\mbox{\tiny $(i)$}}
   \tilde{\eta}^{\mbox{\tiny $(i)$}}
  \right)
  M_{n}(\{\lambda^{\mbox{\tiny $(i)$}},\,
   \tilde{\lambda}^{\mbox{\tiny $(i)$}};\,
   \tilde{\eta}^{\mbox{\tiny $(i)$}}\}),
\end{equation}
with $M_n$ which just transforms under the action of the 
supersymmetric charge $\tilde{Q}$ via a shift of $\tilde{\eta}$.

\subsection{Three-particle amplitudes}\label{subsec:3ptampl}

In the previous section, we have been discussing the construction of
the asymptotic states from the representations of the 
(Super)-Poincar{\'e} group as well as some general properties of the 
amplitudes. As already mentioned, we will consider all states to be 
in the $\tilde{\eta}$-representation. 
The simplest objects that we can
determine just by symmetries are the three-particle amplitudes:
as the Poincar{\'e} invariance fixes them up to an overall constant
\cite{Benincasa:2007xk}, the Super-Poincar{\'e} group fixes the
scattering of three coherent states \cite{ArkaniHamed:2008gz}.
In particular, momentum conservation can be written as
\begin{equation}\eqlabel{eq:momcons}
 \langle i,j\rangle[i,j]\:=\:0,
 \qquad
 \forall\; i,\,j\:=\:1,\,2,\,3,
\end{equation}
implying that either all the $\lambda$'s or the $\tilde{\lambda}$ 
are proportional to each other. Therefore, a given three-particle
amplitude can either depend just on Lorentz invariant combinations
of $\lambda$'s or of $\tilde{\lambda}$'s. Requiring that the 
amplitudes transform correctly under the Lorentz little group and 
supersymmetry as well as that they vanish on the real sheet, the 
explicit form of the three-particle amplitudes turns out to be
\begin{equation}\eqlabel{eq:3ptampl}
 \begin{split}
  &\mathcal{M}_{3}^{\mbox{\tiny $(\mathfrak{1})$}}(1^{+},2^{+},3^{-})\:=\:
   \delta^{(2\times 2)}
   \left(
    \sum_{i=1}^{3}\lambda^{\mbox{\tiny $(i)$}}
    \tilde{\lambda}^{\mbox{\tiny $(i)$}}
   \right)
   \delta^{(1\times\mathcal{N})}
   \left(
    \sum_{i=1}^{3}[i+1,i-1]\tilde{\eta}^{\mbox{\tiny $(i)$}}
   \right)
   \frac{[1,2]^{4-\mathcal{N}}}{[1,2][2,3][3,1]},\\
  &\mathcal{M}_{3}^{\mbox{\tiny $(\mathfrak{2})$}}(1^{-},2^{-},3^{+})\:=\:
   \delta^{(2\times 2)}
   \left(
    \sum_{i=1}^{3}\lambda^{\mbox{\tiny $(i)$}}
    \tilde{\lambda}^{\mbox{\tiny $(i)$}}
   \right)
   \delta^{(2\times\mathcal{N})}
   \left(
    \sum_{i=1}^3\lambda^{\mbox{\tiny $(i)$}}
    \tilde{\eta}^{\mbox{\tiny $(i)$}}
   \right)
   \frac{\langle1,2\rangle^{4-\mathcal{N}}}{
    \langle1,2\rangle\langle2,3\rangle\langle3,1\rangle},
 \end{split}
\end{equation}
where the apex $(k)$ on the left-hand-side indicates the number
of negative helicity multiplets, while $\mathcal{N}$ indicates
the number of supersymmetries -- notice that the expressions above
reproduce the three-particle amplitudes for any 
$\mathcal{N}\,\le\,4$. As already mentioned at the very
 beginning, we will just discuss the planar sector of the 
$\mathcal{N}\,\le\,2$ supersymmetric theories, and thus all the 
amplitudes are understood to be colour ordered.

The three-particle amplitudes \eqref{eq:3ptampl} can be actually 
thought as {\it on-shell forms} \cite{ArkaniHamed:2012nw} by 
associating to them the super phase-space of each particle
\begin{equation}\eqlabel{eq:3ptosform}
 \mathcal{A}_3^{\mbox{\tiny $(\mathfrak{k})$}}\:=\:
 \mathcal{M}_3^{\mbox{\tiny $(\mathfrak{k})$}}
 \prod_{i=1}^{3}
 \frac{d^2\lambda^{\mbox{\tiny $(i)$}}\,
       d^2\tilde{\lambda}^{\mbox{\tiny $(i)$}}}{
       \mbox{vol}\{GL(1)\}}
       d^{\mathcal{N}}\tilde{\eta}^{\mbox{\tiny $(i)$}}.
\end{equation}
This turns out to be useful for gluing the three-particle amplitudes
together, generating higher point on-shell processes.

\begin{figure}[htbp]
 \centering 
  \scalebox{.45}{\includegraphics{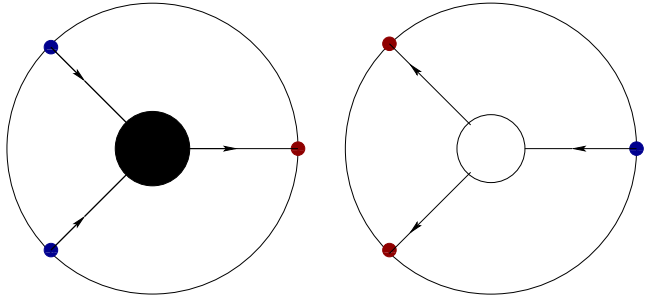}}
  \caption{Decorated on-shell diagrams for three-particle 
   amplitudes: The black and white nodes represent the MHV and 
   $\bar{\mbox{MHV}}$ amplitudes respectively, while the incoming
   (outgoing) arrows represent the negative (positive) helicity
   multiplets. The decoration provides a {\it perfect orientation}
   which {\it sources} and {\it sinks} are associated to and
   identified by small dark blue and dark red nodes respectively
   at the boundary of the graph.}
  \label{fig:3ptampl}
\end{figure}

At diagrammatic level, 
$\mathcal{M}_3^{\mbox{\tiny $(\mathfrak{1})$}}$ 
($\mathcal{A}_3^{\mbox{\tiny $(\mathfrak{1})$}}$) and 
$\mathcal{M}_3^{\mbox{\tiny $(\mathfrak{2})$}}$ 
($\mathcal{A}_3^{\mbox{\tiny $(\mathfrak{2})$}}$) are typically 
depicted as a white and a black trivalent nodes respectively, with 
the three lines departing from their centres representing the 
scattering states. 
Furthermore, we need to graphically distinguish between the 
negative- and positive-helicity multiplets.  
To this purpose, we conventionally {\it decorate} the three-particle
diagrams by associating an incoming/outgoing arrow to the external
lines to represent a negative/positive-helicity multiplet, as in 
Figure \ref{fig:3ptampl}. Such a decoration provides the diagrams 
with a {\it perfect orientation} \cite{Postnikov:2006kva} to
which it is possible to associate external nodes which
the oriented arrows depart from (blue node, named {\it source}) or 
arrive to (red node, named {\it sink})\footnote{Another convention 
which could have been taken is to identify the boundary nodes with 
outgoing/incoming arrow via a white/black node, accordingly with the
three-particle  amplitudes. However, we prefer the convention we 
will use throughout the paper because it remarks the different 
nature of the boundary nodes and the internal vertices}.

As a final comment, in the non-supersymmetric case a further set
of tree-level amplitudes is allowed by Poincar{\'e} invariance
(while supersymmetry forbids it)
\begin{equation}\eqlabel{eq:3ptampl2}
 \begin{split}
  \mathcal{M}_{3}^{\mbox{\tiny $(\mathfrak{0})$}}(1^{+},2^{+},3^{+})\:=\:
   \delta^{(2\times 2)}
   \left(
    \sum_{i=1}^{3}\lambda^{\mbox{\tiny $(i)$}}
    \tilde{\lambda}^{\mbox{\tiny $(i)$}}
   \right)
   \frac{[1,2][2,3][3,1]}{M^2},\\
  \mathcal{M}_{3}^{\mbox{\tiny $(\mathfrak{3})$}}(1^{-},2^{-},3^{-})\:=\:
   \delta^{(2\times 2)}
   \left(
    \sum_{i=1}^{3}\lambda^{\mbox{\tiny $(i)$}}
    \tilde{\lambda}^{\mbox{\tiny $(i)$}}
   \right)
   \frac{\langle1,2\rangle\langle2,3\rangle\langle3,1\rangle}{M^2},
 \end{split}
\end{equation}
where the scale $M^2$ has been introduce to make explicit the
dimensionfulness of the coupling constant of these amplitudes,
{\it i.e.} the dimensionful coupling constant for such an operator
has been replaced by a dimensionful one and the scale $M^2$.
Their coupling constant is typically zero in pure Yang-Mills theory.
However, the amplitudes \eqref{eq:3ptampl2} can appear in
effective field theory -- they correspond to a dimension six
operator $\mbox{tr}\{F^3\}$ -- or also in loop amplitudes where
$M^2$ is actually given by a propagator \cite{Bern:2005hs}. In this
last case, it is important to notice that they become highly
singular: taken by themselves, because of the propagator-like 
factor, they diverge in the complexified momentum space, while
they still vanish on the real-sheet, so they can in principle 
provide a finite quantity just when they get suitably glued to
an object which would be vanishing in the complexified momentum space.
Finally, taken $M^2$ not to have any kinematic 
dependence but just as a full-fledge dimensionful coupling constant,
an eventual theory built up just from the amplitudes 
\eqref{eq:3ptampl2} seem to require the introduction of higher
and higher dimension operators leading to the breakdown of locality
\cite{Benincasa:2011pg}. In the following, we will build the 
planar non-supersymmetric spin-$1$ theory just out of the 
amplitude \eqref{eq:3ptampl} and some comment on the emergence of
\eqref{eq:3ptampl2} will eventually be made at loop level.

\subsection{Building higher point diagrams}\label{subsec:HighDiag}

The isometry group of our space-time defined for us the simplest
scattering process. Now we can use them to build more complicated
on-shell processes. The most natural operation which can be defined
is the gluing of two three-particle amplitudes or, more generally, 
of any two on-shell diagrams at hand, along one leg each. 
The natural prescription is to integrate over the super phase-space 
of the glued legs imposing momentum conservation and summing over 
all the coherent states which are allowed to propagate 
\cite{ArkaniHamed:2012nw}:
\begin{equation}\eqlabel{eq:fusion}
 \mathcal{M}_{m_1+m_2}\:=\:
  \sum_{h\,=\,\mp}
  \int\frac{%
       d^2\lambda^{\mbox{\tiny $(P)$}}\,
       d^2\tilde{\lambda}^{\mbox{\tiny $(P)$}}}{
       \mbox{vol}\{GL(1)\}}
       d^{\mathcal{N}}\tilde{\eta}^{\mbox{\tiny $(P)$}}
    \mathcal{M}_{m_1+1}(-P,\,-h;\,\tilde{\eta}^{\mbox{\tiny $(P)$}})
    \mathcal{M}_{m_2+1}(P,\,h;\,\tilde{\eta}^{\mbox{\tiny $(P)$}}),
\end{equation}
where the $\mathcal{M}$'s represent generic on-shell processes
(they are not necessarily full-fledge amplitudes) and momentum 
conservation has been already implemented. This procedure is 
completely natural if one thinks of the on-shell processes as 
on-shell forms, as in \eqref{eq:3ptosform}: gluing two on-shell 
forms return a higher degree on-shell form. At a diagrammatic
level this gluing prescription generates diagrams with 
a perfect orientation.

As just mentioned, the simplest higher point on-shell process which 
can be built is the four-point one obtained by gluing two 
three-particle amplitudes. It is possible to glue two 
three-particle amplitudes of different or of the same type. 

\begin{figure}[htbp]
 \centering 
  \scalebox{.35}{\includegraphics{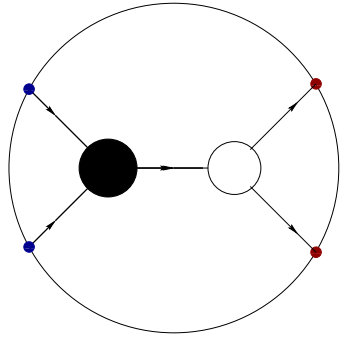}}
  \caption{Simplest four-particle on-shell diagrams: Two on-shell
   $3$-forms are glued by integrating over the super phase-space of
   the internal on-shell leg. In this figure, this on-shell diagram
   appears having boundary nodes (and thus the four un-glued states 
   are fixed): there is just a single coherent state which can 
   propagate, and the integration over its super phase-space leads 
   to a constraint on the external momenta.}
  \label{fig:4ptampl}
\end{figure}

In the first case (Figure \ref{fig:4ptampl}), the integration over
the super phase-space of the glued (intermediate) line returns a 
constraint on the momenta of the ``un-glued states'', and it 
represents a singularity. Furthermore, if the external states are 
fixed, there is just one allowed coherent state propagating.

We can also glue together two three-particle amplitudes of the same 
type. In this case all the spinors of the same type in the whole
(sub)-diagram turn out to be proportional to each other, which 
implies that it does not matter along which channel the four states
are connected to each other. This defines an equivalence operation,
dubbed {\it merger} \cite{ArkaniHamed:2012nw},
which allows to contract two three-particle amplitudes of the same 
type in a four-particle object along a channel and then expand it 
again along a different one. 
Notice that these equivalent four-particle (sub)-processes
are characterised by three ``external'' states having a certain
helicity while the fourth one to having the other, {\it i.e.}
three arrows in the on-shell diagram has a certain direction, while
the last one has the opposite decoration.

\begin{figure}[htbp]
 \centering 
 \[
  \raisebox{-.9cm}{\scalebox{.40}{\includegraphics{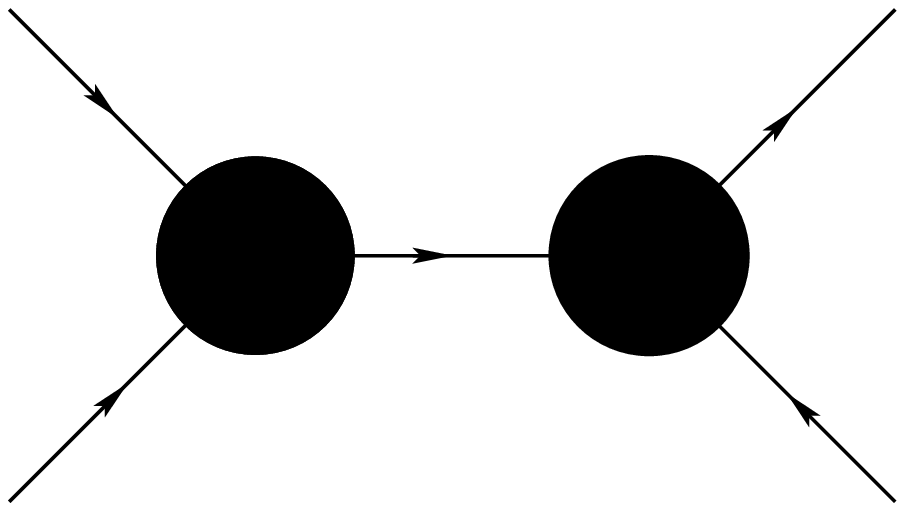}}}
  \qquad\Longleftrightarrow\qquad
  \raisebox{-.9cm}{\scalebox{.40}{\includegraphics{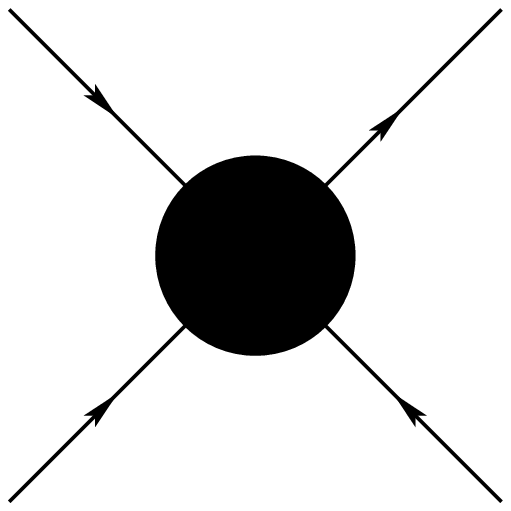}}}
  \qquad\Longleftrightarrow\qquad
  \raisebox{-1.5cm}{\scalebox{.40}{\includegraphics{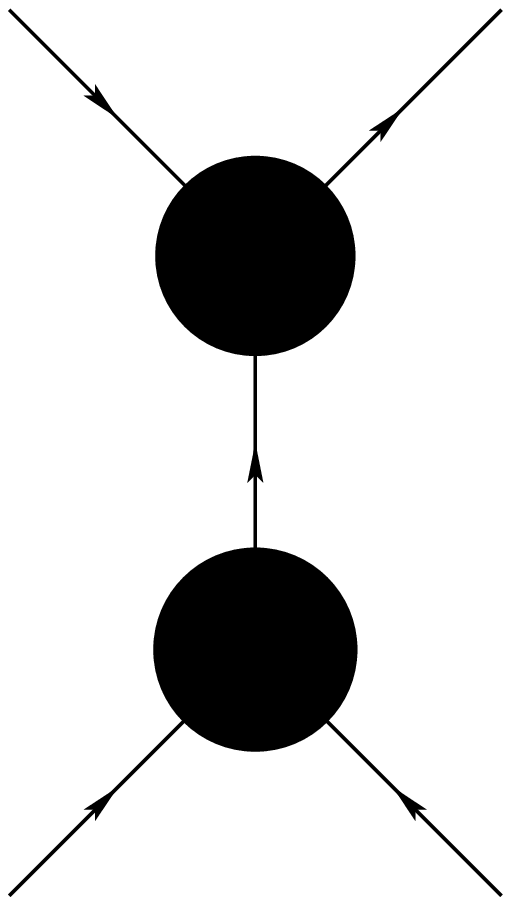}}}
 \]
  \caption{Merger operation: The proportionality among all the 
  $\tilde{\lambda}$'s (all the $\lambda$'s in the case of two 
  $\mathcal{M}_3^{\mbox{\tiny $(1)$}}$) implies that the four
  external states can be connected to each other equivalently 
  in both channels. In each of such channels, just a single
  coherent state can propagate in the internal on-shell leg.}
  \label{fig:4ptmerge}
\end{figure}

So far, there is no relevant difference with the maximally
supersymmetric case: the gluing of two on-shell diagrams is
defined according to the same prescription as well as the merger
operation still holds. The first slight difference arises in this
equivalence relation where just a certain multiplet can propagate
as intermediate state.

\subsection{BCFW bridges and helicity flows}
\label{subsec:BCFWbH}

\begin{figure}[htbp]
 \centering 
 \[
  \raisebox{-1.5cm}{\scalebox{.30}{\includegraphics{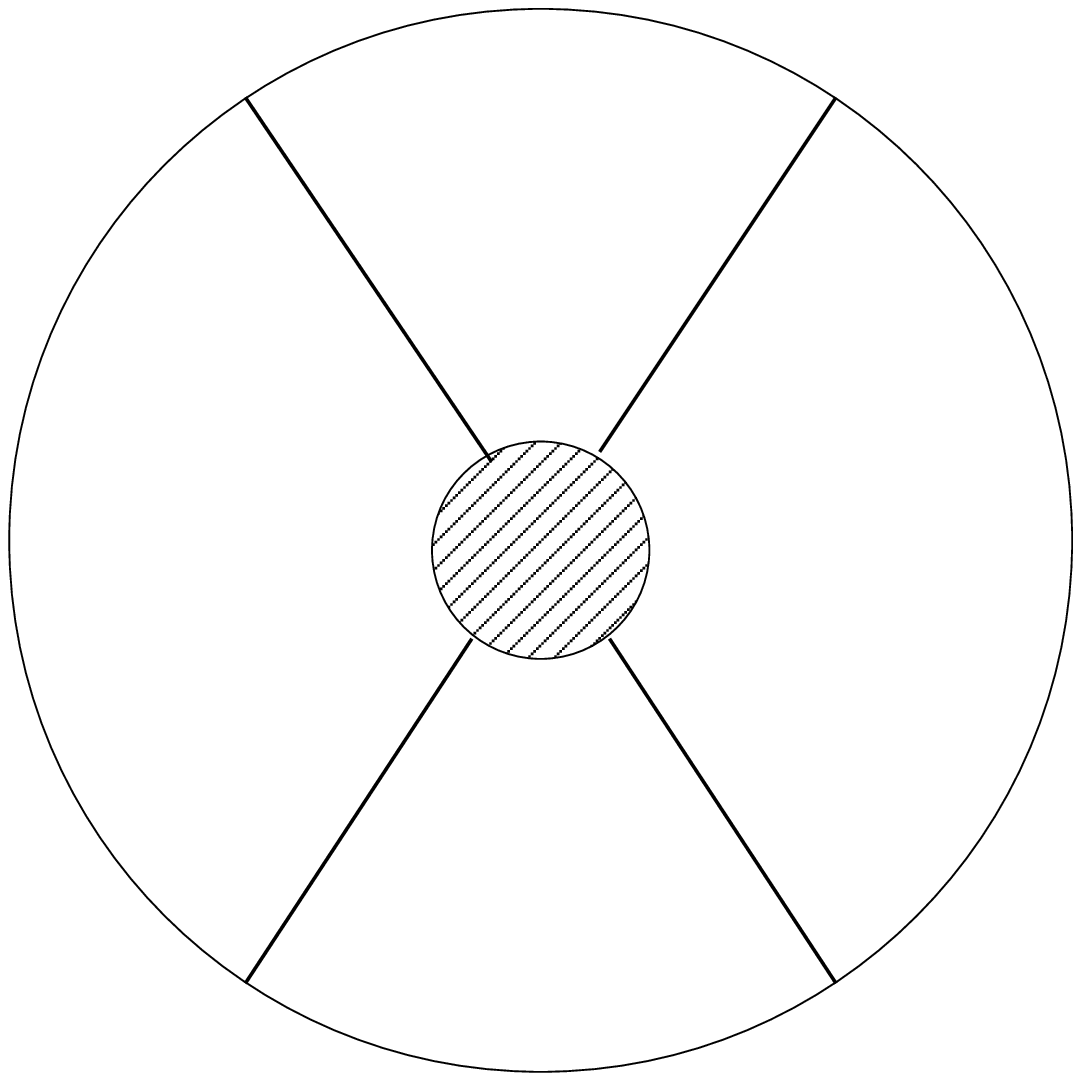}}}
   \hspace{1cm}\Longrightarrow\hspace{1cm}
  \raisebox{-1.5cm}{\scalebox{.30}{\includegraphics{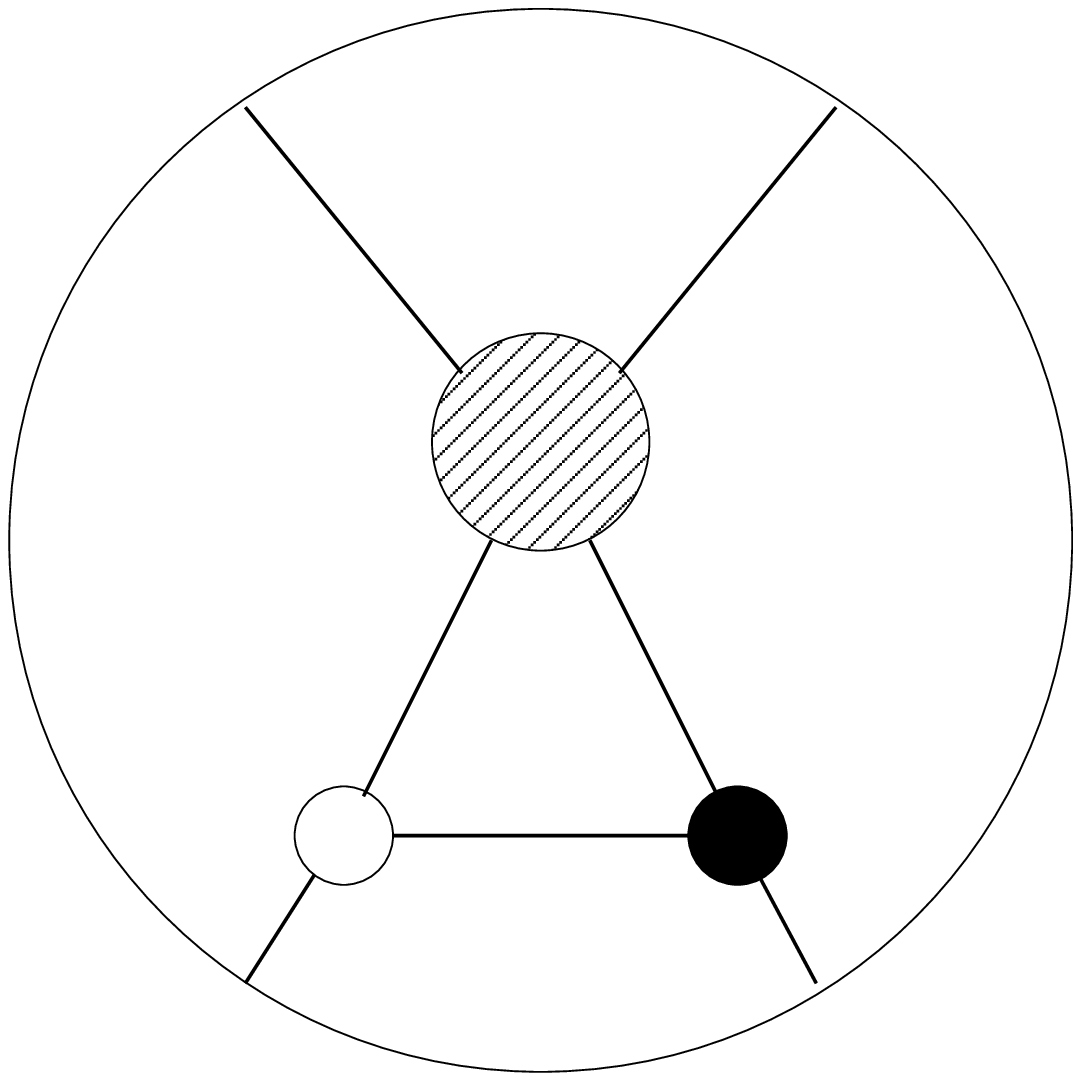}}}
 \]
 \caption{Generating higher degree diagrams via a BCFW bridge: From
          a given on-shell diagram (generically represented on the
          left-hand-side), a higher degree on-shell diagram can
          be generated attaching a {\it BCFW bridge} to two selected
          external states, {\it i.e.} they get connected via two
          three-particle amplitudes of different type which are
          glued to each other (on right-hand-side).}
 \label{fig:BCFWbridge}
\end{figure}

Let us now discuss the general way to generate higher degree 
on-shell processes. Given a certain on-shell diagram 
$\mathcal{M}_n^{\mbox{\tiny $(0)$}}$
\footnote{Notice the difference with the notation used for the 
three particles \eqref{eq:3ptampl}, \eqref{eq:3ptosform}, 
\eqref{eq:3ptampl2} -- and more generally for the N${}^{k}$MHV
amplitudes -- where the apex ${}^{\mbox{\tiny $(\mathfrak{k})$}}$ 
indicates the number of negative helicity external states, while 
here the apex  ${}^{\mbox{\tiny $(d)$}}$ indicates the number
of un-fixed degrees of freedom.}, it is always
possible to single out two external lines and connecting them by 
gluing one three-particle amplitude to each of them, being of
different type, and gluing those three-particle amplitudes between
them. The integration over the delta-functions leaves one degree of
freedom unfixed, mapping 
$\mathcal{M}_n^{\mbox{\tiny $(0)$}}$ to a
(higher degree) differential form
\begin{equation}\eqlabel{eq:BCFWbridge}
 \mathcal{M}_n^{\mbox{\tiny $(0)$}}\qquad\longrightarrow\qquad
 \mathcal{M}_n^{\mbox{\tiny $(1)$}}\:=\:
 dz\,\mu(z)\,\mathcal{M}_n^{\mbox{\tiny $(0)$}}(z),
\end{equation}
where $\mathcal{M}_n^{\mbox{\tiny $(0)$}}(z)$ is nothing but the
BCFW-deformed $\mathcal{M}_n^{\mbox{\tiny $(0)$}}$ 
(Figure \ref{fig:BCFWbridge}) and $\mu(z)$ is a measure induced
by the BCFW bridge itself. 

As prescribed,
when this {\it BCFW bridge} is attached to a given on-shell diagram,
one needs to sum over all the possible coherent states which can
propagate. Typically, one is used to think of the BCFW deformations
as preserving the helicities of the deformed states. Summing over
all the allowed coherent states instead admits the possibility 
that this does not hold. Let us clarify this point. From a given
on-shell diagram, let us single out two external states having
different helicities and attach a BCFW bridge to them. There are two
{\it generally inequivalent} ways to perform such an operation,
which differ from each other dependently on which three-particle 
amplitude of the bridge is associated to each external coherent 
state.

Let us start with associating the anti-holomorphic (holomorphic) 
amplitude to the negative (positive) helicity external coherent 
state:
\begin{equation}\eqlabel{eq:BCFWbridge1}
 \raisebox{-1.7cm}{\scalebox{.33}{\includegraphics{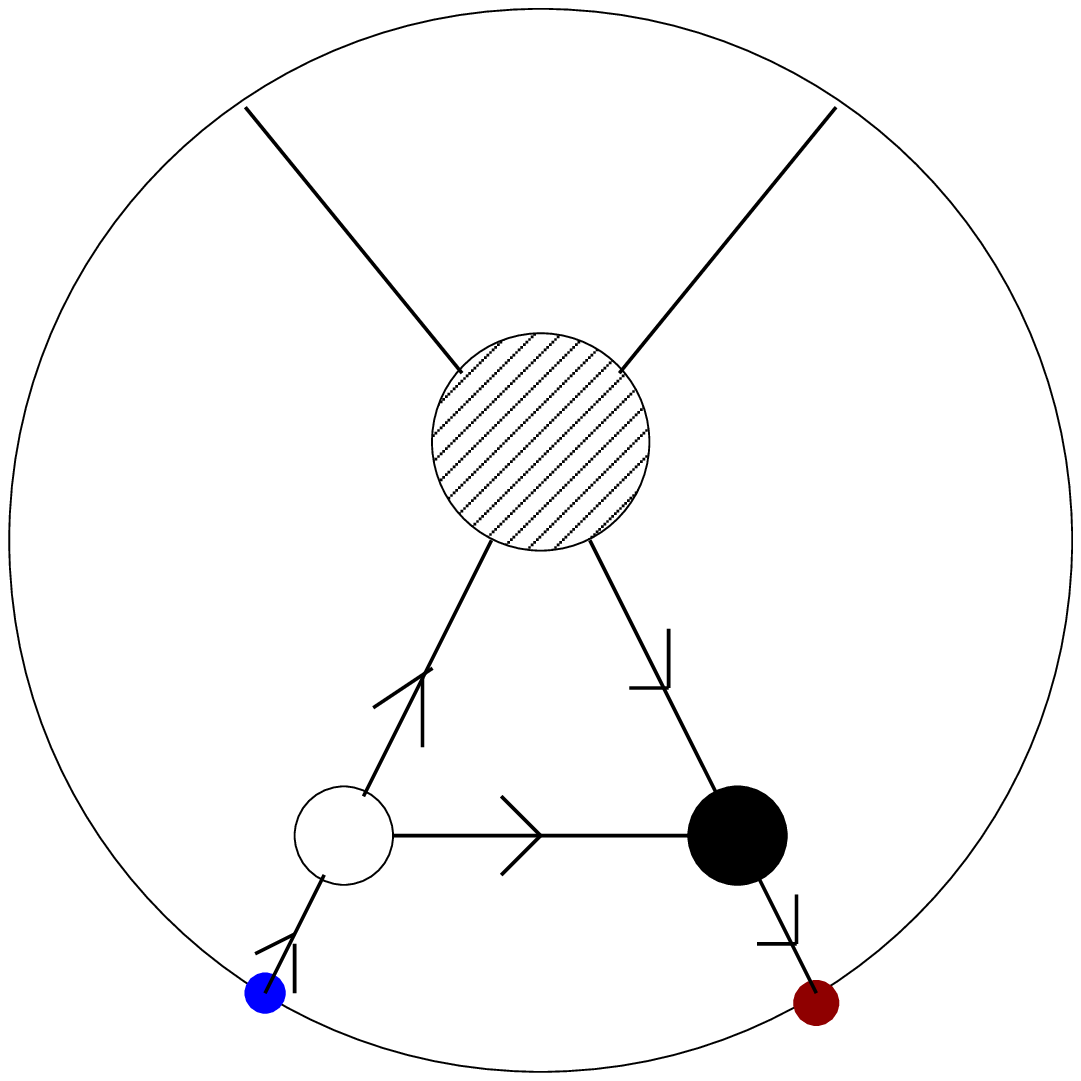}}}
 \:=\:\frac{dz}{z}\mathcal{M}_{n}^{\mbox{\tiny $(0)$}}(z),
 \hspace{2cm}
 \left\{
  \begin{array}{l}
   q_{\mbox{\tiny $ij$}}\:=\:z\lambda^{\mbox{\tiny $(i)$}}
    \tilde{\lambda}^{\mbox{\tiny $(j)$}},\\
   \hat{p}^{\mbox{\tiny $(i)$}}\:=\:
    \lambda^{\mbox{\tiny $(i)$}}
    (
     \tilde{\lambda}^{\mbox{\tiny $(i)$}}-
     z\tilde{\lambda}^{\mbox{\tiny $(j)$}}
    ),\\
   \tilde{\eta}^{\mbox{\tiny $(\hat{i})$}}\:=\:
     \tilde{\eta}^{\mbox{\tiny $(i)$}}-
     z\tilde{\eta}^{\mbox{\tiny $(j)$}},\\
   \hat{p}^{\mbox{\tiny $(j)$}}\:=\:
    (
     \lambda^{\mbox{\tiny $(j)$}}+z\lambda^{\mbox{\tiny $(i)$}}
    )
    \tilde{\lambda}^{\mbox{\tiny $(j)$}}.
  \end{array}
 \right. ,
\end{equation}
where the upper vertical (intermediate) lines need to be thought of
as attached to a putative on-shell diagram and have momenta
$\hat{p}^{\mbox{\tiny $(i)$}}$ and $\hat{p}^{\mbox{\tiny $(j)$}}$,
$q_{\mbox{\tiny $ij$}}$ is the momentum in the horizontal 
intermediate line connecting the two three-particle amplitudes,
while the lower lines are external states with momenta 
$p^{\mbox{\tiny $(i)$}}$ and $p^{\mbox{\tiny $(j)$}}$.
With such a choice, there is only one coherent state which can
propagate in the each intermediate line. As it is straightforward
to see from \eqref{eq:BCFWbridge1}, this BCFW bridge induces a
BCFW deformation on the on-shell diagrams which deforms the 
positive/negative helicity spinor of the negative/positive
helicity multiplet of the given on-shell process. Furthermore,
a well-defined helicity flow can be identified from the
external coherent states to the on-shell diagram the BCFW bridge
is applied to.

Let us now consider a BCFW bridge obtained from the previous one
by exchanging the holomorphicity of the three-particle amplitudes in
the bridge (the helicities of the external states are always kept
fixed). In this case, two possible coherent states are admitted
in the intermediate lines
\begin{equation}\eqlabel{eq:BCFWbridge2}
 \raisebox{-1.8cm}{\scalebox{.33}{\includegraphics{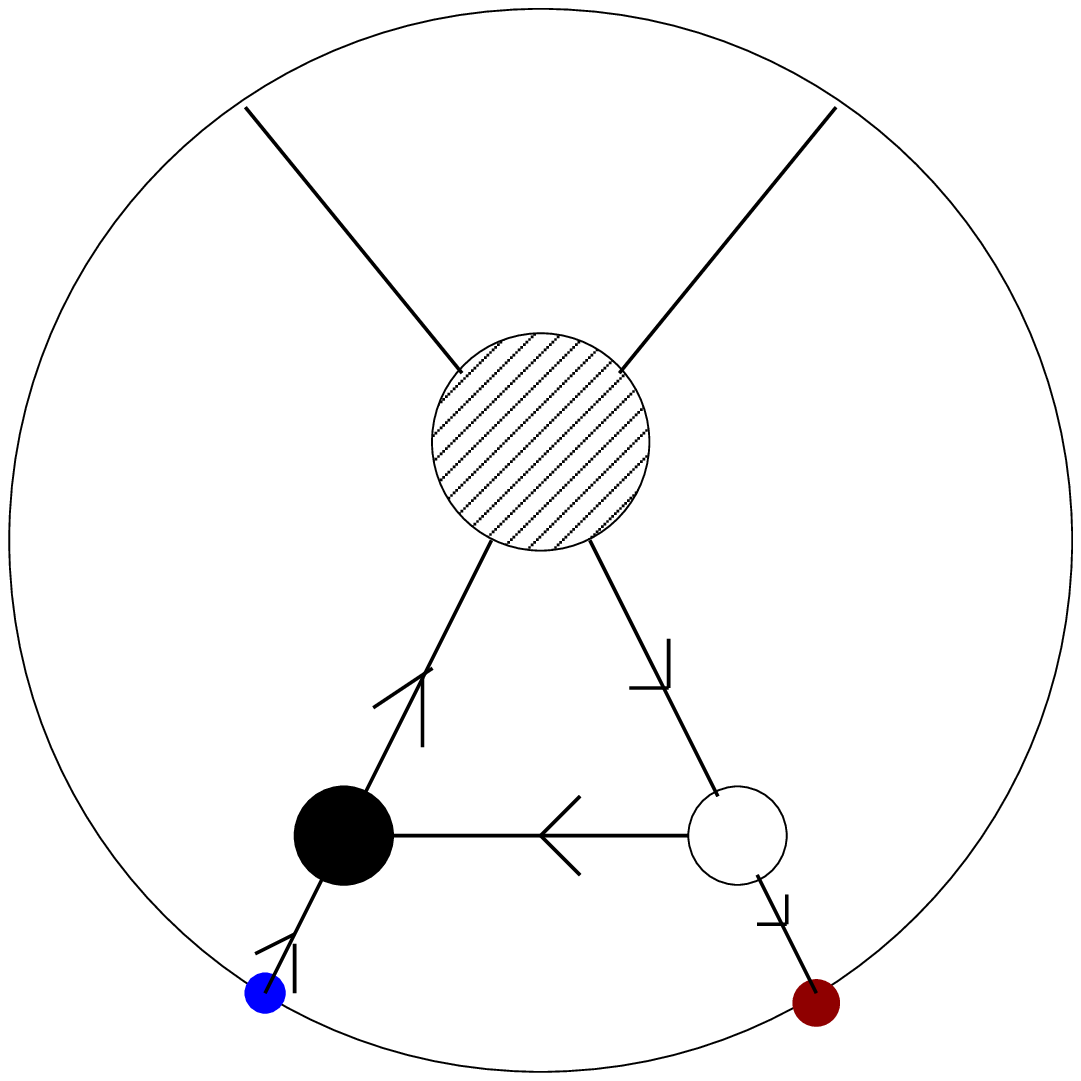}}}
 \quad+\quad
 \raisebox{-1.8cm}{\scalebox{.33}{\includegraphics{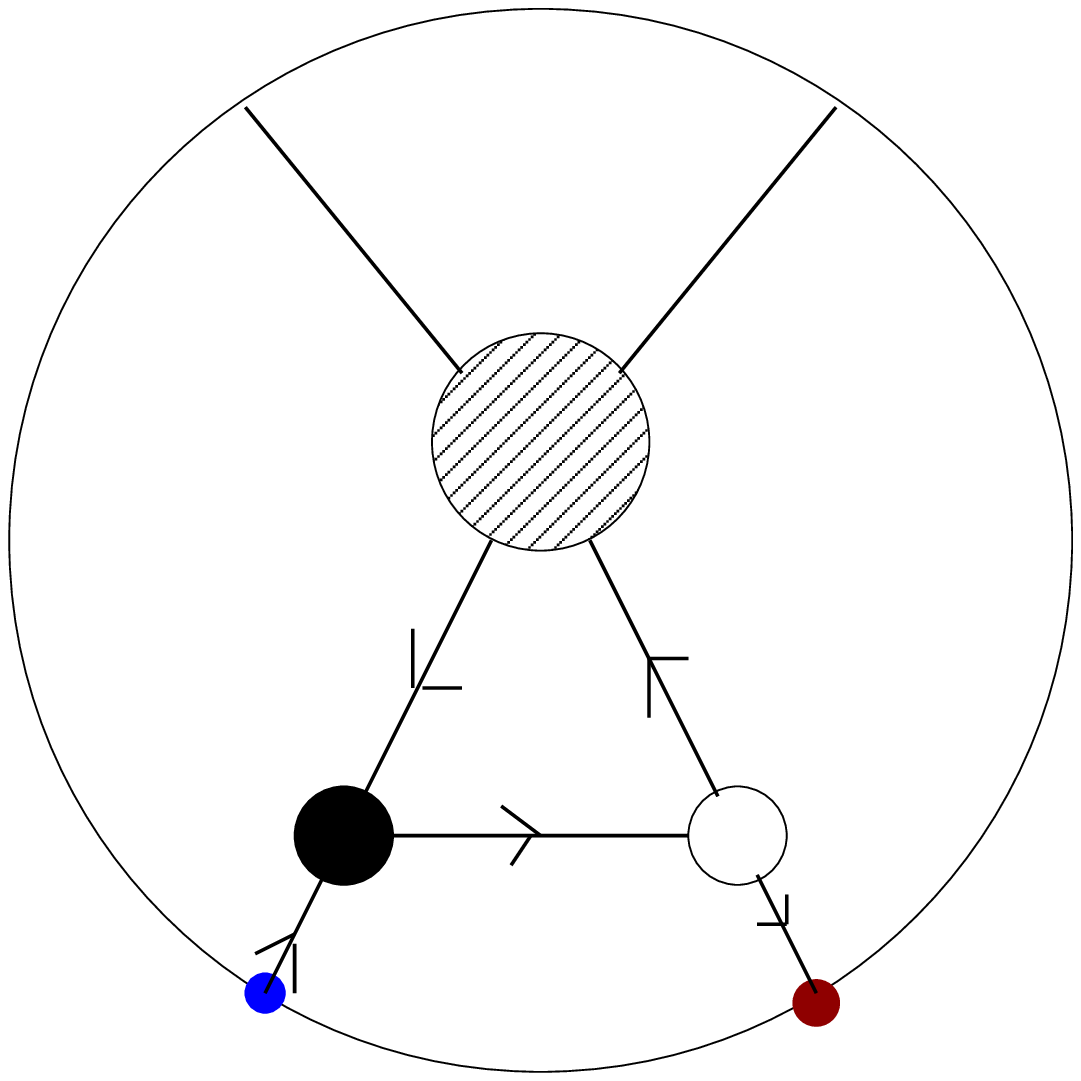}}}
 ,
 \hspace{2cm}
 \left\{
  \begin{array}{l}
   q_{\mbox{\tiny $ij$}}\:=\:z\lambda^{\mbox{\tiny $(j)$}}
    \tilde{\lambda}^{\mbox{\tiny $(i)$}},\\
   \hat{p}^{\mbox{\tiny $(i)$}}\:=\:
    (
     \lambda^{\mbox{\tiny $(i)$}}-z\lambda^{\mbox{\tiny $(j)$}}
    )
    \tilde{\lambda}^{\mbox{\tiny $(i)$}},\\
   \hat{p}^{\mbox{\tiny $(j)$}}\:=\:
    \lambda^{\mbox{\tiny $(j)$}}
    (
     \tilde{\lambda}^{\mbox{\tiny $(j)$}}+
     z\tilde{\lambda}^{\mbox{\tiny $(i)$}}
    )\\
   \tilde{\eta}^{\mbox{\tiny $(\hat{j})$}}\:=\:
     \tilde{\eta}^{\mbox{\tiny $(j)$}}-
     z\tilde{\eta}^{\mbox{\tiny $(i)$}}.
  \end{array}
 \right. \;.
\end{equation}
Let us consider these two contributions separately. As far as the
first one is concerned, it provides the same differential as
\eqref{eq:BCFWbridge1} (even if the on-shell diagram the 
bridge is attached to gets deformed in a different way) and
there is a well-defined helicity flow from the external coherent
states to the on-shell diagram the BCFW bridge is attached to,
as in the previous case
\begin{equation}\eqlabel{eq:BCFWbridge2a}
 \raisebox{-1.7cm}{\scalebox{.33}{\includegraphics{BCFWbridgeB.eps}}}
 \:=\:\frac{dz}{z}\mathcal{M}_{n}^{\mbox{\tiny $(0)$}}(z).
\end{equation}
Thus, this contribution induces the ``conjugate'' BCFW deformation
of \eqref{eq:BCFWbridge1} on the on-shell diagram the bridge is 
attached to. Another feature that can be identified is the existence
of a helicity flow along the intermediate lines in the 
left-hand-side of \eqref{eq:BCFWbridge2a}.

In the second contribution in \eqref{eq:BCFWbridge2}, there is
no helicity flow from the external lines and which connect the
bridge to a putative on-shell diagram, while there is along
the intermediate lines as in \eqref{eq:BCFWbridge2a} but in
opposite direction. Furthermore, differently from 
\eqref{eq:BCFWbridge1} and \eqref{eq:BCFWbridge2a}, the differential
has a different structure
\begin{equation}\eqlabel{eq:BCFWbridge2b}
 \raisebox{-1.7cm}{\scalebox{.33}{\includegraphics{BCFWbridgeC.eps}}}
 \:=\:dz\,z^{3-\mathcal{N}}\mathcal{M}_{n}^{\mbox{\tiny $(0)$}}(z).
\end{equation}
Contrarily to the other cases just discussed, this contribution is
not directly related to a BCFW deformation of the original on-shell
diagram because of the change in helicity of the lines the bridge
is attached to, and the differential shows a multiple pole. 
This for $\mathcal{N}\,\le\,2$. For $\mathcal{N}\,=\,3$ the
differential measure $\mu$ is just equal to $1$, while for the 
maximally supersymmetric case $\mathcal{N}\,=\,4$, also this bridge 
has the same measure $1/z$ as the other two, making it equivalent to 
the previous one in eq \eqref{eq:BCFWbridge2a}. Notice also
that the differential \eqref{eq:BCFWbridge2b} is not helicity blind
for $\mathcal{N}\,\neq\,4$ (the degree of freedom labelled by $z$
transforms not trivially under the little group of 
$p^{\mbox{\tiny $(i)$}}$ and $p^{\mbox{\tiny $(j)$}}$), which 
implies that the helicity configuration of the on-shell diagram
the bridge is attached to is different of the helicity
configuration of the full on-shell diagram.

So far we have been discussing BCFW bridges whose external coherent
states had different helicities. In order to complete this analysis,
let us consider BCFW bridges whose external states have the same
helicity. In this case, no matter how the bridge is formed, there is
just one possible coherent state propagating in each intermediate
line\footnote{In \eqref{eq:BCFWbridge3} we explicitly consider
the BCFW bridge whose external states have both incoming helicity
arrows, {\it i.e.} they have negative helicity. Exactly the same
discussion holds if the external coherent states are taken to
have positive helicity up to a change in the direction of the
helicity arrows in the intermediate lines glued to a putative 
on-shell diagram.}:
\begin{equation}\eqlabel{eq:BCFWbridge3}
 \begin{split}
&\raisebox{-1.7cm}{\scalebox{.33}{\includegraphics{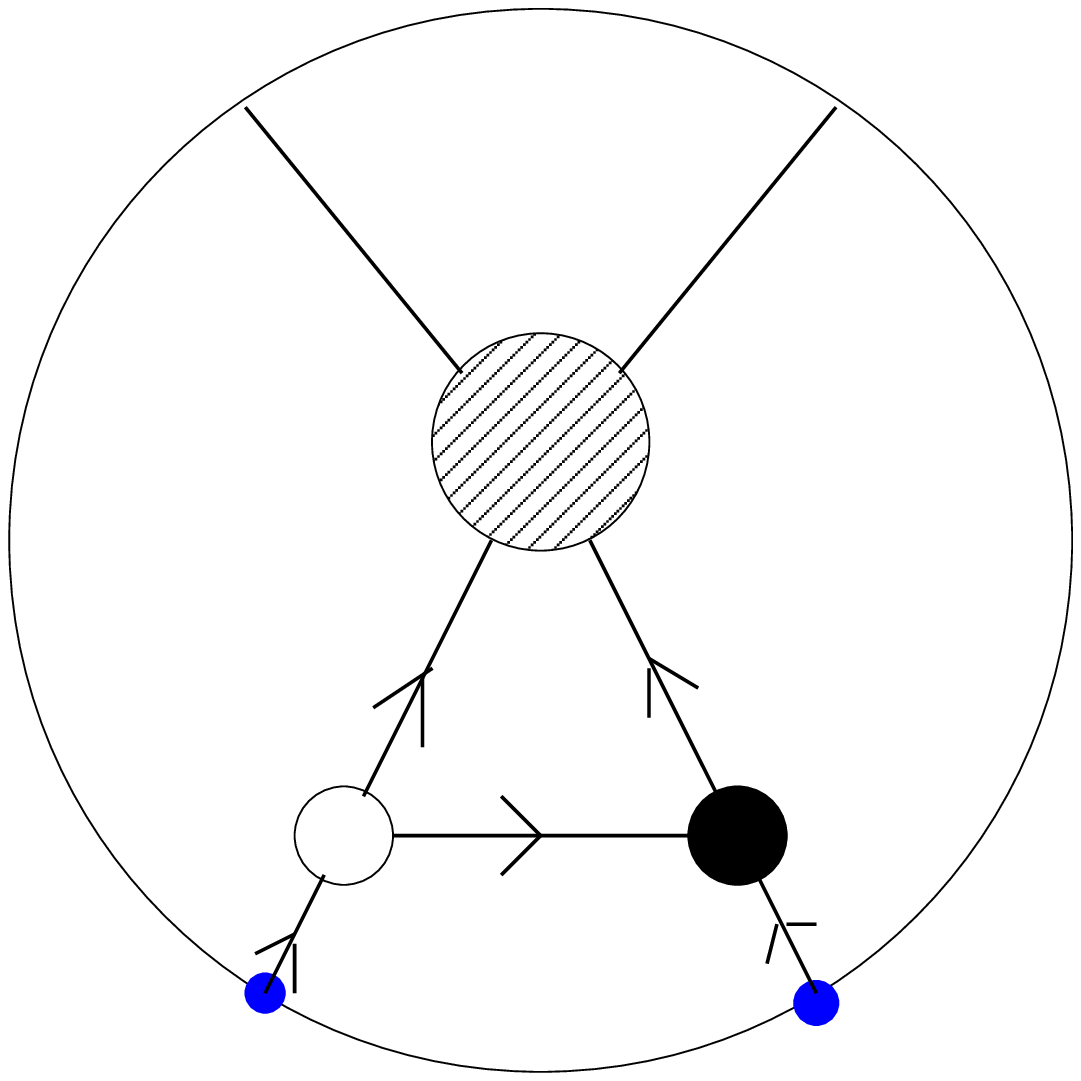}}}
  \:=\:\frac{dz}{z}\mathcal{M}_{n}^{\mbox{\tiny $(0)$}}(z),
  \hspace{2cm}
  \left\{
   \begin{array}{l}
    q_{\mbox{\tiny $ij$}}\:=\:z\lambda^{\mbox{\tiny $(i)$}}
     \tilde{\lambda}^{\mbox{\tiny $(j)$}}\\
    \hat{p}^{\mbox{\tiny $(i)$}}\:=\:
     \lambda^{\mbox{\tiny $(i)$}}
     (
      \tilde{\lambda}^{\mbox{\tiny $(i)$}}-
      z\tilde{\lambda}^{\mbox{\tiny $(j)$}}
     ),\\
    \tilde{\eta}^{\mbox{\tiny $(\hat{i})$}}\:=\:
     \tilde{\eta}^{\mbox{\tiny $(i)$}}-
     z\tilde{\eta}^{\mbox{\tiny $(j)$}}\\
    \hat{p}^{\mbox{\tiny $(j)$}}\:=\:
     (
      \lambda^{\mbox{\tiny $(j)$}}+z\lambda^{\mbox{\tiny $(i)$}}
     )
     \tilde{\lambda}^{\mbox{\tiny $(j)$}}
   \end{array}
  \right. \\
&\raisebox{-1.7cm}{\scalebox{.33}{\includegraphics{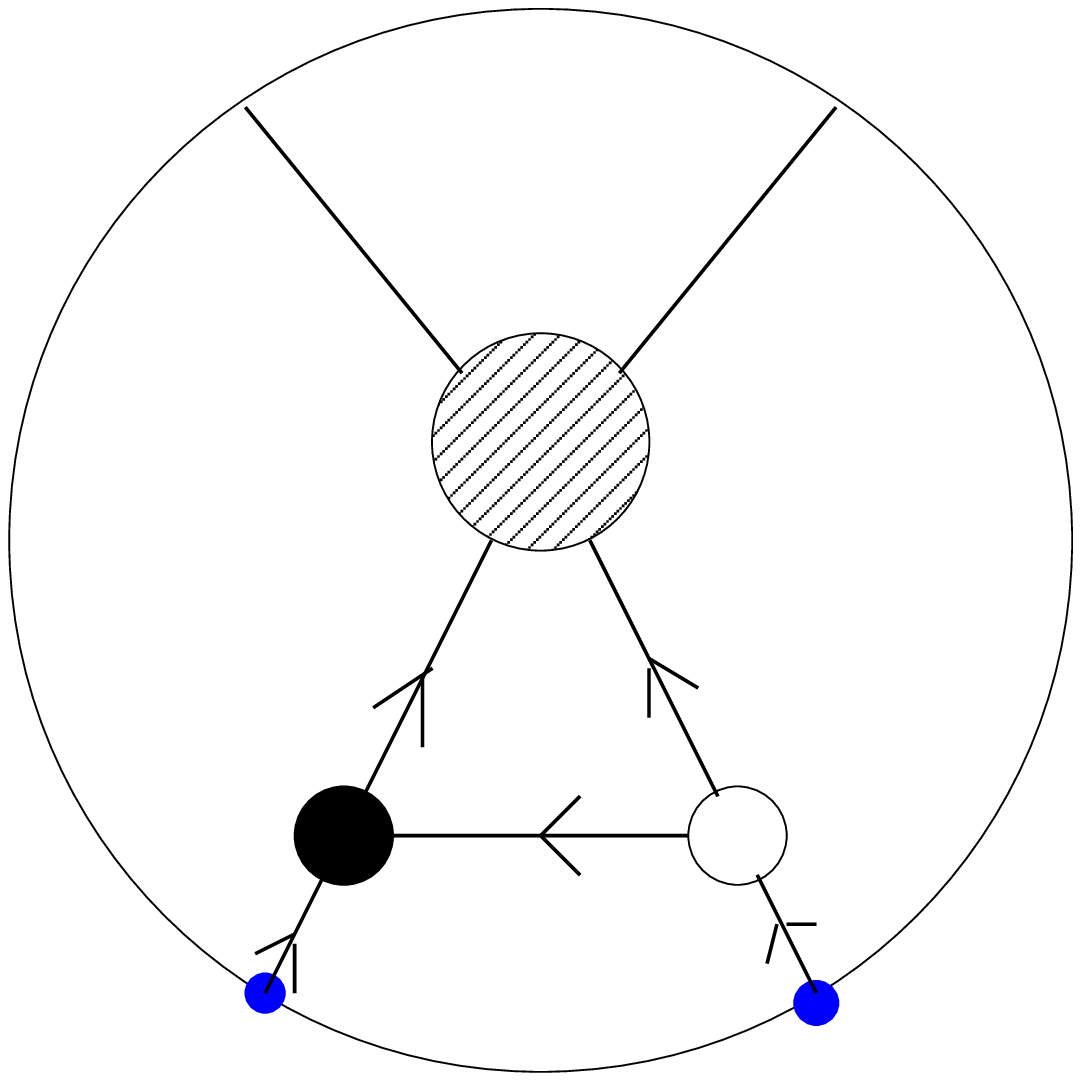}}}
  \:=\:\frac{dz}{z}\mathcal{M}_{n}^{\mbox{\tiny $(0)$}}(z),
  \hspace{2cm}
  \left\{
  \begin{array}{l}
   q_{\mbox{\tiny $ij$}}\:=\:z\lambda^{\mbox{\tiny $(j)$}}
    \tilde{\lambda}^{\mbox{\tiny $(i)$}}\\
   \hat{p}^{\mbox{\tiny $(i)$}}\:=\:
    (
     \lambda^{\mbox{\tiny $(i)$}}-z\lambda^{\mbox{\tiny $(j)$}}
    )
    \tilde{\lambda}^{\mbox{\tiny $(i)$}}\\
   \hat{p}^{\mbox{\tiny $(j)$}}\:=\:
    \lambda^{\mbox{\tiny $(j)$}}
    (
     \tilde{\lambda}^{\mbox{\tiny $(j)$}}+
     z\tilde{\lambda}^{\mbox{\tiny $(i)$}}
    )\\
   \tilde{\eta}^{\mbox{\tiny $(\hat{j})$}}\:=\:
     \tilde{\eta}^{\mbox{\tiny $(j)$}}-
     z\tilde{\eta}^{\mbox{\tiny $(i)$}}
  \end{array}
 \right. ,
 \end{split}
\end{equation}
as well as there is just a helicity flow between the external
lines and the intermediate ones which glue the bridge with a 
putative on-shell diagram, the differential measure is as in
\eqref{eq:BCFWbridge1} and \eqref{eq:BCFWbridge2a}.

It is important to stress here that the existence or not of 
helicity flows plays a very important role in identifying the
physical structure of a theory. For the moment we have learnt
that the presence of a helicity flow between the external lines
of a BCFW bridge and the intermediate ones attached to a putative
on-shell diagram correspond to the differential measure of $1/z$,
while in absence of such a flow, it develops a multiple pole;
when different states can propagate in the intermediate lines,
a helicity flow in them is generated. How deep is this observation? 
In order to understand it, let us discuss the helicity flows
more generally

\subsection{On-shell diagrams and helicity flows}\label{subsec:OShf}

\begin{figure}[htbp]
 \centering 
 \[
  \raisebox{-.9cm}{\scalebox{.40}{\includegraphics{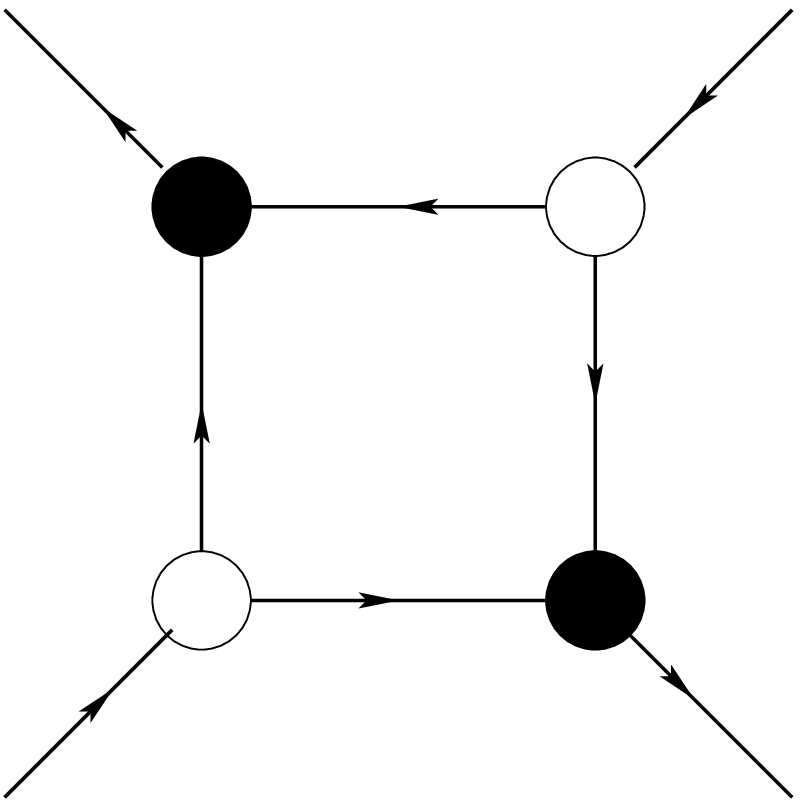}}}
   \hspace{1cm}
  \raisebox{-.9cm}{\scalebox{.40}{\includegraphics{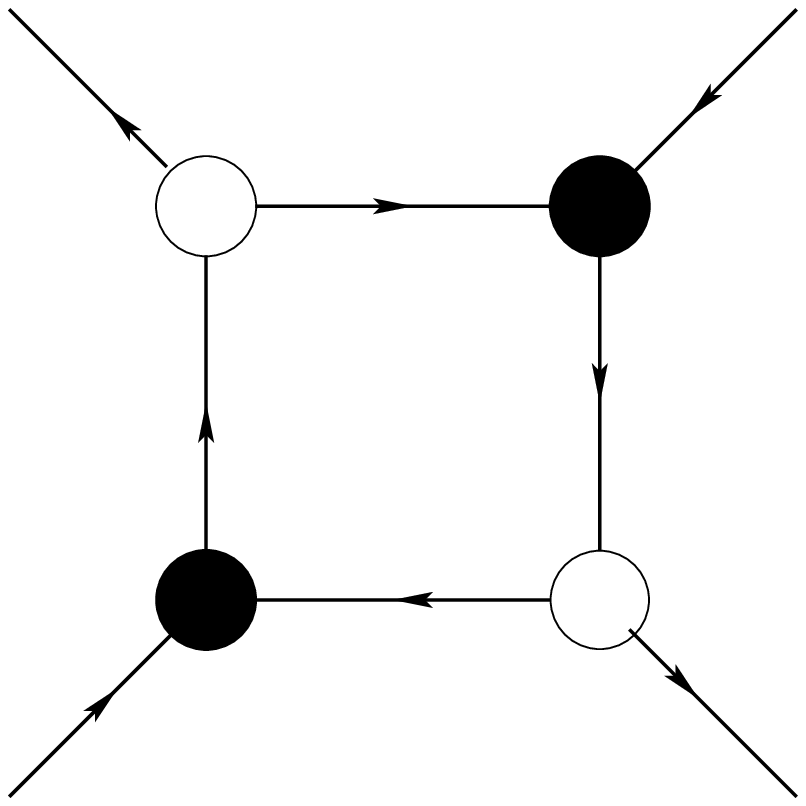}}}
   \hspace{1cm}
  \raisebox{-.9cm}{\scalebox{.40}{\includegraphics{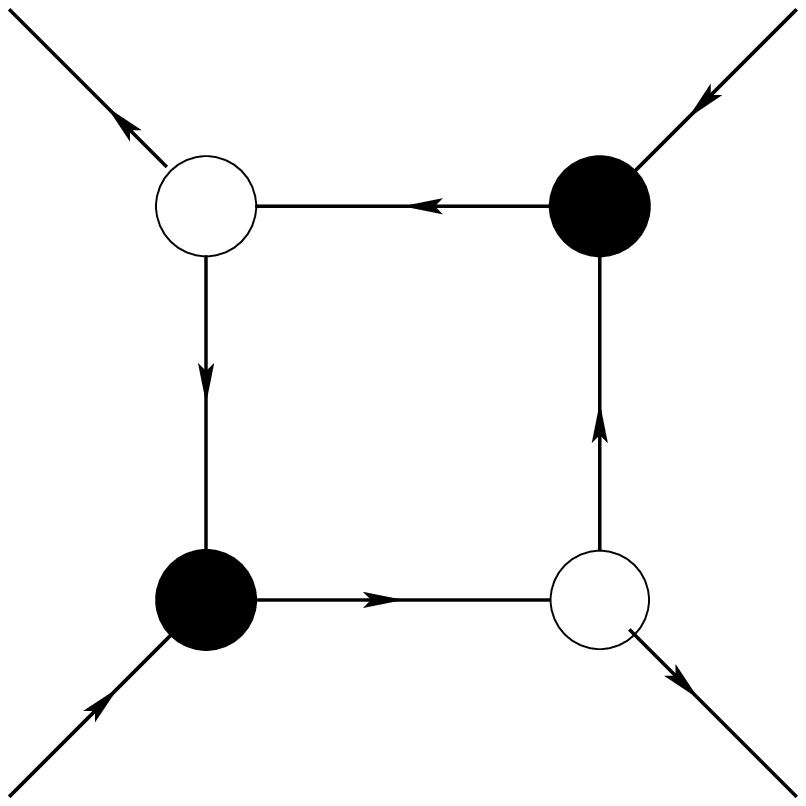}}}
 \]
 \caption{Simplest four-particle on-shell diagrams viewable in
  terms of a BCFW bridge attached to an other (simpler) on-shell 
  diagram.}
 \label{fig:4ptOSdiags}
\end{figure}

Let us consider the simplest (non-singular) on-shell diagrams,
such as in Figure \ref{fig:4ptOSdiags}, where the helicity 
configuration $(-,+,-,+)$ has been chosen for the external
coherent states. Notice that the second and third diagrams actually
come from the sum over the coherent states which can propagate in
the glued intermediate lines once the holomorphicity of the 
three-particle amplitude has been chosen. For the choice in the
first diagram instead, just one coherent state per intermediate line
is allowed.
 
They can be considered as a BCFW bridge attached to a diagram of the
type in Figure \ref{fig:4ptampl}. The latter, providing a 
delta-function on the momenta, fixes the degree of freedom 
introduced by the bridge. Thus, these on-shell diagrams are fully 
localised and return a rational function of the Lorentz invariants. 
Let us notice that each of these on-shell diagrams can
be viewed as generated by a BCFW bridge in two different channels.
In order to be as explicit as possible, we discuss separately
the three on-shell diagrams. Beginning with the first on-shell 
diagram in Figure \ref{fig:4ptOSdiags}, as we just mentioned
it can be seen as a BCFW bridge applied in two different channels
\begin{equation*}
 \raisebox{-.9cm}{\scalebox{.40}{\includegraphics{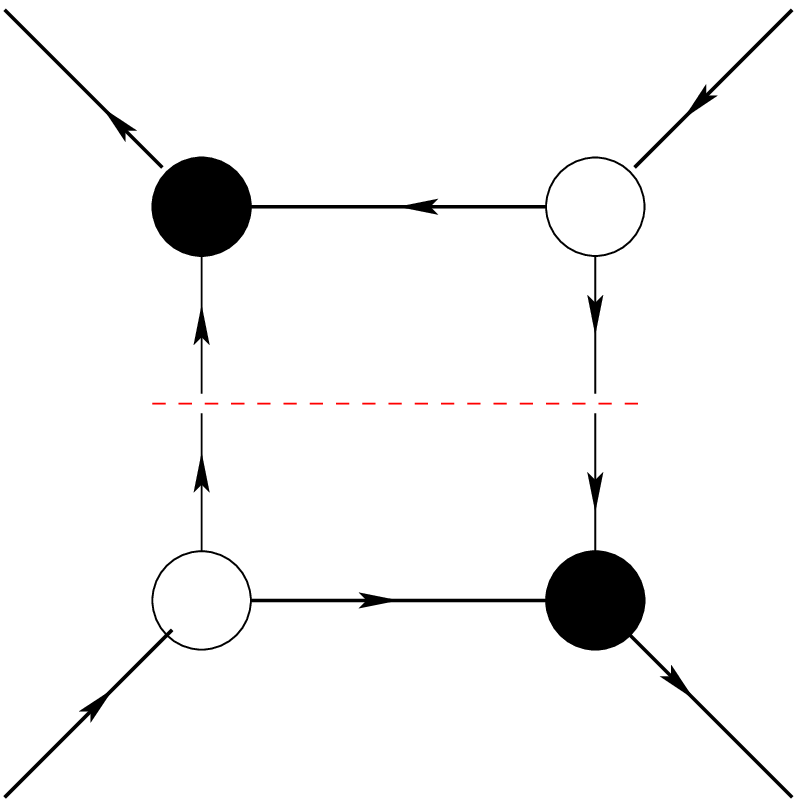}}}
 \hspace{1.2cm}
 \raisebox{-.9cm}{\scalebox{.40}{\includegraphics{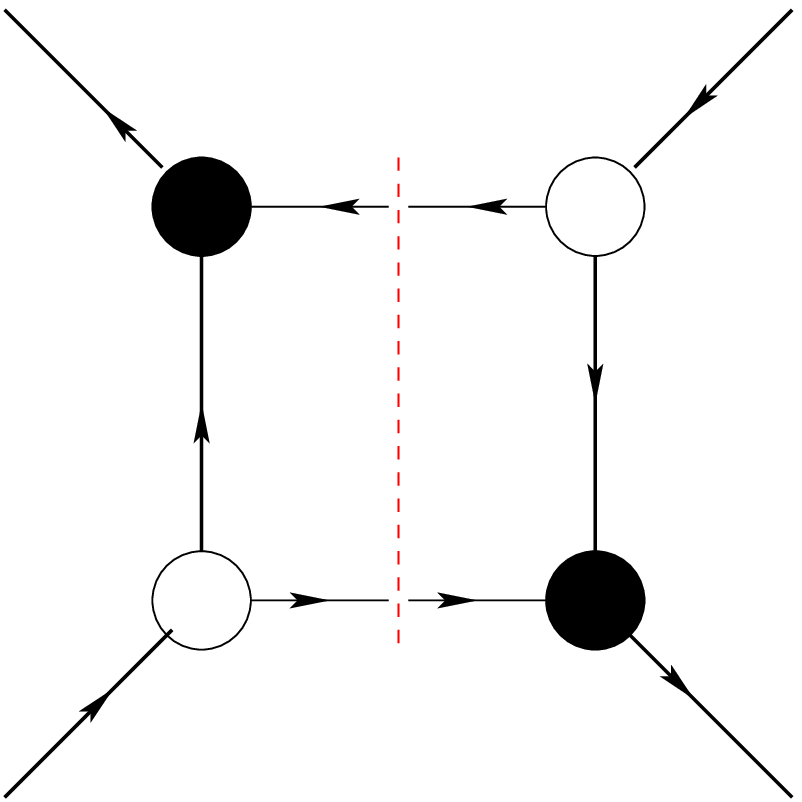}}}
\end{equation*}
Notice that each of them can be viewed in two different ways because
each sub-diagram in which they have been virtually separated by the
dashed red line can be thought as implementing a BCFW bridge.
Following the helicity arrows, it is straightforward to realise that
there exists a helicity flow between the external states and
the intermediate one, while there is no helicity flow in the 
intermediate lines: all the possible BCFW bridges are of the type
\eqref{eq:BCFWbridge1}. 
Furthermore, each single sub-diagram has the same helicity 
configuration $(-,+,-,+)$ as the full diagram and the full
diagram shows helicity flows between all the adjacent states.
This is just the statement that the whole on-shell diagram
contains all and only the singularities (simple poles) related
to the helicity configuration $(-,+,-,+)$ with the correct residues.
More explicitly, the fact that the full diagram shows helicity 
flows between all the adjacent external states implies that the
physical object it represents contains all and only the complex
factorisation in both the $s$- and $t$-channels

\begin{equation*}
 \begin{array}{cccccc}
  {}
 &{}
 &\raisebox{-1.4cm}{\scalebox{.30}{\includegraphics{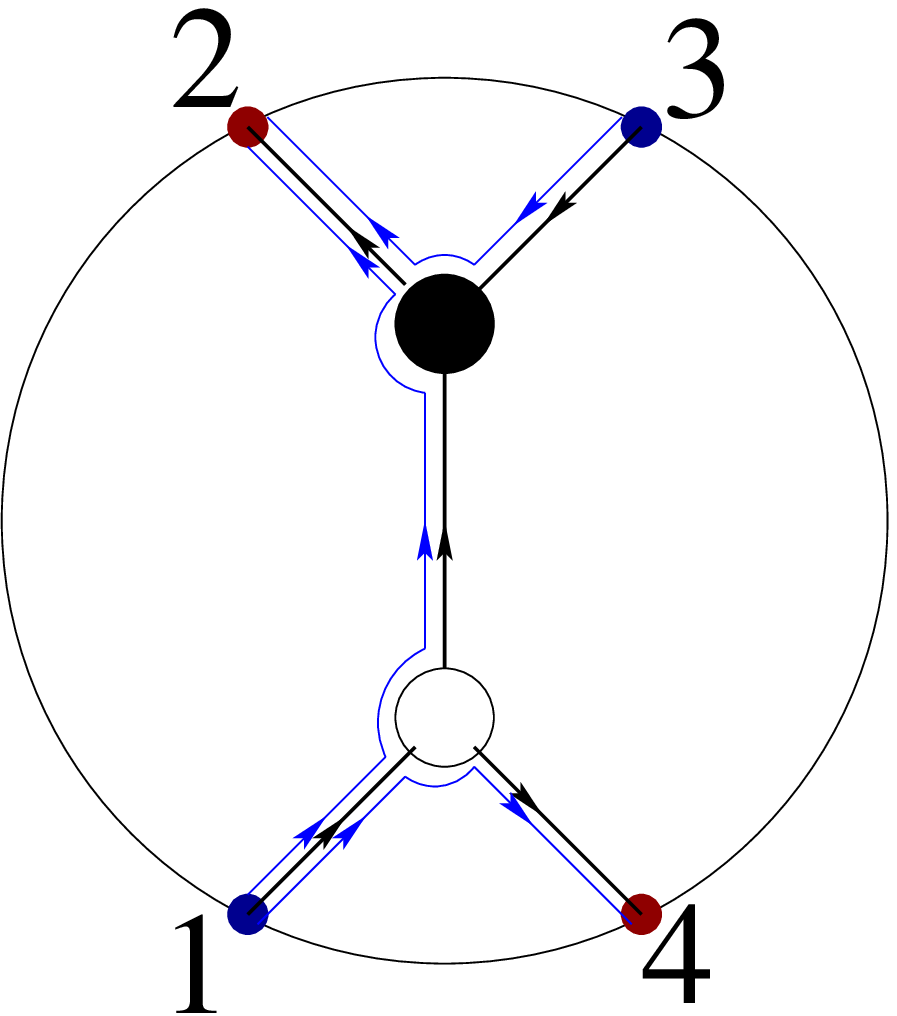}}}
 &+
 &\raisebox{-1.4cm}{\scalebox{.30}{\includegraphics{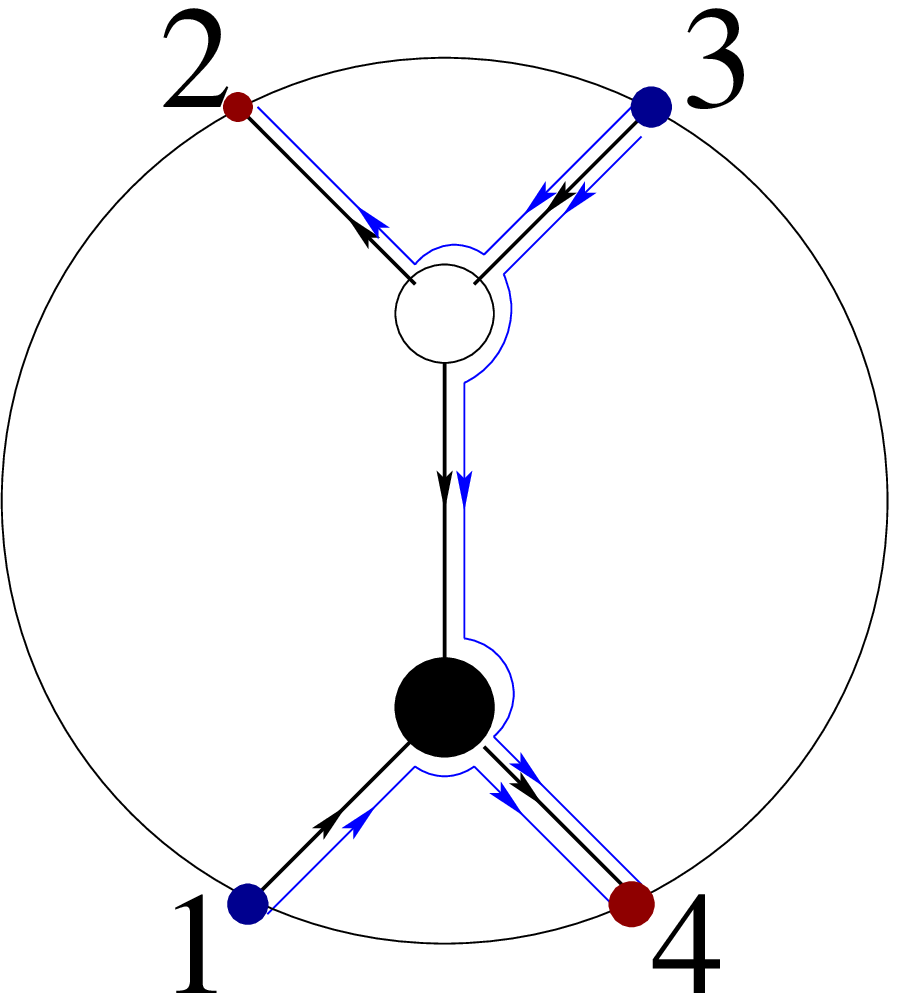}}}
 &+\\
  {\raisebox{-1.6cm}{\scalebox{.30}{\includegraphics{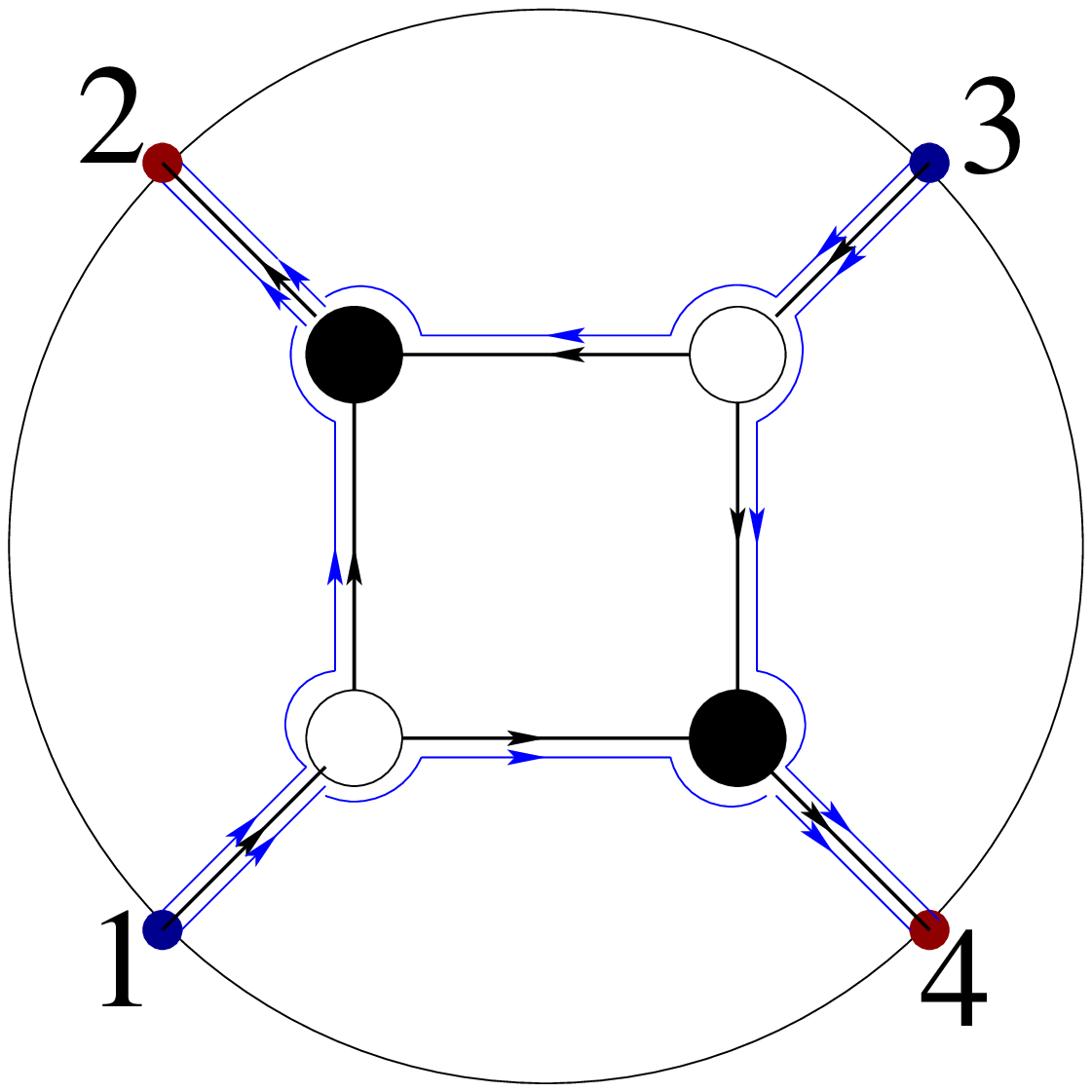}}}}
  &{}
  &
   \begin{array}{ccc}
    {\color{red} 2}  & \longleftarrow  & {\color{blue} 3}\\
    \uparrow         & {}              & {}              \\
    {\color{blue} 1} & \longrightarrow & {\color{red} 4}
   \end{array}
  &{}
  &
   \begin{array}{ccc}
    {\color{red} 2}  & \longleftarrow  & {\color{blue} 3}\\
    {}               & {}              & \downarrow      \\
    {\color{blue} 1} & \longrightarrow & {\color{red} 4}
   \end{array}
  &{}\\
   {}
  &{\Longrightarrow}
  &{}
  &{}
  &{}
  &{}\\
   {\vspace{-.5cm}
   \begin{array}{ccc}
    {\color{red} 2}  & \longleftarrow  & {\color{blue} 3}\\
    \uparrow         & {}              & \downarrow      \\
    {\color{blue} 1} & \longrightarrow & {\color{red} 4}
   \end{array}}
  &{}
  &\raisebox{-1cm}{\scalebox{.30}{\includegraphics{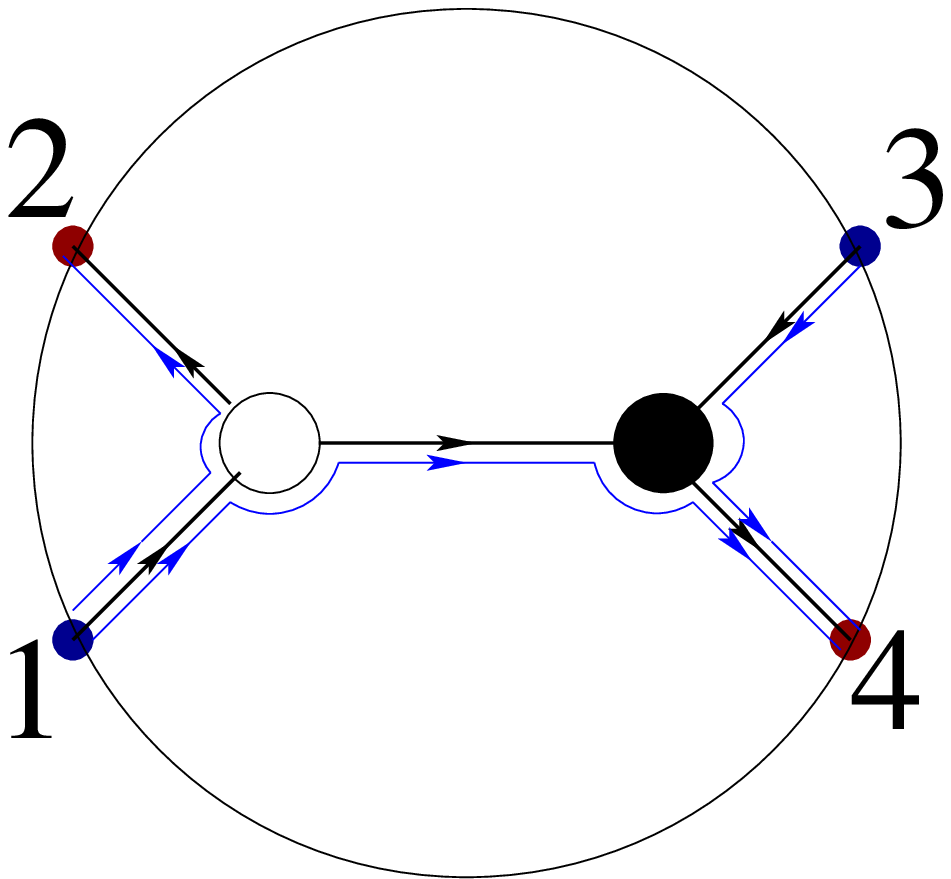}}}
  &+
  &\raisebox{-1cm}{\scalebox{.30}{\includegraphics{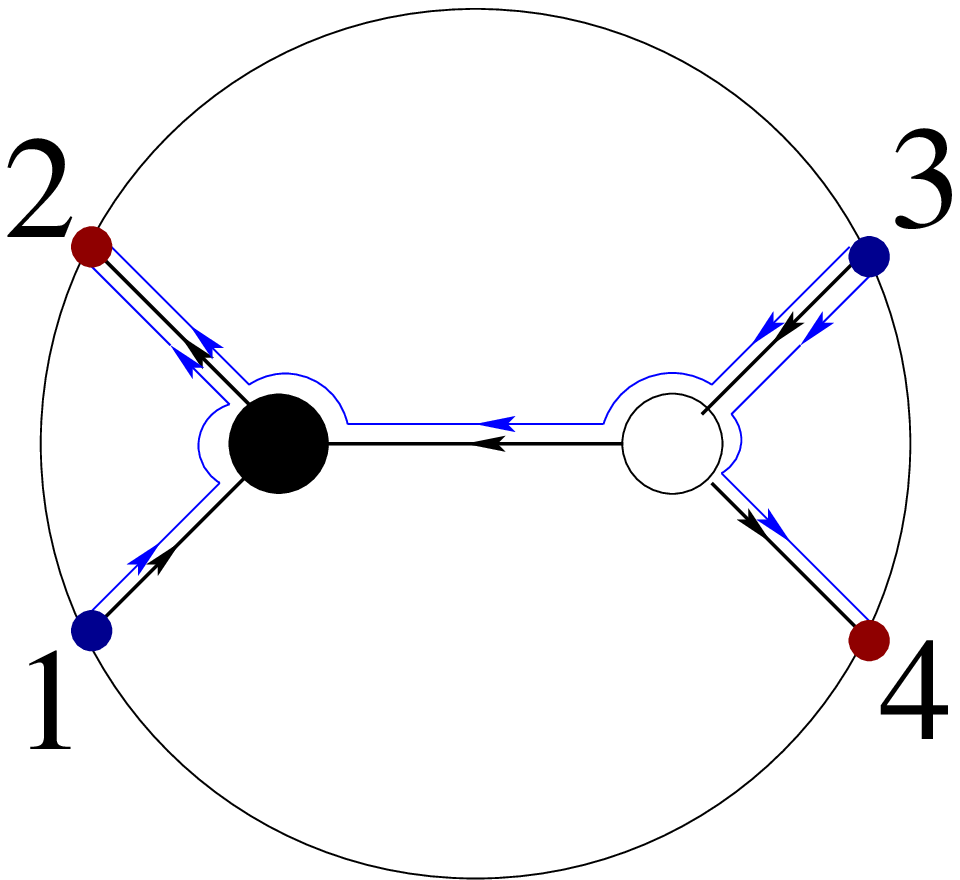}}}
  &{}
  \\
   {}
  &{}
  &
   \begin{array}{ccc}
    {\color{red} 2}  & {}  & {\color{blue} 3}\\
    \uparrow         & {}              & \downarrow      \\
    {\color{blue} 1} & \longrightarrow & {\color{red} 4}
   \end{array}
  &{}
  &
   \begin{array}{ccc}
    {\color{red} 2}  & \longleftarrow  & {\color{blue} 3}\\
    \uparrow         & {}              & \downarrow      \\
    {\color{blue} 1} & {}              & {\color{red} 4}
   \end{array}
  &{}
 \end{array}
\end{equation*}


Let us repeat the same analysis for the second on-shell diagram
of Figure \ref{fig:4ptOSdiags}:
\begin{equation*}
 \raisebox{-.9cm}{\scalebox{.40}{\includegraphics{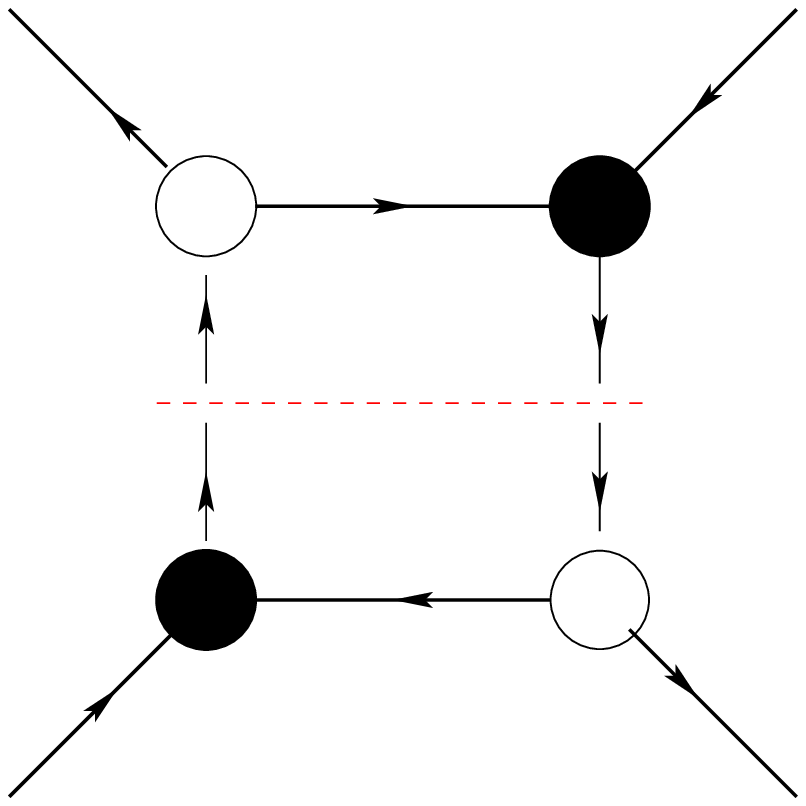}}}
 \hspace{1.2cm}
 \raisebox{-.9cm}{\scalebox{.40}{\includegraphics{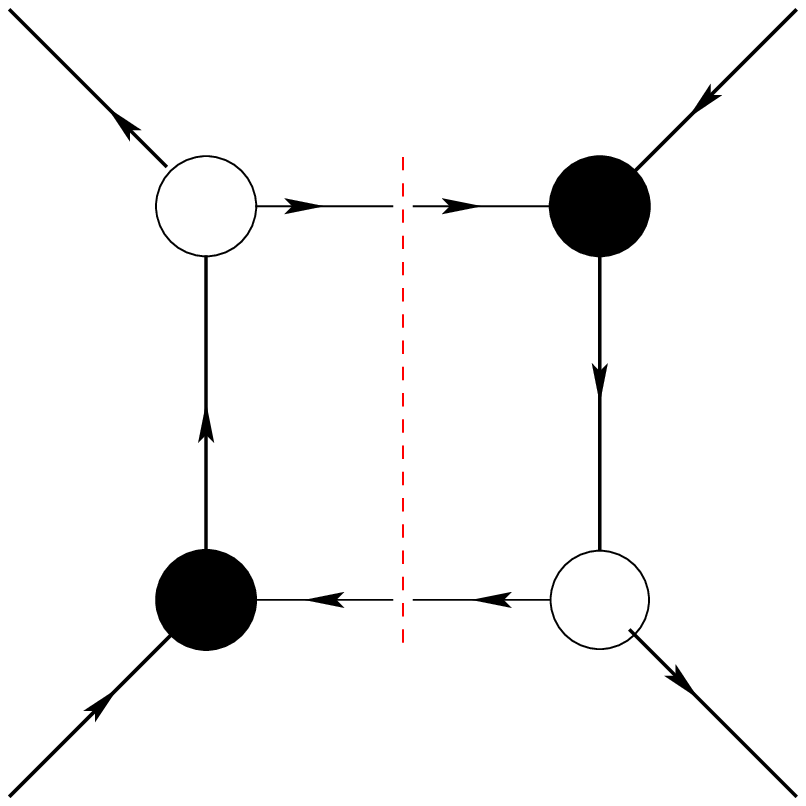}}}
\end{equation*}
In the diagram on the left, it is possible to notice a helicity flow
between the external lines and the intermediate one in the two
possible BCFW bridges, and furthermore there is also a helicity flow along just the intermediate lines. As a consequence, each of the
two sub-diagrams in this channel has the same helicity structure of
the full diagram.

In the picture on the right, the very same diagram is viewed in a
different channel. Here one can instead identify a helicity flow 
along the intermediate lines as well as one involving the two 
external lines of a bridge and the intermediate one connecting them,
while there is no flow from the external lines to the ones attached 
to the other diagram. As a consequence, both of the two sub-diagrams
{\it do not preserve} the helicity structure of the full diagram.
This means that in this channel the on-shell diagram under 
discussion does not contain the right residue to represent a 
full-fledge four-particle amplitude. This information is encoded in 
the  (clockwise) orientation of the helicity flow in the 
intermediate lines in the full diagram.

A similar analysis holds for the last diagram in 
Figure \ref{fig:4ptOSdiags}, where the helicity flow of the 
intermediate lines is counter-clockwise:
\begin{equation*}
 \raisebox{-.9cm}{\scalebox{.40}{\includegraphics{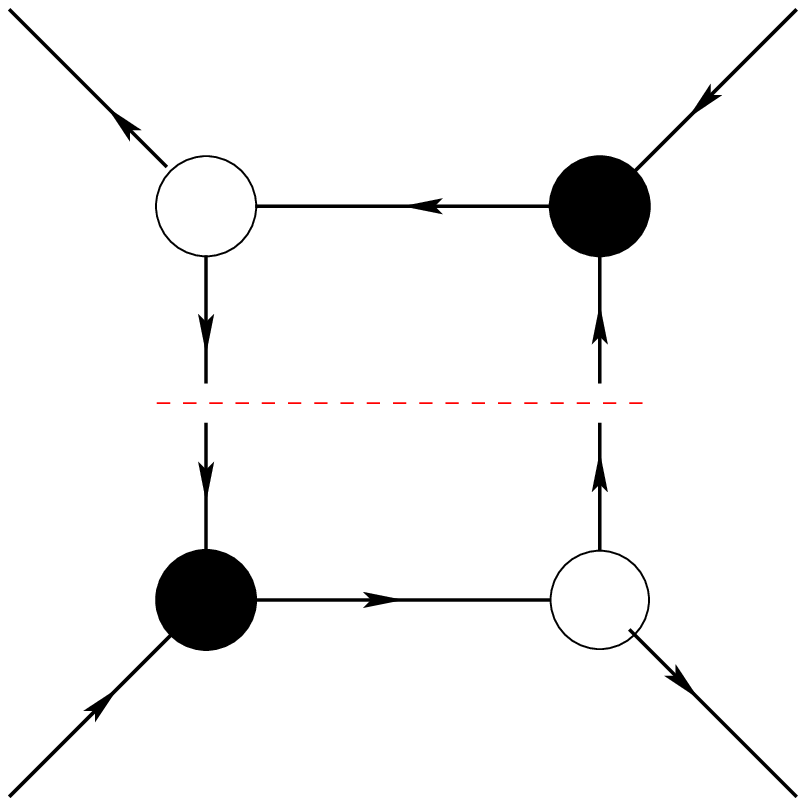}}}
 \hspace{1.2cm}
 \raisebox{-.9cm}{\scalebox{.40}{\includegraphics{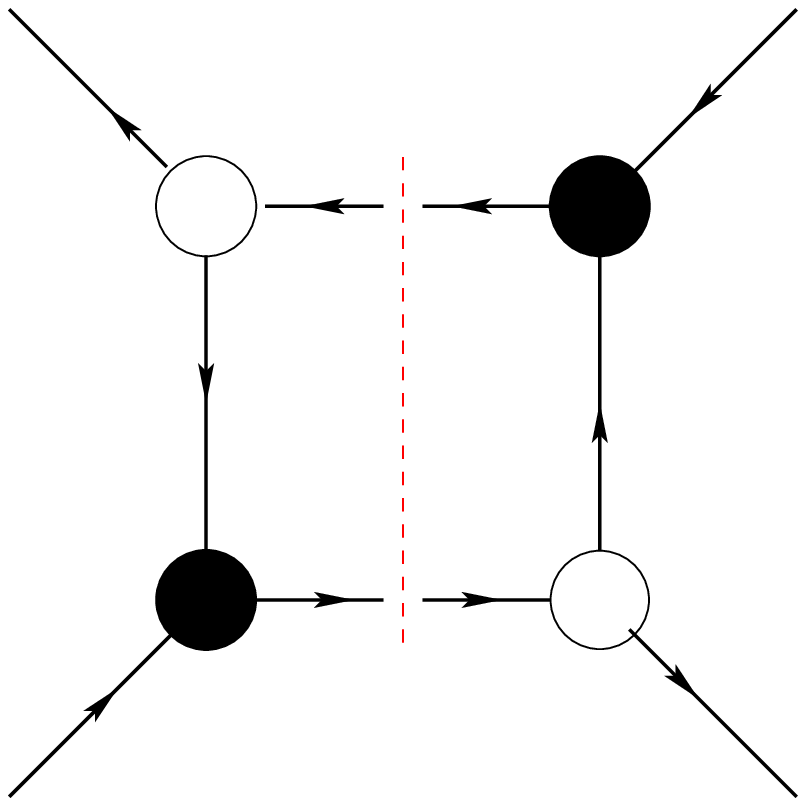}}}
\end{equation*}
Summarising, the two diagrams just analysed (which, we do not have
to forget, need to me summed up) shows two helicity flows between
adjacent coherent states as well as a oriented helicity flows in the
intermediate lines. The former implies that the full on-shell 
diagram contains a pole in the channel where the external helicity
flow appears, while the latter that diagram also has structures
related to different helicity configurations, which boils down to
having higher order poles:
\begin{equation*}
 \begin{split}
&\raisebox{-1.6cm}{\scalebox{.32}{\includegraphics{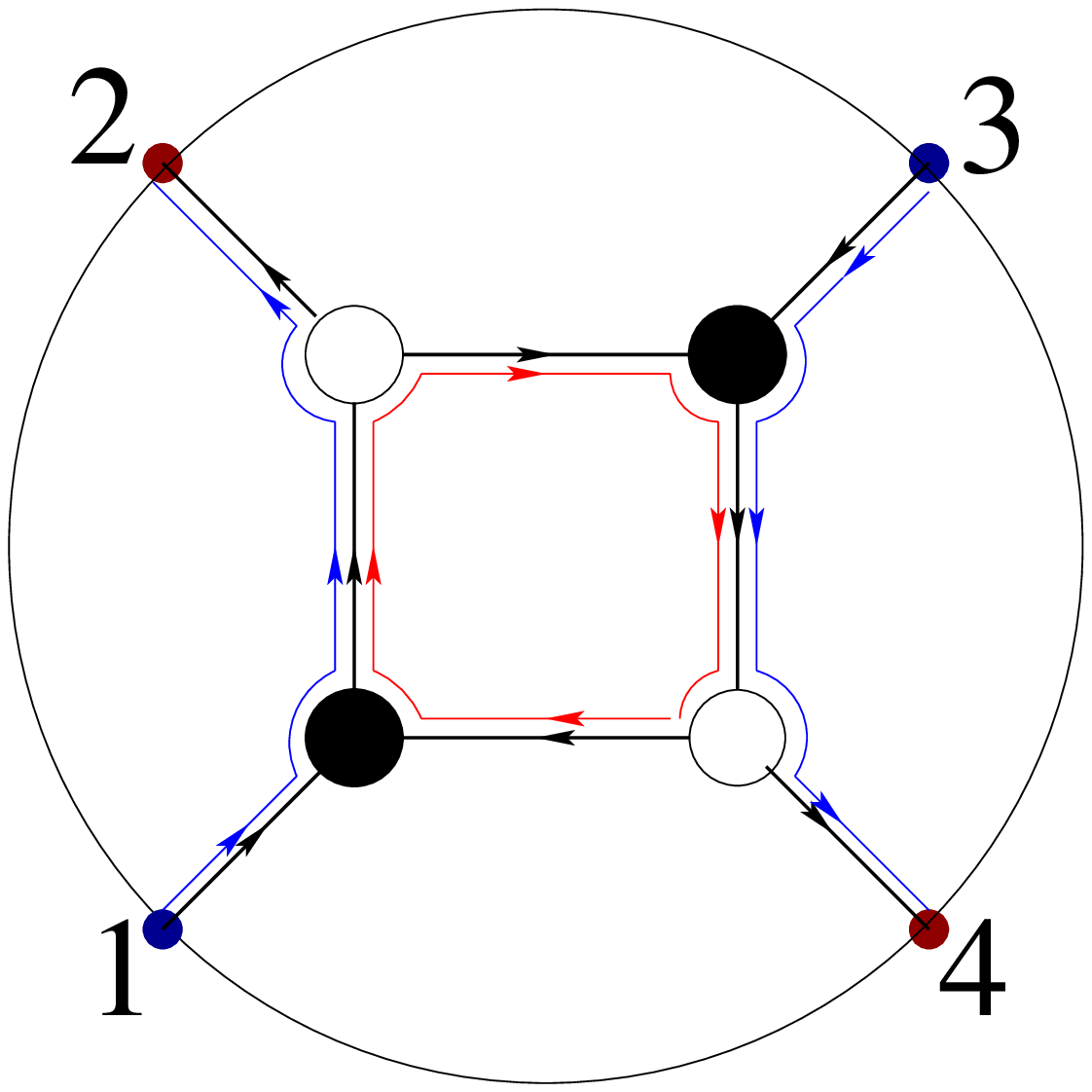}}}
 \;\Longrightarrow\;
  \raisebox{-1.4cm}{\scalebox{.32}{\includegraphics{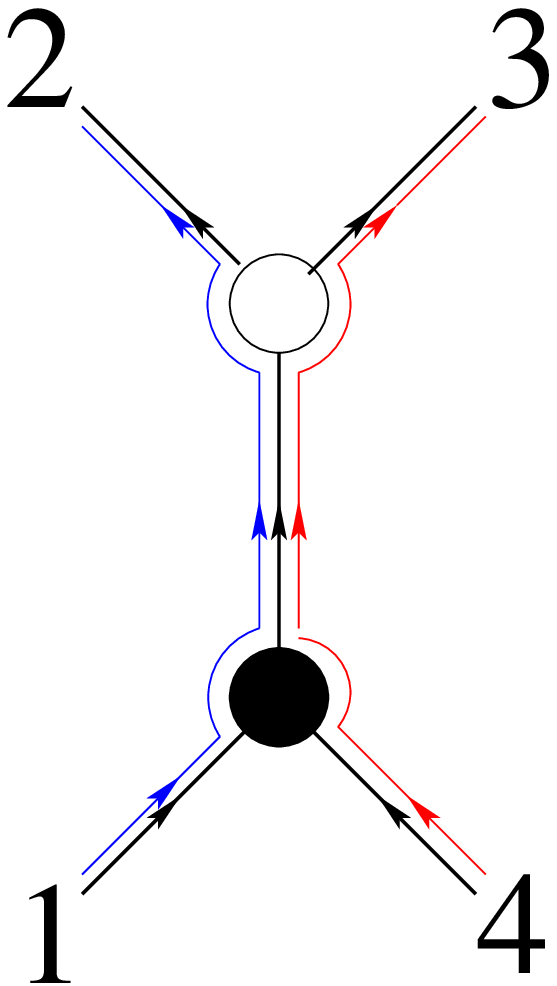}}}
   \;+\;
 \raisebox{-1.4cm}{\scalebox{.32}{\includegraphics{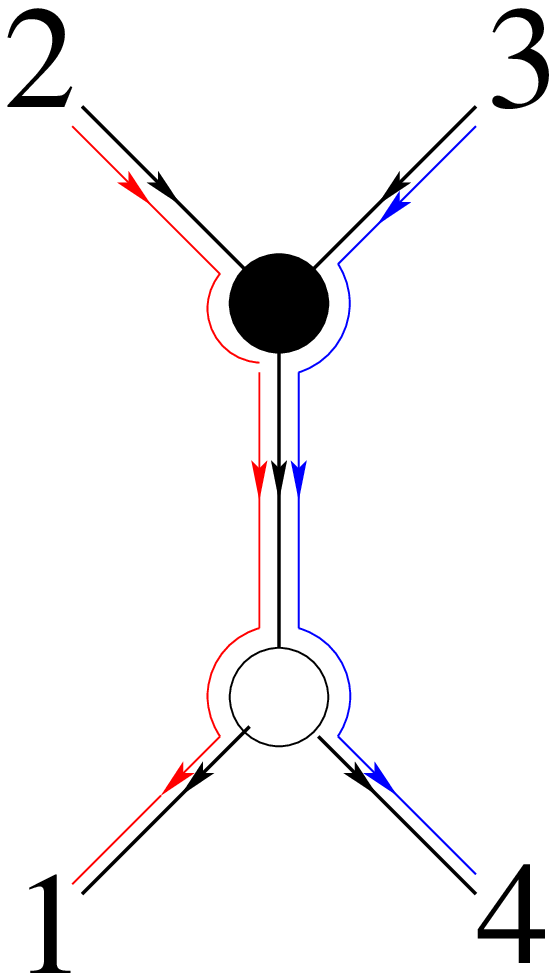}}}
   \;+\;
  \raisebox{-.7cm}{\scalebox{.32}{\includegraphics{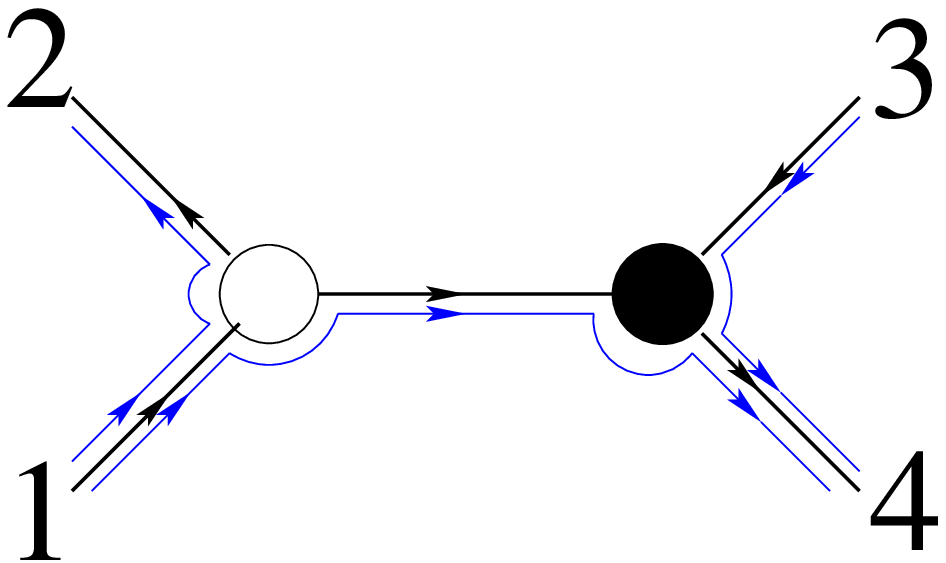}}}
  \;+\;
  \raisebox{-.7cm}{\scalebox{.32}{\includegraphics{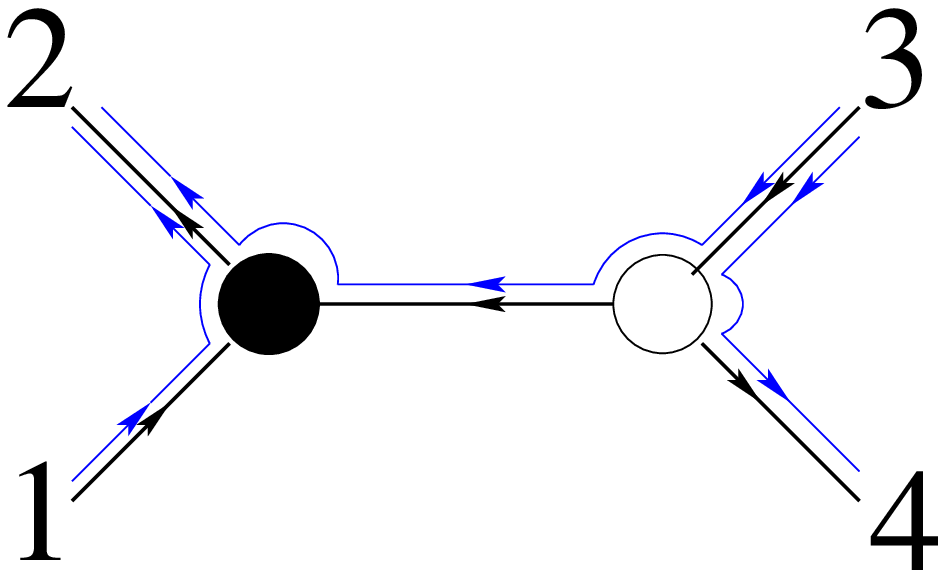}}}
 \\
&\raisebox{-1.6cm}{\scalebox{.32}{\includegraphics{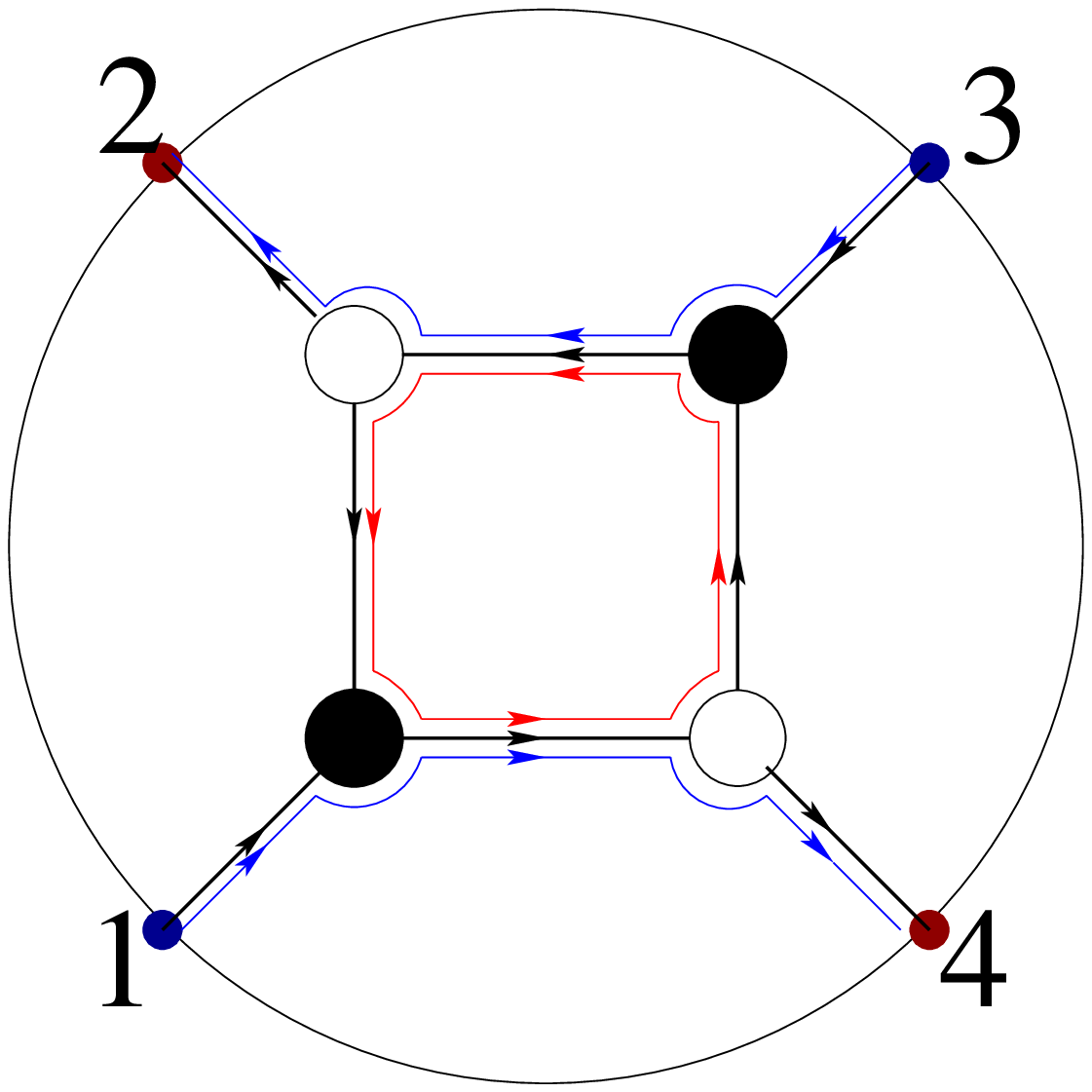}}}
 \;\Longrightarrow\;
 \raisebox{-1.4cm}{\scalebox{.32}{\includegraphics{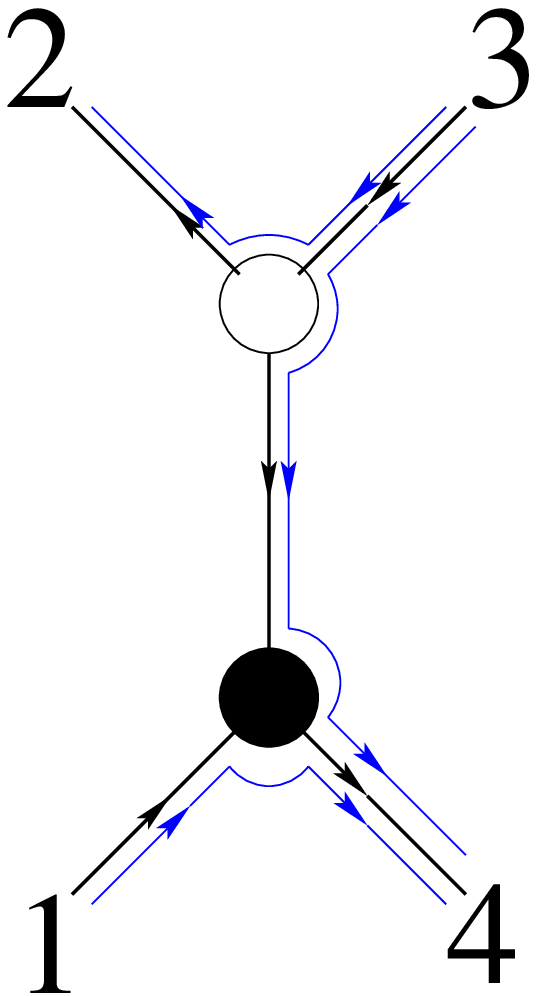}}}
   \;+\;
 \raisebox{-1.4cm}{\scalebox{.32}{\includegraphics{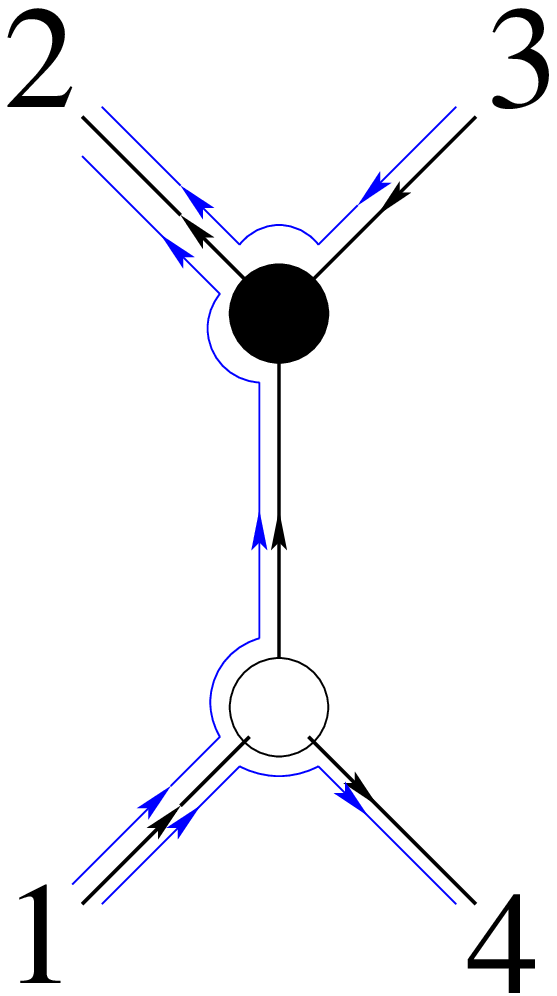}}}
  \;+\;
  \raisebox{-.7cm}{\scalebox{.32}{\includegraphics{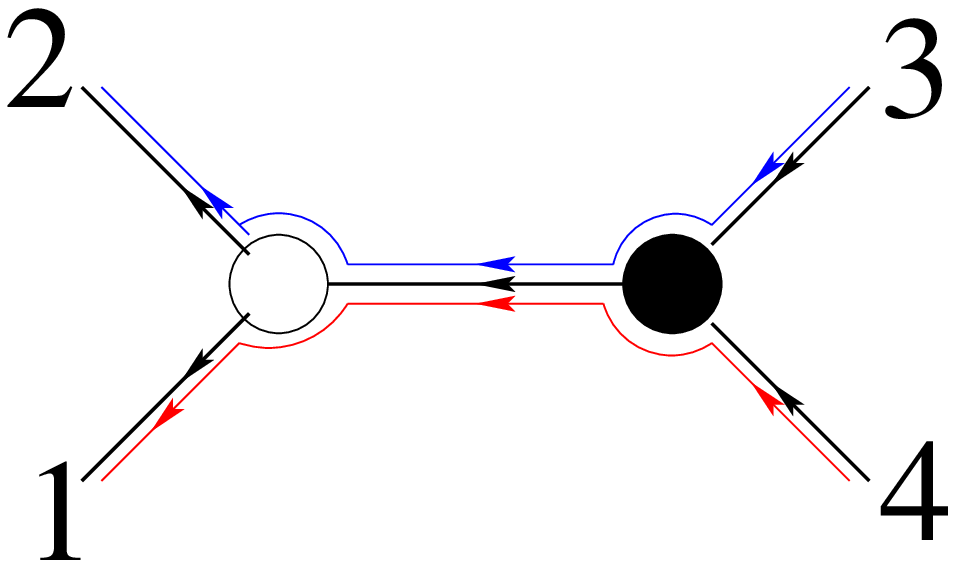}}}
  \;+\;
  \raisebox{-.7cm}{\scalebox{.32}{\includegraphics{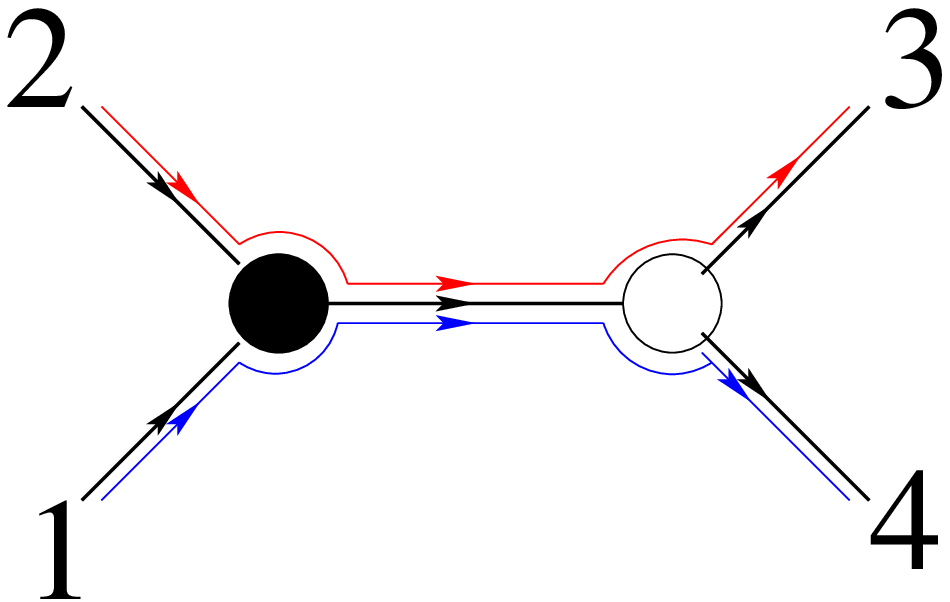}}}
 \end{split}
\end{equation*}
where the red helicity flows indicate the ones characterising a 
different helicity configuration. {\it Individually}, these diagrams
cannot represent the full-fledge four-particle amplitude with 
helicity configuration $(-,+,-,+)$ given that each of them shows
just one factorisation channel. Their sum instead could, provided
that suitable cancellations occur. As we will see later, those
cancellations occur just for $\mathcal{N}\,=\,3$ (in the maximally
supersymmetric case, there is no distinction between them).

\begin{figure}[htbp]
 \centering 
 \[
  \raisebox{-.9cm}{\scalebox{.40}{\includegraphics{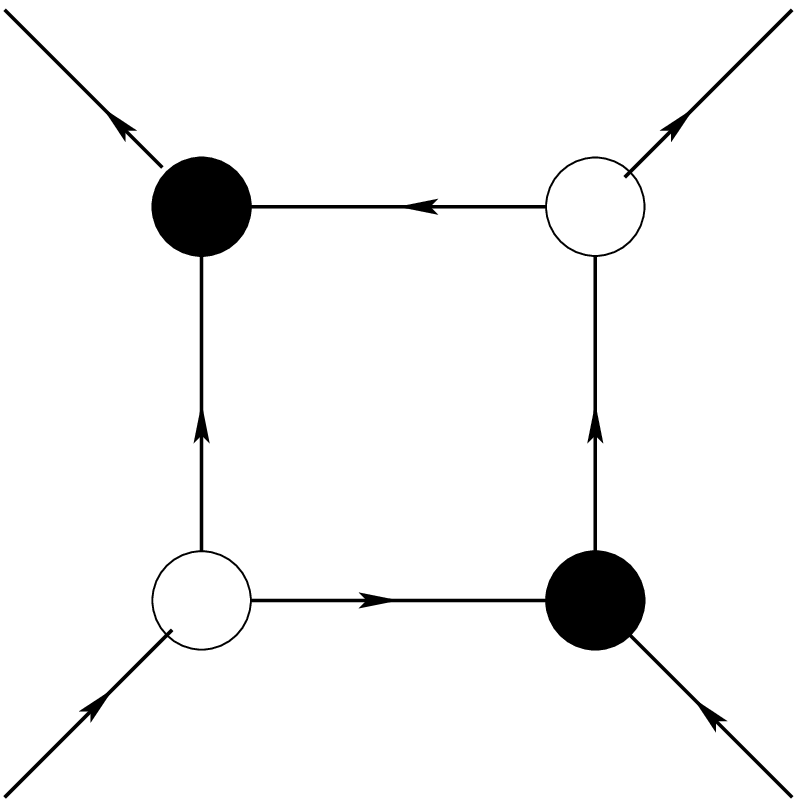}}}
   \hspace{1.5cm}
  \raisebox{-.9cm}{\scalebox{.40}{\includegraphics{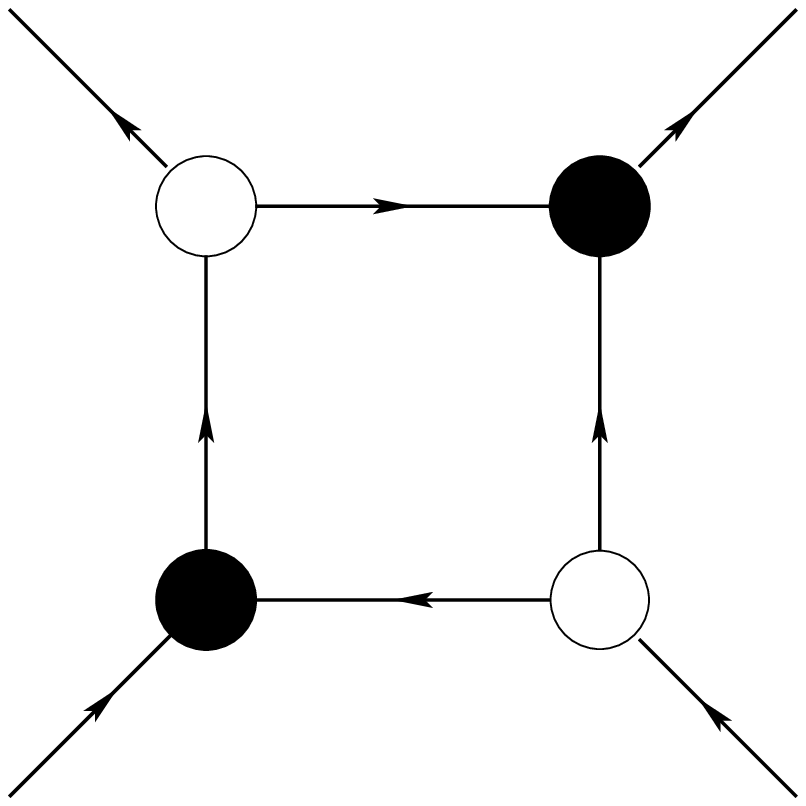}}}
 \]
 \caption{Four-particle on-shell diagrams with external helicity 
          configuration $(-,-,+,+)$}
 \label{fig:4ptdiags2}
\end{figure}

For the sake of completeness, we explicitly consider the fully 
localised four-particle on-shell diagram with the helicity 
configuration $(-,-,+,+)$. Actually, as we will see, this class
of diagram exhibits a property which is absent in the one analysed
so far. It is easy to see that the helicity configuration 
$(-,-,+,+)$ for the external states allows for two diagrams only,
as shown in Figure \ref{fig:4ptdiags2}

Notice that in none of them there is a helicity flow within the
intermediate lines only. 
As for the previous class of on-shell diagrams,
also in this case each diagram can be viewed as two BCFW-{\it like}
bridges applied to another on-shell diagram:
\begin{equation*}
 \begin{split}
  &\raisebox{-.9cm}{\scalebox{.40}{\includegraphics{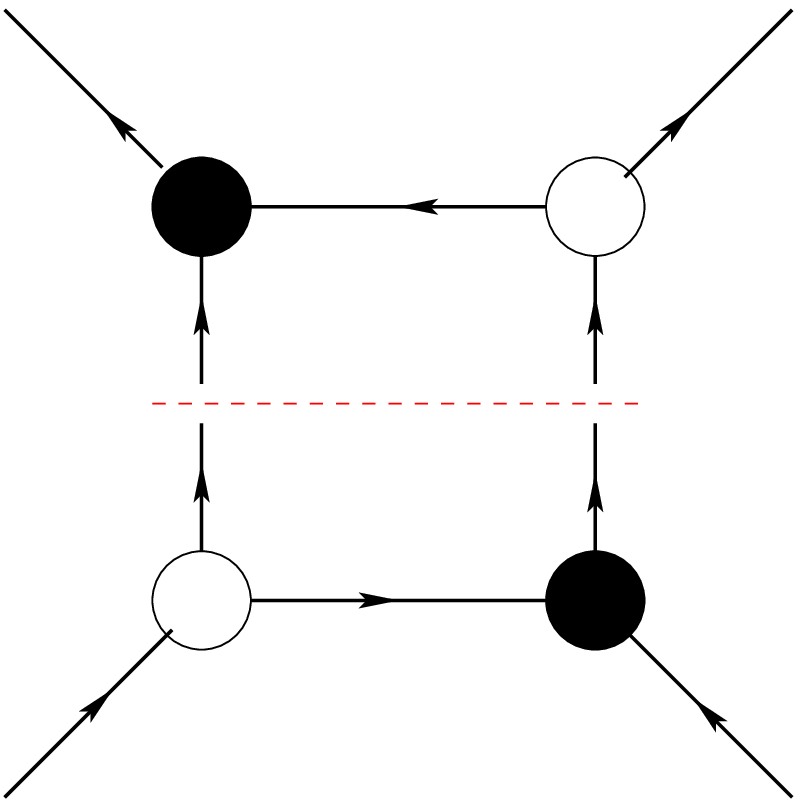}}}
   \hspace{1.2cm}
   \raisebox{-.9cm}{\scalebox{.40}{\includegraphics{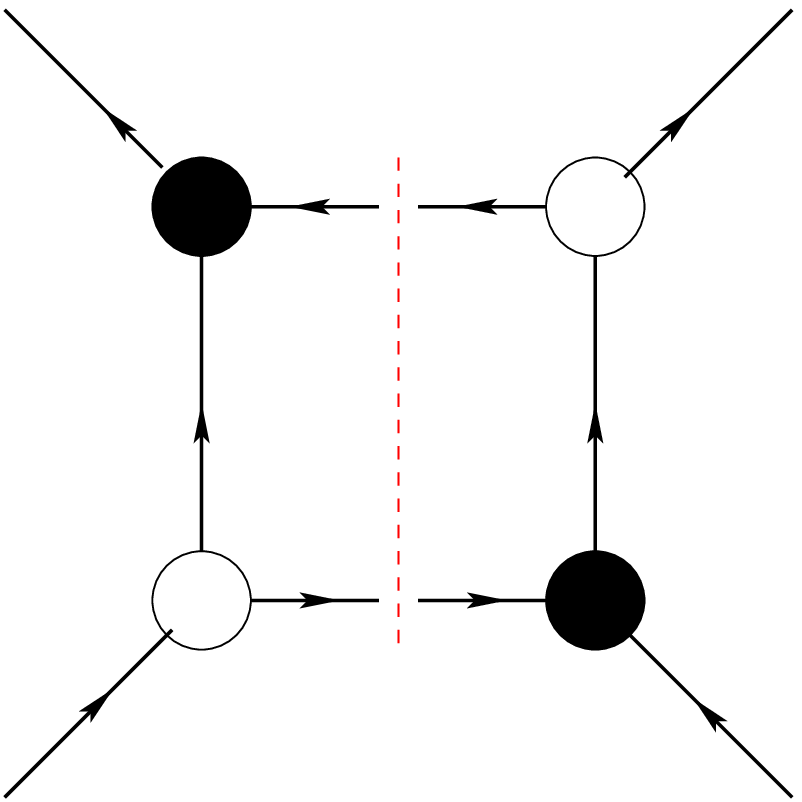}}}
  \\
  \phantom{\ldots}
  \\
  &\raisebox{-.9cm}{\scalebox{.40}{\includegraphics{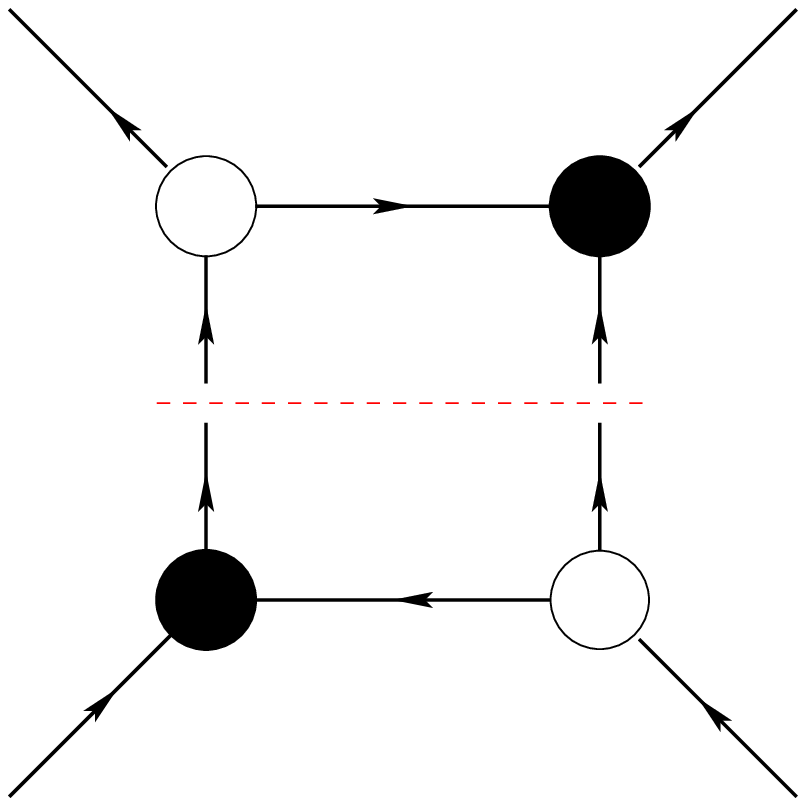}}}
   \hspace{1.2cm}
   \raisebox{-.9cm}{\scalebox{.40}{\includegraphics{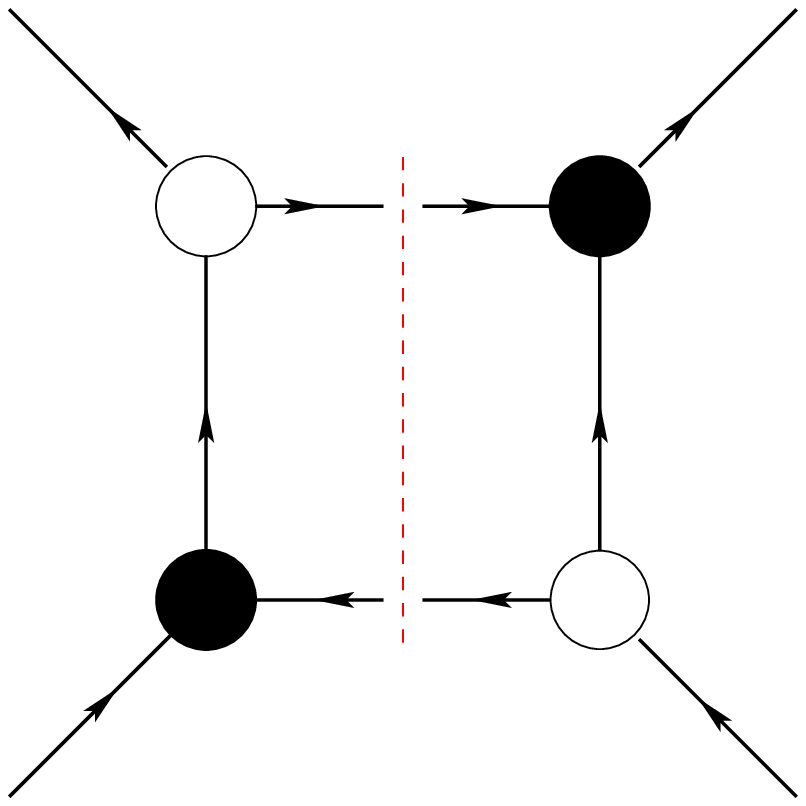}}}
 \end{split}
\end{equation*}
In the on-shell diagrams on the left (both at the top and at the 
bottom) both the sub-diagrams preserve the helicity configuration
of the full diagram and they show a helicity flow between the
external states and the intermediate lines which get glued to
the other diagram. In the on-shell diagrams on the right instead
just one of the two sub-diagrams preserves the helicity configuration
of the full diagram and it is also the only one which shows a
helicity flow between the external lines and the ones which get
glued to the other diagram. 


\begin{equation*}
 \begin{split}
&\raisebox{-1.6cm}{\scalebox{.30}{\includegraphics{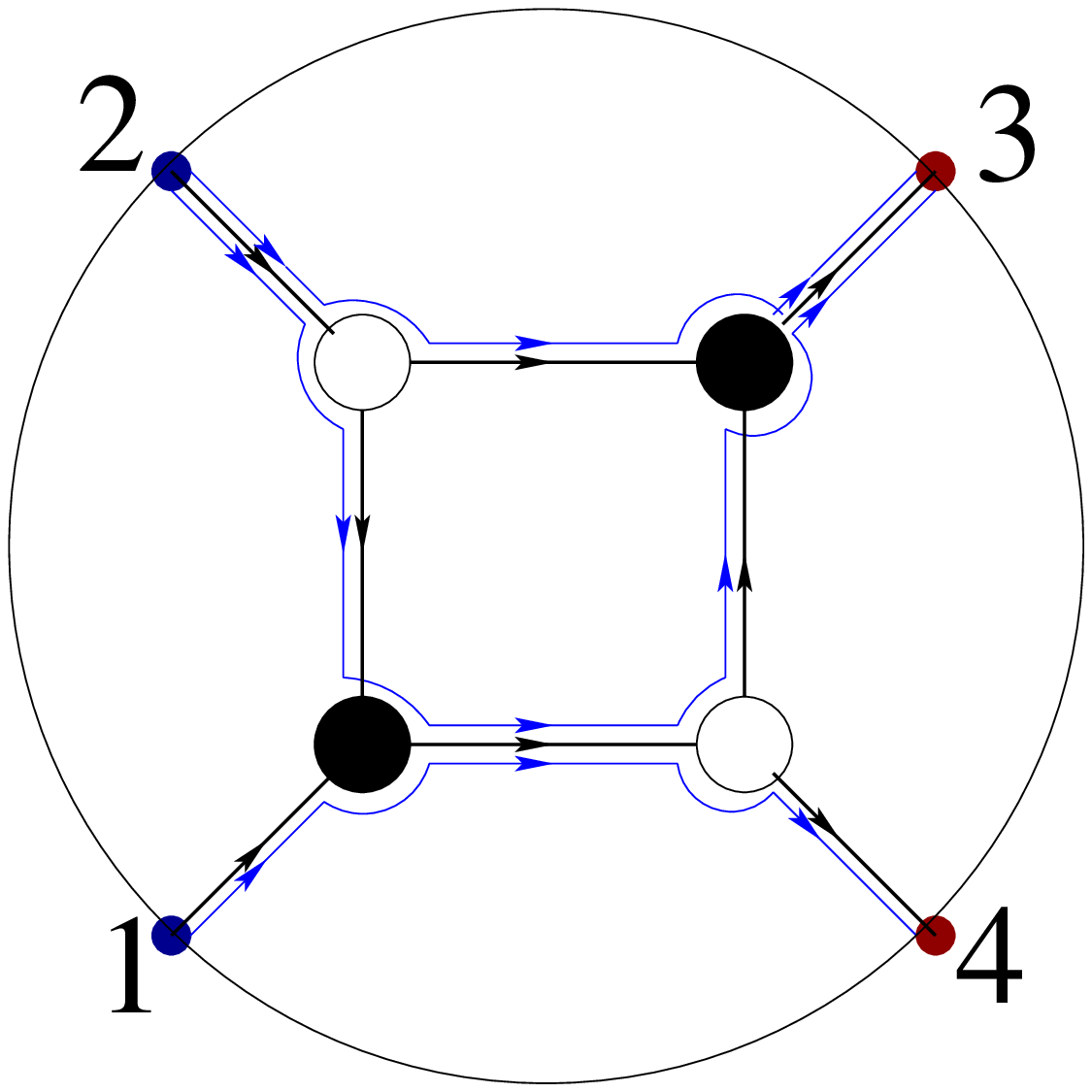}}}
 \;\Longrightarrow\;
  \raisebox{-1.4cm}{\scalebox{.32}{\includegraphics{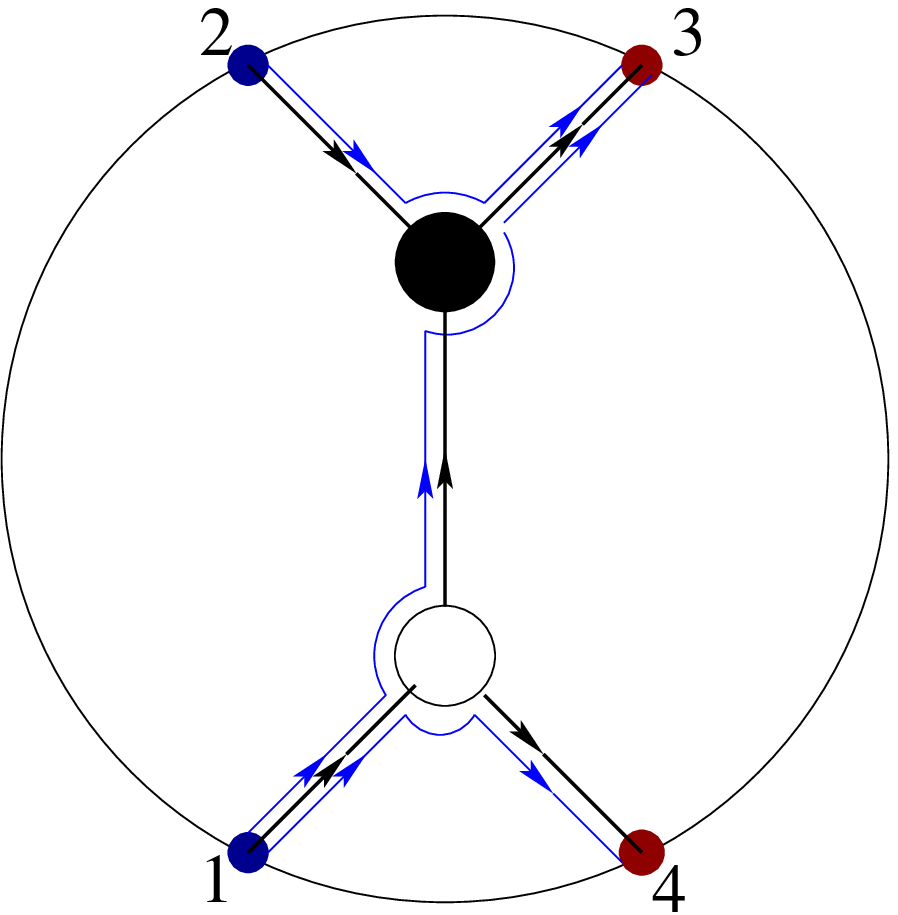}}}
   \;+\;
 \raisebox{-1.4cm}{\scalebox{.32}{\includegraphics{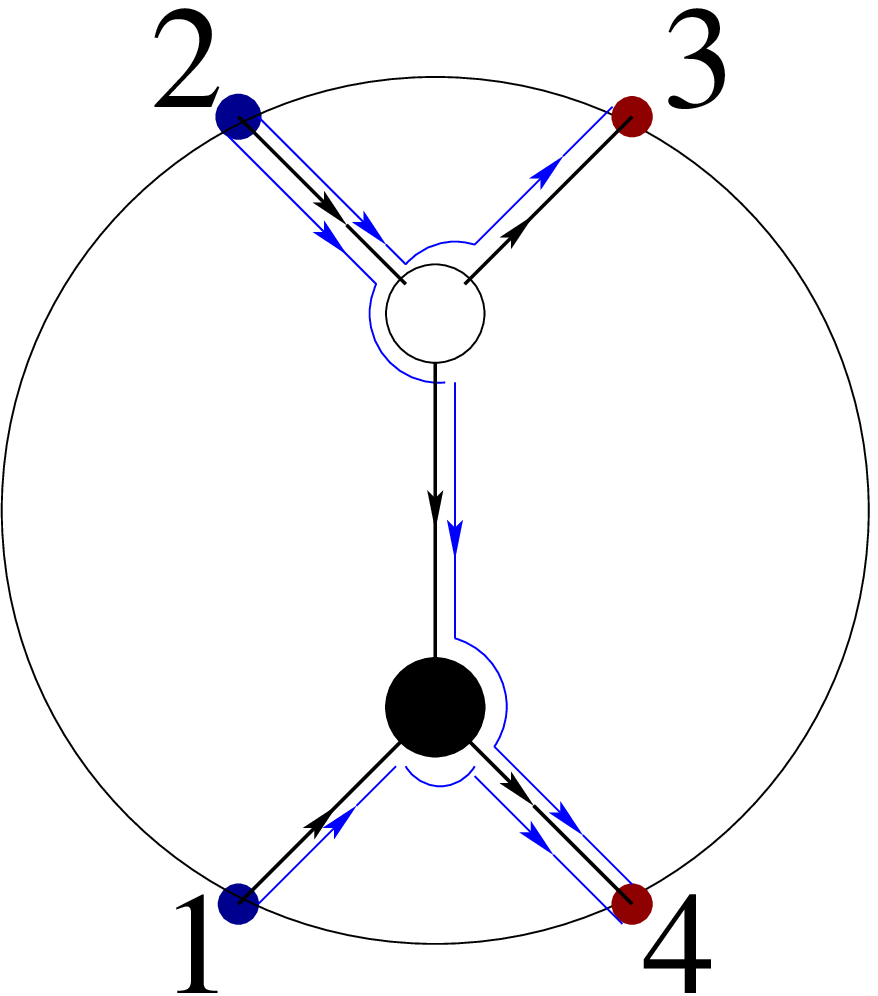}}}
   \;+\;
  \raisebox{-1.4cm}{\scalebox{.32}{\includegraphics{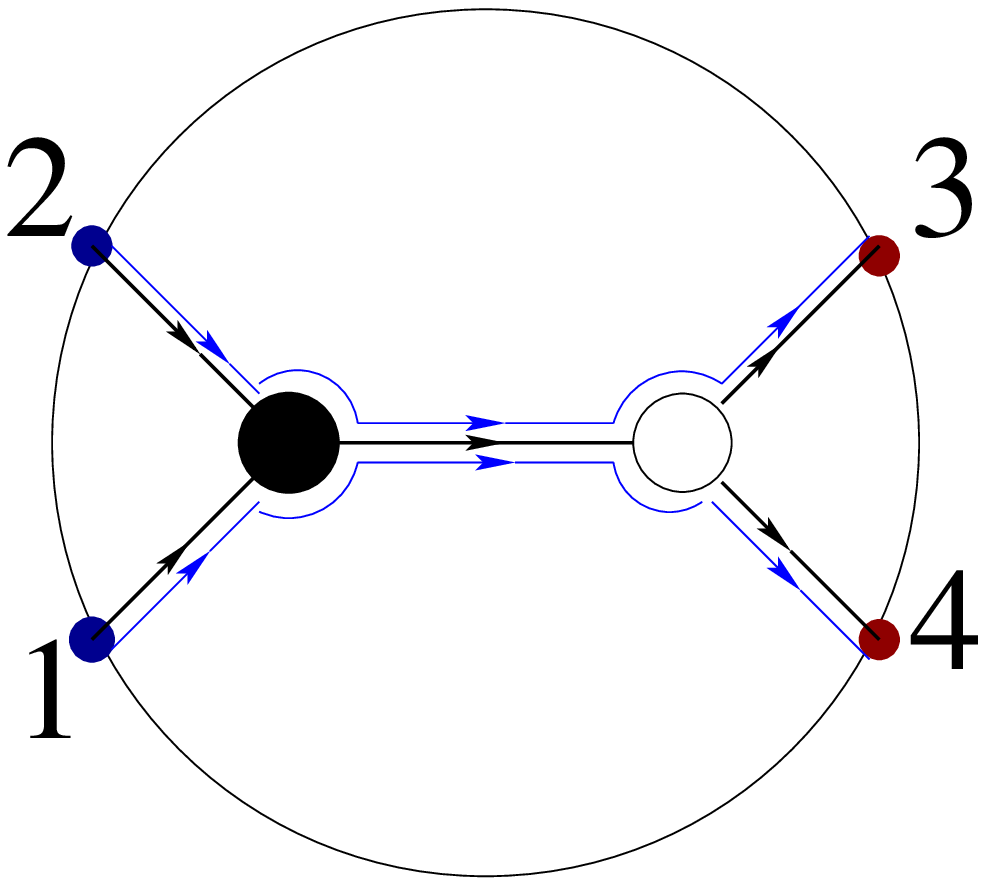}}}
 \\
&\raisebox{-1.6cm}{\scalebox{.30}{\includegraphics{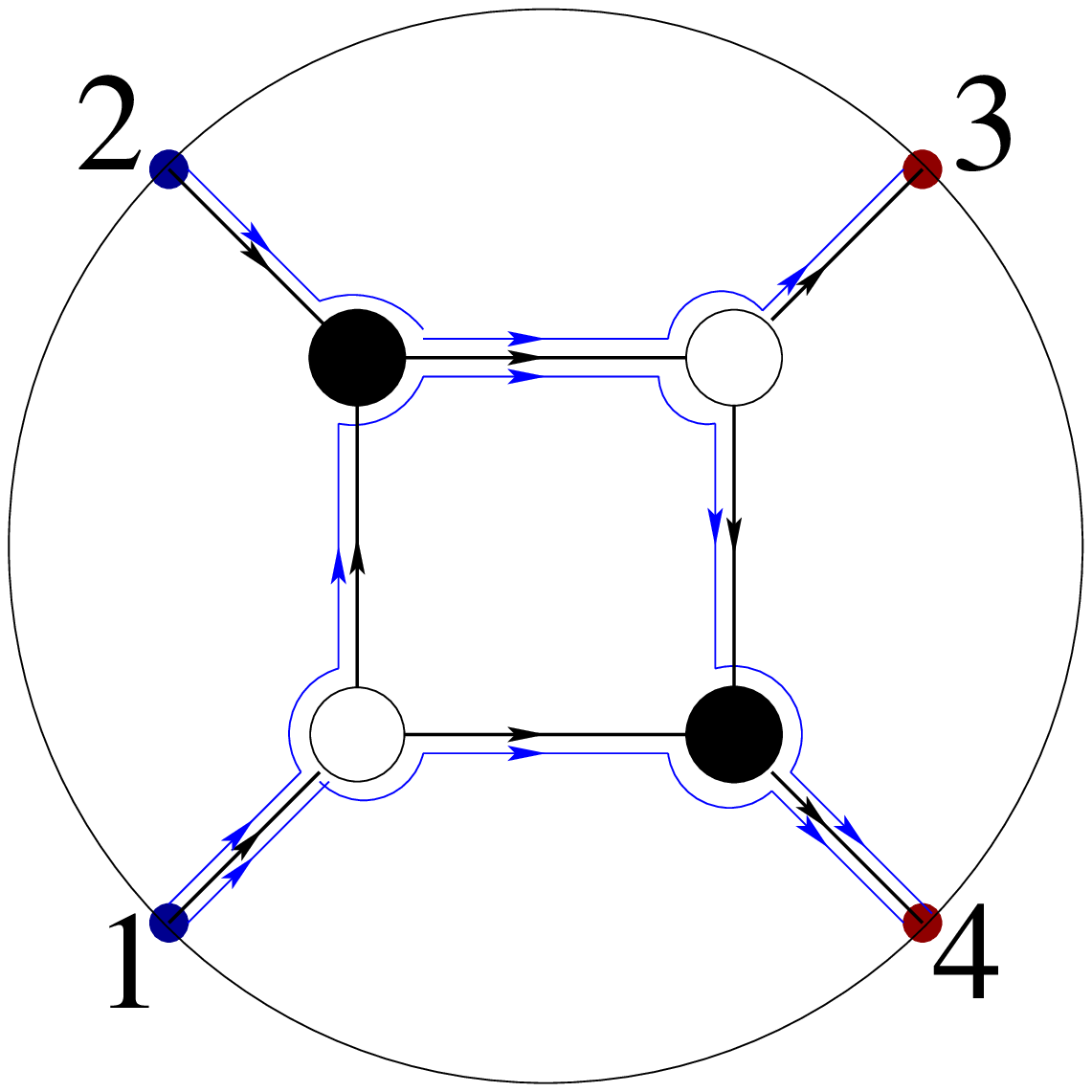}}}
 \;\Longrightarrow\;
  \raisebox{-1.4cm}{\scalebox{.32}{\includegraphics{4ptSing3a.eps}}}
   \;+\;
 \raisebox{-1.4cm}{\scalebox{.32}{\includegraphics{4ptSing3d.eps}}}
   \;+\;
  \raisebox{-1.4cm}{\scalebox{.32}{\includegraphics{4ptSing3c.eps}}}
 \end{split}
\end{equation*}

Furthermore, the diagram at the top and
the one at the bottom are one the mirror of the other, {\it i.e.}
they are {\it topologically equivalent}: they contain the same
complex factorisation channels, as it is made manifest from the
explicit channels above and as it can be deduced from the helicity
flows.

\begin{figure}[htbp]
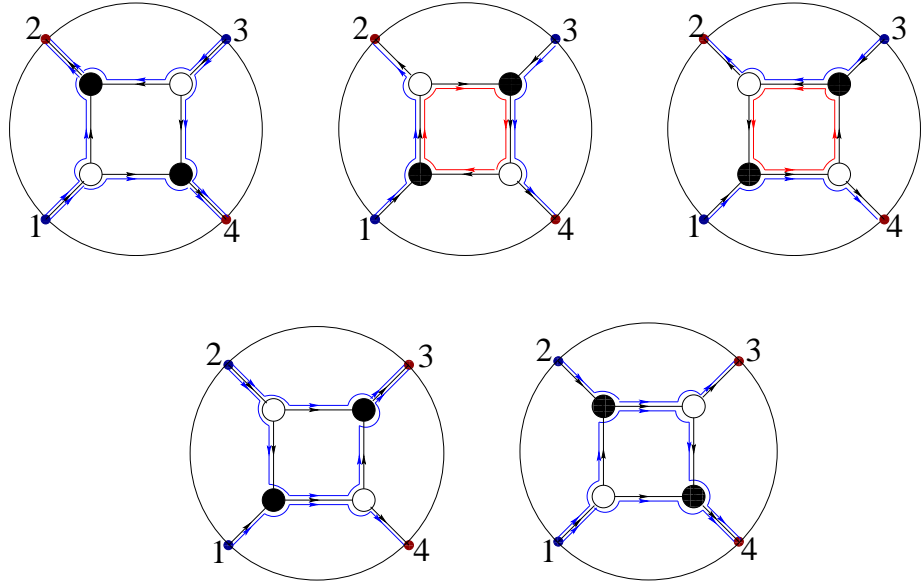

 \centering 
 \begin{equation*}
  \begin{split}
 &\raisebox{-.9cm}{\scalebox{.30}{\includegraphics{4ptHelFlow1.eps}}}
   \hspace{1cm}
  \raisebox{-.9cm}{\scalebox{.30}{\includegraphics{4ptHelFlow2.eps}}}
   \hspace{1cm}
  \raisebox{-.9cm}{\scalebox{.30}{\includegraphics{4ptHelFlow3.eps}}}
 \\
 &\phantom{\ldots}\\
 &\hspace{2.4cm}
  \raisebox{-.9cm}{\scalebox{.30}{\includegraphics{4ptHelFlow4.eps}}}
   \hspace{1cm}
  \raisebox{-.9cm}{\scalebox{.30}{\includegraphics{4ptHelFlow5.eps}}}
  \end{split}
 \end{equation*}
 \caption{Helicity flows. For each possible diagram with external
          helicity configuration $(-,+,-,+)$ (at the top) and 
          $(-,-,+,+)$ (at the bottom) all the helicity flows are 
          shown.}
 \label{fig:4ptHelFlows}
\end{figure}

The lesson we learn from this analysis is that the structure of
the helicity flows reveals the possible equivalence between two
on-shell (four-particle) diagrams related by an exchange of
holomorphicity of their three-particle building blocks. 
In particular, such an equivalence holds just for the helicity 
configuration $(-,-,+,+)$, for which the helicities of the internal 
states are fixed once for all and there is no helicity flow along
the internal lines.

\begin{figure}[htbp]
 \centering 
 \[
  \raisebox{-1.4cm}{\scalebox{.35}{\includegraphics{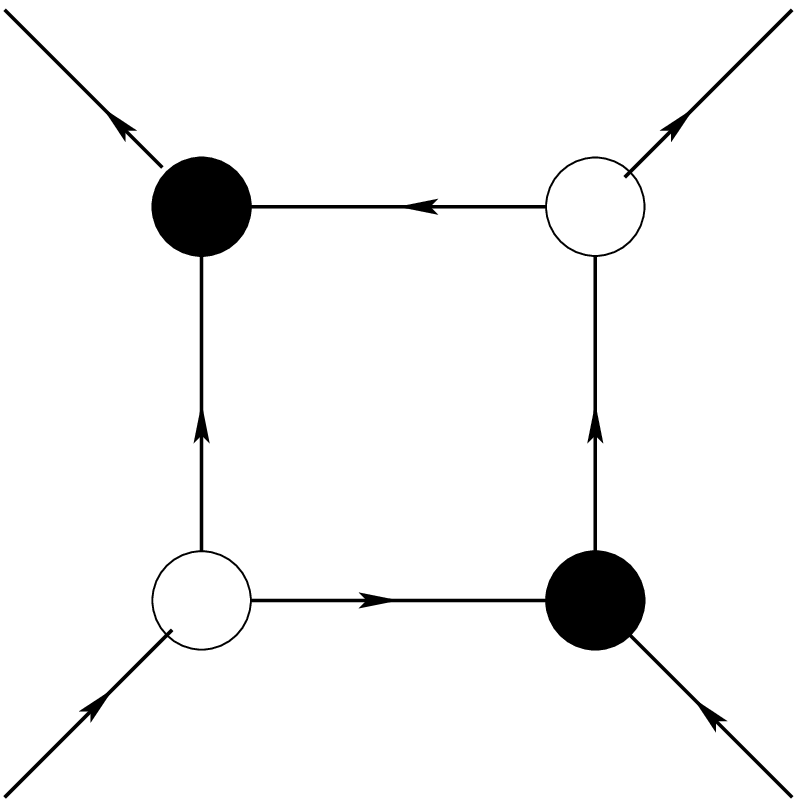}}}
   \qquad\Longleftrightarrow\qquad
  \raisebox{-1.4cm}{\scalebox{.35}{\includegraphics{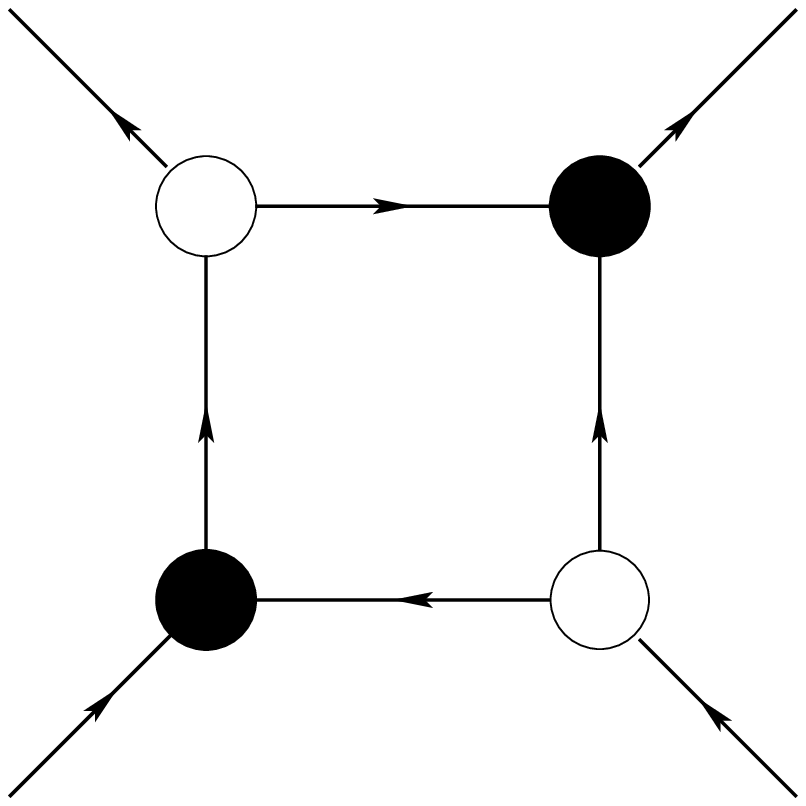}}}
 \]
 \caption{Square move operation. For non-maximally supersymmetric
          theories, it is possible to define an equivalence relation
          between two four-particle fully-localised diagrams if
          and only if the external multiplets with the same helicity
          are adjacent.}
 \label{fig:sqmove}
\end{figure}

Thus, together with the merger operation 
(Figure \ref{fig:4ptmerge}), the decorated diagrammatics enjoys
a further equivalence relation, which is the less/no-supersymmetric
counterpart of the square move in $\mathcal{N}\,=\,4$ SYM 
\cite{ArkaniHamed:2012nw} and involves the coherent states in a 
precise ordering, {\it i.e.} the equal-helicity coherent states have 
to be adjacent (Figure \ref{fig:sqmove}).

The limited validity of this equivalence relation is a further 
crucial difference with the maximally supersymmetric case.

In order to close this section, we need to discuss one further 
operation: the {\it bubble deletion}. 
Generally, the manipulation of a given
on-shell diagram via mergers and square moves can lead to a 
sub-diagram in which two three-particle amplitudes share two lines.
In the maximally supersymmetric case, this bubble could be replaced
by a single intermediate line (the bubble could be deleted) with
the price of generating a $d\log{\zeta}$ term: there is a change
of variables mapping the bubble into a factored-out $d\log{\zeta}$ 
form.
In the less/no-supersymmetric case, we need to distinguish two
cases, depending on whether or not there is an oriented helicity 
flow inside the bubble. In absence of an oriented helicity flow
inside the bubble, the two intermediate states have the same
helicity: there exists a change of variable mapping the bubble
into a single intermediate line with the same helicity arrow as
the states in the bubble with a $d\log{\zeta}$ factor:
\begin{equation}\eqlabel{BubDel}
 \raisebox{-.4cm}{\scalebox{.45}{\includegraphics{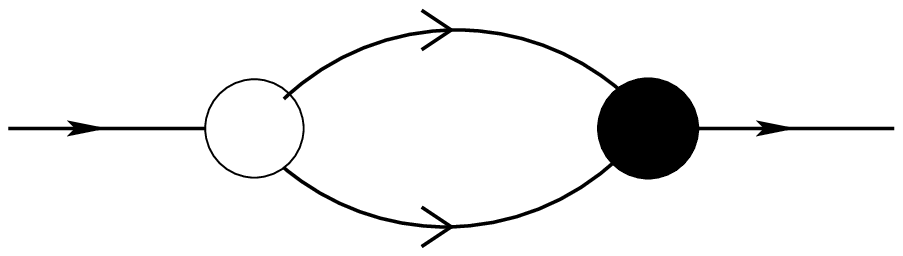}}}
  \qquad = \qquad
 \raisebox{-.0cm}{\scalebox{.45}{\includegraphics{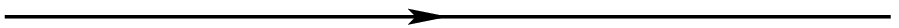}}}\quad d\log{\zeta}
\end{equation}
If there is instead an oriented helicity flow, this is no longer
possible. More precisely, an eventual change of variable which
would allow to map the bubble into a single intermediate line
factoring out the related degree of freedom is given by a 
transcendental equation. At most we can write the following
{\it schematic} relation
\begin{equation}\eqlabel{BubDelw}
 \begin{split}
 &\raisebox{-.4cm}{\scalebox{.45}{\includegraphics{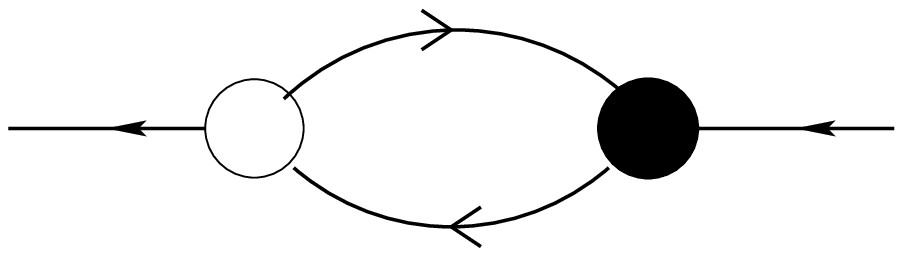}}}
 +
 \raisebox{-.4cm}{\scalebox{.45}{\includegraphics{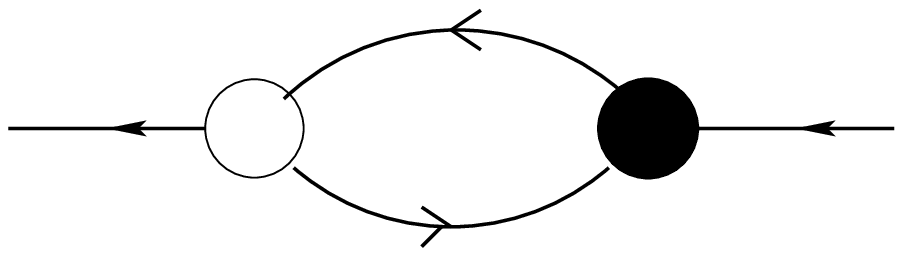}}}
  \quad = \\
 &\hspace{3cm}=\quad
  \raisebox{-.0cm}{\scalebox{.45}{\includegraphics{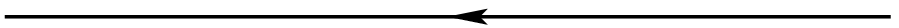}}}
  \,
  \frac{d\zeta}{\zeta}
  \left[
   \frac{1}{(1-\zeta)^{4-\mathcal{N}}}+
   \frac{(-\zeta)^{4-\mathcal{N}}}{(1-\zeta)^{4-\mathcal{N}}}
  \right]
 \end{split}
\end{equation}
Notice that for $\mathcal{N}\,=\,3$ the above bubble deletion
returns a $d\log{\zeta}$, as in $\mathcal{N}\,=\,4$ theory.
Thus, the presence of sub-diagrams
such as the bubbles on the l.h.s of \eqref{BubDelw} is a signal
of the presence of a different structure than the standard
$d\log{\zeta}$, which is typical of $\mathcal{N}\,=\,4$ loop
amplitudes. While the $d\log{\zeta}$-structure is a feature of 
UV-finite contributions to the amplitude, the presence of a
contribution with UV divergencies.

\begin{figure}[htbp]
 \centering 
 \[
  \raisebox{-1.1cm}{\scalebox{.45}{\includegraphics{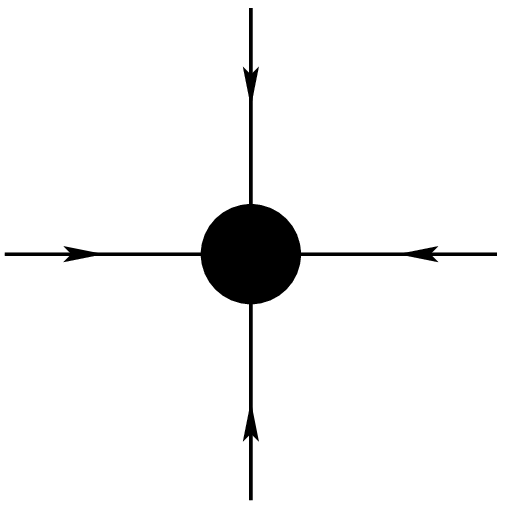}}}
   \qquad\Longleftrightarrow\qquad
  \raisebox{-2cm}{\scalebox{.45}{\includegraphics{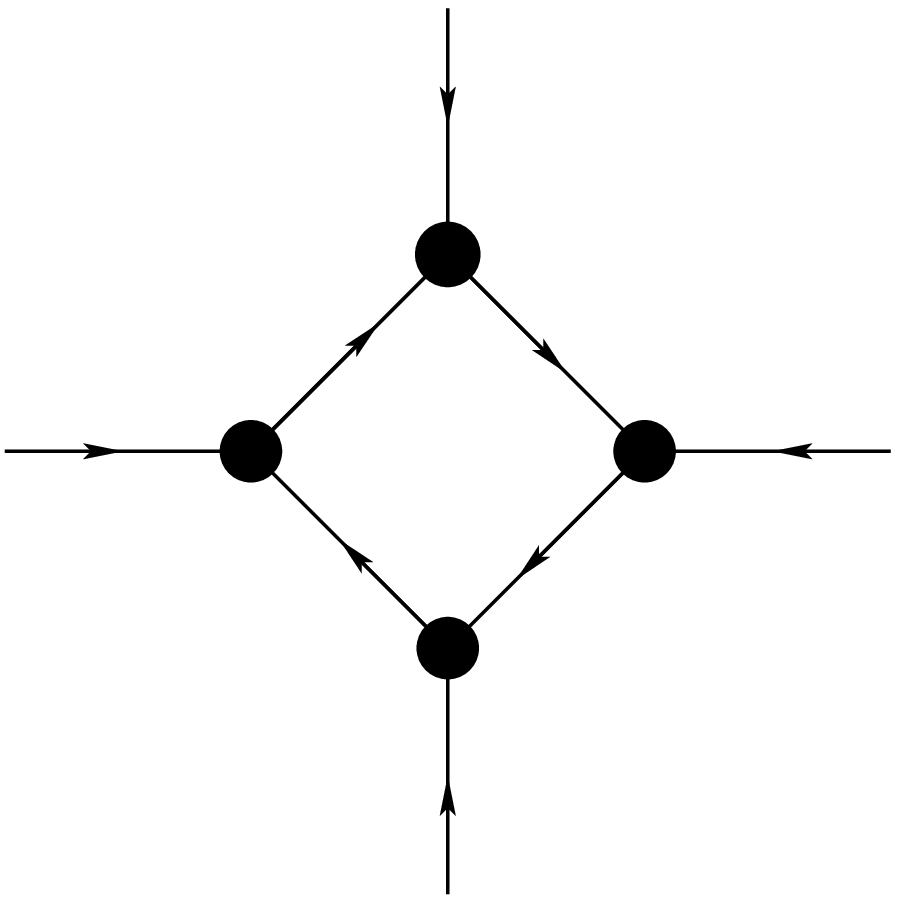}}}
  \:+\:
  \raisebox{-2cm}{\scalebox{.45}{\includegraphics{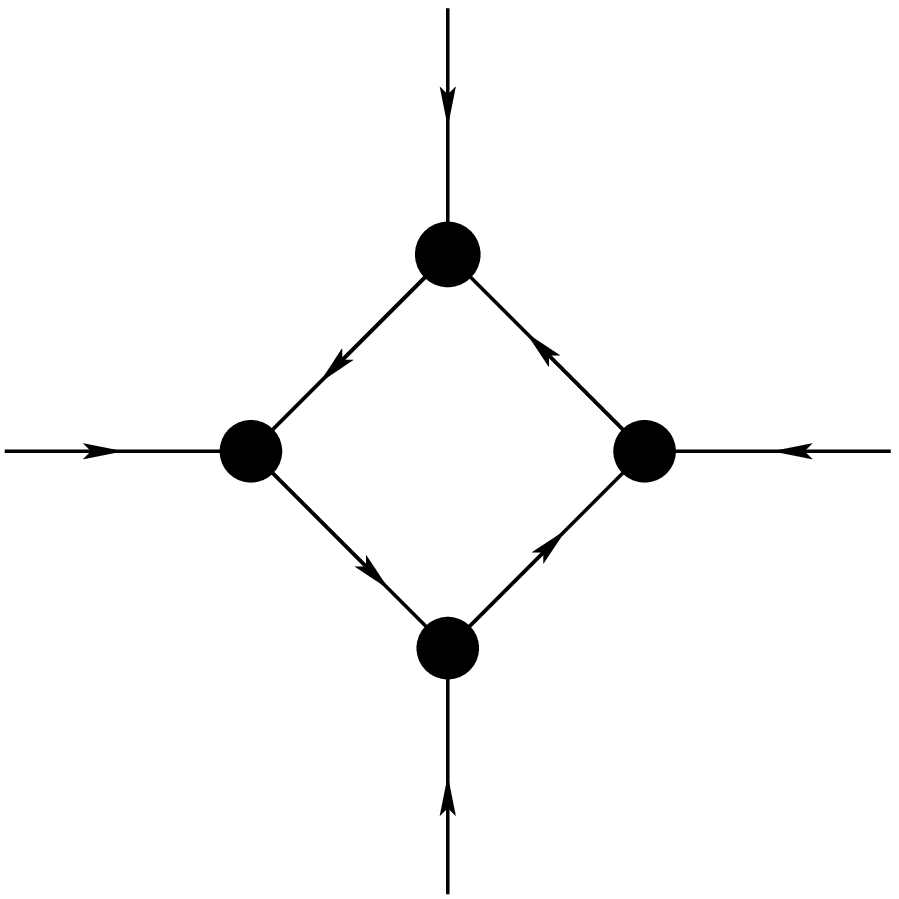}}}  
 \]
 \caption{Blow-up}
 \label{fig:blowup}
\end{figure}

For the sake of completeness, it is worth to mention that the 
decorated diagrams enjoy a further equivalence operation, named
{\it blow up}. Let us consider a black (white) node of valence $v$
whose helicity arrows are all incoming (outgoing). Then, 
because of the proportionality among all the spinors of a same 
type (as for mergers), the $v$-valence node can be open up to a
sum of two cyclic $v$-gons characterised by internal helicity loops,
as in Figure \ref{fig:blowup}.

Notice that if on one side the blow-up can be seen as a sequence
of merger operations, on the other side the fact that the helicity
arrows in the node have the same orientation forces it to open
up as a loop. A {\it tree-like} configuration, as it happens for
the mergers, would be allowed only if one allows also for the
all-equal helicity three-particle amplitudes \eqref{eq:3ptampl2}
(which are anyhow forbidden for $\mathcal{N}\,\neq\,0$).

Our fundamental and physically meaningful objects are the
three-particle amplitudes, which are diagrammatically pictured
as nodes of valence $3$. All the physically non-trivial diagrammatic
operations involve those nodes (recall that any node with valence
higher than $3$ can be recast into combination of nodes of valence
three through mergers and blow-ups). However, one can also perform
trivial operations via nodes of valence $2$: given an edge one can
always insert a valence-$2$ node in the middle of it, and similarly
given the presence of a valence-$2$ note on an edge, it can be
removed by gluing the two edges. The only condition in these trivial
operations is that there should be a well defined helicity flow,
{\it i.e.} the edges in a valence-$2$ node, irrespectively of its 
colour, should have one incoming helicity arrow and one outgoing,
so that it can be inserted on a given edge respecting its helicity
direction and similarly, once the node gets removed, the two
edges can be glued into a single one which respect the helicity
direction of the original ones:
\begin{equation}\eqlabel{eq:2vnode}
 \raisebox{-.1cm}{\scalebox{.60}{\includegraphics{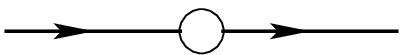}}}
 \qquad\Longleftrightarrow\qquad
 \raisebox{-.1cm}{\scalebox{.60}{\includegraphics{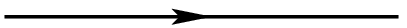}}}
 \qquad\Longleftrightarrow\qquad
 \raisebox{-.1cm}{\scalebox{.60}{\includegraphics{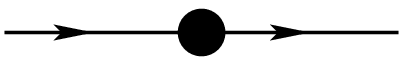}}}
\end{equation}

\subsection{Helicity rules}
\label{subsec:HelRul}

Summarising, the decoration of the on-shell diagrams for keeping
track of the helicities of the coherent states allows to
diagrammatically extract several features of the processes we
are dealing with and, therefore, of a theory. In particular
\begin{itemize}
 \item an oriented helicity flow between two adjacent external
       states encodes the existence of complex factorisations. 
       If such a flow goes through the interior of the diagram,
       as in the last two diagrams in Figure \ref{fig:4ptHelFlows},
       just one of the two possible complex factorisations in a 
       given channel is possible -- in the specific example just
       mentioned, the helicity flow $1\,\longrightarrow\,4$
       going through the interior in the fourth diagram in Figure 
       \ref{fig:4ptHelFlows}, as well as the analogous helicity flow 
       $2\,\longrightarrow\,3$ in the last one both encode the
       existence of the complex factorisation 
       $[4,1]\,\longrightarrow\,0$ 
       ($\langle2,3\rangle\,\longrightarrow\,0$);
 \item equivalence relations hold when the (sub)-diagram shows
       helicity flows between external states only with one of
       them going through the interior of the diagram; they do not
       hold if the helicity flows are just external or if there
       is an internal helicity loop;
 \item the helicity flows between external states which do not
       go through the interior of the diagram are preserved by
       equivalence relations;
 \item the presence of an oriented helicity loop in the intermediate
       lines encodes the presence of a further (high order) 
       singularity;
 \item the presence of an oriented helicity loop in an on-shell 
       bubble encodes a different functional structure than
       just $d\log{\zeta}$, identifies the presence of
       UV-divergent contribution to the integrated amplitude.
\end{itemize}

\subsection{BCFW bridges and M{\"o}bius transformations}
\label{subsec:BCFWbM}

Let us now explore more complicated on-shell diagrams, starting
with the on-shell boxes on which we apply a BCFW bridge. As before,
we are going to consider the external helicity configurations 
$(-,+,-,+)$ and $(-,-,+,+)$. Applying a BCFW bridge returns a higher 
degree differential form (in this concrete case a one-form) for 
still four-particle object.

\begin{figure}[htbp]
 \centering 
 \[
  \raisebox{-1.6cm}{\scalebox{.30}{\includegraphics{4ptHelFlow1.eps}}}
   \quad\Longrightarrow\quad
  \raisebox{-1.6cm}{\scalebox{.23}{\includegraphics{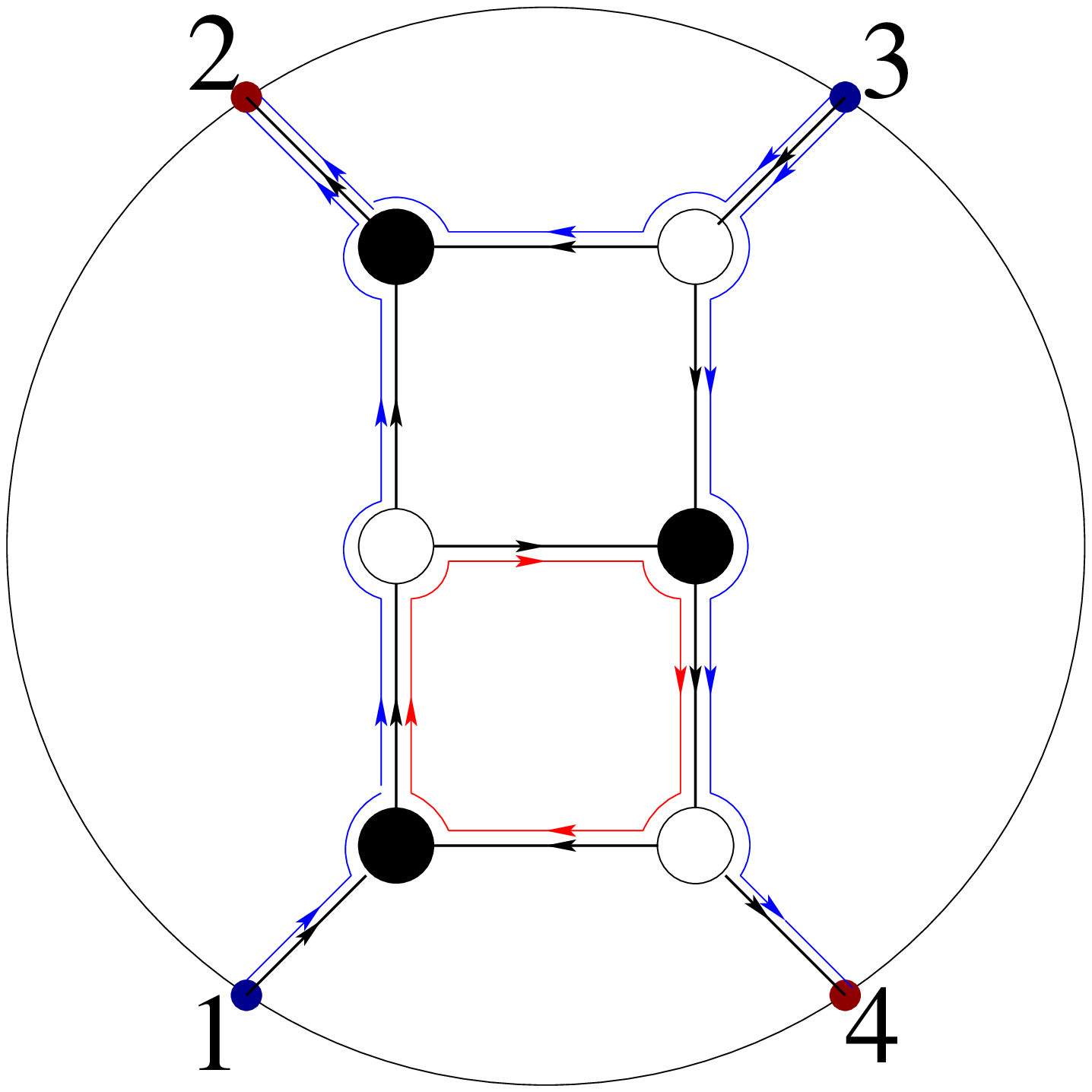}}}
   \quad+\quad
  \raisebox{-1.6cm}{\scalebox{.23}{\includegraphics{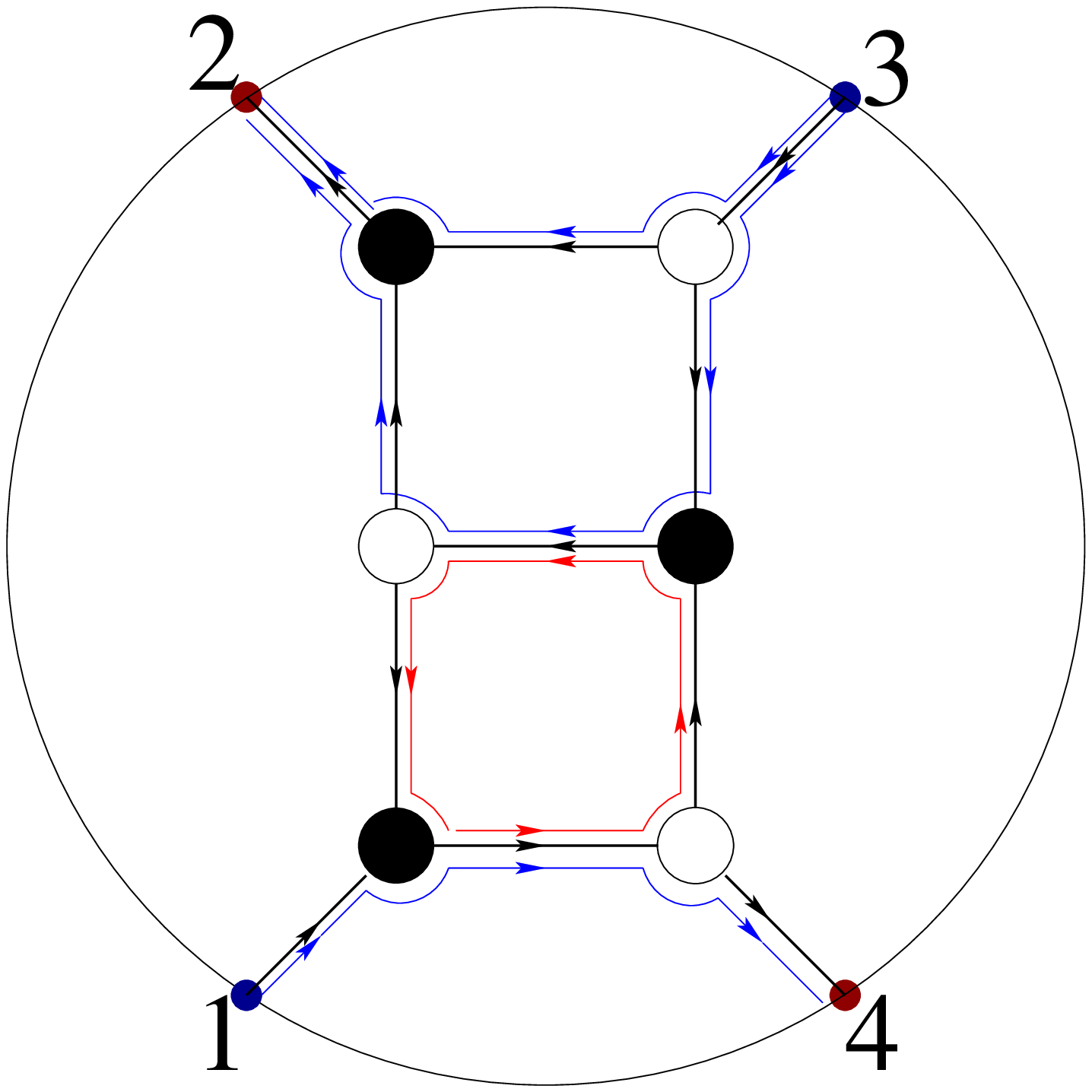}}}
 \]
 \caption{BCFW bridge on a four-particle leading singularity and helicity flow structure.}
 \label{fig:trplcut1}
\end{figure}

Let us begin with the on-shell box with helicity configuration 
$(-,+,-,+)$. As we saw before, it is possible to define two 
inequivalent classes of on-shell boxes: in one, the internal
coherent states are univocally fixed and there are just four
distinct helicity flows between consecutive states; in the other 
one, two sets of states are allowed to propagate (see Figure
\ref{fig:4ptOSdiags}) forming a helicity flow in the intermediate 
lines with different orientation for each set (see the first line in
Figure \ref{fig:4ptHelFlows}). The former, as we showed earlier, 
provides the on-shell representation for the tree-level amplitudes
\begin{equation}\eqlabel{eq:4ptree}
 \raisebox{-1.5cm}{\scalebox{.30}{\includegraphics{4ptHelFlow1.eps}}}
 \hspace{-.1cm}
 \:=\:
  \delta^{\mbox{\tiny $(2\times2)$}}
  \left(
   \sum_{i=1}^4\lambda^{\mbox{\tiny $(i)$}}
   \tilde{\lambda}^{\mbox{\tiny $(i)$}}
  \right)
  \delta^{\mbox{\tiny $(2\times\mathcal{N})$}}
  \left(
   \sum_{i=1}^4\lambda^{\mbox{\tiny $(i)$}}
   \tilde{\eta}^{\mbox{\tiny $(i)$}}
  \right)
  \frac{\langle1,3\rangle^{4-\mathcal{N}}}{
   \langle1,2\rangle\langle2,3\rangle\langle3,4\rangle
   \langle4,1\rangle}
\end{equation}
Let us apply a BCFW bridge on its external lines labelled
by $1$ and $2$. In particular, for our discussion we choose the 
on-shell box which actually represents the full tree-level 
four-particle amplitude 
$\mathcal{M}_4^{\mbox{\tiny tree}}(1^{-},2^{+},3^{-}, 4^{+})$. 
Let us choose the BCFW bridge which associates
the holomorphic three-particle amplitude to the state $1$ while
the anti-holomorphic one to the state $2$, as depicted in Figure
\ref{fig:trplcut1}. Summing over all the allowed states in the
intermediate lines, the resulting object can be expressed as the
sum of two on-shell diagrams (r.h.s. of Figure \ref{fig:trplcut1}).
Let us focus on the first of these two diagrams -- for the present
discussion the differences between these two diagrams are not
relevant. It is easy to see that the upper box has helicity flows
just on its external states, as the original on-shell box, while
the box at the bottom shows a clockwise helicity flows along just
its intermediate lines. This implies that {\it no equivalence
relation holds}.
Generating, as we did, via a BCFW bridge on the states $1$
and $2$, the explicit form is given as
\begin{equation}\eqlabel{eq:3cut1}
 \raisebox{-1.6cm}{\scalebox{.23}{\includegraphics{TripleCut1b.eps}}}
 \:=\:
  \mathcal{M}_4^{\mbox{\tiny tree}}
  \,\frac{dz}{z}
 \frac{\left(1+\frac{\langle3,4\rangle}{\langle3,1\rangle}z\right)^{4-\mathcal{N}}}{1+\frac{\langle4,2\rangle}{\langle1,2\rangle}z},
\end{equation}
where $\mathcal{M}_4^{\mbox{\tiny tree}}$ indicates the r.h.s. of
\eqref{eq:4ptree}.

The very same diagram can be also seen as a BCFW bridge
applied on the states labelled by $3$ and $4$ of an on-shell
diagram with an internal clockwise helicity flow:
\begin{equation}\eqlabel{eq:3cut2}
 \raisebox{-1.6cm}{\scalebox{.23}{\includegraphics{TripleCut1b.eps}}}
 \:=\:
  \mathcal{M}_4^{\mbox{\tiny tree}}
  \left(-\frac{t}{u}\right)^{4-\mathcal{N}}
  \frac{dz'}{z'}
 \frac{1}{
  \left(1+\frac{\langle3,4\rangle}{\langle2,4\rangle}z'\right)^{
   4-\mathcal{N}}
  \left(1+\frac{\langle1,3\rangle}{\langle1,2\rangle}z'\right)}.
\end{equation}
These two expressions are related by a M{\"o}bius transformation
\begin{equation}\eqlabel{eq:MobT1}
 \frac{\langle1,3\rangle}{\langle1,2\rangle}z'\:=\:
  -\frac{1+\frac{\langle4,2\rangle}{\langle1,2\rangle}z}{1+\frac{\langle3,4\rangle}{\langle3,1\rangle}z},
\end{equation}
which actually can be written as just an inversion in the following
variables
\begin{equation}\eqlabel{eq:MobT2}
 \zeta'\:=\:-\frac{u}{s}\frac{1}{\zeta},
 \qquad
 \frac{\langle4,2\rangle}{\langle1,2\rangle}z\:=\:
 \frac{\zeta}{1-\zeta},\quad
 \frac{\langle1,3\rangle}{\langle1,2\rangle}z'\:=\:
 \frac{\zeta'}{1-\zeta'},
\end{equation}
where actually all those change of variables are M{\"o}bius 
transformations. 
Such variables are useful to identify the structure of the
on-shell diagram being the ones which are returned by a bubble
deletion and which makes the $d\log{\zeta}$-structure (and,
consequently, a departure from it) manifest.

The on-shell form above has four special points: 
$z\,=\,0,\,\infty,\,-\langle1,2\rangle/\langle4,2\rangle,\,
      -\langle3,1\rangle/\langle3,4\rangle$
(or, in the $\zeta$-parametrisation, 
$\zeta\,=\,0,\,1,\,\infty,\,-u/t$, with $0$ being
a fixed point for the transformation $z\,\longrightarrow\,\zeta$).
Its integration over a circle $\gamma_{\mbox{\tiny $0$}}$ around 
the point $z\,=\,0$ or a circle $\gamma_{\mbox{\tiny P}}$ around
the simple pole $z\,=\,z_{\mbox{\tiny $P$}}\,\equiv\,-
\langle1,2\rangle/\langle4,2\rangle$, returns two leading 
singularities. More precisely, the integration around 
$\gamma_{\mbox{\tiny $0$}}$ provides the leading singularity
corresponding to the tree amplitude itself, while the one around
$\gamma_{\mbox{\tiny P}}$ return the $s$-channel contribution
to the second leading singularity.

Some comments are now in order. The on-shell diagram we are 
considering, under the parametrisation \eqref{eq:3cut1}, can be
equivalently look at as a {\it wrong} BCFW deformation of the
four-particle amplitude a tree level or as a contribution to the
triple cut of the one-loop amplitude, with the coefficient of the
scalar triangle in a Passarino-Veltman expansion related to the
non-vanishing residue at infinity 
\cite{ArkaniHamed:2008gz, Benincasa:2013faa}. Very interestingly,
the M{\"o}bius transformation \eqref{eq:MobT1} mapping
\eqref{eq:3cut1} into \eqref{eq:3cut2}, maps the multiple pole
at infinity into a pole of the same order at finite location: the 
so-called boundary term for a {\it wrong} BCFW deformation can then
be obtained as a residue of a multiple pole in a BCFW deformation
applied on the residue of the simple pole at finite location
under the original BCFW deformation! More explicitly, given a 
BCFW representation with a boundary term, the latter can be obtained
from the residue of the pole at finite location by applying a 
BCFW deformation to it a reading off the residue of the multiple
pole. Thus, the boundary term can be thought as already encoded
into the term which can be computed recursively. At first sight
this seems to be possible just because the class of theories
we are studying always admits a BCFW representation. The exploration
of the utility of the on-shell diagrammatics and M{\"o}bius 
transformations to understand the boundary terms on general grounds
goes beyond the aim of the present paper and we leave it for future
work. In the present context, it provides a clearer picture.

Let us briefly turn to the second diagram on the right-hand-side
of Figure \ref{fig:trplcut1}. It is easy to see from the helicity
flows that the upper box allows for a square-move so that
the diagram simplifies:
\begin{equation}\eqlabel{eq:3cut3}
\raisebox{-1.6cm}{\scalebox{.23}{\includegraphics{TripleCut1c.eps}}}
 \;\Longleftrightarrow\;
\raisebox{-1.6cm}{\scalebox{.23}{\includegraphics{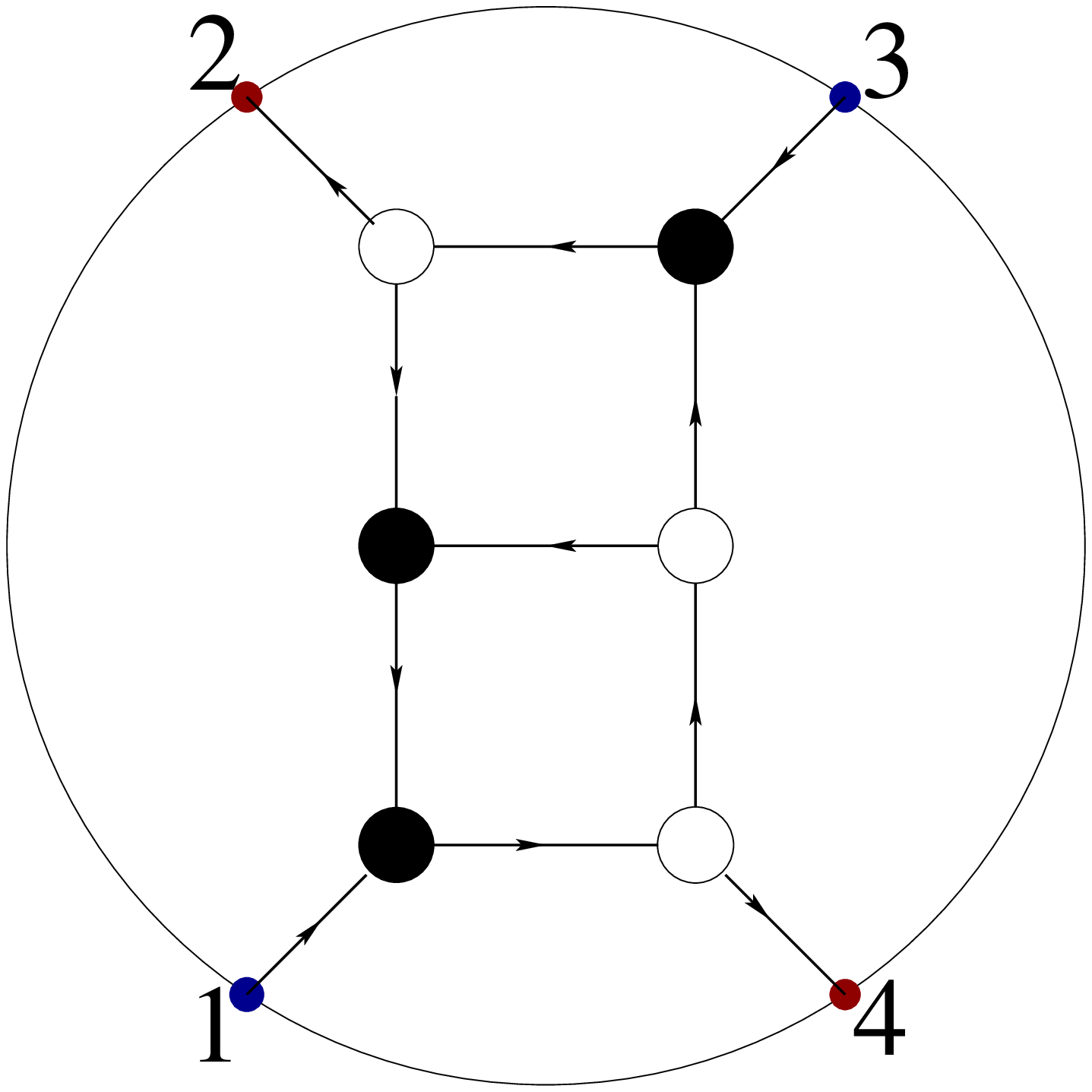}}}
 \;\Longleftrightarrow\;
\raisebox{-1.6cm}{\scalebox{.30}{\includegraphics{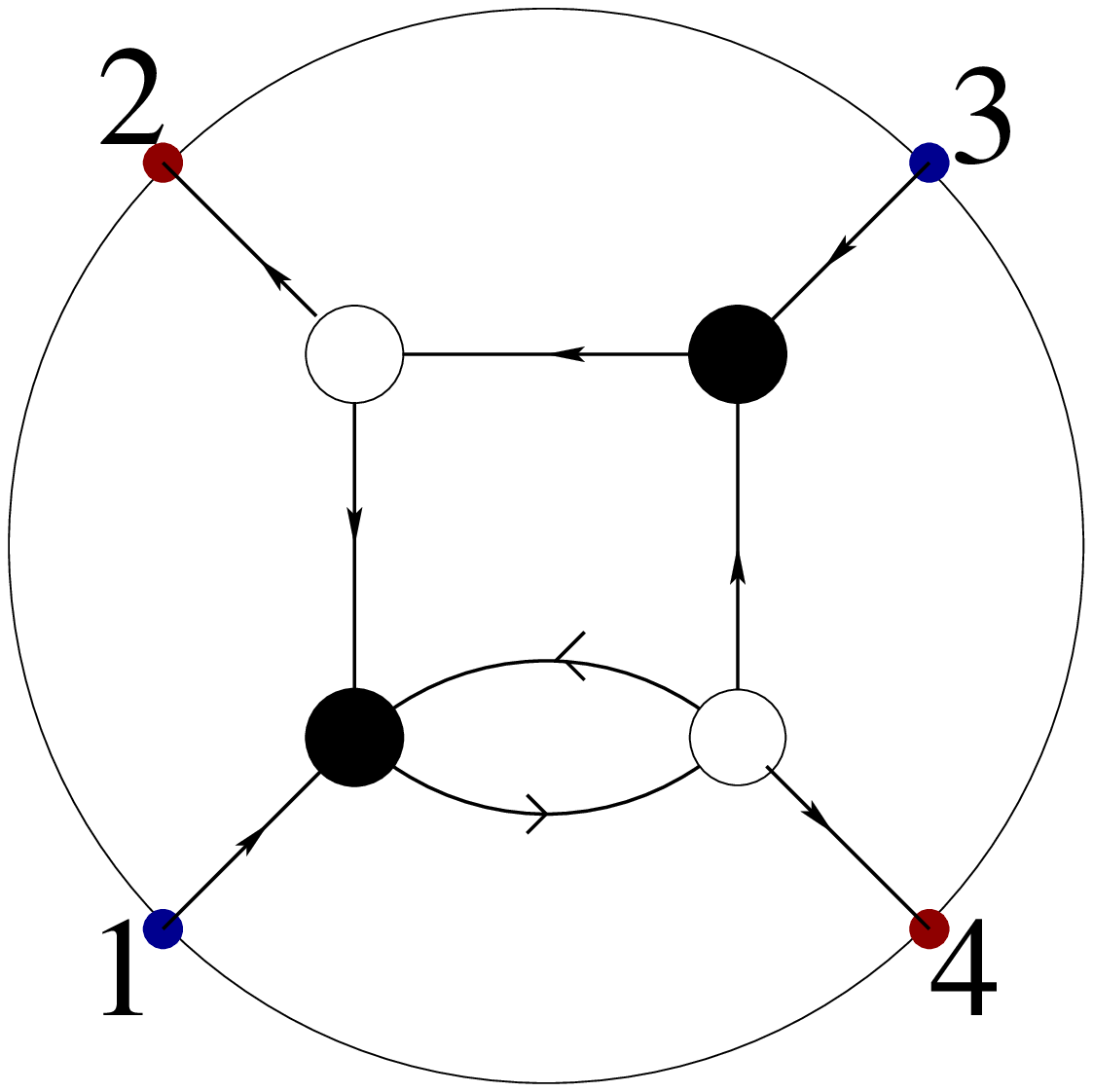}}}
\end{equation}
The diagram at the very right shows a bubble with an internal 
counter-clockwise helicity flow. In general, the presence of such
an oriented helicity bubble is a consequence of the existence of an
oriented helicity flow in the original diagram. As pointed out in
\eqref{BubDelw} in the previous section, one can actually replace
the bubble by an oriented line at the price of obtaining a 
differential which has a richer structure than the simple 
$d\log{\zeta}$. Specifically, one obtains:
\begin{equation}\eqlabel{eq:3cut4}
 \raisebox{-1.6cm}{\scalebox{.23}{\includegraphics{TripleCut1c.eps}}}
 \;\Longleftrightarrow\;
 \raisebox{-1.6cm}{\scalebox{.30}{\includegraphics{4ptHelFlow3.eps}}}
 \quad
 \frac{d\zeta}{\zeta}
 \sum_{k=0}^{4-\mathcal{N}}
 \begin{pmatrix}
  4-\mathcal{N}\\
  k
 \end{pmatrix}
 \frac{(-1)^k}{(1-\zeta)^k},
\end{equation}
where the right-hand-side has been written in a convenient way 
to highlight all the structures emerging from this bubble deletion.
It is easy to see that for $\mathcal{N}\,=\,3,\,4$ the differential
of a logarithm appears -- $d\log{[\zeta/(1-\zeta)]}$ and 
$d\log{\zeta}$ respectively --, while for $\mathcal{N}\,\le\,2$ 
new singularities appear. Namely, in the latter case, one can write 
the differential as
\begin{equation}\eqlabel{eq:3cut5}
 \frac{d\zeta}{\zeta}\mu(\zeta)\:=\:
 d\left[
  \log{\frac{\zeta}{1-\zeta}}+\frac{(-1)^{2-\mathcal{N}}}{1-\zeta}
  \sum_{k=0}^{2-\mathcal{N}}\frac{(-1)^k}{k+1}\frac{1}{(1-\zeta)^k}
 \right].
\end{equation}

As a final remark, notice that the on-shell diagram under analysis
shows the existence of a M{\"o}bius transformation mapping the 
singularities of the leading singularity for the helicity 
configuration $(-,-,+,+)$ into the singularities of (a contribution
to) a leading singularity for $(-,+,-,+)$, and such a transformation
has the same form as \eqref{eq:MobT1}. Thus, M{\"o}bius 
transformations can relate special points of amplitudes with
different helicity configurations.

\section{From decorated to un-decorated diagrams and back}
\label{sec:DecUndec}

The decorated on-shell diagrammatics as defined in the previous
sections is a bit redundant: the existence of equivalence relations,
such as mergers, square moves and blow-ups, imply the need of 
defining equivalence classes for the on-shell processes. In the
maximally supersymmetric case, this issue is solved through
the permutation group \cite{ArkaniHamed:2012nw}: an equivalence
class of on-shell diagrams is defined by a given permutation.

At first sight, the decorated on-shell diagrams might seem to loose
this nice structure because of the perfect orientation itself. 
Luckily, this is not the case and the equivalence classes can be
generically defined via permutations. More precisely, the 
equivalence classes are identified by permutations {\it and} 
helicity flows.

Given a decorated on-shell diagram, one can consider its 
counter-part without perfect orientation and read-off the related
permutation. Then, the equivalence class which the original
diagram with perfect orientation belongs to is defined as the
sub-set of diagrams in the permutation previously found which share
the same helicity flows (once sources and sinks are put back in).

Let us illustrate how the combination of permutation and
helicity flows works, let us discuss some example. In order to
be self-contained, let us briefly recall how permutation are
associated to un-oriented on-shell diagrams (for a more extensive
discussion see \cite{ArkaniHamed:2012nw}).

\subsection{Un-oriented on-shell diagrams and permutations}
\label{subsec:OSperm}

Given the fundamental three-particle diagrams whose external
states are labelled from $1$ to $3$ clockwise, it is assigned
a path directed from $i$ to $i+1$ for the white nodes, and from
$i$ to $i-1$ for the black ones:
\begin{equation}\eqlabel{Perm3pt}
 \begin{split}
  &\raisebox{-1.5cm}{\scalebox{.35}{\includegraphics{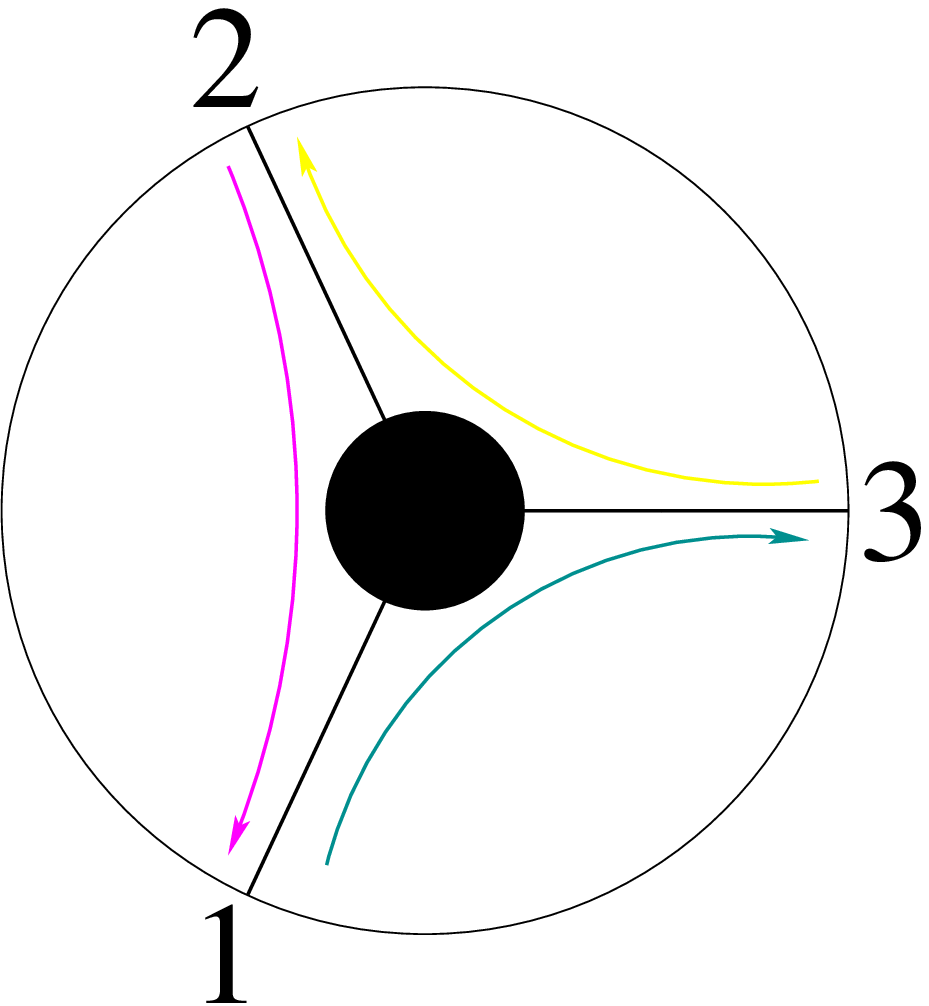}}}
   \quad
   \Longleftrightarrow
   \quad
   \begin{pmatrix}
    i \\
    \downarrow\\
    i-1
   \end{pmatrix}
   \quad
   \Longleftrightarrow
   \quad
   \begin{pmatrix}
    1          & 2          & 3\\
    \downarrow & \downarrow & \downarrow \\
    3          & 1          & 2
   \end{pmatrix}
   \\
  &\raisebox{-1.5cm}{\scalebox{.35}{\includegraphics{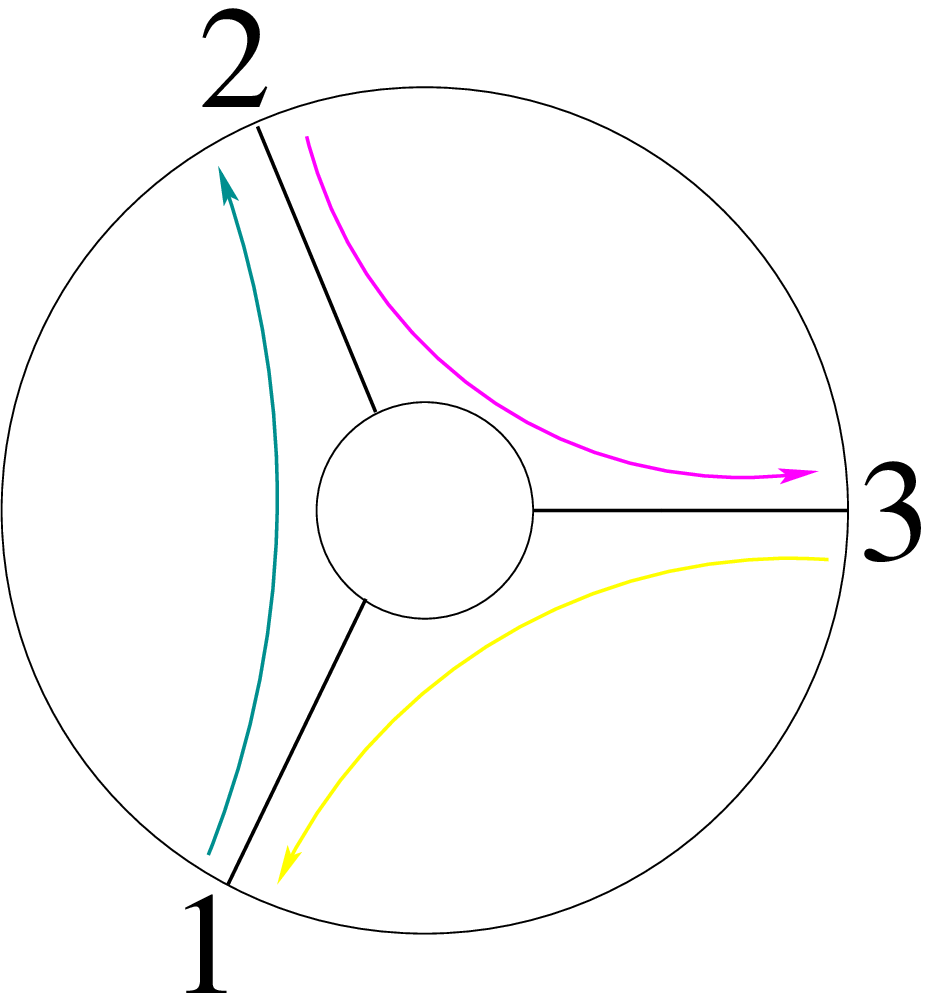}}}
   \quad
   \Longleftrightarrow
   \quad
   \begin{pmatrix}
    i \\
    \downarrow\\
    i+1
   \end{pmatrix}
   \quad
   \Longleftrightarrow
   \quad
   \begin{pmatrix}
    1          & 2          & 3\\
    \downarrow & \downarrow & \downarrow \\
    2          & 3          & 1
   \end{pmatrix}
 \end{split}
\end{equation}
In general, in more complex diagrams, following the paths defined
above, a {\it decorated} permutation can be associated to the 
canonical one, defined as the map
\begin{equation}\eqlabel{PermMap}
 \sigma:\;\{n\}\:\longrightarrow\:\quad\{2n\},
 \qquad
 \sigma(i)\:\in\:[i,\,i+n],
\end{equation}
where $\{n\}$ and $\{2n\}$ respectively indicates a sequence of
$n$ and $2n$ labels. The map has two fixed points $\sigma(i)\,=\,i$
and $\sigma(i)\,=\,i+n$, which can be easily displayed in the
identity diagram:
\begin{equation}\eqlabel{eq:PermId}
   \raisebox{-1.8cm}{\scalebox{.45}{\includegraphics{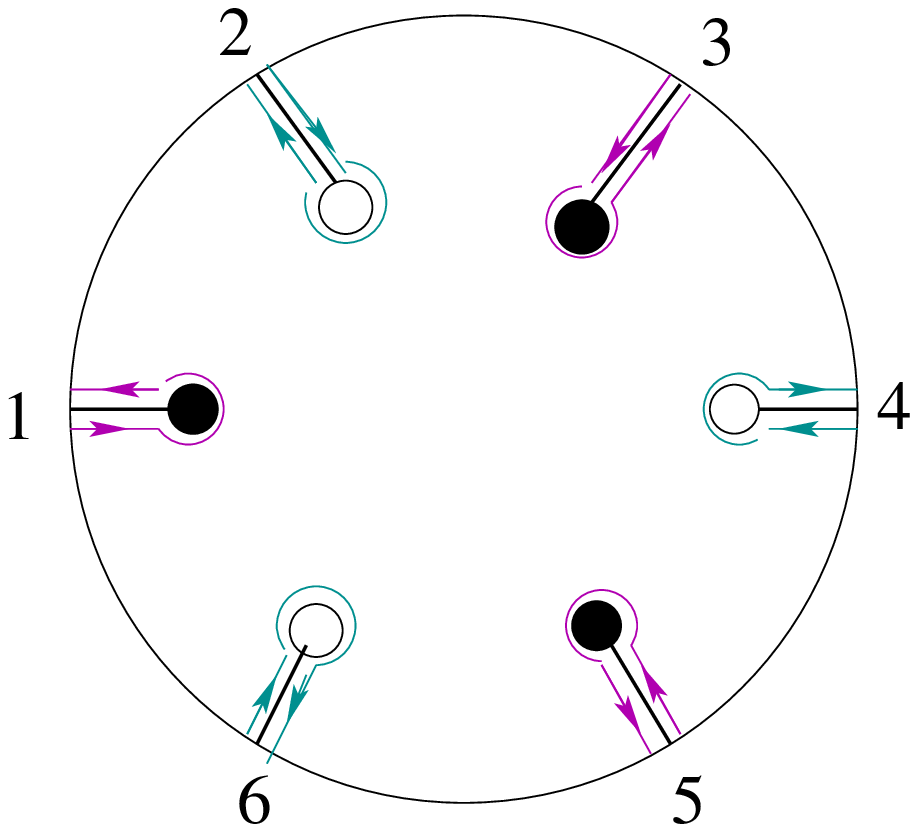}}} 
   \quad\Longleftrightarrow\quad
   \begin{pmatrix}
    1 & 2 & 3 & 4 & 5 & 6 \\
    \downarrow & \downarrow & \downarrow & \downarrow & \downarrow &
      \downarrow \\
    1 & 2 & 3 & 4 & 5 & 6
   \end{pmatrix}
   \quad\Longleftrightarrow\quad
   {\color{red} \{1, 8, 3, 10, 5, 12\}}
\end{equation}
where the set in red indicates the related decorated permutation.

Thus, according to such a prescription, the permutation related
to the following four- and six- particle on-shell processes are
\begin{equation}\eqlabel{eq:PermNpt}
 \begin{split}
   \raisebox{-1.6cm}{\scalebox{.30}{\includegraphics{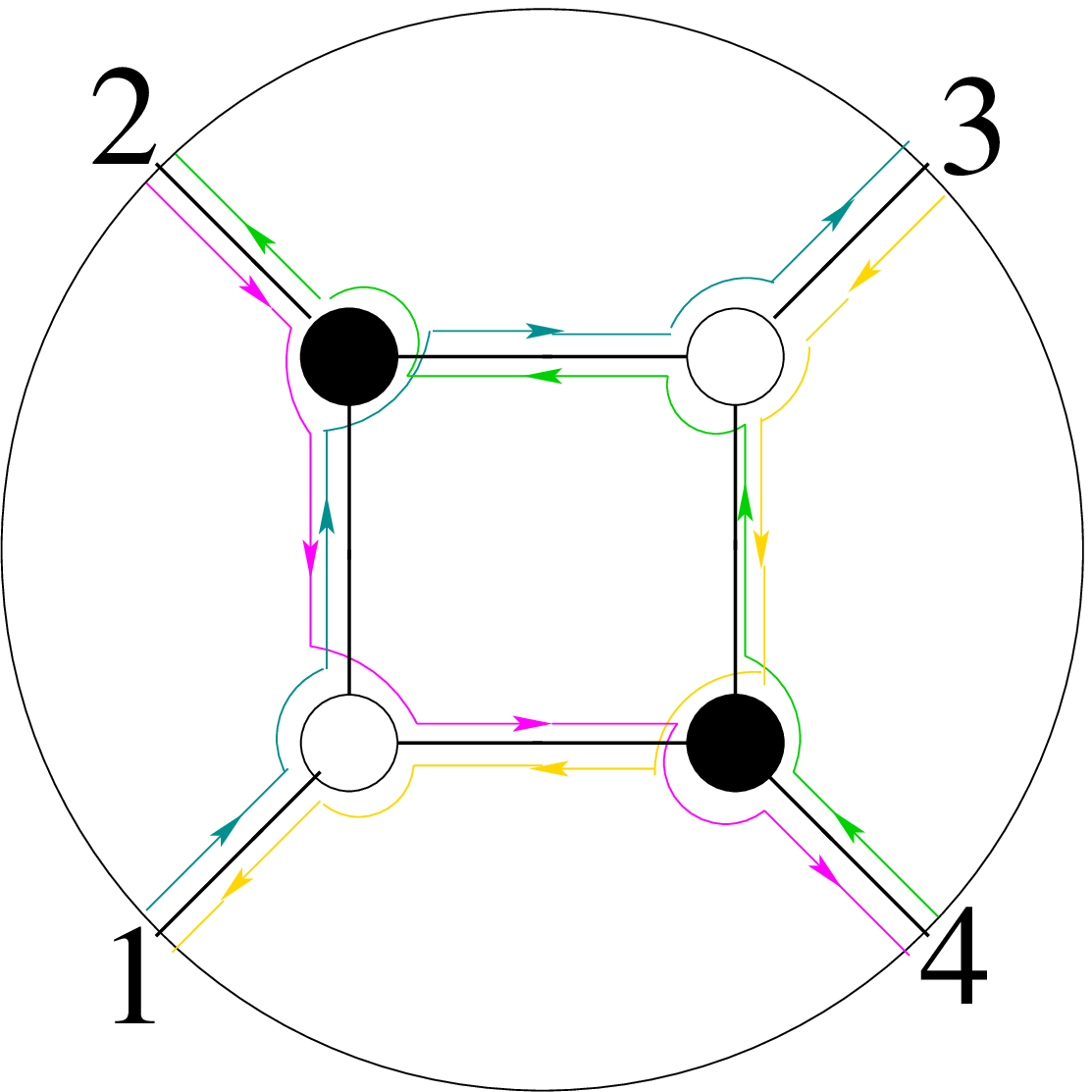}}}
   \quad
   \Longleftrightarrow
   \quad
  &\begin{pmatrix}
    1 & 2 & 3 & 4\\
    \downarrow & \downarrow & \downarrow & \downarrow \\
    3 & 4 & 1 & 2	
   \end{pmatrix}
   \quad
   \Longleftrightarrow
   \quad
   \raisebox{-1.6cm}{\scalebox{.30}{\includegraphics{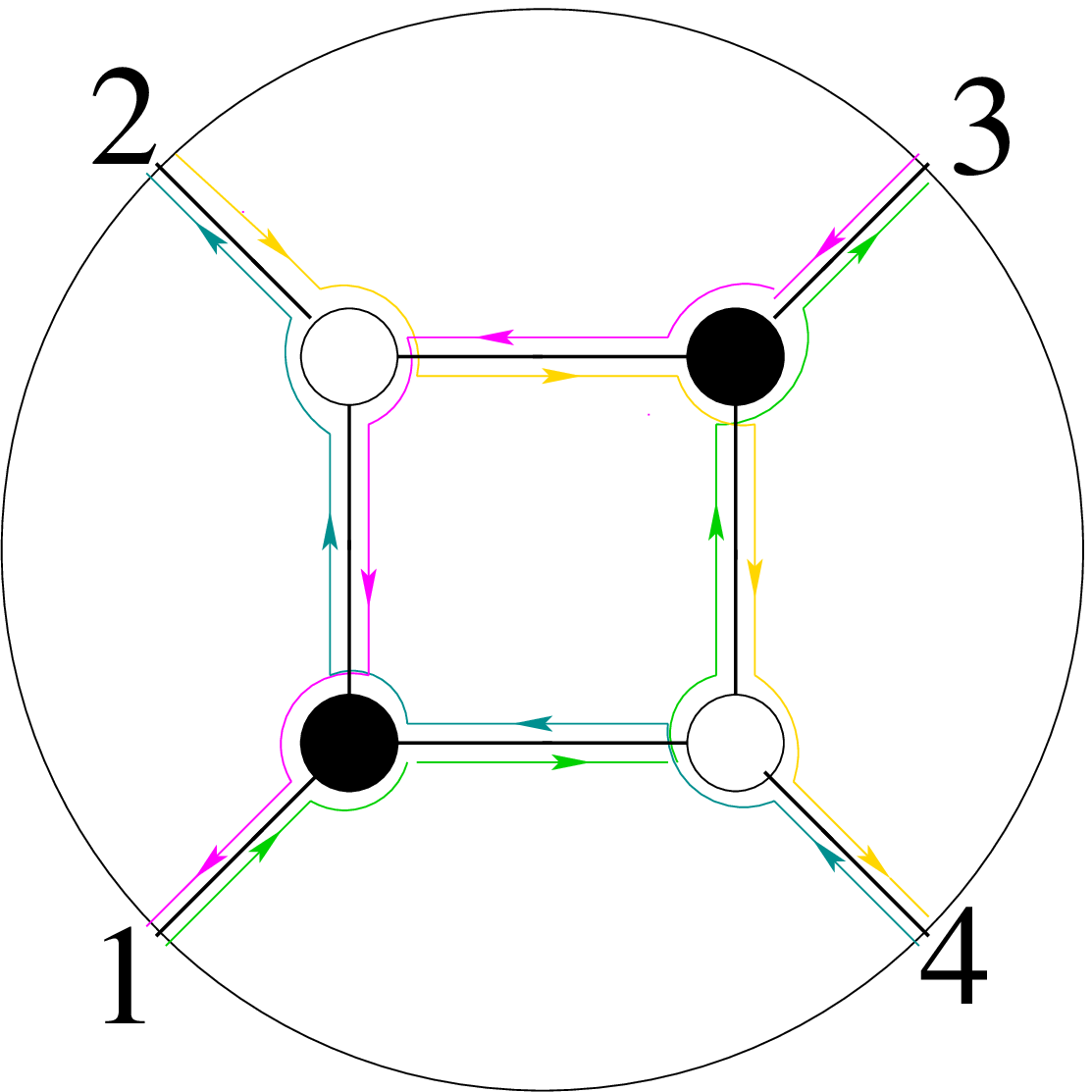}}}
 \end{split}
\end{equation}

\begin{figure}[htbp]
 \centering
 \[
\raisebox{-1.8cm}{\scalebox{.75}{\includegraphics{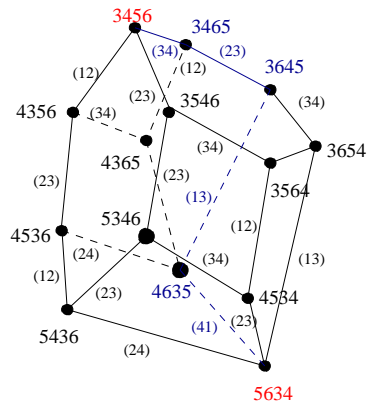}}} 
 \]
 \caption{Bridge polytope $Br_{2,4}$: 
          It is a characterised by vertices representing a 
          decorated permutation of $\{1,\,2,\,3,\,4\}$, while
          the edges represent a transposition mapping a 
          permutation into the other. The top decorated permutation
          $\{3,\,4,\,5,\,6\}$ represents the on-shell box, 
          while the bottom one $\{5,\,6,\,3,\,4\}$ represents the 
          identity. Any path along the edges between these
          two vertices represent a BCFW decomposition of the 
          on-shell box. In blue a possible path is highlighted.}
 \label{fig:bridgehedron}
\end{figure}

A particular type of permutation is the adjacent transposition.
Diagrammatically it is implemented by the BCFW bridges. Therefore,
performing a transposition is equivalent to adding a BCFW bridge.
Vice versa, given a (decorated) permutation $\sigma$, its 
decomposition into adjacent transpositions can be viewed 
as a particular path along the edges of a polytope, named
{\it bridge polytope} $Br_{k,n}$, whose vertices and edges are, 
respectively, (decorated) permutations and transpositions (BCFW 
bridges), leading to the identity \cite{Williams:2015twa}. The
bridge polytope is built as a recursive algorithm:
\begin{equation}\eqlabel{brdec}
 \sigma\:=\:(i,j)\,\circ\,\sigma'\;|\;i,\,j\in\,[1,\,n],\;
    \sigma(i)\,<\,\sigma(j),\;
    \forall\,k\,\in\,[i,\,j]\,:\,
     \sigma(k)\,=\,k\,\wedge\,
     \sigma(k)\,=\,k+n,
\end{equation}
until $\sigma'$ becomes the identity.
In Figure 
\ref{fig:bridgehedron}, the bridge polytope $Br_{2,4}$ is shown,
representing the BCFW decompositions of the (reduced) on-shell box
to the identity.

The blue path highlighted in Figure \ref{fig:bridgehedron} 
represents the following sequence of BCFW bridges:
\begin{equation*}
 \begin{array}{ccccc}
   \raisebox{-1.6cm}{\scalebox{.25}{\includegraphics{Perm4pt.eps}}}
   &
   \overset{(34)}{\longleftrightarrow}
   &
   \raisebox{-1.6cm}{\scalebox{.25}{\includegraphics{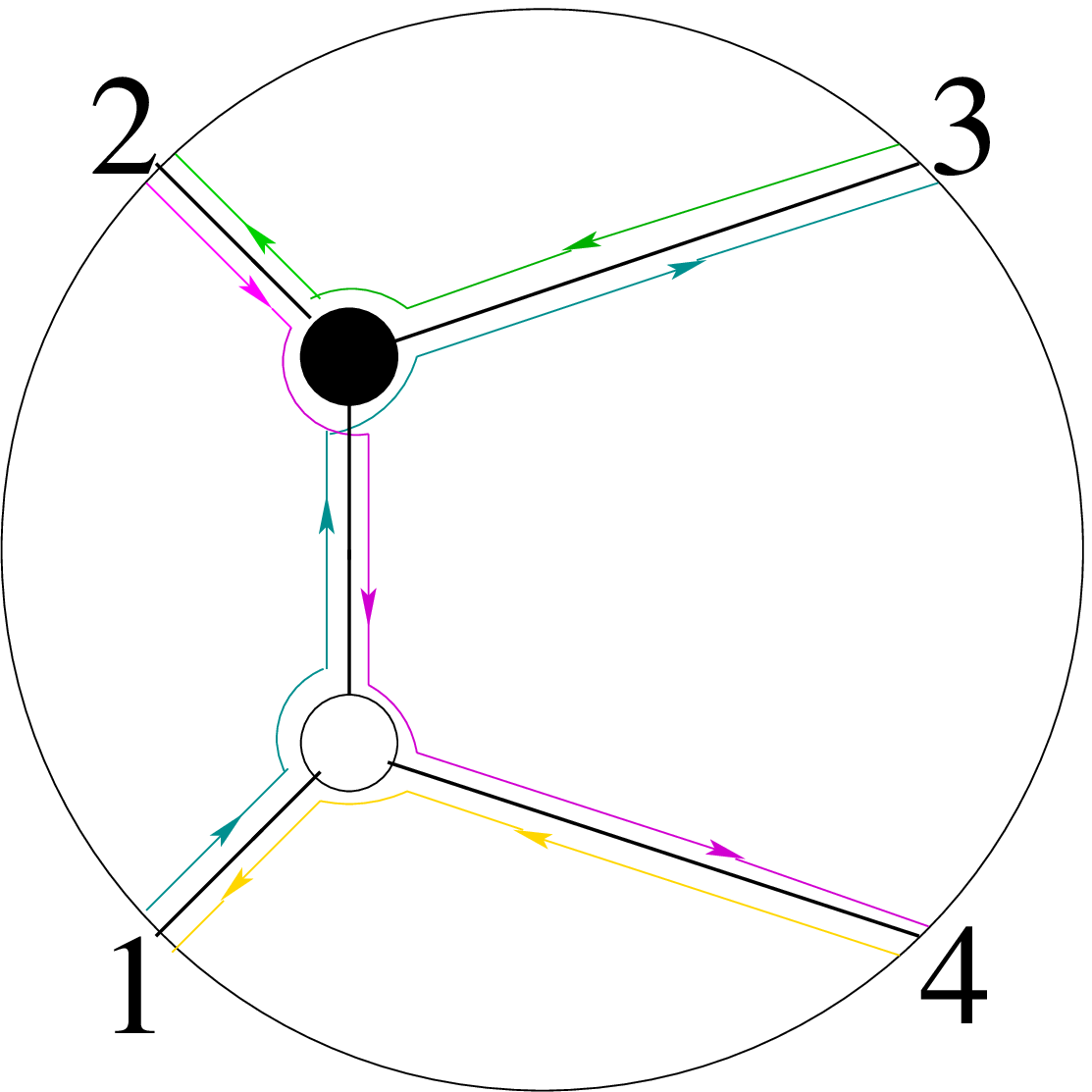}}}
   &
   \overset{\mathbb{I}}{\longleftrightarrow}
   &
   \raisebox{-1.6cm}{\scalebox{.25}{\includegraphics{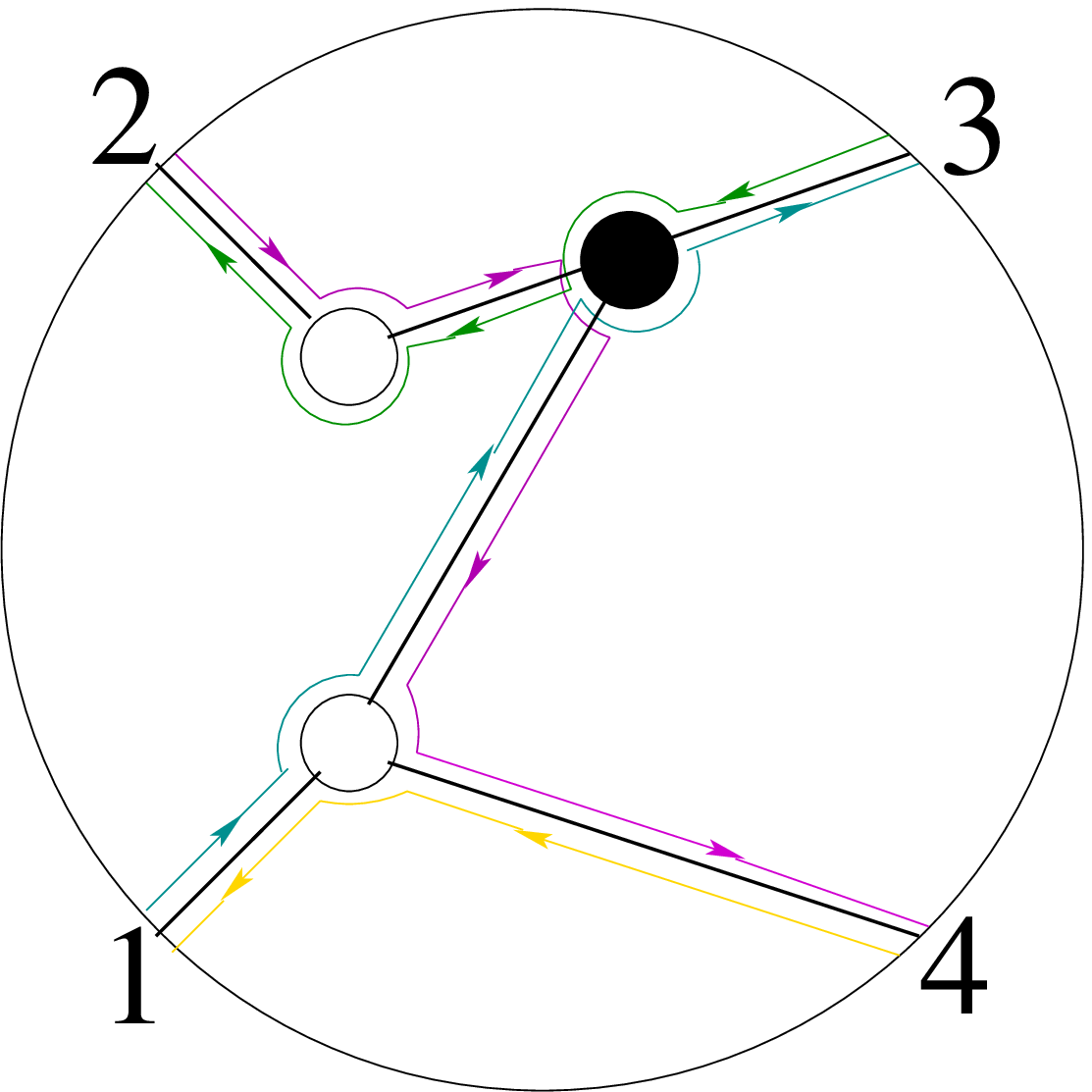}}}   
   \\
   {\color{red} \{3456\}} & {} & {\color{blue} \{3465\}} & {} &
   {\color{blue} \{3465\}}
   \\
   {} & {} & {} & {} & {}
   \\
   {}
   &{}
   &{}
   &{}
   &
   \left\updownarrow\right. {\mbox{\footnotesize $(23)$}}\\
   {} & {} & {} & {} & {}   
   \\
   \raisebox{-1.6cm}{\scalebox{.25}{\includegraphics{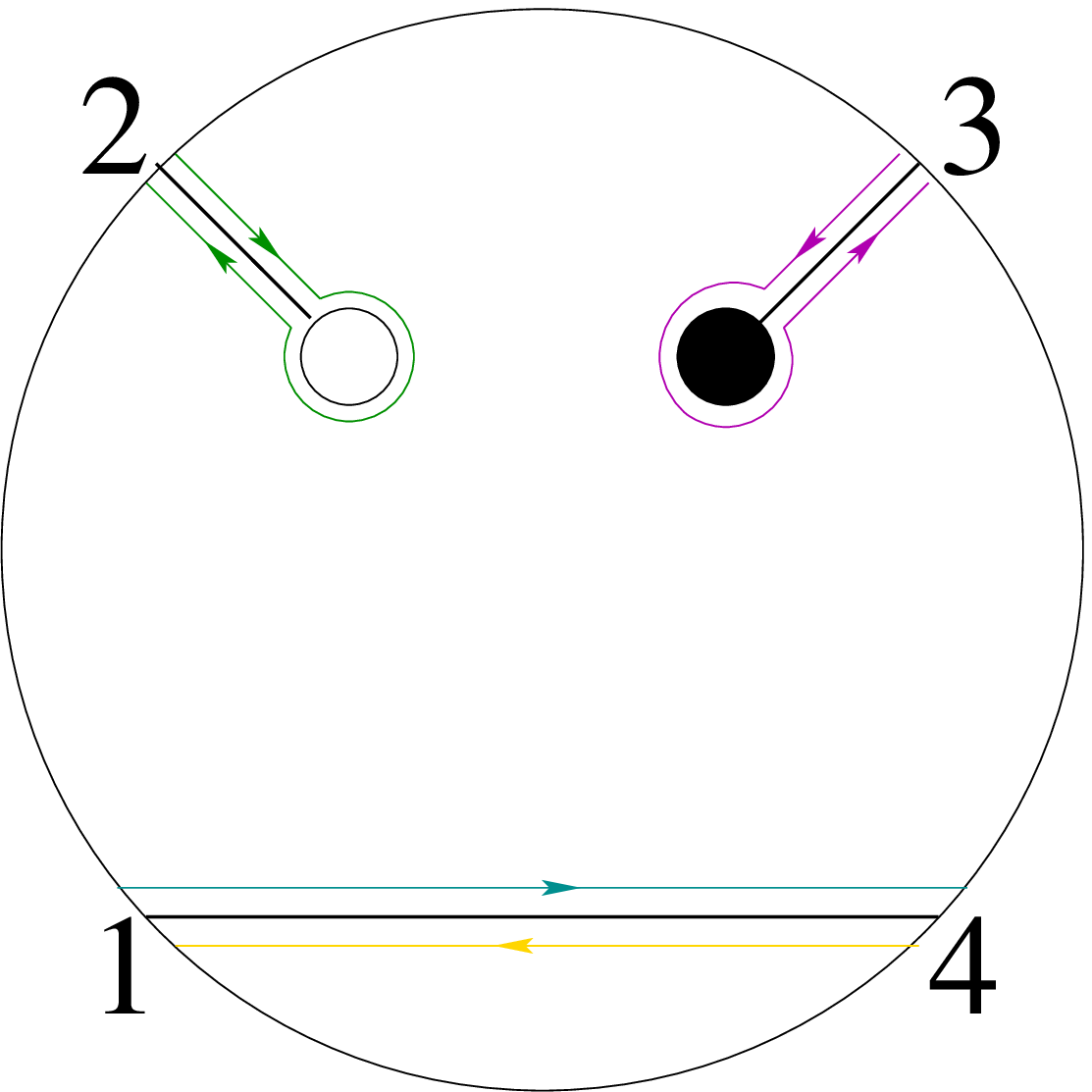}}} 
   &
   \overset{(13)}{\longleftrightarrow}
   &
   \raisebox{-1.6cm}{\scalebox{.25}{\includegraphics{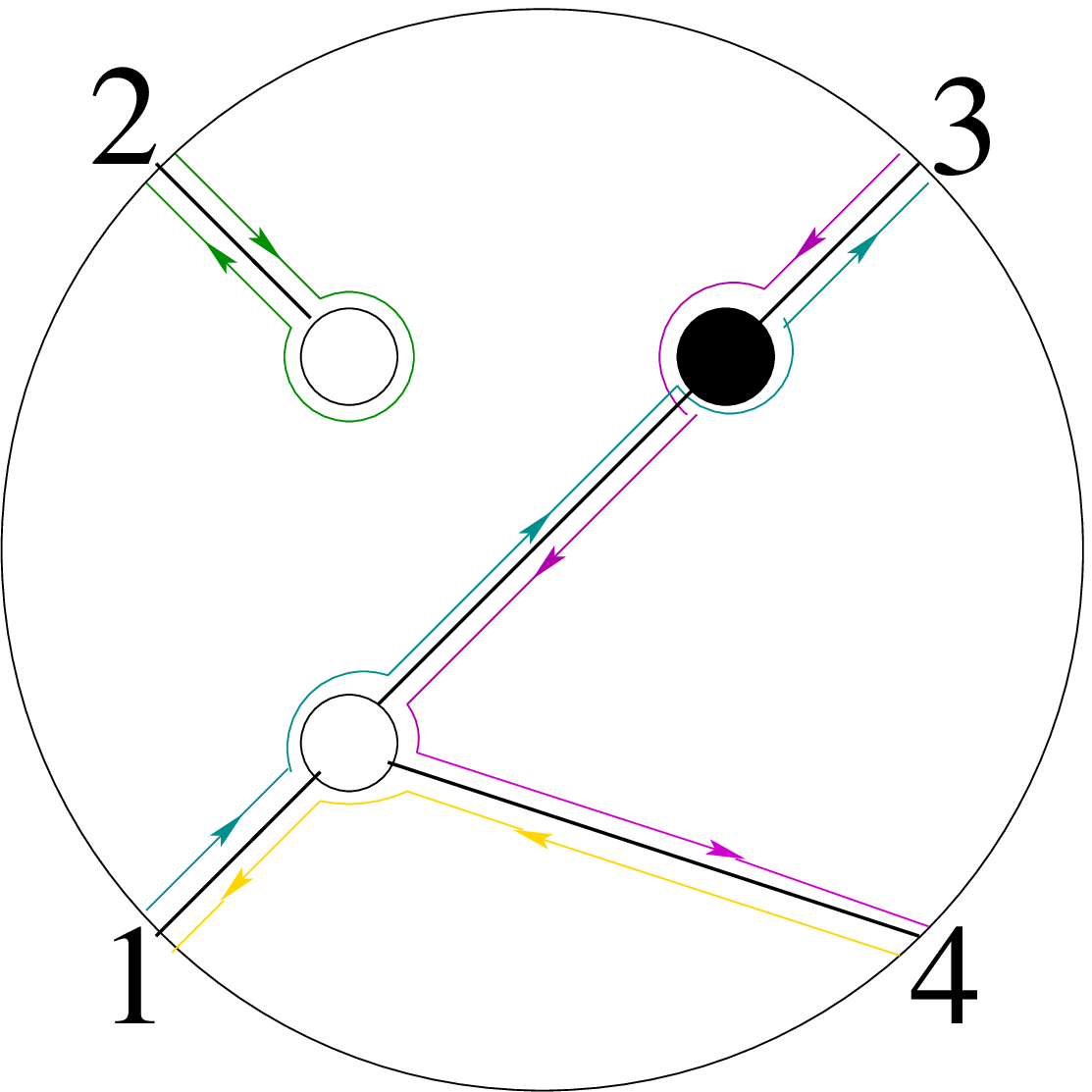}}}
   &
   \overset{\mathbb{I}}{\longleftrightarrow}
   &
   \raisebox{-1.6cm}{\scalebox{.25}{\includegraphics{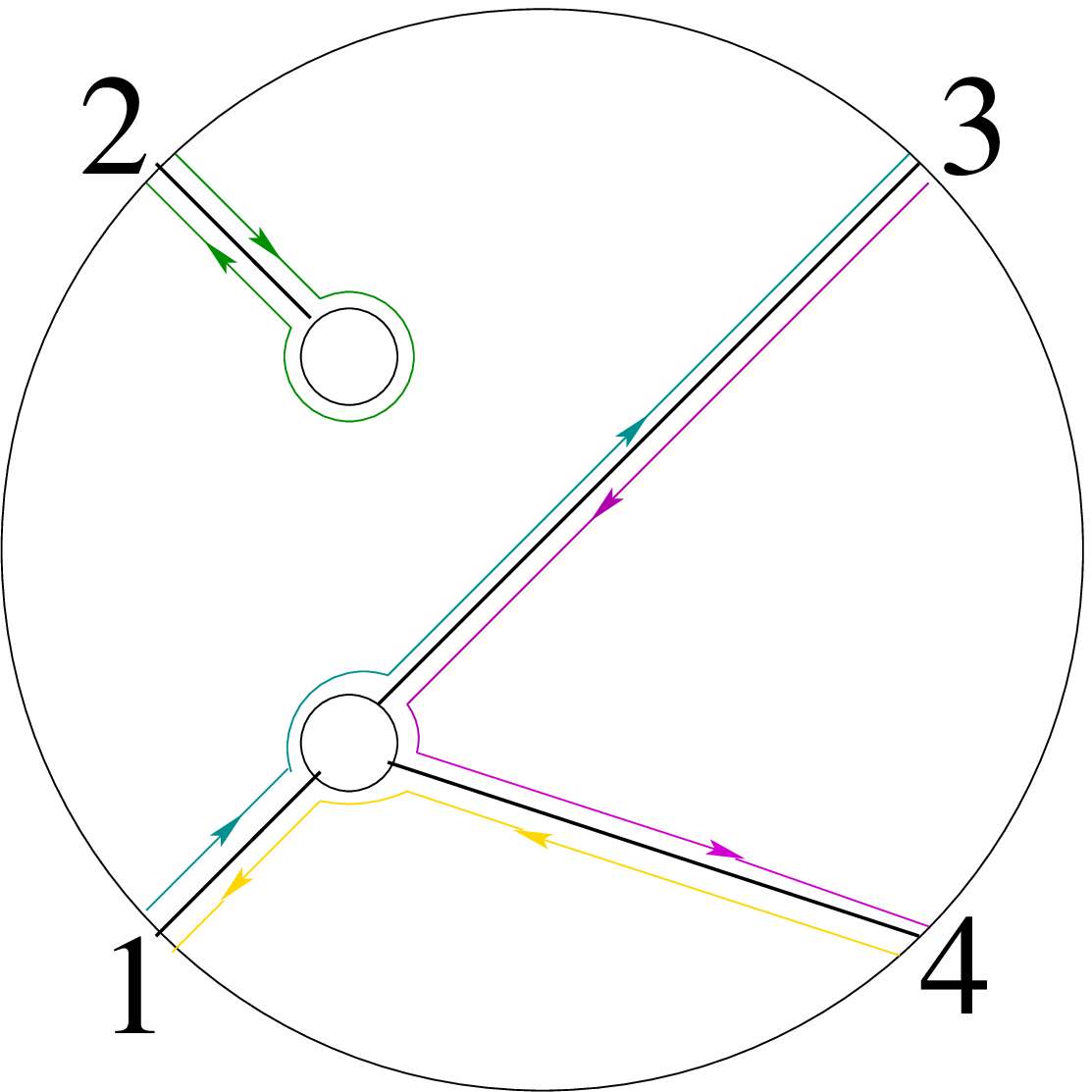}}}\\
  {\color{blue} \{4635\}} & {} & {\color{blue} \{3645\}} & {} &
   {\color{blue} \{3645\}}
  \\
  {} & {} & {} & {} & {}
   \\
   \left\updownarrow\right. {\mbox{\footnotesize $\mathbb{I}$}}   
   &{}
   &{}
   &{}
   &
   {}
   \\
   {} & {} & {} & {} & {}   
   \\
   \raisebox{-1.6cm}{\scalebox{.25}{\includegraphics{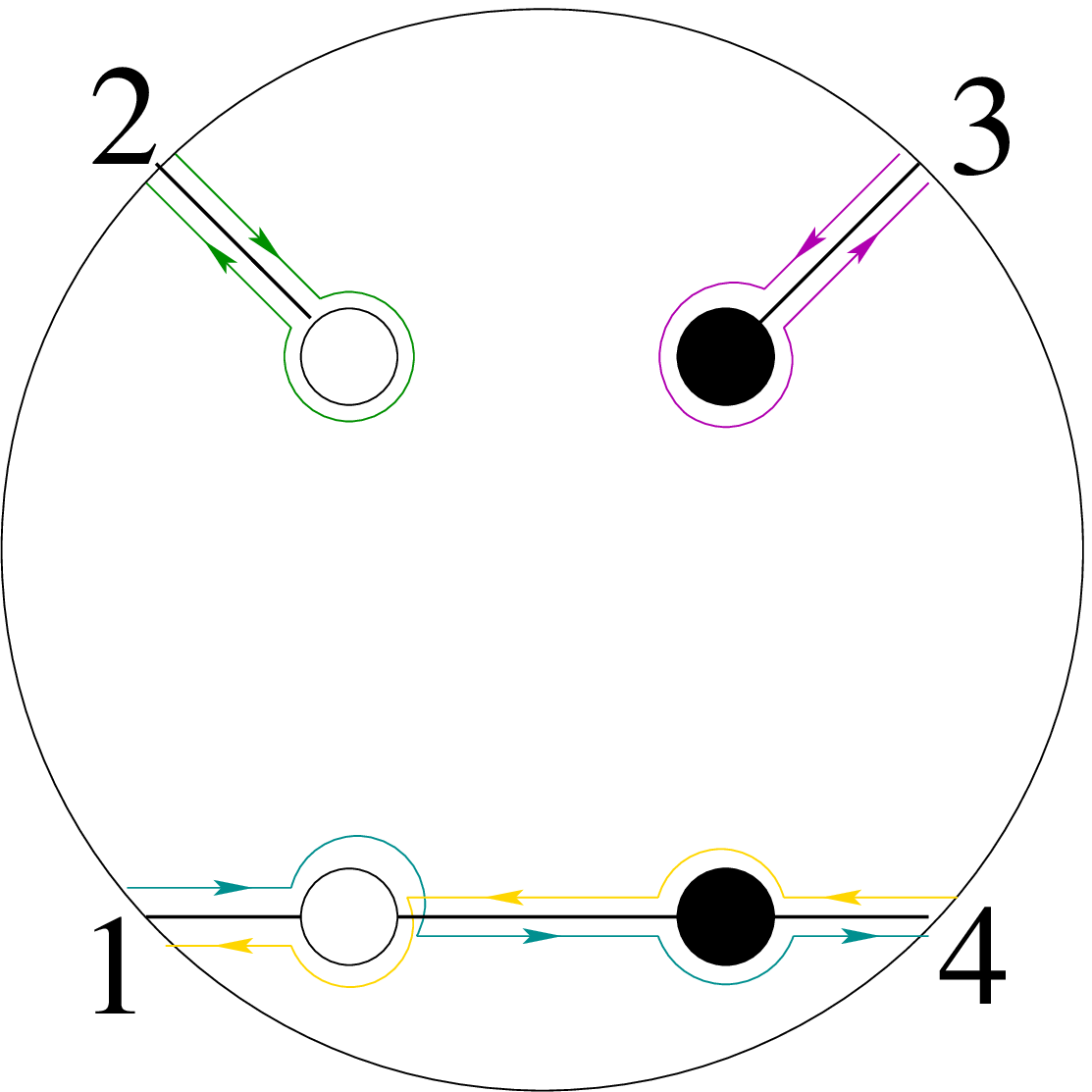}}} 
   &
   \overset{(41)}{\longleftrightarrow}
   &
   \raisebox{-1.6cm}{\scalebox{.25}{\includegraphics{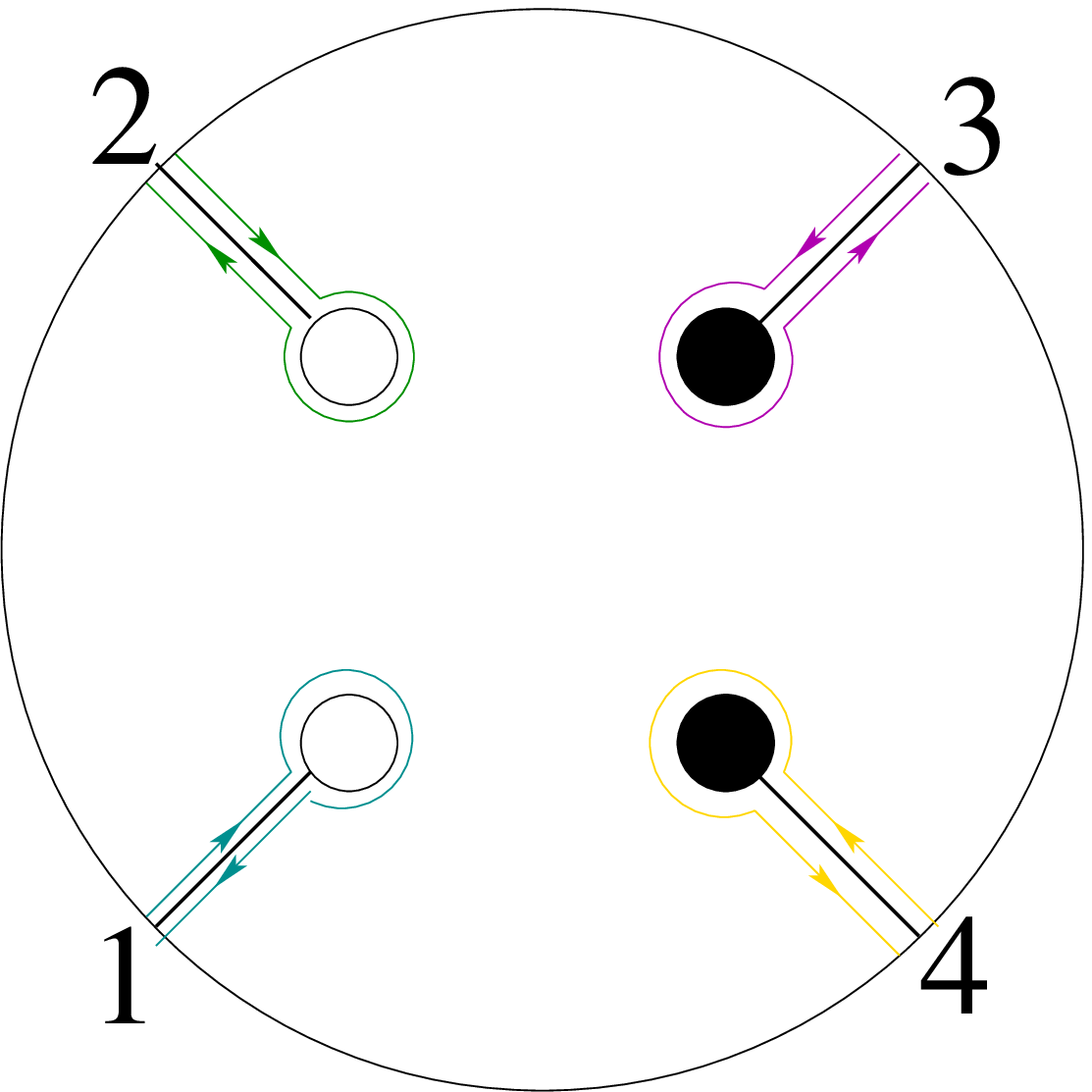}}}
   &{} & {}\\
   {\color{blue} \{4635\}} & {} & {\color{red} \{5634\}} & {} &
   {}
 \end{array}
\end{equation*}
where we displayed the trivial transformations (at step 3 and 7) 
which make manifest the action of the BCFW bridges.

\subsection{Equivalence classes of perfectly-oriented diagrams}
\label{subsec:OSeqcls}

Having reminded how permutation works, we can now discuss how,
together with the helicity flows, they define equivalence classes of
perfectly-oriented diagrams. For this purpose, let us focus on
the simplest example: the on-shell box \eqref{eq:PermNpt}. Once
the labels are fixed (and we do it as in \eqref{eq:PermNpt}), there 
is just a single equivalence class for (un-oriented) on-shell boxes
and it is labelled by $\{3,\,4,\,5,\,6\}$. Let us now associate
sources and sinks, and thus a set of helicity flows, to an on-shell 
box. Specifically, choosing $4,\,1$ as sources and $2,\,3$ as sinks,
it is easy to see that both the un-decorated diagrams in the 
permutation
$\{3,\,4,\,5,\,6\}$ enjoy the same helicity flows 
${\color{blue} 1}\,\longrightarrow\,{\color{red} 2}$ and
${\color{blue} 4}\,\longrightarrow\,{\color{red} 3}$ along the
external lines, as well as the internal 
$\{{\color{blue} 1}{\color{red} 2}\}\,\equiv\,
 \{{\color{blue} 4}{\color{red} 3}\}$ :
\begin{equation}\eqlabel{eq:PermDec1}
\raisebox{-1.8cm}{\scalebox{.35}{\includegraphics{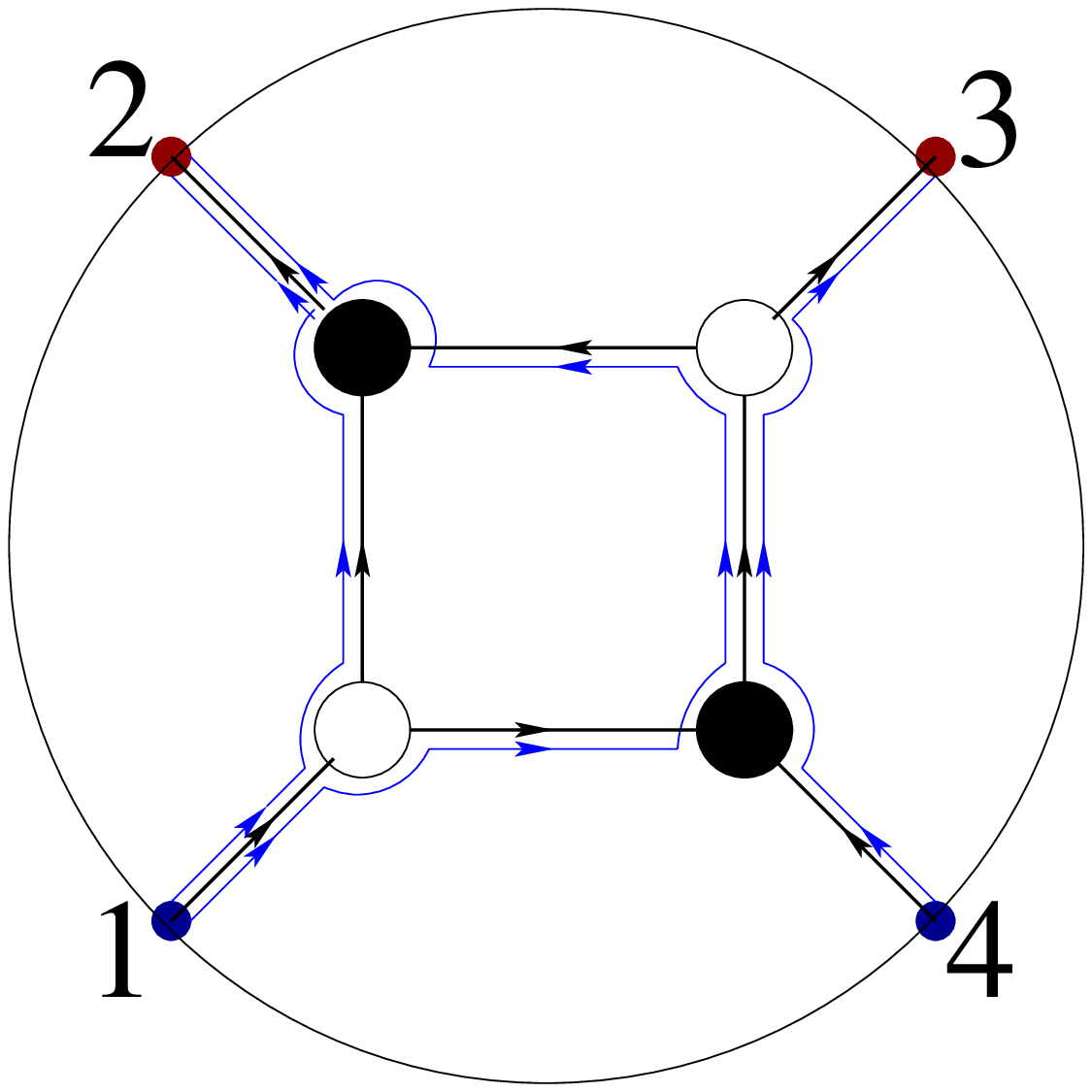}}}
 \qquad\Longleftrightarrow\qquad
\raisebox{-1.8cm}{\scalebox{.35}{\includegraphics{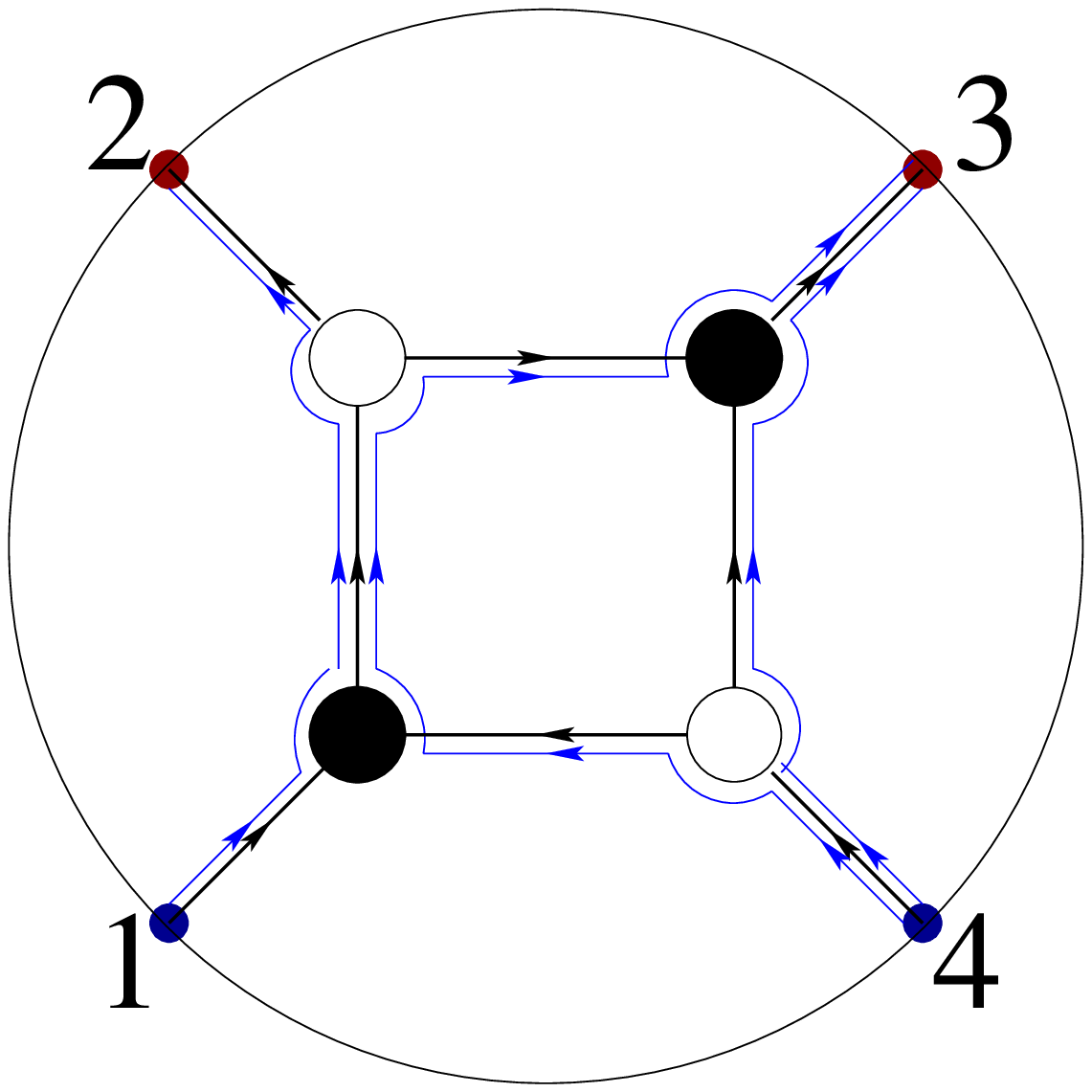}}}
\end{equation}
Such an equivalence class will be denoted as 
$\{{\color{blue} 3},\,{\color{red} 4},\,{\color{red} 5},\,
   {\color{blue} 6}\}$. In principle, the permutation 
$\{3,\,4,\,5,\,6\}$ contains three more equivalence classes of this 
type:
$\{{\color{blue} 3},\,{\color{blue} 4},\,{\color{red} 5},\,
   {\color{red} 6}\}$, 
$\{{\color{red} 3},\,{\color{blue} 4},\,{\color{blue} 5},\,
   {\color{red} 6}\}$ and
$\{{\color{blue} 3},\,{\color{blue} 4},\,{\color{red} 5},\,
   {\color{red} 6}\}$.

Let us now take $1,\,3$ as sources and $2,\,4$ as sinks. In this
case, the helicity flow structure turns out to be different, with 
one diagram showing external helicity flows between the adjacent
boundary nodes 
(${\color{blue} 1}\,\longrightarrow\,{\color{red} 2}$,
 ${\color{red} 2}\,\longleftarrow\,{\color{blue} 3}$, 
 ${\color{blue} 3}\,\longrightarrow\,{\color{red} 4}$,
 ${\color{red} 4}\,\longleftarrow\,{\color{blue} 1}$),
while the other one being actually the same of the contribution
of two diagrams characterised by an internal helicity loop and
by just two external helicity flows
\begin{equation}\eqlabel{eq:PermDec2}
 \raisebox{-1.8cm}{\scalebox{.35}{\includegraphics{4ptHelFlow1.eps}}}
  ;\qquad
 \raisebox{-1.8cm}{\scalebox{.35}{\includegraphics{4ptHelFlow2.eps}}}
 \:+\:
 \raisebox{-1.8cm}{\scalebox{.35}{\includegraphics{4ptHelFlow3.eps}}} 
\end{equation}
The sum above has overall the same external helicity flows, while
the differently oriented internal loops should cancel in order
to belong to the same equivalence class of the first diagram.
This is generically not the case -- it occurs just for 
$\mathcal{N}\,=\,3$ -- and thus the above diagrams belongs
to different equivalence classes, each of them containing just
a single element. We denote them 
$({\color{blue} 1},\,{\color{red} 2},\,{\color{blue} 3},\,
  {\color{red} 4})_{\mbox{\tiny ext}}$,
$({\color{blue} 1},\,{\color{red} 2},\,{\color{blue} 3},\,
  {\color{red} 4})_{\mbox{\tiny R}}$ and
$({\color{blue} 1},\,{\color{red} 2},\,{\color{blue} 3},\,
  {\color{red} 4})_{\mbox{\tiny L}}$
respectively -- where the subscripts ${}_{\mbox{\tiny ext}}$,
${}_{\mbox{\tiny R}}$ and ${}_{\mbox{\tiny L}}$ make reference
to the fact that the related on-shell diagram has just an
external, internal right-``chiral'' or internal left-``chiral''
helicity flow.

It is interesting to notice that the fact that the permutation
contains different equivalence classes is just the statement
that all of them are related by Ward identities. As we
will 
extensively discuss in the companion paper \cite{Benincasa:2015osg}, 
this picture will be manifest on the positive Grassmannian, where
all these diagrams correspond to the same point on $G(2,4)$.
In a nutshell, on-shell diagrams belonging to the same permutation
are geometrically equivalent and are distinguished just by a 
{\it measure} which is characteristic of the particular 
equivalence class inside a given permutation.

For the time being, it is interesting to observe that a similar 
picture of the reduced on-shell diagram as a composition of
BCFW bridges ({\it i.e.} adjacent transposition) holds. In 
particular such decomposition can still be represented as a path
along the edges of a bridge polytope $Br_{k,n}$ from a given 
permutation (representing the diagram) to the identity. The main
difference with the case discussed in Section 
\ref{subsec:OSperm} is due to the fact that endowing the on-shell
diagrams with sources and sinks restricts the edges of the bridge
polytope {\it allowed} in a path towards the 
identity. This restriction is due to the requirement that the
sources and sinks stay fixed along the path, {\it i.e.} the 
{\it boundary} states have the same helicity as one acts with a BCFW
bridge. As a concrete example, let us consider two different 
choices for the sources/sinks: 
$({\color{blue} 1},\,{\color{red} 2},\,{\color{red} 3},\,{\color{blue} 4})$,
and 
$({\color{blue} 1},\,{\color{red} 2},\,{\color{blue} 3},\,{\color{red} 4})$. Beginning with the first case, there are just two
paths along the edges of the bridge polytope which are consistent
with the helicity flows
\begin{equation*}
\raisebox{-1.8cm}{\scalebox{.33}{\includegraphics{4ptHelFlow6.eps}}}
\quad\longleftrightarrow\quad
 \raisebox{-2.6cm}{\scalebox{.75}{\includegraphics{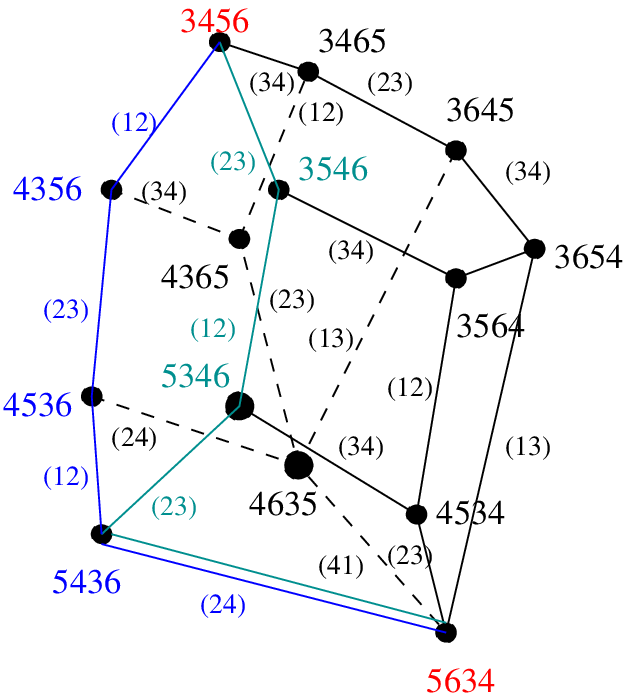}}}
\longleftrightarrow\quad
\raisebox{-1.8cm}{\scalebox{.33}{\includegraphics{4ptHelFlow7.eps}}}
\end{equation*}
where the decomposition paths are highlighted on the bridge polytope
$Br_{2,4}$, in blue for the decomposition leading to the diagram on
the left and in cyan for the one leading to the diagram on the 
right. These paths are a reflection of the
physical singularities of the on-shell process they are providing
the decomposition of. If one takes any of the other edges, the
price one pays is the generation of a (typically helicity-dependent)
Jacobian: the related BCFW bridge is of type 
\eqref{eq:BCFWbridge2b}. Each of the diagrams in the equivalence 
class
$({\color{blue} 1},\,{\color{red} 2},\,{\color{red} 3},\,
  {\color{blue} 4})$ 
correspond to one of the two paths on $Br_{2,4}$.

\begin{figure}[htbp]
 \centering
 \[
   \raisebox{-1.8cm}{\scalebox{.33}{\includegraphics{4ptHelFlow1.eps}}} \quad\longleftrightarrow
   \raisebox{-2.6cm}{\scalebox{.75}{\includegraphics{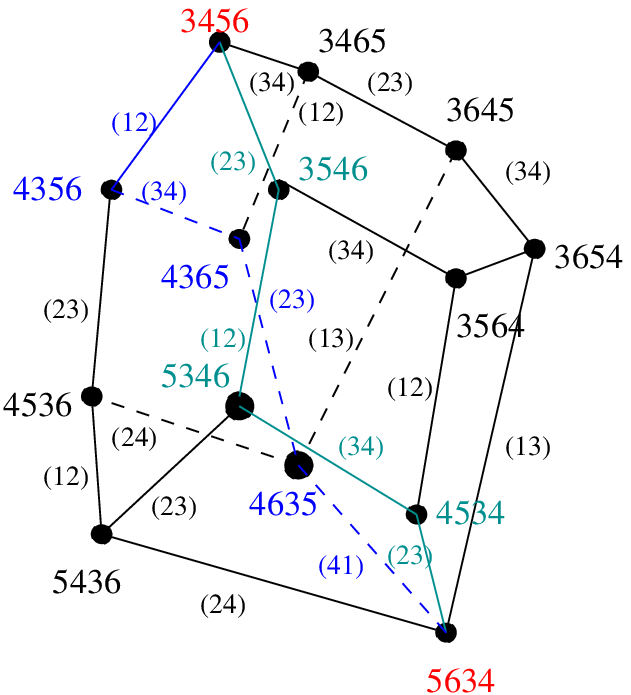}}} \longleftrightarrow\quad
   \raisebox{-1.8cm}{\scalebox{.33}{\includegraphics{4ptHelFlow2.eps}}} \quad   
 \]
 \caption{Bridge polytopes for perfectly-oriented diagrams. The 
          sources and sinks are taken to be 
          $({\color{blue} 1},\,{\color{red} 2},\,
            {\color{blue} 3},\,{\color{red} 4})$. The 
          helicity-preserving paths provide a decomposition
          of either of the on-shell diagram with just external 
          helicity flows $({\color{blue} 1},\,{\color{red} 2},\,
          {\color{blue} 3},\,{\color{red} 4})_{\mbox{\tiny ext}}$
          (in blue), or of the one with the right-chiral
          $({\color{blue} 1},\,{\color{red} 2},\,
            {\color{blue} 3},\,{\color{red} 4})_{\mbox{\tiny R}}$
          internal helicity loops (in cyan).}
 \label{fig:bridgehedronHel}
\end{figure}


Let us now turn to the configuration of sources/sinks 
$({\color{blue} 1},\,{\color{red} 2},\,{\color{blue} 3},\,
{\color{red} 4})$. In this case, as we saw earlier, one can 
distinguish three equivalence classes, each of them containing
a single element. The paths corresponding to the 
{\it helicity-preserving} BCFW decompositions on $Br_{2,4}$ have
been shown in Figure \ref{fig:bridgehedronHel}.  Also in this case,
there are just possible helicity preserving paths possible
one corresponding to $({\color{blue} 1},\,{\color{red} 2},\,
{\color{blue} 3},\,{\color{red} 4})_{\mbox{\tiny ext}}$ and the 
other one to  $({\color{blue} 1},\,{\color{red} 2},\,
{\color{blue} 3},\,{\color{red} 4})_{\mbox{\tiny R}}$.
Let us remark that the existence of just two helicity
preserving paths is just a feature of the particular choices of
sinks and sources (one example where there are more decompositions
allowed which respect the helicity flows is given by 
$({\color{blue} 1},\,{\color{blue} 2},\,{\color{red} 3},\,
  {\color{red} 4})$).

In summary, also for a decorated diagram, the bridge polytope 
provides its BCFW decomposition with the particular feature that
just a subset of the edges are helicity preserving. If one starts
from the identity and endows it with a given choice of sinks
and sources, then the requirement that any transposition is 
decoration preserving leads along all the possible helicity
preserving edges of the related bridge polytope.

\subsection{More on on-shell diagrams, perfect orientations and 
            helicity flows}
\label{subsec:OSpohf}

In the previous sections we saw how the need of introducing new
data, namely to distinguish among the coherent multiplets, led 
naturally to endow the diagrams with perfect orientations which
are intimately related to the singularity structure of the
particular on-shell process through the helicity flows that the
perfect orientation itself generates.

If we abstract ourselves for a moment from our context and consider 
a generic un-oriented bipartite diagram which it is possible to
associated a perfect orientation to, for such a perfectly 
oriented diagram, the inversion of a flow returns another perfect 
orientation as well as any other perfect orientation can be obtained
from the original one by reversing the direction of a flow in the 
latter \cite{Postnikov:2008mp}.

Such a lemma holds for our decorated diagrams as well -- at the
end of the day they are perfectly oriented bipartite diagrams --
with a peculiarity: In our case, the flows dictated by a given
perfect orientation contains information about the helicity of the
states as well as the singularity structure of the on-shell process.
Thus, the reversal of the direction of a helicity flow is 
associated to a Jacobian which transforms not trivially under the
little group if the helicity flow which gets reversed involves
external states, while it is helicity blind if the helicity flow
is internal.

\begin{figure}[htbp]
 \centering
  \raisebox{-1.8cm}{\scalebox{.70}{\includegraphics{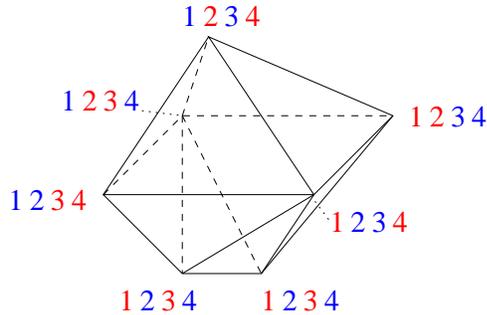}}} 
 \caption{Polytope generated by all possible helicity-flow reversals
          for a fully-localised on-shell box. Each vertex of the
          polytope represents a given perfect orientation, while
          each edge represents the flow-reversal connecting them, to
          which a Jacobian is associated.}
 \label{fig:Opolytope}
\end{figure}

For the sake of definiteness, let us consider the fully-localised
on-shell box with the configuration 
$({\color{blue} 1} ,\,{\color{red} 2},\,{\color{blue} 3},\,
  {\color{red} 4})$ 
for the sources and sinks. This decorate diagram shows four helicity
flows, all of them involving a pair of external states. It is 
possible to pick any of those flows and reverse it, mapping a sink 
into a source and a source into a sink and generating a 
helicity-dependent Jacobian as in the following example:
\begin{equation}\eqlabel{eq:revhelflow}
 \raisebox{-1.3cm}{\scalebox{.25}{\includegraphics{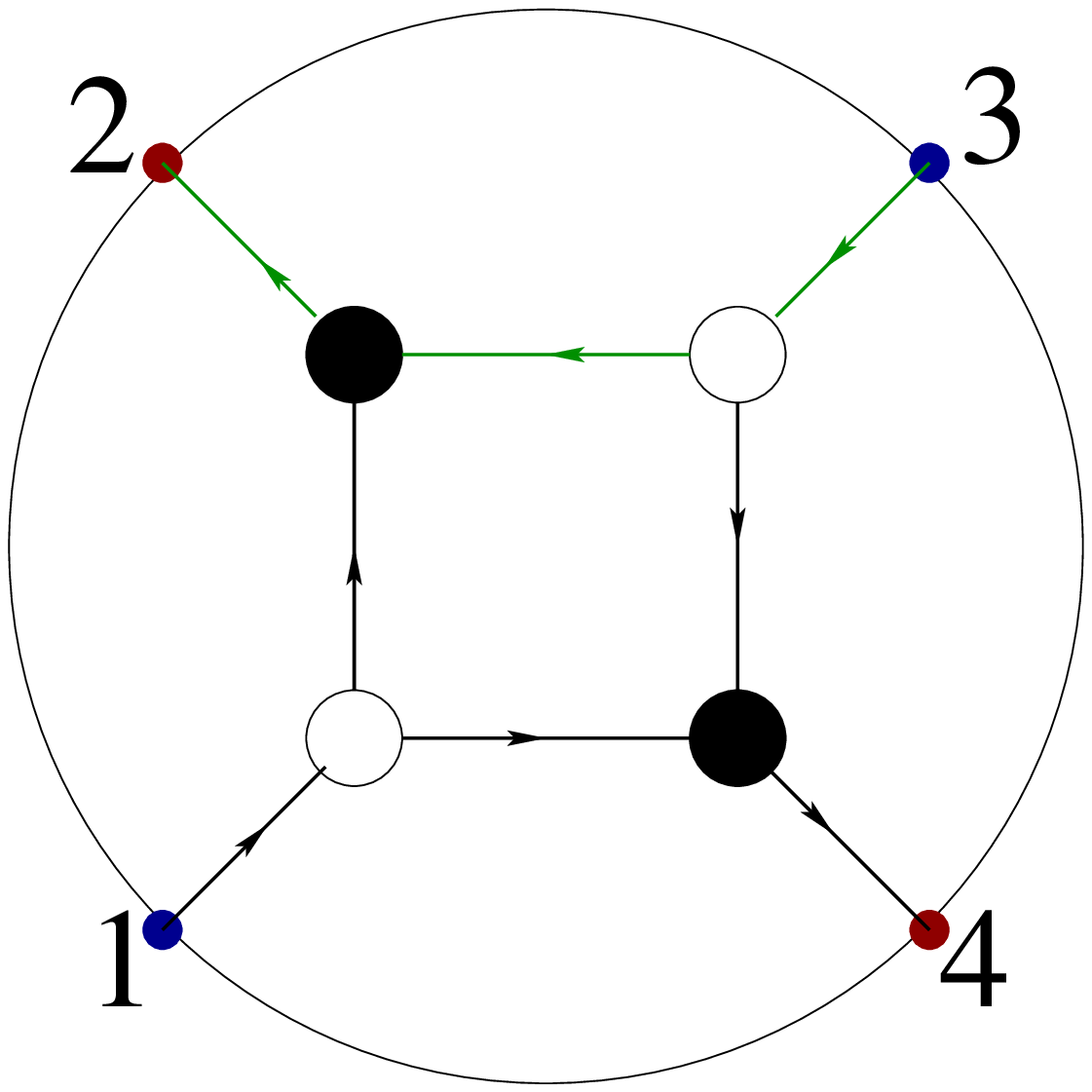}}}
  \quad
  \overset{({\color{red} 2}, {\color{blue} 3})}{\longrightarrow} 
  \quad
 \raisebox{-1.3cm}{\scalebox{.25}{\includegraphics{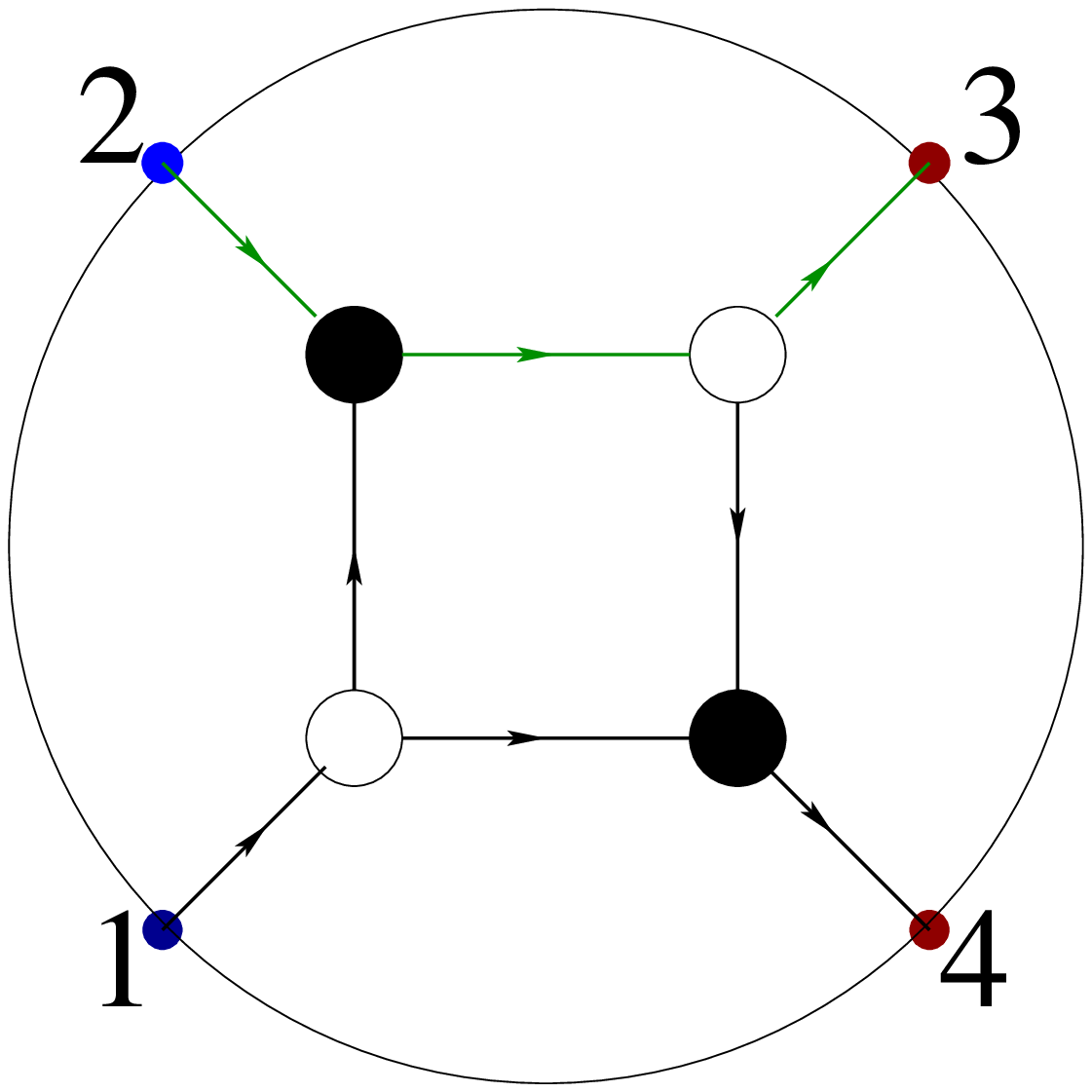}}}
 \;J_{23},
 \qquad
 J_{23}\:=\:
  \left(
   \frac{\langle1,3\rangle}{\langle1,2\rangle}
  \right)^{4-\mathcal{N}}
\end{equation}
In this way one can generate all the possible perfect orientations
for the fully-localised on-shell box, {\it i.e.} all the possible
configurations for the sources and sinks (helicity configuration
for the external states) as well as the two diagrams with internal
flow. This seven configurations can be seen as the vertices of 
a polytope whose edges represent the flow reversals which a 
Jacobian is associated to (see Figure \ref{fig:Opolytope}).

Notice that the edges of the polytope in Figure \ref{fig:Opolytope}
are nothing but the Ward identities relating the tree-level 
four-particle amplitudes (which are all MHV). Some sort of special
points are the two at the bottom of the polytope, which have
the same configuration of the sources and sinks but which 
distinguish themselves for the clockwise/counter-clockwise
internal helicity flow. The Jacobian related to the edges connecting
any other vertex to them has a helicity-dependent factor of the
form of $J_{23}$ in \eqref{eq:revhelflow} and another one which is
helicity blind and depend on the Mandelstam variables. The edge
connecting these two vertices instead is the only one whose
Jacobian is purely helicity blind. In \eqref{eq:revhelflow} the
configuration 
$({\color{blue} 1},\,{\color{red} 2},\,{\color{blue} 3},\,
  {\color{red} 4})$
has been taken as a reference so that, differently from the other
vertices, when one goes back to it taking any path along the 
edges of the polytope, the overall Jacobian is the identity. 
In a more democratic way, one can associate to it the Jacobian
factor $\langle1,3\rangle^{-(4-\mathcal{N})}$, so that the
helicity-dependent term in any Jacobian has the form 
$\langle{\color{blue} i},
 {\color{blue} j}\rangle^{-(4-\mathcal{N})}$,
{\it i.e.}, it depends on the sources only.

As a further comment, notice that the polytope related to the
fully-localised on-shell box with the white and black nodes 
exchanged, can be obtained from the one in Figure 
\ref{fig:Opolytope}, by contracting the two vertices labelled
by $({\color{red} 1},\,{\color{blue} 2},\,{\color{red} 3},\,
     {\color{blue} 4})$ to a single one,
and opening up the vertex 
$({\color{blue} 1},\,{\color{red} 2},\,{\color{blue} 3},\,
     {\color{red} 4})$ into two. This is a reflection of the
fact that these two configurations do not satisfy a square move,
while all the others do.

Finally, such a polytope can obviously be defined for any higher
point on-shell process. When the helicity flows involve external
states, they always show a source and a sink. This mean that any
helicity-flow reversal maps a given on-shell process in another
one in the same N${}^k$MHV-sector. For $k\,=\,2$, the polytope
provides with the well-known Ward identities among tree-level 
amplitudes as well as new ones with {\it non-local}
on-shell processes\footnote{An example is provided by the lower
part of the polytope in Figure \ref{fig:Opolytope}, where the
{\it non-local} on-shell processes, {\it i.e.} having a singularity
which do not correspond to any factorisation channel in the 
full-fledge amplitude, are given by the bottom vertices
$({\color{red} 1},\,{\color{blue} 2},\,{\color{red} 3},\,
  {\color{blue} 4})$}. For $k\,>\,2$, the helicity-flow reversals
do not in general map an amplitude into another amplitude, so that
the polytopes typically describe relation among individual
on-shell processes.

\section{On-shell processes and scattering amplitudes}
\label{sec:OSampl}

So far we have been discussing generic on-shell processes in
order to illustrate some important features of the decorated
diagrammatics as well as how special points can be related to each
other. Let us now turn to actual scattering amplitudes. 
The objects of interest are actually {\it the integrands} in the 
perturbative expansion. 
The on-shell diagrammatics then generates an expansion
in terms of $4L$-forms which, upon (regulated\footnote{In this 
paper we will not discuss the issue of the 
regularisation of these integrals. For a general 
{\it on-shell} proposal which does not rely on any further 
assumption respect to the ones in this paper, we refer to 
\cite{Benincasa:2014zpa, Benincasa:2015zpb}. 
For $\mathcal{N}\,=\,4$ SYM, a natural regularisation is to
introduce a mass, moving onto the Higgs branch of the theory
\cite{Alday:2009zm, Henn:2010bk}
but breaking dual conformal invariance, or moving the external
states slightly off-shell  and preserving such a symmetry 
\cite{Bourjaily:2013mma}. Finally, there
are, again in the maximally supersymmetric context, other
approaches which, using integrability arguments, introduce
spectral parameters as regulator \cite{Ferro:2012xw, Ferro:2013dga},
or a deformation directly into the Grassmannian description
\cite{Ferro:2014gca} (see also \cite{Bargheer:2014mxa} for a 
discussion on the deformed Grassmannian).}) integration on the 
Lorentz real sheet, returns the physical amplitude. Actually, the 
on-shell diagrammatics provides an object from which a great
deal of physical information can be extracted, depending on the
path/manifold on which one integrates it. The general form can
be written as
\begin{equation}\eqlabel{eq:PertExp}
 \mathcal{M}_{n}(\{\lambda^{\mbox{\tiny $(i)$}},\,
  \tilde{\lambda}^{\mbox{\tiny $(i)$}};\,h_i\})\:=\:
 \sum_{L=0}^{\infty}
 \mathcal{M}_{n}^{\mbox{\tiny $(L)$}}
  (\{\lambda^{\mbox{\tiny $(i)$}},\,
   \tilde{\lambda}^{\mbox{\tiny $(i)$}};\,h_i\},\,\{z_l\})
   \bigwedge_{l=1}^{4L}dz_l,
\end{equation}
where the expansion coefficients 
 $\mathcal{M}_{n}^{\mbox{\tiny $(L)$}}
  (\{\lambda^{\mbox{\tiny $(i)$}},\,
   \tilde{\lambda}^{\mbox{\tiny $(i)$}};\,h_i\},\,\{z_l\})$
are rational function of both the Lorentz invariant combination
of the spinors and of the parameters $\{z_l\}$. It is easy to see
that the $L=0$ term is a $0$-form and, therefore, receives
contributions from fully localised on-shell diagrams, while for
$L\,\ge\,1$ the on-shell diagrams must have $4L$ degrees of freedom
left unfixed (for a given $L$) and these terms can generate
branch cuts upon integration of such degrees of freedom over
a suitable manifold. The $L$-th order terms is then the seed from 
which the $L$-loop amplitude can be extracted (upon a suitable 
integration).

In the case of $\mathcal{N}\,=\,4$ SYM it was shown that scattering
amplitudes satisfy an all-loop recursion relation 
\cite{ArkaniHamed:2010kv}, which can be understood as the solution
of a {\it differential equation} encoding all the singularities
of the amplitudes at a given loop $L$, which gets {\it integrated}
via a BCFW bridge \cite{ArkaniHamed:2012nw}:
\begin{equation}\eqlabel{N4SYMrr}
 \begin{split}
  &\partial
   \left[
    \raisebox{-1.5cm}{\scalebox{.55}{\includegraphics{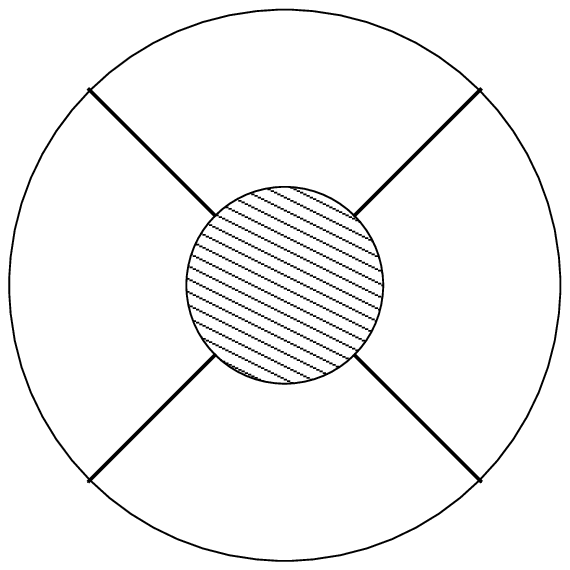}}} 
   \right]\:=\:
   \sum_{L,\,R}\;
   \raisebox{-1cm}{\scalebox{.55}{\includegraphics{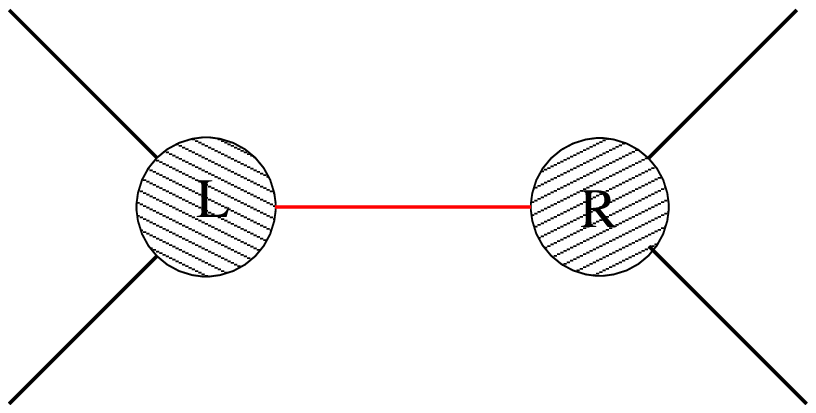}}}+
   \sum_k
   \raisebox{-1.3cm}{\scalebox{.55}{\includegraphics{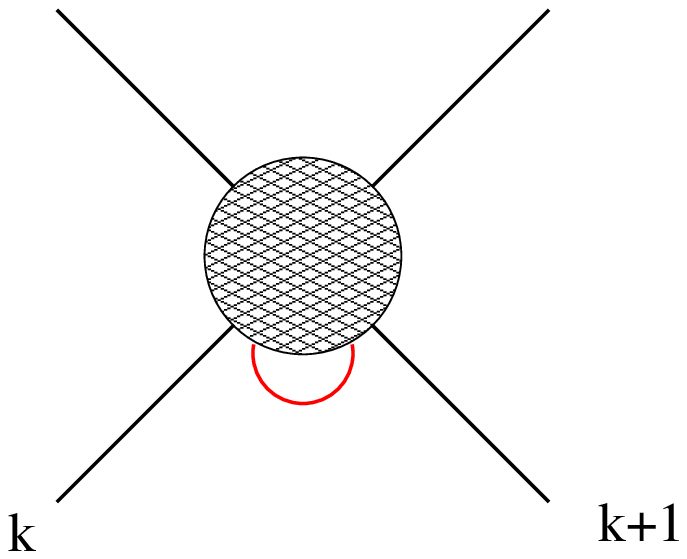}}} \\
  &\hspace{7cm} \Downarrow\\
 &\hspace{.5cm}
  \raisebox{-1.5cm}{\scalebox{.55}{\includegraphics{N4SYMrrLHS.eps}}}
  \:=\:
  \sum_{k\in\mathcal{P}}\;\mathcal{I}_k
  \raisebox{-2.3cm}{\scalebox{.55}{\includegraphics{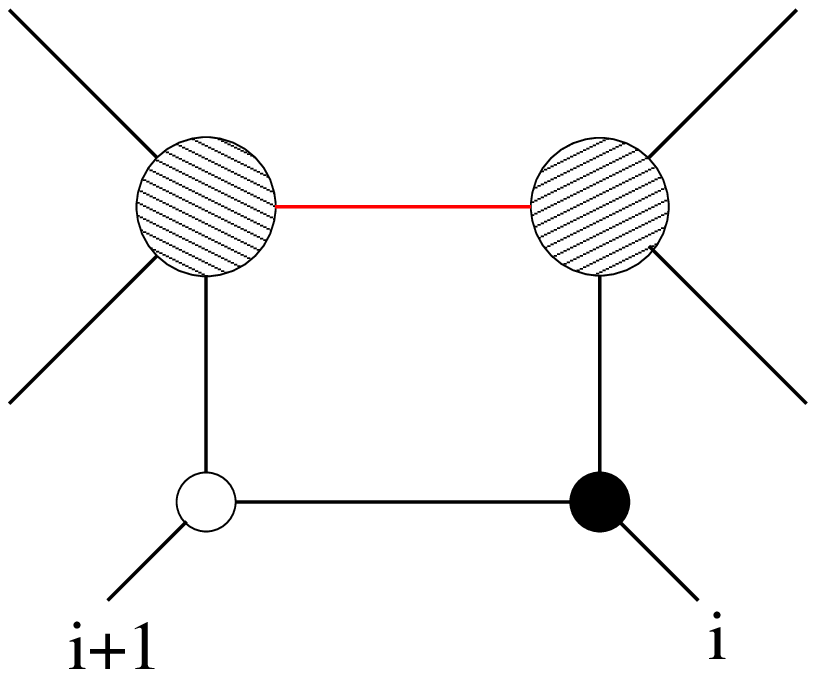}}}\mathcal{J}_k+
 \raisebox{-2.5cm}{\scalebox{.55}{\includegraphics{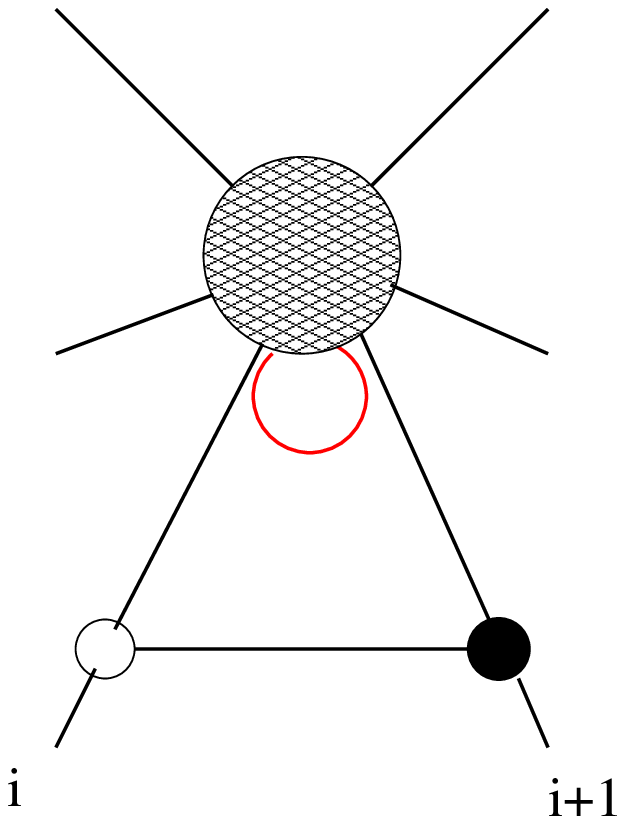}}} 
 \end{split}
\end{equation}
where the first term in the first line indicate all the possible
factorisation channels, with the sub-amplitudes labelled by $L$ and
$R$ being $l$- and $(L-l)$-loop amplitudes with 
$n_{\mbox{\tiny $L$}}+1$ and $n_{\mbox{\tiny $R$}}+1$ external
states ($n_{\mbox{\tiny $L$}}+n_{\mbox{\tiny $R$}}\,=\,n$),
while the second term represents the forward-limit of a 
$(L-1)$-loop amplitude with $n+2$ external states 
\cite{CaronHuot:2010zt}. In the second line, the BCFW bridge
select a sub-set of the singularities, with $\mathcal{P}$ indicating
the set of factorisation channels with the particles in the bridge
belonging to different sub-amplitudes, and $\mathcal{I}_k$ and 
$\mathcal{J}_k$ indicating a partition of all the external particles
but the ones labelled by $i$ and $i+1$ which respects colour ordering
($\mathcal{I}_k\,\cup\,\mathcal{J}_k\:=\:\{1,\ldots,n\}\,\backslash\,
  \{i,i+1\}$, with 
  dim$\{\mathcal{I}_k\,\cup\,\mathcal{J}_k\}\,=\,n-2$).
The validity of the recursive relation \eqref{N4SYMrr} has a very
neat fully-diagrammatic proof \cite{ArkaniHamed:2012nw}.

In the less/no-supersymmetric case the story turns out to be
similar but not quite, due to the presence of additional structures
than the simple $d\log{\zeta}$ which characterises the maximally
supersymmetric case. The first concern which arises is related
to the definition itself of the {\it integrand} due to the possible
presence of external bubbles and external tadpoles. However, for
$\mathcal{N}\,\ge\,1$ it has been shown that the sum over the
full multiplet(s) makes these contributions vanish 
\cite{CaronHuot:2010zt} and, thus, the object {\it integrand} stays 
well-defined, with the loop singularities -- {\it i.e.} the ones
corresponding to a loop-degree-of-freedom dependent propagator
going on-shell -- which still can be interpreted as a forward
limit of a lower-loop amplitude. Furthermore, for 
$\mathcal{N}\,\ge\,1$, the amplitudes are cut-constructible \cite{Bern:1994zx, Bern:1994cg}, which
suggests the validity of a loop recursive structure. More
subtle is the $\mathcal{N}\,=\,0$ case.


\subsection{Fully localised diagrams: The tree level structure}
\label{sec:TreeStr}

Let us begin with considering just fully-localised on-shell 
diagrams. The validity of the recursion relations of the 
amplitude under a particular {\it BCFW bridging} can be shown
diagrammatically along the same lines of the diagrammatic proof
\cite{ArkaniHamed:2012nw} that all the correct factorisation 
channels are contained in \eqref{N4SYMrr}.

The starting point is given by the two sets of four-particle 
amplitudes for which the factorisation channels are manifest and
follow the helicity lines

\begin{equation}\eqlabel{eq:treeRRind0}
 \begin{split}
 &\partial\left[
 \raisebox{-1.2cm}{\scalebox{.25}{\includegraphics{4ptHelFlow1.eps}}}
 \right]
 \:=\:
  \raisebox{-1cm}{\scalebox{.25}{\includegraphics{4ptSing2a.eps}}}
  \;+\;
  \raisebox{-1cm}{\scalebox{.25}{\includegraphics{4ptSing2b.eps}}}
  \;+\;
  \raisebox{-.9cm}{\scalebox{.25}{\includegraphics{4ptSing2c.eps}}}
  \;+\;
  \raisebox{-.9cm}{\scalebox{.25}{\includegraphics{4ptSing2d.eps}}}
 \\
 &\partial\left[
 \raisebox{-1.2cm}{\scalebox{.25}{\includegraphics{4ptHelFlow4.eps}}}
 \right]
 \:=\:
  \raisebox{-1cm}{\scalebox{.25}{\includegraphics{4ptSing3a.eps}}}
   \;+\;
 \raisebox{-1cm}{\scalebox{.25}{\includegraphics{4ptSing3d.eps}}}
   \;+\;
  \raisebox{-.9cm}{\scalebox{.25}{\includegraphics{4ptSing3c.eps}}}
 \end{split}
\end{equation}
Notice that the {\it differential equations} above are 
{\it generally} integrated to return a scattering amplitude just by 
those BCFW bridges which {\it do not} create internal helicity loops
and preserve the helicity flows on the right-hand-side of 
\eqref{eq:treeRRind0}. A special case is provided by the 
$\mathcal{N}\,=\,3$ SYM for which the sum of the two diagrams
with internal loops also provides the correct tree amplitudes. As we
mentioned earlier, this is the consequence of the fact
that, for $\mathcal{N}\,=\,3$, an internal helicity loop corresponds
to the presence of a {\it simple} pole rather than a {\it multiple}
one as for $\mathcal{N}\,\le\,2$, which disappears upon summing 
between the two diagrams with differently-oriented helicity-loops.

Then, one can assume as induction hypothesis the following relation
for the $n$-particle amplitude\footnote{In formula 
\eqref{eq:treeRRind1} neither the external states in the sets
$\mathcal{I}_k$'s and $\mathcal{J}_k$'s have been endowed with a
definite helicity to be as general as possible: For most of out
statements the exact helicity multiplet will not matter. An explicit 
assignment will be provided in case subtleties arise.}:
\begin{equation}\eqlabel{eq:treeRRind1}
 \begin{split}
 \raisebox{-1.1cm}{\scalebox{.45}{\includegraphics{N4SYMrrLHS.eps}}}
  \:&=\:
  \sum_{k\in\mathcal{P}}\;\mathcal{I}_k
  \raisebox{-1.9cm}{\scalebox{.45}{\includegraphics{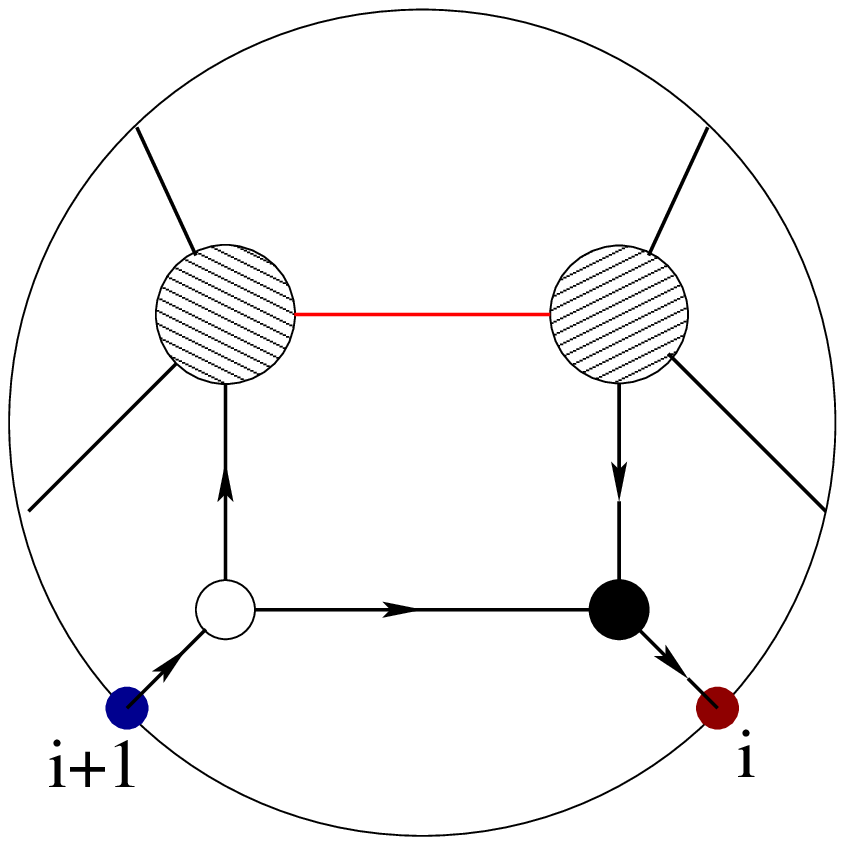}}}\mathcal{J}_k\:=\:
  \sum_{k\in\tilde{\mathcal{P}}}\;
  \tilde{\mathcal{I}}_k
  \raisebox{-2cm}{\scalebox{.45}{\includegraphics{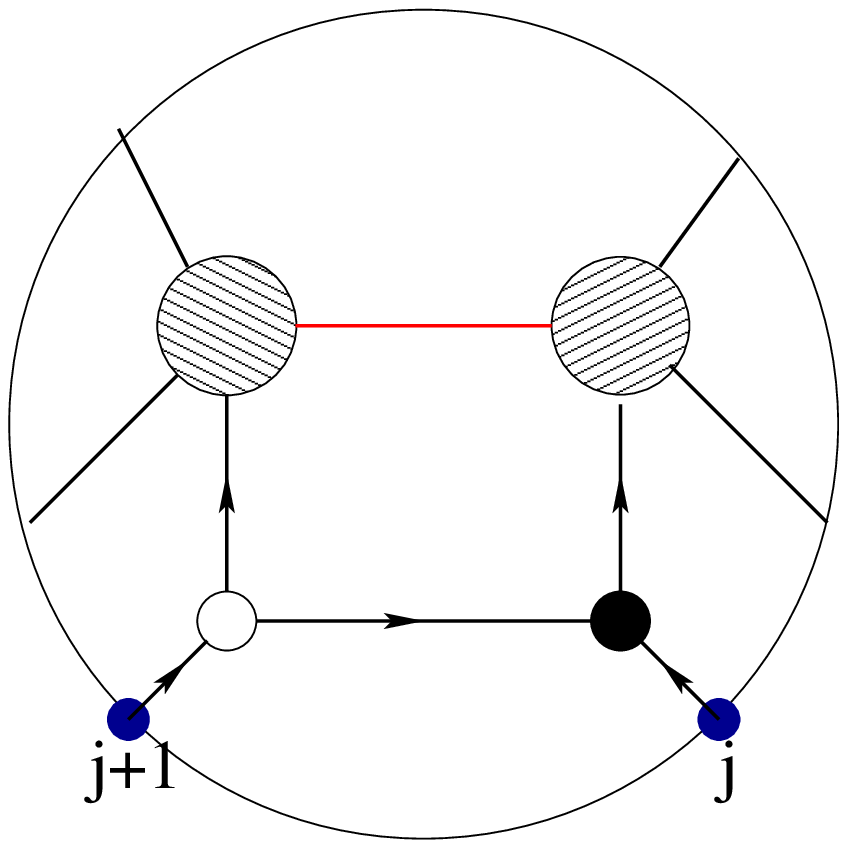}}}\tilde{\mathcal{J}}_k\\
  &=\:
  \sum_{k\in\hat{\mathcal{P}}}\;
  \hat{\mathcal{I}}_k
  \raisebox{-2cm}{\scalebox{.45}{\includegraphics{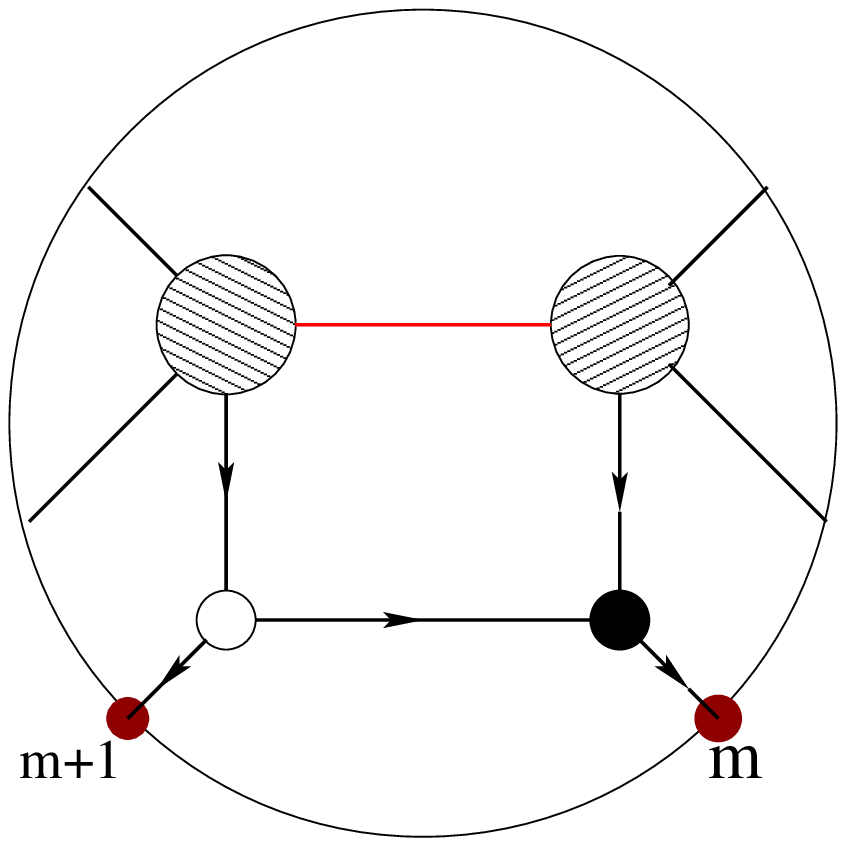}}}\hat{\mathcal{J}}_k
 \end{split}
\end{equation}
where 
$\mathcal{I}_k\,\cup\,\mathcal{J}_k\,=\,\{1,\,\ldots,\,n\}\,\backslash\,\{i,\,i+1\}$, 
$\tilde{\mathcal{I}}_k\,\cup\,\tilde{\mathcal{J}}_k\,=\,\{1,\,\ldots,\,n\}\,\backslash\,\{j,\,j+1\}$, 
$\hat{\mathcal{I}}_k\,\cup\,\hat{\mathcal{J}}_k\,=\,\{1,\,\ldots,\,n\}\,\backslash\,\{m,\,m+1\}$, and the BCFW bridge needs to be
chosen in such a way that no helicity loop is generated in the
interior of the diagram. With this assumption, one now need to show 
that the above formula is valid for a higher number of particles, 
{\it i.e.} that for higher number of particles it contains all and 
only the correct collinear and multi-particle factorisation channels.

Actually, the full discussion follows closely the one
for the maximally supersymmetric case, given that it is part
of our inductive assumption the way that a BCFW bridge can be
taken. This means the helicity flows are always preserved and
at no stage internal helicity loop can be generated.

The representations \eqref{eq:treeRRind1} make all
those channels having the particle pairs $\{i,\,i+1\}$,
$\{j,\,j+1\}$ and $\{m,\,m+1\}$ on different sub-amplitudes 
manifest and, thus, these factorisation are trivially included.

Let us consider the factorisation channel $\mathcal{K}$, where
$\mathcal{K}$ is a set of external states such that either
$\mathcal{K}\:\subset\:\mathcal{I}_k$ or 
$\mathcal{K}\:\subset\:\mathcal{J}_k$\footnote{Since now on,
unless otherwise specified, we will indicate with 
$\mathcal{I}_k$ any of the three sets in the recursive formulas
\eqref{eq:treeRRind1} on the amplitude on the left, and
similarly for the amplitude on the right where we will generically
use $\mathcal{J}_k$.}. In this case, one has two classes
of terms characterised by two sub-amplitudes are bridged by the 
initial bridge itself. The sum over these two classes of terms
because of our induction hypothesis provides a single full-fledge
sub-amplitude. Explicitly:
\begin{equation}\eqlabel{eq:Kfact}
 \begin{array}{ccc}
   \mathcal{I}_k
    \raisebox{-1.6cm}{\scalebox{.40}{\includegraphics{RRmp.eps}}}
   \mathcal{J}_k
  &
   {}
  &
   \raisebox{-1.6cm}{\scalebox{.40}{\includegraphics{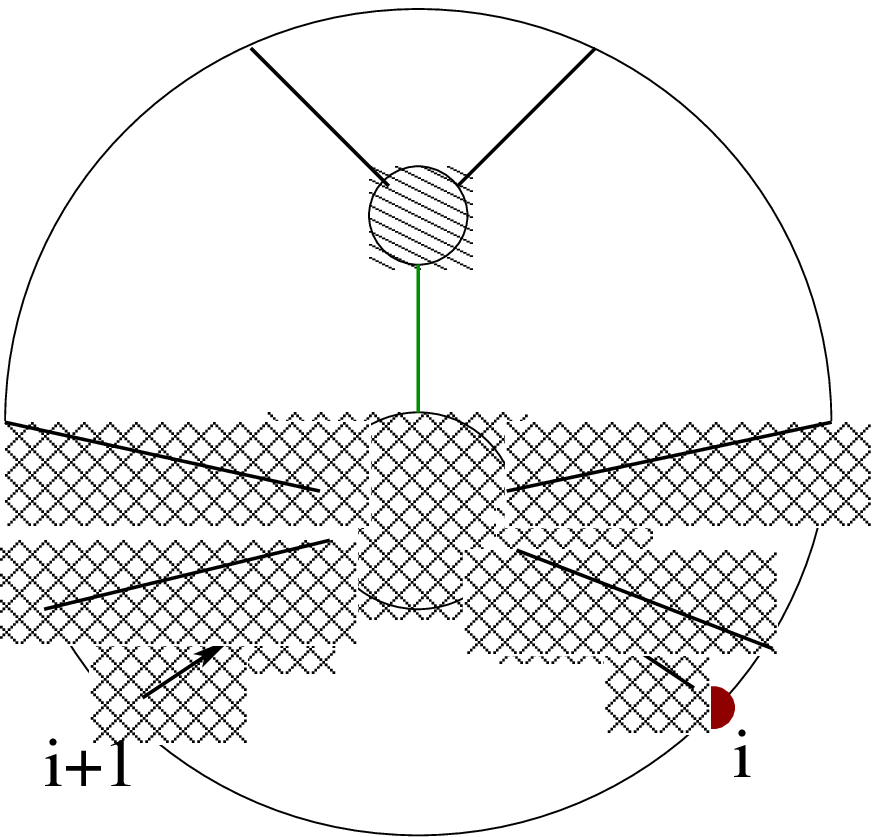}}}\\
  {} & {} & {} 
  \\
   \downarrow & {} & \Uparrow
  \\
   {} & {} & {}\\
   \hspace{-.2cm}
   \sum_{k'}
   \hspace{-.5cm}
    \begin{array}{l}
     \vspace{.4cm}\hspace{.2cm}\mathcal{K}\\
     \phantom{\mathcal{K}}\\
     \mathcal{I}_{k'}^{\mbox{\tiny $(a)$}}
    \end{array}
   \hspace{-.2cm}
   \raisebox{-1.6cm}{\scalebox{.40}{\includegraphics{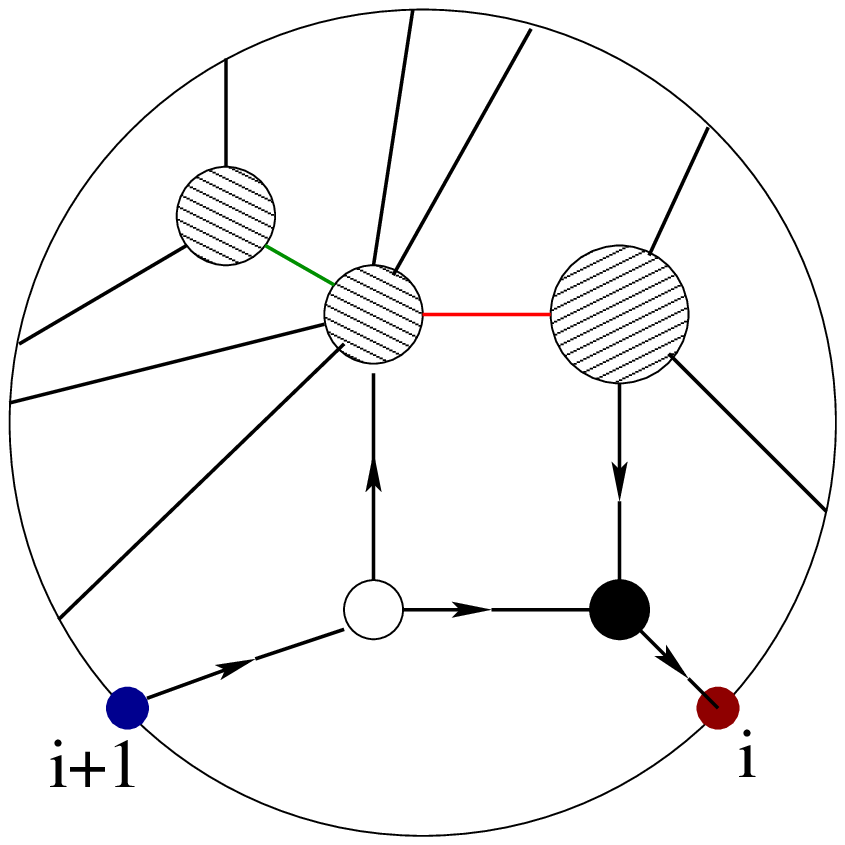}}}
   \;\mathcal{J}_{k'}
  &\hspace{-.1cm}+
  &\hspace{-.1cm}\sum_{k''}
   \mathcal{I}_{k''}
   \raisebox{-1.6cm}{\scalebox{.40}{\includegraphics{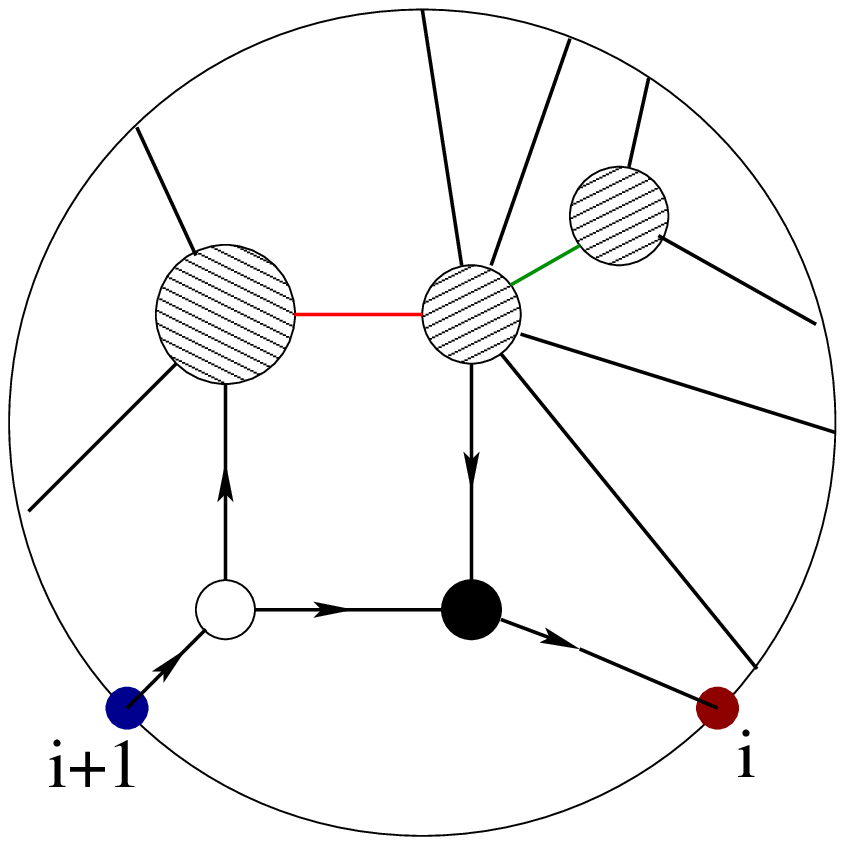}}}
   \begin{array}{l}
    \hspace{-.4cm}\vspace{1cm}\mathcal{K}\\
    \phantom{\mathcal{K}}\\
    \hspace{-.4cm}\vspace{.2cm}\mathcal{J}_{k''}^{\mbox{\tiny $(b)$}}
   \end{array}
  \end{array}
\end{equation}

As far as the factorisation in the $(i,\,i+1)$-channel is concerned,
the diagrammatics offers a very straightforward way to show that
it is also contained in the recursive formula\footnote{Both this
argument on the collinear factorisation in the $(i,i+1)$-channel
and the previous one for the $\mathcal{K}$-channel is the fully
diagrammatic version of the argument used in \cite{Schuster:2008nh}
to discuss the breaking down of the BCFW recursion relations
in pure Yang-Mills and in gravity, and in 
\cite{Benincasa:2011kn} to constrain the therein proposed 
generalised on-shell recursion relations.}. Notice that for each
of the two possible complex factorisations, there is just a
single diagram which can contribute, which is characterised by
a three-particle amplitude in the factorisation channel. 
For the sake of concreteness, let us fix the sink/sources for the 
states in the bridge to be $({\color{red} i},\,{\color{blue} i+1})$,
then irrespectively of whether the three-particle amplitude which 
connects particle $i$ and $i+1$ has a sink or sources, there are
always well defined helicity flows among the external states:
\begin{equation}\eqlabel{eq:RRcoll}
 \begin{array}{ccc}
  \raisebox{-1.6cm}{\scalebox{.40}{\includegraphics{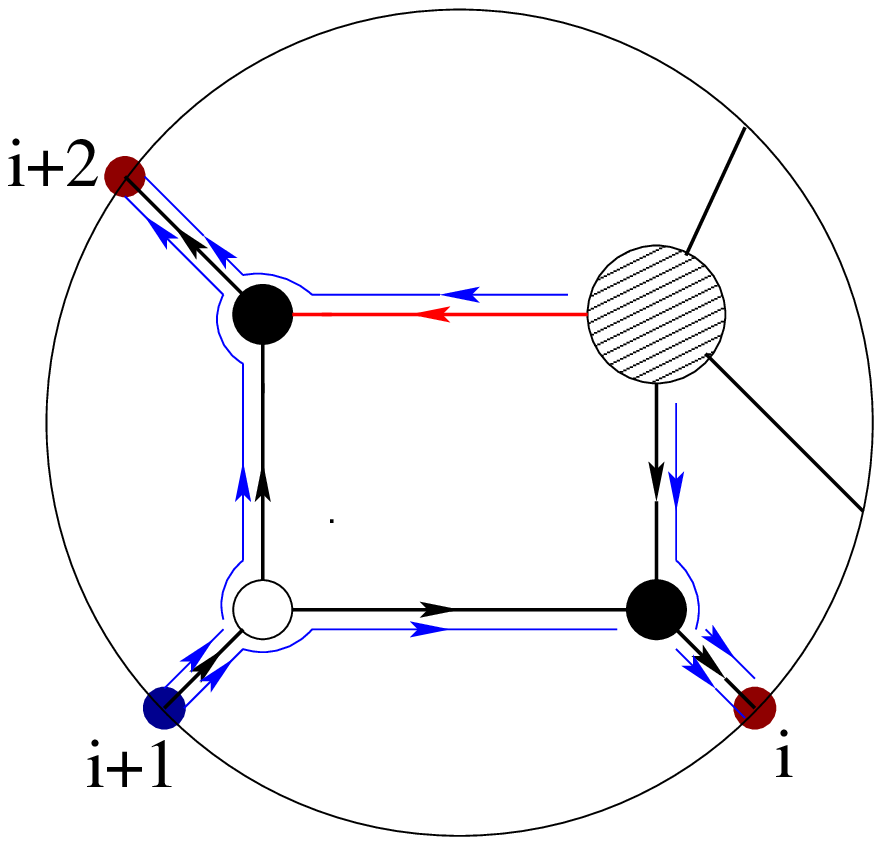}}} 
 &
  \qquad\longrightarrow\qquad
 &
  \raisebox{-1.6cm}{\scalebox{.40}{\includegraphics{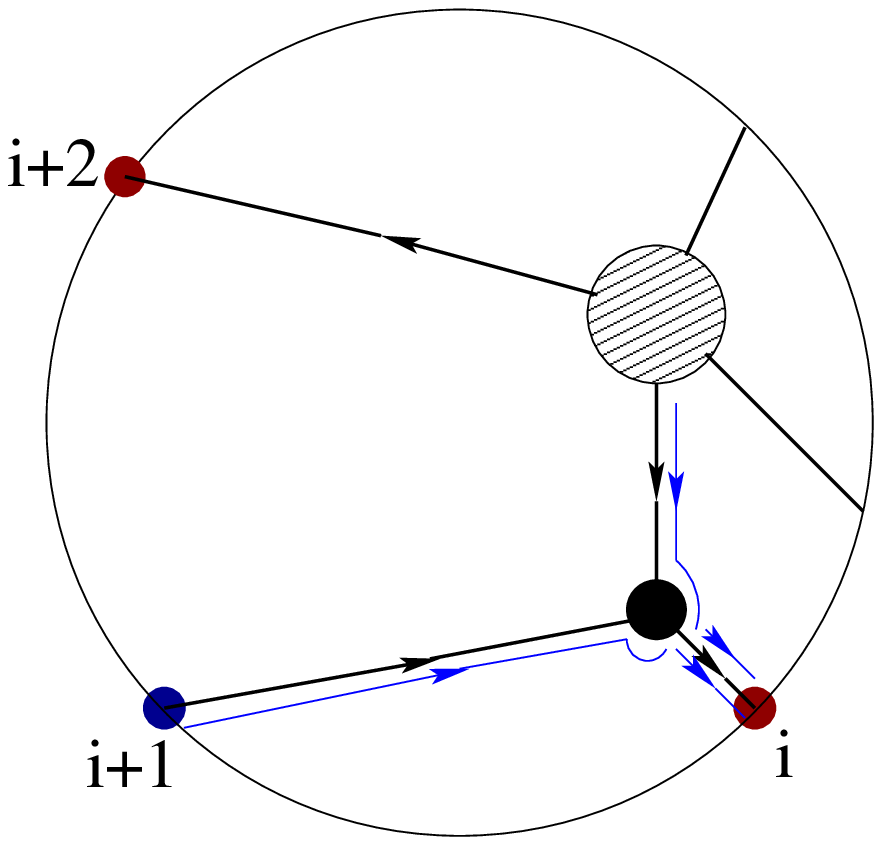}}}   
 \\
 {} & {} & {}
 \\
  \raisebox{-1.6cm}{\scalebox{.40}{\includegraphics{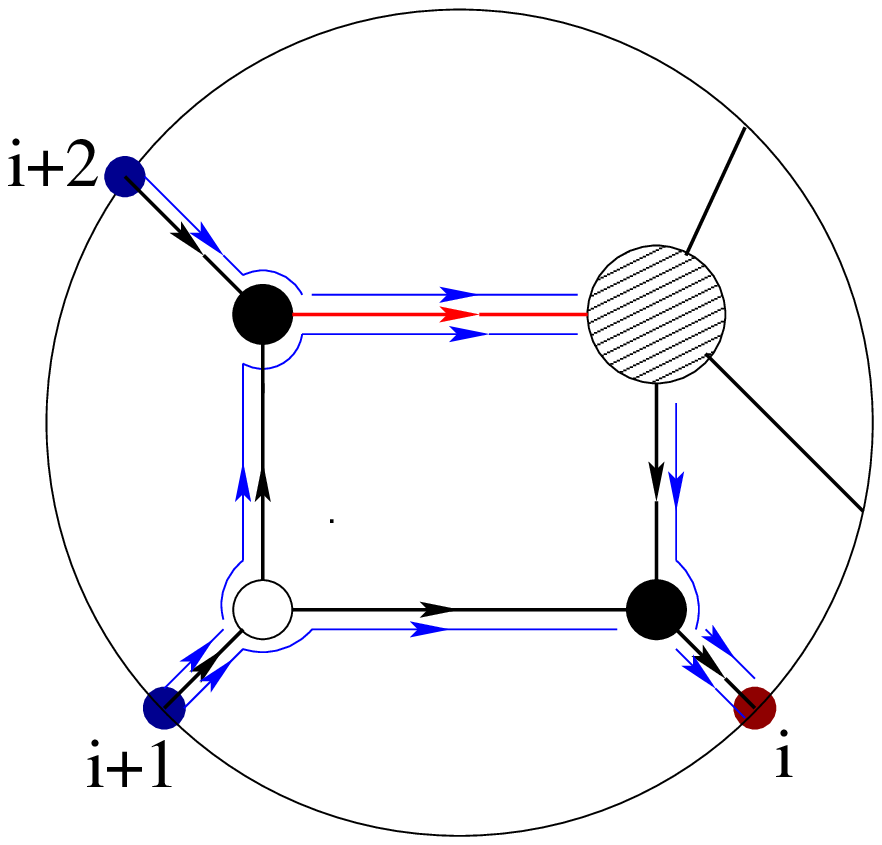}}}
 &
  \qquad\longrightarrow\qquad
 &\quad
  \raisebox{-1.6cm}{\scalebox{.40}{\includegraphics{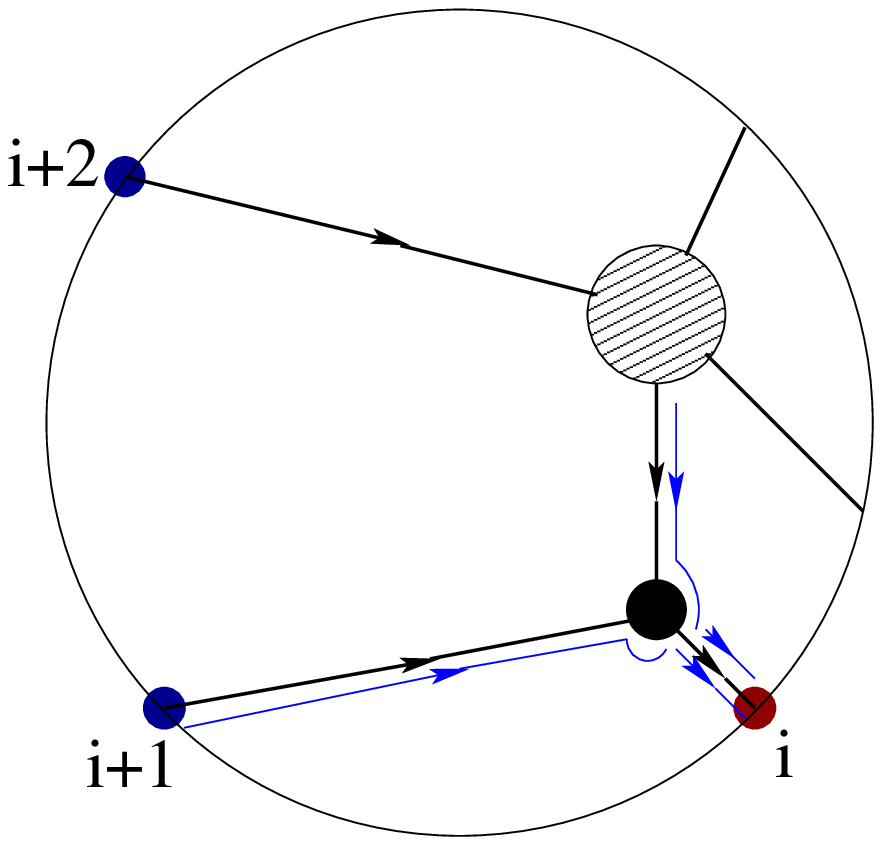}}} 
 \end{array}
\end{equation}
The very same argument applies when the three-particle amplitude is
$\mathcal{M}_3^{\mbox{\tiny $(1)$}}$. These two possibilities
correspond to the two possible complex factorisation channels that 
one can have in a collinear limit 
\cite{Schuster:2008nh, Benincasa:2011kn}. Indeed, not necessarily
an amplitude must factorise under both. However, are exactly 
the helicity flows which tell us whether both are allowed or
just one of them.

Finally, there are further poles that {\it in principle} can arise and which are non-local. However
they always appear in pairs and they cancel, leaving just the physical singularities. This can be
easily see diagrammatically as follows
\begin{equation}\eqlabel{eq:RRnlpoles}
 \raisebox{-1.6cm}{\scalebox{.40}{\includegraphics{RRmp.eps}}}\quad\Longrightarrow\quad
 \raisebox{-1.6cm}{\scalebox{.40}{\includegraphics{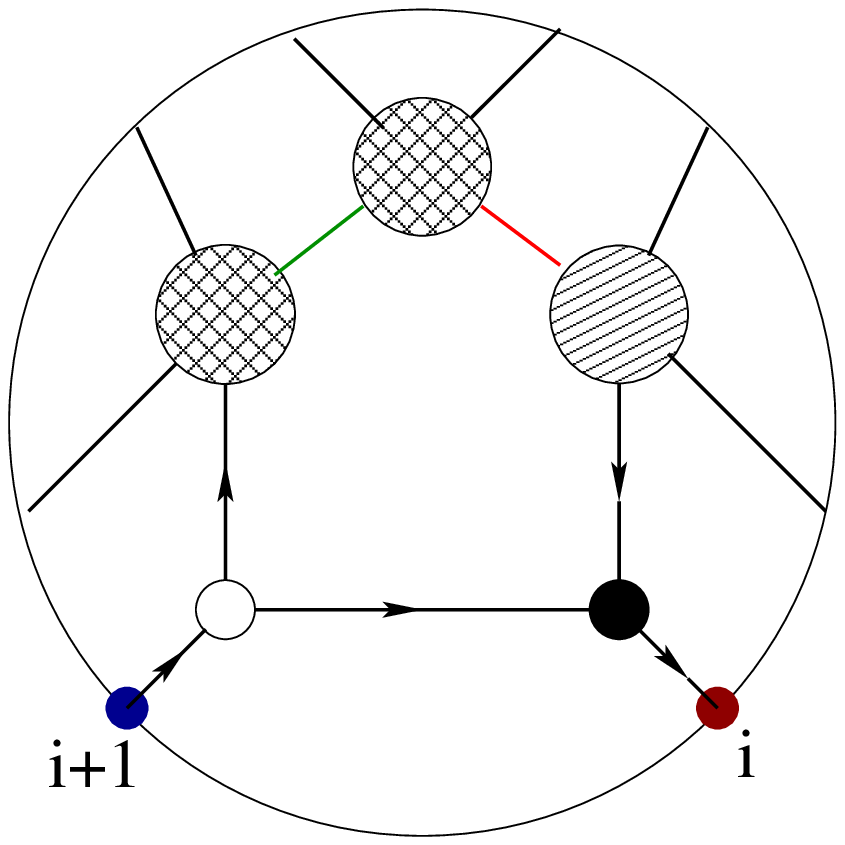}}}\quad+\quad
 \raisebox{-1.6cm}{\scalebox{.40}{\includegraphics{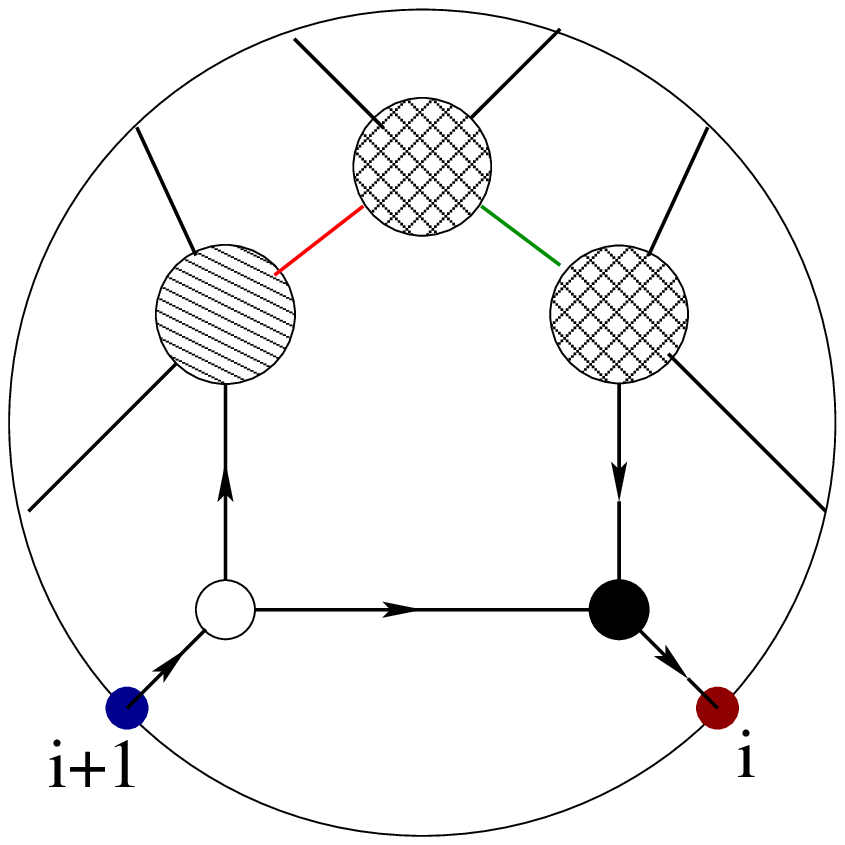}}}
\end{equation}
where the green line indicates the non-local pole, which in the first diagram is generated from the 
sub-amplitude on the left while in the second diagram from the sub-amplitude on the right.

A couple of comments are now in order. Our starting point --
{\it i.e.} the four-point relations in which all the
factorisation channels are manifest \eqref{eq:treeRRind0} -- 
shows just external helicity flows. As we already discussed,
in the case the other possible relation is considered,
there are two contributions each of which show two 
helicity flows in a single channel as well as an internal
helicity loop. This internal helicity loop corresponds to a further
singularity which disappears upon their summation just for
$\mathcal{N}\,=\,3$. Except that for this particular theory,
these diagrams provide an object with a different
singularity structure than a scattering amplitude and, therefore,
they cannot be used as a starting point for an inductive argument.
Furthermore, their presence at any step of the inductive procedure
would invalidate the proof. However, as stated earlier, they can
never arise: This is a consequence of the fact that the
inductive hypothesis just contain full-fledge sub-amplitudes which
themselves can be just expressed in terms of diagrams with
no internal helicity loop!


\subsubsection{On the boundary terms}
\label{subsubsec:BT}

It is instructive to further explore the fully localised diagrams
by comparing the different ones with a given external helicity 
configuration. 

Given a fixed helicity configuration for the external states and
singled out two of them, we define as boundary term the difference 
between the two ways of integrating the on-shell differential 
equation for the amplitude. If this boundary term is non-vanishing,
a singularity has been missed by at least one of the BCFW bridge.
In the case of planar gauge theories, such a boundary term is 
related just to a specific BCFW bridge.

As a starting point, let us consider the four-particle amplitudes
both in the $(-,-,+,+)$ and $(-,+,-,+)$ helicity configurations.
The boundary terms $\Delta\mathcal{M}_{4}^{\mbox{\tiny tree}}$
are then given by
\begin{equation}\eqlabel{eq:Bt}
 \begin{split}
  &\Delta\mathcal{M}_{4}^{\mbox{\tiny tree}}(-,-,+,+)\:=\:
\raisebox{-1.1cm}{\scalebox{.25}{\includegraphics{4ptHelFlow5.eps}}}
    -
\raisebox{-1.1cm}{\scalebox{.25}{\includegraphics{4ptHelFlow4.eps}}}
\\
  &\Delta\mathcal{M}_{4}^{\mbox{\tiny tree}}(-,+,-,+)\:=\:
\raisebox{-1.1cm}{\scalebox{.25}{\includegraphics{4ptHelFlow1.eps}}}
   -
\raisebox{-1.1cm}{\scalebox{.25}{\includegraphics{4ptHelFlow2.eps}}}
   -
\raisebox{-1.1cm}{\scalebox{.25}{\includegraphics{4ptHelFlow3.eps}}}
 \end{split}
\end{equation}
and they can also be seen as a measure of the inequivalence of
a square move. As we saw earlier, the two terms in the first line of
\eqref{eq:Bt} are actually always ({\it i.e.} for any $\mathcal{N}$) 
equal -- as a quick analysis of the helicity flows shows. As
far as the boundary term in the second line is concerned, while
the first diagram shows all the complex factorisation in both the
$s$- and $t$-channels, the second and third diagrams shows the 
two complex factorisations in just one channel ($t$ and $s$, 
respectively) while the presence of oriented internal helicity loops
reveal the presence of a extra poles
\begin{equation}\eqlabel{eq:Bt2}
 \begin{split}
 \Delta\mathcal{M}_4^{\mbox{\tiny tree}}(-,+,-,+)\:&=\:
    \mathcal{M}_4^{\mbox{\tiny tree}}(-,+,-,+)(-1)^{4-\mathcal{N}}
    \sum_{k=1}^{3-\mathcal{N}}
    \begin{pmatrix}
     4-\mathcal{N}\\
     k
    \end{pmatrix}
    \frac{s^kt^{4-\mathcal{N}-k}}{u^{4-\mathcal{N}}}\:=\\
  &=\:\mathcal{M}_4^{\mbox{\tiny tree}}(-,+,-,+)
  \left[
   \varepsilon_{\mathcal{N},3}(4-\mathcal{N})\frac{st}{u^2}-
   \delta_{\mathcal{N},0}2\frac{s^2t^2}{u^4}
  \right].
 \end{split}
\end{equation}
The $\mathcal{N}\,=\,3$ case is the only one in which the poles
related to the internal loops are simple pole and cancel when
the two contributions get summed. For $\mathcal{N}\,\le\,2$, these
poles become higher order a no cancellation occurs. More precisely,
the poles do not disappear, however, while
the theory is supersymmetric, cancellations occur in such a way that
just double poles appear\footnote{Here the order of the pole is
actually refer to the function obtained dividing \eqref{eq:Bt2} by the tree-level amplitude}. In particular, this 
boundary term appears to be universal, up to an overall constant, in its kinematic 
structure ({\it i.e.} it does not depend on $\mathcal{N}$ as long as
$\mathcal{N}\,\le\,2$). 
Just for $\mathcal{N}\,=\,0$, an extra term appears
showing a fourth order pole. 

If we instead look at higher point localised diagrams, it is
interesting to notice that in the MHV sector the boundary term 
defined through diagrams containing {\it at most} one internal
helicity loop can be constructed from the one we just discussed
by adding inverse soft factors
\begin{equation}\eqlabel{eq:BtMHV}
 \Delta\mathcal{M}_{n}^{\mbox{\tiny tree}}
  (1^{-},2^{+},3^{-},4^{+},\ldots,n^{+})\:=\:
 \Delta\mathcal{M}_{4}^{\mbox{\tiny tree}}
  (1^{-},2^{+},3^{-},n^{+})\otimes
 \bigotimes_{k=0}^{n-5}\mbox{Soft}_{+}^{\mbox{\tiny tree}}(n-1-k),
\end{equation}
with $\mbox{Soft}_{+}^{\mbox{\tiny tree}}(n-1-k)$ being the soft 
factor for the positive helicity multiplet labelled by $n-1-k$, 
{\it i.e.} three-particle 
amplitude suitably attached to the lower-point amplitude. 
This is just a consequence of the 
fact that the full-fledge amplitudes in the MHV sector can be 
constructed by inverse soft procedure (they are always given by a 
single on-shell diagram irrespectively of the number of external 
states), as well as in the definition of $\Delta\mathcal{M}_{n}$ 
diagrams with at most one internal helicity loop are considered.
Exactly the same construction holds for the $\bar{\mbox{MHV}}$
sector
\begin{equation}\eqlabel{eq:BtAMHV}
 \Delta\mathcal{M}_{n}^{\mbox{\tiny tree}}
  (1^{-},2^{+},3^{-},4^{+},5^{-}\ldots,n^{-})\:=\:
 \Delta\mathcal{M}_{4}^{\mbox{\tiny tree}}
  (1^{-},2^{+},3^{-},4^{+})\otimes
 \bigotimes_{k=5}^{n}\mbox{Soft}_{-}^{\mbox{\tiny tree}}(k).
\end{equation}

\subsection{On-shell $1$-forms and $2$-forms: the one-loop 
structure}
\label{subsec:1loopStr}

Let us now turn to more interesting diagrams, beginning with those
ones which can be obtained from the localised diagrams via the
application of a BCFW bridge, as in Figure \ref{fig:trplcut1}. This is the same class of diagrams we discussed in Section 
\ref{subsec:BCFWbM}, where we analysed its pole structure and 
emphasised mostly the tree level information. Actually, the very
same information is directly connected with the loop structure of
the theory. Specifically, this class of diagram corresponds, in
a more standard language, to a triple cut of a one-loop amplitude.
More precisely, it corresponds to one of the two families of 
solutions of the triple-cut equations. Let us analyse it in some
detail. The one-form $\mathcal{M}_{4}^{\mbox{\tiny $(1)$}}(z)$ 
generated by BCFW-bridging the tree-level four-particle amplitude 
$\mathcal{M}_4(-,+,-,+)$ can be written as
\begin{equation}\eqlabel{eq:OS3cut1}
 \begin{split}
  \mathcal{M}_4^{\mbox{\tiny $(1)$}}(z)
  \:&=\:\mathcal{M}_{4}^{\mbox{\tiny tree}}\frac{dz_{12}}{z_{12}
  \left(1+\frac{\langle4,2\rangle}{\langle4,1\rangle}z_{12}\right)}
  \left[
   \left(
    1+\frac{\langle2,3\rangle}{\langle1,3\rangle}z_{12}
   \right)^{4-\mathcal{N}}
   +
   \left(
    -\frac{\langle2,3\rangle}{\langle1,3\rangle}z_{12}
   \right)^{4-\mathcal{N}}
  \right]\:=\\
  &=\:\mathcal{M}_{4}^{\mbox{\tiny tree}}\frac{dz_{34}}{z_{34}
  \left(1+\frac{\langle3,1\rangle}{\langle4,1\rangle}z_{34}\right)}
  \left[
   \frac{\left(-\frac{s}{u}\right)^{4-\mathcal{N}}+
      \left(-\frac{t}{u}\right)^{4-\mathcal{N}}
       \left(1+\frac{\langle3,1\rangle}{\langle4,1\rangle}
    z_{34}\right)^{4-\mathcal{N}}}{\left(1+\frac{\langle2,3\rangle}{
    \langle2,4\rangle}z_{34}\right)^{4-\mathcal{N}}}
  \right],
 \end{split}
\end{equation}
where the second line represents the parametrisation of this
diagram looked as a BCFW bridge of the two helicity loop 
four-particle diagrams. This on-shell one-form shows three singular
points: 
$z_{12}\,=\,0,\,\infty,\,-\langle4,1\rangle/\langle4,2\rangle$,
or, equivalently in the $z_{34}$ parametrisation, 
$z_{34}\,=\,0,\,-\langle2,4\rangle/\langle2,4\rangle,\,
 -\langle4,1\rangle/\langle4,2\rangle$.
As mentioned in Section \ref{subsec:BCFWbM}, the integration on a 
$T^1$ around the points $z_{12}\,=\,0$ 
($z_{34}\,=\,-\langle4,1\rangle/\langle4,2\rangle$)
and $z_{12}\,=\,-\langle4,1\rangle/\langle4,2\rangle$ 
($z_{34}\,=\,0$)
returns the two leading singularities, {\it i.e.} this integration
localises the on-shell form on diagrams which has either external
helicity flows only or internal helicity loops. The one-form
\eqref{eq:OS3cut1} shows also a multiple pole which is at infinity
in the $z_{12}$ parametrisation or at finite location in the 
$z_{34}$-one. The related residue is nothing but the boundary term
of the tree-level amplitude we started with, but it also provides
the coefficient of the scalar triangle in the $(1,2)$-channel.

Let us now turn to the following on-shell two-forms with four
external states:
\begin{equation}\eqlabel{eq:OS2cut}
 \begin{split}
  \mathcal{M}^{\mbox{\tiny $(2)$}}_{4}(\zeta)\:&=\: 
  \raisebox{-1.25cm}{\scalebox{.30}{\includegraphics{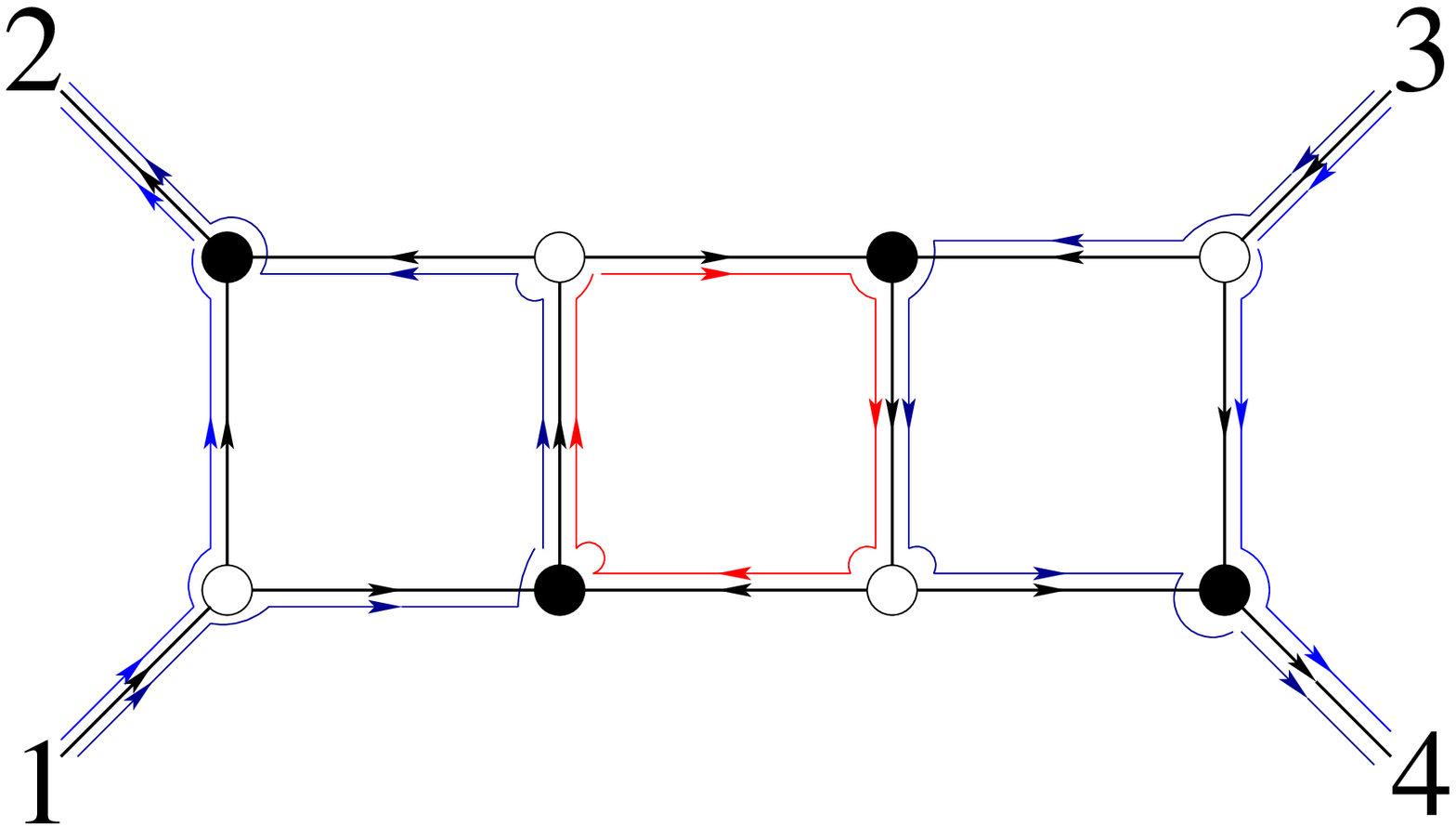}}}
  +
  \raisebox{-1.25cm}{\scalebox{.30}{\includegraphics{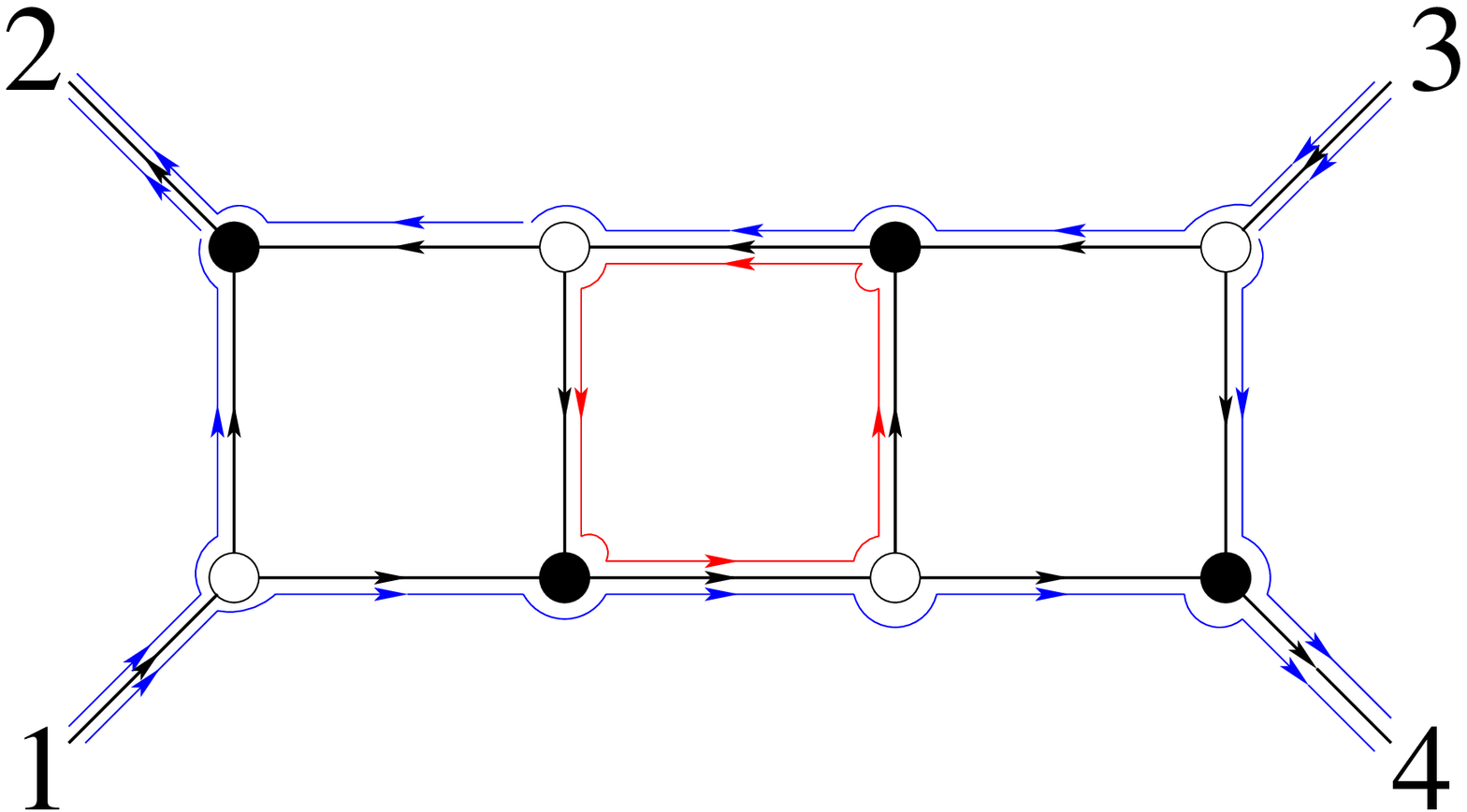}}}\:=
  \\
  &=\:\mathcal{M}_4^{\mbox{\tiny tree}}\,
   \frac{d\zeta_{12}}{\zeta_{12}}\wedge
   \frac{d\zeta_{34}}{\zeta_{34}}
   \frac{
    \left(-\frac{t}{u}\right)^{4-\mathcal{N}}
    +
    \left(-\frac{s}{u}\right)^{4-\mathcal{N}}
    \left(1-\zeta_{12}\right)^{4-\mathcal{N}}
    \left(1-\zeta_{34}\right)^{4-\mathcal{N}}
   }{
    \left[
     -\frac{t}{u}-\frac{s}{u}\left(1-\zeta_{12}\right)
     \left(1-\zeta_{34}\right)
    \right]^{4-\mathcal{N}}
   },
 \end{split}
\end{equation}
which can be looked at as two BCFW bridges applied onto the two
{\it chiral} boxes, {\it i.e.} one of the two leading singularities,
with $\zeta_{12}$ and $\zeta_{34}$ being the free degrees of 
freedom induced by the bridging in the $(1,2)$- and $(3,4)$-channel
respectively\footnote{The degrees of freedom parametrised by
$\zeta_{12}$ and $\zeta_{34}$ are related to the original BCFW 
parameters (which we typically indicate with) $z$ by a M{\"o}bius
transformation \eqref{eq:MobT2}.}. The helicity flows highlight
the equivalence operation which are allowed: while the second
contribution \eqref{eq:OS2cut} is rigidly fixed (the helicity flows
among the external states are all outer to the diagram), in the
first one, it is possible to apply first a square move on each
of the most external boxes and perform a bubble deletion on 
clockwise on-shell bubbles. 
The presence of such on-shell bubble after equivalence operations 
can be predicted because of the presence of the clockwise 
{\it chiral} box. In any case, in both terms, the presence of
the {\it chiral} boxes implies the presence of a multiple pole.
As usual, a big deal of physical information can be extracted from 
the (multivariate) residues of the singularities.


Differently from the case analysed earlier, now the integration
contours are $2$-cycles in $\mathbb{CP}^2$ and, in principle,
the multivariate residues can depend on the ordering in which the
integration is performed\footnote{For a discussion about the
use of multivariate residues to compute (maximal) unitarity cuts, 
see \cite{Sogaard:2013fpa}.}. 

Using the homogeneous coordinates $[w_0,\,w_{12},\,w_{34}]$ for
$\mathbb{CP}^2$, which are related to our local coordinates 
$(\zeta_{12},\,\zeta_{34})\,\in\,\mathbb{C}^2$ via the map
$[w_0,\,w_{12},\,w_{34}]\,\longrightarrow\,
 [\zeta_{12}(w),\,\zeta_{34}(w)]$ with 
$\zeta_{i,i+1}\,=\,w_{i,i+1}/w_0$, the two-form 
$\mathcal{M}_{4}^{\mbox{\tiny $(2)$}}$ in
\eqref{eq:OS2cut} is well-defined everywhere except on
the following hypersurfaces:
\begin{equation}\eqlabel{eq:hypsing}
 w_0\,=\,0,\quad w_{12}\,=\,0,\quad w_{34}\,=\,0,\quad 
  -\frac{s}{u}(w_0-w_{12})(w_0-w_{34})-\frac{t}{u}w_0^2\,=\,0,
\end{equation}
with $w_0\,=\,0$ being the point at infinity. The singularities
can be seen as the discrete set of points given by two divisors
defined out of the hypersurfaces \eqref{eq:hypsing}. Let us 
begin with defining the following divisors 
\begin{equation}\eqlabel{eq:divs1}
 D_{1}\,=\,
  \left\{
   w_0 w_{12}\,=\,0
  \right\}
 \;\mbox{and}\;
 D_{2}\,=\,
  \left\{
   w_{34}
   \left(
    -\frac{s}{u}(w_0-w_{12})(w_0-w_{34})-\frac{t}{u}w_0^2
   \right)^{4-\mathcal{N}}\,=\,0\,
  \right\}.
\end{equation}
The intersection $D_1\,\cap\,D_2$ is a discrete set whose elements
are singular points for the two-form under analysis:
\begin{equation}\eqlabel{eq:divint1}
 D_1\,\cap\,D_2\:=\:
 \left\{
  \left.[w_0,\,w_{12},\,w_{34}]\,\in\,\mathbb{CP}^2\,\right|\,
  [0,\,1,\,0],\,[1,\,0,\,0],\,
  \left[1,\,0,\,-\frac{u}{s}\right]
 \right\}.
\end{equation}
Such a set contains one point, the first one in \eqref{eq:divint1}, 
at infinity and the origin as single poles, as well as a further
point (the third one) as a multiple pole.

The two-form \eqref{eq:OS2cut} is invariant under the label
exchange $(1,2)\,\longleftrightarrow\,(3.4)$, so that two more
divisors, and therefore their intersection, is obtained from
\eqref{eq:divs1} and \eqref{eq:divint1} by such a label
exchange.

The two intersections so defined expose two poles at infinity,
the origin and two multiple poles. It is not difficult to see that
the residue of both poles at infinity corresponds to the leading 
singularity with no internal helicity loop. 
In the same way, the residues at the origin and at the multiple
pole respectively provides the other
leading singularity and the scalar triangle coefficient. 
Furthermore, the global residue theorem \cite{Griffiths:1978gh} 
implies that the sum of the residues of the two-form for each 
divisor intersection above vanishes, {\it i.e.} the global residue
theorem relates the leading singularities and (one) coefficient of
the scalar triangle, returning the tree level relations discussed
in the previous subsection.

A further choice for the divisors can be done by taking 
\begin{equation}\eqlabel{eq:divs2}
 D'_{1}\,=\,
  \left\{
   w_{12}\,w_{34}\,=\,0
  \right\}
 \;\mbox{and}\;
 D'_{2}\,=\,
  \left\{
   w_{0}
   \left(
    -\frac{s}{u}(w_0-w_{12})(w_0-w_{34})-\frac{t}{u}w_0^2
   \right)^{4-\mathcal{N}}\,=\,0\,
  \right\},
\end{equation}
whose intersection is made of two points at infinity and two
multiple poles
\begin{equation}\eqlabel{eq:divint2}
 D'_1\,\cap\,D'_2\:=\:
 \left\{
  \left.[w_0,\,w_{12},\,w_{34}]\,\in\,\mathbb{CP}^2\,\right|\,
  [0,\,0,\,1],\,[0,\,1,\,0],\,
  \left[1,\,0,\,-\frac{u}{s}\right],\,
  \left[1,\,-\frac{u}{s},\,0\right]
 \right\}.
\end{equation}
The residues at these points are equal in pairs and, when applying
the global residue theorem, they acquire opposite sign so that the
overall sign over the residues for this intersection is also zero.
However, contrarily to the previous case no interesting physical
relation is encoded.

The $2$-cycles described above are also straightforward to analyse
directly in the local coordinates $(\zeta_{12},\,\zeta_{34})$ in
\eqref{eq:OS2cut}. In particular:
\begin{itemize}
 \item $\gamma_{\mbox{\tiny LS}}^{\mbox{\tiny $(1)$}}\,=\,
       \{(\zeta_{23},\,\zeta_{41})\,\in\hat{\mathbb{C}}^2\,|\,
         \zeta_{23}\,=\,0,\,\zeta_{41}\,=\,0\}$ returns one of
       the leading singularities:
       \begin{equation}\eqlabel{eq:OS2cutcont1}
        \begin{split}
         \oint_{\gamma_{\mbox{\tiny LS}}^{\mbox{\tiny $(1)$}}}\,
          \mathcal{M}_{4}^{\mbox{\tiny $(2)$}}(\zeta)\:&=\:
          \raisebox{-1.25cm}{\scalebox{.25}{\includegraphics{4ptHelFlow2.eps}}}\hspace{-.2cm}\quad+\:
          \raisebox{-1.25cm}{\scalebox{.25}{\includegraphics{4ptHelFlow3.eps}}}\hspace{-.5cm}\qquad=\\
         &=\:\mathcal{M}_{4}^{\mbox{\tiny tree}}
          \left[
           \left(-\frac{t}{u}\right)^{4-\mathcal{N}}+
           \left(-\frac{s}{u}\right)^{4-\mathcal{N}}
          \right]
        \end{split}
       \end{equation}
 \item $\gamma_{\mbox{\tiny LS}}^{\mbox{\tiny $(2)$}}\,=\,
       \{(\zeta_{23},\,\zeta_{41})\,\in\hat{\mathbb{C}}^2\,|\,
         \zeta_{23}\,=\,\infty,\,\zeta_{41}\,=\,\infty\}$ returns 
       the second leading singularity:
       \begin{equation}\eqlabel{eq:OS2cutcont2}
        \oint_{\gamma_{\mbox{\tiny LS}}^{\mbox{\tiny $(2)$}}}\,
        \mathcal{M}_{4}^{\mbox{\tiny $(2)$}}(\zeta)\:=\:
        \raisebox{-1.25cm}{\scalebox{.25}{\includegraphics{4ptHelFlow1.eps}}}\:=\:\mathcal{M}_{4}^{\mbox{\tiny tree}}.
       \end{equation}
       Notice that both in this case and in the previous one,
       the integration order does not matter.
 \item $\gamma_{\mbox{\tiny $\triangle$}}^{\mbox{\tiny $(1)$}}\,=\,
       \{(\zeta_{23},\,\zeta_{41})\,\in\hat{\mathbb{C}}^2\,|\,
         \zeta_{23}\,=\,0,\,\zeta_{41}\,=\,-u/t;\:
         \zeta_{41}\,\succ\,\zeta_{23}\}$, where
       $\zeta_{41}\,\succ\,\zeta_{23}$ establishes a lexicographical
       order. Actually, one can integrate over $\zeta_{41}$ first
       by taking $\zeta_{23}$ fixed and then integrating over
       $\zeta_{23}$: The result stays unchanged and is the 
       coefficient of a triangle integral in the $t$-channel.
       \begin{equation}\eqlabel{eq:OScutcont3}
        \begin{split}
         \oint_{\gamma_{\triangle}^{\mbox{\tiny $(1)$}}} 
         \mathcal{M}_{4}^{\mbox{\tiny $(2)$}}(\zeta)\:&=\:
         \raisebox{-1.25cm}{\scalebox{.30}{\includegraphics{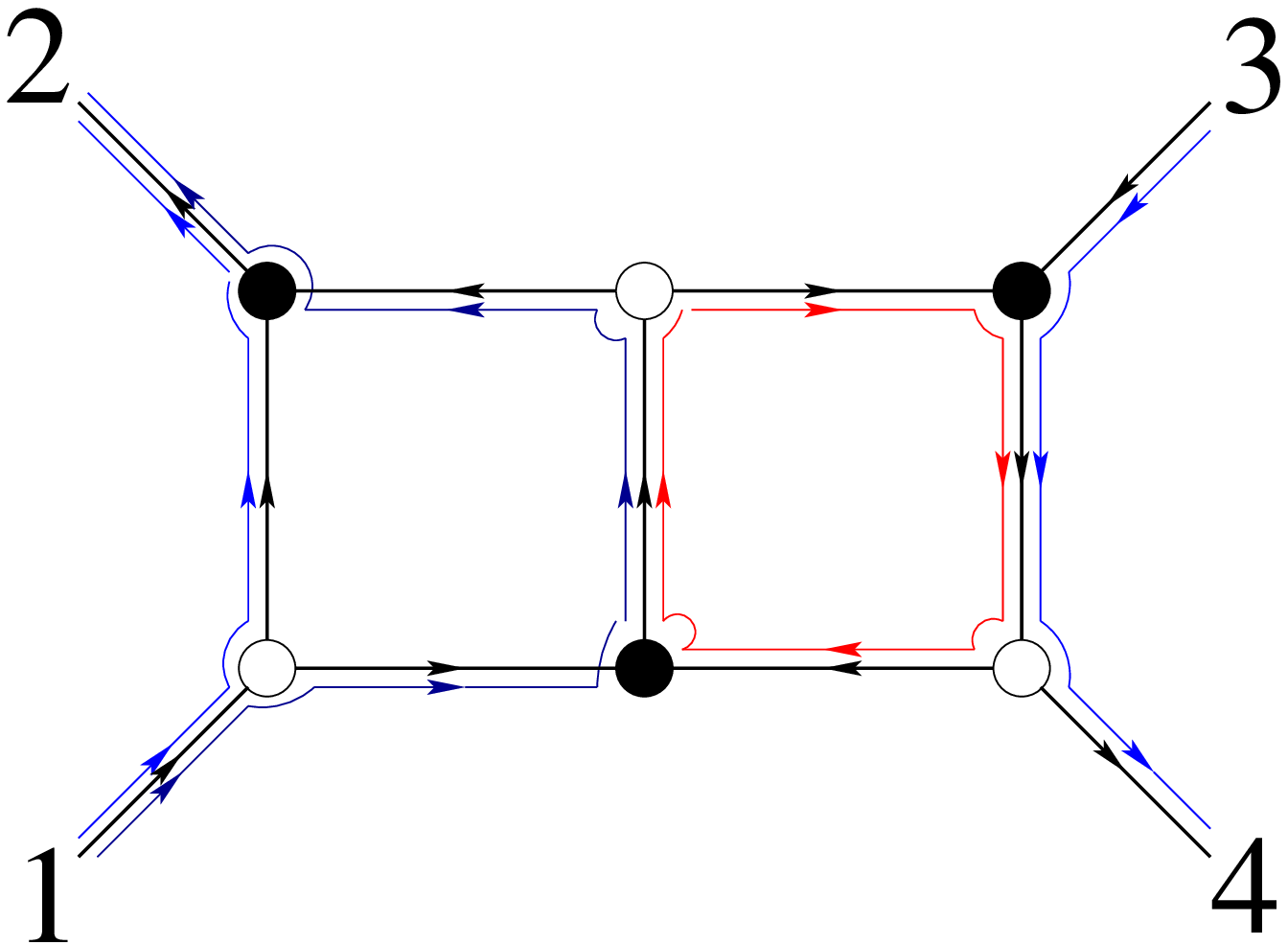}}}+
        \raisebox{-1.25cm}{\scalebox{.30}{\includegraphics{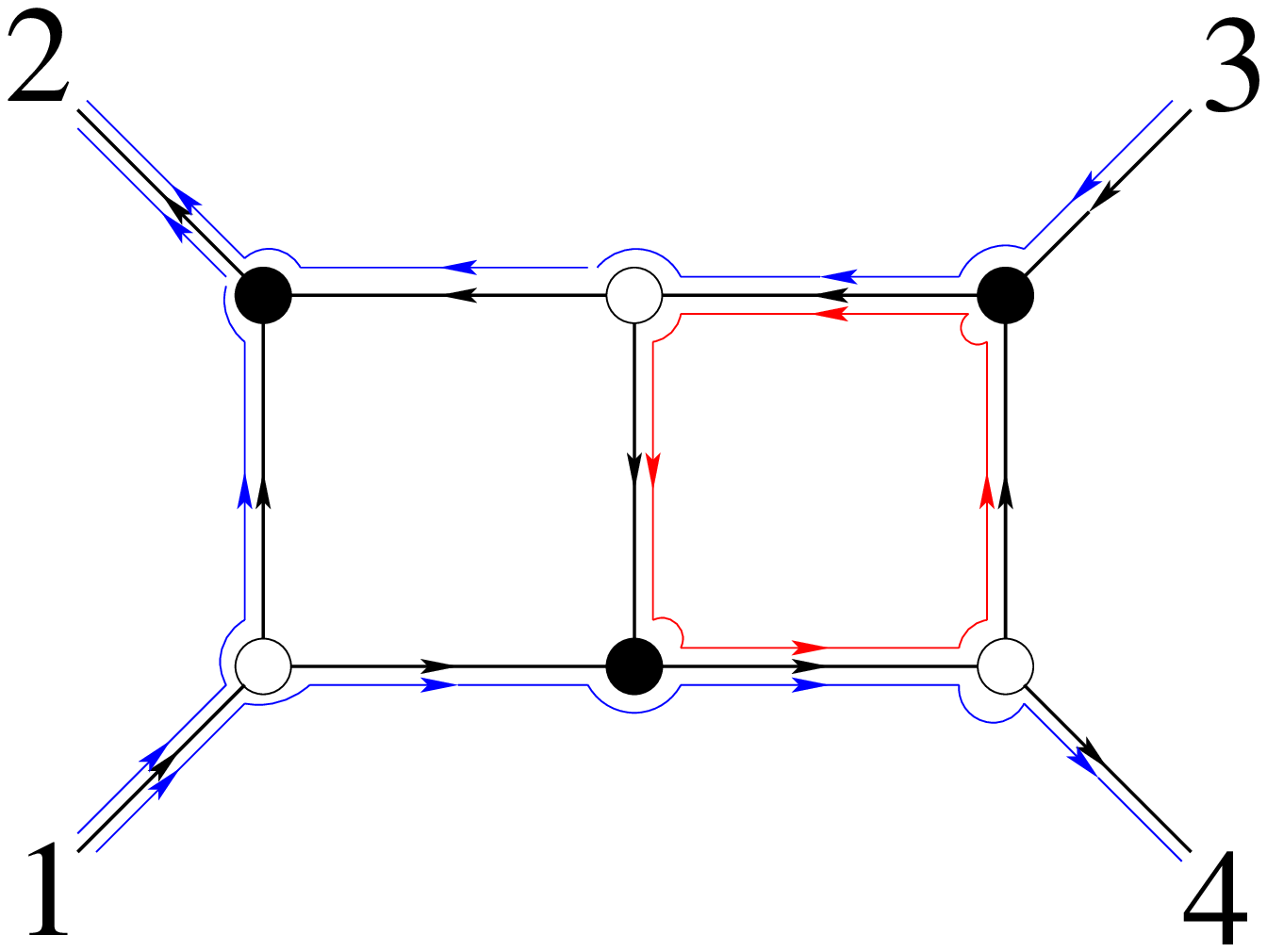}}}
        \:=\\
        &=\mathcal{M}_{4}^{\mbox{\tiny tree}}
         \left[
          \varepsilon_{\mathcal{N},3}(4-\mathcal{N})\frac{st}{u^2}
          -\delta_{\mathcal{N},0}2\frac{s^2 t^2}{u^4}
         \right]
        \end{split}
       \end{equation}
 \item $\gamma_{\mbox{\tiny $\triangle$}}^{\mbox{\tiny $(2)$}}\,=\,
       \{(\zeta_{23},\,\zeta_{41})\,\in\hat{\mathbb{C}}^2\,|\,
         \zeta_{41}\,=\,0,\,\zeta_{23}\,=\,-u/t;\:
         \zeta_{23}\,\succ\,\zeta_{41}\}$ returns again the
       coefficient of a triangle integral -- the same discussion
       above applies up to the label exchange 
       $23\,\longleftrightarrow\,41$.
\end{itemize}
The extraction of the bubble coefficient is a bit more subtle and
it is related to the multiple pole defined by the last hypersurface
in \eqref{eq:hypsing} with no intersection with any of the other
hypersurfaces.

Thus, the residues of the poles in the $2$-form \eqref{eq:OS2cut}
encodes all the physical information contained in a double cut.
In order to understand how the differences between the 
supersymmetric theories and pure Yang-Mills arise, it is
instructive to rewrite \eqref{eq:OS2cut} as follows:
\begin{equation}\eqlabel{eq:OS2cut2}
 \begin{split}
  \mathcal{M}_4^{\mbox{\tiny $(2)$}}(\zeta)\:&=\:
  \frac{d\zeta_{12}}{\zeta_{12}}\wedge\frac{d\zeta_{34}}{\zeta_{34}}
  \left\{
   1-\varepsilon_{\mathcal{N},3}
     \left[
      \frac{(4-\mathcal{N})\left(-\frac{t}{u}\right)}{
       -\frac{t}{u}-\frac{s}{u}\left(1-\zeta_{12}\right)
      \left(1-\zeta_{34}\right)}-
     \right.
  \right.\\
  &\left.\left.
    -\frac{(4-\mathcal{N})\left(-\frac{t}{u}\right)^2}{
     \left[-\frac{t}{u}-\frac{s}{u}\left(1-\zeta_{12}\right)
     \left(1-\zeta_{34}\right)\right]^2}
   \right]+
   \delta_{\mathcal{N},0}
   \left[
    \frac{2\left(-\frac{t}{u}\right)^2\left(-\frac{s}{u}\right)}{
    \left[-\frac{t}{u}-\frac{s}{u}\left(1-\zeta_{12}\right)
     \left(1-\zeta_{34}\right)\right]^3}-
   \right.\right.\\
  &-\left.\left.
    \frac{2\left(-\frac{t}{u}\right)^3\left(-\frac{s}{u}\right)}{
    \left[-\frac{t}{u}-\frac{s}{u}\left(1-\zeta_{12}\right)
     \left(1-\zeta_{34}\right)\right]^4}
   \right]
  \right\}.
 \end{split}
\end{equation}
The expression above makes manifest that the difference in
between the structure of supersymmetric case and pure Yang-Mills 
is encoded in the different behaviour with the {\it composite}
pole, which is a direct consequence of the different structure in
the {\it boundary} terms discussed in Section \ref{subsubsec:BT}.

\subsection{Forward amplitudes and singularities}
\label{subsec:FwAmpl}

One of the key observations which allowed to prove the all-loop 
integrand BCFW recursion relation in $\mathcal{N}\,=\,4$ SYM has 
been the correct and non-ambiguous interpretation of the 
singularities characteristic of the loop amplitudes. Specifically, 
when a BCFW deformation of a loop amplitude is performed, also the
loop propagators acquire a BCFW-parameter dependence and the
related residues are given by a single cut. In general,
single cuts can be a highly ill-defined procedure, except for
the $\mathcal{N}\,\ge\,1$ massless supersymmetric planar gauge 
theories and the $\mathcal{N}\,\ge\,2$ massive ones 
\cite{CaronHuot:2010zt}. In those theories, the residues of 
this singularity are interpreted as forward amplitudes of one
perturbative order less.

In this section, we analyse the one-loop four-particle case in
some detail, including the non-supersymmetric case. The
general claim is that the on-shell diagrammatics provides a 
natural definition for the forward limit also in those cases where
the limit is singular, such as $\mathcal{N}\,=\,0$ Yang-Mills, where
some subtleties need to be taken care of. 
As a first step, we consider the following quantity
\begin{equation}\eqlabel{eq:1loopInt}
 \mathcal{M}_{4}^{\mbox{\tiny $(1)$}}\:=\:
\raisebox{-1.8cm}{\scalebox{.40}{\includegraphics{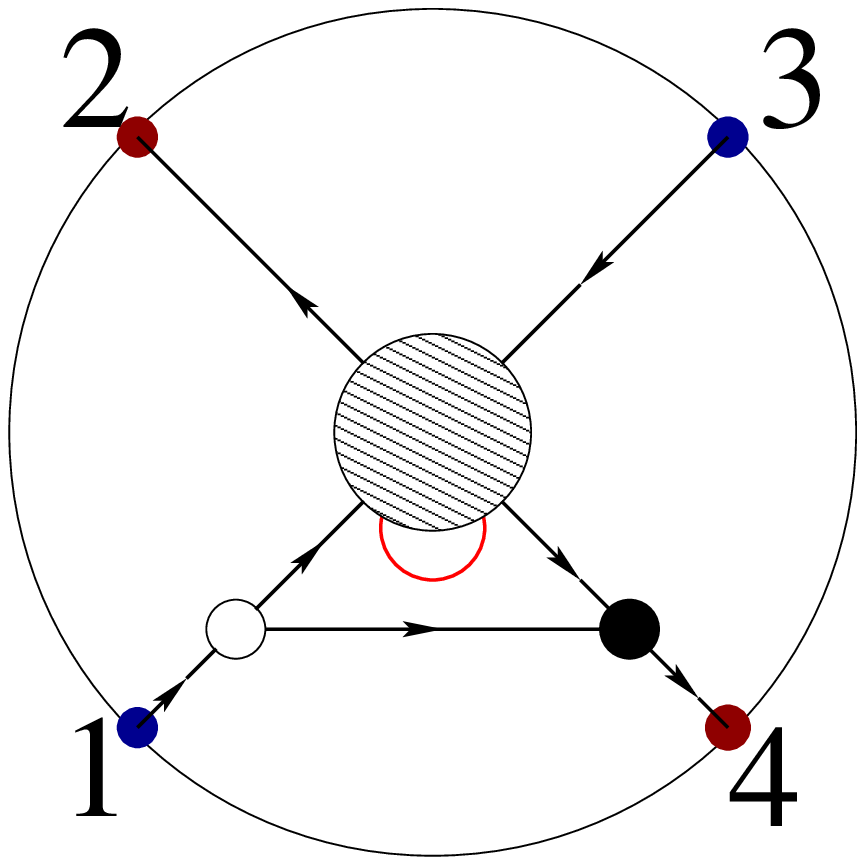}}}
 \:=\:
 \frac{dz_{41}}{z_{41}}\wedge
 \frac{d^2\lambda^{\mbox{\tiny $(AB)$}}
       d^2\tilde{\lambda}^{\mbox{\tiny $(AB)$}}}{
       \mbox{Vol}\{GL(1)\}}
 \,\mathcal{M}_6^{\mbox{\tiny tree}}
 ({\color{red} AB},{\color{red} -AB};\,z_{41})
\end{equation}
where the singularity singled out by the BCFW bridge is the
forward limit of the tree-level six-particle amplitude, while
$z_{41}$ is the parameter related to the $(4,1)$ BCFW bridge. In the 
forward (red) line of \eqref{eq:1loopInt} one has to sum over all the
multiplets which can propagate. In order to study the contribution 
to this singularity let us explicitly consider the (NMHV) 
six-particle amplitudes contributing {\it before} that they are 
taken to be forward:
\begin{equation}\eqlabel{eq:6ptampl}
 \begin{split}
  &\mathcal{M}_6^{\mbox{\tiny tree}}
   ({\color{red} A},{\color{red} B})
   \:=\hspace{-.5cm}
\raisebox{-1.8cm}{\scalebox{.35}{\includegraphics{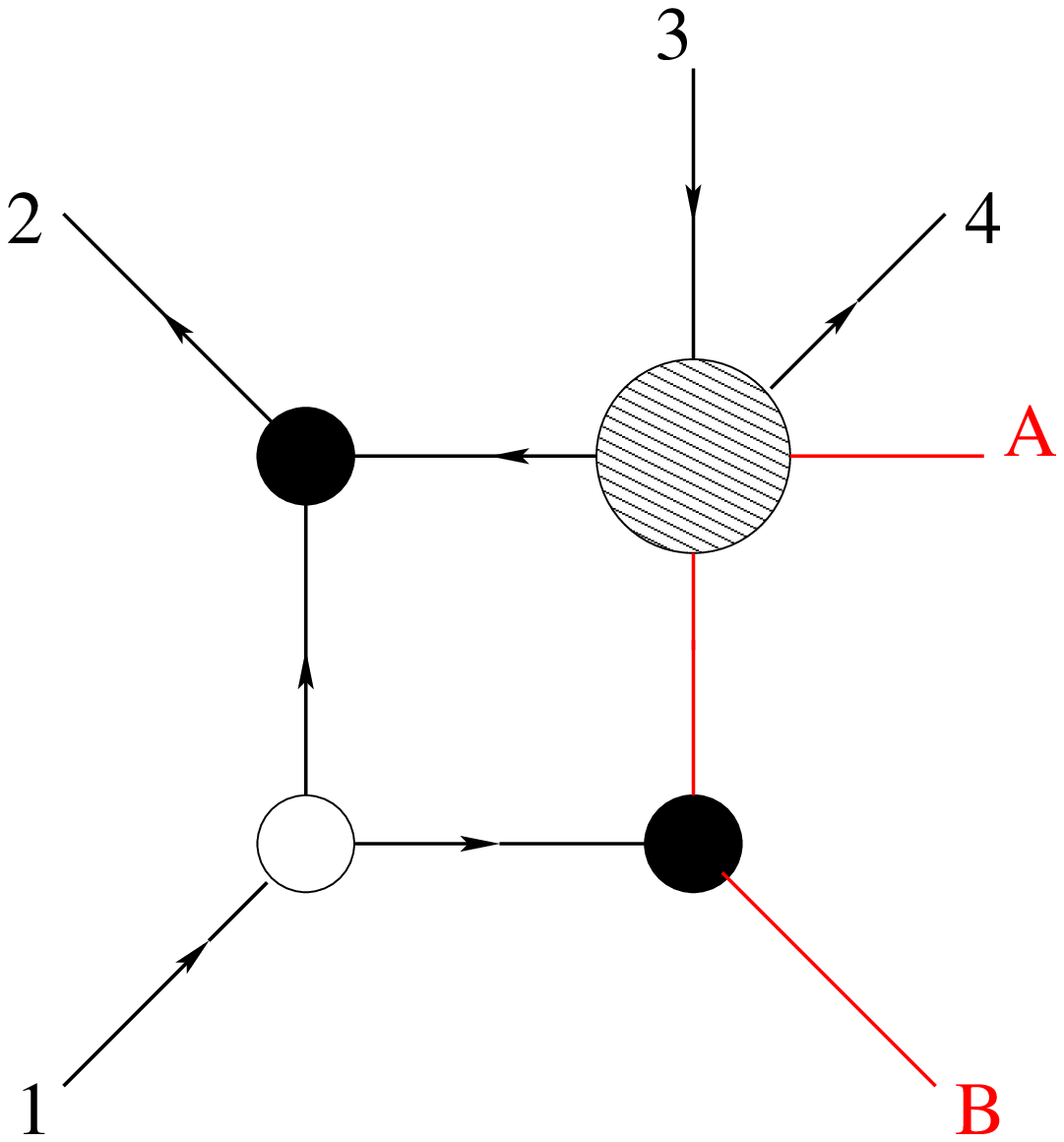}}}
+
\raisebox{-2cm}{\scalebox{.35}{\includegraphics{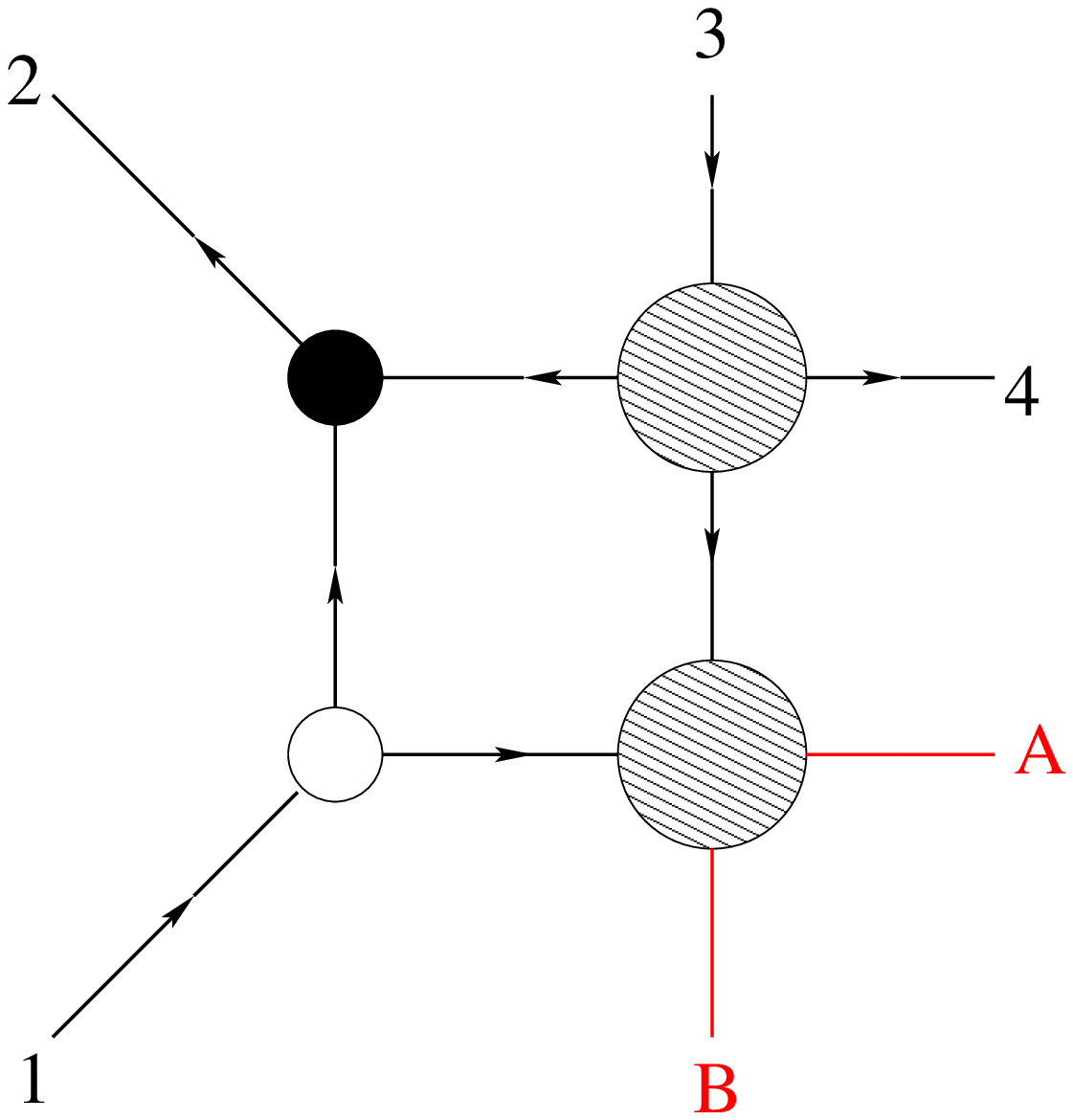}}}
+
\raisebox{-2.2cm}{\scalebox{.35}{\includegraphics{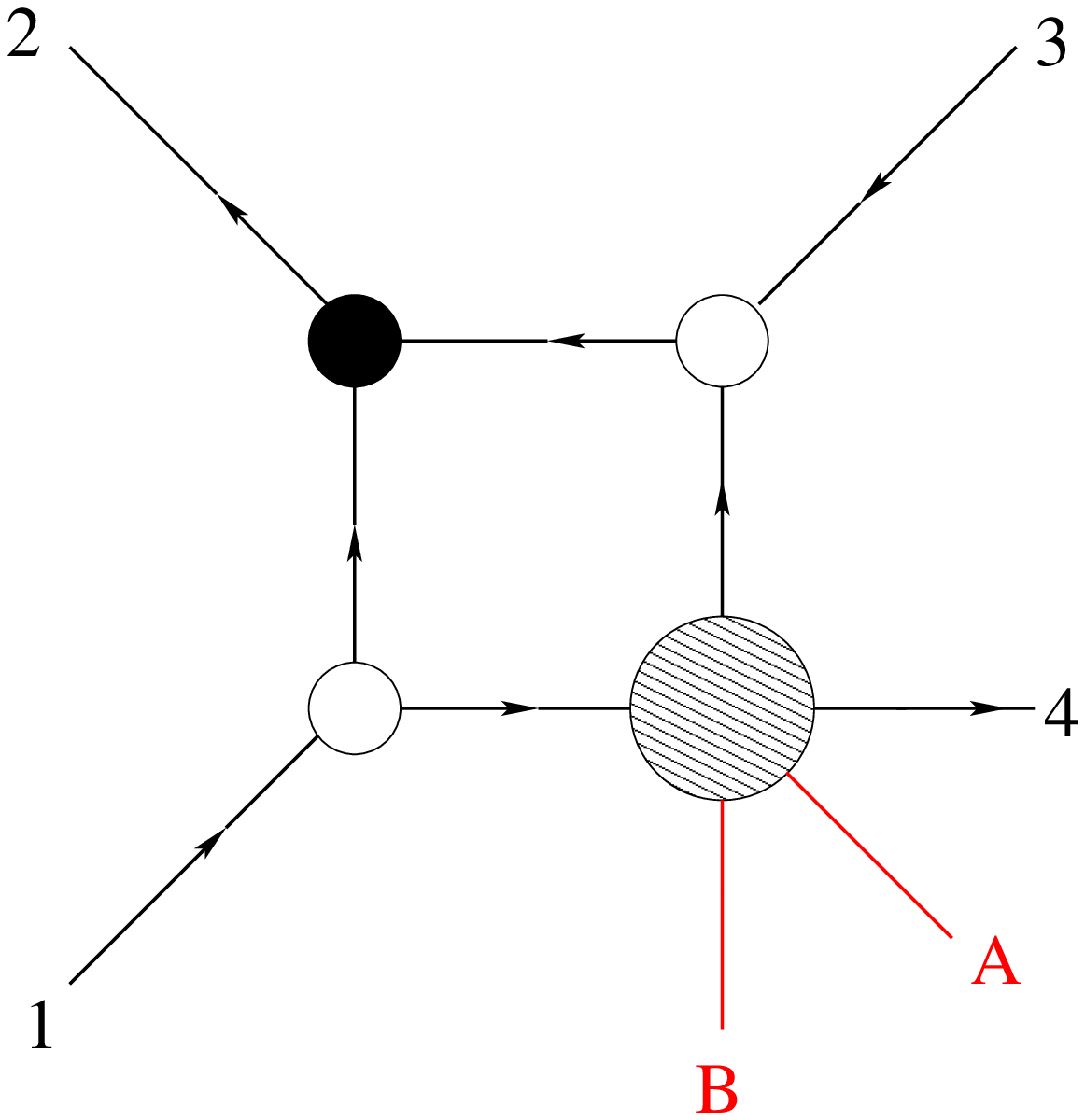}}}
   \\
  &
   \hspace{.2cm}=
\raisebox{-1.6cm}{\scalebox{.22}{\includegraphics{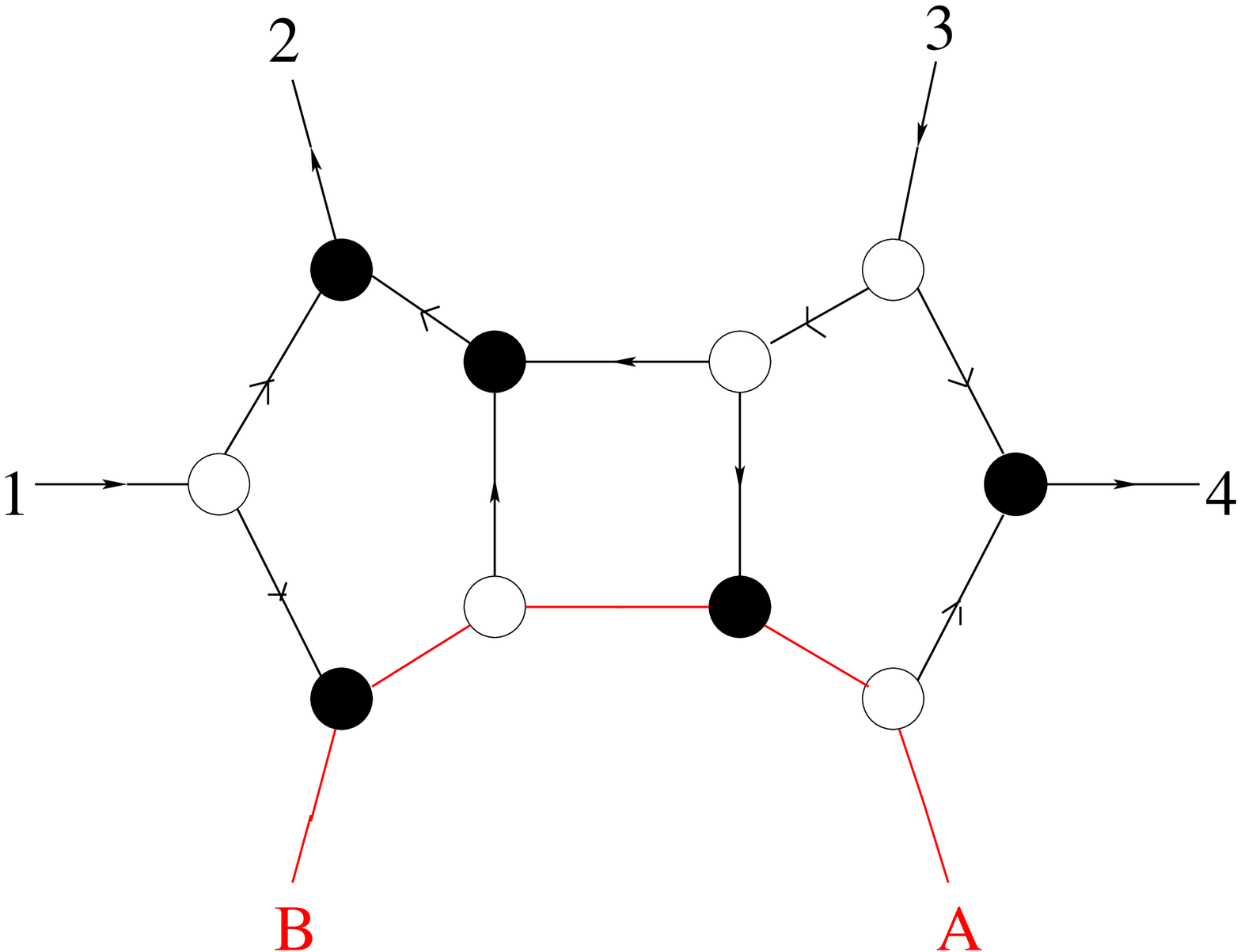}}}
+
\raisebox{-1.6cm}{\scalebox{.22}{\includegraphics{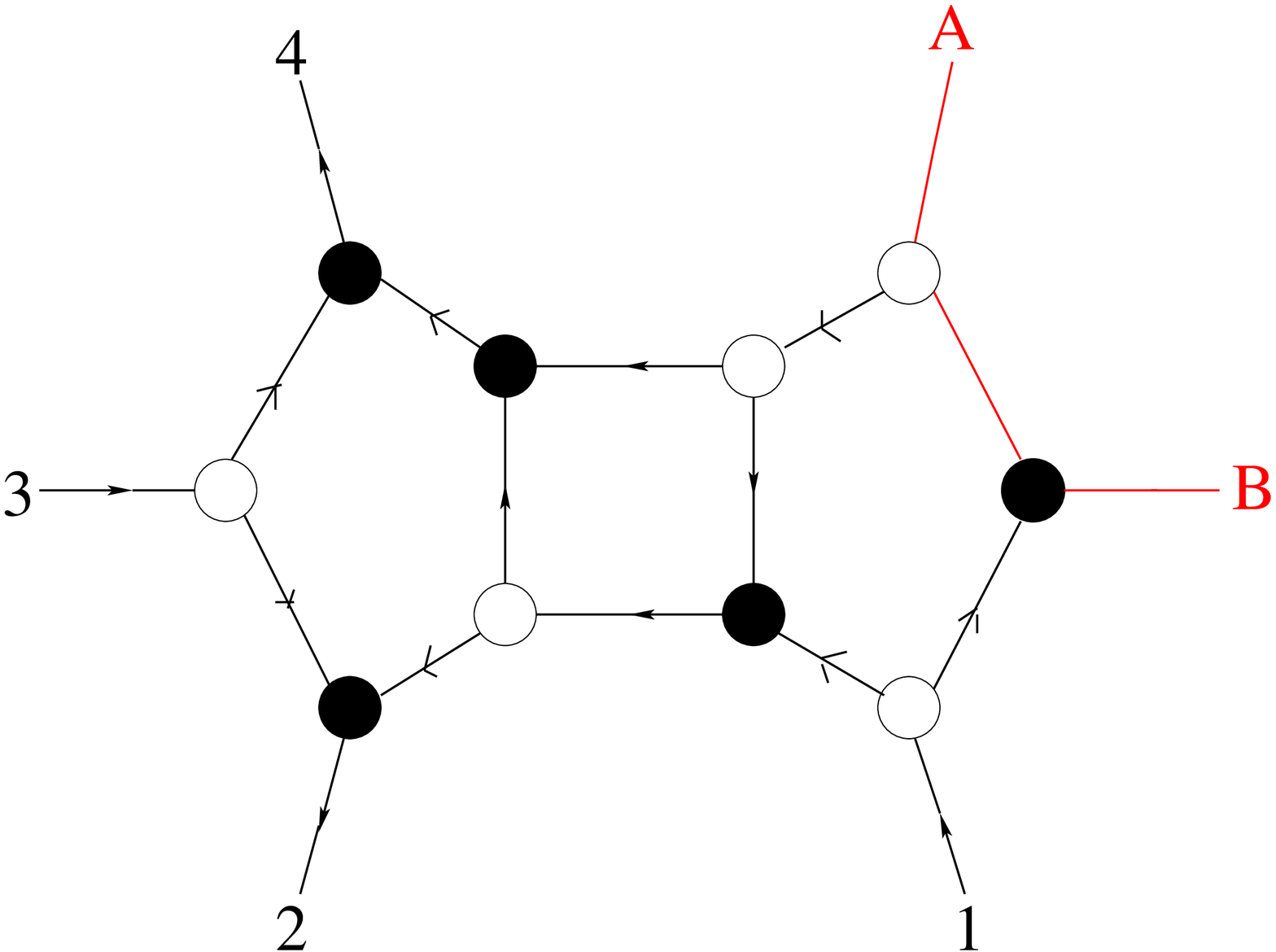}}}
+
\raisebox{-1.6cm}{\scalebox{.22}{\includegraphics{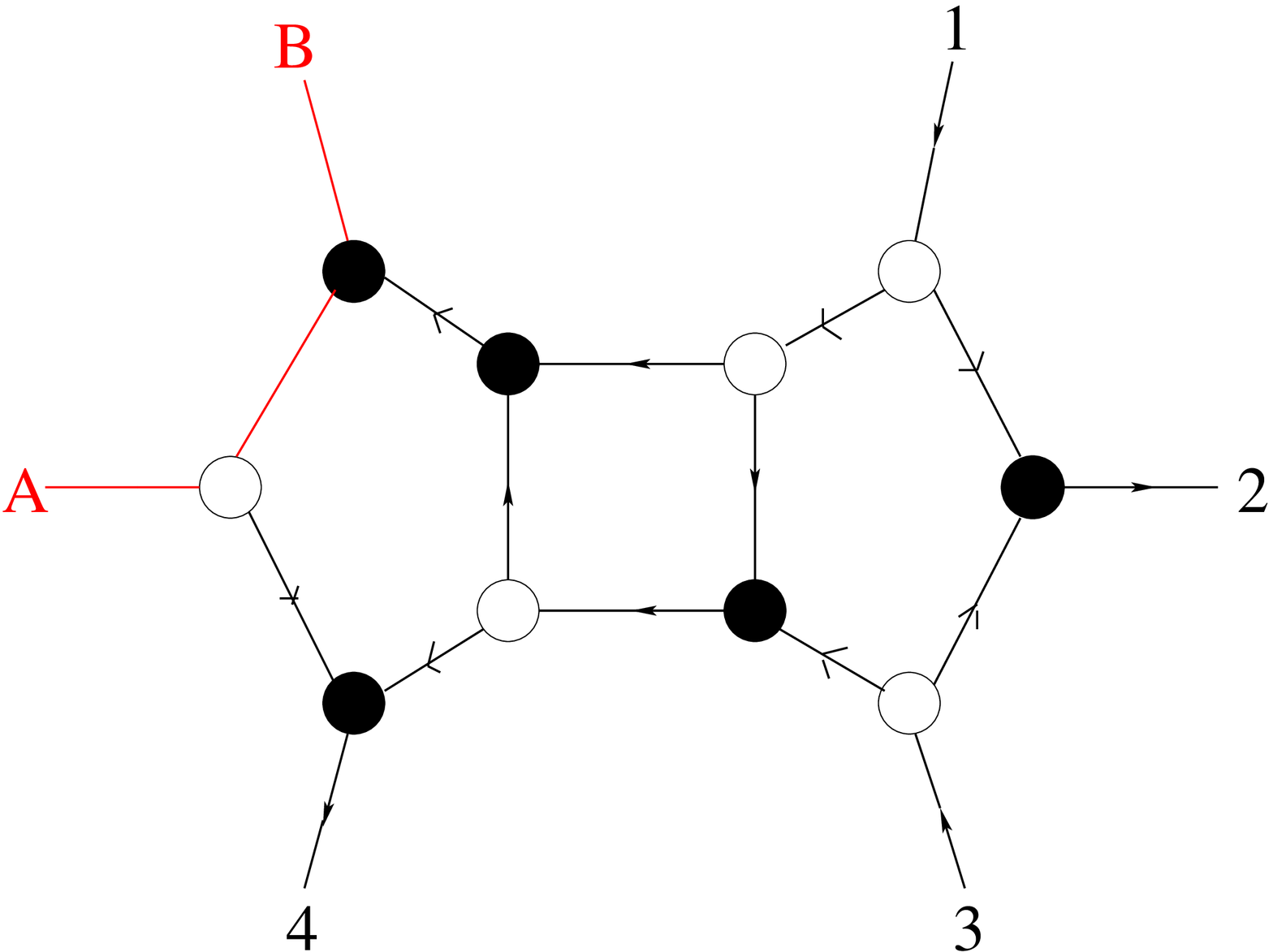}}}
 \end{split}
\end{equation}
where the second line is the complete on-shell diagrammatic 
representation of the tree-level six-particle amplitude. With the
choice of the external coherent states $(1,2,3,4)$ in 
\eqref{eq:6ptampl}, the internal states are fixed, except for
the ones in red. When the states ${\color{red} A}$ and 
${\color{red} B}$ are taken to be forward, an internal helicity
loop is generated, where both the orientations are allowed.

Notice that, in principle, when the amplitude above is taken to
be forward, the first term in \eqref{eq:6ptampl}  turns out to be 
well-defined while the second and third ones show a singularity in
$1/(p^{\mbox{\tiny $(1)$}})^2$ and $1/(p^{\mbox{\tiny $(4)$}})^2$
respectively\footnote{Here we are just indicating the type of
singularity rather than the exact power, shown by the second and 
third diagram in \eqref{eq:6ptampl} when they are taken to be 
forward.}. Let us analyse all the terms in some detail.

\subsubsection{The non-singular term}
\label{subsubsec:NonSing}

Let us begin with the first term in \eqref{eq:6ptampl}, which
we take to be forward
\begin{equation}\eqlabel{eq:M6treefw}
\raisebox{-1.4cm}{\scalebox{.25}{
 \includegraphics{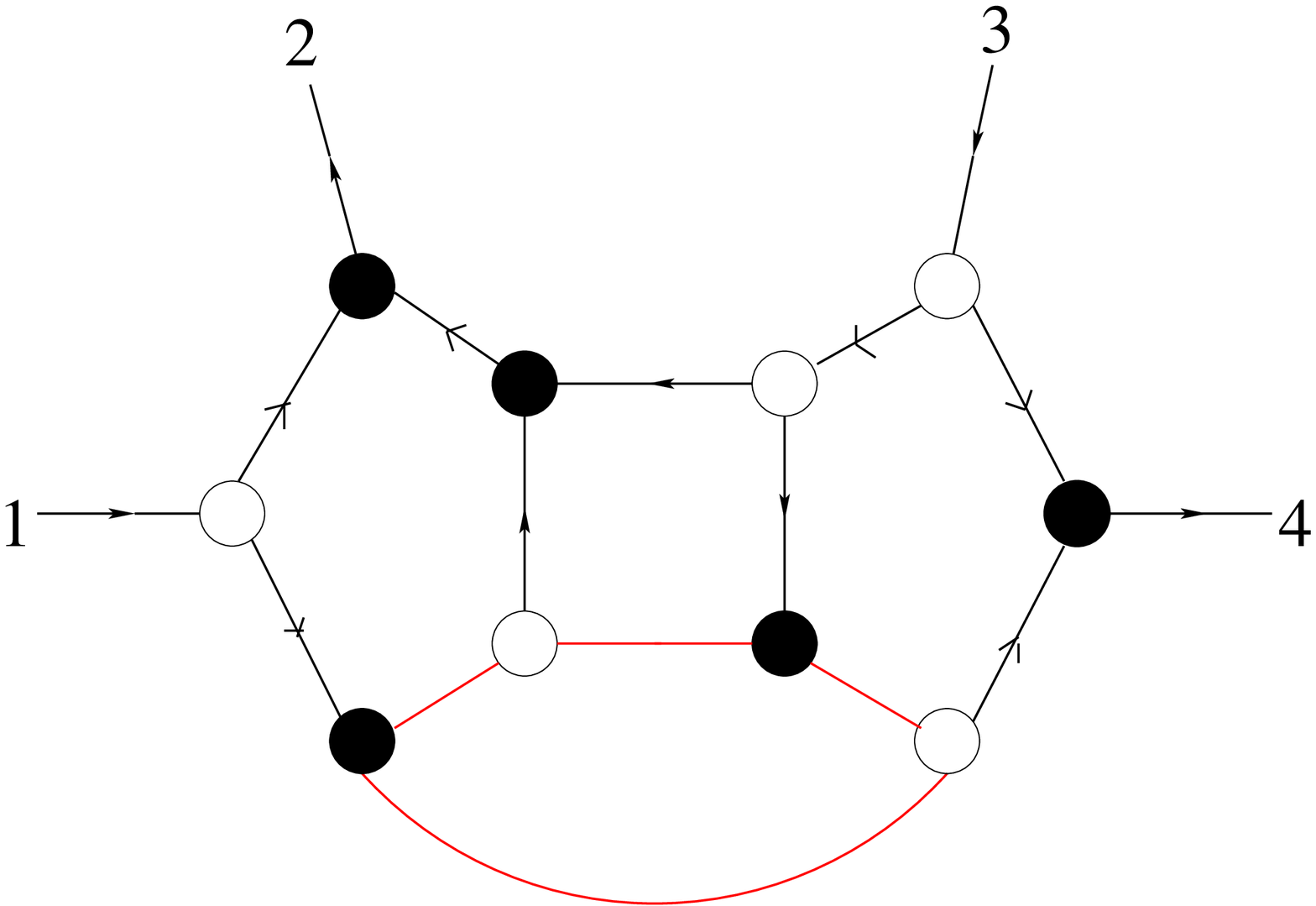}}}
\quad\Longrightarrow\quad
\raisebox{-2.0cm}{\scalebox{.25}{\includegraphics{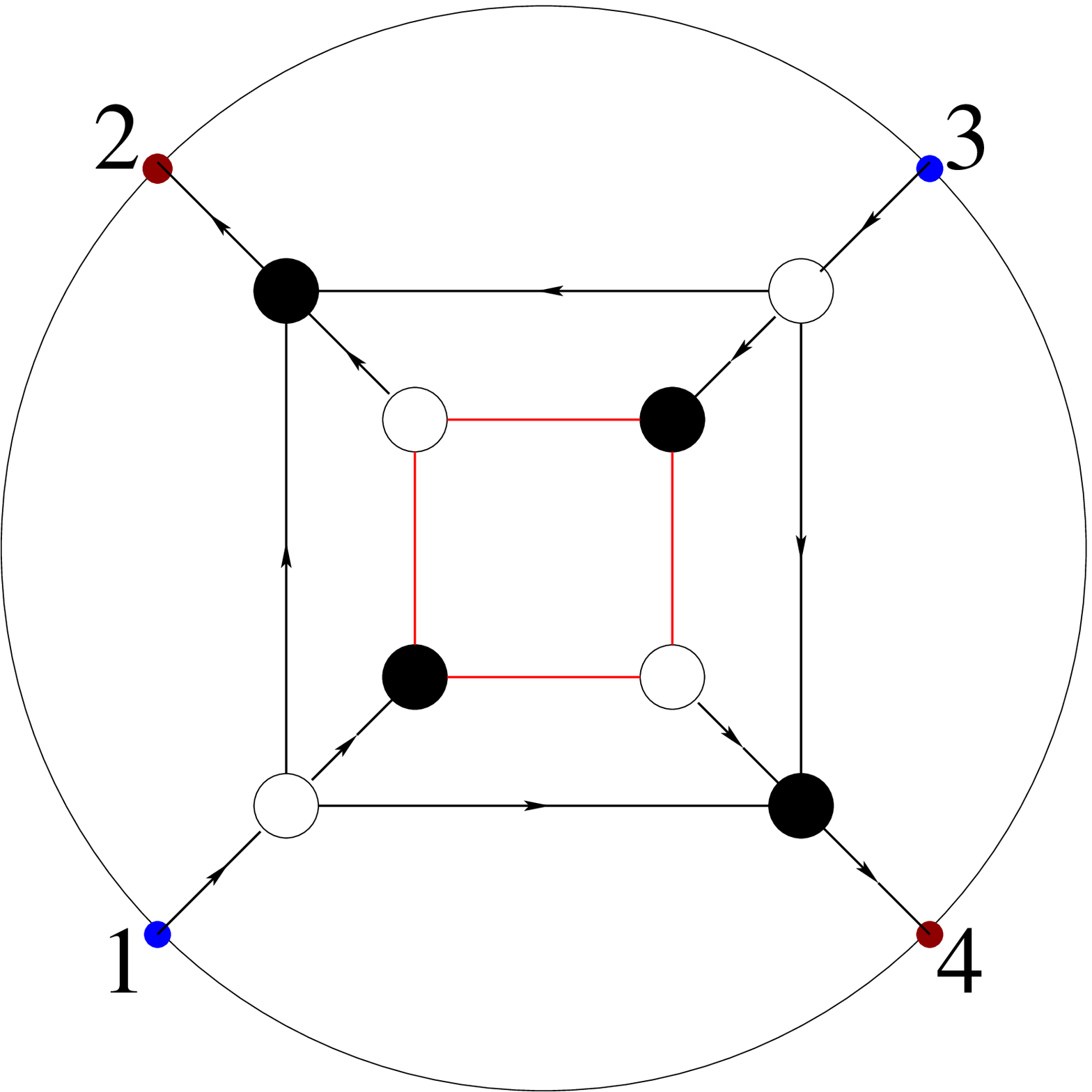}}},
\end{equation}
where on the right-hand-side we re-introduced the BCFW-bridge of
\eqref{eq:1loopInt}, so to obtain the full-fledge contribution to
the one-loop integrand. The red un-decorated internal lines
emphasise the relation to the forward limit as well as they are
kept un-decorated because they allow for both the clockwise and
counter-clockwise helicity loops, {\it i.e.} the sum over all
the multiplet is allowed. Furthermore, notice that such a term
can be seen as the application of four BCFW-bridge to an on-shell
$0$-form with internal helicity loop. Finally, the only internal
helicity loops are related to the forward limit.

With in mind the preliminary discussion 
about the one-loop structure in Section \ref{subsec:1loopStr},
it is easy to see, even  diagrammatically, that the on-shell diagram
on the right-hand-side in \eqref{eq:M6treefw} contains all the 
{\it cut-constructible} information about our one-loop amplitude,
which can be extracted by performing an integration over suitable
contours. Explicitly, the on-shell form encoded in \eqref{eq:M6treefw}
can be written as

\begin{equation}\eqlabel{M4L1a}
 \raisebox{-1.85cm}{\scalebox{.25}{\includegraphics{1loopM4A.eps}}}
 \:=\:
 \mathcal{M}_4^{\mbox{\tiny tree}}
 \bigwedge_{i=1}^4\frac{d\zeta_{i,i+1}}{\zeta_{i,i+1}}\;
 \frac{
  \left[
   \mathcal{J}_{s}(\{\zeta\})
  \right]^{4-\mathcal{N}}+
  \left[
    \mathcal{J}_{t}(\{\zeta\})
  \right]^{4-\mathcal{N}}}{
  \left[
   \mathcal{J}(\{\zeta\})
  \right]^{4-\mathcal{N}}},
\end{equation}
where
\begin{equation}\eqlabel{eq:M4L1aJac}
 \mathcal{J}\:=\:\mathcal{J}_{s}+\mathcal{J}_{t},\quad
 \mathcal{J}_{s}\:=\:-\frac{s}{u}(1-\zeta_{12})(1-\zeta_{34}),\quad
 \mathcal{J}_{t}\:=\:-\frac{t}{u}(1-\zeta_{23})(1-\zeta_{41})
\end{equation}
and, as usual, we performed a M{\"o}bius transformation from
the ``natural'' BCFW-parameters $z_{i,i+1}$ to $\zeta_{i,i+1}$
via \eqref{eq:MobT2}.

For $\mathcal{N}\,=\,3$, the ratio in \eqref{M4L1a} reduces to
the unity leaving the one-loop integrand to be a wedge product of 
$d$logs, as for $\mathcal{N}\,=\,4$: as expected, the 
$\mathcal{N}\,=\,3$ multiplets reorganise themselves to form
the single $\mathcal{N}\,=\,4$ one. For the $\mathcal{N}\,=\,4$ 
itself, the on-shell form \eqref{M4L1a} needs a factor $1/2$ due
to the fact that it is given by a sum over two multiplets while
$\mathcal{N}\,=\,4$ has just one.

For $\mathcal{N}\,\le\,2$, the ratio is non-trivial and thus the
singularity structure is reacher because of the multiple singularity
represented by the denominator in \eqref{M4L1a}. The integrations
over $\{\zeta_{23}\,=\,0,\;\zeta_{41}\,=\,0\}$ and
$\{\zeta_{12}\,=\,0,\;\zeta_{34}\,=\,0\}$ return an on-shell
two-form, such as the one discussed in Section 
\ref{subsec:1loopStr}, representing the double cut in the $s$-
and $t$-channel respectively.

\subsubsection{The singular terms}
\label{eq:Sing}

We are now ready to move on the the discussion of the other two
terms, which at first sight look to be singular making the
forward limit not well-defined. Let us start our discussion
focusing on just one of the two term in \eqref{eq:6ptampl}, given that the analysis
of {\it each single} terms goes similarly.

Let us consider the second diagram in \eqref{eq:6ptampl} 
{\it before} the forward limit is taken:
\begin{equation}\eqlabel{eq:M6treebffw2}
 \begin{split}
  &\raisebox{-1.6cm}{\scalebox{.22}{
   \includegraphics{6ptForwardB2.eps}}}
   \:=\:
   \delta^{\mbox{\tiny $(2\times2)$}}
   \left(
    \sum_{k=1}^4\lambda^{\mbox{\tiny $(k)$}}\tilde{\lambda}^{\mbox{\tiny $(k)$}}+
     \lambda^{\mbox{\tiny $(A)$}}\tilde{\lambda}^{\mbox{\tiny $(A)$}}+
     \lambda^{\mbox{\tiny $(B)$}}\tilde{\lambda}^{\mbox{\tiny $(B)$}}
    \right)\times
  \\
  &\hspace{.5cm}\times
   \delta^{\mbox{\tiny $(2\times\mathcal{N})$}}
   \left(
    \sum_{k=1}^4\lambda^{\mbox{\tiny $(k)$}}
     \tilde{\eta}^{\mbox{\tiny $(k)$}}+
    \lambda^{\mbox{\tiny $(A)$}}\tilde{\eta}^{\mbox{\tiny $(A)$}}+
    \lambda^{\mbox{\tiny $(B)$}}\tilde{\eta}^{\mbox{\tiny $(B)$}}
   \right)
   \delta^{\mbox{\tiny $(1\times\mathcal{N})$}}
   \left(
    [4,3]\tilde{\eta}^{\mbox{\tiny $(2)$}}+[2,4]\tilde{\eta}^{\mbox{\tiny $(3)$}}+
    [3,2]\tilde{\eta}^{\mbox{\tiny $(4)$}}
   \right)\\
  &\hspace{.5cm}\times
   \frac{[2,4]^{4-\mathcal{N}}
    \langle\mathcal{K},1\rangle^{4-\mathcal{N}}}{
    \langle A,B\rangle \langle1|A+B|4]P^2_{AB1}
    \langle A|B+1|2]\langle B,1\rangle
    [2,3][3,4]},
   \qquad \mathcal{K}\:=\:A,\,B,
 \end{split}
\end{equation}
with $\mathcal{K}\:=\:A,\,B$ depending on the helicity configuration
for $A$ and $B$, which, for our purposes, need to have opposite
helicities.
 
First of all, notice that \eqref{eq:M6treebffw2} not only shows singularities of the
full six-particle amplitude, but also two {\it non-local} ones. In the forward limit 
this term is indeed singular. Taking 
$p^{\mbox{\tiny $(B)$}}\,\longrightarrow\,-p^{\mbox{\tiny $(A)$}}$
the singularity are given by the following terms in the denominator
of \eqref{eq:M6treebffw2}:
\begin{equation}\eqlabel{eq:M41loopDiv1}
 \langle A,B\rangle \hspace{2cm}
 \langle1|A+B|4] \hspace{2cm}
 P^2_{AB1}.
\end{equation}
Notice that considering the further degree-of-freedom introduced by
the $(4,1)$ BCFW-bridge {\it does not} cure such divergencies: this class
of divergencies is reminiscent of the bubbles in the external lines,
with the difference that the singularity does not appear to be exactly 
$1/(p^{\mbox{\tiny $(1)$}})^4$ but rather it has a $1/(p^{\mbox{\tiny $(1)$}})^2$
factor (whose origin is the last term in \eqref{eq:M41loopDiv1}), a further singularity
which coincides with the holomorphic collinear limit in the $(A,B)$-channel\footnote{In 
the complexified momentum space, a collinear limit $P_{\mbox{\tiny $ij$}}^2\,\equiv\,
\langle i,j\rangle[i,j]\,\longrightarrow\,0$ can be taken in two different ways, by 
sending the either holomorphic inner product $\langle i,j\rangle$ or the 
anti-holomorphic one $[i,j]$ to zero. We refer to the former as holomorphic collinear limit. In the present case -- the singularity $\langle A,B\rangle$ -- the on-shell diagram
under analysis is the only one of the on-shell representation of the tree-level 
six-particle amplitude containing the singularity $\langle A,B\rangle$ and thus it
resembles its factorisation property in this channel.}, and finally a {\it non-local} 
one (the second two term of \eqref{eq:M41loopDiv1}).

Furthermore, indicating the momentum of the internal lines 
between two consecutive external states with $l_{i,i+1}$, it is
interesting to focus on the following
\begin{equation}\eqlabel{eq:ls}
 \begin{split}
  &l_{dA}\:=\:\lambda^{\mbox{\tiny $(A)$}}
   \left(
    \tilde{\lambda}^{\mbox{\tiny $(A)$}}+
    \frac{\langle B,1\rangle}{\langle A,1\rangle}
     \tilde{\lambda}^{\mbox{\tiny $(B)$}}
   \right),
   \qquad
   l_{AB}\:=\:\frac{\langle B,1\rangle}{\langle A,1\rangle}
   \lambda^{\mbox{\tiny $(A)$}}\tilde{\lambda}^{\mbox{\tiny $(B)$}},
  \\
  &\phantom{\ldots}\\
  &l_{B1}\:=\:-\frac{\langle A,B\rangle}{\langle A,1\rangle}
   \lambda^{\mbox{\tiny $(1)$}}\tilde{\lambda}^{\mbox{\tiny $(B)$}},
   \hspace{2.2cm}
   l_{1a}\:=\:\lambda^{\mbox{\tiny $(1)$}}
   \left(
    \tilde{\lambda}^{\mbox{\tiny $(1)$}}+
    \frac{\langle A,B\rangle}{\langle A,1\rangle}
    \tilde{\lambda}^{\mbox{\tiny $(B)$}}
   \right),
 \end{split}
\end{equation}
where $l_{dA}$ and $l_{1a}$ indicate the momenta between an
external state ($A$ and $1$ respectively) and an internal node.
In the forward limit:
\begin{equation}\eqlabel{eq:ls2}
 \raisebox{-1.6cm}{\scalebox{.25}{
   \includegraphics{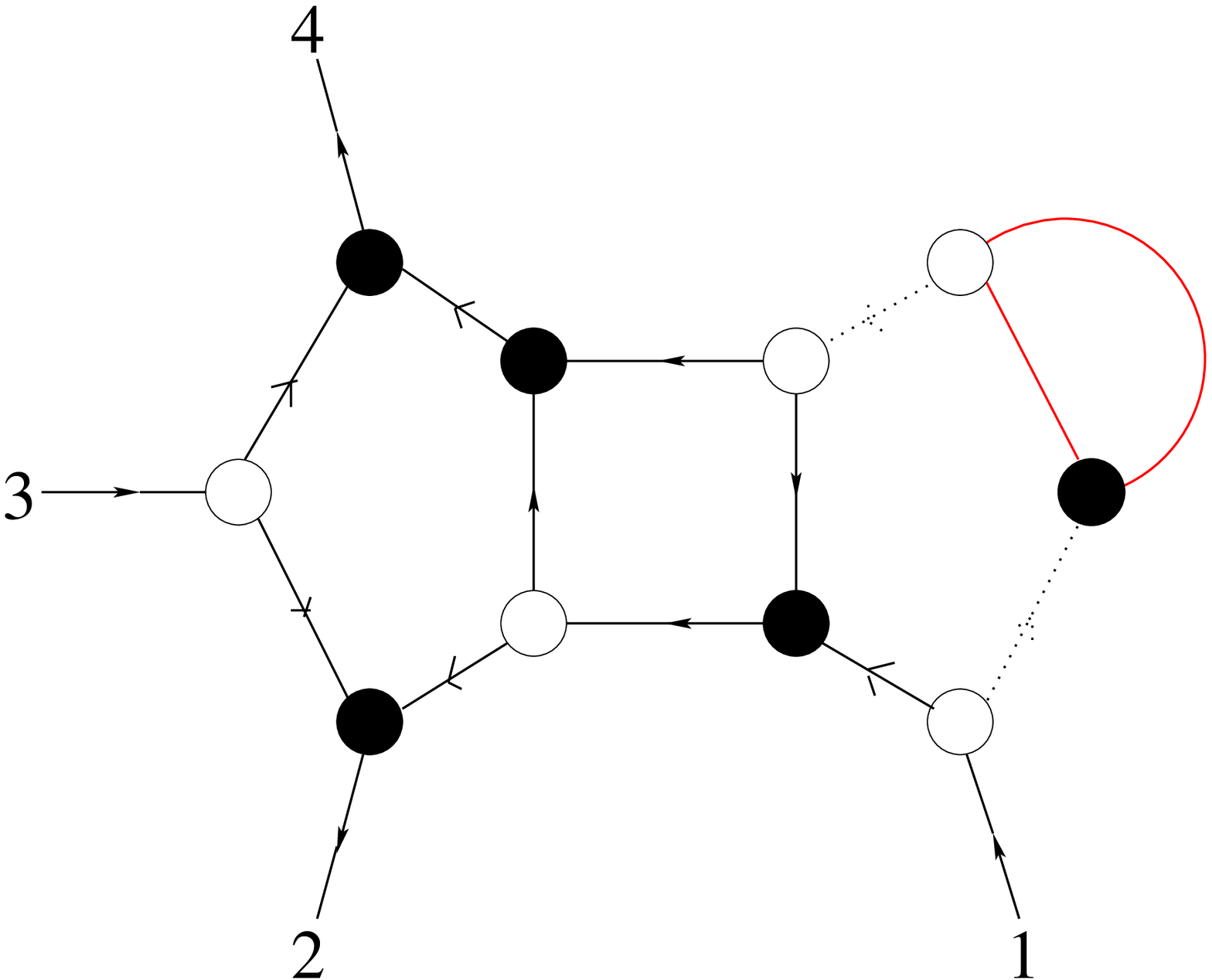}}}
 \hspace{1.8cm}
 \begin{array}{cl}
    l_{dA}\,\longrightarrow\,0, 
  & l_{AB}\,\longrightarrow\,-p^{\mbox{\tiny $(A)$}},\\
    {} & {}\\
    l_{B1}\,\longrightarrow\,0,
  & l_{1a}\,\longrightarrow\,p^{\mbox{\tiny $(1)$}},
 \end{array}
\end{equation}
with $l_{dA}$ and $l_{B1}$ becoming soft\footnote{The soft lines $l_{dA}$ and $l_{B1}$
are indicated in \eqref{eq:ls2} with dotted lines.} and the five-particle
sub-diagram factorising into a tree-level four-particle amplitude
and a soft factor.

As for a first analysis, we can take a {\it quasi}-forward limit of our diagram	,
{\it i.e.} infinitesimally away from being forward, in such a way
that the momentum conservation keeps being preserved:
\begin{equation}\eqlabel{eq:qflim1}
 \begin{split}
  &\lambda^{\mbox{\tiny $(B)$}}\:\longrightarrow\:
    -\lambda^{\mbox{\tiny $(A)$}}
    +\epsilon
    \lambda^{\mbox{\tiny $(4)$}},
   \qquad
    \tilde{\lambda}^{\mbox{\tiny $(B)$}}\:\longrightarrow\:
    \tilde{\lambda}^{\mbox{\tiny $(A)$}},
   \hspace{2cm}
    \tilde{\eta}^{\mbox{\tiny $(B)$}}\:\longrightarrow\:
    \tilde{\eta}^{\mbox{\tiny $(A)$}}
   \\
 &\phantom{\ldots}\\
 &\lambda^{\mbox{\tiny $(i)$}}\:\longrightarrow\:
   \lambda^{\mbox{\tiny $(i)$}},\hspace{2.4cm}
  \tilde{\lambda}^{\mbox{\tiny $(4)$}}\:\longrightarrow\:
   \tilde{\lambda}^{\mbox{\tiny $(4)$}}-\epsilon\tilde{\lambda}^{\mbox{\tiny $(A)$}},
  \hspace{.9cm}
  \tilde{\eta}^{\mbox{\tiny $(4)$}}\:\longrightarrow\: 
   \tilde{\eta}^{\mbox{\tiny $(4)$}}-\epsilon\tilde{\eta}^{\mbox{\tiny $(A)$}}.
 \end{split}
\end{equation}
Notice that such a deformation preserves both momentum and super-momentum conservation.

%
Thus, in this {\it quasi}-forward limit, the term under analysis 
becomes
\begin{equation}\eqlabel{eq:M41Ldiv2}
 \begin{split}
  \raisebox{-1.6cm}{\scalebox{.22}{
   \includegraphics{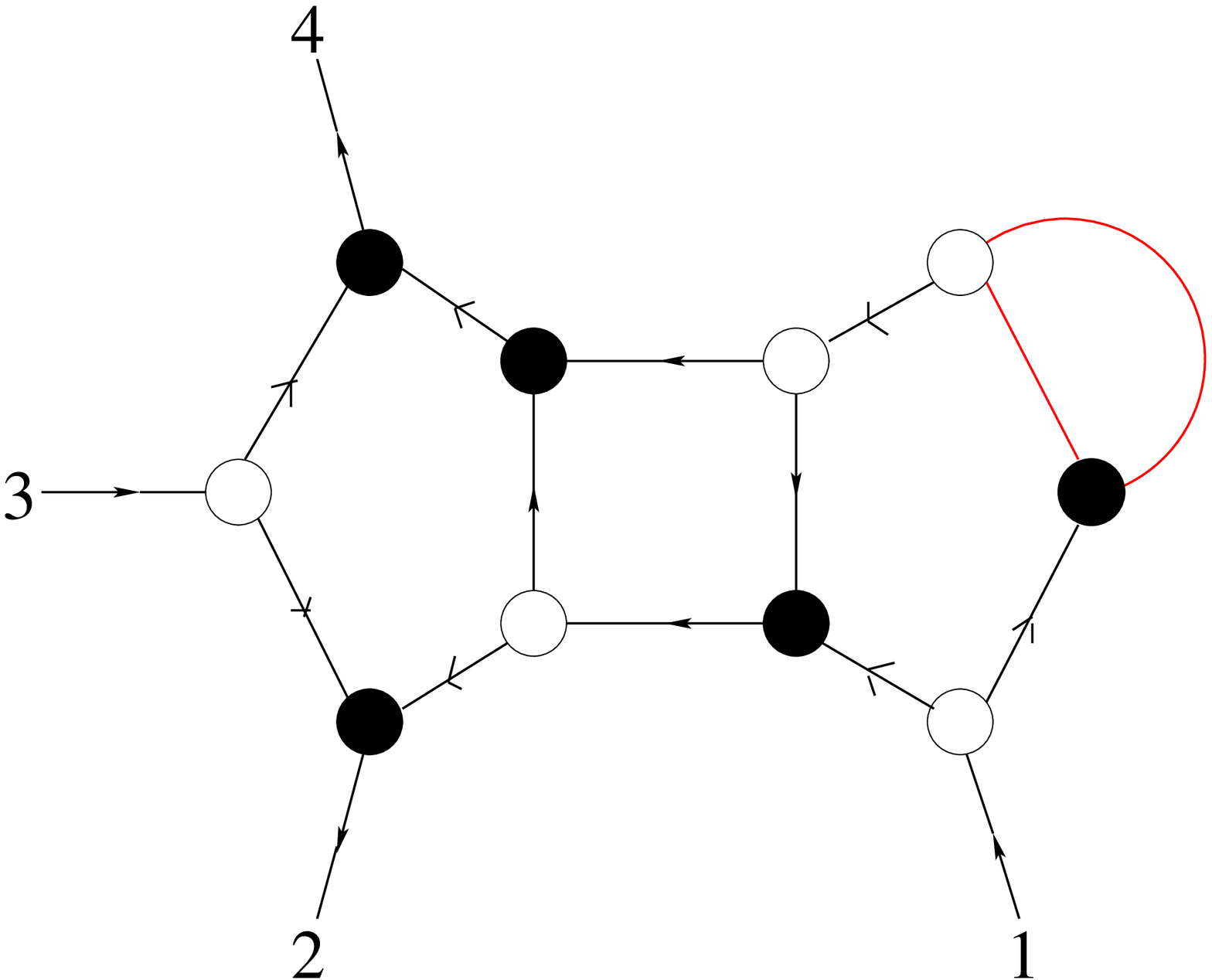}}}
   \:=&\:
   \mathcal{M}_4^{\mbox{\tiny tree}}\,
   \frac{d^2\lambda^{\mbox{\tiny $(A)$}}d^2\tilde{\lambda}^{\mbox{\tiny $(A)$}}}{
    \mbox{Vol}\{GL(1)\}}\,
   \frac{[4,1]}{[4,A][A,1]}
   \left(
    \frac{\langle A,1\rangle}{\langle 4,1\rangle}
   \right)^{2-\mathcal{N}}\,\frac{\epsilon^{\mathcal{N}-3}}{\langle A,4\rangle}\times\\
 &\hspace{.5cm}\times
   \frac{
   \left[ 
    1+(-1)^{\mathcal{N}}
    \left(
     1-\epsilon\frac{\langle 4,1\rangle}{\langle A,1\rangle}
    \right)^{4-\mathcal{N}}
   \right]}{
   \left(
    1+\epsilon\frac{\langle A,4\rangle[A,2]}{\langle A,1\rangle[1,2]}
   \right)
   \left(
    1-\epsilon\frac{\langle4,1\rangle}{\langle A,1\rangle}
   \right)
   \left(
    1-\epsilon\frac{[3,A]}{[3,4]}
   \right)},
 \end{split}
\end{equation}
where the factor $\epsilon^{\mathcal{N}}$ comes from the integration
over $\tilde{\eta}^{\mbox{\tiny $(A)$}}$, {\it i.e.} from the sum over
components of a multiplet, while the two terms are given exactly
by the sum over the two multiplets.

For $\mathcal{N}\,=\,4$, this term is of order 
$\mathcal{O}(\epsilon)$ and thus vanishes, as expected. 
For $\mathcal{N}\,=\,3$, the behaviour is analogous to the previous
case due to the cancellation between the two multiplets, which is
reflected in the $(-1)^{\mathcal{N}}$ in the numerator of 
\eqref{eq:M41Ldiv2}. The same holds for the other divergent diagram.
As a consequence in these cases, the forward limit is well-defined and it is
completely given by \eqref{eq:M6treefw}.

As far as the less supersymmetric theories ($\mathcal{N}\,\in\:]0,2]$) are concerned, a 
single pole in the parameter $\epsilon$ appears
\begin{equation}\eqlabel{eq:N12div}
 \begin{split}
  \raisebox{-1.6cm}{\scalebox{.22}{
   \includegraphics{6ptForwardB3.eps}}}
   \:&\overset{\epsilon\,\longrightarrow\,0}{=}\:
   \mathcal{M}_4^{\mbox{\tiny tree}}\,
   \frac{d^2\lambda^{\mbox{\tiny $(A)$}}d^2\tilde{\lambda}^{\mbox{\tiny $(A)$}}}{
    \mbox{Vol}\{GL(1)\}}\,
   \frac{[4,1]}{[4,A][A,1]\langle A,4\rangle}\times
  \\
  &\hspace{-.8cm}\times\left[
    \frac{4-\mathcal{N}}{\epsilon}+
    (4-\mathcal{N})
    \frac{
     \langle A|
      [1,2]\lambda^{\mbox{\tiny $(1)$}}\tilde{\lambda}^{\mbox{\tiny $(3)$}}+
      [3,4]\lambda^{\mbox{\tiny $(4)$}}\tilde{\lambda}^{\mbox{\tiny $(2)$}}|A]}{
     \langle A,1\rangle[1,2][3,4]}+\mathcal{O}(\epsilon)
   \right],
 \end{split}
\end{equation}
where the leading term resembles the divergence of an external bubble, while the
coefficient of order $\mathcal{O}(1)$ in the square brackets is helicity-dependent
because the expansion parameter itself $\epsilon$, as defined in \eqref{eq:qflim1},
scales as $\lambda^{\mbox{\tiny $(A)$}}$ and $\tilde{\lambda}^{\mbox{\tiny $(4)$}}$.
Notice that for $\mathcal{N}\,=\,1$ the behaviour $\epsilon^{-2}$ suggested by the
overall factor in \eqref{eq:M41Ldiv2} is actually enhanced to $\epsilon^{-1}$
because of cancellation among the two multiplets.

Finally, in the non-supersymmetric case no cancellation occurs whatsoever so that
the leading term in the small-$\epsilon$ expansion is of order 
$\mathcal{O}(\epsilon^{-3})$ and all the lower order poles are present.

Some comments are now in order. The individual terms in the on-shell representation
of the tree-level six-particle amplitude \eqref{eq:6ptampl} are non-local as they
show poles which are not factorisation channels and thus they are not present in the
final amplitude. Interestingly enough, when one takes the forward limit such 
non-localities disappear just in one of these terms, namely in the only one which
appears to be well-defined in this limit, at least at the leading order $\mathcal{O}(\epsilon^0)$. 
In particular, one of these spurious poles is reached in the forward limit, contributing
to the {\it forward singularity} as a factor of order $\mathcal{O}(\epsilon^{-1})$. 
Furthermore, this same pole is common to both singular contributions to the forward 
amplitude. This means that if we treat each singular on-shell diagram separately as we 
did above, our regulated expansion \eqref{eq:N12div} would depend crucially on such 
non-locality. 
Therefore, in order to see its effects cancelling out, one has to consider both term at 
once. Let us see this in more detail. The pole in question is $\langle 1|A+B|4]$ which 
would correspond to the following factorisation of our diagrams:
\begin{equation}
 \begin{split}
  &\raisebox{-1.8cm}{\scalebox{.30}{\includegraphics{6ptForwardB.eps}}}
    \qquad\longrightarrow\qquad
   \raisebox{-1.8cm}{\scalebox{.30}{\includegraphics{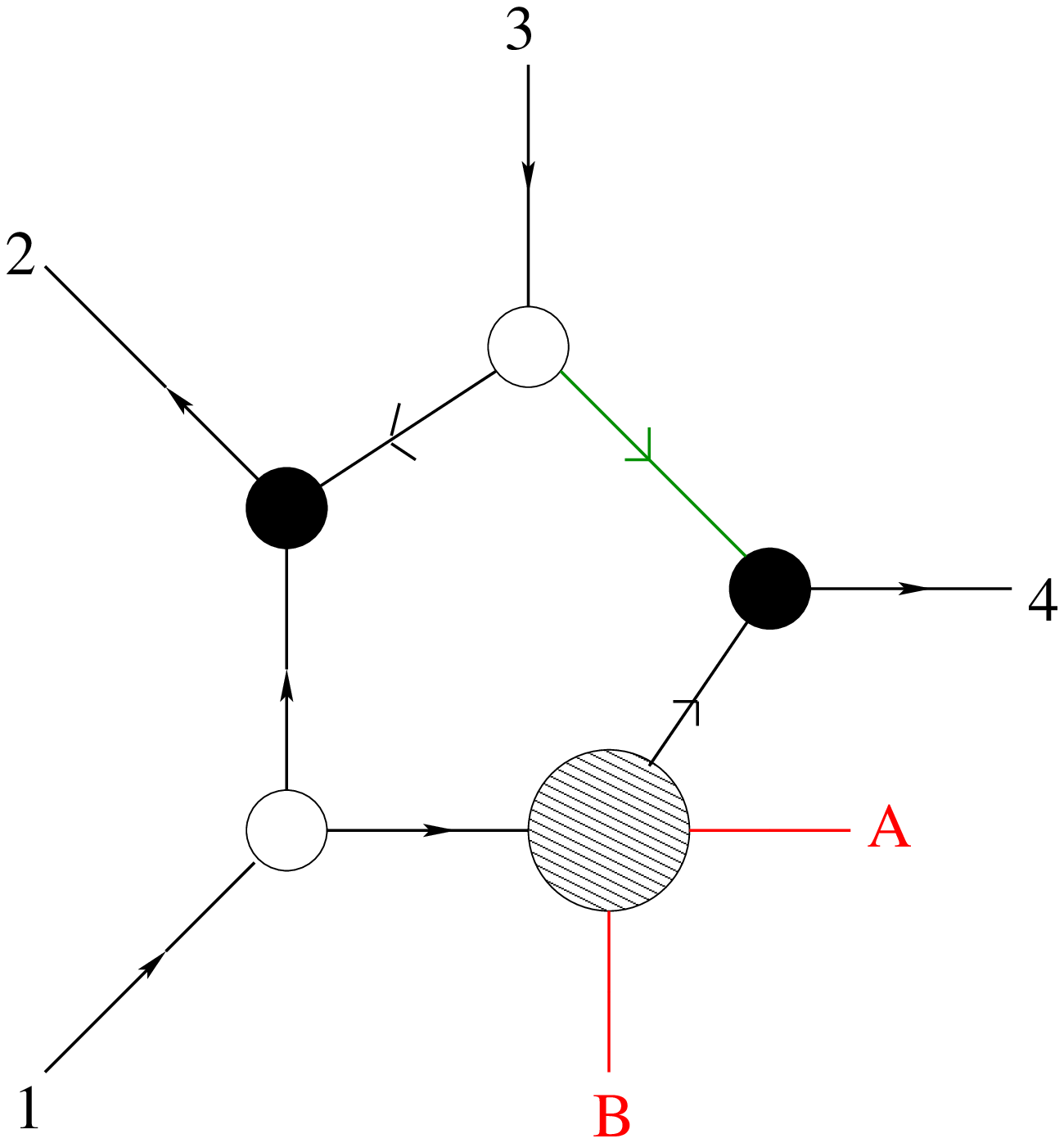}}}\\
  &\raisebox{-1.8cm}{\scalebox{.30}{\includegraphics{6ptForwardC.eps}}}
    \qquad\longrightarrow\qquad
   \raisebox{-1.8cm}{\scalebox{.30}{\includegraphics{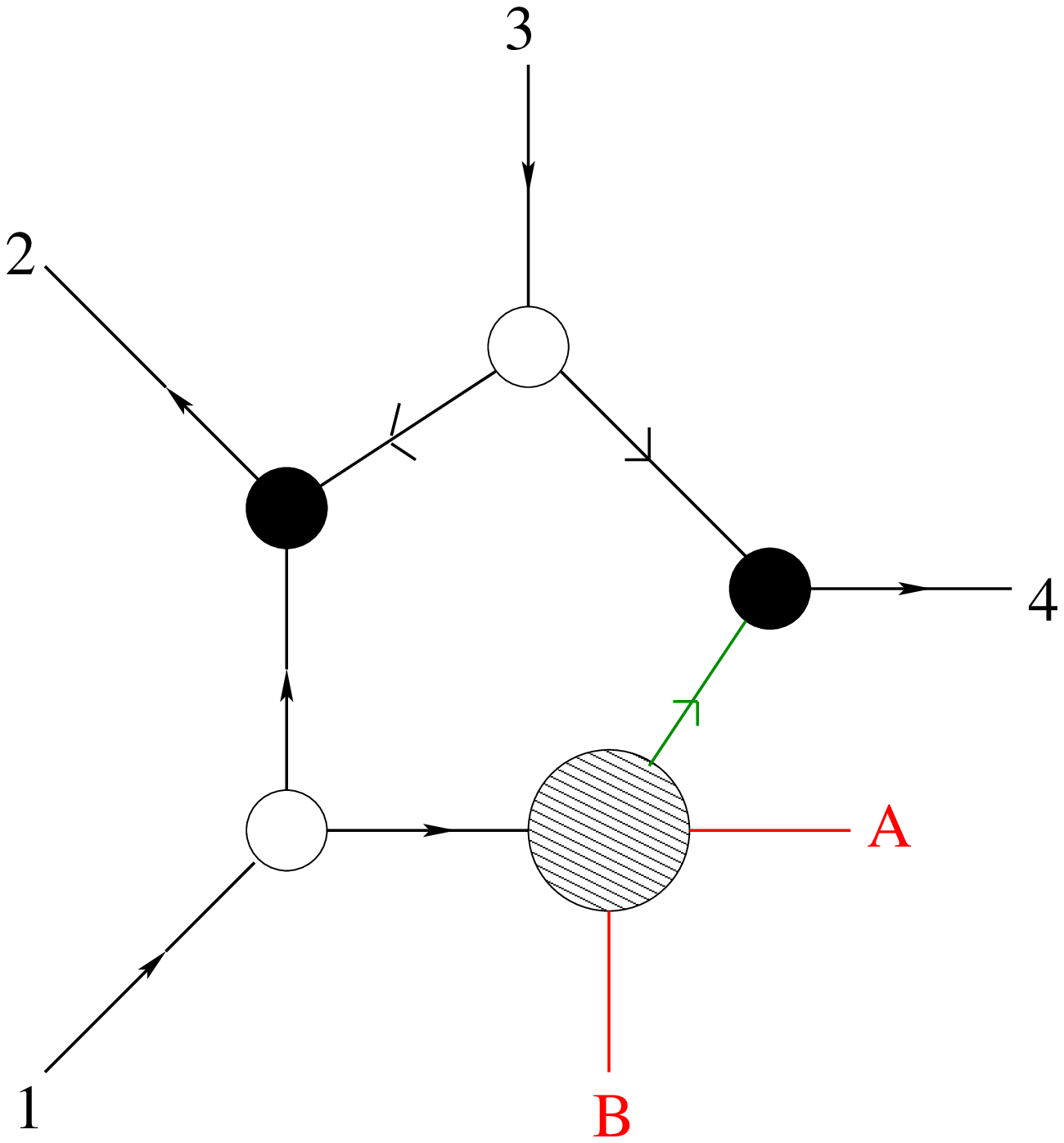}}}
 \end{split}
\end{equation}
which turn out to have different sign leading to the cancellation of such a
contribution -- the green lines above emphasise how these spurious factorisation 
emerge. As a consequence, the behaviour in our {\it quasi}-forward limit is
enhanced of one power in $\epsilon$.

The second non-local term $\langle A|B+1|2]$ is instead common to the on-shell diagram we have been
analysing and to the non-singular term \eqref{eq:M6treefw}. In the {\it quasi}-forward limit it is of
order $\mathcal{O}(\epsilon^0)$ and, at this order, it becomes local. The non-localities might eventually
reappear at higher order in $\epsilon$, not affecting the one-loop result.

Finally, our {\it quasi}-forward limit can regulate the diagram we discussed above,
but it is not able to do the same with the second singular term. The reason is that
the latter shows the collinear singularity $[A,B]$ (rather than $\langle A,B\rangle$),
which would stay $\epsilon$-independent under the deformation \eqref{eq:qflim1}.
Therefore a more general way to stay ``{\it quasi}-forward'' is needed in order
to treat consistently and simultaneously both contributions.

\subsubsection{{\it Quasi}-forward limits and singular terms}
\label{subsubsec:qflst}

Let us go back to the on-shell representation \eqref{eq:6ptampl} for the six-particle
amplitude and let us consider the second and third terms all together. The explicit
expression for one of them is given by \eqref{eq:M6treebffw2}, while for the other one
it can be easily be obtained from \eqref{eq:M6treebffw2} through the label-shift
$i\,\longrightarrow\,i-2$:
\begin{equation}\eqlabel{eq:M6treebffw3}
 \begin{split}
  &\raisebox{-1.6cm}{\scalebox{.22}{
   \includegraphics{6ptForwardC2.eps}}}
   \:=\:
   \delta^{\mbox{\tiny $(2\times2)$}}
   \left(
    \sum_{k=1}^4\lambda^{\mbox{\tiny $(k)$}}\tilde{\lambda}^{\mbox{\tiny $(k)$}}+
     \lambda^{\mbox{\tiny $(A)$}}\tilde{\lambda}^{\mbox{\tiny $(A)$}}+
     \lambda^{\mbox{\tiny $(B)$}}\tilde{\lambda}^{\mbox{\tiny $(B)$}}
    \right)\times
  \\
  &\hspace{.5cm}\times
   \delta^{\mbox{\tiny $(2\times\mathcal{N})$}}
   \left(
    \sum_{k=1}^4\lambda^{\mbox{\tiny $(k)$}}
     \tilde{\eta}^{\mbox{\tiny $(k)$}}+
    \lambda^{\mbox{\tiny $(A)$}}\tilde{\eta}^{\mbox{\tiny $(A)$}}+
    \lambda^{\mbox{\tiny $(B)$}}\tilde{\eta}^{\mbox{\tiny $(B)$}}
   \right)
   \delta^{\mbox{\tiny $(1\times\mathcal{N})$}}
   \left(
    [B,A]\tilde{\eta}^{\mbox{\tiny $(4)$}}+[4,B]\tilde{\eta}^{\mbox{\tiny $(A)$}}+
    [A,4]\tilde{\eta}^{\mbox{\tiny $(B)$}}
   \right)\\
  &\hspace{.5cm}\times
   \frac{[4,\mathcal{Q}]^{4-\mathcal{N}}
    \langle 1,3\rangle^{4-\mathcal{N}}}{
    \langle 1,2\rangle \langle3|4+A|B]P^2_{AB4}
    \langle 1|A+B|4]\langle 2,3\rangle
    [4,A][A,B]},
   \qquad \mathcal{Q}\:=\:B,\,A,
 \end{split}
\end{equation}

As we observed at the end of the previous section, if we take the {\it quasi}-forward
limits as in \eqref{eq:qflim1}, neither $[A,B]$ nor $P_{\mbox{\tiny $AB4$}}^2$ are
mapped into poles in the parameter $\epsilon$ and thus this diagram stays ill-defined.
In order to be able to treat simultaneously both \eqref{eq:M6treebffw2} and 
\eqref{eq:M6treebffw3}, we define that {\it quasi-forward} limit as
\begin{equation}\eqlabel{eq:qflim2}
 \begin{split}
  &\lambda^{\mbox{\tiny $(A)$}}(\epsilon)\:=\:\lambda^{\mbox{\tiny $(A)$}},\qquad
   \tilde{\lambda}^{\mbox{\tiny $(A)$}}(\epsilon)\:=\:
   \tilde{\lambda}^{\mbox{\tiny $(A)$}}+
    \epsilon\frac{\langle1,4\rangle}{\langle A,4\rangle}\tilde{\lambda}^{\mbox{\tiny $(1)$}},
   \qquad
   \tilde{\eta}^{\mbox{\tiny $(A)$}}(\epsilon)\:=\:
   \tilde{\eta}^{\mbox{\tiny $(A)$}}+
    \epsilon\frac{\langle1,4\rangle}{\langle A,4\rangle}\tilde{\eta}^{\mbox{\tiny $(1)$}},\\
  &\lambda^{\mbox{\tiny $(B)$}}(\epsilon)\:=\:
    -\lambda^{\mbox{\tiny $(A)$}}-\epsilon\frac{[4,1]}{[A,1]}\lambda^{\mbox{\tiny $(4)$}},
   \qquad
   \tilde{\lambda}^{\mbox{\tiny $(B)$}}(\epsilon)\:=\:\tilde{\lambda}^{\mbox{\tiny $(A)$}},\qquad
   \tilde{\eta}^{\mbox{\tiny $(B)$}}(\epsilon)\:=\:\tilde{\eta}^{\mbox{\tiny $(A)$}}\\
  &\lambda^{\mbox{\tiny $(4)$}}(\epsilon)\:=\:\lambda^{\mbox{\tiny $(4)$}},\qquad
   \tilde{\lambda}^{\mbox{\tiny $(4)$}}(\epsilon)\:=\:
   \tilde{\lambda}^{\mbox{\tiny $(4)$}}+
    \epsilon\frac{[4,1]}{[A,1]}\tilde{\lambda}^{\mbox{\tiny $(A)$}},
   \qquad
   \tilde{\eta}^{\mbox{\tiny $(4)$}}(\epsilon)\:=\:
   \tilde{\eta}^{\mbox{\tiny $(4)$}}+
    \epsilon\frac{[4,1]}{[A,1]}\tilde{\eta}^{\mbox{\tiny $(A)$}},\\
  &\lambda^{\mbox{\tiny $(1)$}}(\epsilon)\:=\:
    \lambda^{\mbox{\tiny $(1)$}}-\epsilon\frac{\langle1,4\rangle}{\langle A,4\rangle}\lambda^{\mbox{\tiny $(A)$}},
   \qquad
   \tilde{\lambda}^{\mbox{\tiny $(1)$}}(\epsilon)\:=\:\tilde{\lambda}^{\mbox{\tiny $(1)$}},\qquad
   \tilde{\eta}^{\mbox{\tiny $(1)$}}(\epsilon)\:=\:\tilde{\eta}^{\mbox{\tiny $(1)$}},
 \end{split}
\end{equation}
which again respects both momentum and super-momentum conservation -- notice that 
the deformation parameter $\epsilon$ is now defined to be helicity-blind. With such a
deformation applied to both \eqref{eq:M6treebffw2} and \eqref{eq:M6treebffw3}, the singularities in the
Lorentz invariants are mapped into poles in the regularisation parameter $\epsilon$ and we can now freely
integrate over the Grassmann variable $\tilde{\eta}^{\mbox{\tiny $(A)$}}$ related to the forward line.

\subsection{Forward amplitudes and the one-loop integrand}\label{subsec:1loopInt}

In the previous section we discussed in some detail the structure of the forward amplitudes for 
less/no-supersymmetric theories, which is related to the residue of the pole characterising the loop integrand.
As in \eqref{eq:1loopInt}, let us consider the one-loop integrand as generated via the BCFW-bridge in the 
$(4,1)$-channel. The full-fledge contribution to the one-loop integrand from the forward amplitudes is given by:
\begin{equation}\eqlabel{eq:1lInt}
 \left.\mathcal{M}_4^{\mbox{\tiny $(1)$}}\right|_{\mbox{\tiny $(4,1)$}}\:=\:
  \raisebox{-1.8cm}{\scalebox{.23}{\includegraphics{1loopM4A.eps}}}+
  \raisebox{-1.8cm}{\scalebox{.23}{\includegraphics{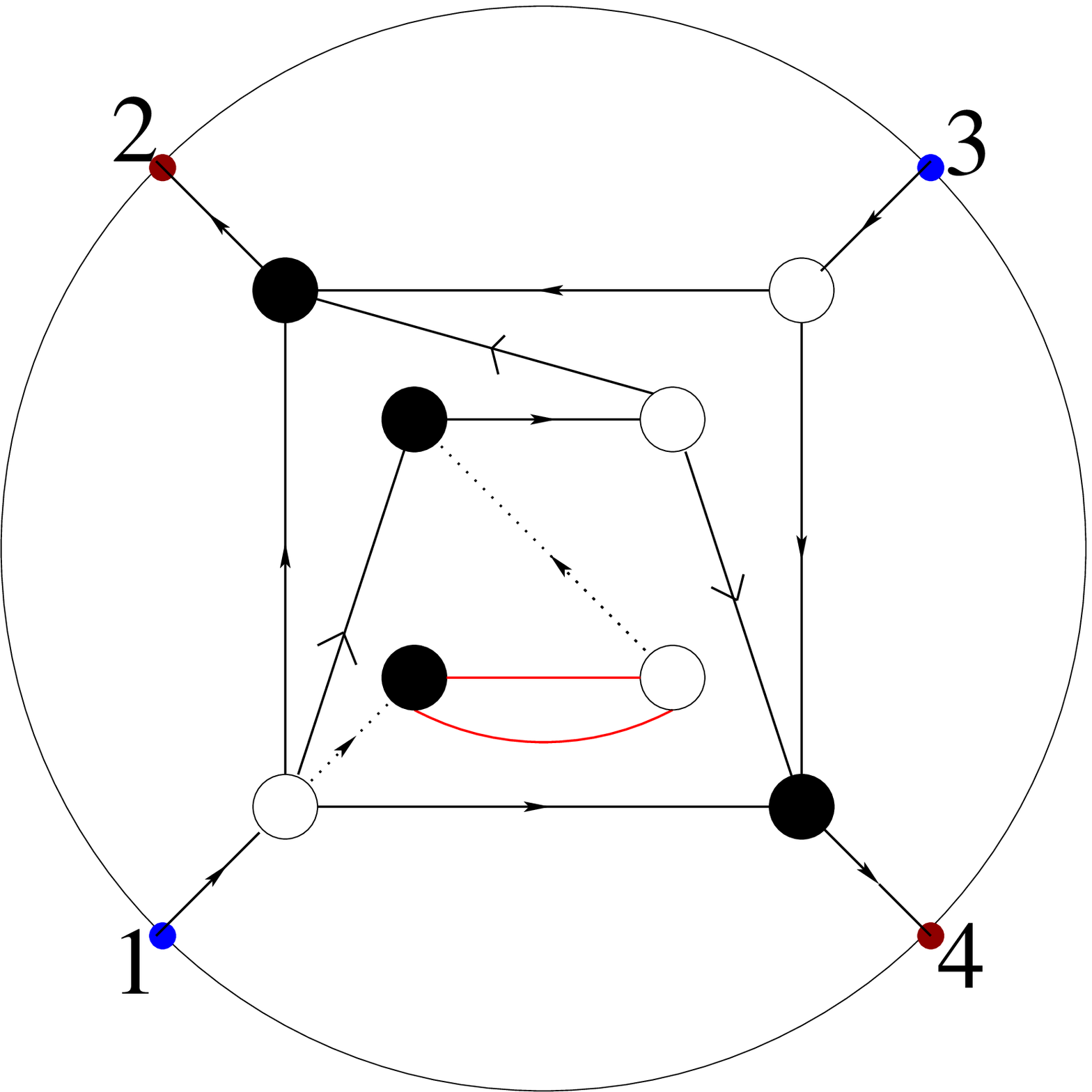}}}+
  \raisebox{-1.8cm}{\scalebox{.23}{\includegraphics{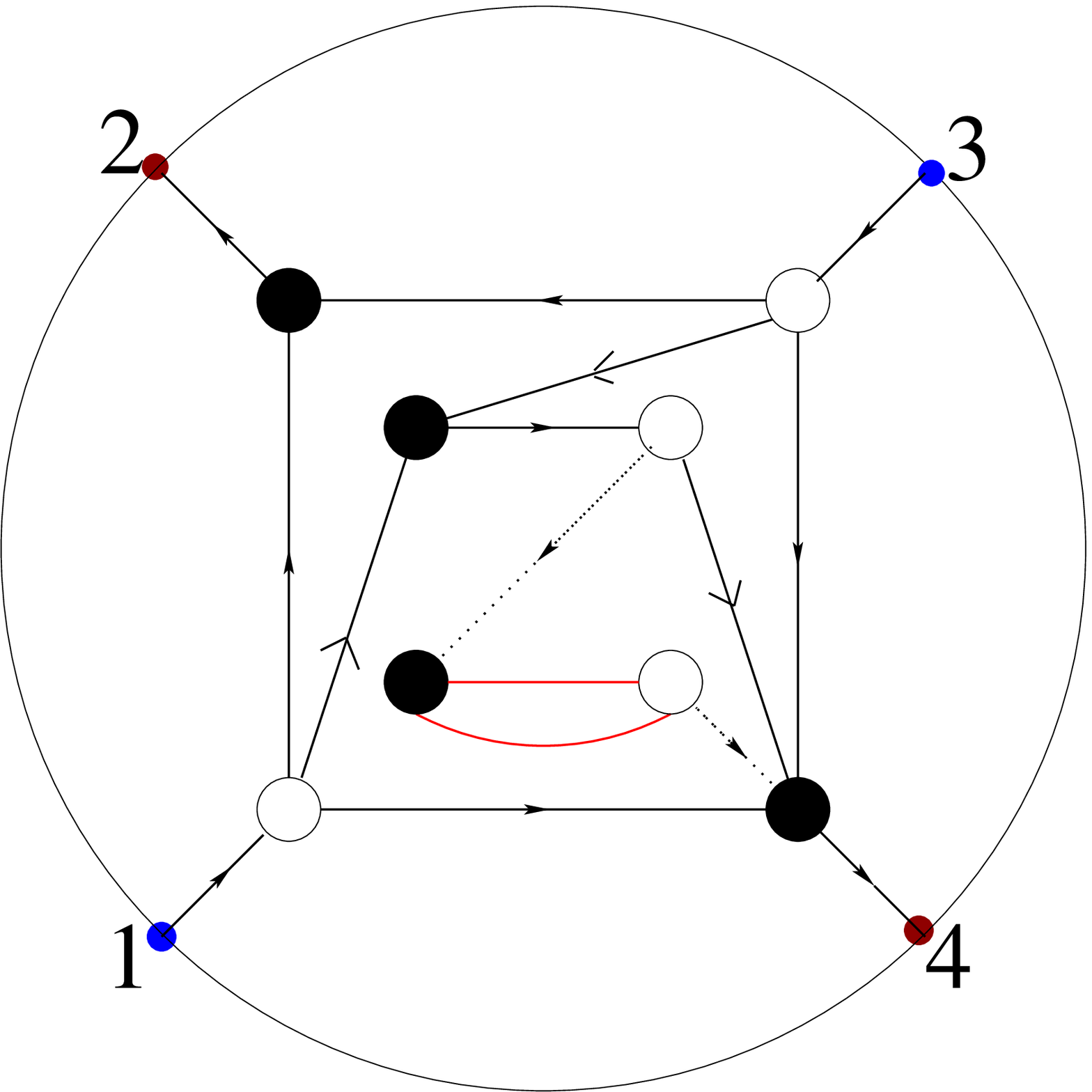}}}
\end{equation}
where the dotted lines, as before, identify the soft lines, and the on-shell diagrams are thought to be suitably
regularised. Notice that just the first diagram above can be generated by any BCFW bridges, while the other two
have a more peculiar structure. This suggests that the diagrammatic expression \eqref{eq:1lInt}, which has
been obtained by {\it assuming} that the residue of the pole in the one loop integrand under a BCFW deformation
in the $(4,1)$-channel is given by a tree-level six-particle forward amplitude, is incomplete. In some sense,
this might have been expected given that a BCFW deformation in the $(i,i+1)$-channel can see at most the external
bubbles related to the lines $i$ and $i+1$. Therefore, one can catch just part of these singularities. 
For supersymmetric theories, this is not really an issue, at least as long as we stick to one-loop, given that 
these terms are expected to be of order $\mathcal{O}(\epsilon)$, $\epsilon$ being a regularisation parameter. 
However, one would like to see this from a full-fledge on-shell perspective. What we are observing here is that both
the second and third term in \eqref{eq:1lInt} individually show non-local contribution to the forward singularity
and the overall leading behaviour turns out to be $\mathcal{O}(\epsilon^{-1})$ for $\mathcal{N}\,=\,1,\,2$ 
supersymmetric theories and $\mathcal{O}(\epsilon^{-3})$ for pure Yang-Mills in our quasi-forward regularisation. 
However, both terms show a common non-local pole whose contribution to the forward singularity cancels upon 
summation, so that the real behaviour turns out to be $\mathcal{O}(\epsilon^{0})$ and $\mathcal{O}(\epsilon^{-2})$
in the supersymmetric and non-supersymmetric cases respectively. This does not still coincide with the general
argument that supersymmetry is sufficient to cancel the external bubble divergencies \cite{CaronHuot:2010zt}. 

The divergencies on line-$1$, which are caught in the second term \eqref{eq:1lInt}, can in principle be singled-out
by a BCFW bridge in the $(1,2)$-channel as well. It is not difficult to understand that these two are not 
equivalent. For an explicit comparison, let us consider the on-shell diagrams generated by a BCFW bridge in the 
$(1,2)$-channel:
\begin{equation}\eqlabel{eq:1lInt2}
 \left.\mathcal{M}_4^{\mbox{\tiny $(1)$}}\right|_{\mbox{\tiny $(1,2)$}}\:=\:
  \raisebox{-1.8cm}{\scalebox{.23}{\includegraphics{1loopM4A.eps}}}+
  \raisebox{-1.8cm}{\scalebox{.23}{\includegraphics{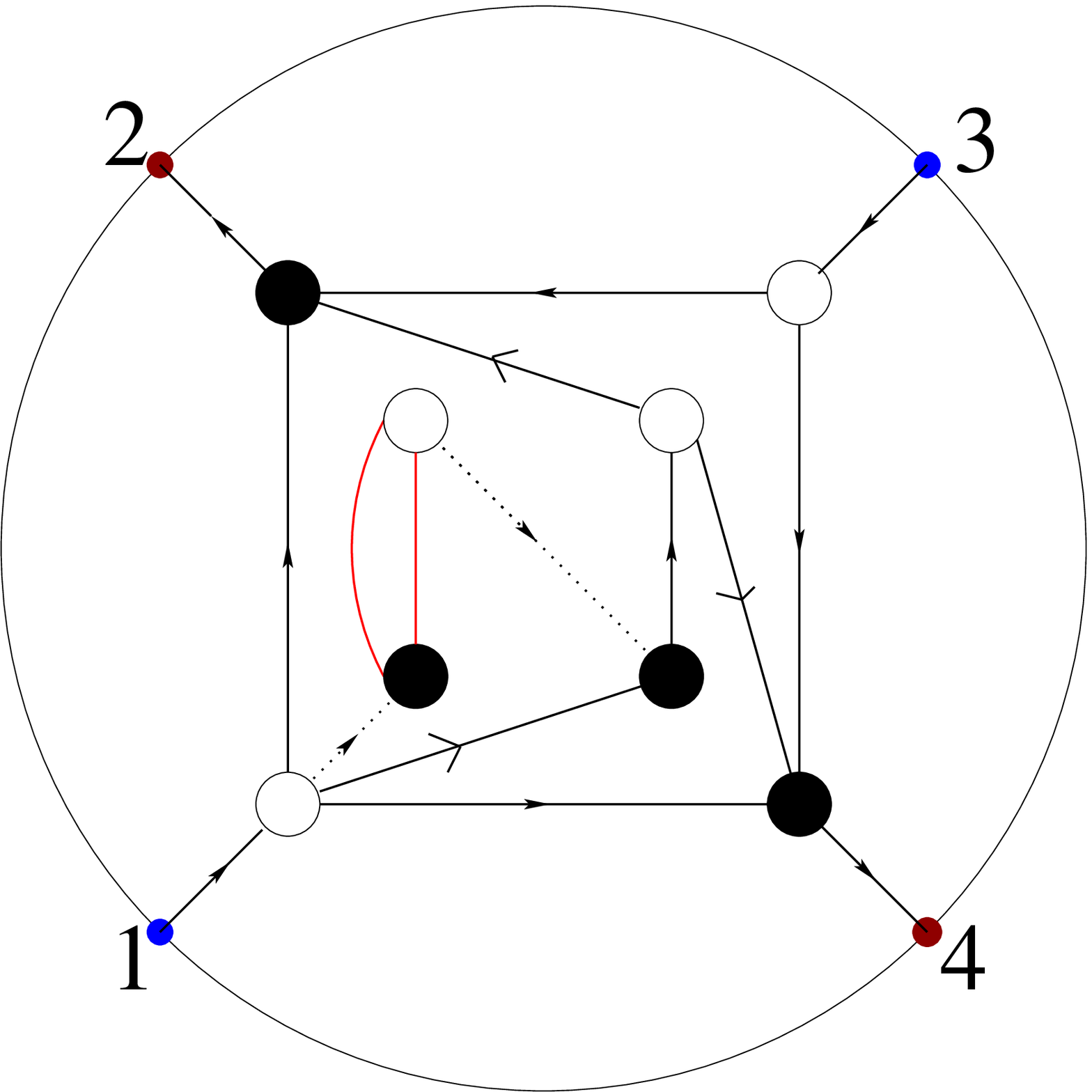}}}+
  \raisebox{-1.8cm}{\scalebox{.23}{\includegraphics{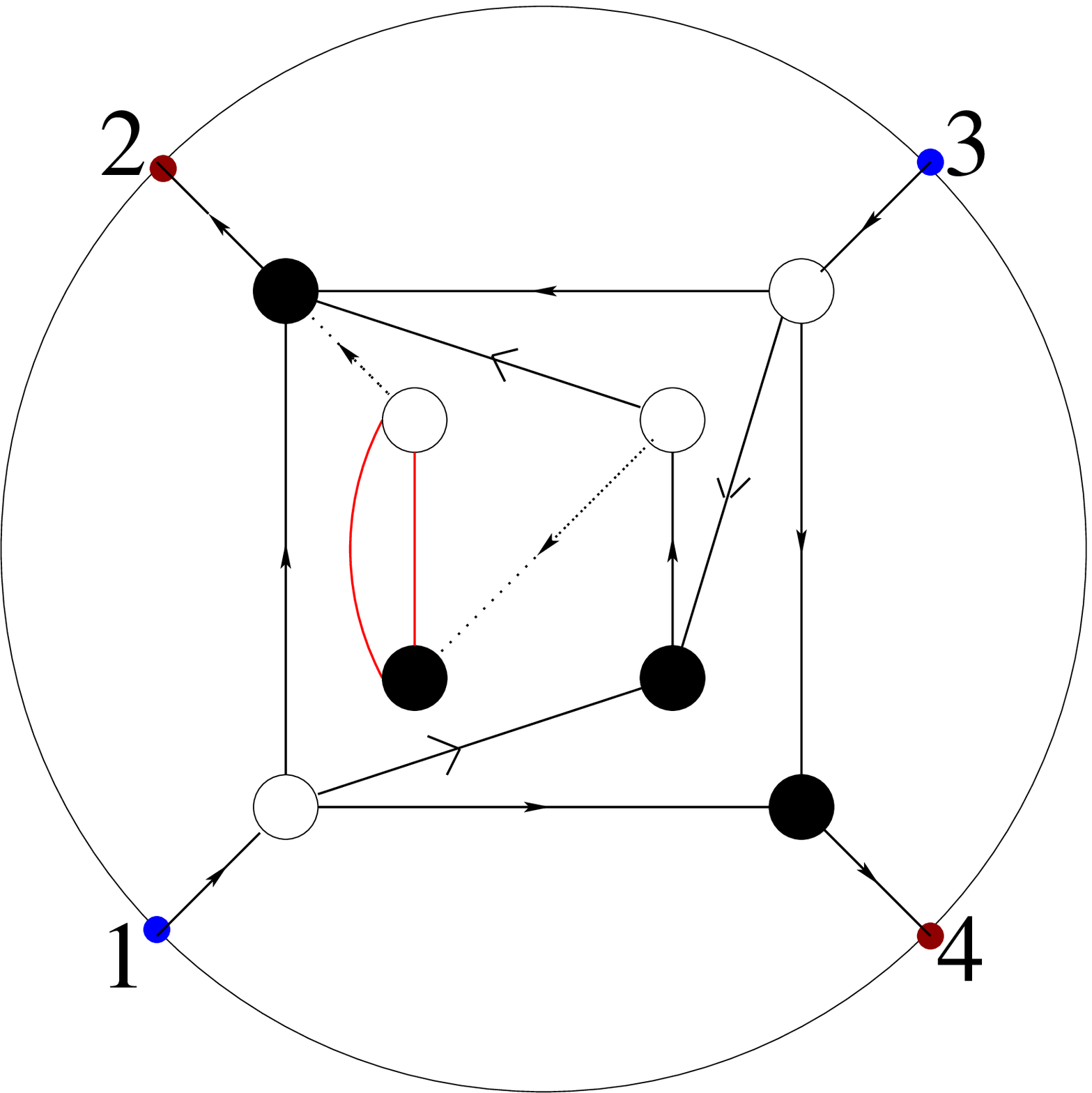}}}
\end{equation}
and consider the second term in both \eqref{eq:1lInt} and  \eqref{eq:1lInt2}, which can be recasted in a more
transparent form for our present purpose:
\begin{equation}\eqlabel{eq:1lInt3}
 \begin{split}
  &\raisebox{-1.8cm}{\scalebox{.23}{\includegraphics{1loopM4B.eps}}}\longleftrightarrow
   \raisebox{-1.8cm}{\scalebox{.15}{\includegraphics{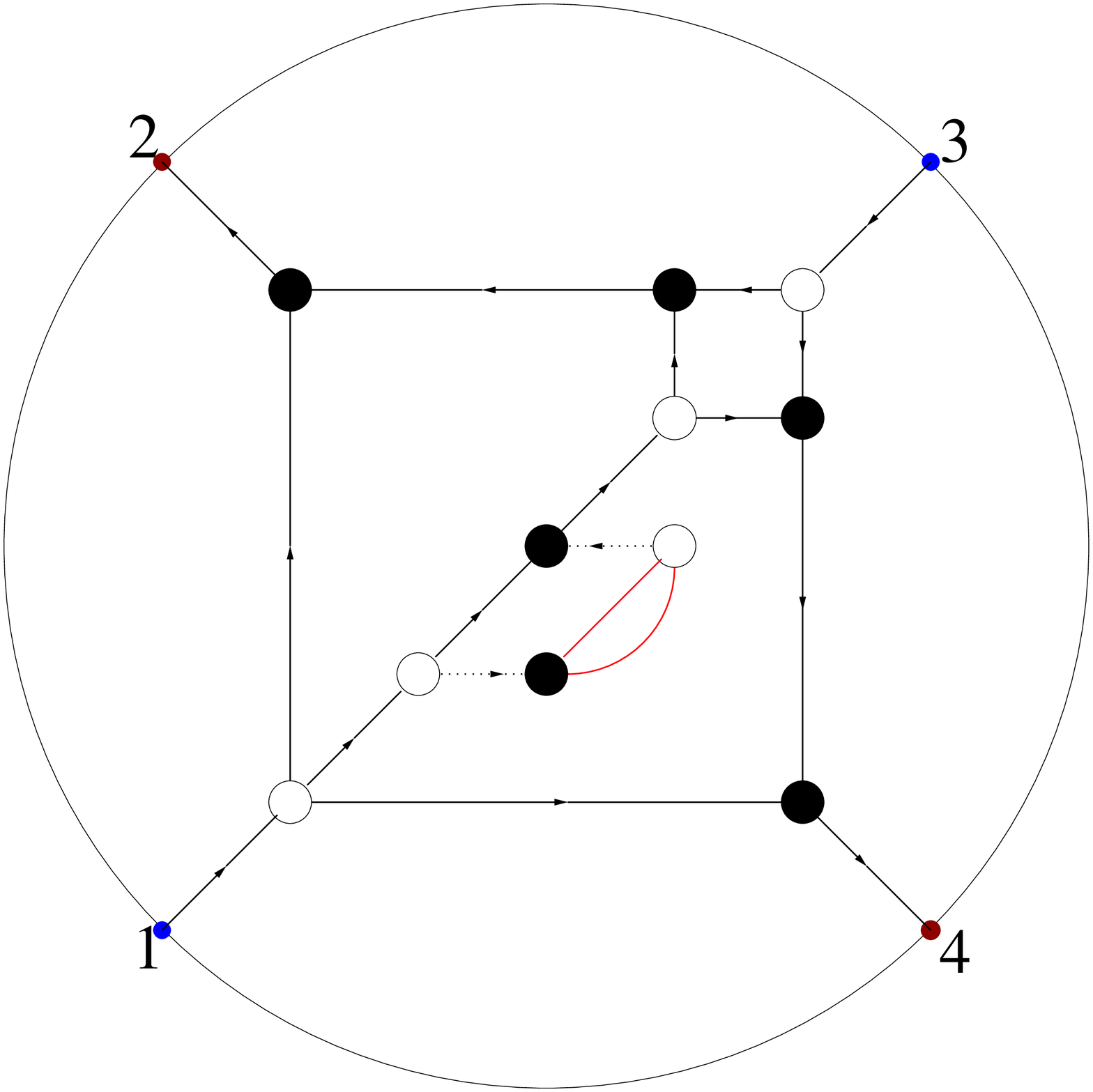}}}
  \\
  &\raisebox{-1.8cm}{\scalebox{.23}{\includegraphics{1loopM4E.eps}}}\longleftrightarrow
   \raisebox{-1.8cm}{\scalebox{.15}{\includegraphics{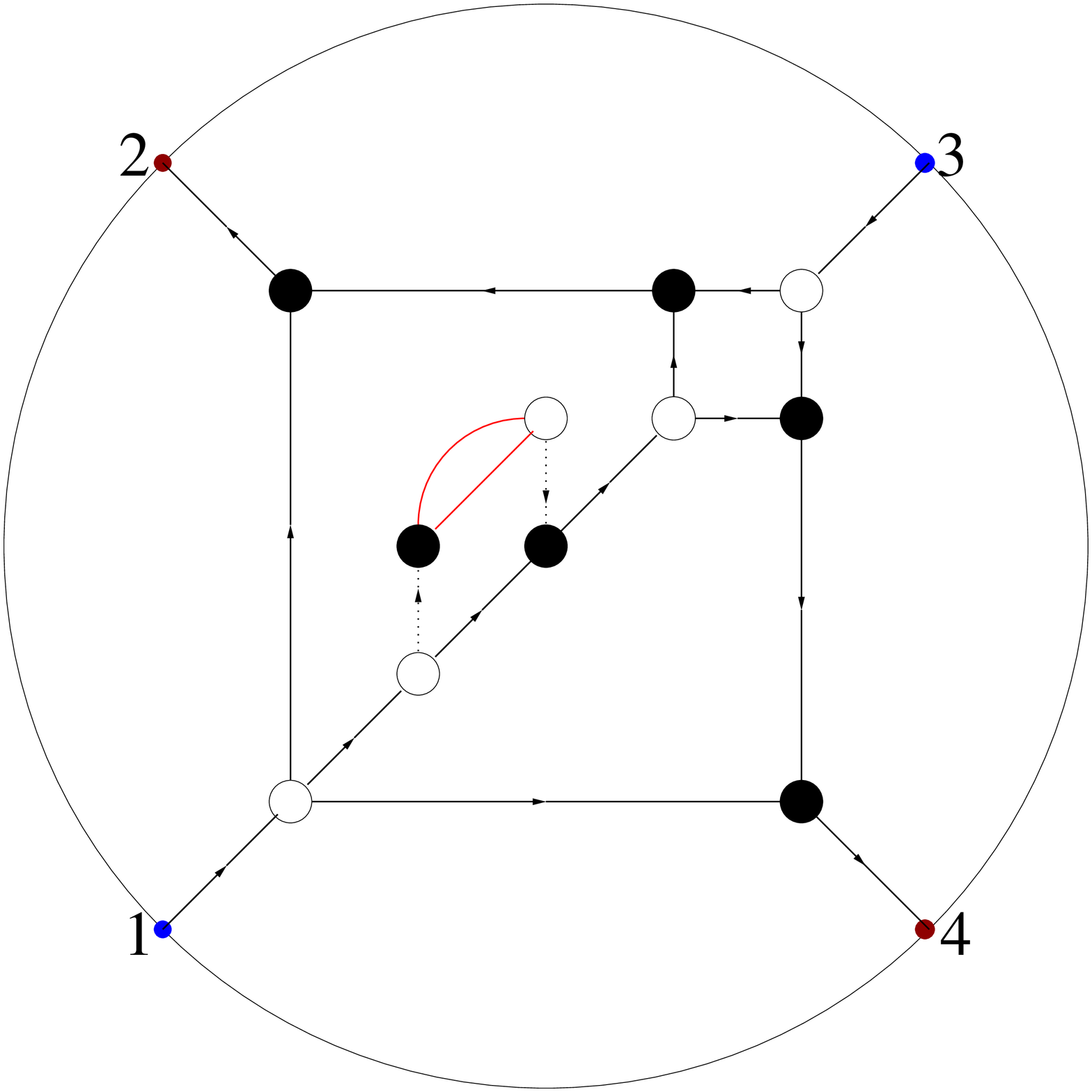}}}
 \end{split}
\end{equation}
where it becomes manifest how both terms can be seen as a two BCFW-bridges attached to a tree-level four-particle 
amplitude (the up-right on-shell box) with an on-shell bubble on the particle-$1$ line. What distinguish in between
the two lines in \eqref{eq:1lInt3} is {\it how} the on-shell bubble is attached on the particle-$1$ line.
It is important to stress the on-shell bubble structure, which is a bit peculiar: it appears as a loop which
is {\it softly} attached to a particle line and the side of the particle line is attached to affects the spinor
dependence of the ``soft terms''. Parametrising the right-hand-side of both lines in \eqref{eq:1lInt3} in such
a way that the two terms can be compared, it is easy to see that their sum provides with a further cancellation
returning an overall behaviour of the (quasi)-forward limits as $\mathcal{O}(\epsilon)$ for $\mathcal{N}\,=\,1,\,2$ 
and $\mathcal{O}(\epsilon^{-1})$ for $\mathcal{N}\,=\,0$.

Some comments are now in order. Even for the $\mathcal{N}\,=\,3,4$ cases, for which from the start the forward limit
was well-defined, seeing the one-loop integrand as the result of the integration of the singularity equation via
a BCFW bridge does not contain the full structure of the theory at this order. Luckily enough, for supersymmetric
theories these terms turn out to vanish. However, while for $\mathcal{N}\,=\,3,4$ all these terms vanish
individually\footnote{Indeed, for each {\it individual} diagram, one keeps into account both the summation among
the components of the loop multiplet as well as between all the multiplet which can propagate.}, so no real 
issue occurs, in the less-supersymmetric theories this is not the case: some cancellation occurs between the two
singular terms caught by a given BCFW bridge -- in particular, the contribution to the forward singularity of the
common non-local pole -- but still a single BCFW bridge does not seem capable to catch the full structure of the
singularity along a given particle line nor the singularities along all the particle lines. Diagrammatically,
this is shown by the appearance of on-shell bubbles just on the deformed lines and by the fact the way the on-shell
bubble is attached to a given line $i$ changes depending on whether one is considering a BCFW bridge in the
$(i,i+1)$- or in the $(i-1,i)$-channel, as it occurs in \eqref{eq:1lInt3}.

These missing contributions should be related to a boundary term in the integrand: under a given BCFW deformation of
the one-loop integrand, the loop structure along the undeformed external lines remain blind to it. They can be 
recovered by applying to the eventual boundary term other BCFW bridges, in a sort of multi-step procedure.
Diagrammatically, this is equivalent to complete \eqref{eq:1lInt} on symmetry basis to get:
 \begin{align}\eqlabel{eq:1lIntFin}
  \mathcal{M}_4^{\mbox{\tiny $(1)$}}\:=\:
  &\raisebox{-1.8cm}{\scalebox{.23}{\includegraphics{1loopM4A.eps}}}+
   \raisebox{-1.8cm}{\scalebox{.15}{\includegraphics{1loopM4F.eps}}}+
   \raisebox{-1.8cm}{\scalebox{.15}{\includegraphics{1loopM4G.eps}}}+ \nonumber\\
  &\raisebox{-1.8cm}{\scalebox{.15}{\includegraphics{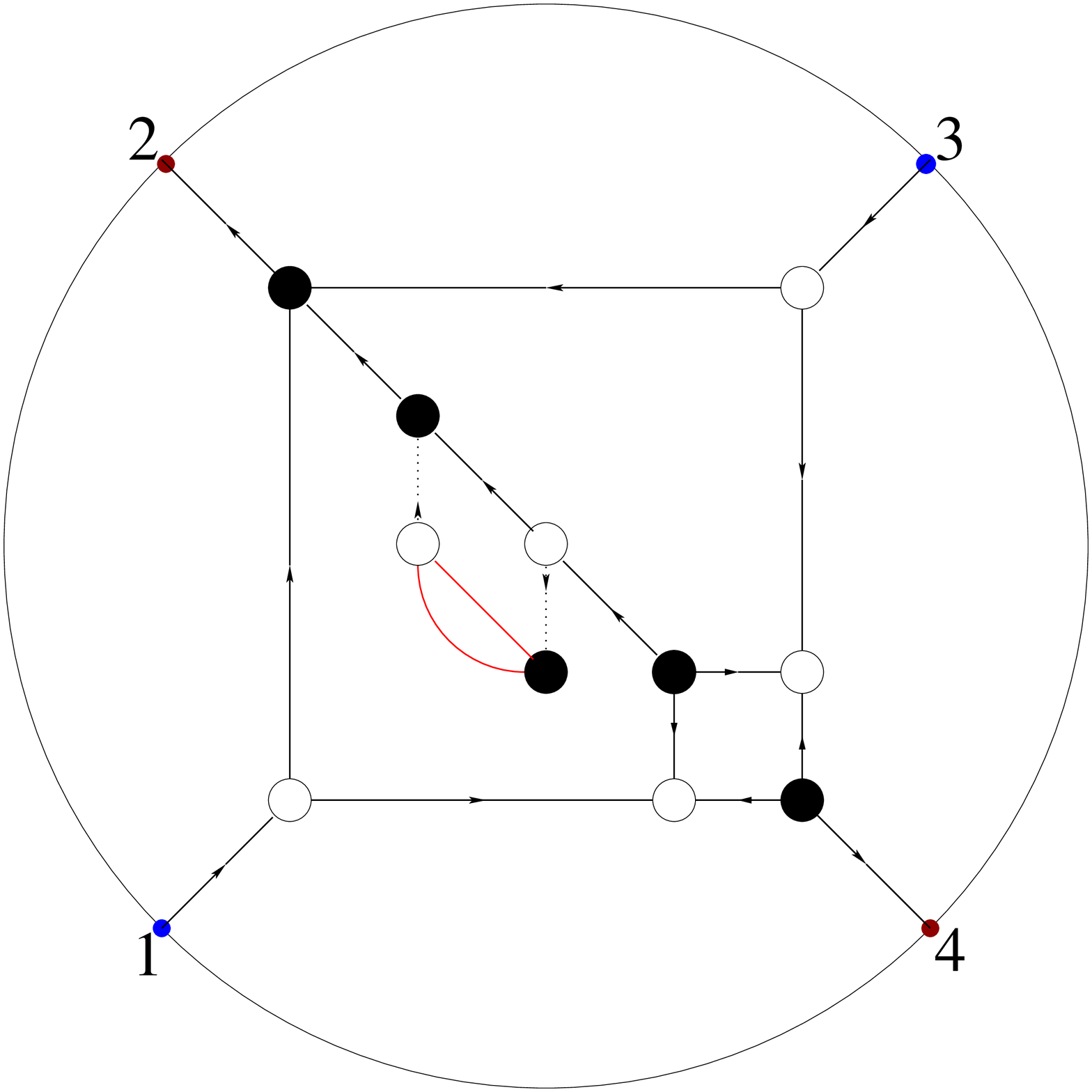}}}+
   \raisebox{-1.8cm}{\scalebox{.15}{\includegraphics{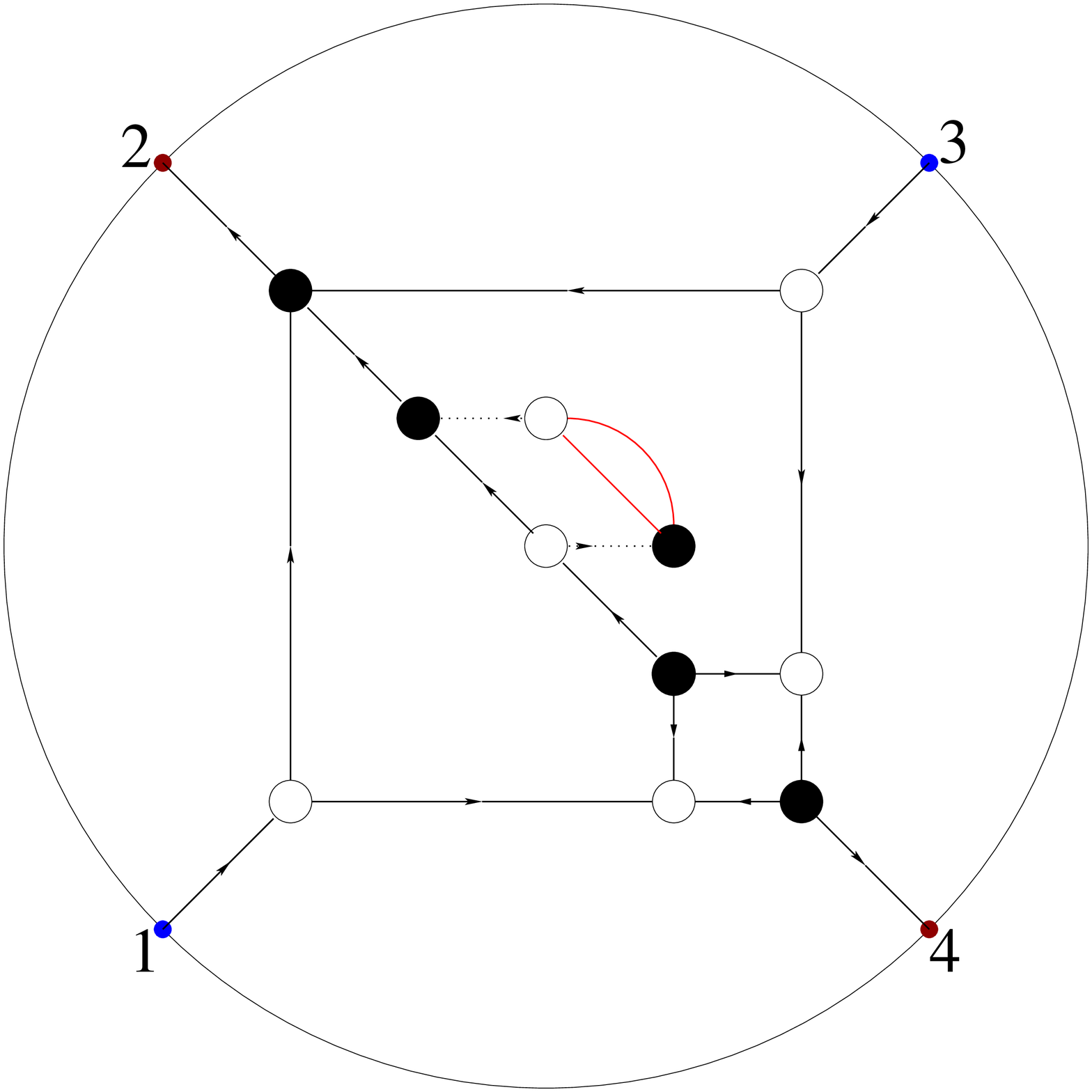}}}+
   \raisebox{-1.8cm}{\scalebox{.15}{\includegraphics{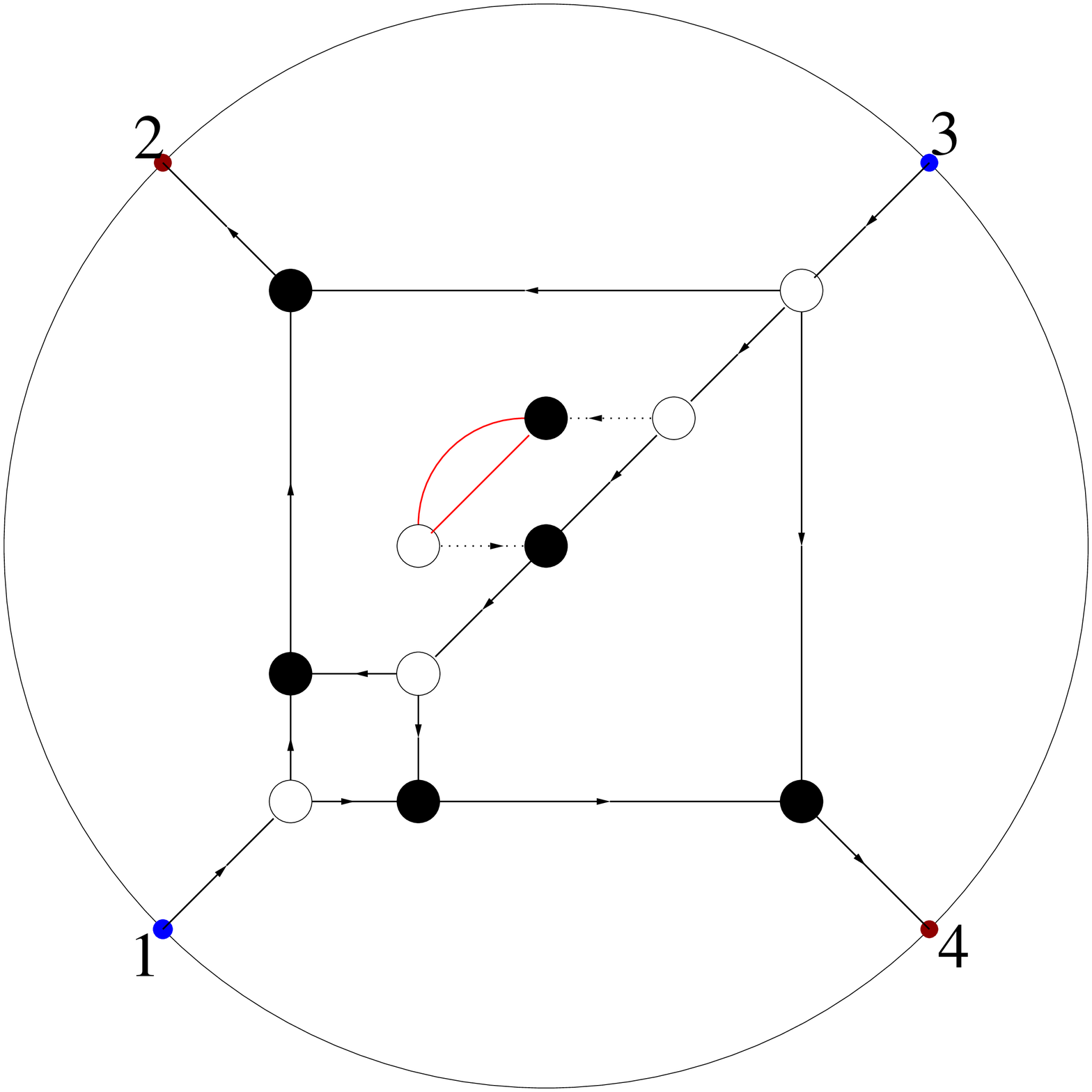}}}+\displaybreak[1]\\
  &\raisebox{-1.8cm}{\scalebox{.15}{\includegraphics{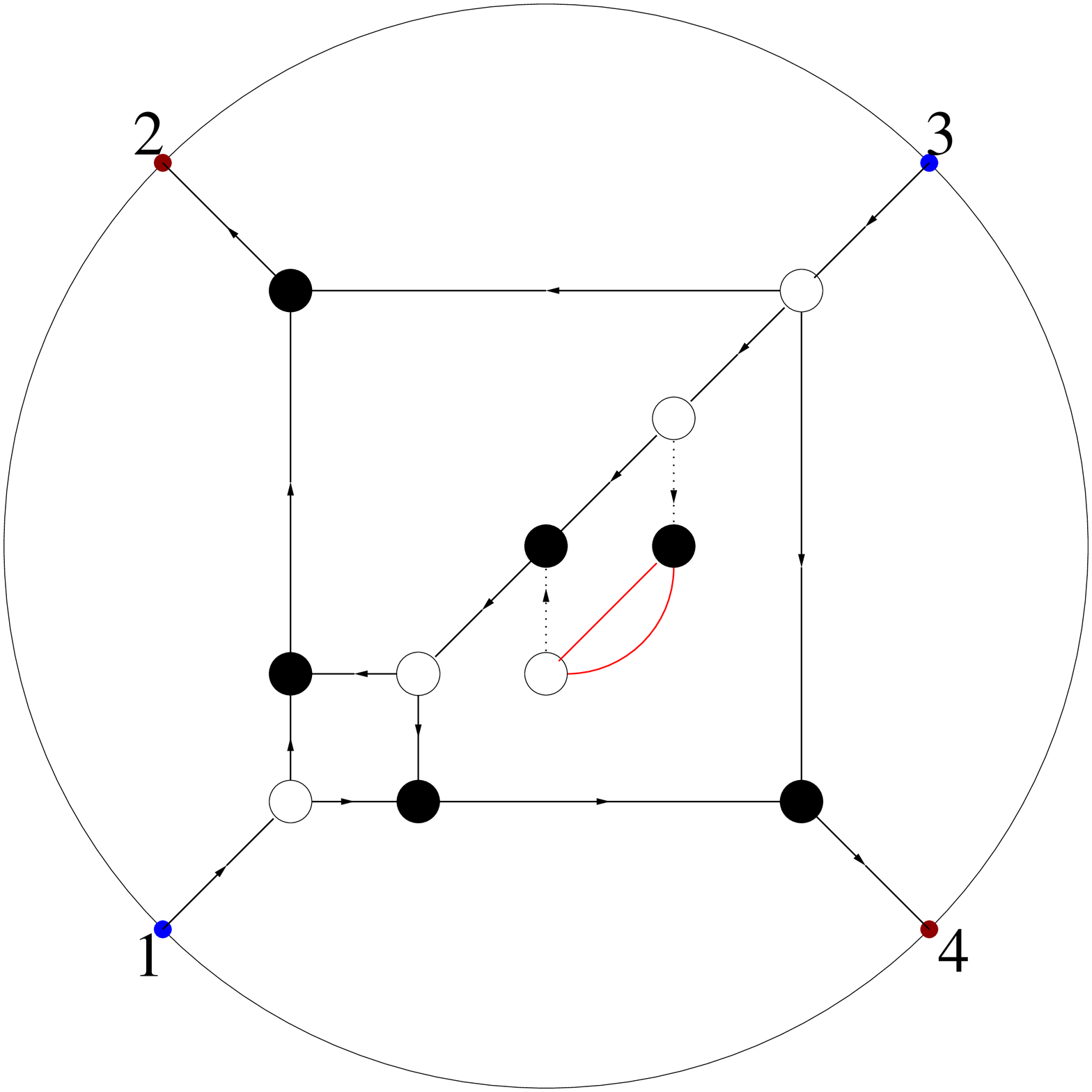}}}+
   \raisebox{-1.8cm}{\scalebox{.15}{\includegraphics{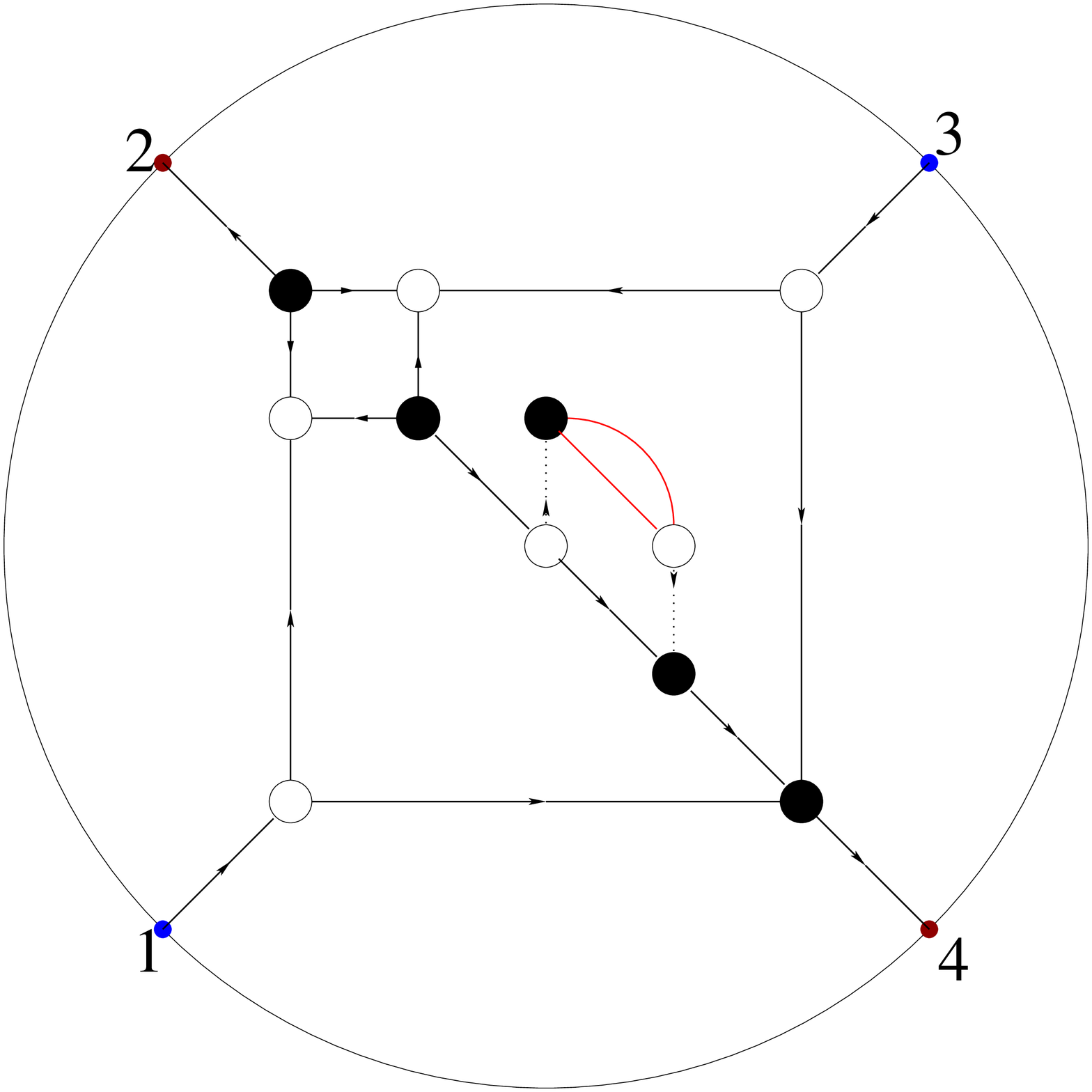}}}+
   \raisebox{-1.8cm}{\scalebox{.15}{\includegraphics{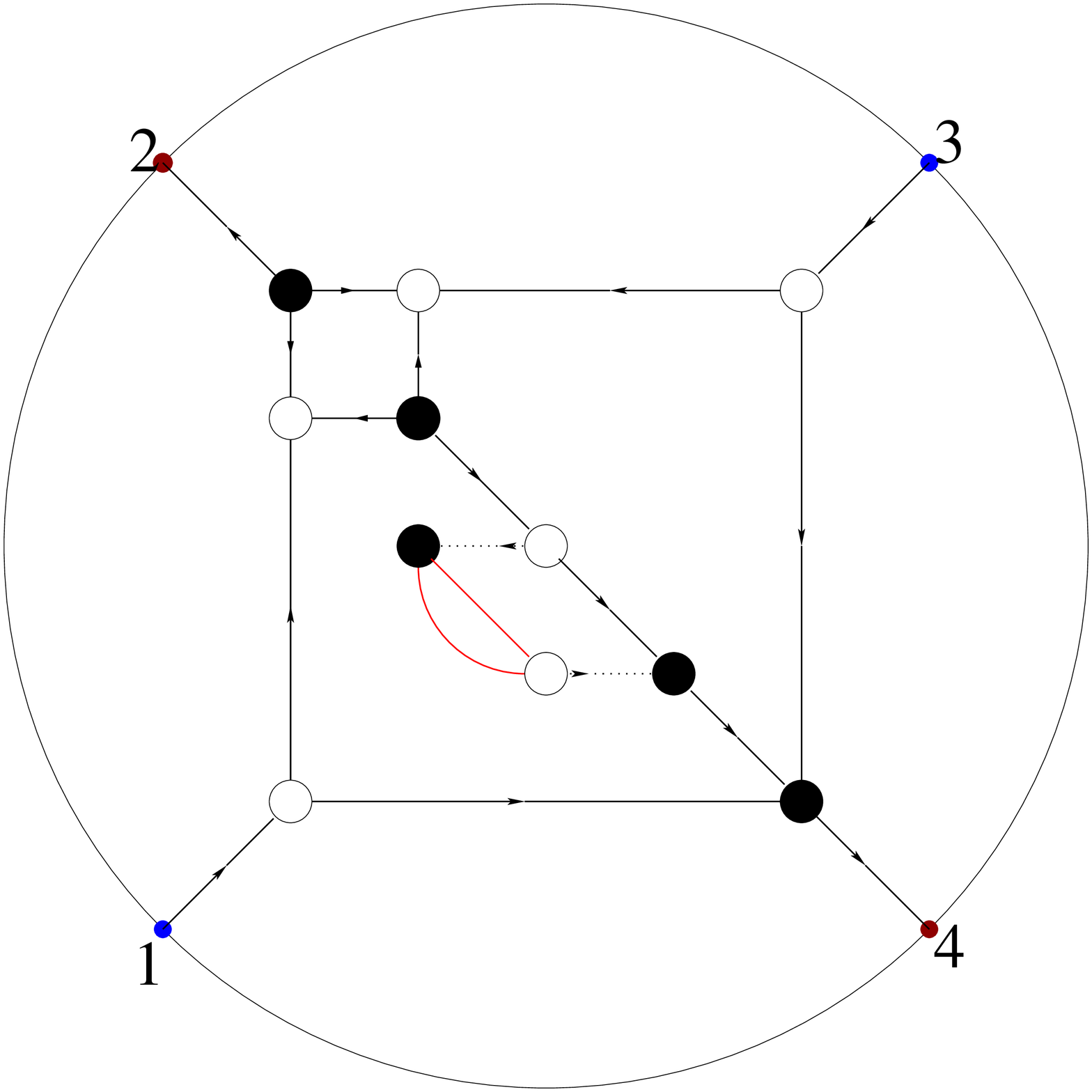}}} \nonumber
 \end{align}

In the supersymmetric case, this issue is important just to have a full-fledge on-shell proof of the fact that
the singular terms in the forward limit do not contribute even at integrand level once a suitable regularisation
is introduced. The cut-constructible structure is instead completely encoded in the non-singular term (the first
in \eqref{eq:1lIntFin}), which can be obtained by any suitable BCFW bridge as its very symmetric structure 
explicitly shows.

For pure Yang-Mills instead having a good control on the divergent term is crucial for constructing higher loops
on-shell diagrams. One can imagine that a contribution to a two-loop integrand -- {\it i.e.} an 8-form -- can be
 given by suitably gluing together two one-loop on-shell diagrams. In this case a finite contribution can be given
by a divergent term (in the regularisation parameter) of one of two diagrams and an order $\mathcal{O}(\epsilon)$
term for the other one. Furthermore, these terms are important for studying in detail the UV-divergent structure.

\subsection{Rational terms and the one-loop integrand}\label{subsec:Rats}

In order to complete our discussion about the four-particle integrand at one-loop we need to understand how
the information about the rational terms, {\it i.e.} those contributions to the one-loop amplitude which
are not characterised by branch-cuts, are encoded from this on-shell point of view. In general, these terms
are not caught by four-dimensional unitarity, but rather by considering unitarity in $D$-dimensions
\cite{Bern:1995db, Brandhuber:2005jw, Anastasiou:2006jv, Britto:2006fc, Anastasiou:2006gt, Britto:2008vq,
Britto:2008sw, Feng:2008ju, Badger:2008cm, NigelGlover:2008ur}. Furthermore, the gluon loops are always dealt
with by using a decomposition in terms of $\mathcal{N}\,=\,4$ and $\mathcal{N}\,=\,1$ multiplets as well as a 
scalar, so that the rational terms are computed from the latter contribution only. The massless scalars 
in $D\,=\,4-2\epsilon$ are actually equivalent to a four-dimensional massive scalar. As a consequence,
helicity configurations such as the all-plus and the mostly plus one becomes non-trivial 
\cite{Bern:1991aq, Bern:1995db}, differently from what happens at tree level for which they vanish.

By itself, the quasi-forward regularisation scheme we have been using to make sense of the in principle ill-defined
forward limit, is not capable to catch this type of contribution. The reason is easy to understand: this scheme
regularises an integrand which can be defined in principle but has some pathologies, while it does not give
rise to new degrees of freedom and thus to new integrands. Therefore, in order to be able to detect the missing
terms, we consider the contribution of some massive scalar in the forward lines, in a similar fashion as in
dimensional regularisation. Indeed, this is not satisfactory at all, but allows us to provide a first realisation of
such contribution in terms of on-shell processes. For a complete satisfactory treatment one would need to
propose a full-fledge regularisation scheme for the on-shell diagrams and, thus, for the {\it integrand} which
is capable to take care of the forward terms and catch the non-cut-constructible terms at once. We leave this
to future work.

Let us turn to those configurations whose tree-level amplitude vanish. In these cases the one-loop amplitude
is completely given by a rational term, and therefore, thinking of it in terms of the decomposition mentioned above,
it is fully equivalent to having just a massive scalar running in the loop. From our perspective, this amplitude
is determined by the forward limit of a tree-level-six particle amplitude. Let us consider some explicit example.

\subsubsection{The all-plus four-gluon integrand}\label{subsubsec:PPPPint}

We begin with simplest example. As already mentioned, this class of amplitudes does not have neither factorisation
channels nor branch cuts. However, {\it under a suitable regularisation} it is possible to mapping in an object --
our integrand -- which do have a pole structure. In what follows, we {\it assume} the existence of a regularisation
scheme which reduces the problem to considering a four-dimensional massive scalar running in the loop, as it
happens in dimensional regularisation. The introduction of a (regularising) mass allows for non-vanishing on-shell
$0$-forms and thus there is a concrete sense in which one can think of the residues of the poles in the integrand 
we want to compute as forward amplitudes. Specifically, the full integrand is completely determined by such a
singularity and it can be written in same fashion of the MHV pure gluonic amplitudes:
\begin{equation}\eqlabel{eq:M41lallp}
 \mathcal{M}_4^{\mbox{\tiny $(1)$}}(1^{+},2^{+},3^{+},4^{+})\:=\:
  \raisebox{-1.8cm}{\scalebox{.45}{\includegraphics{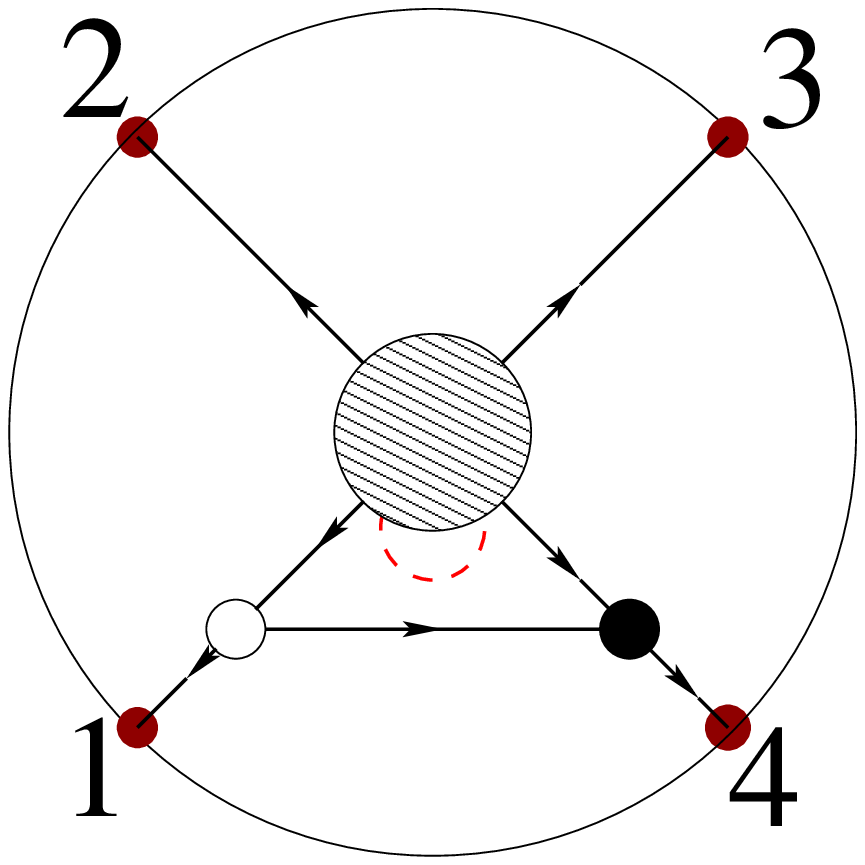}}}\:=\:
  \raisebox{-2.0cm}{\scalebox{.25}{\includegraphics{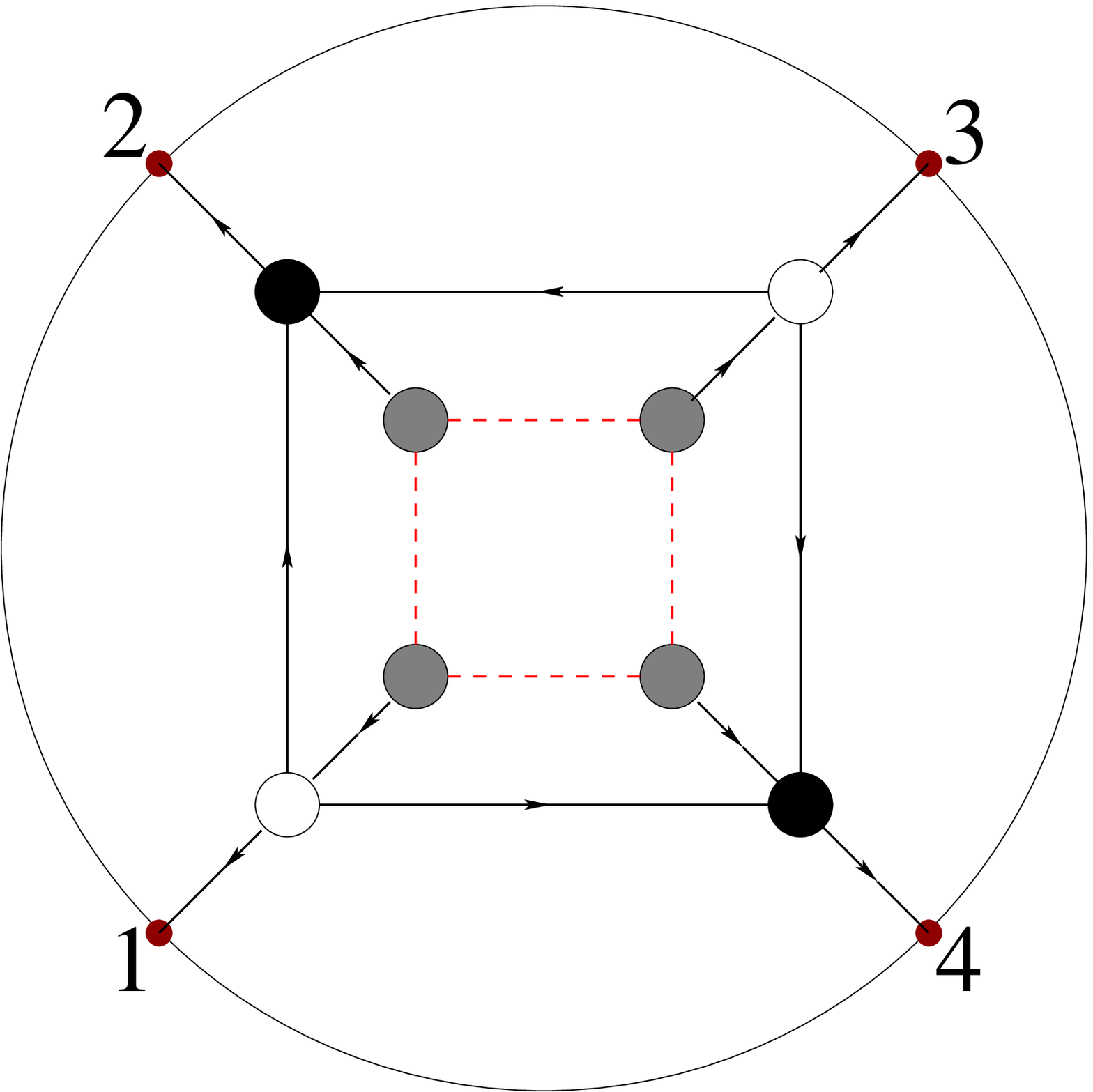}}},
\end{equation}
where the dashed red lines indicate the (forward) massive scalars while the grey nodes are the three-point 
amplitudes with one massless gluon and two massive scalars:
\begin{equation}\eqlabel{eq:3ptGSS}
 \begin{split}
   \mathcal{M}_3(1^0,2^{+},3^0)\:&\equiv\:
   \raisebox{-1.2cm}{\scalebox{.33}{\includegraphics{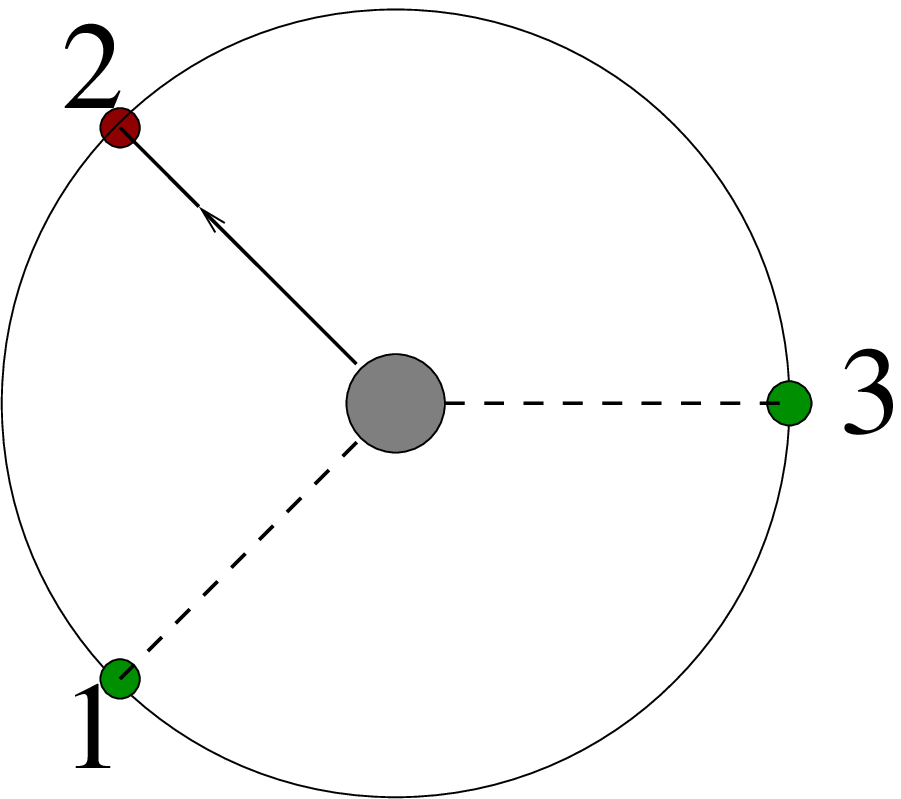}}}\:=\\
  &=\:\delta^{\mbox{\tiny $(2\times2)$}}
   \left(
    \sum_{i=1}^3\lambda^{\mbox{\tiny $(i)$}}\tilde{\lambda}^{\mbox{\tiny $(i)$}}+
    m^2\sum_{j=3}^1\frac{q_j\tilde{q}_j}{\langle j,q_j\rangle[j,q_j]}
   \right)
   m^2\frac{[2,1]}{\langle2|p^{\mbox{\tiny $(1)$}}|1]},\\
    \mathcal{M}_3(1^0,2^{-},3^0)\:&\equiv\:
   \raisebox{-1.2cm}{\scalebox{.33}{\includegraphics{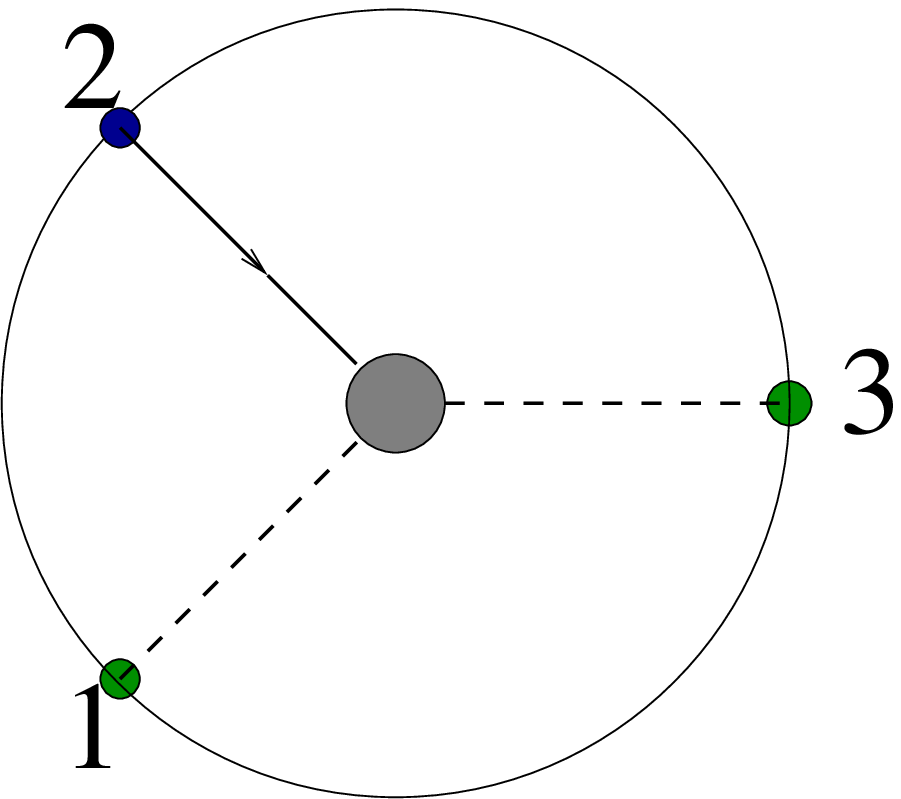}}}\:=\\
  &=\:\delta^{\mbox{\tiny $(2\times2)$}}
   \left(
    \sum_{i=1}^3\lambda^{\mbox{\tiny $(i)$}}\tilde{\lambda}^{\mbox{\tiny $(i)$}}+
    m^2\sum_{j=3}^1\frac{q_j\tilde{q}_j}{\langle j,q_j\rangle[j,q_j]}
   \right)
   m^2\frac{\langle3,2\rangle}{\langle3|p^{\mbox{\tiny $(3)$}}|2]}.
 \end{split}
\end{equation}
The momenta of the massive scalars are represented as the sum of two light-like bispinors with
$q_j$ and $\tilde{q}_j$ being the reference spinors related to particle $j$.

Following the diagrammatic reasoning in Section \ref{sec:TreeStr}, it is possible to show that tree-level
amplitudes with massive scalars and at least two external massless gluons can be fully expressed in terms
of on-shell processes built out of the three-particle amplitudes \eqref{eq:3ptampl} for pure gluons 
($\mathcal{N}\,=\,0$) and the gluon-scalar amplitudes \eqref{eq:M41lallp}. In the case we are interested now, the 
residue of the loop propagator is the forward limit of the six-particle amplitude 
$\mathcal{M}_6^{\mbox{\tiny tree}}(1^{+},2^+,3^+,4^+,{\color{red} A},{\color{red} B})$:
\begin{equation}\eqlabel{eq:GSSfw}
 \raisebox{-1.8cm}{\scalebox{.34}{\includegraphics{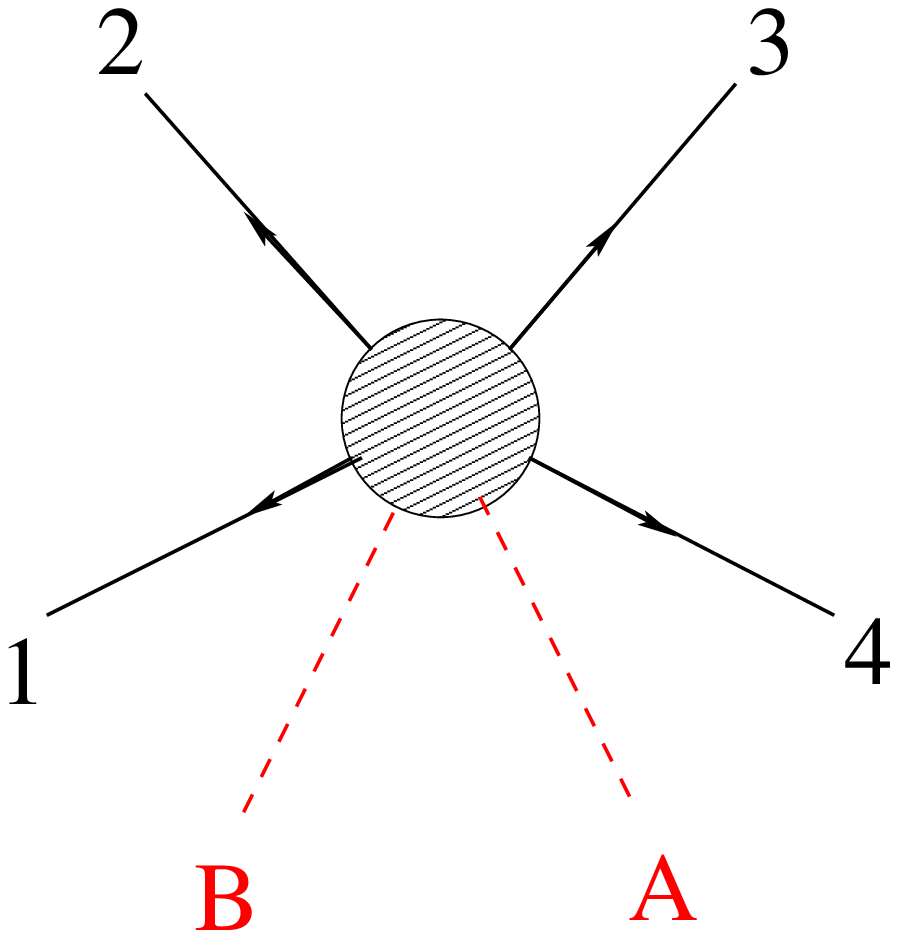}}}\longrightarrow\:
 \raisebox{-1.6cm}{\scalebox{.22}{\includegraphics{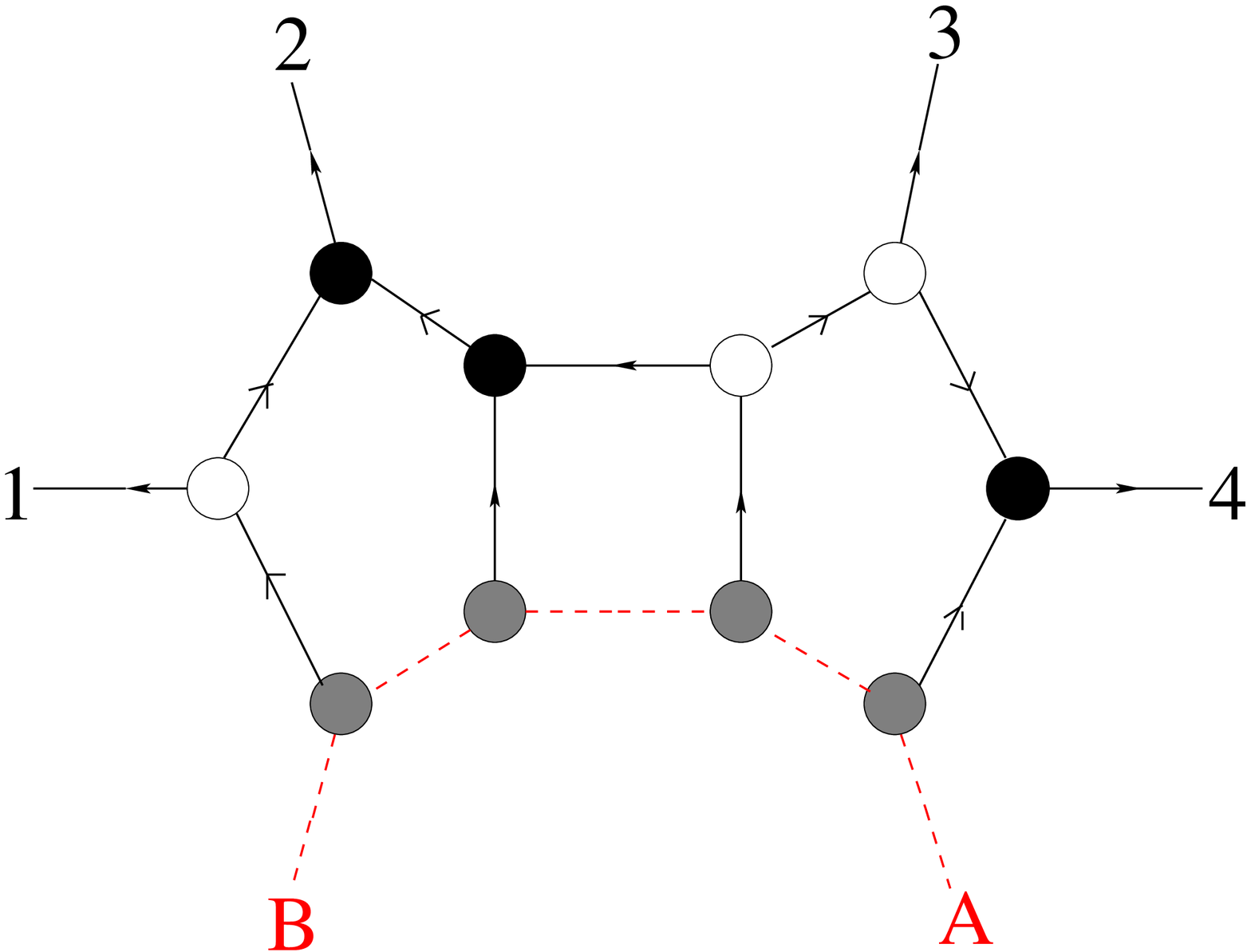}}}\:\longrightarrow\:
 \raisebox{-1.0cm}{\scalebox{.22}{\includegraphics{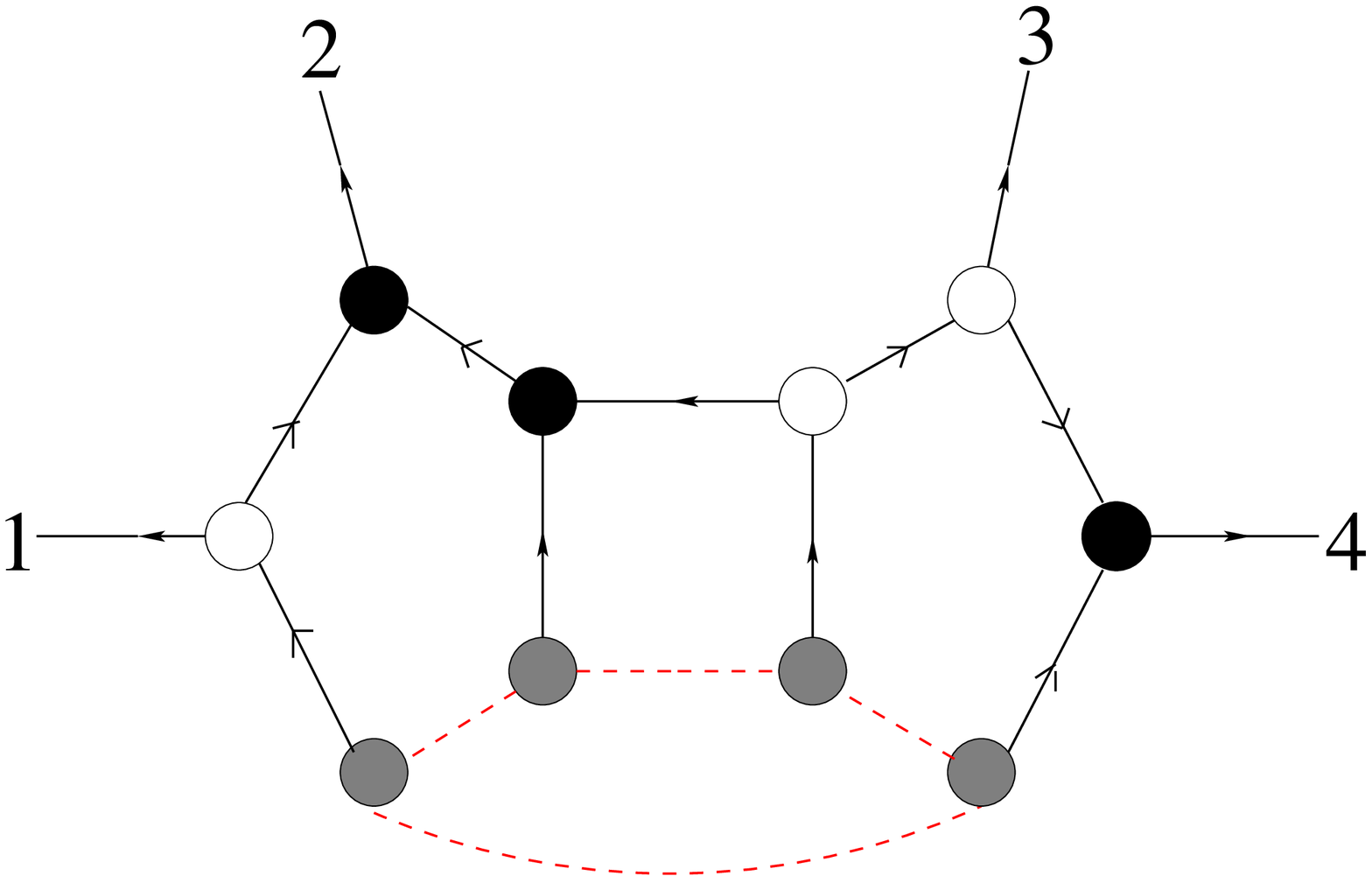}}}.
\end{equation}
Notably, the six-particle amplitude of interest can be represented by a two on-shell processes one of which 
turns out to be well-defined in the forward limit, while the other one vanishes:
\begin{equation}\eqlabel{eq:GSSfw2}
 \begin{split}
  \raisebox{-1.1cm}{\scalebox{.23}{\includegraphics{6ptFwSSm3.eps}}}\:=\:
  &\frac{d^2\lambda^{\mbox{\tiny $(A)$}}d^2\tilde{\lambda}^{\mbox{\tiny $(A)$}}}{\mbox{Vol}\{GL(1)\}}
  \frac{m^4\frac{st}{\langle1,2\rangle\langle2,3\rangle\langle3,4\rangle\langle4,1\rangle}}{
             \left(2p^{\mbox{\tiny $(A)$}}p^{\mbox{\tiny $(1)$}}\right)
             \left(2p^{\mbox{\tiny $(A)$}}p^{\mbox{\tiny $(4)$}}\right)
             \left(P_{12}^2-2p^{\mbox{\tiny $(A)$}}P_{12}\right)}\\
  &p^{\mbox{\tiny $(A)$}}\:=\:\tau\lambda^{\mbox{\tiny $(A)$}}\tilde{\lambda}^{\mbox{\tiny $(A)$}}+
    m^2\frac{q\tilde{q}}{\tau\langle A,q\rangle[A,q]},
 \end{split}
\end{equation}
which coincides with the single-cut computation in \cite{NigelGlover:2008ur}. Notice that the phase-space
of the massive forward state coincides with the one of its light-like projection because the on-shell 
condition fixes its further degree of freedom to be $m^2/(\langle A,q\rangle[A,q])$. Attaching the BCFW bridge
in the $(4,1)$-channel and using the merger operation, one can easily get to the on-shell diagram at the very
right of \eqref{eq:M41lallp}: (the integrand of) the all-plus four-particle amplitude can be seen as four
BCFW bridges attached to an on-shell $0$-form. Explicitly, let us apply the BCFW bridge in the $(4,1)$-channel
to the $3$-form \eqref{eq:GSSfw2}:
\begin{equation}\eqlabel{eq:M41lpppp}
 \begin{split}
   \raisebox{-1.6cm}{\scalebox{.20}{\includegraphics{1loopM4pppp.eps}}}\:=\:
  &\frac{dz_{41}}{z_{41}}\,
   \frac{d^2\lambda^{\mbox{\tiny $(A)$}}d^2\tilde{\lambda}^{\mbox{\tiny $(A)$}}}{\mbox{Vol}\{GL(1)\}}
   \frac{m^4\frac{st}{\langle1,2\rangle\langle2,3\rangle\langle3,4\rangle\langle4,1\rangle}}{
             \left[2p^{\mbox{\tiny $(A)$}}p^{\mbox{\tiny $(1)$}}(z)\right]
             \left[2p^{\mbox{\tiny $(A)$}}p^{\mbox{\tiny $(4)$}}(z)\right]
             \left[P_{12}^2(z)-2p^{\mbox{\tiny $(A)$}}P_{12}(z)\right]}\\
  &p^{\mbox{\tiny $(1)$}}(z)\:=\:
   \lambda^{\mbox{\tiny $(1)$}}
   \left(
    \tilde{\lambda}^{\mbox{\tiny $(1)$}}-z_{41}\tilde{\lambda}^{\mbox{\tiny $(4)$}}
   \right),\quad
   p^{\mbox{\tiny $(4)$}}(z)\:=\:
   \left(
    \lambda^{\mbox{\tiny $(4)$}}+z_{41}\lambda^{\mbox{\tiny $(1)$}}
   \right)
   \tilde{\lambda}^{\mbox{\tiny $(4)$}},
 \end{split}
\end{equation}
where $p^{\mbox{\tiny $(A)$}}$ is given in \eqref{eq:GSSfw2}. It is straightforward to notice that the 
usual off-shell loop momentum $l$ is related to the above parametrisation via 
$l\,=\,p^{\mbox{\tiny $(A)$}}+z_{41}\lambda^{\mbox{\tiny $(1)$}}\tilde{\lambda}^{\mbox{\tiny $(4)$}}$,
with the pole $z_{41}\,=\,0$ clearly related to the massive on-shell condition $(l^2-m^2)\,=\,0$.
Seeing instead our on-shell process as generated by four BCFW bridges applied to the (internal) $0$-form on-shell 
box:
\begin{equation}\eqlabel{eq:M41lpppp2}
 \begin{split}
  &\raisebox{-1.6cm}{\scalebox{.20}{\includegraphics{1loopM4pppp.eps}}}\:=\:
   \frac{m^4}{\langle1,2\rangle\langle2,3\rangle\langle3,4\rangle\langle4,1\rangle}
   \bigwedge_{i=1}^4\frac{dz_{i,i+1}}{z_{i,i+1}(1+a_{i,i+1}z_{i,i+1})},\\
  &\left\{
    \begin{array}{l}
     \tilde{\lambda}^{\mbox{\tiny $(1)$}}(z)\:=\:
      \tilde{\lambda}^{\mbox{\tiny $(1)$}}-z_{41}\tilde{\lambda}^{\mbox{\tiny $(4)$}}
                                          -z_{12}\tilde{\lambda}^{\mbox{\tiny $(2)$}},\\
     \lambda^{\mbox{\tiny $(2)$}}(z)\:=\:
      \lambda^{\mbox{\tiny $(2)$}}+z_{12}\lambda^{\mbox{\tiny $(1)$}}+z_{23}\lambda^{\mbox{\tiny $(3)$}},\\
     \tilde{\lambda}^{\mbox{\tiny $(3)$}}(z)\:=\:
      \tilde{\lambda}^{\mbox{\tiny $(3)$}}-z_{23}\tilde{\lambda}^{\mbox{\tiny $(2)$}}
                                          -z_{34}\tilde{\lambda}^{\mbox{\tiny $(4)$}},\\
     \lambda^{\mbox{\tiny $(4)$}}(z)\:=\:
      \lambda^{\mbox{\tiny $(4)$}}+z_{34}\lambda^{\mbox{\tiny $(3)$}}+z_{41}\lambda^{\mbox{\tiny $(1)$}}
    \end{array}
   \right.
   \qquad
   \begin{array}{cc}
    a_{12}\:=\:\frac{\langle1,3\rangle}{\langle2,3\rangle},& 
    a_{23}\:=\:\frac{\langle1,3\rangle}{\langle1,2\rangle},\\
    {} & {} \\
    a_{34}\:=\:\frac{\langle3,1\rangle}{\langle4,1\rangle},& 
    a_{41}\:=\:\frac{\langle3,1\rangle}{\langle3,4\rangle},
   \end{array}
 \end{split}
\end{equation}
with the poles in $z_{i,i+1}\,=\,0$ corresponding to $(l^2-m^2)\,=\,0$, $((l-p^{\mbox{\tiny $(1)$}})^2-m^2)\,=\,0$,
$((l+p^{\mbox{\tiny $(4)$}})^2-m^2)\,=\,0$, $((l-P_{\mbox{\tiny $12$}})^2-m^2)\,=\,0$.

Such an on-shell four form shows a $d\log$ structure, which reflects the fact that the maximal cut fixes 
completely this amplitude \cite{Brandhuber:2005jw}.

A comment is now in order. The on-shell diagrammatic representation of the integrands corresponding to the
rational terms has been possible because we {\it assumed} that a suitable regularisation scheme for on-shell
diagram would also induce a mass, as it occurs for dimensional regularisation (where one then also integrates over 
such a mass to obtain a number -- up to order $\mathcal{O}(\epsilon)$).

\subsubsection{The UHV four-gluon integrand}\label{subsubsec:MPPPint}

Let us now turn to the UHV amplitudes, which at tree level also vanish. As in the previous case,
the relevant contribution is equivalent to a massive four-dimensional scalar running in the loop. The residue
related to the pole induced by a BCFW bridge in the $(4,1)$-channel is given by the six-particle amplitude
$\mathcal{M}_6^{\mbox{\tiny tree}}(1^+,2^+,3^-,4^+,{\color{red} A^0},{\color{red} B^0})$, with the momenta
of the scalars ${\color{red} A}$ and  ${\color{red} B}$ taken to be forward:
\begin{equation}\eqlabel{eq:M4ppmp}
 \begin{split}
  &\raisebox{-1.7cm}{\scalebox{.45}{\includegraphics{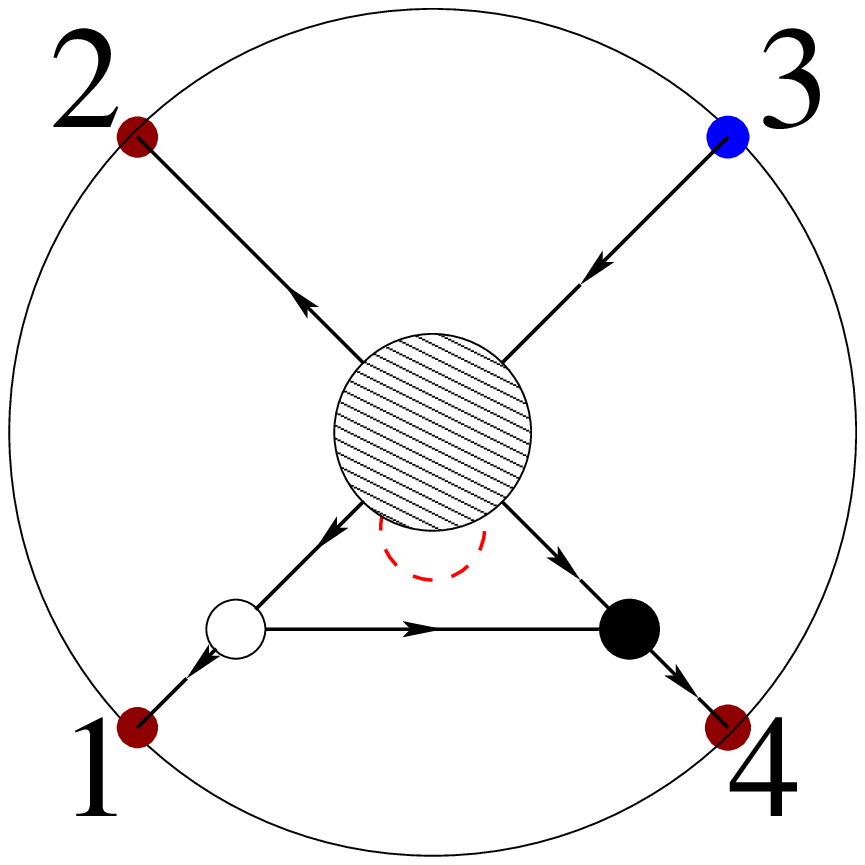}}}\:=\:
   \raisebox{-1.9cm}{\scalebox{.25}{\includegraphics{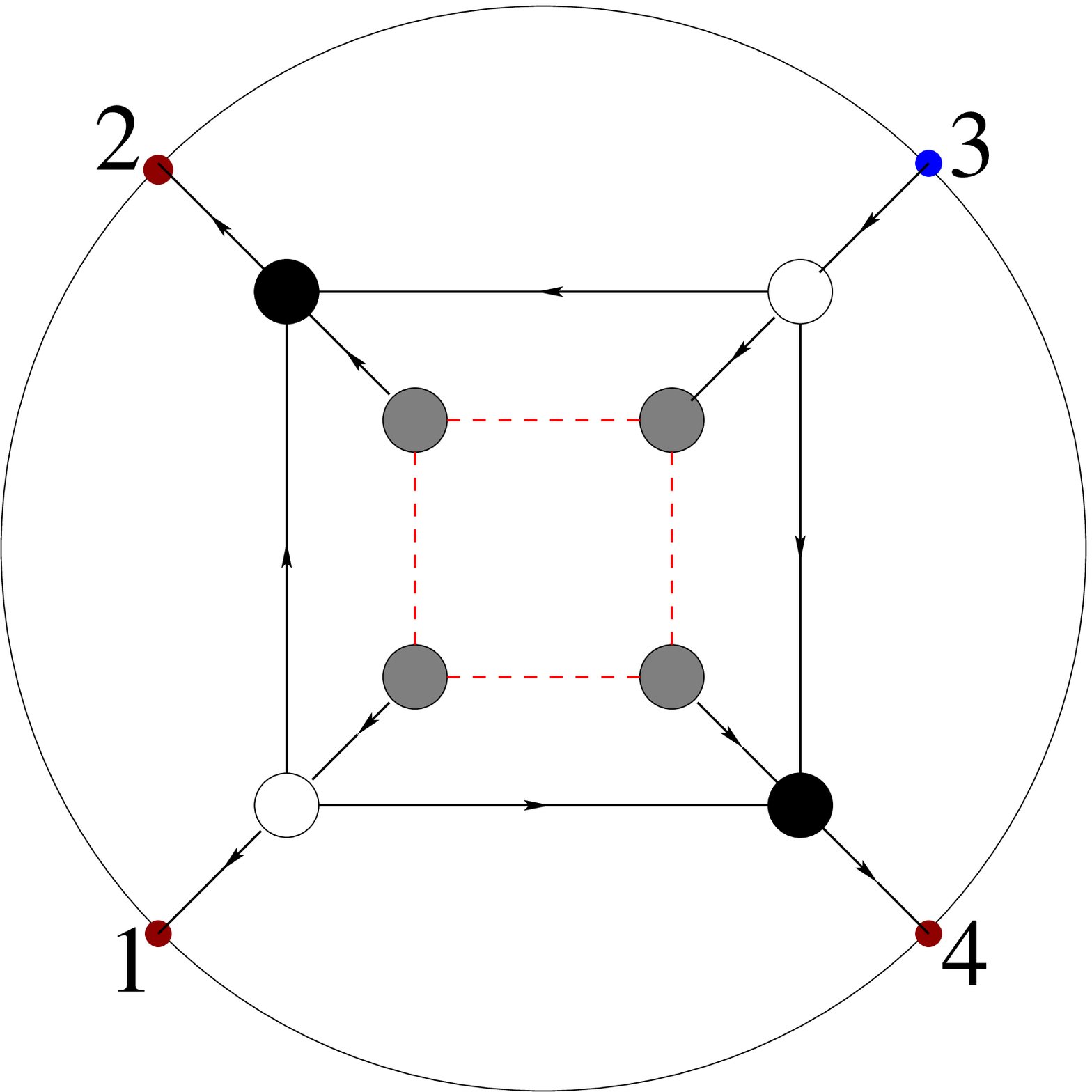}}}+
   \raisebox{-1.9cm}{\scalebox{.17}{\includegraphics{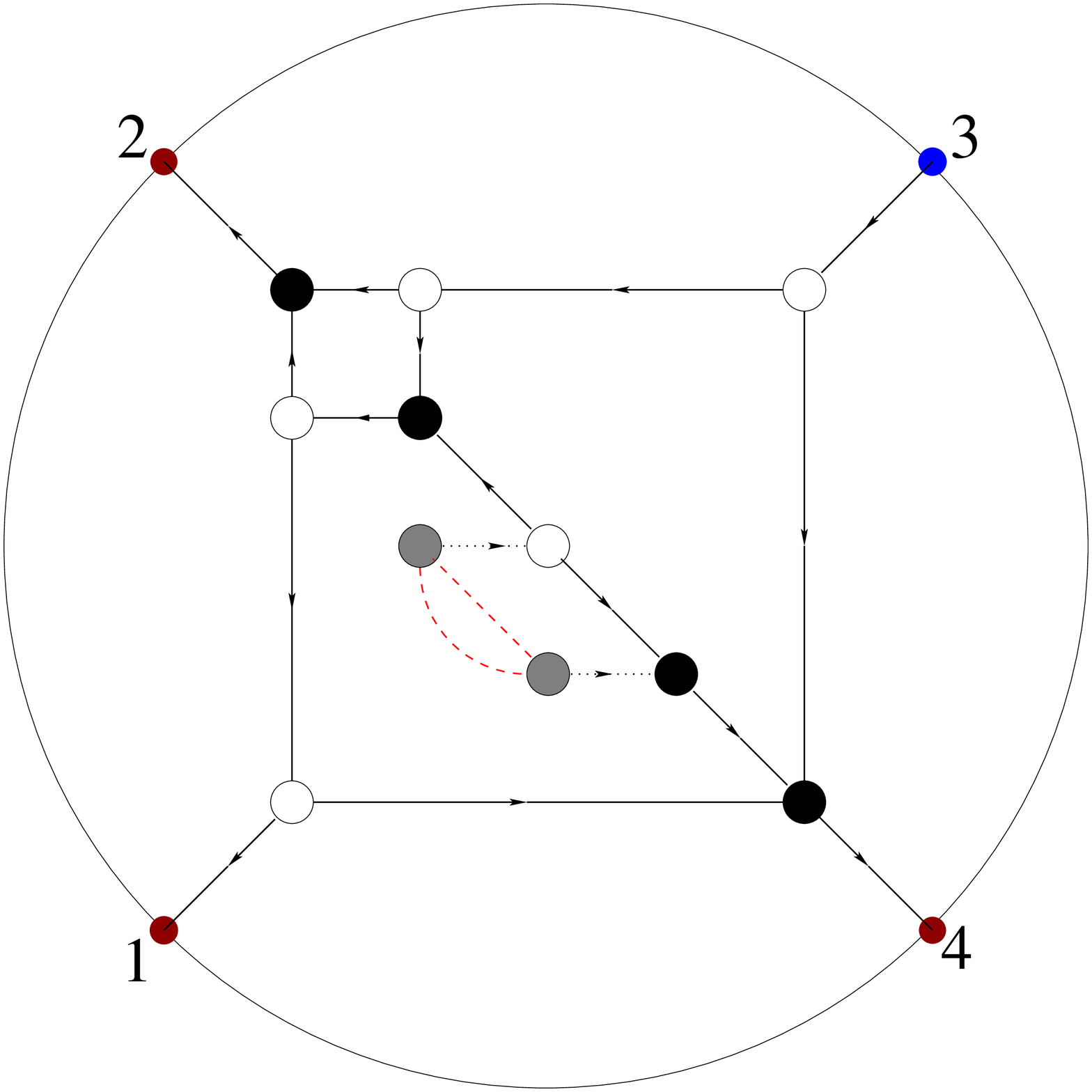}}}+\\
  &\raisebox{-1.9cm}{\scalebox{.17}{\includegraphics{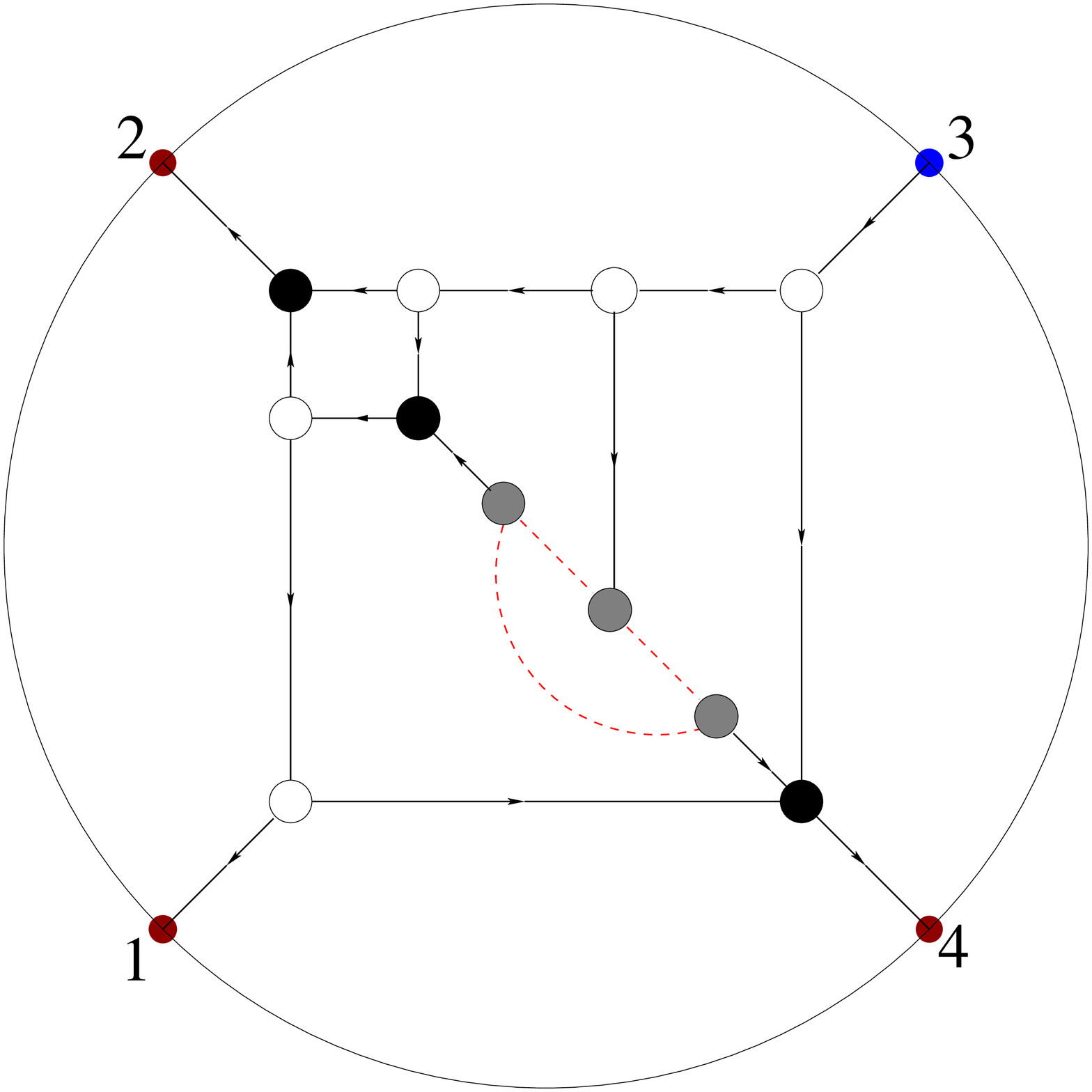}}}+
   \raisebox{-1.9cm}{\scalebox{.17}{\includegraphics{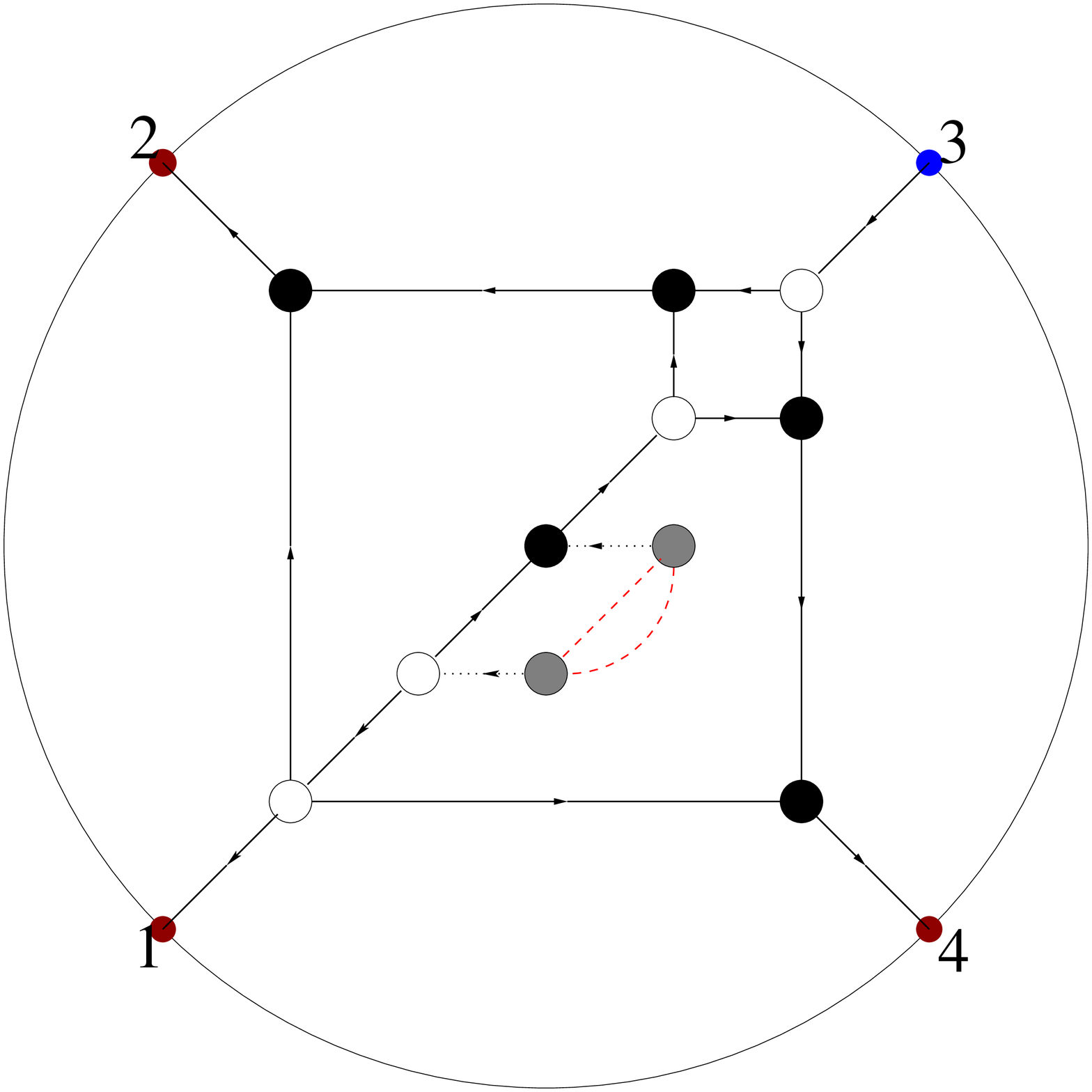}}}+
   \raisebox{-1.9cm}{\scalebox{.17}{\includegraphics{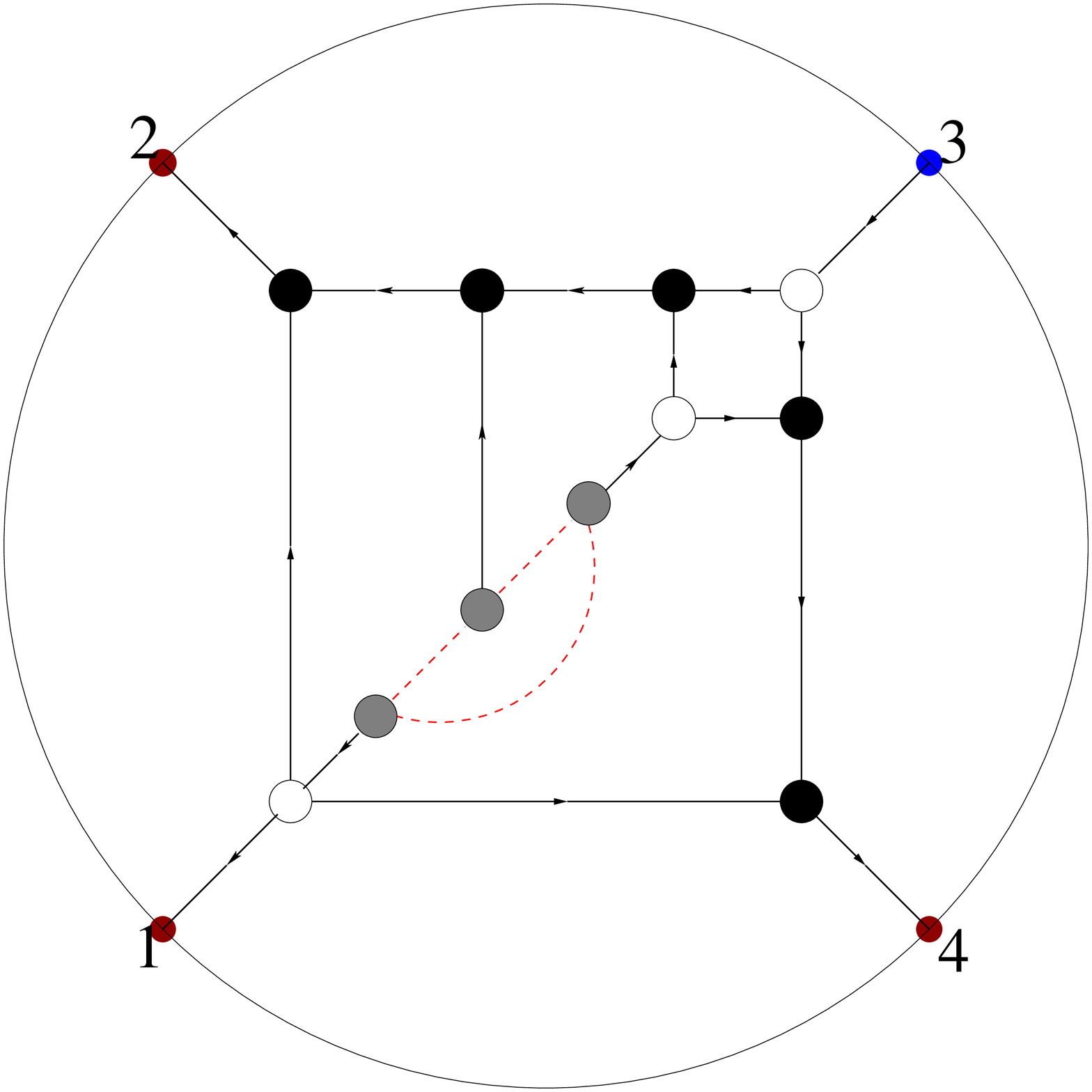}}}
 \end{split}
\end{equation}
The first diagram on the right-hand-side has the usual structure of four BCFW bridges attached to an on-shell
$0$-form which has the same helicity structure of the (integrand of the) amplitude we want to compute. This diagram
represents a finite quantity:
\begin{equation}\eqlabel{eq:M4ppmpFin}
 \begin{split}
   \raisebox{-1.9cm}{\scalebox{.25}{\includegraphics{1loopM4ppmp.eps}}}\:=\:
  &\frac{m^2[2,4]^2}{\langle1,2\rangle[2,3][3,4]\langle4,1\rangle}\frac{1}{u}\times\\
  &\hspace{-3cm}\times
   \bigwedge_{i=1}^4\frac{dz_{i,i+1}}{z_{i,i+1}}\,
   \frac{\mathcal{J}^{-1}}{(1+a_{34}z_{34})^2(1+a_{23}z_{23})^2}
   \left[
    m^2-\frac{st}{2u}\prod_{i=1}^4(1+a_{i,i+1}z_{i,i+1})\mathcal{J}^{-1}
   \right],
 \end{split}
\end{equation}
where the coefficients $a_{i,i+1}$, given in \eqref{eq:M41lpppp2} , and the Jacobian $\mathcal{J}$ are functions of 
the Lorentz invariants:
\begin{equation}\eqlabel{eq:M4ppmpFin2}
   \mathcal{J}\:=\:
   -\frac{s}{u}(1+a_{23}z_{23})(1+a_{41}z_{41})-\frac{t}{u}(1+a_{12}z_{12})(1+a_{34}z_{34})
\end{equation}
The integration along the contours 
$\gamma_{i,i+1}\,=\,\left\{z_{j,j+1}\,=\,0\:\forall\,j\,\neq\,i,\;\&\;z_{i,i+1}\,|\,\mathcal{J}\,=\,0\right\}$
provide the contributions from the triple cuts, in agreement with \cite{Brandhuber:2005jw}.

The second and fourth term in \eqref{eq:M4ppmp} instead shows the same structure of the diagrams
in \eqref{eq:1lInt} and they are divergent as well, making the forward limit we are considering ill-defined
unless a suitable regularisation procedure is introduced. Finally, the UHV integrand shows a new structure
which is encoded in the third and fifth diagram in the right-hand-side of \eqref{eq:M4ppmp}. Such terms need also
a suitable regularisation given that they are ill-defined: the corresponding terms in the forward six-particle 
amplitude in the left-hand-side of \eqref{eq:M4ppmp} purely contribute to the forward singularity.

In order to regularise these terms one can generalise the quasi-forward limit discussed previously to the
massive scalars. Specifically, the idea is to deform the momenta of the massive scalars on the tree-level 
six-particle amplitude {\it before} the forward limit is taken. The deformation is chosen in a BCFW fashion, so that both momentum conservation and on-shell conditions are preserved:

{\footnotesize
\begin{equation}\eqlabel{eq:MassQF}
 \begin{split}
  &\tilde{\lambda}^{\mbox{\tiny $(4)$}}(\epsilon)\,=\,
    \tilde{\lambda}^{\mbox{\tiny $(4)$}}+
    \epsilon
    \left[
     \frac{\langle B,r_B\rangle}{\langle 4,r_B\rangle}\tilde{\lambda}^{\mbox{\tiny $(B)$}}+
     \frac{m^2\,\tilde{\lambda}^{\mbox{\tiny $(r_B)$}}}{\langle4|B|r_B]}
    \right],
   \qquad
   \lambda^{\mbox{\tiny $(1)$}}(\epsilon)\,=\,
    \lambda^{\mbox{\tiny $(1)$}}+
    \epsilon
    \left[
     \frac{[A,r_A]}{[1,r_A]}\lambda^{\mbox{\tiny $(A)$}}+
     \frac{m^2\,\lambda^{\mbox{\tiny $(r_A)$}}}{\langle r_A|A|1]}
    \right],\\
  &\lambda^{\mbox{\tiny $(B)$}}(\epsilon)\,=\,
    \lambda^{\mbox{\tiny $(B)$}}-
    \epsilon\frac{\langle B,r_B\rangle}{\langle4,r_B\rangle}\lambda^{\mbox{\tiny $(4)$}},
   \hspace{2.7cm}
   \tilde{\lambda}^{\mbox{\tiny $(A)$}}(\epsilon)\,=\,
    \tilde{\lambda}^{\mbox{\tiny $(A)$}}-
    \epsilon\frac{[A,r_A]}{[1,r_A]}\tilde{\lambda}^{\mbox{\tiny $(1)$}},\\
  &\lambda^{\mbox{\tiny $(r_B)$}}(\epsilon)\,=\,
    \lambda^{\mbox{\tiny $(r_B)$}}+
    \epsilon\frac{\langle r_B,B\rangle}{\langle4,B\rangle}\lambda^{\mbox{\tiny $(4)$}},
   \hspace{2.5cm}
   \tilde{\lambda}^{\mbox{\tiny $(r_A)$}}(\epsilon)\,=\,
    \tilde{\lambda}^{\mbox{\tiny $(r_A)$}}+
    \epsilon\frac{[r_A,A]}{[1,A]}\tilde{\lambda}^{\mbox{\tiny $(1)$}},
 \end{split}
\end{equation}
} 
\noindent
where $r_A$ and $r_B$ label the reference spinors for the massive momenta $p^{\mbox{\tiny $(A)$}}$
and $p^{\mbox{\tiny $(B)$}}$ respectively. Then, one can take the quasi-forward limit 
$\lambda^{\mbox{\tiny $(B)$}}\,\longrightarrow\,-\lambda^{\mbox{\tiny $(A)$}}$,
$\tilde{\lambda}^{\mbox{\tiny $(B)$}}\,\longrightarrow\,\tilde{\lambda}^{\mbox{\tiny $(A)$}}$,
$\lambda^{\mbox{\tiny $(r_B)$}}\,\longrightarrow\,-\lambda^{\mbox{\tiny $(r_A)$}}$,
$\tilde{\lambda}^{\mbox{\tiny $(r_B)$}}\,\longrightarrow\,\tilde{\lambda}^{\mbox{\tiny $(r_A)$}}$.
As in the massless case, the propagators which originally were divergent are now mapped in poles in the
parameter $\epsilon$. It can be easily checked that the singularities in the non-local poles in an individual
on-shell diagram cancel upon summation in \eqref{eq:M4ppmp}. In order to complete the integrand, one would need
to apply a multi-step BCFW algorithm, as for the massless case.

\subsection{Higher point one-loop integrands}\label{subsec:HP}

So far we analysed in detail the one-loop structure of four-particle amplitudes. We learnt that the forward
amplitude related to a given BCFW bridging can {\it in principle} contain ill-defined terms. Upon the
quasi-forward regularisation outlined above, poles in the regularisation parameters appear both related to
physical singularities and to non-local poles. The latter cancel upon summation among diagrams, some of which
are obtained by a multi-step BCFW algorithm (or equivalently by symmetry). For $\mathcal{N}\,\neq\,0$, supersymmetry
guarantees further cancellations so that the originally ill-defined terms become of order $\mathcal{O}(\epsilon)$.
For $\mathcal{N}\,=\,0$, these terms are of order $\mathcal{O}(\epsilon^{-1})$. In both cases, the finite diagram
contains all the cut-constructible information as well as, for $\mathcal{N}\,=\,0$, the integrand related to the 
rational terms upon a suitable mass-deformation.

These results extend also to higher number of external states. In order to provide a complete proof of this 
statement, we need to prove that the same cancellations occurs for a larger number of particles as well as
the finite contributions returned from the start contains all and only the correct singularities. 

As a first step, let us consider the following $n$-particle forward term:
\begin{equation}\eqlabel{eq:MnFw}
 \raisebox{-1.4cm}{\scalebox{.20}{\includegraphics{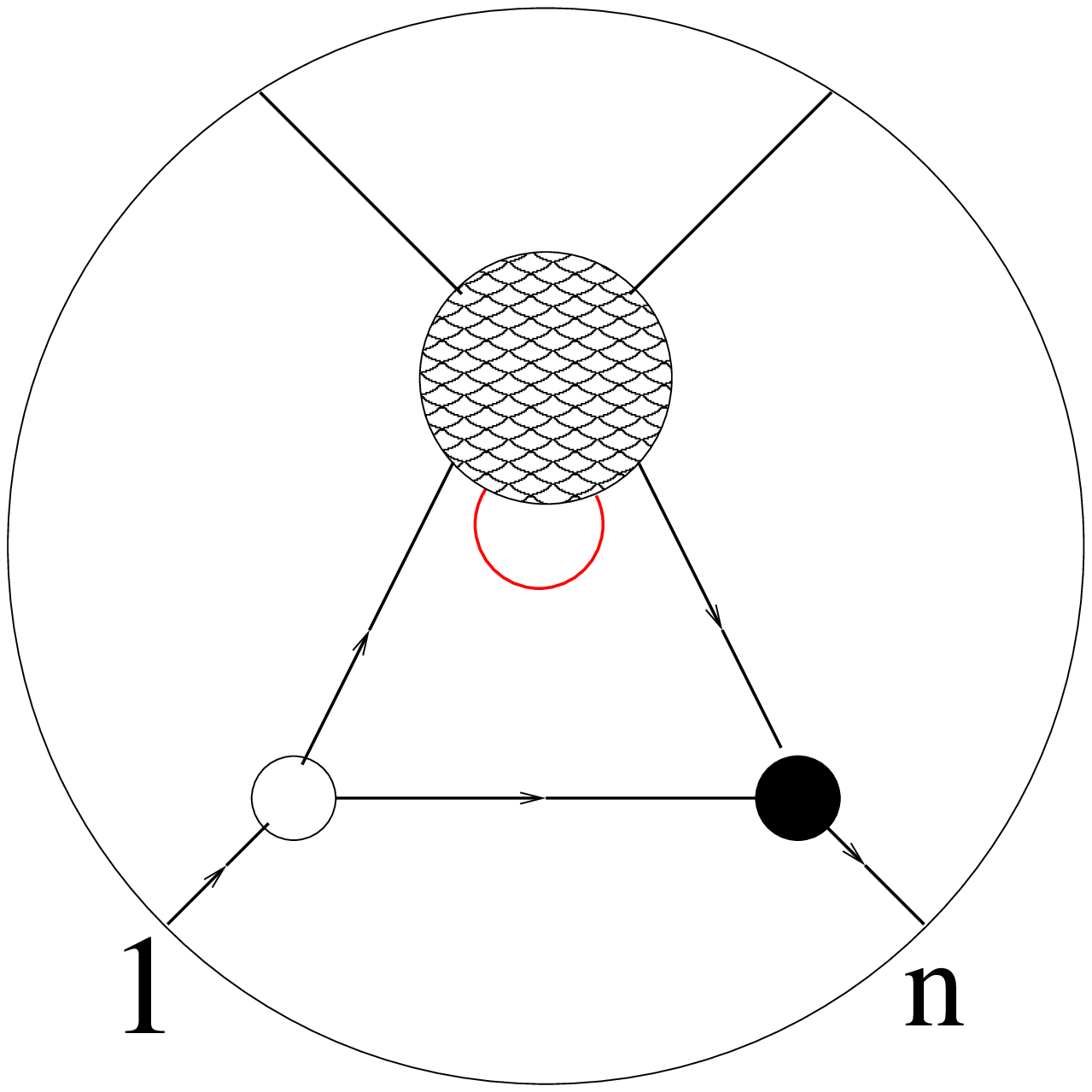}}}
  \:\Longrightarrow\:
 \raisebox{-1.4cm}{\scalebox{.20}{\includegraphics{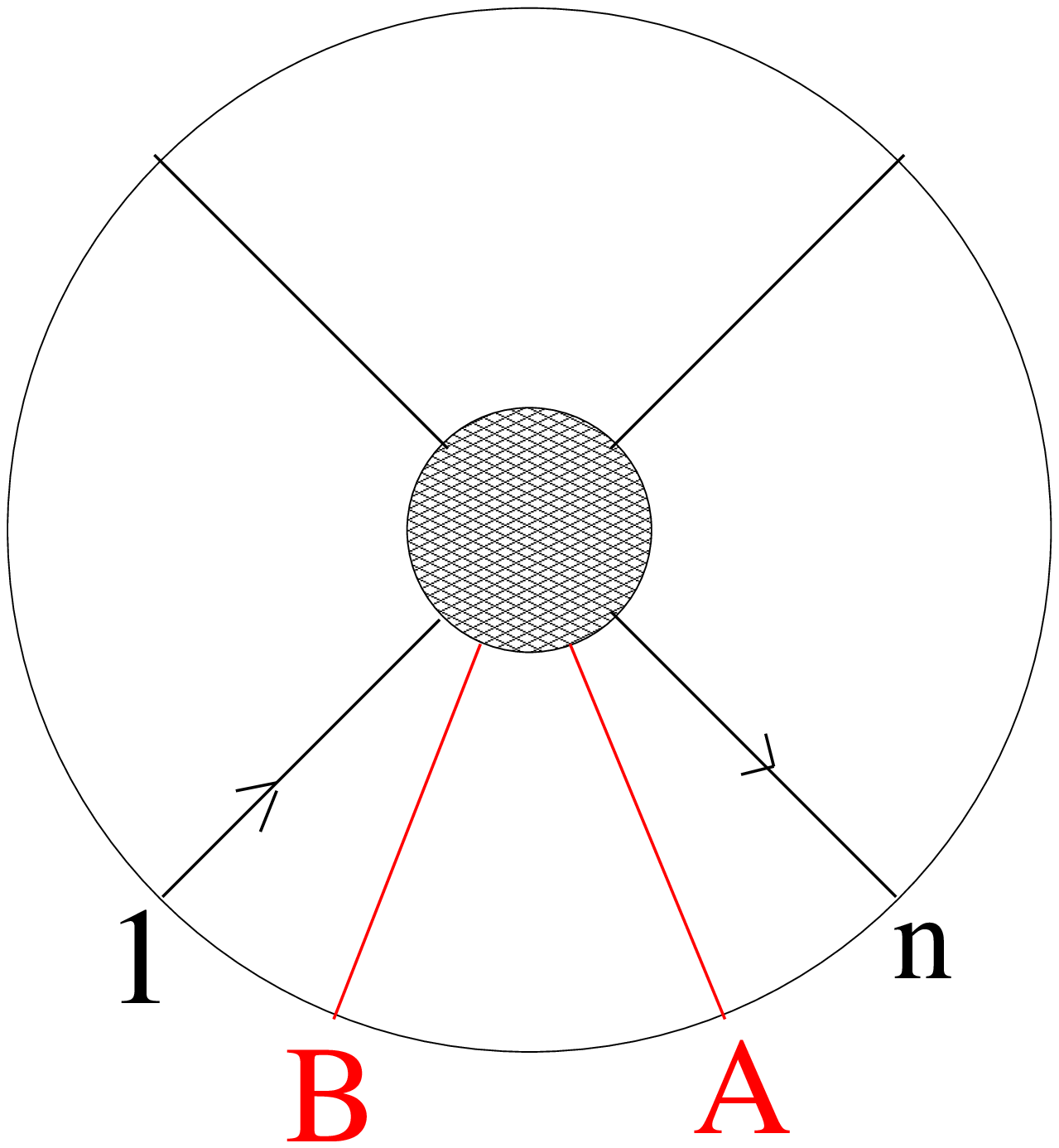}}}=
 \raisebox{-1.3cm}{\scalebox{.20}{\includegraphics{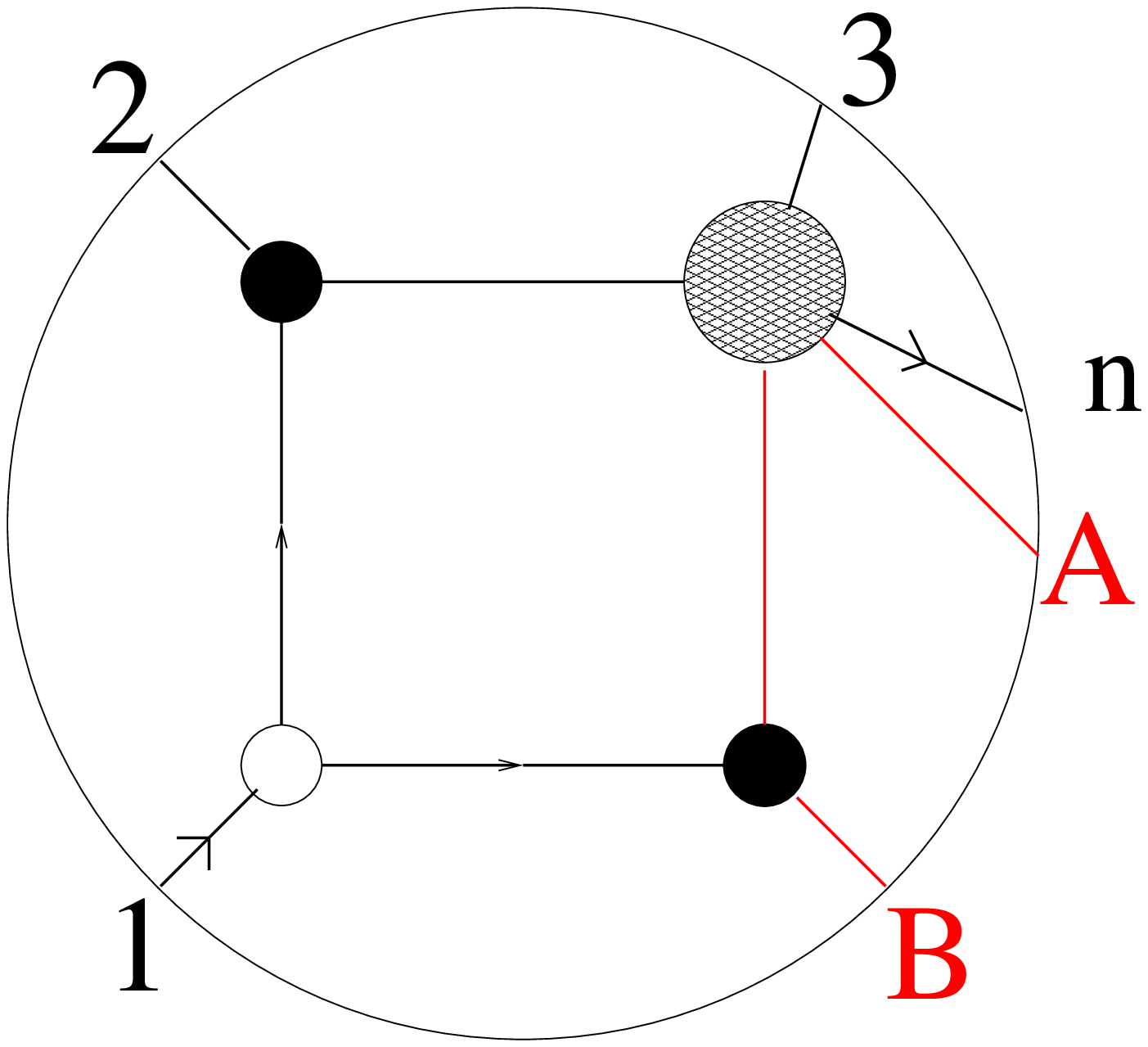}}}+
 \sum_{k}
 \raisebox{-1.3cm}{\scalebox{.20}{\includegraphics{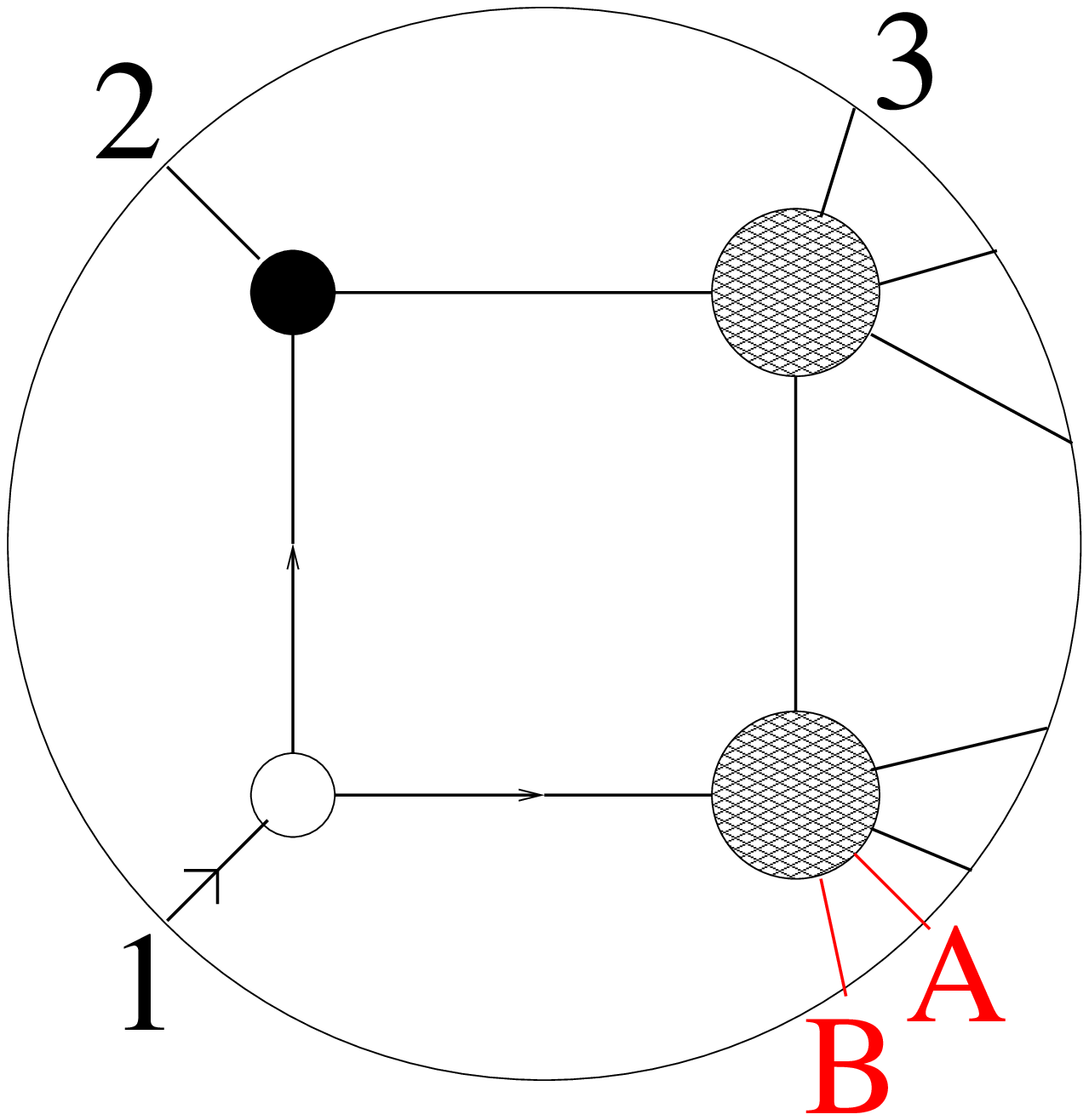}}}
 \begin{array}{l}
  \hspace{-.2cm}\mathcal{I}_k\\
  {}\\
  \hspace{-.2cm}\mathcal{J}_k
 \end{array}
\end{equation}
with $\mathcal{I}_k\,\cup\,\mathcal{J}_k\,=\,\{4,\,\ldots,\,n\}$. As the momenta of particle $A$ and $B$ are taken
to be forward, the first term in the right-hand-side of \eqref{eq:MnFw} turns out to be completely regular, while
all the other terms are not well defined. More precisely, the terms in the sum in  \eqref{eq:MnFw} can be written
as two sets of terms one which is ill-defined in the forward limit, while the other is well-defined, provided that
$\mbox{dim}\{\mathcal{J}_k\}\,\ge\,2$ -- for $\mbox{dim}\{\mathcal{J}_k\}\,<\,2$ all the terms in which it can be
represented are ill-defined in the forward limit. Upon the quasi-forward regularisation, each of such terms' leading 
behaviour is of order $\mathcal{O}(\epsilon^{\mathcal{N}-3})$ for $\mathcal{N}$ even and
$\mathcal{O}(\epsilon^{\mathcal{N}-2})$ for $\mathcal{N}$ odd, with the divergencies due also to non-local poles.
However, such non-local poles are not singularities of the full tree-level $(n+2)$-particle amplitude. In fact,
in a neighbourhood of such a pole, the following factorisations occur:
\begin{equation}\eqlabel{eq:MnFwCanc}
 \begin{split}
  &\raisebox{-1.3cm}{\scalebox{.20}{\includegraphics{1LMnFw3.eps}}}
   \begin{array}{l}
    \hspace{-.2cm}\mathcal{I}_{k'}\\
    {}\\
    \hspace{-.2cm}\mathcal{J}_{\bar{k}}
   \end{array}
   \hspace{-.2cm}\Longrightarrow\:
   \raisebox{-1.3cm}{\scalebox{.20}{\includegraphics{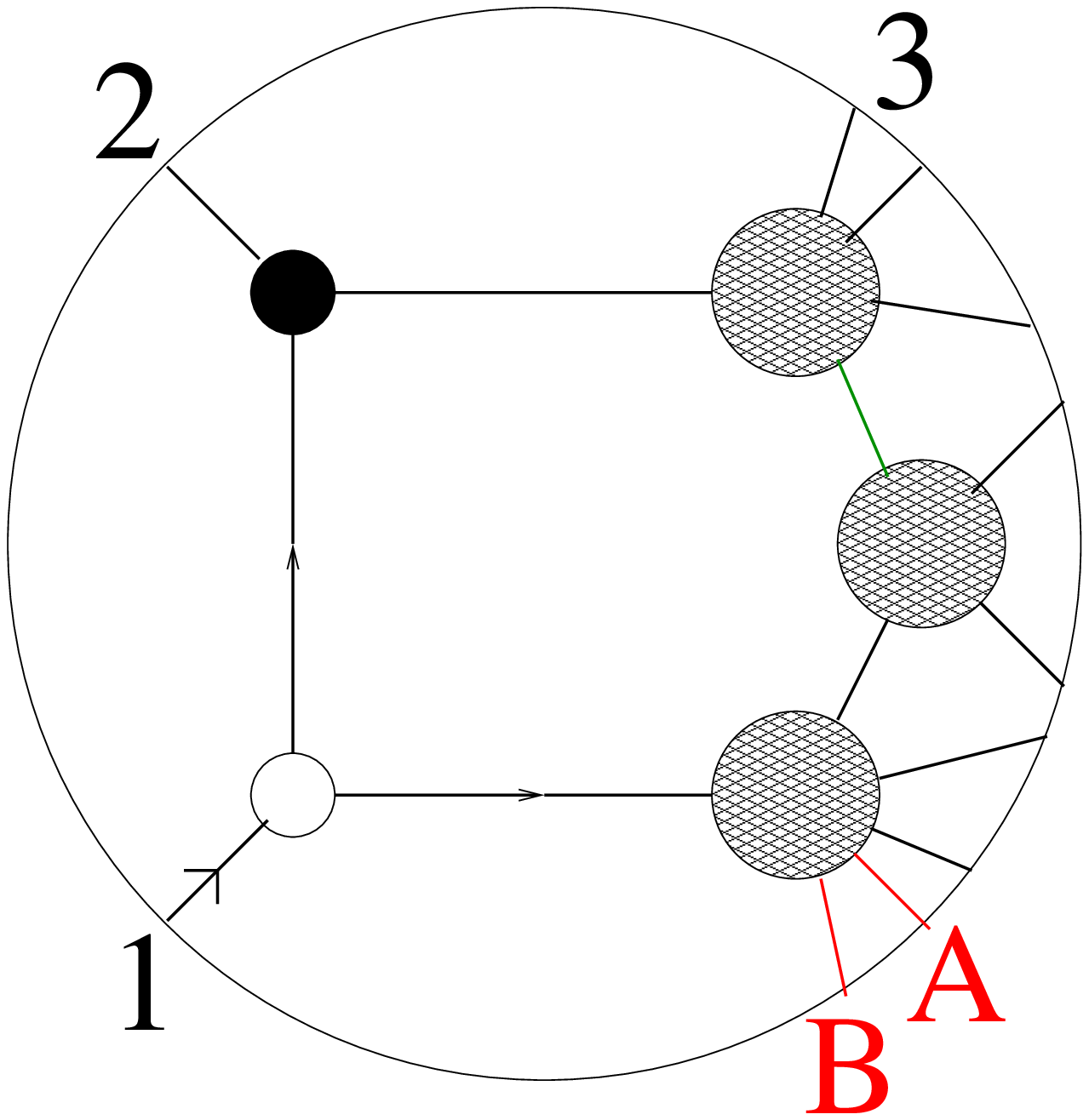}}}
   \begin{array}{l}
    \hspace{-.2cm}\vspace{.2cm}\mathcal{I}_{\bar{k}}\\
    \hspace{-.1cm}\mathcal{K}\\
    \hspace{-.2cm}\mathcal{J}_{\bar{k}}
   \end{array}
   \hspace{1.5cm}
   \begin{array}{l}
    \mathcal{I}_{\bar{k}}\,\cup\,\mathcal{K}\:=\: \mathcal{I}_{k'}\\
    {}\\
    \mathcal{I}_{k'}\,\cup\,\mathcal{J}_{\bar{k}}\,=\,\{4,\ldots,n\}
   \end{array}
   \\
  &\raisebox{-1.3cm}{\scalebox{.20}{\includegraphics{1LMnFw3.eps}}}
   \begin{array}{l}
    \hspace{-.2cm}\mathcal{I}_{\bar{k}}\\
    {}\\
    \hspace{-.2cm}\mathcal{J}_{k''}
   \end{array}
   \hspace{-.2cm}\Longrightarrow\:
   \raisebox{-1.3cm}{\scalebox{.20}{\includegraphics{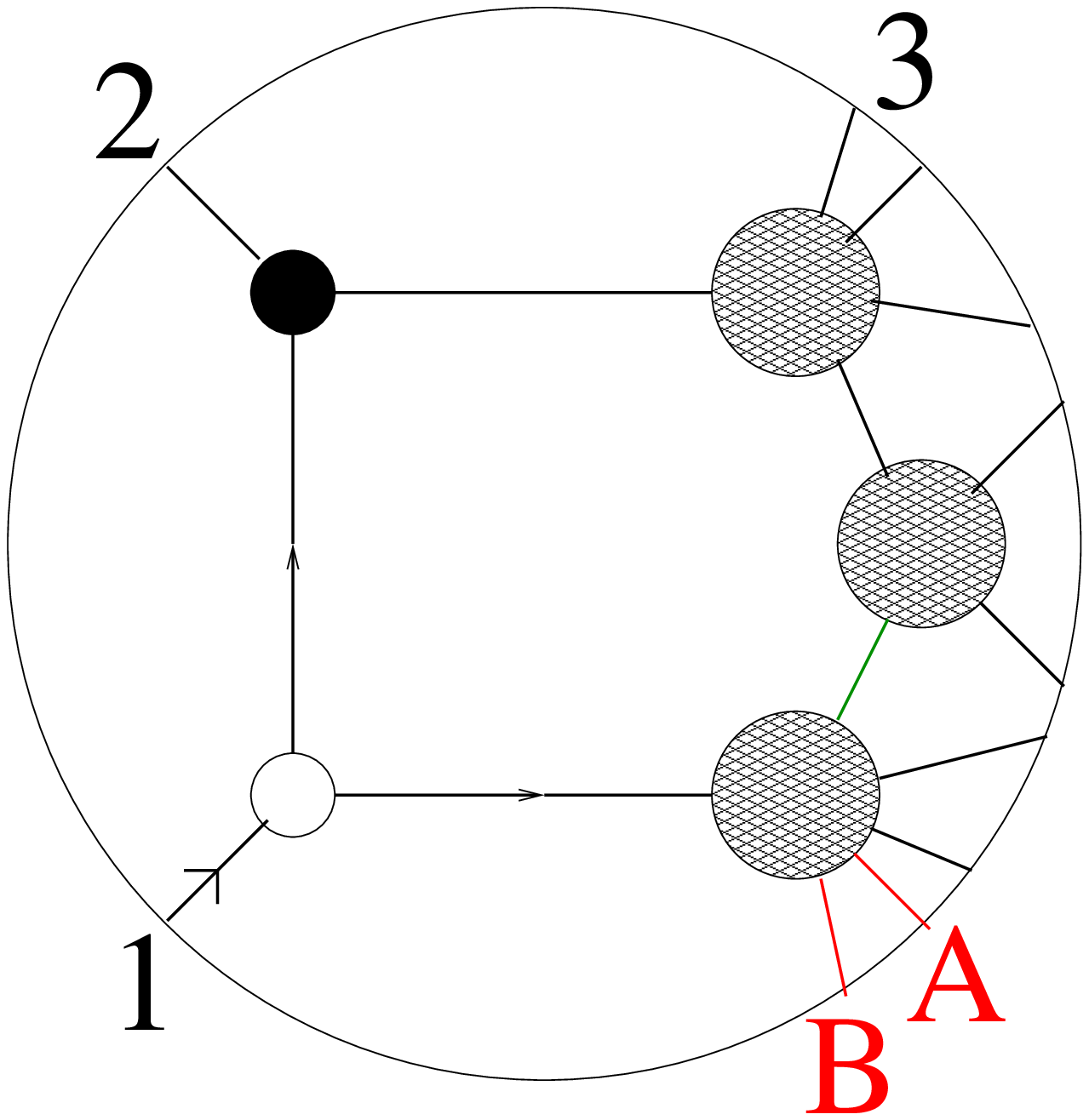}}}
   \begin{array}{l}
    \hspace{-.2cm}\vspace{.2cm}\mathcal{I}_{\bar{k}}\\
    \hspace{-.1cm}\mathcal{K}\\
    \hspace{-.2cm}\mathcal{J}_{\bar{k}}
   \end{array}
   \hspace{1.5cm}
   \begin{array}{l}
    \mathcal{J}_{\bar{k}}\,\cup\,\mathcal{K}\:=\: \mathcal{J}_{k''}\\
    {}\\
    \mathcal{I}_{\bar{k}}\,\cup\,\mathcal{J}_{k''}\,=\,\{4,\ldots,n\}
   \end{array}
 \end{split}
\end{equation}
where $\{\mathcal{I}_{k'},\,\mathcal{J}_{\bar{k}}\}$ and $\{\mathcal{I}_{\bar{k}},\,\mathcal{J}_{k''}\}$ are two
different partitions of $\{4,\ldots,n\}$, while the green interior line points out the factorisation which leads to 
the diagrams on the right-hand-side. The two factorisation diagrams are equal up to an overall sign, so that the 
related pole cancels out upon summation. Hence, when the {\it quasi-forward} limit is taken, the pole in the
regularisation parameter due to the individual diagrams on the left-hand side in \eqref{eq:MnFwCanc} cancels.
Therefore, the contribution coming from the originally ill-defined diagrams in the sum are 
\begin{equation}\eqlabel{eq:MnFwFin}
 \raisebox{-1.3cm}{\scalebox{.20}{\includegraphics{1LindMnRHSc.eps}}}\;\Longrightarrow\;
 \raisebox{-1.3cm}{\scalebox{.20}{\includegraphics{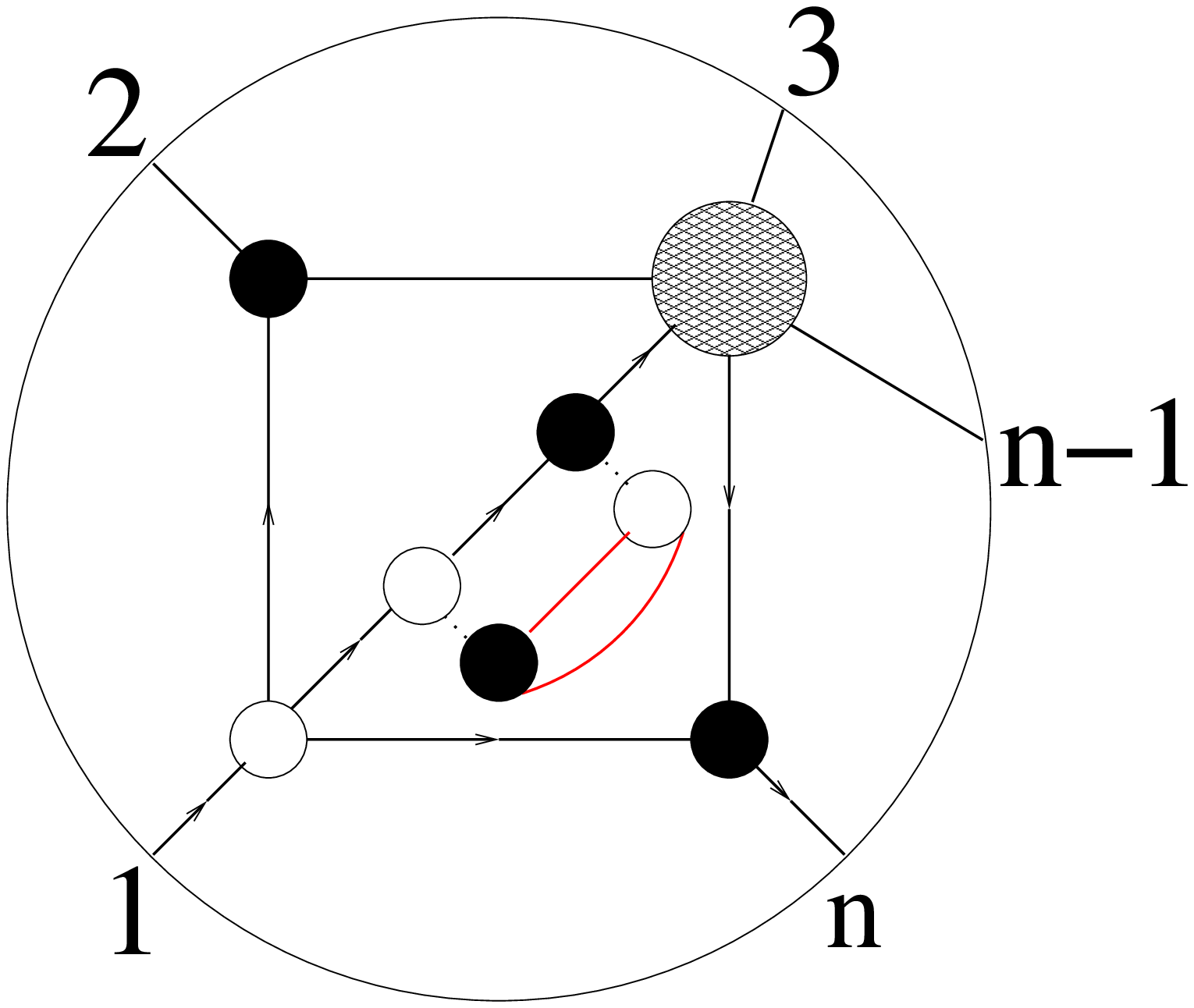}}}\:+\:
 \raisebox{-1.3cm}{\scalebox{.20}{\includegraphics{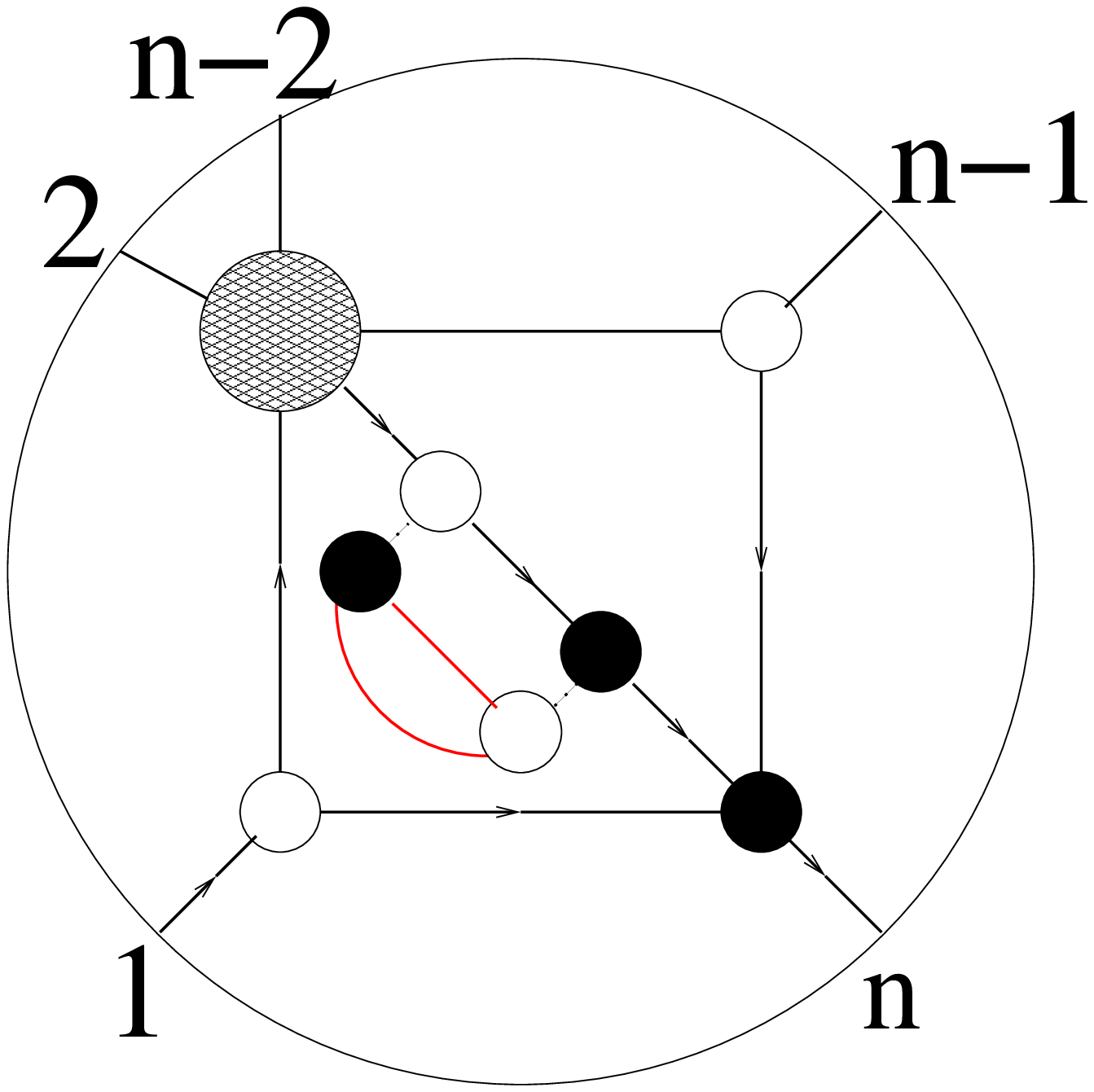}}}
\end{equation}
As in the four-particle case, such diagrams show a forward singularity of the order 
$\mathcal{O}(\epsilon^{\mathcal{N}-2})$ for $\mathcal{N}$ even and $\mathcal{O}(\epsilon^{\mathcal{N}-1})$ for 
$\mathcal{N}$ odd: The $\epsilon^{\mathcal{N}}$ factor comes from the sum over the components of a multiplet 
running in the ``forward lines''/on-shell bubbles, an extra cancellation occurs for $\mathcal{N}$ odd when
summing the two multiplets and, finally, the factor $\epsilon^{-2}$ comes from a collinear singularity
(either $\langle A,B\rangle$ or $[A,B]$) and a ``bubble divergence'' 
$P^2_{ABi}\,\sim\,(p^{\mbox{\tiny $(i)$}})^2$. The other terms of the type in \eqref{eq:MnFwFin} can again
be generated via a multi-step BCFW algorithm and, upon summation, the behaviour of the originally ill-defined
diagrams is enhanced to $\mathcal{O}(\epsilon^{\mathcal{N}-1})$ for $\mathcal{N}$ even and 
$\mathcal{O}(\epsilon^{\mathcal{N}-1})$ for $\mathcal{N}$ odd. Thus, the cancellations we observed at four-particle 
level actually extend to an arbitrary number $n$ of external states.

We now need to prove that the finite contributions singled out by a given BCFW bridge contain all and only
the physical singularities. The proof can be formulated inductively on the same lines of the all-loop one for 
$\mathcal{N}\,=\,4$ Super Yang-Mills, with the crucial difference that the helicity flows now provide a guide for
reading off the singularities. The induction hypothesis is given by the 
following relation:
\begin{equation}\eqlabel{eq:MnIndHyp}
  \raisebox{-1.6cm}{\scalebox{.22}{\includegraphics{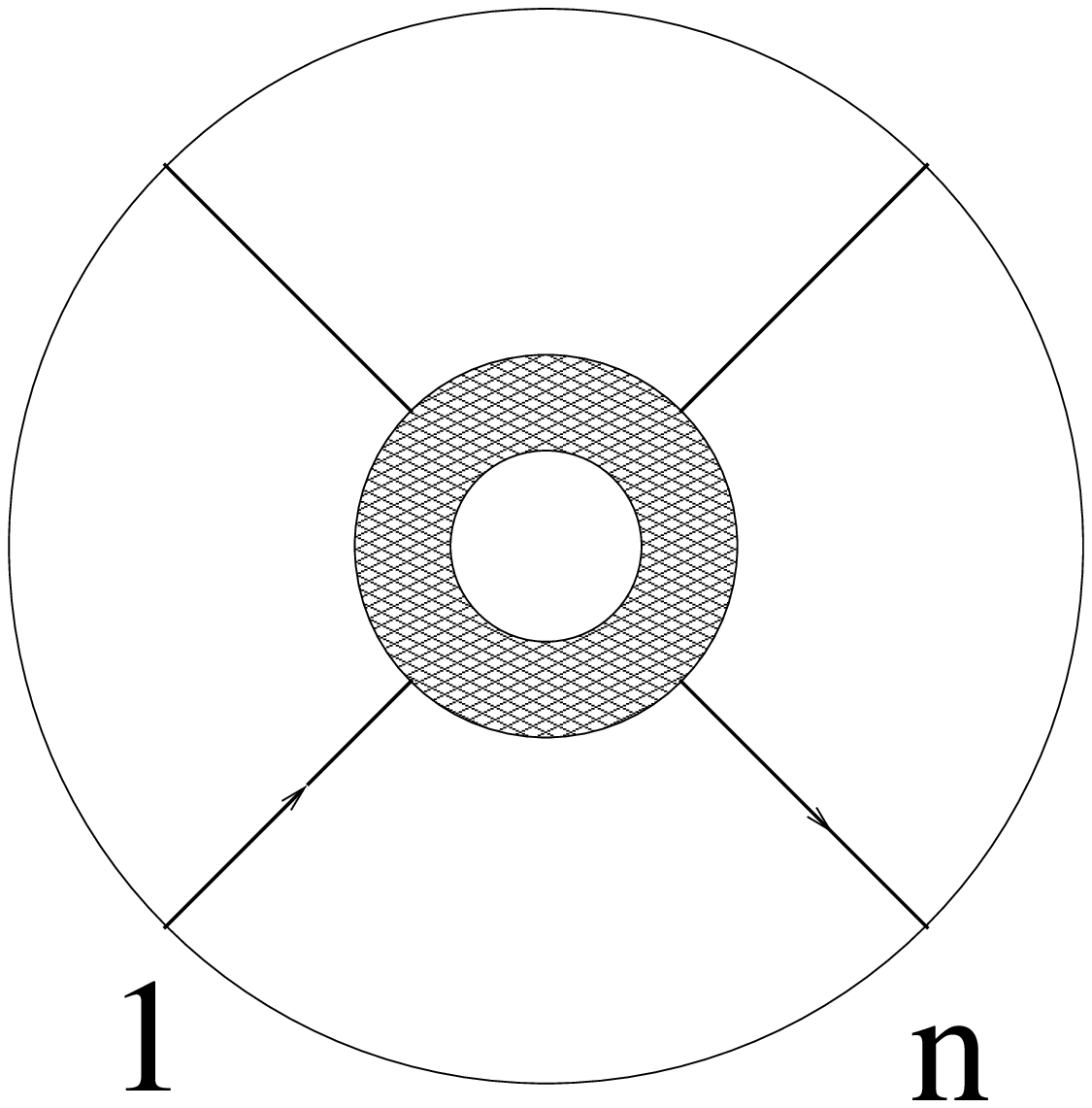}}}
  \begin{array}{l}
   {} \\
   \hspace{-.1cm}=\:\sum_k
  \end{array}
    \begin{array}{l}
     \vspace{1.5cm}\mathcal{I}_k
    \end{array}
   \hspace{-.6cm}\raisebox{-1.7cm}{\scalebox{.20}{\includegraphics{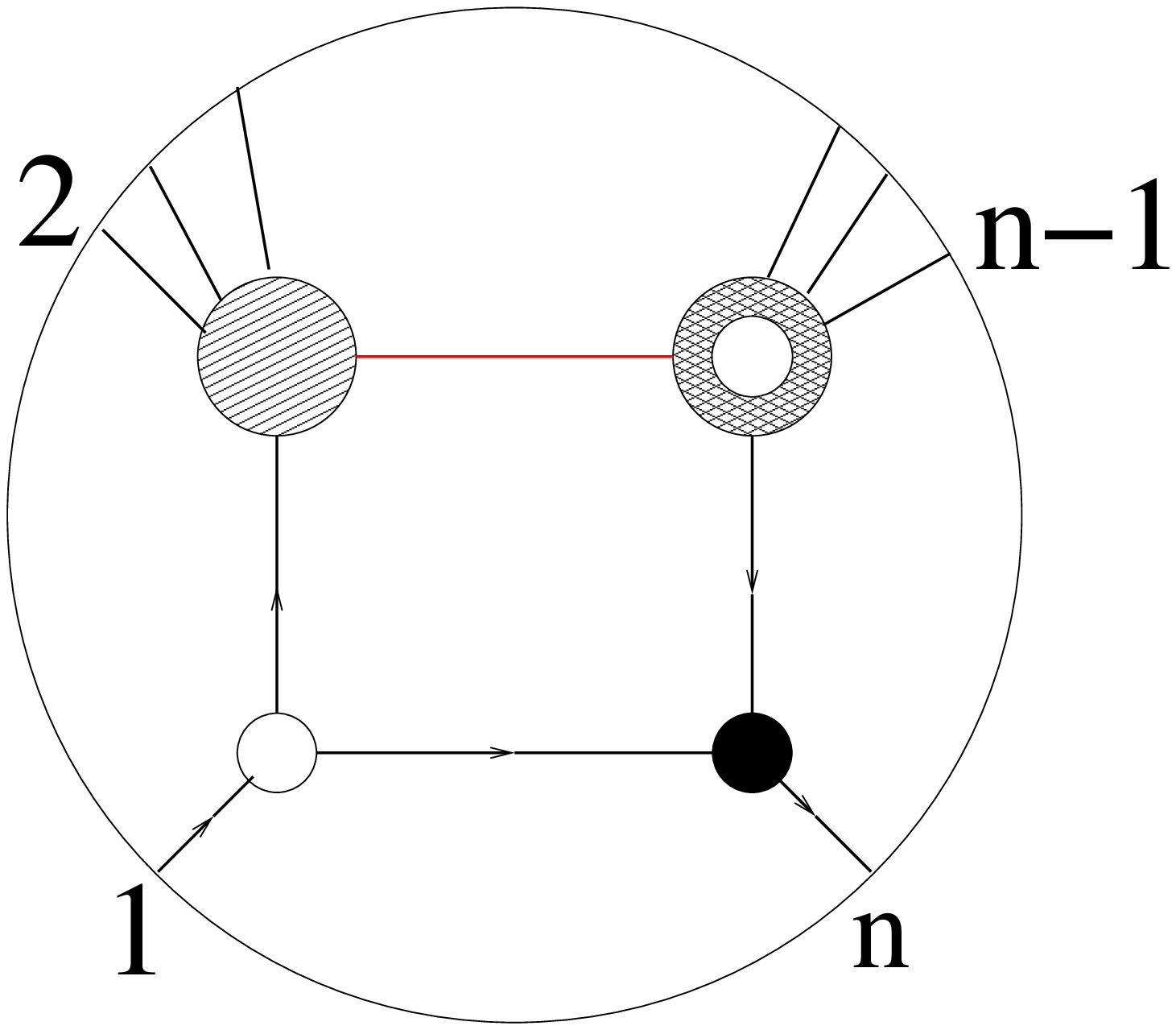}}}
    \begin{array}{l}
     \vspace{1.3cm}\hspace{-1cm}\mathcal{J}_k
    \end{array}
   \begin{array}{l}
   {} \\
   \hspace{-.3cm}+\:\sum_k
  \end{array}
   \begin{array}{l}
     \vspace{1.5cm}\mathcal{I}_k
    \end{array}
   \hspace{-.6cm}\raisebox{-1.7cm}{\scalebox{.20}{\includegraphics{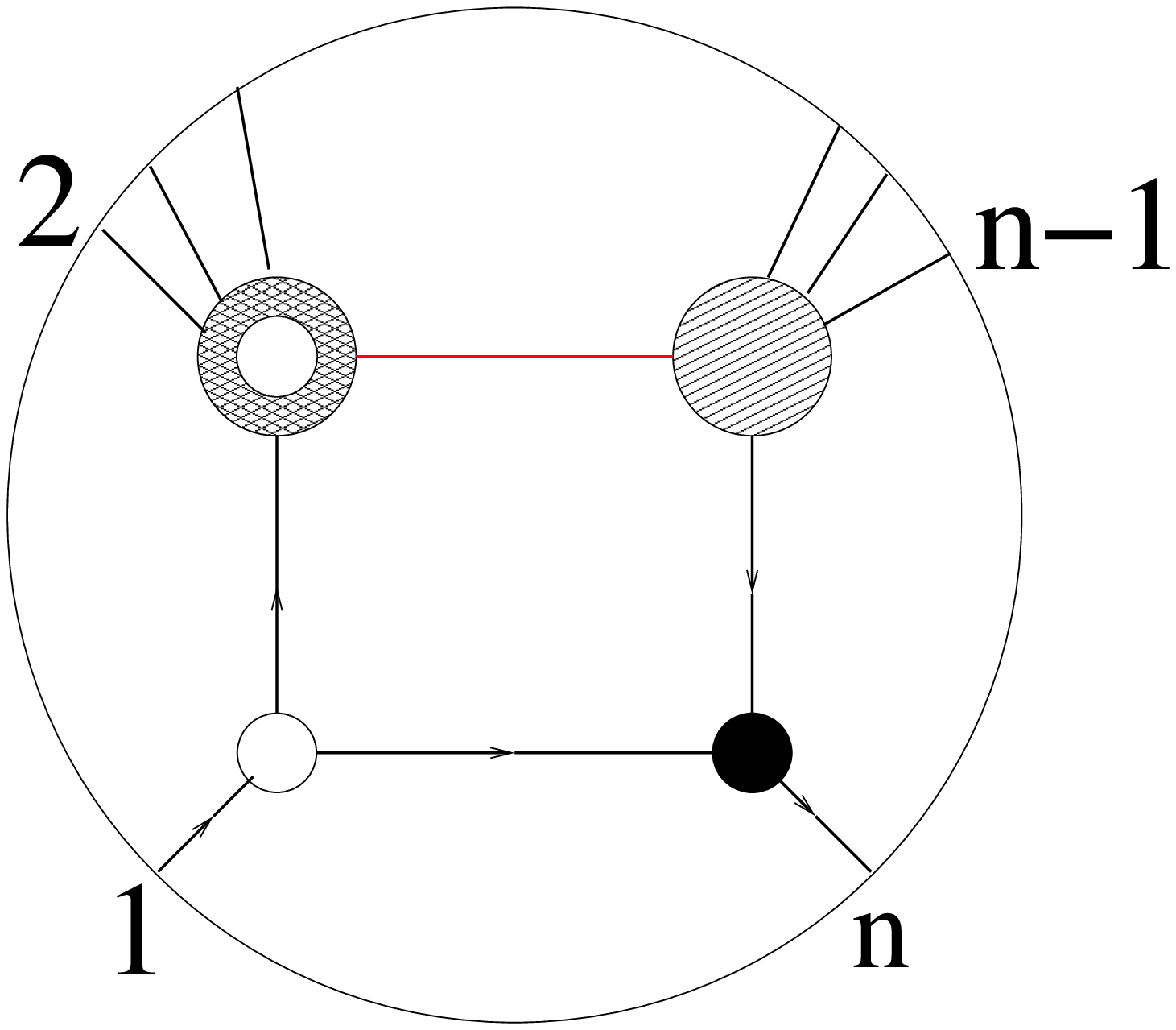}}}
   \begin{array}{l}
     \vspace{1.3cm}\hspace{-1cm}\mathcal{J}_k
    \end{array}
    \begin{array}{l}
     {} \\
     +
    \end{array}
   \raisebox{-1.7cm}{\scalebox{.20}{\includegraphics{1LindMnRHSc.eps}}}
\end{equation}
Having demonstrated in the previous sections that, upon the quasi-forward regularisation, cancellations occur in
the four-point one-loop integrand among divergent terms so that they turn out to behave as $\mathcal{O}(\epsilon)$ 
for $\mathcal{N}\,=\,1,\,2$ and $\mathcal{O}(\epsilon^{-1})$ for $\mathcal{N}\,=\,0$, the induction hypothesis
\eqref{eq:MnIndHyp} implies that all the one-loop sub-amplitudes in the two sets of factorisation channels
do have this same behaviour and in no inverse powers of $\epsilon$ ({\it i.e.} new divergencies) can be added in
the factorisation diagrams by the BCFW bridge. Furthermore, 
as we just stressed, a single BCFW bridge is not able to capture all the divergent terms
so that via either a multi-step BCFW algorithm or symmetry argument, one can obtain all the missing terms. 
In this sense, one {\it would not} be allowed to write \eqref{eq:MnIndHyp}, not even in the four-particle case.
However, 
if we consider the completion of the divergent structure as part of the regularisation scheme, \eqref{eq:MnIndHyp} 
can be indeed written for $\mathcal{N}\,=\,1,\,2$ theories once we prove that these terms are of order 
$\mathcal{O}(\epsilon)$ in the regularisation parameter. Thus,
the idea is to always generate the higher point integrand from the lower ones, regularising it at each stage of
the recursive procedure via the quasi-forward scheme. The $\mathcal{N}\,=\,0$ case is more subtle due to the
presence of the $\mathcal{O}(\varepsilon^{-1})$ terms as well as of the would-be rational contributions, and we 
will discuss it separately.


\subsubsection{One-loop integrand structure for $\mathcal{N}\,=\,1,\,2$ supersymmetric theories}
\label{subsubsec:1lRRN12}

Let us therefore focus on the $\mathcal{N}\,=\,1,\,2$ theories, taking \eqref{eq:MnIndHyp} as induction hypothesis,
where all the one-loop (sub)-diagrams are understood to be regularised under the quasi-forward regularisation scheme
with the $\mathcal{O}(\epsilon)$ behaviour for the on-shell bubble diagrams, and analysing the one-loop
amplitudes with $n'\,>\,n$ external states. 

At first, we discuss the factorisation channels. The channels with the bridged particle on different sub-amplitudes
are manifestly shown:
\begin{equation}\eqlabel{eq:MnFact1n}
 \begin{split}
   &\begin{array}{l}
     \vspace{2.0cm}\mathcal{I}_k
    \end{array}
    \hspace{-.6cm}\raisebox{-1.3cm}{\scalebox{.20}{\includegraphics{1LindMnRHSa.eps}}}
    \begin{array}{l}
     \vspace{1.8cm}\hspace{-1cm}\mathcal{J}_k
    \end{array}
    \:\Longrightarrow\:
    \begin{array}{l}
     \vspace{1.9cm}\mathcal{I}_k
    \end{array}
    \hspace{-.7cm}\raisebox{-1.3cm}{\scalebox{.20}{\includegraphics{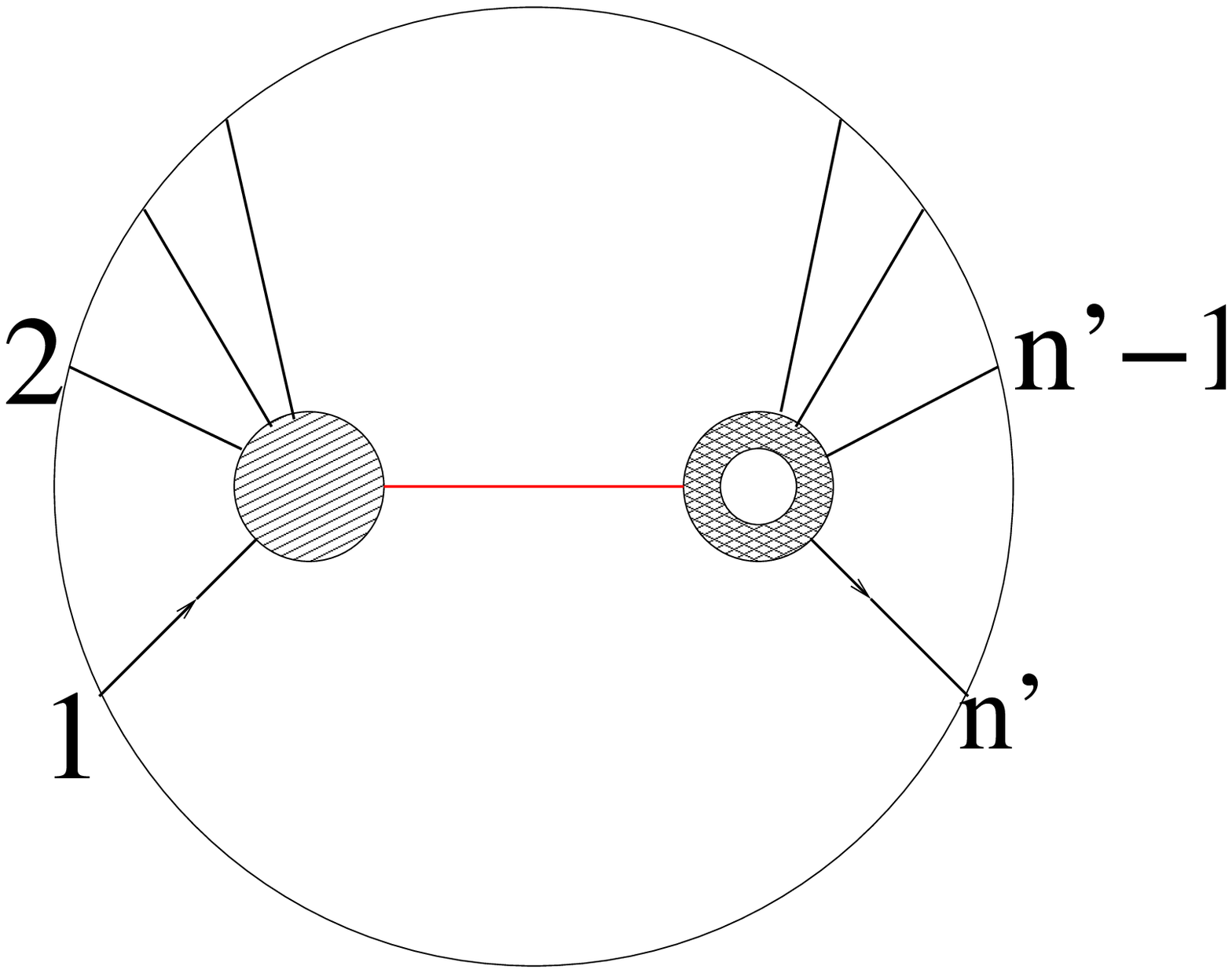}}}
    \begin{array}{l}
     \vspace{1.7cm}\hspace{-1cm}\mathcal{J}_k
    \end{array}\\
   &\begin{array}{l}
     \vspace{2.0cm}\mathcal{I}_k
    \end{array}
    \hspace{-.6cm}\raisebox{-1.3cm}{\scalebox{.20}{\includegraphics{1LindMnRHSb.eps}}}
    \begin{array}{l}
     \vspace{1.8cm}\hspace{-1cm}\mathcal{J}_k
    \end{array}
    \:\Longrightarrow\:
    \begin{array}{l}
     \vspace{1.9cm}\mathcal{I}_k
    \end{array}
    \hspace{-.7cm}\raisebox{-1.3cm}{\scalebox{.20}{\includegraphics{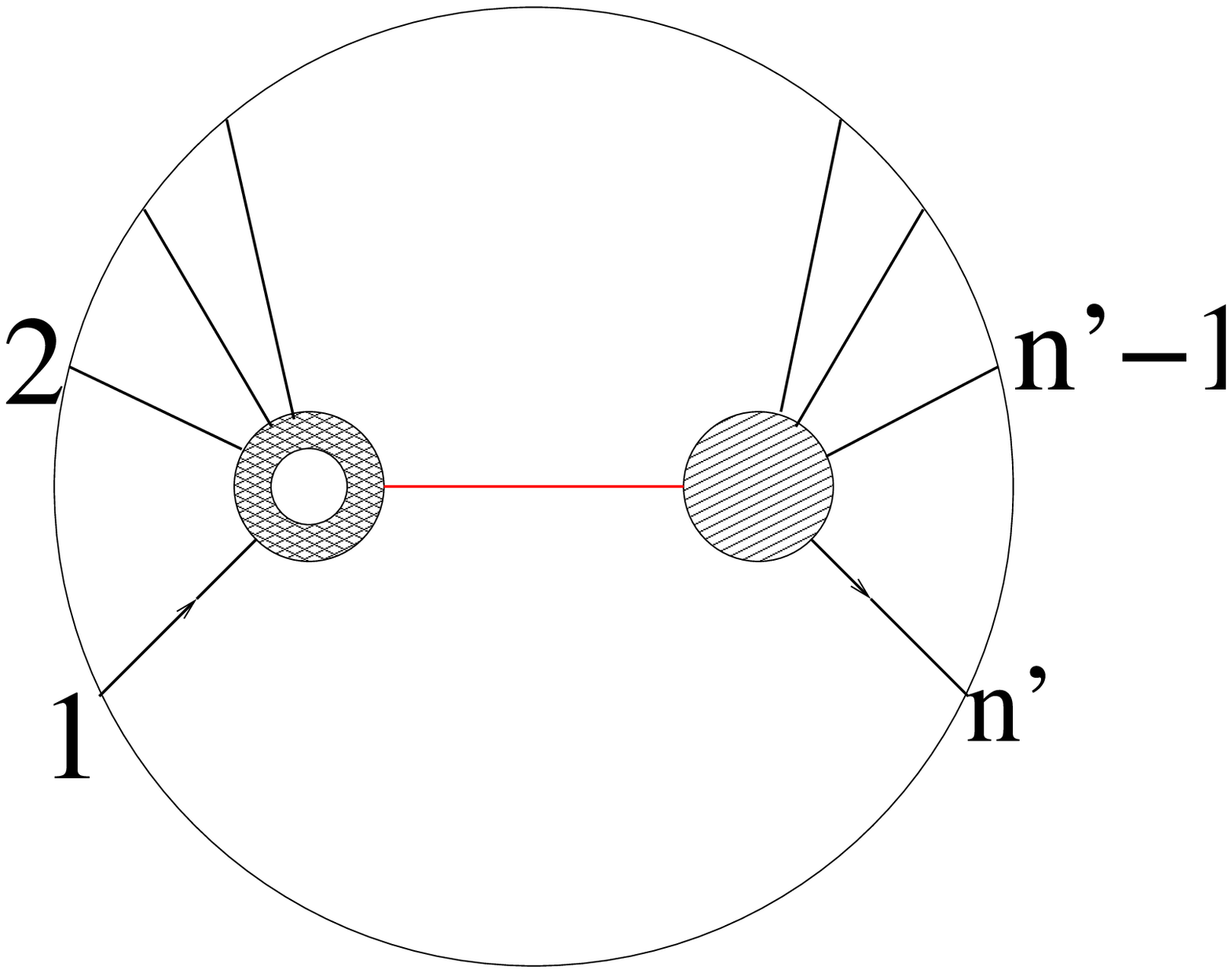}}}
    \begin{array}{l}
     \vspace{1.7cm}\hspace{-1cm}\mathcal{J}_k
    \end{array}
 \end{split}
\end{equation}
Notice that the helicities of the bridged particles are preserved. This occurs with any bridge with helicity 
multiplets $(\mp,\mp)$ as well as $(-,+)$ with the (negative)-positive helicity multiplet on the (anti)-holomorphic
three-particle amplitude. The leftover possible helicity choice of a bridge typically allows both multiplets to
propagate in the internal lines, generating the helicity loops and, therefore, extra singularities. Furthermore,
one term of this sum is not helicity preserving with respect to the bridged particles.

Let us move on to factorisation channels where the bridged particles are allowed to be on the same sub-amplitude.
Let us label a generic factorisation channel with $\mathcal{K}$ and, for the time being, consider the tree-level
contribution to the factorising amplitude:
\begin{equation}\eqlabel{eq:MnFactMult}
   \begin{array}{l}
    \vspace{2.0cm}\mathcal{I}_k
   \end{array}
   \hspace{-.6cm}\raisebox{-1.3cm}{\scalebox{.20}{\includegraphics{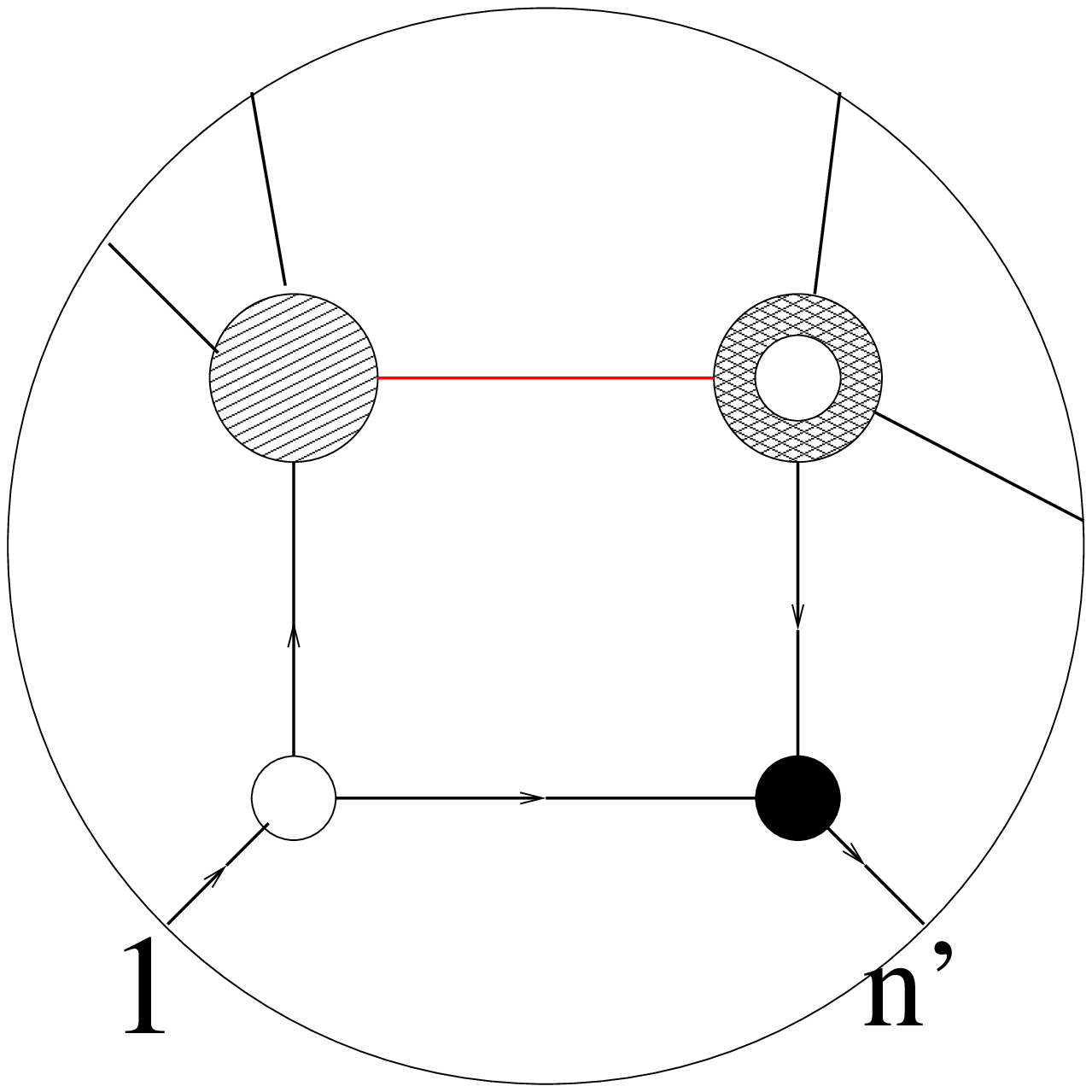}}}
   \begin{array}{l}
    \vspace{1.8cm}\hspace{-.5cm}\mathcal{J}_k
   \end{array}
   \:\Longrightarrow\:
   \sum_{k'}
   \begin{array}{l}
    \vspace{.6cm}\hspace{.3cm}\mathcal{K}\\
    \vspace{.6cm}\hspace{-.2cm}\mathcal{I}_{\bar{k}}^{\mbox{\tiny $(a)$}}
   \end{array}
   \hspace{-.35cm}\raisebox{-1.3cm}{\scalebox{.20}{\includegraphics{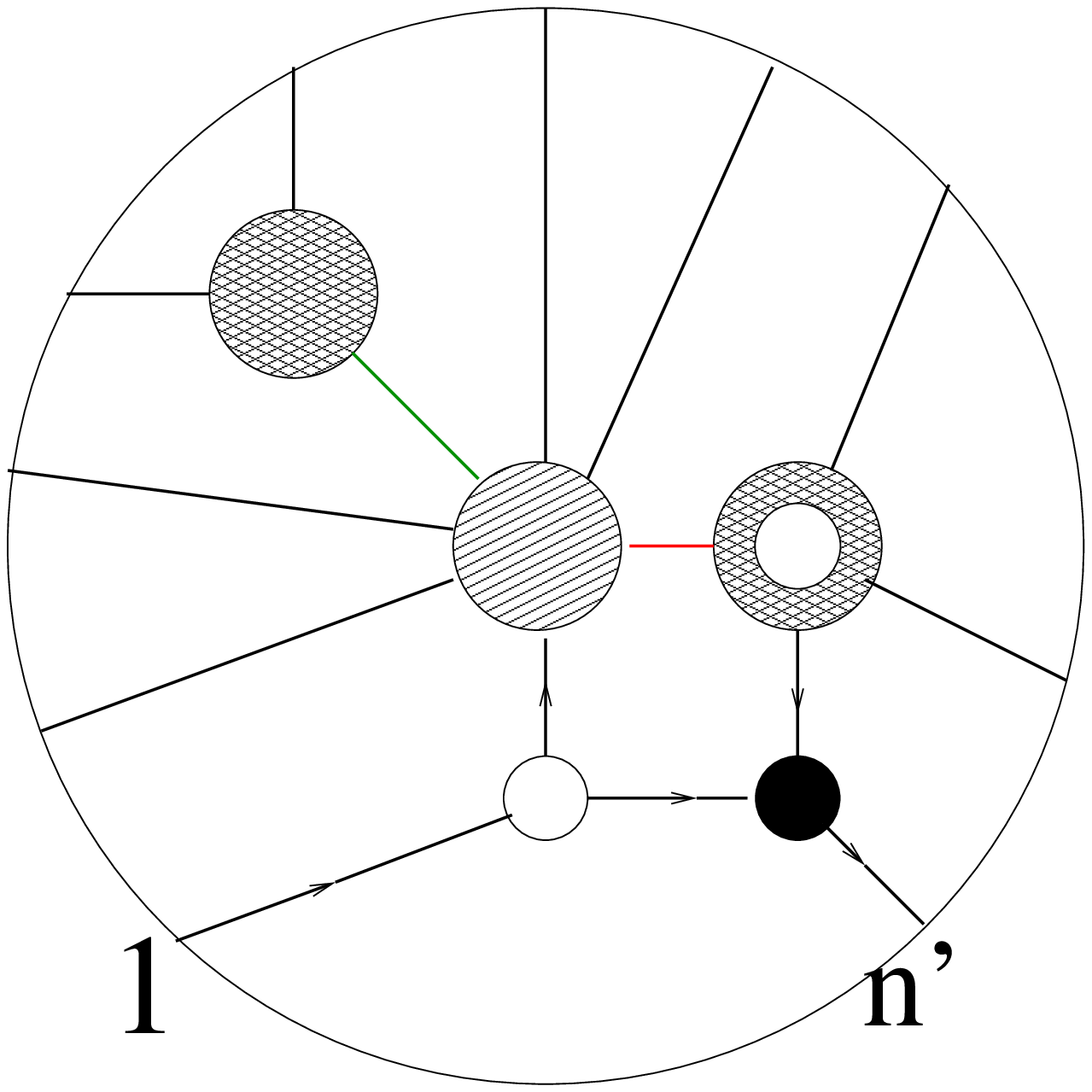}}}
   \begin{array}{l}
    \vspace{.6cm}\hspace{-1.2cm}\mathcal{I}_{k'}^{\mbox{\tiny $(b)$}}\\
    \vspace{1.5cm}\hspace{-.1cm}\mathcal{J}_{k'}
   \end{array}
   +
   \sum_{k'}
   \begin{array}{l}
    \vspace{.65cm}\hspace{.8cm}\mathcal{J}_{k'}^{\mbox{\tiny $(a)$}}\\
    \vspace{1.5cm}\hspace{-.2cm}\mathcal{I}_{k'}
   \end{array}
   \hspace{-1.4cm}\raisebox{-1.3cm}{\scalebox{.20}{\includegraphics{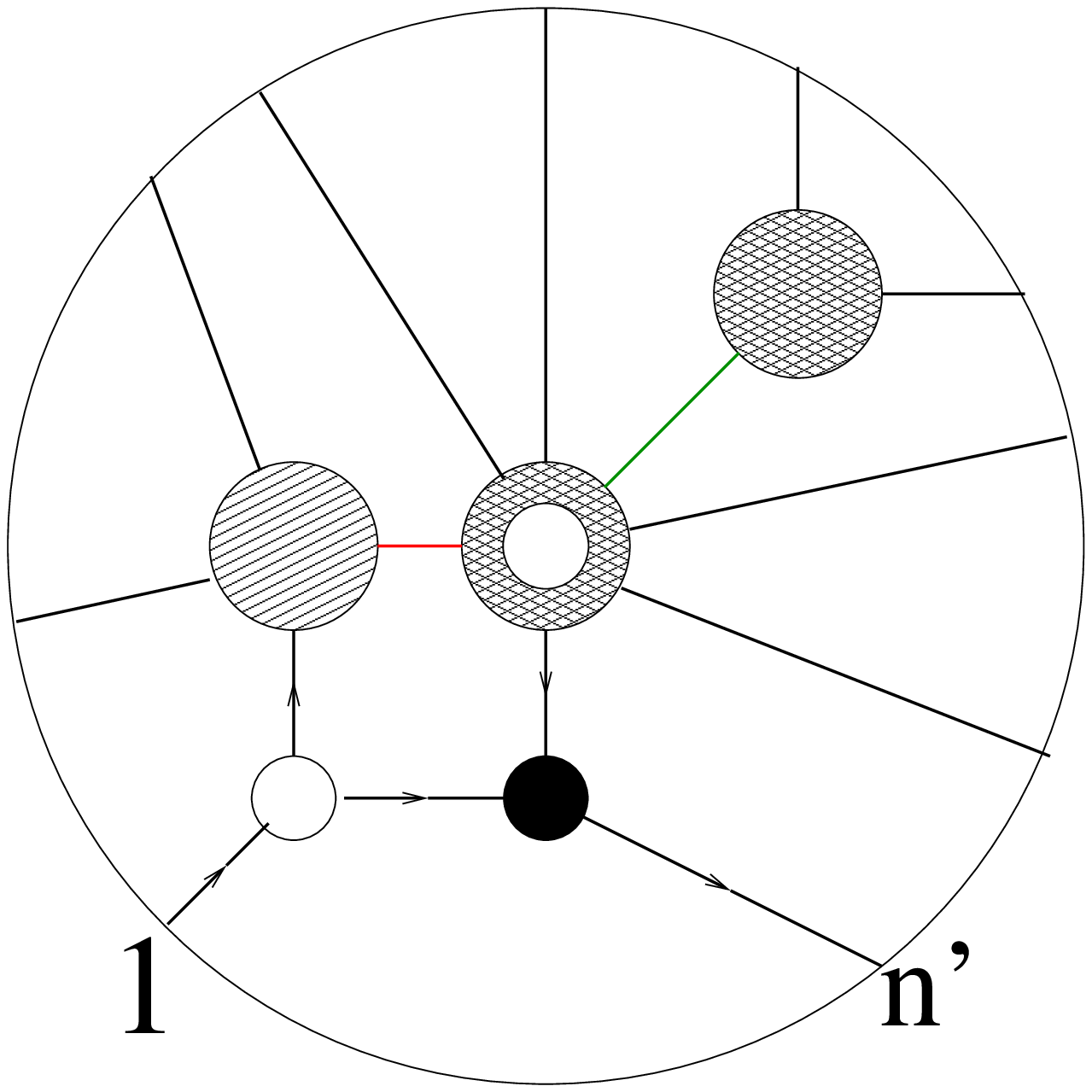}}}
   \begin{array}{l}
    \vspace{.6cm}\hspace{-.4cm}\mathcal{K}\\
    \vspace{.6cm}\hspace{-.1cm}\mathcal{J}_{\bar{k}}^{\mbox{\tiny $(b)$}}
   \end{array}
\end{equation}
where the factorisation channel is again identified by the helicity flows in the tree-level sub-amplitude.
Similarly, for the other set of factorisation
diagrams. As far as the forward diagram is concerned, in this class of factorisation channels, a contribution
it provides is given by
\begin{equation}\eqlabel{eq:MnFactMultFw}
 \raisebox{-1.3cm}{\scalebox{.20}{\includegraphics{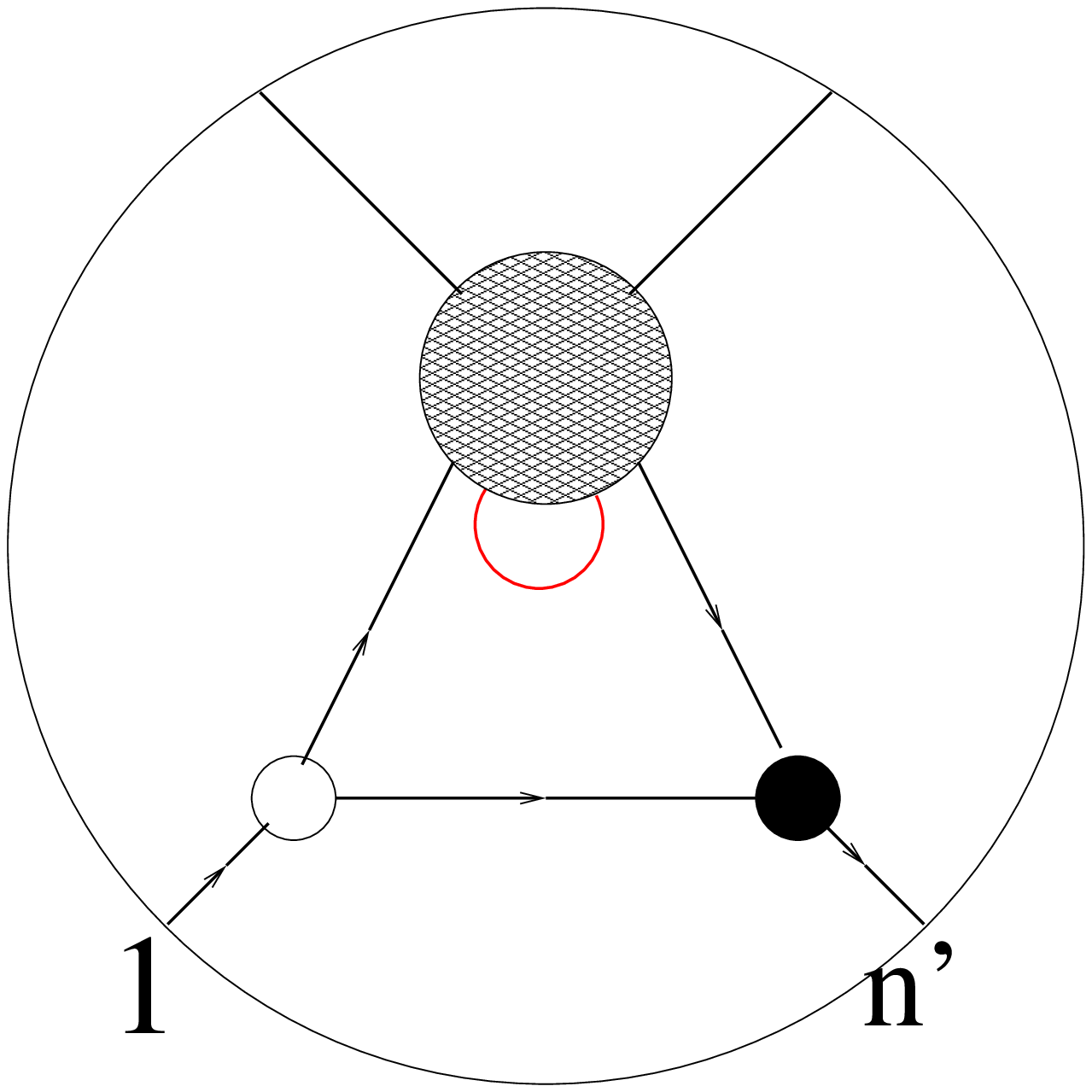}}}\:\Longrightarrow\:
 \raisebox{-1.3cm}{\scalebox{.20}{\includegraphics{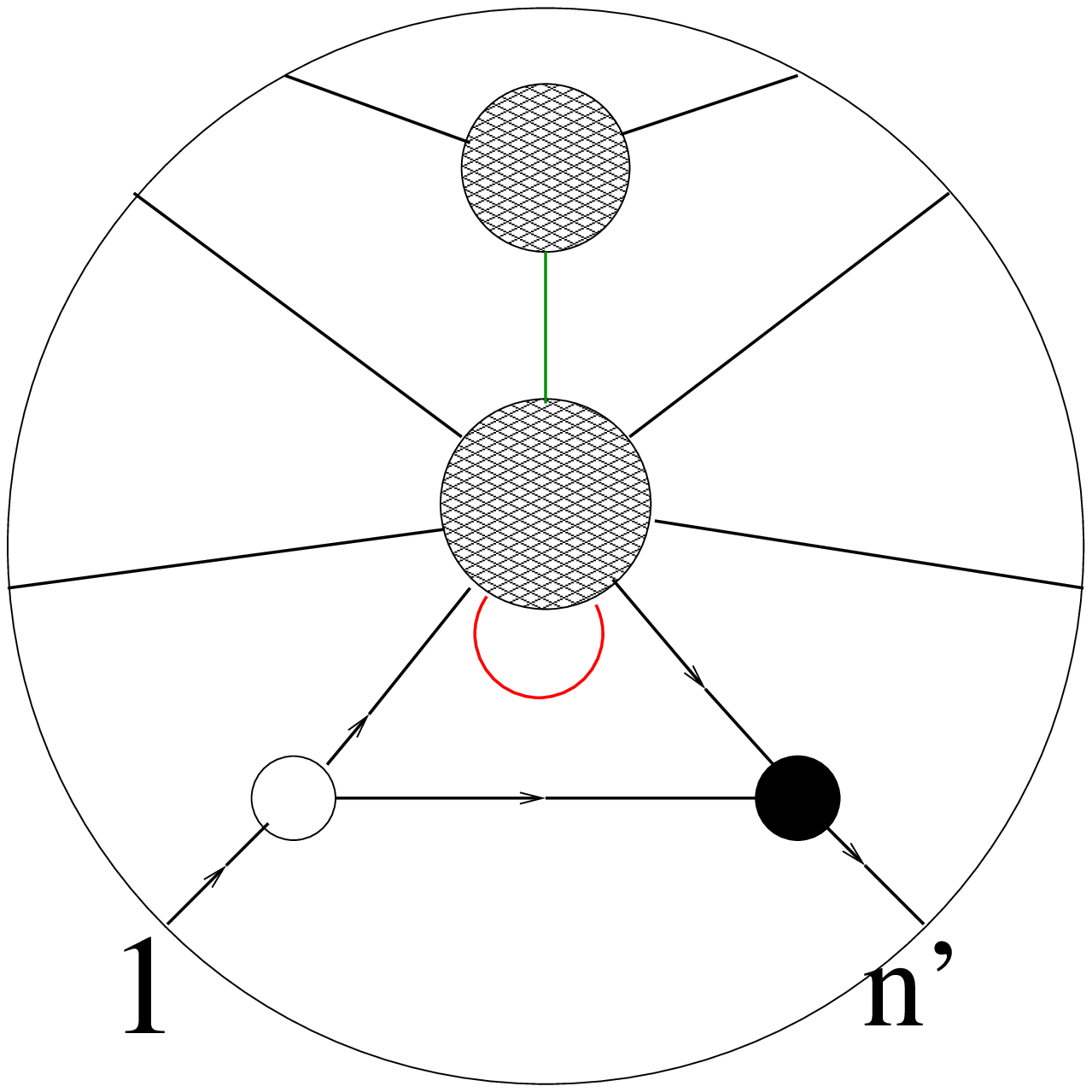}}}
 \begin{array}{l}
  \vspace{2.8cm}\hspace{-1.6cm}\mathcal{K}
  \end{array}
\end{equation}
Recollecting all these diagrams, it easy to see that, by virtue of the induction hypothesis, they sum up to a 
factorisation diagram with the sub-amplitudes are given by a tree-level sub-amplitude with the set $\mathcal{K}$
of external states, and a one-loop sub-amplitude

{\footnotesize
\begin{equation*}
 \begin{split}
  &\sum_{k'}
   \begin{array}{l}
    \vspace{.6cm}\hspace{.3cm}\mathcal{K}\\
    \vspace{.6cm}\hspace{-.2cm}\mathcal{I}_{\bar{k}}^{\mbox{\tiny $(a)$}}
   \end{array}
   \hspace{-.35cm}\raisebox{-1.3cm}{\scalebox{.20}{\includegraphics{1LMnFact1nc.eps}}}
   \begin{array}{l}
    \vspace{.6cm}\hspace{-1.2cm}\mathcal{I}_{k'}^{\mbox{\tiny $(b)$}}\\
    \vspace{1.5cm}\hspace{-.1cm}\mathcal{J}_{k'}
   \end{array}
   \hspace{-.2cm}+
   \sum_{k'}
   \begin{array}{l}
    \vspace{.65cm}\hspace{.8cm}\mathcal{J}_{k'}^{\mbox{\tiny $(a)$}}\\
    \vspace{1.5cm}\hspace{-.2cm}\mathcal{I}_{k'}
   \end{array}
   \hspace{-1.4cm}\raisebox{-1.3cm}{\scalebox{.20}{\includegraphics{1LMnFact1nc3.eps}}}
   \begin{array}{l}
    \vspace{.6cm}\hspace{-.4cm}\mathcal{K}\\
   \vspace{.6cm}\hspace{-.1cm}\mathcal{J}_{\bar{k}}^{\mbox{\tiny $(b)$}}
   \end{array}
   \hspace{-.2cm}+
   \sum_{k'}
   \begin{array}{l}
    \vspace{.6cm}\hspace{.3cm}\mathcal{K}\\
    \vspace{.6cm}\hspace{-.2cm}\mathcal{I}_{\bar{k}}^{\mbox{\tiny $(a)$}}
   \end{array}
   \hspace{-.35cm}\raisebox{-1.3cm}{\scalebox{.20}{\includegraphics{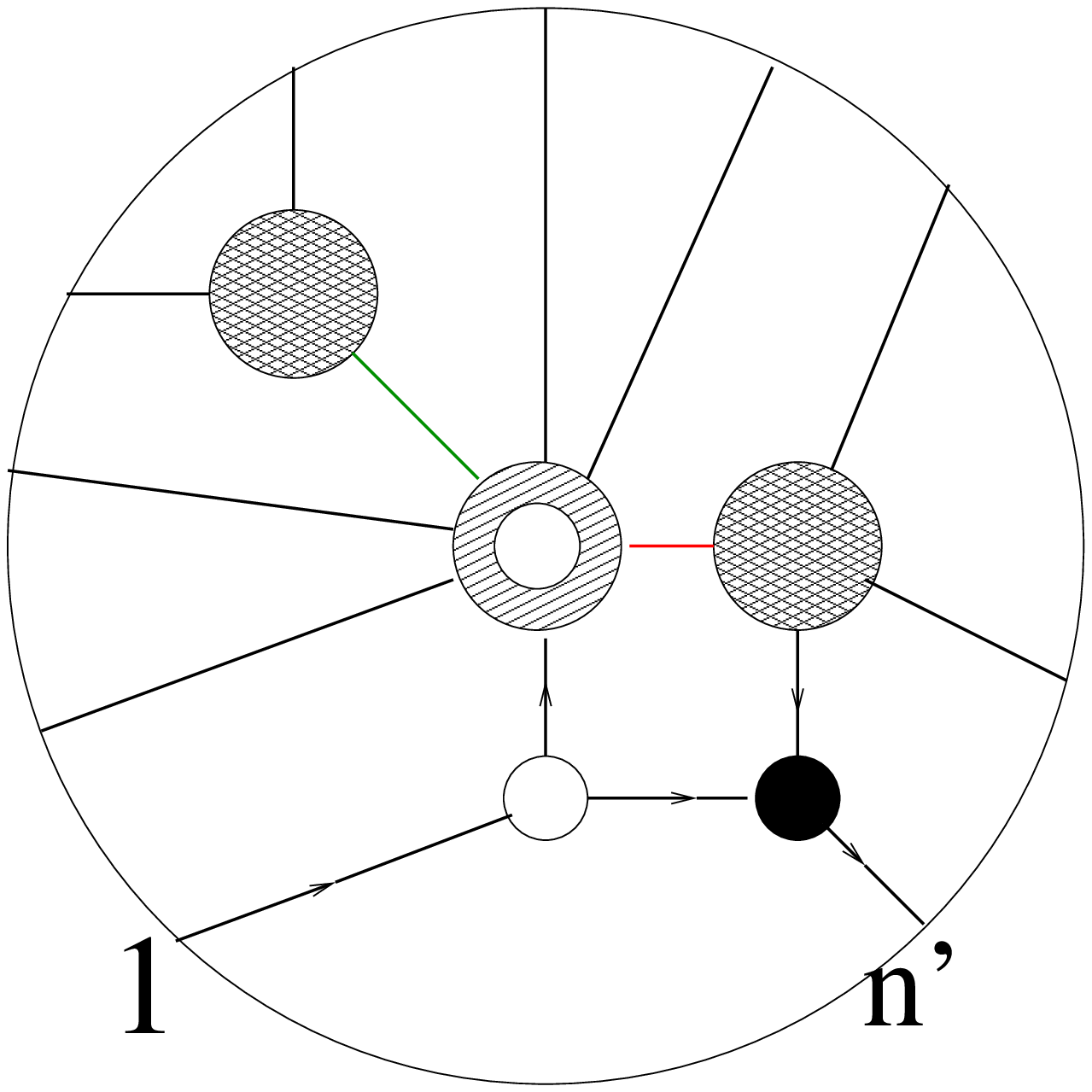}}}
   \begin{array}{l}
    \vspace{.6cm}\hspace{-1.2cm}\mathcal{I}_{k'}^{\mbox{\tiny $(b)$}}\\
    \vspace{1.5cm}\hspace{-.1cm}\mathcal{J}_{k'}
   \end{array}
   \hspace{-.2cm}+
  \\
  &\hspace{2cm}
   +\sum_{k'}
   \begin{array}{l}
    \vspace{.65cm}\hspace{.8cm}\mathcal{J}_{k'}^{\mbox{\tiny $(a)$}}\\
    \vspace{1.5cm}\hspace{-.2cm}\mathcal{I}_{k'}
   \end{array}
   \hspace{-1.4cm}\raisebox{-1.3cm}{\scalebox{.20}{\includegraphics{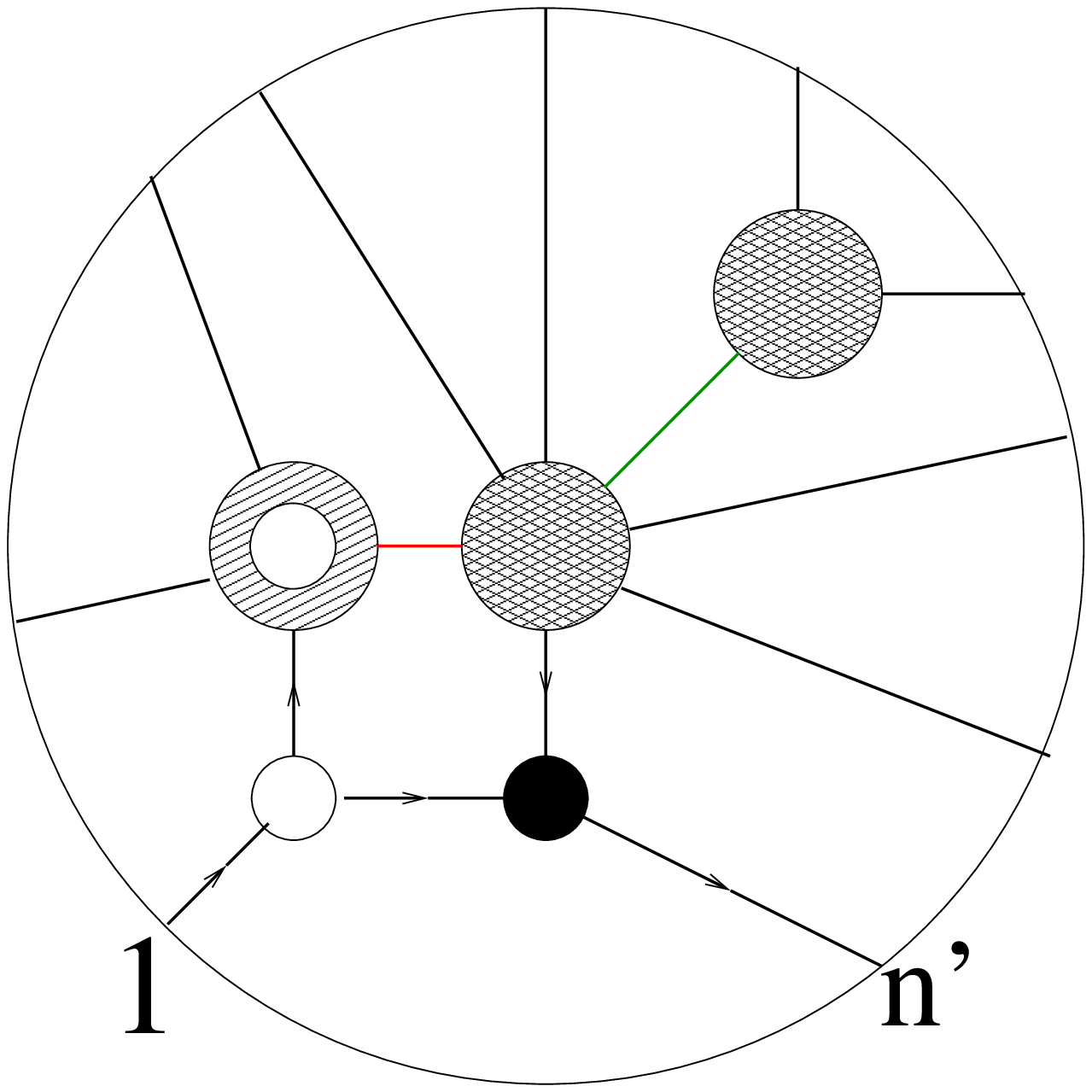}}}
   \begin{array}{l}
    \vspace{.6cm}\hspace{-.4cm}\mathcal{K}\\
   \vspace{.6cm}\hspace{-.1cm}\mathcal{J}_{\bar{k}}^{\mbox{\tiny $(b)$}}
   \end{array}
   \hspace{-.2cm}+
   \begin{array}{l}
    \hspace{-.2cm}\vspace{.5cm}\mathcal{I}_{\bar{k}}^{\mbox{\tiny $(a)$}}
   \end{array}
   \hspace{-.3cm}\raisebox{-1.3cm}{\scalebox{.20}{\includegraphics{1LMnFactFw2.eps}}}
   \begin{array}{l}
    \vspace{.6cm}\hspace{-1.6cm}\mathcal{K}\\
    \vspace{1.8cm}\hspace{-.2cm}\mathcal{J}_{\bar{k}}^{\mbox{\tiny $(b)$}}
   \end{array}
   \hspace{-.3cm}=
   \begin{array}{l}
    \vspace{.8cm}\hspace{.3cm}\mathcal{K}\\
    \vspace{.4cm}\hspace{-.2cm}\mathcal{I}_{\bar{k}}^{\mbox{\tiny $(a)$}}
   \end{array}
   \hspace{-.35cm}\raisebox{-1.3cm}{\scalebox{.20}{\includegraphics{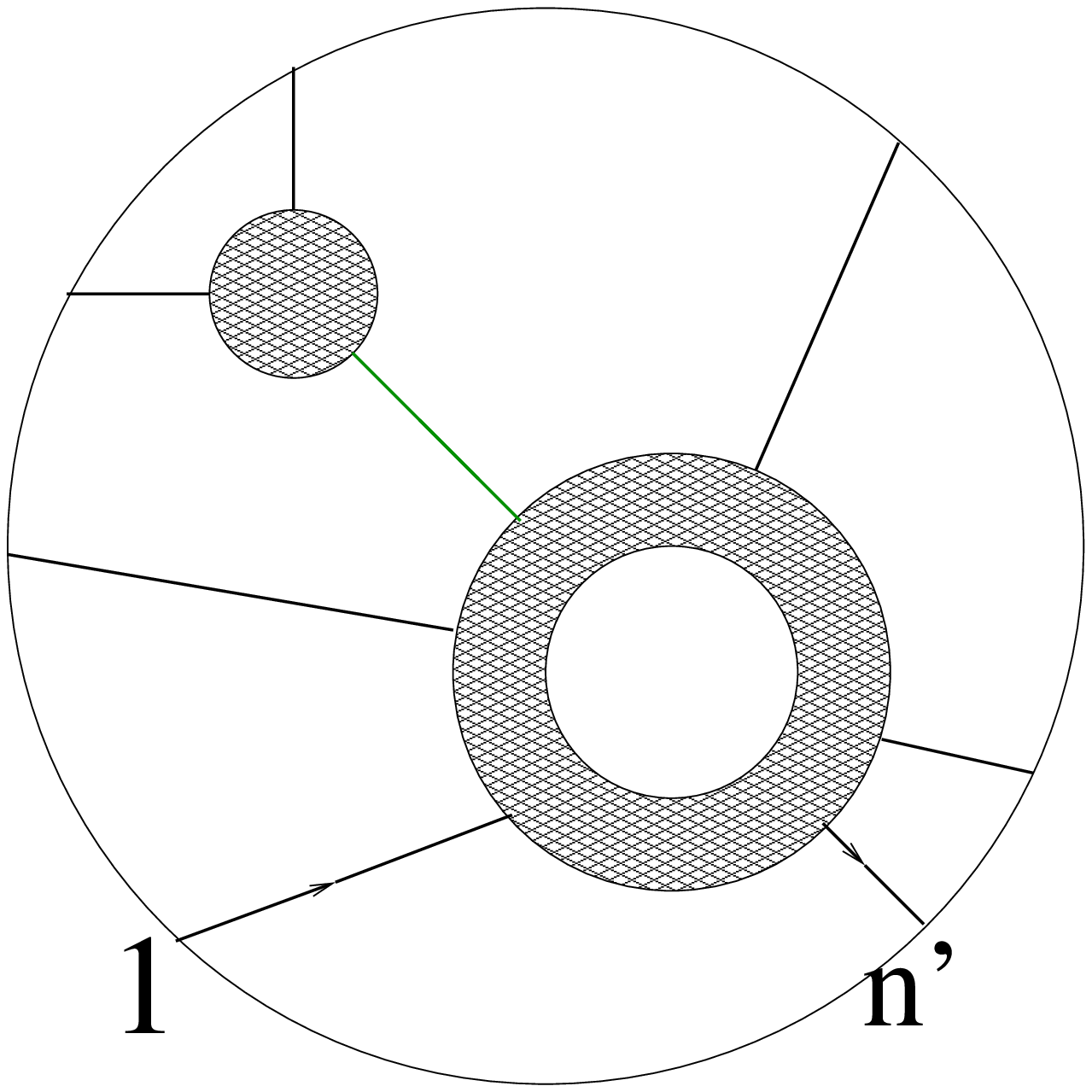}}}
   \begin{array}{l}
    \vspace{.6cm}\hspace{-1.2cm}\phantom{\mathcal{I}_{k''}}\\
    \vspace{1.5cm}\hspace{-.1cm}\mathcal{J}_{\bar{k}}^{\mbox{\tiny $(b)$}}
   \end{array}
 \end{split}
\end{equation*}
}
In the same fashion, but without contribution from the forward diagrams, one obtains
\begin{equation*}
 \sum_{k'}
  \begin{array}{l}
   \vspace{.6cm}\hspace{.3cm}\mathcal{K}\\
   \vspace{.6cm}\hspace{-.2cm}\mathcal{I}_{\bar{k}}^{\mbox{\tiny $(a)$}}
  \end{array}
  \hspace{-.35cm}\raisebox{-1.3cm}{\scalebox{.20}{\includegraphics{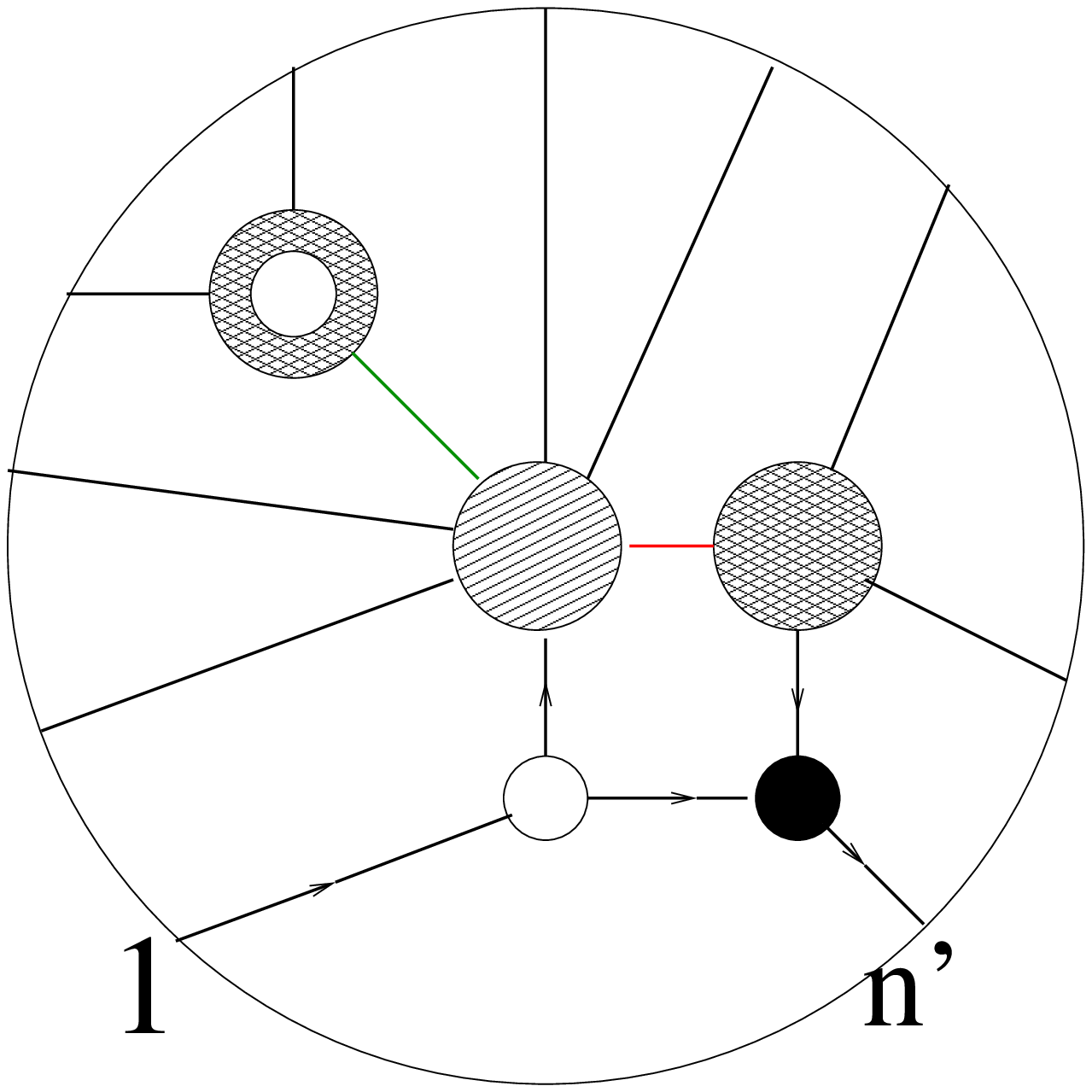}}}
  \begin{array}{l}
   \vspace{.6cm}\hspace{-1.2cm}\mathcal{I}_{k'}^{\mbox{\tiny $(b)$}}\\
   \vspace{1.5cm}\hspace{-.1cm}\mathcal{J}_{k'}
  \end{array}
  \hspace{-.2cm}+
 \sum_{k'}
  \begin{array}{l}
   \vspace{.65cm}\hspace{.8cm}\mathcal{J}_{k'}^{\mbox{\tiny $(a)$}}\\
   \vspace{1.5cm}\hspace{-.2cm}\mathcal{I}_{k'}
  \end{array}
  \hspace{-1.4cm}\raisebox{-1.3cm}{\scalebox{.20}{\includegraphics{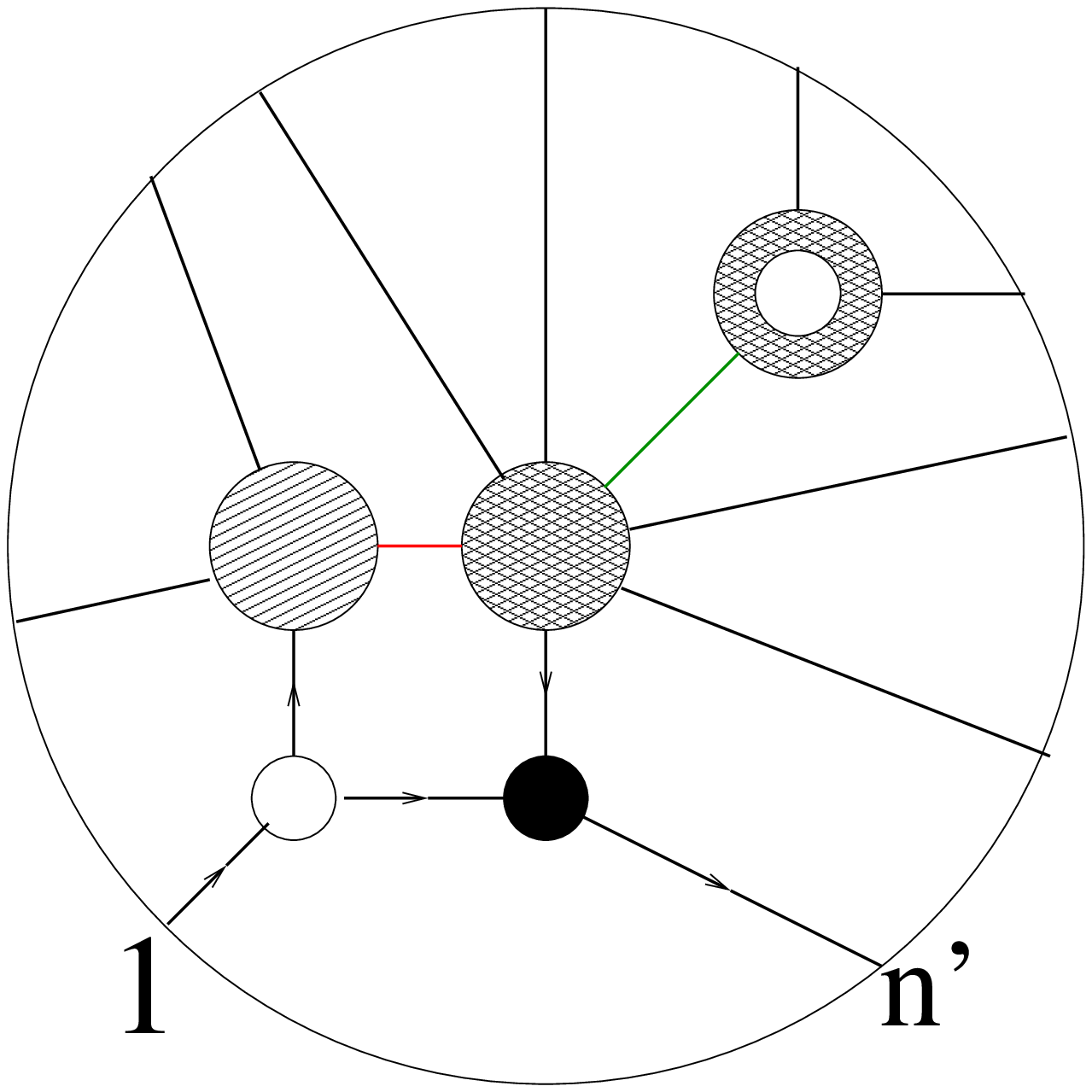}}}
  \begin{array}{l}
   \vspace{.6cm}\hspace{-.4cm}\mathcal{K}\\
   \vspace{.6cm}\hspace{-.1cm}\mathcal{J}_{\bar{k}}^{\mbox{\tiny $(b)$}}
  \end{array}
  \hspace{-.2cm}=\:
  \begin{array}{l}
   \vspace{.65cm}\hspace{.8cm}\phantom{\mathcal{J}_{k'}^{\mbox{\tiny $(a)$}}}\\
   \vspace{1.5cm}\hspace{-.2cm}\mathcal{I}_{\bar{k}}^{\mbox{\tiny $(a)$}}
  \end{array}
  \hspace{-1.3cm}\raisebox{-1.3cm}{\scalebox{.20}{\includegraphics{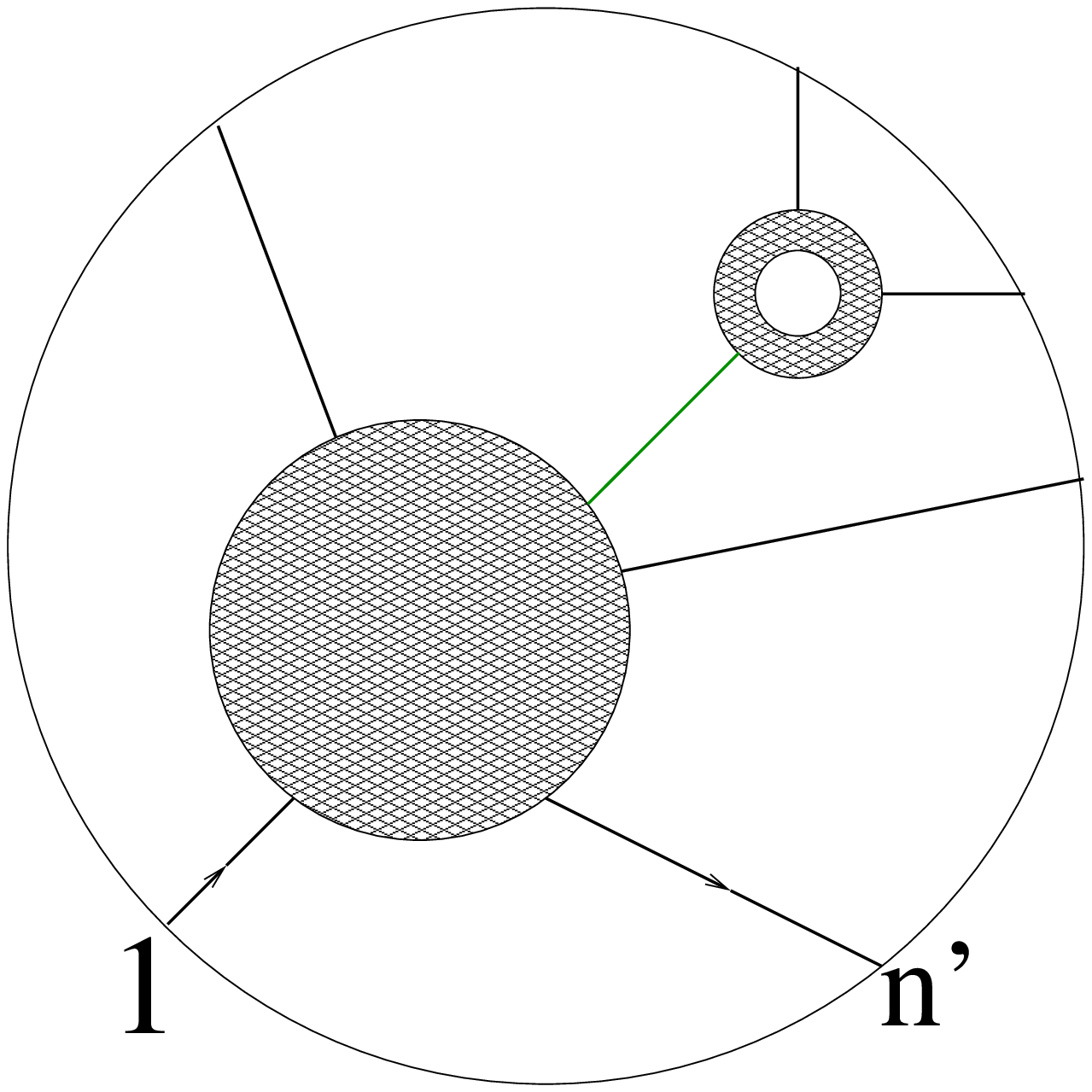}}}
  \begin{array}{l}
   \vspace{.6cm}\hspace{-.4cm}\mathcal{K}\\
   \vspace{.6cm}\hspace{-.1cm}\mathcal{J}_{\bar{k}}^{\mbox{\tiny $(b)$}}
  \end{array}
\end{equation*}
where we used the induction hypothesis in order to write down the right-hand-side.

The collinear factorisation involving the bridged particles is a bit more subtle. There is just one diagram
which can contribute for each of the two ways in which such a factorisation can occur. The first one, which
correspond to the anti-holomorphic two particle factorisation is given by
\begin{equation}\eqlabel{eq:MnFactCollA}
 \raisebox{-1.3cm}{\scalebox{.25}{\includegraphics{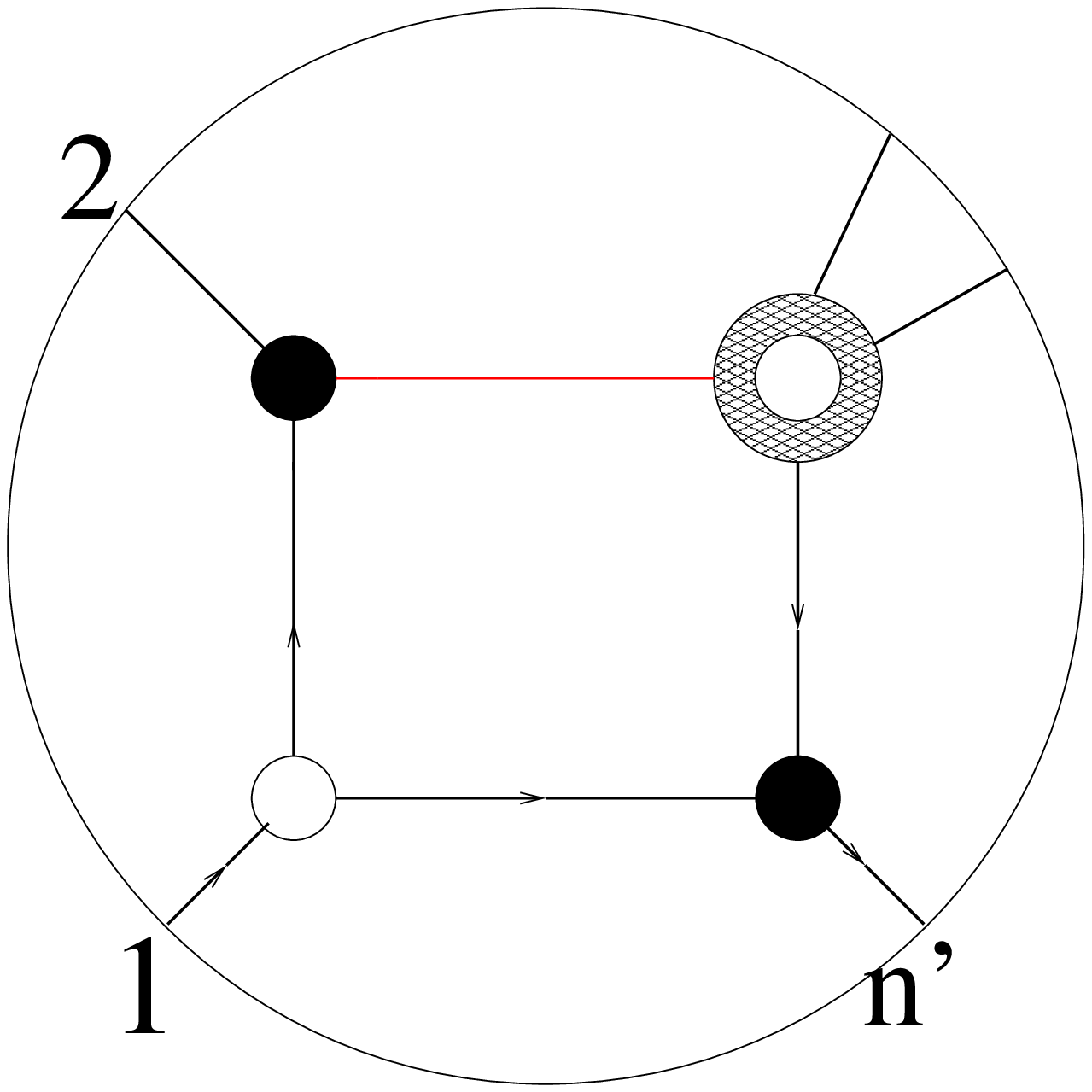}}}\:\Longrightarrow\:
 \raisebox{-1.3cm}{\scalebox{.25}{\includegraphics{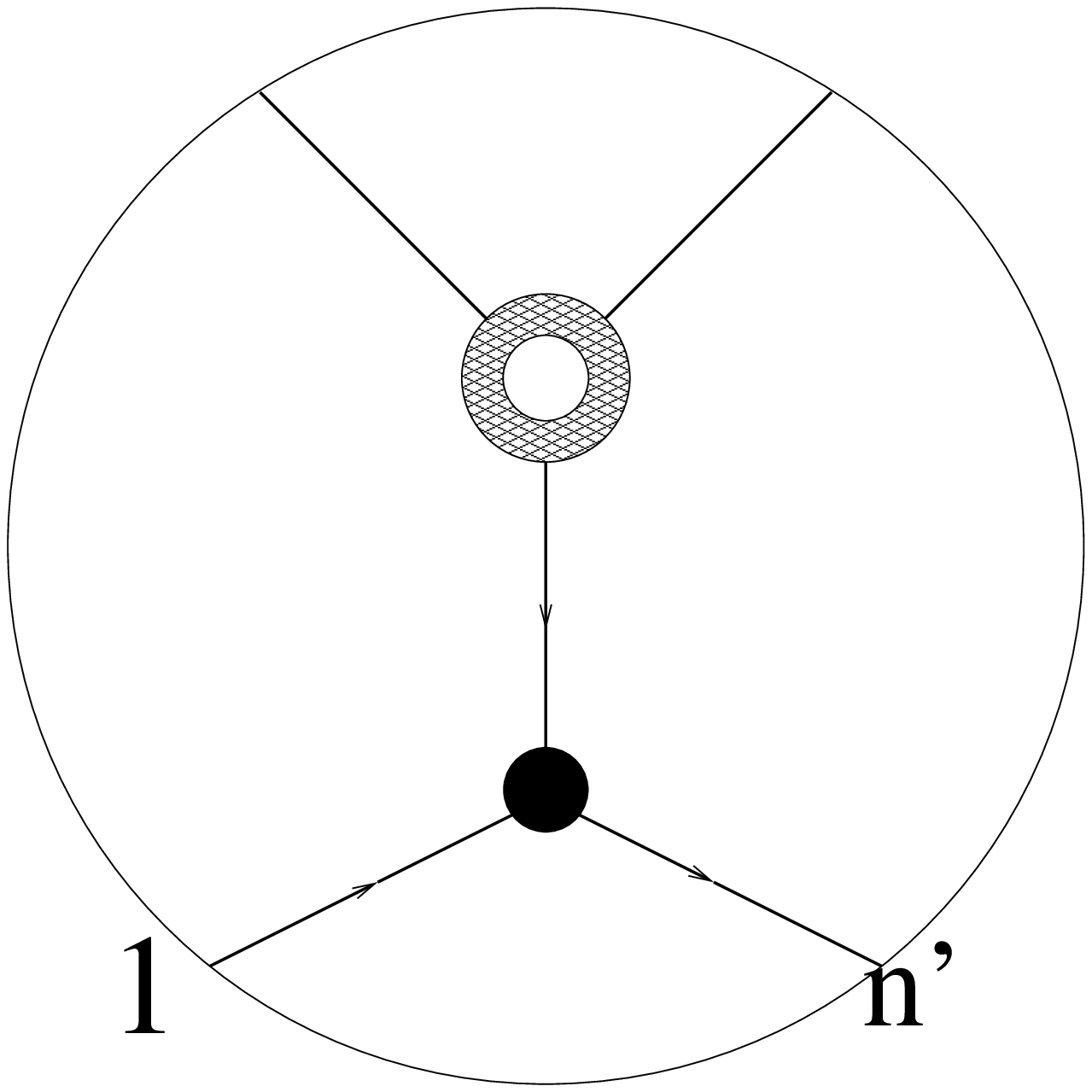}}}
\end{equation}
Notice that, with the current choice of the helicity configuration for the bridged particles and irrespectively
of the helicity of particle $2$, the helicity flow from the external state $2$ to the $(n'-2)$-particle 
sub-amplitude  is preserved: Such a channel is always a collinear singularity of our amplitude.
Similarly, the holomorphic one arises from
\begin{equation}\eqlabel{eq:MnFactCollH}
 \raisebox{-1.3cm}{\scalebox{.25}{\includegraphics{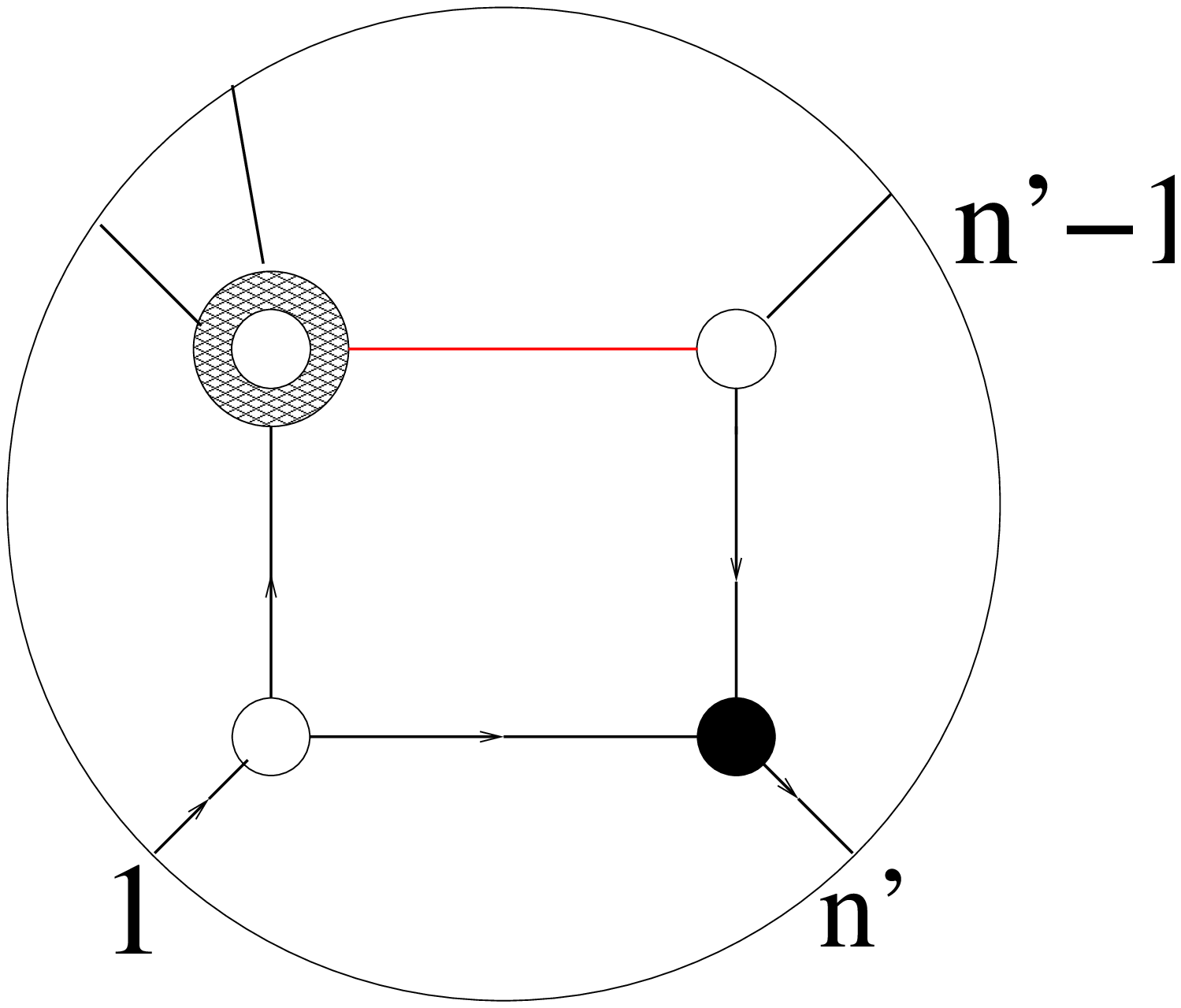}}}\:\Longrightarrow\quad
 \raisebox{-1.3cm}{\scalebox{.25}{\includegraphics{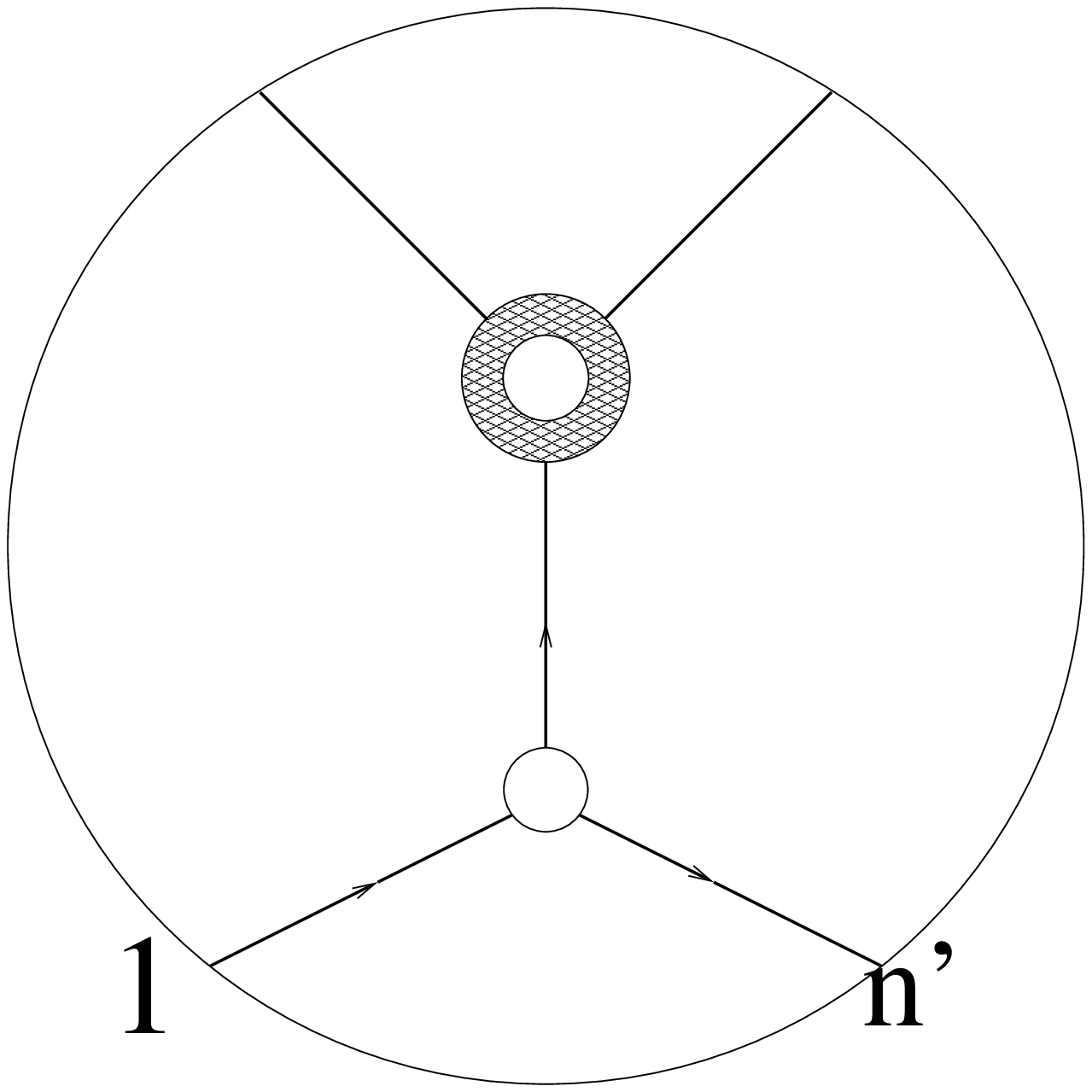}}}
\end{equation}
As in the previous case, the helicity flow between the external state $n'$ and the $(n'-2)$-particle 
sub-amplitude  is preserved, so that the singularity diagram on the right-hand-side is effectively a
factorisation channel for the full amplitude.

A comment is now in order. Let us consider the helicity configuration $(-,-)$ for the bridged particles. While
the discussion for the anti-holomorphic factorisation goes as in \eqref{eq:MnFactCollA}, something different
happens for the holomorphic one. First, the diagram of the type of the one at the left-hand-side in 
\eqref{eq:MnFactCollH} is present if and only if particle $(n'-1)$ has positive helicity. In this case
\begin{equation}\eqlabel{eq:MnFactCollH2}
 \raisebox{-1.3cm}{\scalebox{.25}{\includegraphics{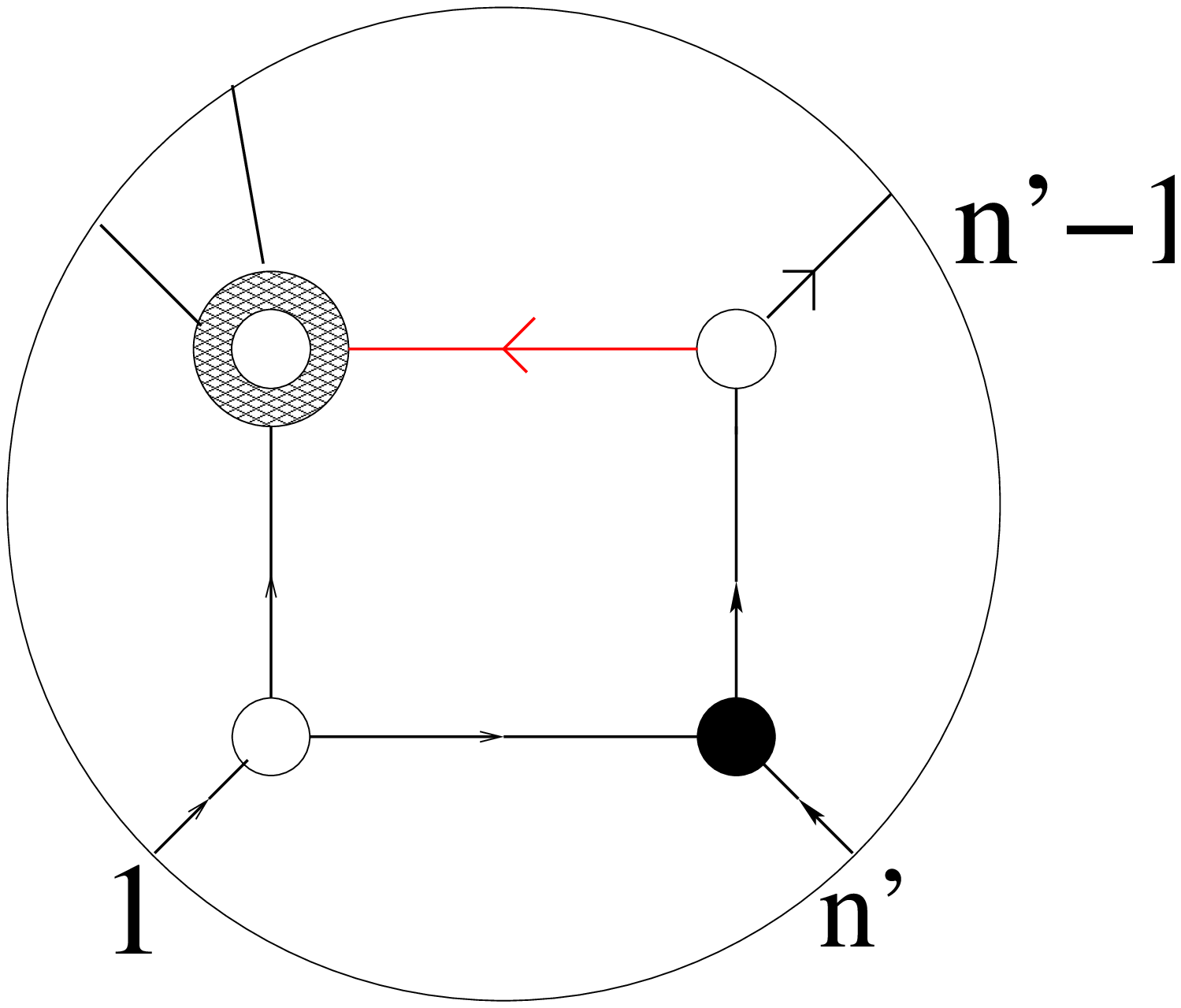}}}\:\Longrightarrow\quad
 \raisebox{-1.55cm}{\scalebox{.25}{\includegraphics{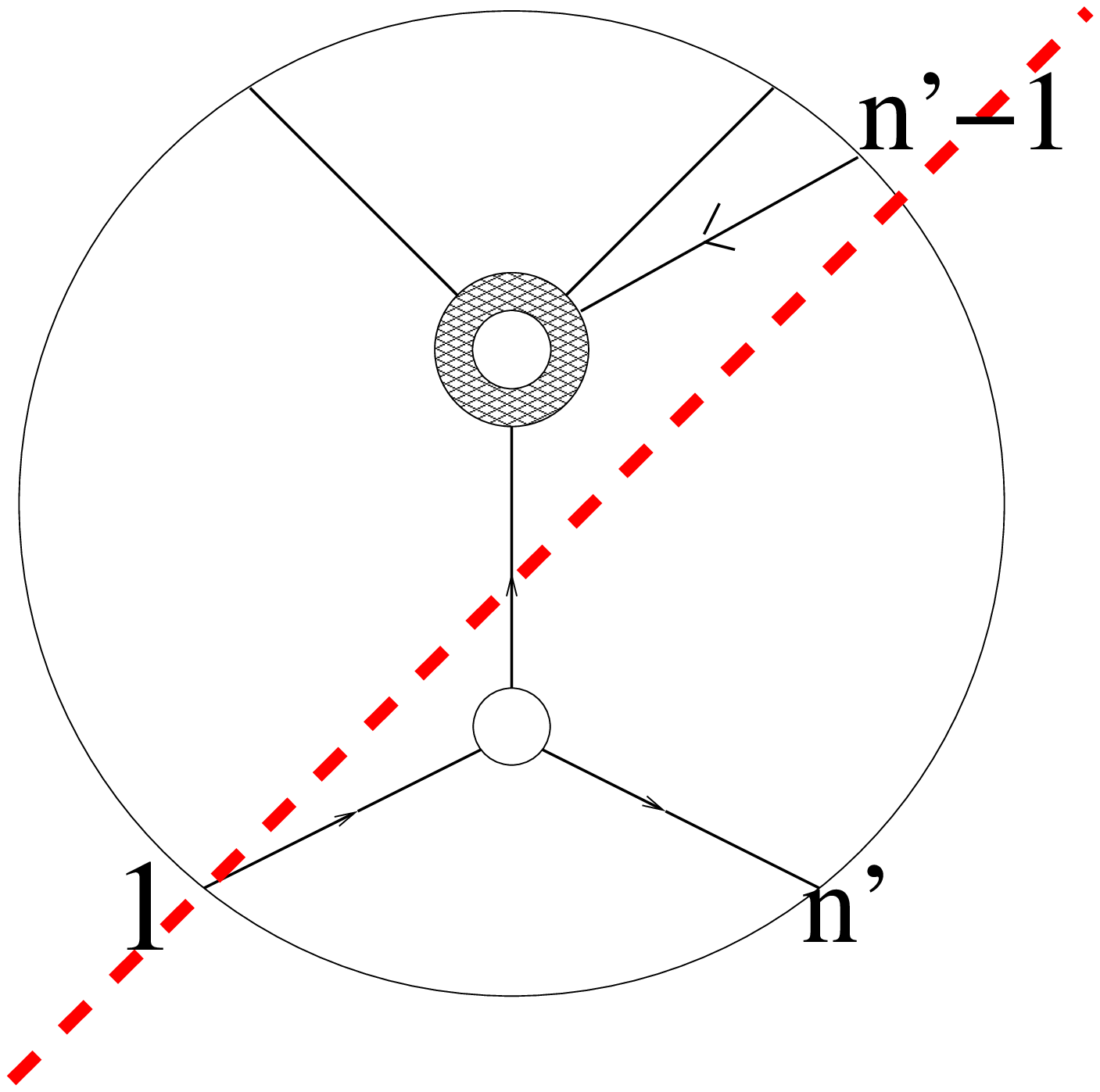}}}
\end{equation}
There is no helicity flow along the lines $(n'-1)$ and $n'$ and thus there is no holomorphic factorisation in the 
$(n',1)$ channel.

Finally, from the very same argument that led to prove the cancellation of non-local poles in 
Section \ref{sec:TreeStr}, it follows that the factorisation channels are all and only the one described.

Let us move on the forward diagram. As for the factorisation channels, a given BCFW bridge makes manifest a forward
limit\footnote{It is important to stress that such a statement can be made because we are considering all the 
on-shell diagrams as regularised in the quasi-forward scheme, in which any other contribution is of order
$\mathcal{O}(\epsilon)$ and the simple bridging operation cannot make such a behaviour worse.}:
\begin{equation}\eqlabel{eq:MnFwLm}
 \raisebox{-1.3cm}{\scalebox{.20}{\includegraphics{1LMnFactFw.eps}}}\:\Longrightarrow\quad
 \raisebox{-1.3cm}{\scalebox{.20}{\includegraphics{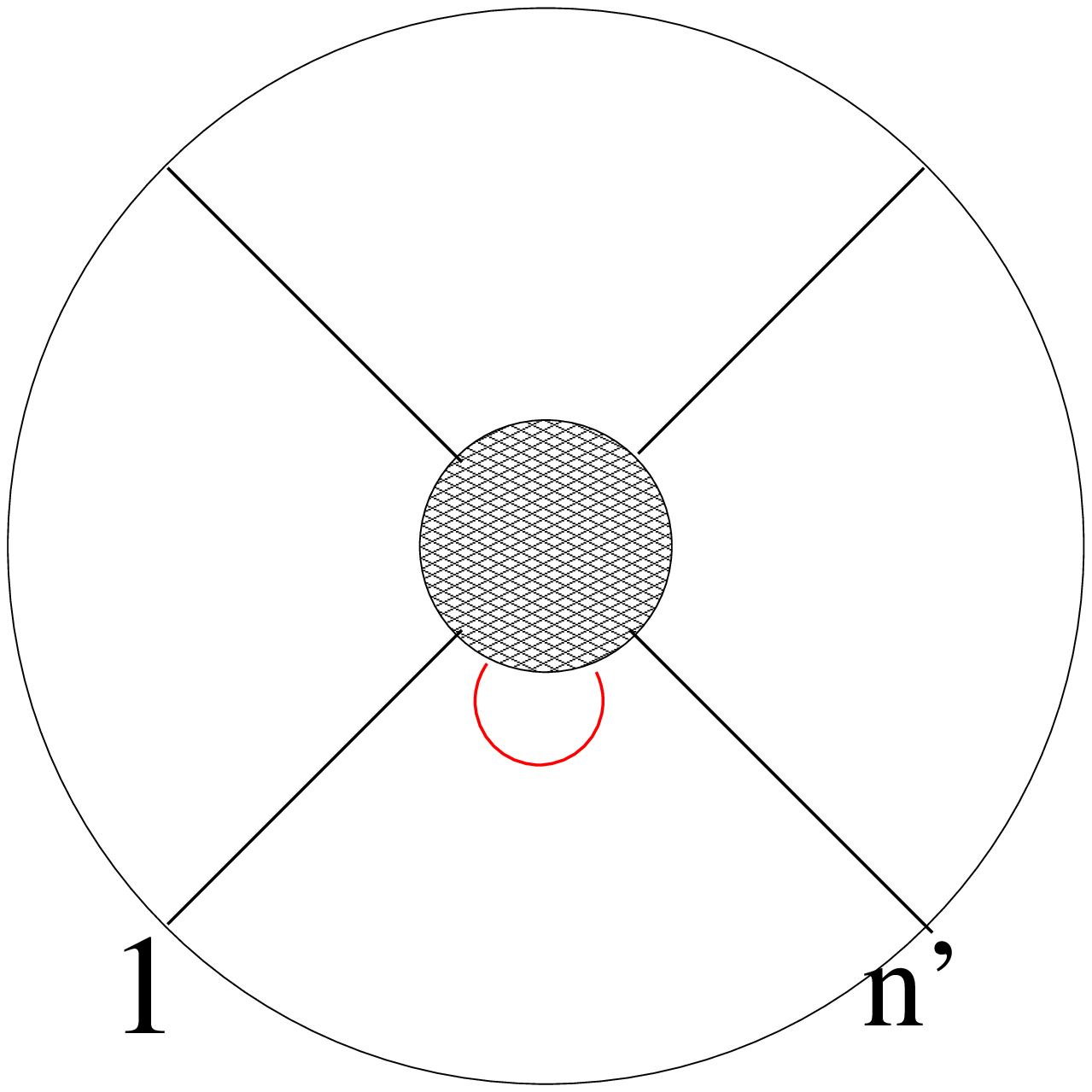}}}
\end{equation}
The other forward channels are a bit more subtle. Let us consider the forward singularity between particle $i$
and $i+1$. Such a singularity is contained in the following factorisation terms:
\begin{equation}\eqlabel{eq:MnFwLm2}
 \raisebox{-1.3cm}{\scalebox{.20}{\includegraphics{1LMnFact.eps}}}\:\Longrightarrow\:
 \raisebox{-1.3cm}{\scalebox{.20}{\includegraphics{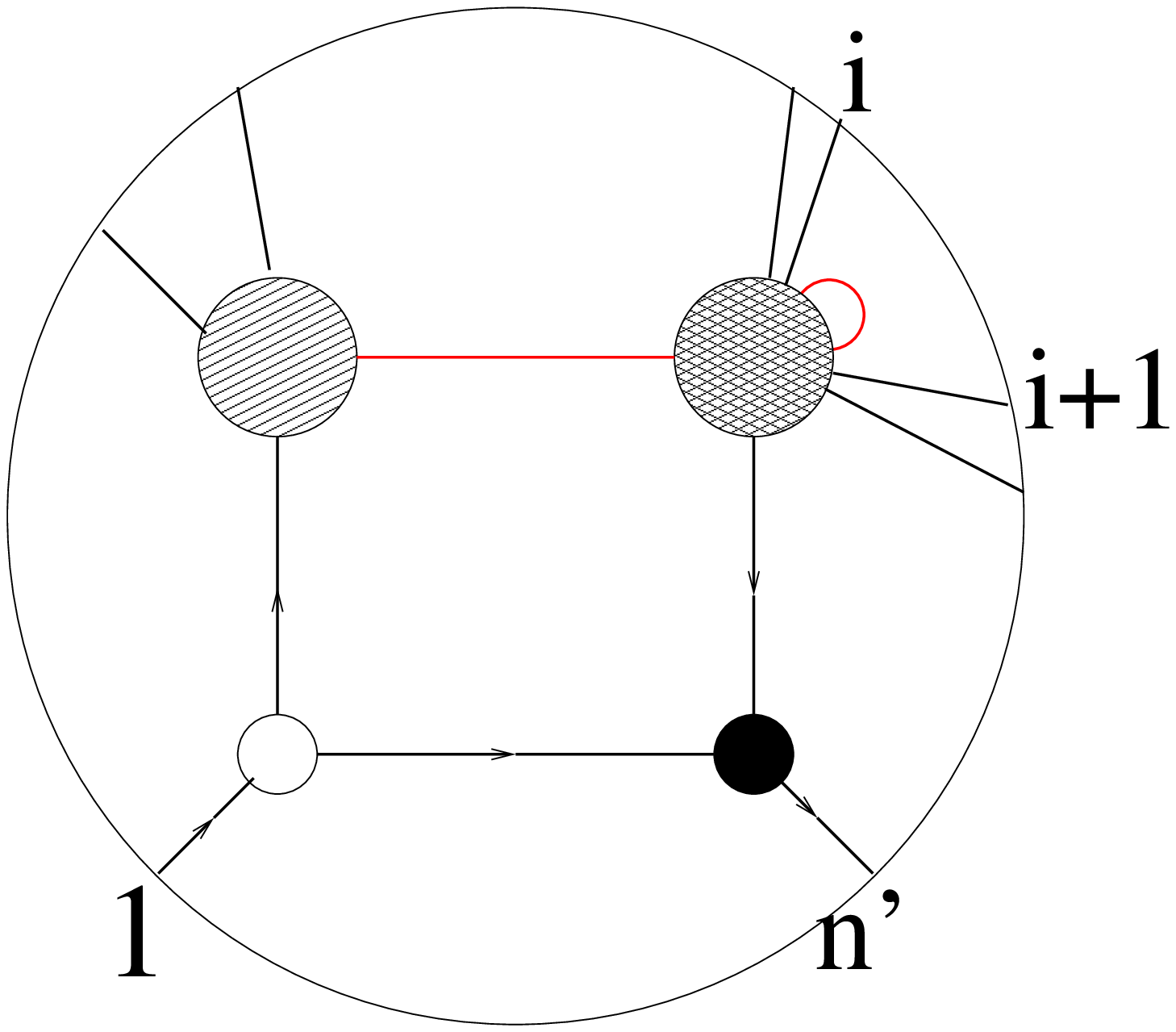}}};
 \quad
 \raisebox{-1.3cm}{\scalebox{.20}{\includegraphics{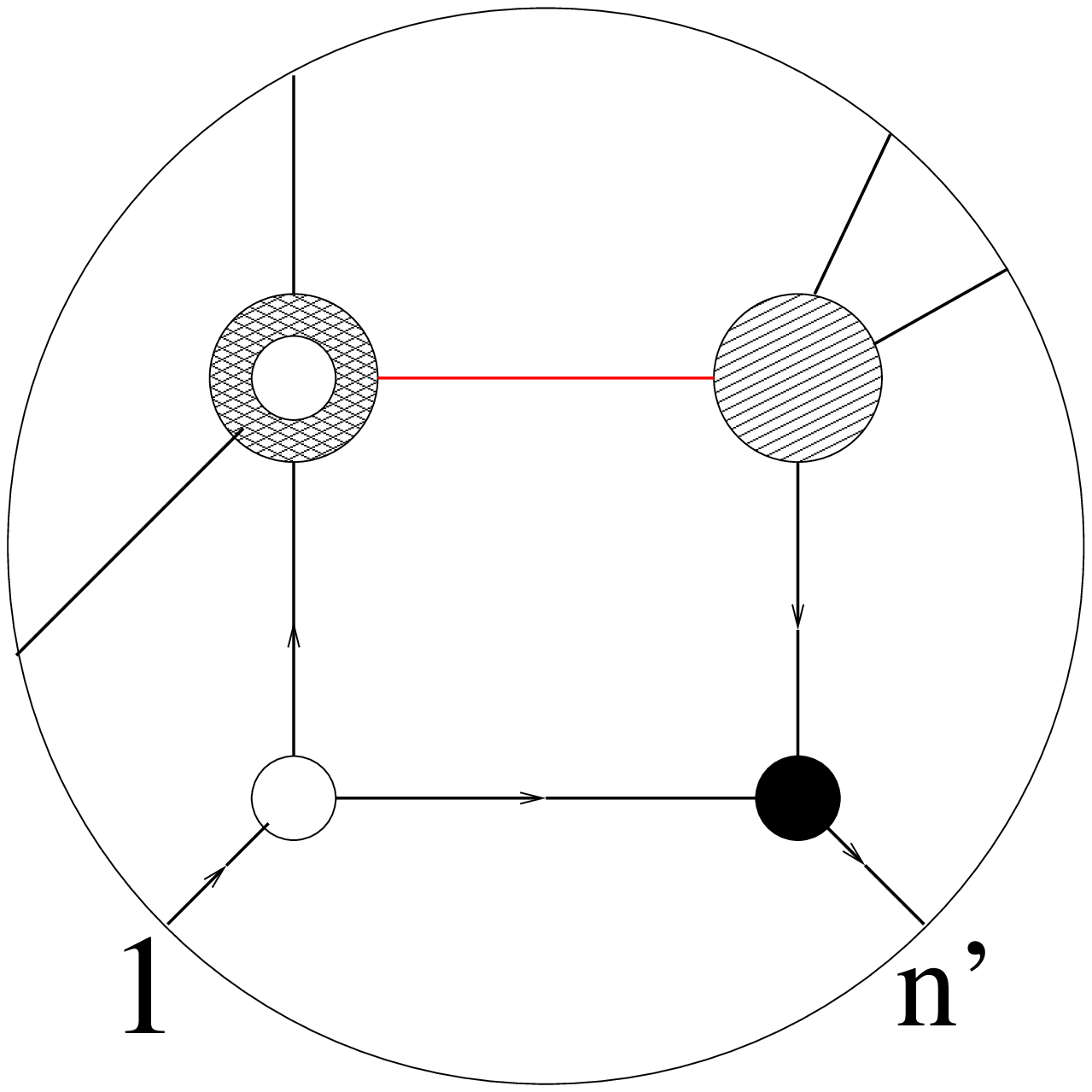}}}\:\Longrightarrow\:
 \raisebox{-1.3cm}{\scalebox{.20}{\includegraphics{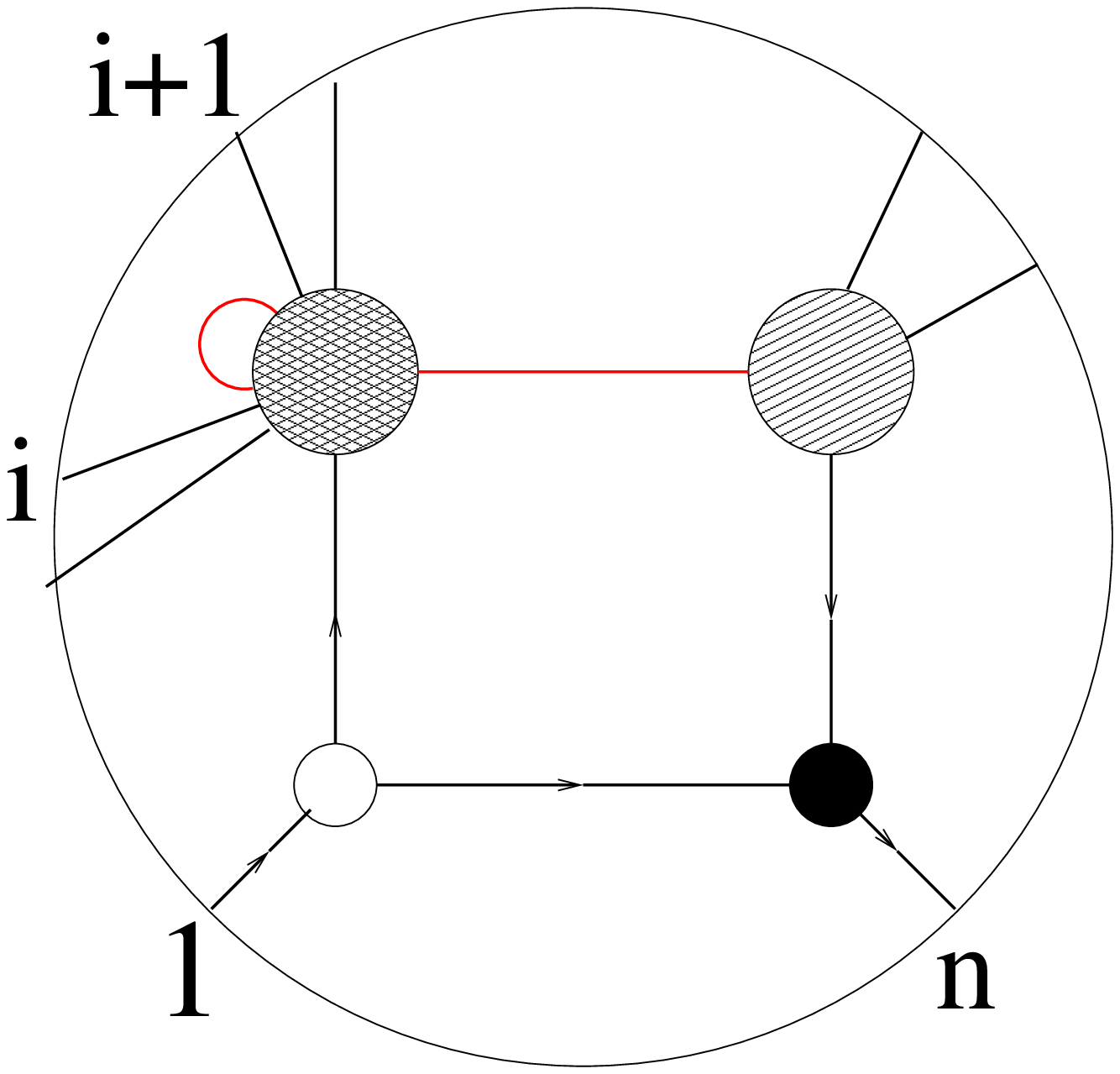}}}
\end{equation}
as well as 
\begin{equation}\eqlabel{eq:MnFwLm3}
 \raisebox{-1.3cm}{\scalebox{.20}{\includegraphics{1LMnFactFw.eps}}}\:\Longrightarrow\quad
 \raisebox{-1.3cm}{\scalebox{.20}{\includegraphics{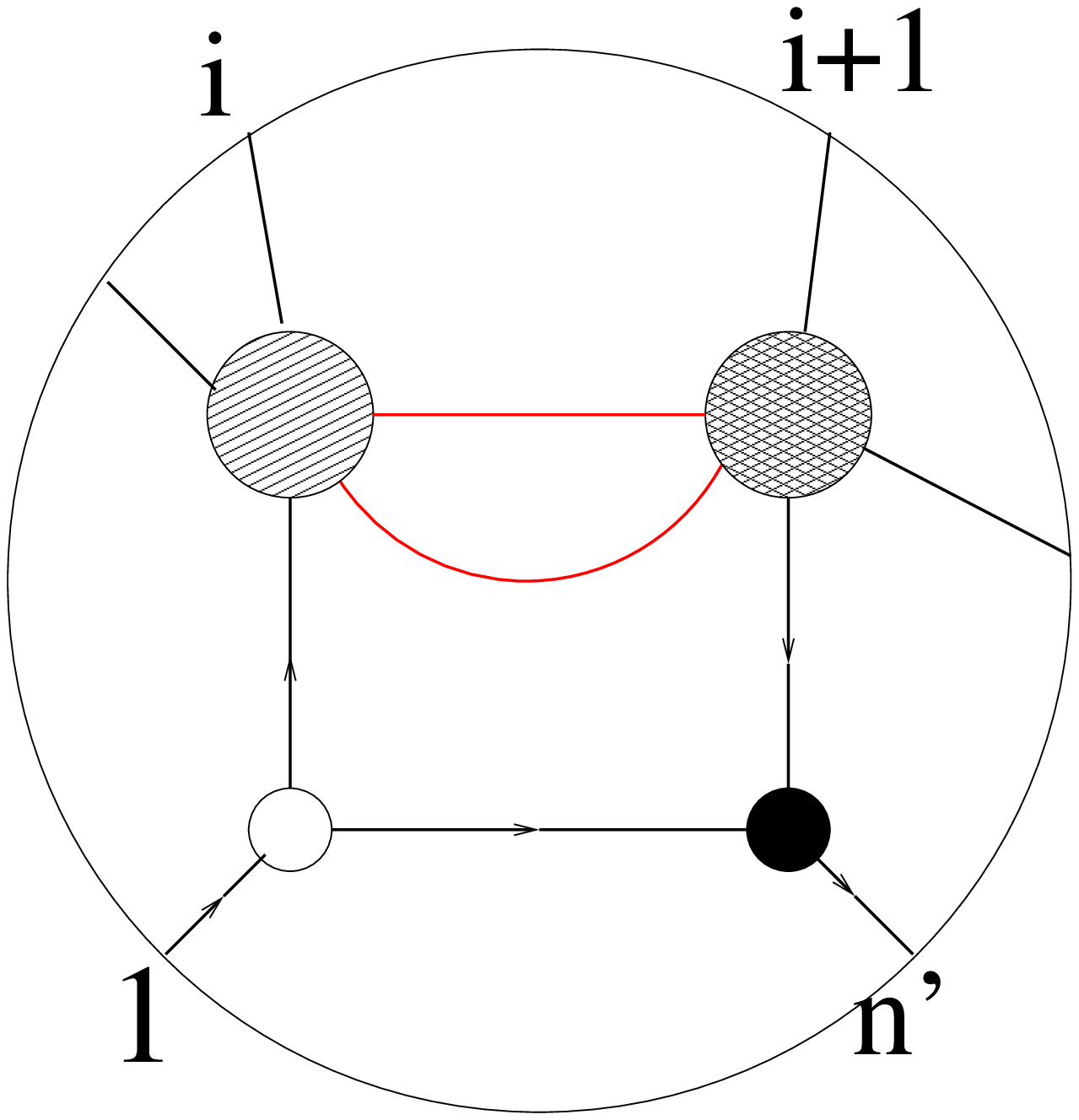}}}
\end{equation}
Thus, all the physical forward singularities are guaranteed by the induction hypothesis. Finally, 
in order to prove that there is no other type of forward-channel implied by the BCFW formula for arbitrary
$n'$ external states, we need to show that any other possible forward singularity is actually spurious and
thus it needs to cancel. The cancellation mechanism turns out to be quite similar to the one which eliminates
all the non-local poles for the factorisation channels, {\it i.e.} the same forward singularity is contained
by two different diagrams and its residue is the same but with different sign:
\begin{equation}\eqlabel{eq:MnFwLm4}
 \raisebox{-1.3cm}{\scalebox{.20}{\includegraphics{1LMnFactB.eps}}}\:\Longrightarrow\:
 \raisebox{-1.3cm}{\scalebox{.20}{\includegraphics{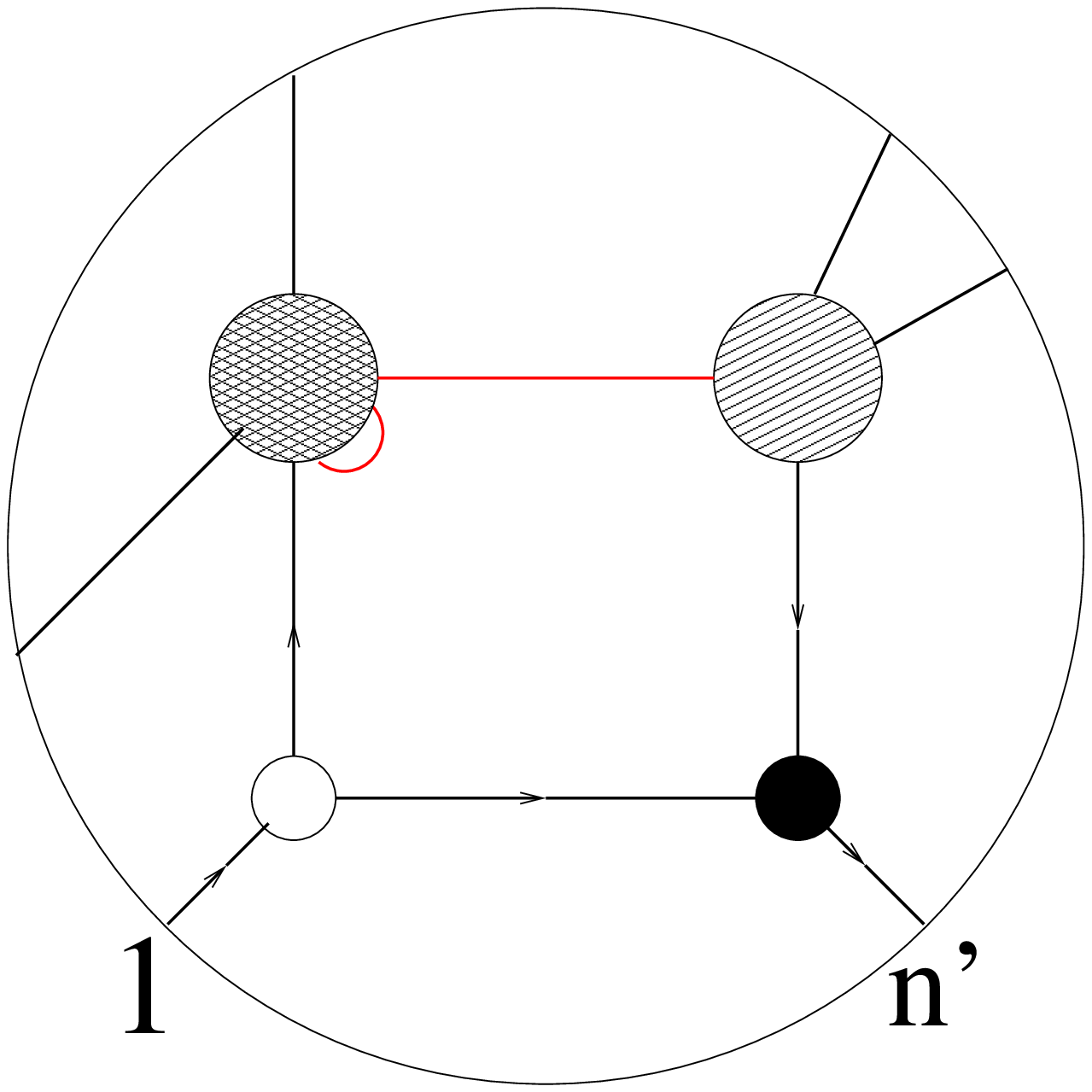}}}\:\Longleftarrow\:
 \raisebox{-1.3cm}{\scalebox{.20}{\includegraphics{1LMnFactFw.eps}}}
\end{equation}
and similarly for the other factorisation-channel where the tree-level and one-loop sub-amplitudes are
exchanged.

\subsubsection{One-loop integrand structure for pure Yang-Mills}
\label{subsubsec:1lRRpYM}

Differently from the supersymmetric theories, pure Yang-Mills turns out to have a richer structure, having both
rational terms and diagrams with bubbles on the external legs. As we discussed in the four-particle case at 
one-loop, one can generate these terms in the quasi-forward regularisation with a massive deformation.
Schematically, the on-shell diagrams representing a generic four-particle amplitude at one-loop can be
organised in three classes of contributions:
\begin{equation}\eqlabel{eq:1LpYMM4}
 \mathcal{M}_4^{\mbox{\tiny $(1L)$}}\:=\:
  \tilde{\mathcal{M}}_4^{\mbox{\tiny $(1L)$}}
  \,+\,
  \hat{\mathcal{M}}_{4}^{\mbox{\tiny $(1L)$}}(m^2)\,+\,
  \frac{1}{\epsilon}\sum_{r}
  \left[
   \tilde{\mathcal{M}}_{4,\,r}^{\mbox{\tiny $(1L)$}}+
   \hat{\mathcal{M}}_{4,\,r}^{\mbox{\tiny $(1L)$}}(m^2)
  \right]\epsilon^r
\end{equation}
where $\tilde{\mathcal{M}}_4$ resembles the on-shell diagram with only massless states and which is well-defined
in the forward limit and encodes the cut-constructible information, 
$\hat{\mathcal{M}}_{4}^{\mbox{\tiny $(1L)$}}(m^2)$ is such that it vanishes as
$m^2\,\longrightarrow\,0$ and encodes the information on the rational terms, and finally the last set of 
contributions come from the on-shell bubbles upon quasi-forward regularisation, with $\epsilon$ being the
regularisation parameter. All these sets of terms are actually on-shell four-forms, despite the fact that we refer 
to $\hat{\mathcal{M}}_{4}^{\mbox{\tiny $(1L)$}}(m^2)$ as encoding the rational terms. 
The sets of contributions which have been indicated with $\tilde{\mathcal{M}_4}$ and $\tilde{\mathcal{M}}_{4,r}$
contain all the information coming from considering just massless gluons in the forward lines, while
$\hat{\mathcal{M}_4}(m^2)$ and $\hat{\mathcal{M}}_{4,r}(m^2)$ take into account massive scalars only.
For all-plus helicity amplitudes just the second term is present, for the UHV ones the first term is absent, while 
all of them are present for MHV configurations. 

We can now proceed again via induction and take again \eqref{eq:MnIndHyp} as induction hypothesis, where {\it also}
massive scalars need to be considered as propagating in the forward lines. All the sub-amplitudes in the
factorisation channels and the forward term are understood to be regularised in the quasi-forward 
scheme\footnote{It is important here to remember that in this scheme we include both a quasi-forward deformation
such as \eqref{eq:qflim2} and \eqref{eq:MassQF}, and the completion of the recursive expression via multi-step
BCFW algorithm.}. The analysis of the singularity structure for the $n'$-particle amplitude at one-loop ($n'\,>\,n$)
goes exactly as for $\mathcal{N}\,=\,1,\,2$ in Section \eqref{subsubsec:1lRRN12}, showing that the
recursive relation contains all and only the physical factorisation channels as well as the forward singularities.
However, some comments are in order. First of all, the $n$-particle integrand resulting from the recursion relation
can be naturally reorganised according to the following structure
\begin{equation}\eqlabel{eq:1LpYMMn}
 \mathcal{M}_n^{\mbox{\tiny $(1L)$}}\:=\:
 \tilde{\mathcal{M}}_n^{\mbox{\tiny $(1L)$}}
  \,+\,
  \hat{\mathcal{M}}_n^{\mbox{\tiny $(1L)$}}(m^2)\,+\,
  \frac{1}{\epsilon}\sum_{r}
  \left[
   \tilde{\mathcal{M}}_{n,\,r}^{\mbox{\tiny $(1L)$}}+
   \hat{\mathcal{M}}_{n,\,r}^{\mbox{\tiny $(1L)$}}(m^2)
  \right]\epsilon^r
\end{equation}
where
\begin{equation}\eqlabel{eq:1LpYMMnCC}
 \tilde{\mathcal{M}}_n^{\mbox{\tiny $(1L)$}}\:=\:
 \sum_{k\in\mathcal{P}^{\mbox{\tiny $(1,n)$}}}
  \left[
   \tilde{\mathcal{M}}_{\mathcal{I}_k}^{\mbox{\tiny $(1L)$}}\otimes\mathcal{M}_{\mathcal{J}_k}^{\mbox{\tiny tree}}
   \:+\:
   \mathcal{M}_{\mathcal{I}_k}^{\mbox{\tiny tree}}\otimes\tilde{\mathcal{M}}_{\mathcal{J}_k}^{\mbox{\tiny $(1L)$}}
  \right]+\mathcal{M}_{n+2}^{\mbox{\tiny fw}}
\end{equation}
contains the contributions of the massless gluons in the forward lines which are well-defined from the start, and
encode the cut-constructible information;
\begin{equation}\eqlabel{eq:1LpYMMnRat}
 \hat{\mathcal{M}}_n^{\mbox{\tiny $(1L)$}}(m^2)\:=\:
  \sum_{k\in\mathcal{P}^{\mbox{\tiny $(1,n)$}}}
   \left[
    \hat{\mathcal{M}}_{\mathcal{I}_k}^{\mbox{\tiny $(1L)$}}(m^2)\otimes
    \mathcal{M}_{\mathcal{J}_k}^{\mbox{\tiny tree}}
    \:+\:
    \mathcal{M}_{\mathcal{I}_k}^{\mbox{\tiny tree}}\otimes
    \hat{\mathcal{M}}_{\mathcal{J}_k}^{\mbox{\tiny $(1L)$}}(m^2)
   \right]+\mathcal{M}_{n+2}^{\mbox{\tiny fw}}(m^2),
\end{equation}
contains the contributions of the massive scalars in the forward line which are well-defined from the start, and
encode the information on the rational terms; and
\begin{equation}\eqlabel{eq:1LpYMMnDiv}
 \begin{split}
  &\tilde{\mathcal{M}}_{n,r}^{\mbox{\tiny $(1L)$}}\:=\:
   \sum_{k\in\mathcal{P}^{\mbox{\tiny $(1,n)$}}}
    \left[
     \tilde{\mathcal{M}}_{\mathcal{I}_k,r}^{\mbox{\tiny $(1L)$}}\otimes\mathcal{M}_{\mathcal{J}_k}^{\mbox{\tiny tree}}
     \:+\:
     \mathcal{M}_{\mathcal{I}_k}^{\mbox{\tiny tree}}\otimes\tilde{\mathcal{M}}_{\mathcal{J}_k}^{\mbox{\tiny $(1L)$}}
    \right]+\mathcal{M}_{n+2,r}^{\mbox{\tiny fw}},\\
  &\hat{\mathcal{M}}_{n,r}^{\mbox{\tiny $(1L)$}}(m^2)\:=\:
   \sum_{k\in\mathcal{P}^{\mbox{\tiny $(1,n)$}}}
    \left[
     \hat{\mathcal{M}}_{\mathcal{I}_k,r}^{\mbox{\tiny $(1L)$}}(m^2)\otimes\mathcal{M}_{\mathcal{J}_k}^{\mbox{\tiny tree}}
     \:+\:
     \mathcal{M}_{\mathcal{I}_k}^{\mbox{\tiny tree}}\otimes\hat{\mathcal{M}}_{\mathcal{J}_k,r}^{\mbox{\tiny $(1L)$}}(m^2)
    \right]+\mathcal{M}_{n+2,r}^{\mbox{\tiny fw}}(m^2)
 \end{split}
\end{equation}
are understood to have been completed via multi-step BCFW algorithm, and 
encodes the $(r-1)$-terms in the quasi-forward expansion and, in principle, takes contributions from both the
massless gluons and the massive scalars in the (quasi)-forward lines. In all the formulae
\eqref{eq:1LpYMMnCC}, \eqref{eq:1LpYMMnRat} and \eqref{eq:1LpYMMnDiv}, $\mathcal{P}^{\mbox{\tiny $(1,n)$}}$
represents the set of factorisation channels singled out by the BCFW bridge in the $(1,n)$-channel, 
$\mathcal{I}_k$ and $\mathcal{J}_k$ are such that $\mathcal{I}_k\,\cup\,\,\mathcal{J}_k\,=\,\{2,\ldots,n-1\}$
and represent all the possible partitions of $\{2,\ldots,n-1\}$ in $\mathcal{P}^{\mbox{\tiny $(1,n)$}}$, and
finally the apex ${}^{\mbox{\tiny (fw)}}$ indicates the terms coming from the highest-degree forward term in the 
recursion relation.

It is easy to notice that the overall recursion relation is actually a direct sum of three recursion relations,
one for each ``sector'': \eqref{eq:1LpYMMnCC} provides a recursion relation for the cut-constructible terms,
\eqref{eq:1LpYMMnRat} for the ``rational'' ones and finally \eqref{eq:1LpYMMnDiv} for the on-shell bubbles --
Notice that, in principle, \eqref{eq:1LpYMMnDiv} contains an order $\mathcal{O}(\epsilon^0)$ term. However, it
does not contain cut-constructible information and can be easily isolated, as we did, because it is sourced by
those on-shell diagrams which are ill-defined before quasi-forward regularisation.


\subsection{Higher loop integrands}\label{subsec:HL}

In order to extend the previous analysis to higher loops, at least for $\mathcal{N}\,=\,1,\,2$, we need to 
ensure that the potentially problematic terms keep being (at least) of order $\mathcal{O}(\epsilon)$ at all loops. 
As far as pure Yang-Mills is concerned, the nice structure \eqref{eq:1LpYMMn} observed at one loop does not 
hold in general at higher loops: the terms of order $\mathcal{O}(\epsilon)$ can combine with the 
$\mathcal{O}(\epsilon^{-1})$ ones giving rise to physically meaningful contributions and thus mixing the various 
sectors which appear instead decoupled at one-loop level.

The potentially problematic terms arise from the following structures
\begin{equation}\eqlabel{eq:ProbTerms}
 \raisebox{-1.3cm}{\scalebox{.20}{\includegraphics{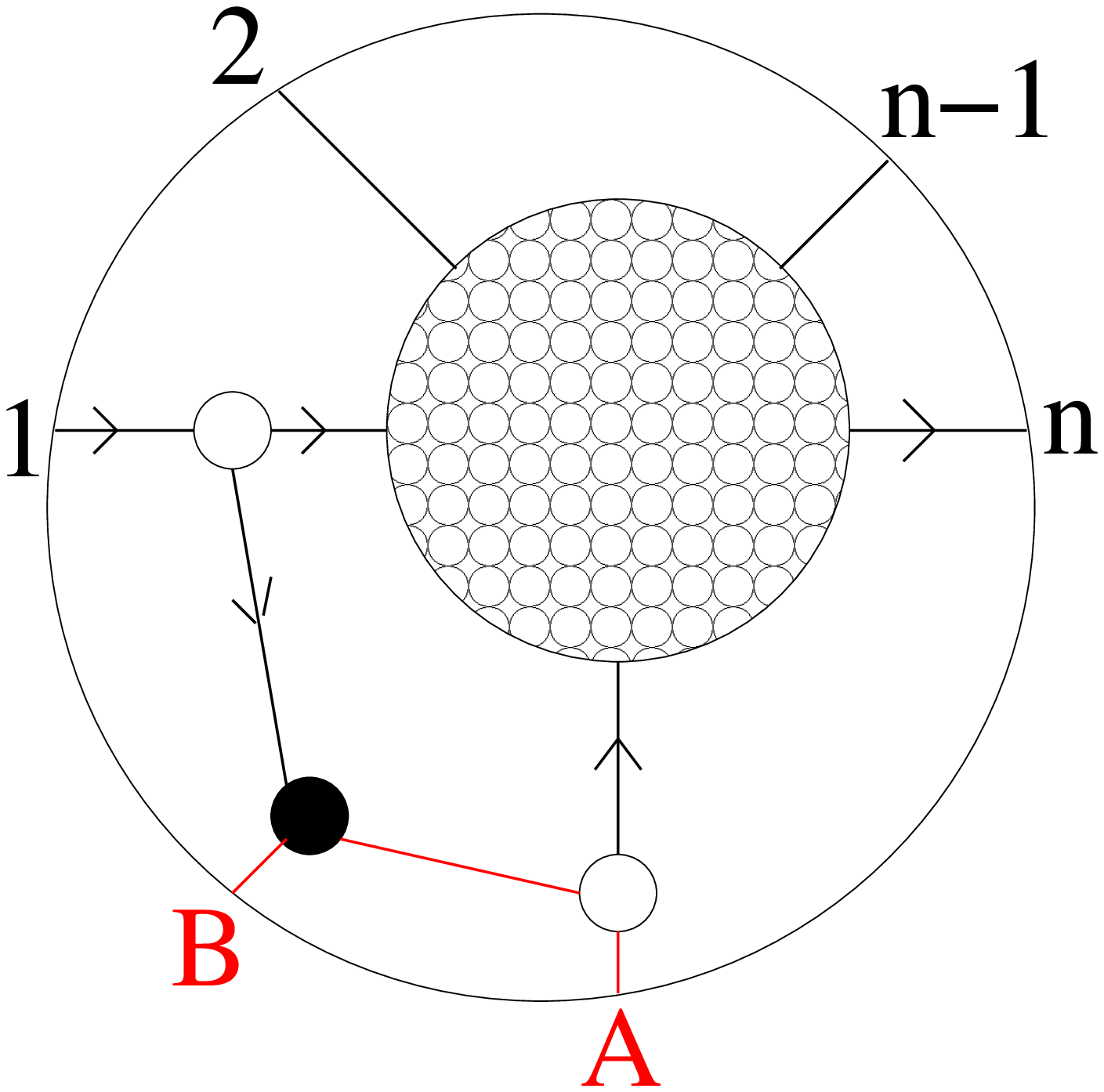}}}
 \hspace{2cm}
 \raisebox{-1.3cm}{\scalebox{.20}{\includegraphics{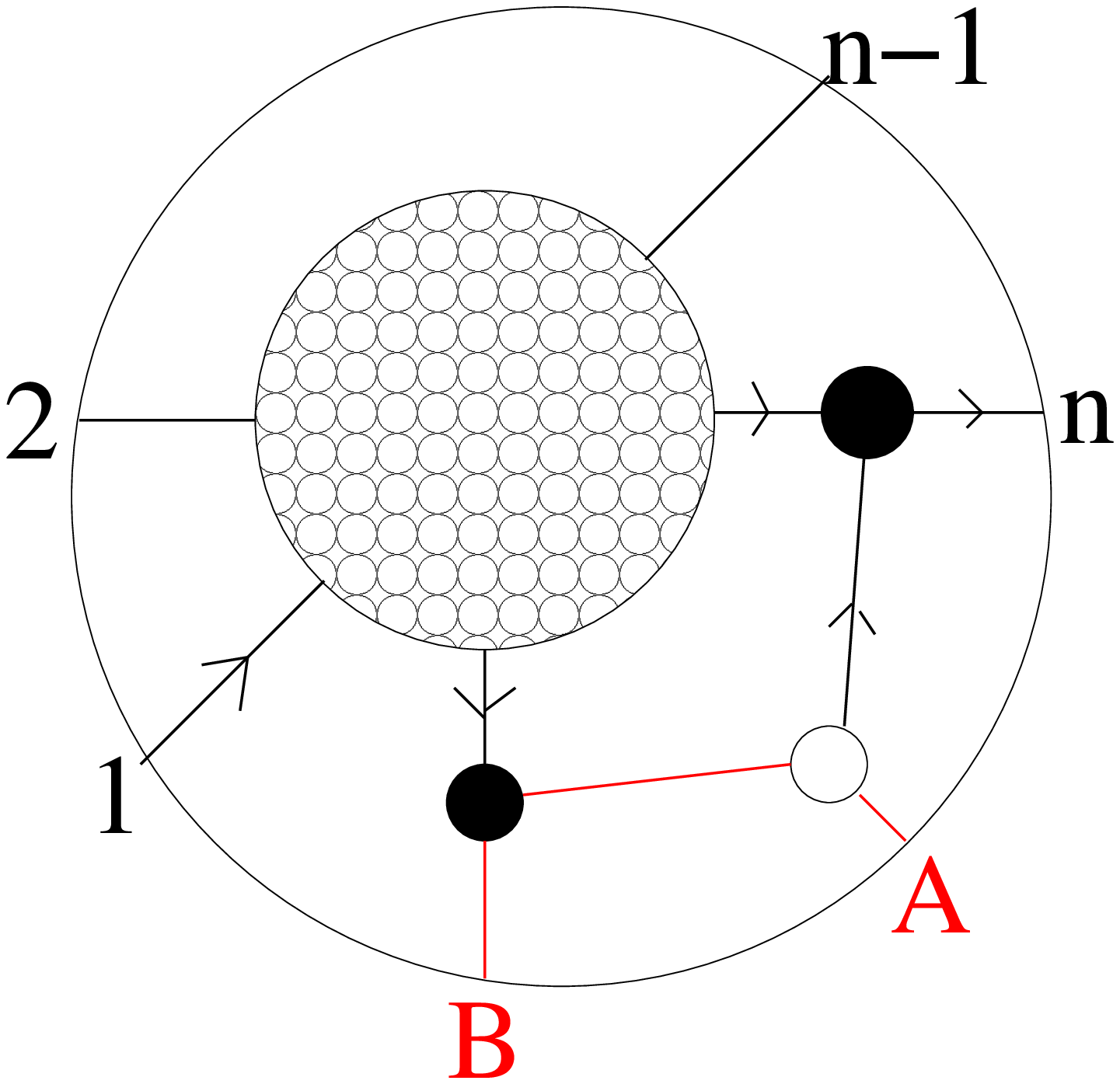}}}
\end{equation}
where the big blobs represent $(n+1)$-particle amplitudes at $(L-1)$-loop order, and the red lines labelled by
$A$ and $B$ are the ones which are taken to be forward. Notice that these two diagrams share the same non-local
pole
\begin{equation*}
 \raisebox{-1.3cm}{\scalebox{.20}{\includegraphics{HighLoopDivs3.eps}}}
 \:\Longrightarrow\:
 \raisebox{-1.3cm}{\scalebox{.20}{\includegraphics{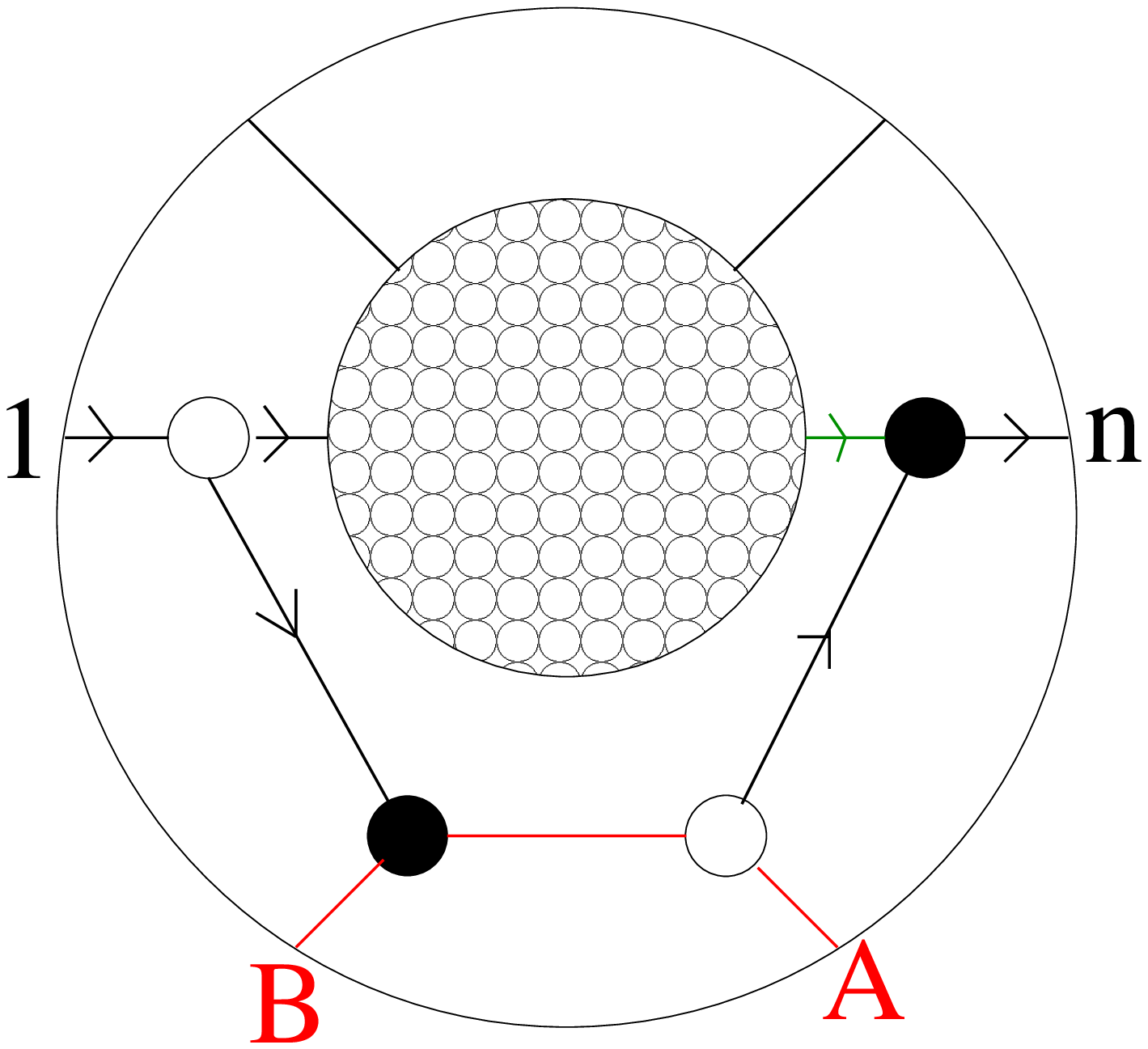}}}
 \:\Longleftrightarrow\:
 \raisebox{-1.3cm}{\scalebox{.20}{\includegraphics{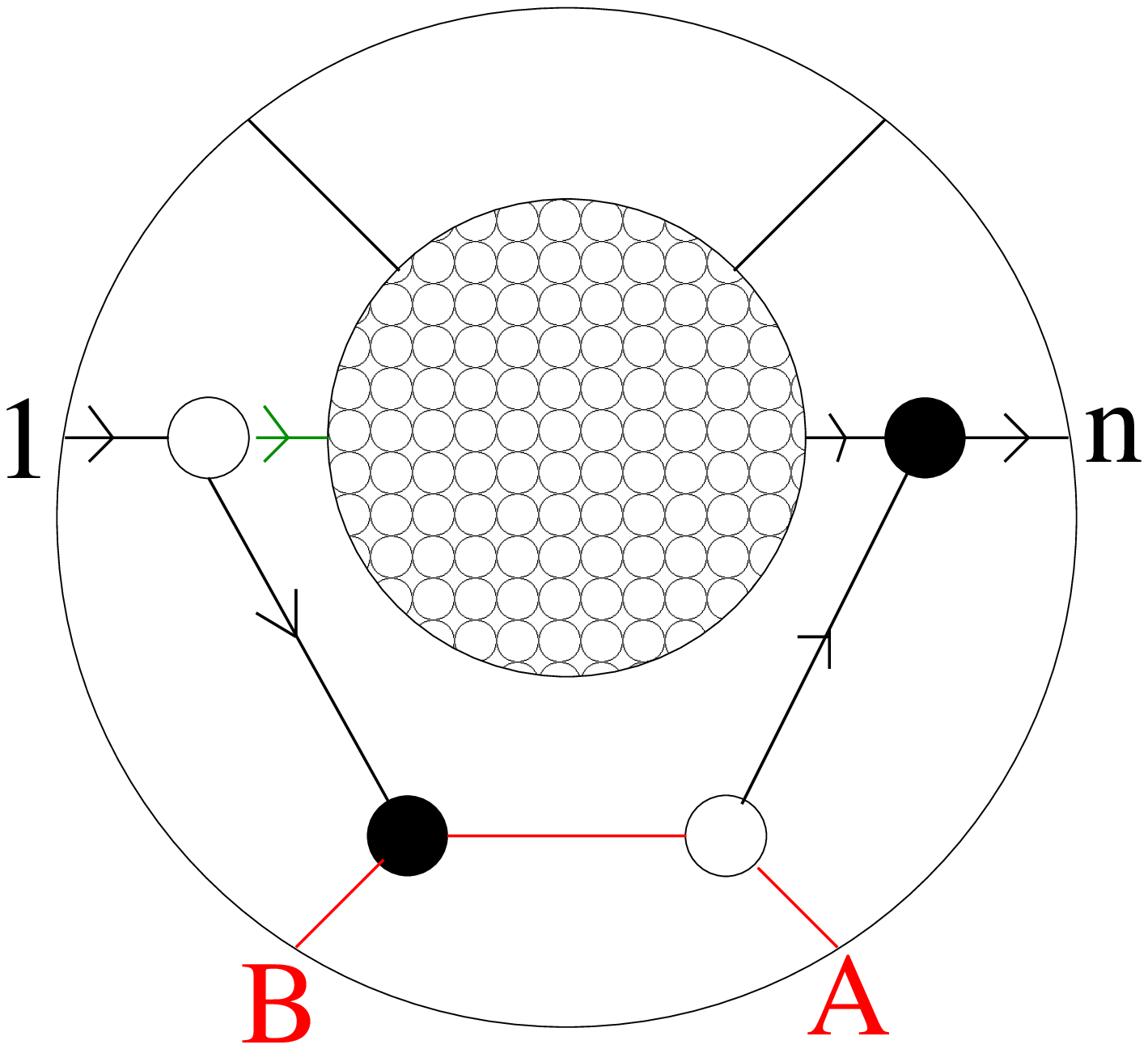}}}
 \:\Longleftarrow\:
 \raisebox{-1.3cm}{\scalebox{.20}{\includegraphics{HighLoopDivs.eps}}}
\end{equation*}
where the green line emphasises how such factorisations arise. This singularity disappears upon summation.
In the quasi-forward regularisation for the individual diagrams above, the pole is mapped into a pole in the
regularisation parameter. Thus, when the two quasi-forward diagrams are summed, the pole in $\epsilon$ related
to such a channel disappears as well: The sum of the two diagrams in \eqref{eq:ProbTerms} is better behaved than
the individual diagrams. Also, in this limit two intermediate legs become soft -- of order $\mathcal{O}(\epsilon)$ 
-- so that the $(n+1)$-particle sub-amplitude at $(L-1)$-loop factorises in a soft factor of order 
$\mathcal{O}(\epsilon^{-2})$ and an $n$-particle sub-amplitude at $(L-1)$-loop. Therefore, at the leading order in 
the $\epsilon$-expansion:
\begin{equation}\eqlabel{eq:ProbTermsQF}
  \raisebox{-1.3cm}{\scalebox{.20}{\includegraphics{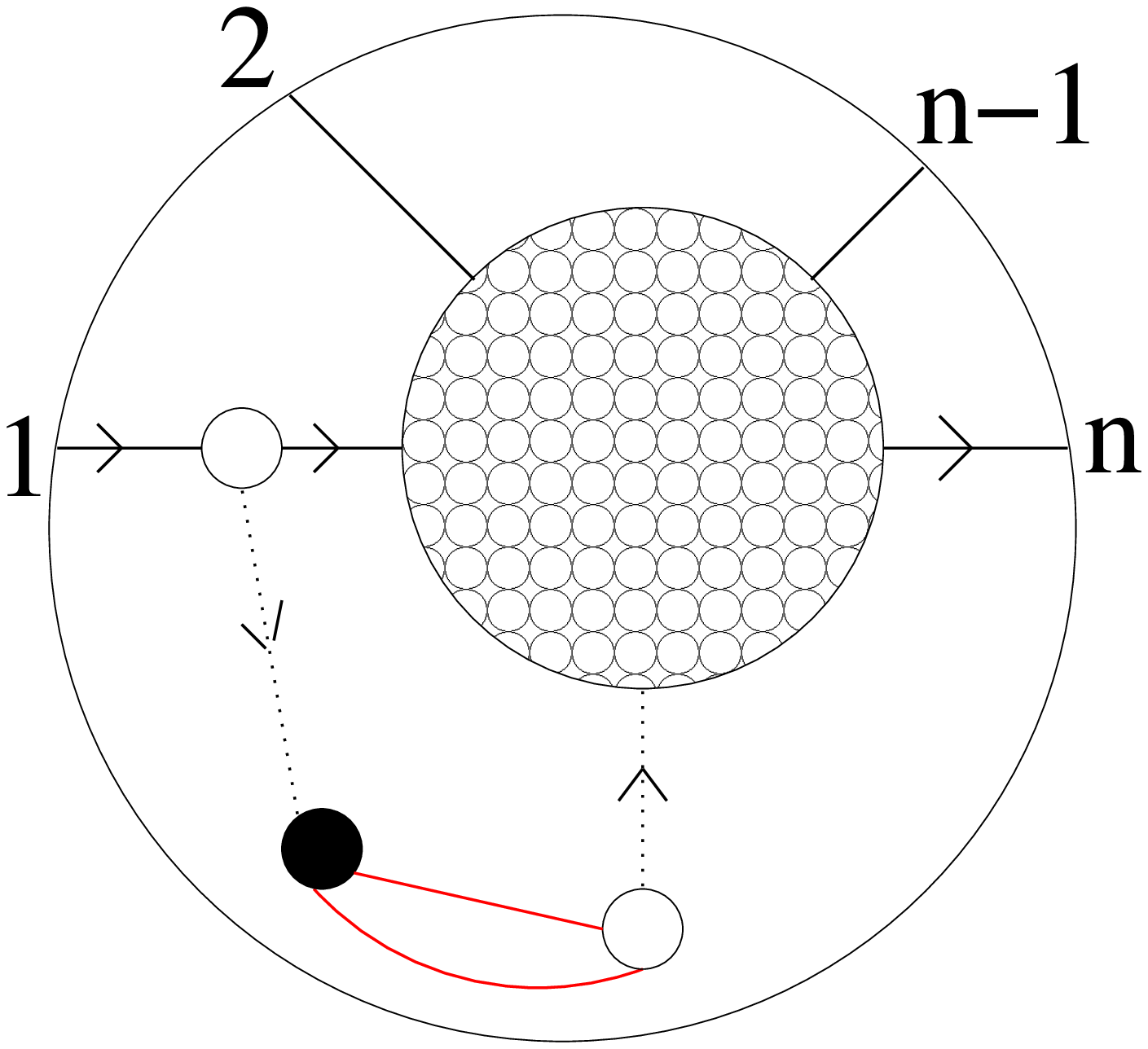}}}\:+\:
  \raisebox{-1.3cm}{\scalebox{.20}{\includegraphics{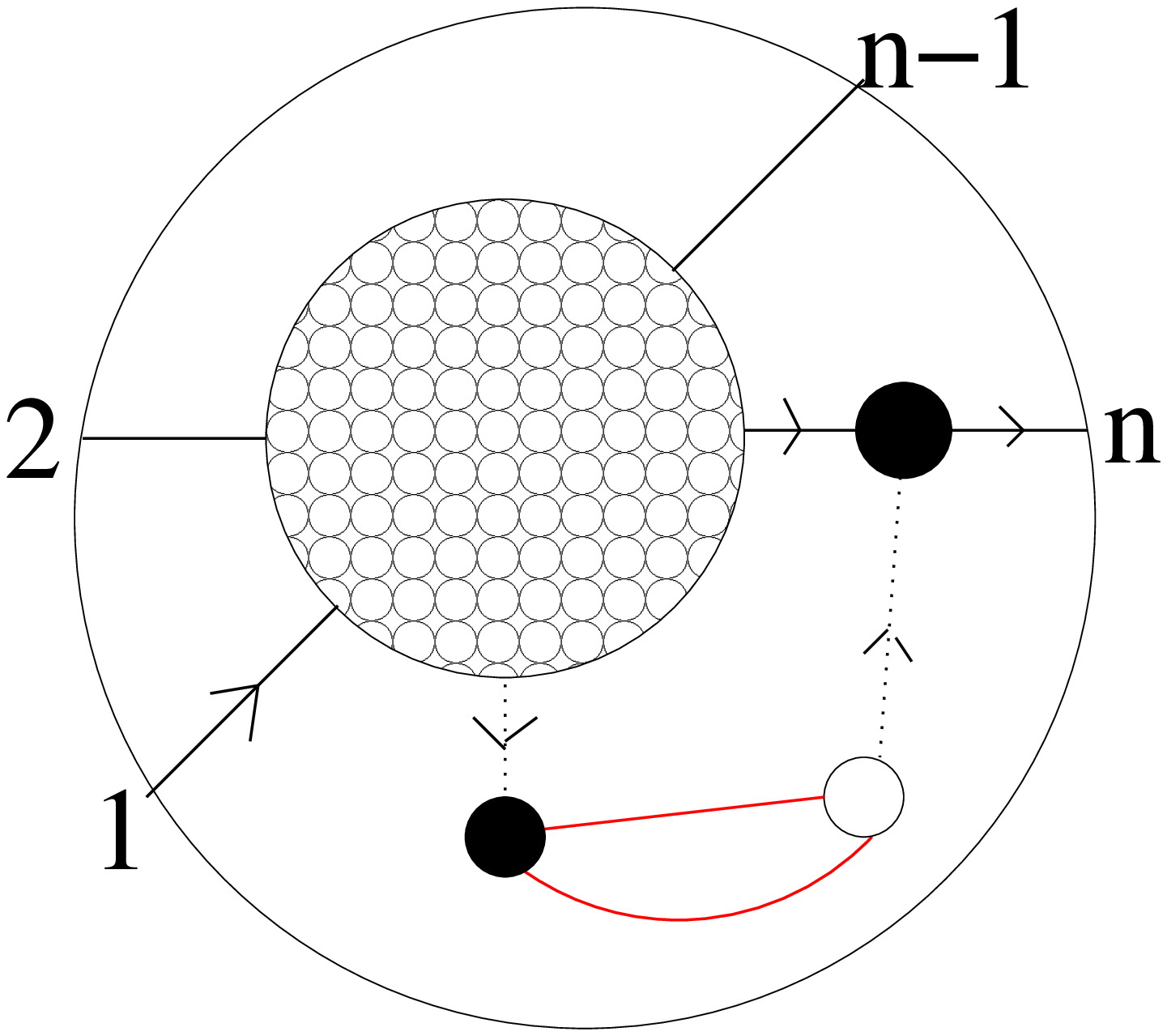}}}\:\sim\:
  \epsilon^{\mathcal{N}-2}\times\left[1+(-1)^{\mathcal{N}}\right]
\end{equation}
where the dotted internal lines represent the soft states. Notice that the factor $\epsilon^{\mathcal{N}}$ comes
from the integration over the component of a single multiplet, the term in the square brackets represents the
sum over the two multiplets, the extra powers $\epsilon^{-2}$ is a result of a collinear and a soft singularity 
-- the cancellation of the non-local pole discussed above is already taken into account.

Upon completion via the multi-step BCFW algorithm/symmetry, also diagrams of the following type are introduced
\begin{equation}\eqlabel{eq:ProbTermsQF2}
 \raisebox{-1.3cm}{\scalebox{.20}{\includegraphics{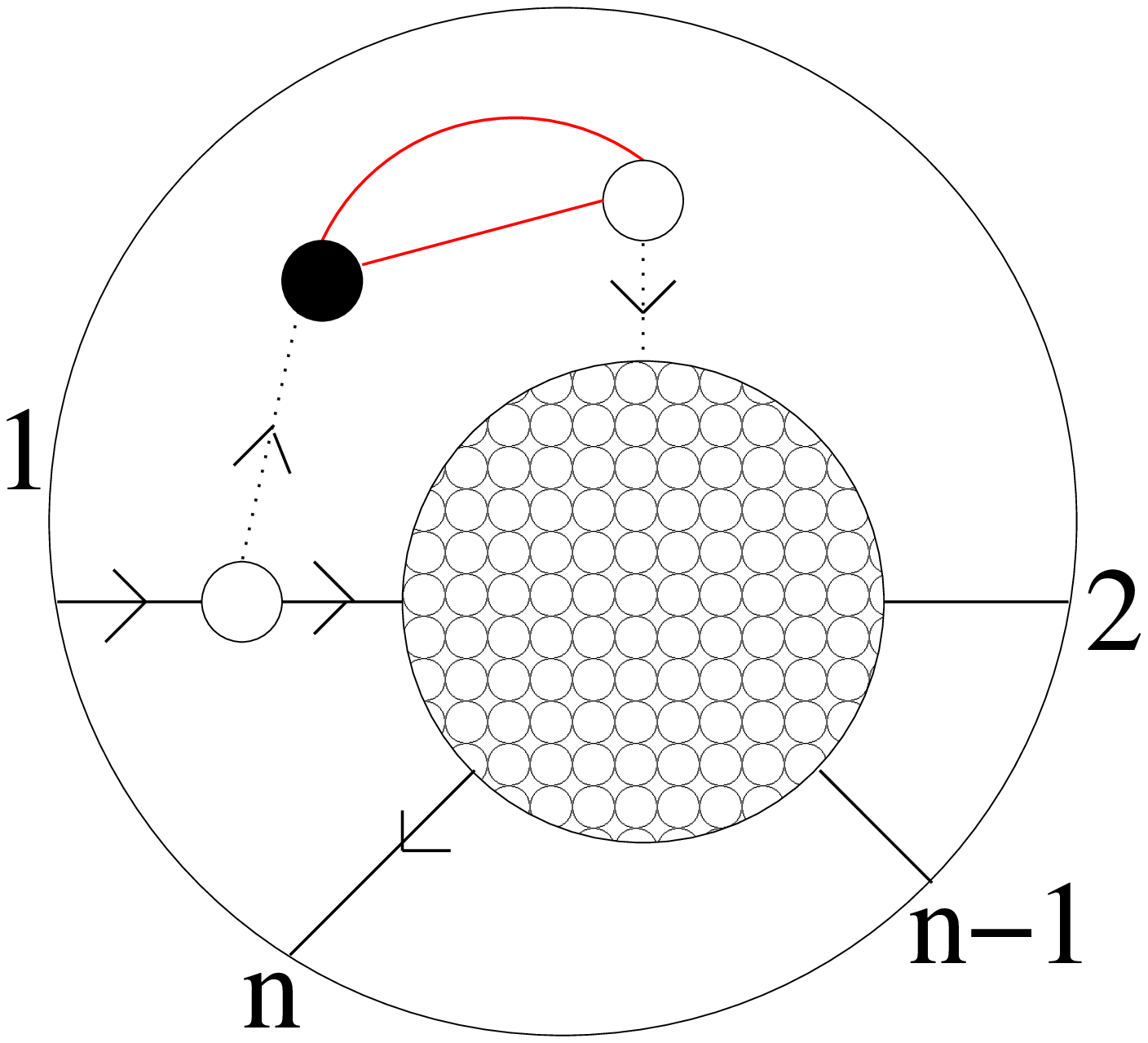}}}\hspace{2cm}
 \raisebox{-1.3cm}{\scalebox{.20}{\includegraphics{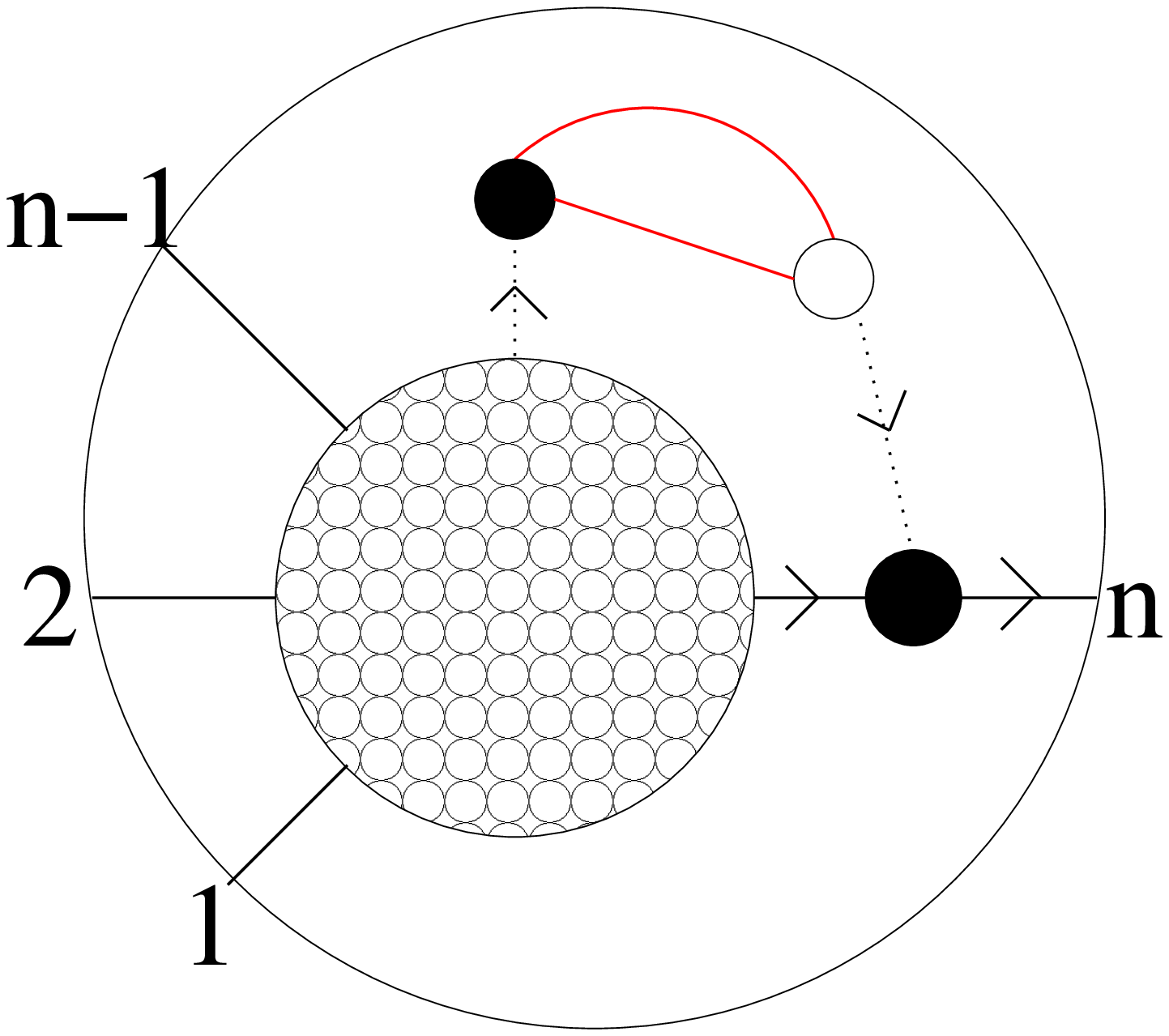}}}
\end{equation}
which share a forward singularity with \eqref{eq:ProbTerms}. The overall behaviour gets therefore enhanced of
one power to $\mathcal{O}(\epsilon)$ for supersymmetric theories and 
$\mathcal{O}(\epsilon^{-1})$ for pure Yang-Mills. Notice that this is the same mechanism which we 
described in detail at one-loop. Furthermore, in a hypothetical recursion relation at higher loops, one can find 
factorisation terms such as
\begin{equation*}
 \raisebox{-1.3cm}{\scalebox{.20}{\includegraphics{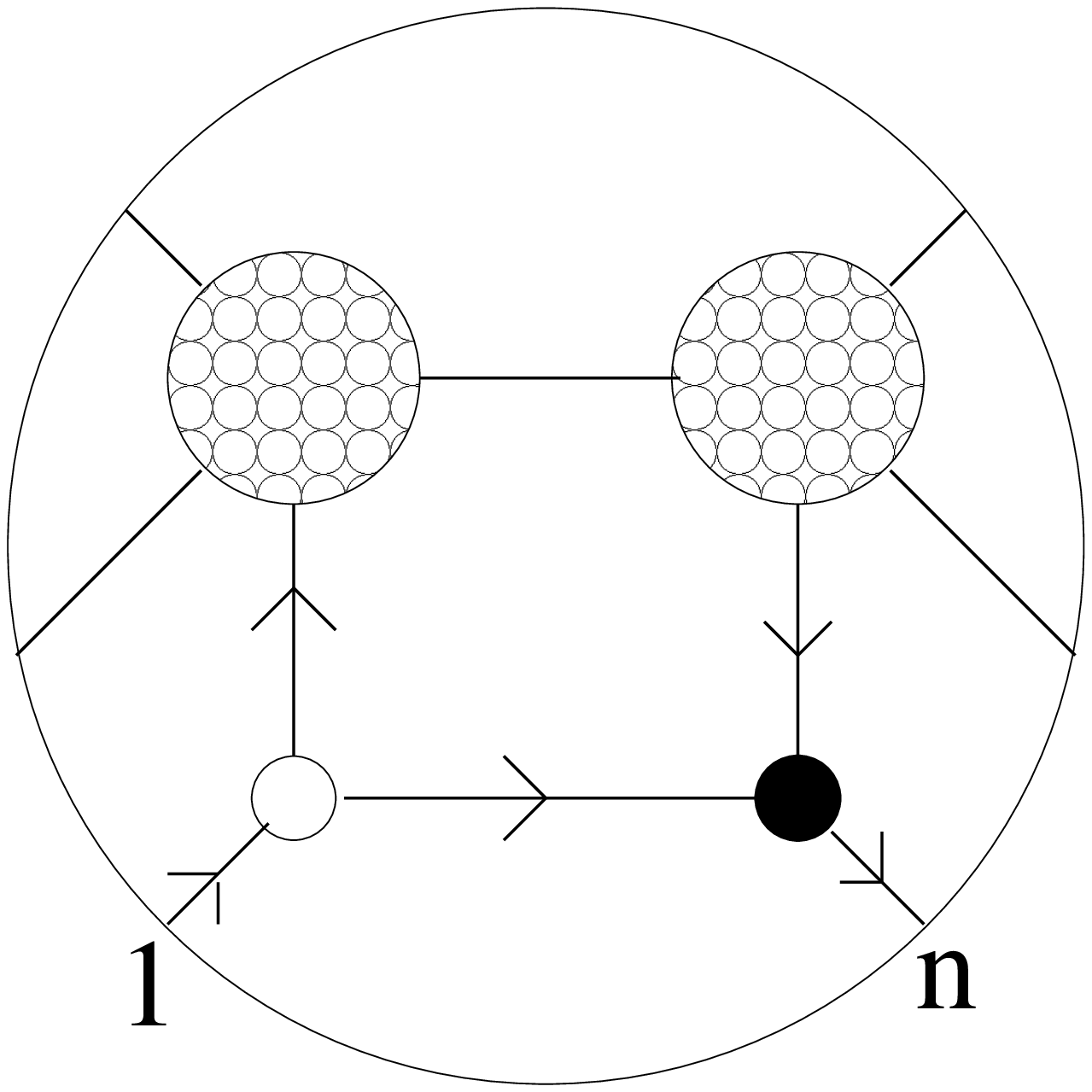}}}
\end{equation*}
where the two sub-amplitudes are at $l$- and $(L-l)$-loop. In the supersymmetric case, the products of the
two sub-amplitudes in a factorisation channel produces terms which are at least of order 
$\mathcal{O}(\epsilon^{0})$ which are just given by the produces of those terms which are well-defined from the
start in the forward limit. Therefore, in the supersymmetric case, one can apply to all loops the same inductive 
reasoning used to prove the validity of the recursion relations at tree- and one-loop level, given that
no pole in the regularisation parameter can be generated.

Different is instead the case of pure Yang-Mills which shows already at one-loop a pole in the regularisation
parameter. In principle, if one tries to glue two loop amplitudes in pure Yang-Mills, one could generate higher
order poles in the regularisation parameters as well as physically meaningful contribution can arise from
the product of the singular term of one sub-amplitude and the $\mathcal{O}(\epsilon)$ term of the other 
sub-amplitude, coming from the contribution which is well-defined from the very beginning.


 
\section{Conclusion}\label{sec:Concl}

In this paper we discussed the on-shell diagrammatics for less/no-supersymmetric theories and the possibility
of representing the (integrand of the) amplitudes in terms of on-shell processes at all loop orders. 

For the class of theories we are interested in, the physical states can be packed into two coherent multiplets 
which, grouping states with the same helicity sign, can be labelled by the related helicity sign itself. From a
diagrammatic point of view, this is reflected in a direction assignment to the lines in the diagrams. With such
a prescription, the three-particle amplitudes are endowed with two incoming and one outgoing helicity arrows
if they are holomorphic, and one incoming and two outgoing if they are anti-holomorphic. As a consequence, any
more complicated on-shell process built by suitably gluing them is characterised by a perfect orientation, which
actually has a physical meaning. First of all, the presence of a physical perfect orientation restricts the 
equivalence operations on a diagram. More precisely, while the merger/expansion involving two three-particle 
amplitudes of the same type holds as in the maximally supersymmetric case, the square move is admitted for
a specific helicity flow structure: it holds if and only if the on-shell box shows helicity flows between external
state and one of them goes around the full box. The path of the helicity flows in the on-shell box is a reflection
of its singularity structure. In the case just mentioned, the existence of helicity flows between external states 
and with one of them characterised by a particular path is the diagrammatic codification of the fact that the
on-shell diagram in question undergoes a complex factorisation under both limit in a channel, while in the other
just one of the two complex factorisations is allowed. Such an on-shell diagram is just a representation of
a tree-level four-particle amplitude with the same helicity states being consecutive.

In the case the coherent states with the same helicity are not consecutive, the on-shell box is characterised
either by direct helicity flows between consecutive states or by two helicity flows between consecutive states
in a given channel and an internal helicity loop. The presence of the latter is a manifestation of the presence
of a higher order pole. When an internal helicity loop is admitted, both orientation 
(clockwise and counter-clockwise) are allowed: this correspond to the fact that one has to sum over both the
multiplets. Interestingly, in on-shell diagrams contributing to a loop integrand (what we have called
on-shell $4L$-forms), the presence of such internal helicity loops in the interior of the diagram is 
related to the presence of a more complex structure than the $d\log$ which appear in the maximally supersymmetric
theory. In particular, it indicates the presence of UV divergencies.

The existence of these equivalence relations implies that also for the decorated on-shell diagrams there is 
a degree of redundancy, and equivalence classes can be defined. In the context of $\mathcal{N}\,=\,4$ SYM,
the equivalence classes are defined via (decorated) permutations. In the less/no-supersymmetric case, they
are defined combining permutations with the helicity flows. Interestingly, a given permutation can contain
different equivalence classes, which is the statement that such equivalence classes are related to
each other by Ward identities. More generically, given a certain on-shell diagram with the number of sink and
sources fixed ({\it i.e.} the diagram belongs to a fixed N${}^k$MHV-sector) and a given perfect orientation, one
can obtain any other perfect orientation by helicity flows reversal. Such an operation produces a Jacobian:
this is the statement that, once the N${}^k$MHV-sector is fixed, the on-shell processes can be related by Ward
identities as well.

The on-shell diagrams so defined allows to provide a representation for scattering amplitudes also for 
$\mathcal{N}\,\le\,2$ (S)YM theories. Making this statement more precise, for $\mathcal{N}\,=\,1,\,2$
it is possible to prove via induction that the on-shell diagrams provide a full-fledge representation for the
integrand at all-loops, and such a representation relies on the possibility of reconstructing the integrand
of the amplitudes from factorisation and forward singularities. For an individual diagram, the forward limit
is well defined just for $\mathcal{N}\,=\,3,\,4$, because of the presence of non-local poles: While supersymmetry
is enough to kill both local and non-local divergencies in $\mathcal{N}\,=\,3,\,4$ SYM theories, it is not the case 
for $\mathcal{N}\,\le\,2$. The problematic terms have the topology of a lower-level amplitude with one of the states
softly connected to an on-shell bubble. Some of these non-localities which characterise such diagrams cancel upon 
summation of the on-shell diagrams in a given representation. This can be seen explicitly by treating the forward 
limit through the introduction of a quasi-forward deformation such that both the on-shell condition and momentum 
conservation are preserved. In this way the terms which become singular in the forward limit are mapped into poles 
in the deformation parameter: One can then explicitly see how the sum over the potentially problematic terms becomes
of order $\mathcal{O}(\epsilon^0)$ for $\mathcal{N}\,=\,1,\,2$ and $\mathcal{O}(\epsilon^{-2})$ for pure Yang-Mills.
However, we observed that, while any BCFW bridge captures the so-called cut-constructible part of the amplitude,
it does not capture all the inequivalent diagrams with external on-shell bubbles. The missing terms can be
reconstructed either via a multi-step BCFW algorithm or, equivalently, via symmetry. Once also these terms are added,
the behaviour improves to $\mathcal{O}(\epsilon)$ for $\mathcal{N}\,=\,1,\,2$ and $\mathcal{O}(\epsilon^{-1})$ for 
pure Yang-Mills. While for the supersymmetric case this is an all-loop statement and, thus, one can recursively
reconstruct the integrand at all loop in terms of on-shell processes, with the only condition that
no BCFW bridge has an internal helicity flow, for pure Yang-Mills this holds only at one loop. Staying at one loop,
in order to obtain the terms related to the so-called rational terms, we introduce a mass-deformation in
the forward lines and, in particular, considered the contribution of a massive scalar. The quasi-forward 
regularisation, together with the mass-deformation, allows 
to reconstruct the integrand containing the cut-constructible information as well as the rational terms and the
divergent pieces (which should vanish upon integration).

However, the treatment of pure Yang-Mills needs to be considered just preliminary. First of all, the way that
we obtained the part of the integrand which encodes the rational terms is {\it ad hoc}: In a sense, we mimic
what happens in dimensional regularisation where, dividing the $(4-2\epsilon)$-dimensional loop momentum
in a direct product of a $4$-dimensional and a $-2\epsilon$-dimensional one, a sort of effective mass is generated.
Furthermore, we also uses the ``decomposition'' of the gluon into a $\mathcal{N}\,=\,4$ component, $4$  
$\mathcal{N}\,=\,1$ components and a scalar, with now the massive scalar being the only contribution to the
non-cut-constructible part. Furthermore, in this way the bipartite nature of the on-shell diagrammatic 
gets broken. It would be desirable to have a regularisation procedure at integrand level which can take care
of the forward limit and generate the ``rational terms'' at once. One possibility might be to generalise
our quasi-forward deformation in such a way that rather than preserving the on-shell condition of the
forward lines, would make them square to some new parameter $m^2$. However, also this prescription can be 
applied only to those diagrams that can be already defined (before the forward limit is taken), while
it will not be able to generate those amplitudes such as the all-plus and UHV ones.
It is therefore compelling to find
a suitable regularisation scheme for the on-shell diagrams which can preserve as many of their properties as 
possible. One the regularisation is under control, it is tempting to try to face the renormalisation issue and 
formulate a sort of on-shell renormalisation group.

\section*{Acknowledgement}

It is a pleasure to thank the CERN theory division for hospitality while this work was in progress as 
well as Freddy Cachazo, David Gordo and Henrik Johansson for insightful discussions and the organisers of the 
{\it Iberian Strings 2015} workshop where some results were presented. I would like also to thank the MiTP for
hospitality and partial support during the workshop {\it Stringy geometry}. A special acknowledgement is for the
Theory Group at the Universidade de Santiago de Compostela for hospitality during the final stage of this 
work. I would also like to thank the developers of SAGE \cite{sage}, Maxima \cite{maxima} and 
Xfig \cite{xfig}. This work is supported in part by Plan Nacional de Altas Energ{\'i}as (FPA2012-32828), 
and the Spanish MINECO's Centro de Excelencia Severo Ochoa Programme under grants SEV-2012-0249.

\appendix

\section{BCFW-{\it like} integrands from integral basis at one loop}
\label{app:BCFWinteg}

According to the common wisdom on loop amplitudes, they are computed
by looking for a scalar integral basis and using {\it generalised
unitarity} to compute their coefficients 
\cite{Bern:1994zx, Bern:1994cg, Bern:1995db, Bern:1996je, Bern:1996ja, Bern:1997sc, Bern:2004cz}
which are just rational functions of the Lorentz invariants. At
one loop, such a basis can either be determined by the 
Passarino-Veltman reduction 
\cite{Passarino:1978jh, vanNeerven:1983vr, Bern:1993kr}, or
also by general arguments on the amplitude singularity structure
\cite{ArkaniHamed:2008gz}: In a four-dimensional space-time and
for fixed external momenta, the leading singularity is determined
by sending on-shell four internal propagators in complexified
momentum space. Thus, a putative scalar integral basis of a 
one-loop amplitude can be thought to be given by the sum of all
those scalar $k$-gon integrals with $k\,\le\,4$. One can actually
proceed iteratively by writing down the one-loop amplitude as a sum of all the scalar 
boxes reproducing all the leading singularities.
Then, one can perform a triple cut on this ansatz. If it provides
a consistent result, the ansatz is complete. Otherwise one needs
to add all the scalar triangles which allows the ansatz to be
consistent with the triple cuts. Then, one can check this corrected
ansatz by performing a double cut. Again, if the ansatz turns out
to be consistent with the double cut, the basis can be considered
as determined. Otherwise, it would be necessary to add all those
bubble integrals which are consistent with the double cuts.
Actually, in this case, one would need also to add a rational
term which cannot be determined by $k$-cuts with $k\,\ge\,2$.
Those terms can be actually fixed if one implements generalised
unitarity in dimensional regularisation as zeroth-order terms
in the regulator expansion \cite{Bern:1995db, Brandhuber:2005jw, 
Anastasiou:2006jv, Britto:2006fc, Anastasiou:2006gt, Britto:2008vq,
Britto:2008sw, Feng:2008ju, Badger:2008cm, NigelGlover:2008ur}.

Therefore, the minimal scalar integral basis for a general 
one-loop amplitude is given by
\begin{equation}\eqlabel{eq:PVbasis}
 M_{n}^{\mbox{\tiny $(1)$}}\:=\:
    \sum_{i\in\mathcal{S}_4}\mathcal{C}_4^{\mbox{\tiny $(i)$}}{\raisebox{-.8cm}{\scalebox{.30}{\includegraphics{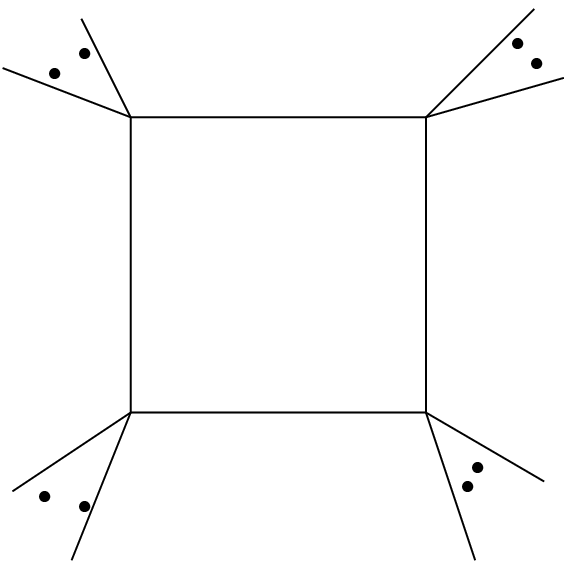}}}}+
    \sum_{i\in\mathcal{S}_3}\mathcal{C}_3^{\mbox{\tiny $(i)$}}{\raisebox{-.8cm}{\scalebox{.30}{\includegraphics{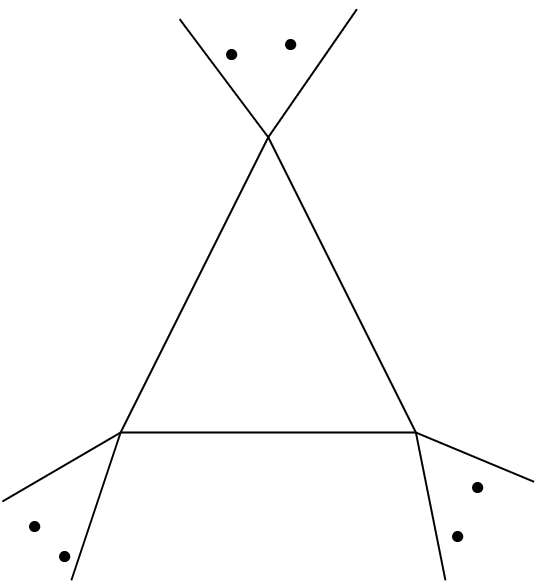}}}}+
    \sum_{i\in\mathcal{S}_2}\mathcal{C}_2^{\mbox{\tiny $(i)$}}{\raisebox{-.4cm}{\scalebox{.30}{\includegraphics{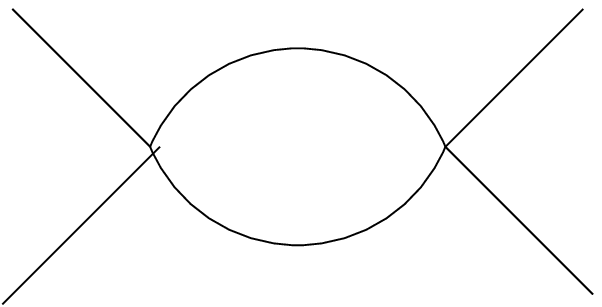}}}}+
    \mathcal{R}^{\mbox{\tiny $(1)$}}
\end{equation}
For the sake of simplicity, let us restrict ourself to a 
four-particle colour-ordered amplitude and let us play with the 
integrals. Fixing the loop momentum $l$ to run from particle $1$ to 
particle $2$, the box integral is given by
\begin{equation}\eqlabel{eq:Box4}
 \mathcal{I}_4\:\equiv\:
  \raisebox{-1.2cm}{\scalebox{.50}{\includegraphics{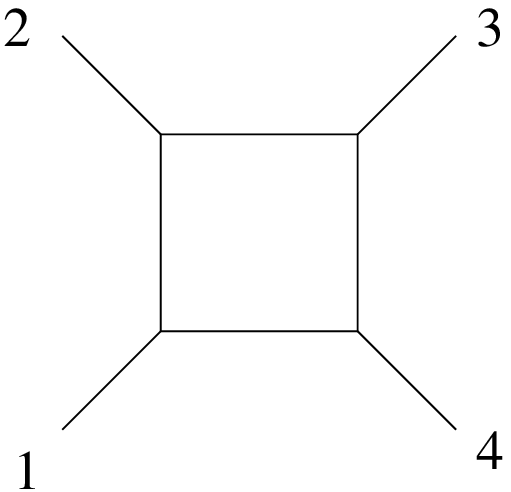}}}
  \:=\:
  \int\,d^4l\,
  \frac{st\,\delta^{\mbox{\tiny $(4)$}}
        \left(
         \sum_{k=1}^4p^{\mbox{\tiny $(i)$}}
        \right)}{l^2(l-p^{\mbox{\tiny $(1)$}})^2
       (l+p^{\mbox{\tiny $(2)$}})^2(l-P_{14})^2},
\end{equation}
where both the Mandelstam variables $s$ and $t$ and the 
momentum-conserving $\delta$-function have been included into the
definition. Using the momentum-conserving $\delta$-function,
the scalar box above can be written as 
\begin{equation}\eqlabel{eq:Box4a}
 \begin{split}
  \mathcal{I}_4\:=\:st\int\,\prod_{i=1}^4
  \frac{d^4l_{\mbox{\tiny $i,i+1$}}}{l_{\mbox{\tiny $i,i+1$}}^2}
  \delta^{\mbox{\tiny $(4)$}}
   \left(
    p^{\mbox{\tiny $(1)$}}-l_{12}-l_{41}
   \right)
  &\delta^{\mbox{\tiny $(4)$}}
   \left(
    p^{\mbox{\tiny $(2)$}}+l_{12}+l_{23}
   \right)
  \delta^{\mbox{\tiny $(4)$}}
   \left(
    p^{\mbox{\tiny $(3)$}}-l_{23}-l_{34}
   \right)\times\\
  &\times\delta^{\mbox{\tiny $(4)$}}
   \left(
    p^{\mbox{\tiny $(4)$}}+l_{34}+l_{41}
   \right),
 \end{split}
\end{equation}
with $l_{12}\,\equiv\,l$, 
$l_{23}\,\equiv\,-(l+p^{\mbox{\tiny $(2)$}})$,
$l_{34}\,\equiv\,l-P_{14}$,
$l_{41}\,\equiv\,-(l-p^{\mbox{\tiny $(1)$}})$. All the 
$l_{\mbox{\tiny $i,i+1$}}$'s can be written as a sum over two
light-like vectors
\begin{equation}\eqlabel{eq:li}
 l_{\mbox{\tiny $i,i+1$}}\:=\:\mu_{\mbox{\tiny $i,i+1$}}+
  z_{\mbox{\tiny $i,i+1$}}q_{\mbox{\tiny $i,i+1$}},
 \qquad 
 \mu_{\mbox{\tiny $i,i+1$}}^2\:=\:0\:=\:
 q_{\mbox{\tiny $i,i+1$}}^2,
\end{equation}
$q_{\mbox{\tiny $i,i+1$}}$'s being fixed reference spinors.
The momenta $l_{\mbox{\tiny $i,i+1$}}$ are therefore parametrised
in terms of the light-like momenta $\mu_{\mbox{\tiny $i,i+1$}}$,
encoding three degrees of freedom each, and 
$z_{\mbox{\tiny $i,i+1$}}$.
In these new variables, the scalar box becomes
\begin{equation}\eqlabel{eq:Box4b}
 \begin{split}
  \mathcal{I}_4\:=\:st\int\prod_{i=1}^4
  \frac{dz_{\mbox{\tiny $i,i+1$}}}{z_{\mbox{\tiny $i,i+1$}}}
  \int\prod_{i=1}^4d^4\mu_{\mbox{\tiny $i,i+1$}}
  \delta^{\mbox{\tiny $(4)$}}
  \left(
   \mu_{\mbox{\tiny $i,i+1$}}^2
  \right)
  \delta^{\mbox{\tiny $(4)$}}
  \left(
   p^{\mbox{\tiny $(i)$}}(z)-\mu_{\mbox{\tiny $i,i+1$}}-
   \mu_{\mbox{\tiny $i-1,i$}}
  \right),
 \end{split}
\end{equation}
where $p^{\mbox{\tiny $(i)$}}(z)\,\equiv\, p^{\mbox{\tiny $(i)$}}\,
+\,(-1)^i z_{\mbox{\tiny $i,i+1$}}\,q_{\mbox{\tiny $i,i+1$}}\,
+\,(-1)^i z_{\mbox{\tiny $i-1,i$}}\,q_{\mbox{\tiny $i-1,i$}}$.
The reference four-vectors can be conveniently chosen in such a 
way that $[p^{\mbox{\tiny $(i)$}}(z)]^2\,=\,0$. There are two,
in principle equivalent, straightforward choices which satisfy such 
a condition:
$q_{\mbox{\tiny $i,i+1$}}\,=\,
 \lambda^{\mbox{\tiny $(i)$}}\tilde{\lambda}^{\mbox{\tiny $(i+1)$}}$
or
$q_{\mbox{\tiny $i,i+1$}}\,=\,
\lambda^{\mbox{\tiny $(i+1)$}}\tilde{\lambda}^{\mbox{\tiny $(i)$}}$
-- notice that with such a choices the variables 
$z_{\mbox{\tiny $i,i+1$}}$'s are taken to transform not trivially
under the little group of particles $i$ and $i+1$.

Performing the integration over $\mu_{\mbox{\tiny $i,i+1$}}$,
one obtains the following form for the scalar box
\begin{equation}\eqlabel{eq:Box4c}
 \begin{split}
  \mathcal{I}_4\:&=\:\delta^{\mbox{\tiny $(4)$}}
  \left(
   \sum_{i=1}^4 p^{\mbox{\tiny $(i)$}}
  \right)
  \int\prod_{i=1}^4
  \frac{dz_{\mbox{\tiny $i,i+1$}}}{z_{\mbox{\tiny $i,i+1$}}}
  \frac{st}{s(z)t(z)}\:=\\
 &=\:\delta^{\mbox{\tiny $(4)$}}
  \left(
   \sum_{i=1}^4 p^{\mbox{\tiny $(i)$}}
  \right)
  \int\prod_{i=1}^4
  \frac{dz_{\mbox{\tiny $i,i+1$}}}{z_{\mbox{\tiny $i,i+1$}}
  (1+a_{\mbox{\tiny $i,i+1$}}z_{\mbox{\tiny $i,i+1$}})},
 \end{split}
\end{equation}
where the $a_{\mbox{\tiny $i,i+1$}}$'s are rational functions of
the Lorentz invariant spinor inner products, whose explicit form 
depends on the choices of the reference vectors/bispinors
$q_{\mbox{\tiny $i,i+1$}}$. Notice that the form of the integrand
of \eqref{eq:Box4c} resembles the one which is obtained from
the on-shell diagrammatics, where the $d\log{\zeta}$ structure
is manifest
\begin{equation}\eqlabel{eq:Box4d}
 \mathcal{I}_4\:=\:\delta^{\mbox{\tiny $(4)$}}
  \left(
   \sum_{i=1}^4 p^{\mbox{\tiny $(i)$}}
  \right)
  \int\bigwedge_{i=1}^4d\log{\zeta_{\mbox{\tiny $i,i+1$}}}
  \:\equiv\:
  \delta^{\mbox{\tiny $(4)$}}
  \left(
   \sum_{i=1}^4 p^{\mbox{\tiny $(i)$}}
  \right)
  \int\bigwedge_{i=1}^4
  d\log{\frac{a_{\mbox{\tiny $i,i+1$}}z_{\mbox{\tiny $i,i+1$}}}{
  1+a_{\mbox{\tiny $i,i+1$}}z_{\mbox{\tiny $i,i+1$}}}}
\end{equation}
and the definition of $p^{\mbox{\tiny $(i)$}}(z)$'s recalls 
a composite BCFW deformation.

Let us now move to the scalar triangles, proceeding in a similar
fashion. As an explicit example, let us consider the following
integral
\begin{equation}\eqlabel{eq:Tr4}
 \mathcal{I}_3\:\equiv\:
 \raisebox{-1.2cm}{\scalebox{.50}{\includegraphics{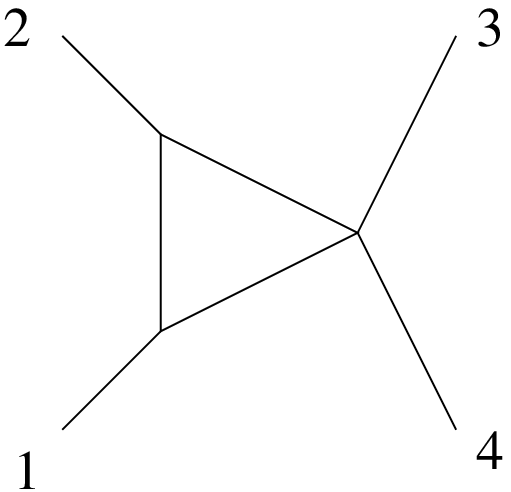}}}
  \:=\:
  \int\,d^4l\,
  \frac{s\,\delta^{\mbox{\tiny $(4)$}}
        \left(
         \sum_{k=1}^4p^{\mbox{\tiny $(i)$}}
        \right)}{l^2(l-p^{\mbox{\tiny $(1)$}})^2
       (l+p^{\mbox{\tiny $(2)$}})^2},
\end{equation}
where the loop momentum $l$ is again taken to run from particle $1$
to particle $2$. As in the scalar box case, using the 
momentum-conserving $\delta$-function:
\begin{equation}\eqlabel{eq:Tr4a}
 \mathcal{I}_3\:=\:s\int\prod_{i=4}^2
 \frac{d^4l_{\mbox{\tiny $i,i+1$}}}{l_{\mbox{\tiny $i,i+1$}}^2}
  \delta^{\mbox{\tiny $(4)$}}
   \left(
    p^{\mbox{\tiny $(1)$}}-l_{12}-l_{41}
   \right)
  \delta^{\mbox{\tiny $(4)$}}
   \left(
    p^{\mbox{\tiny $(2)$}}+l_{12}+l_{23}
   \right)
  \delta^{\mbox{\tiny $(4)$}}
   \left(
    P_{34}+l_{41}-l_{23}
   \right).
\end{equation}
Taking now the very same parametrisation \eqref{eq:li} for the
$l_{\mbox{\tiny $i,i+1$}}$'s, and integrating over the
light-like vectors/bispinors $\mu_{\mbox{\tiny $i,i+1$}}$'s,
the scalar triangle becomes
\begin{equation}\eqlabel{eq:Tr4b}
 \begin{split}
  \mathcal{I}_3\:&=\:\delta^{\mbox{\tiny $(4)$}}
   \left(
    \sum_{i=1}^4 p^{\mbox{\tiny $(i)$}}
   \right)
   \int\prod_{i=4}^2
   \frac{dz_{\mbox{\tiny $i,i+1$}}}{z_{\mbox{\tiny $i,i+1$}}}
   \int\frac{d\xi}{\xi}\,
   \frac{s}{s(z)}\:=\\
   &=\:\delta^{\mbox{\tiny $(4)$}}
   \left(
    \sum_{i=1}^4 p^{\mbox{\tiny $(i)$}}
   \right)
   \int
   \frac{dz_{\mbox{\tiny $41$}}}{z_{\mbox{\tiny $41$}}
   (1+a_{\mbox{\tiny $41$}}z_{\mbox{\tiny $41$}})}
   \frac{dz_{\mbox{\tiny $23$}}}{z_{\mbox{\tiny $23$}}
   (1+a_{\mbox{\tiny $23$}}z_{\mbox{\tiny $23$}})}
   \frac{dz_{12}}{z_{12}}
   \int\frac{d\xi}{\xi}
 \end{split}
\end{equation}
where $\xi$ is related to $\mu_{12}$ via the $\delta$-function 
which ``localises'' it over $\mu_{12}^2\,=\,0$, so that 
$\mu_{12}\,=\,\xi\,\nu\tilde{\nu}$ with $\xi$, $\nu_{a}$
and $\tilde{\nu}_{\dot{a}}$ parametrising the three degrees of
freedom of $\mu_{12}$, as well as
$p^{\mbox{\tiny $(i)$}}(z)\,\equiv\, p^{\mbox{\tiny $(i)$}}\,
+\,(-1)^i z_{\mbox{\tiny $i,i+1$}}\,q_{\mbox{\tiny $i,i+1$}}\,
+\,(-1)^i z_{\mbox{\tiny $i-1,i$}}\,q_{\mbox{\tiny $i-1,i$}}$
and 
$P_{34}(z)\:=\:P_{34}+z_{41}q_{41}-z_{23}q_{23}$.
As for the box, the reference $q$'s are taken in such a way
that $[p^{\mbox{\tiny $(i)$}}(z)]^2\,=\,0$.

Notice that also the scalar triangles have a $d\log$ structure
for their integrand
\begin{equation}\eqlabel{eq:Tr4c}
 \begin{split}
  \mathcal{I}_3\:&=\:\delta^{\mbox{\tiny $(4)$}}
   \left(
    \sum_{i=4}^2 p^{\mbox{\tiny $(i)$}}
   \right)
   \int d\log{\xi}\wedge
   \bigwedge_{i=4}^2d\log{\zeta_{\mbox{\tiny $i,i+1$}}}
   \:\equiv\\
  &\equiv\:
   \delta^{\mbox{\tiny $(4)$}}
   \left(
    \sum_{i=4}^2 p^{\mbox{\tiny $(i)$}}
   \right)
   \int d\log{z_{12}} \wedge d\log{\xi}\wedge
   d\log{\frac{a_{\mbox{\tiny $41$}}z_{\mbox{\tiny $41$}}}{
   1+a_{\mbox{\tiny $41$}}z_{\mbox{\tiny $41$}}}}
   \wedge
   d\log{\frac{a_{\mbox{\tiny $23$}}z_{\mbox{\tiny $23$}}}{
   1+a_{\mbox{\tiny $23$}}z_{\mbox{\tiny $23$}}}}.
 \end{split}
\end{equation}
For planar theories, the integration variables among the different
elements of the basis can be identified. Comparing how we 
parametrised $\mathcal{I}_4$ and $\mathcal{I}_3$, 
the $z_{\mbox{\tiny $i,i+1$}}$'s in the two integrals are
exactly the same for $i\,=\,4,\ldots\,2$, while $z_{34}$ and
$\xi$ need to be related by a change of variable, which
can be easily found if we consider 
$P_{34}(z)\:=p^{\mbox{\tiny $(3)$}}(z)+p^{\mbox{\tiny $(4)$}}(z)$,
with $p^{\mbox{\tiny $(i)$}}(z)$ as defined in \eqref{eq:Box4b} --
which just amounts to add and subtract $z_{34}q_{34}$ in the
definition of $P_{34}(z)$ in the above paragraph -- and it is
given by
\begin{equation}\eqlabel{eq:xiz34}
 \xi\:=\:\frac{b(1+a_{34}z_{34})}{
  -\frac{s}{u}(1+a_{23}z_{23})(1+a_{41}z_{41})-\frac{t}{u}(1+a_{12}z_{12})(1+a_{34}z_{34})},
\end{equation}
where $b$, as the $a_{\mbox{\tiny $i,i+1$}}$'s, is a rational 
function of the Lorentz invariants and depends on the choices
of the reference $q_{\mbox{\tiny $i,i+1$}}$'s.

The expression as $d{\log}$ of the integrand of the other scalar
triangles can be obtained by cyclic permutations of the indices.

Finally, let us take a look at the scalar bubble
\begin{equation}\eqlabel{eq:Bbl4}
 \begin{split}
  \mathcal{I}_2\:\equiv\:
  \raisebox{-1.2cm}{\scalebox{.50}{\includegraphics{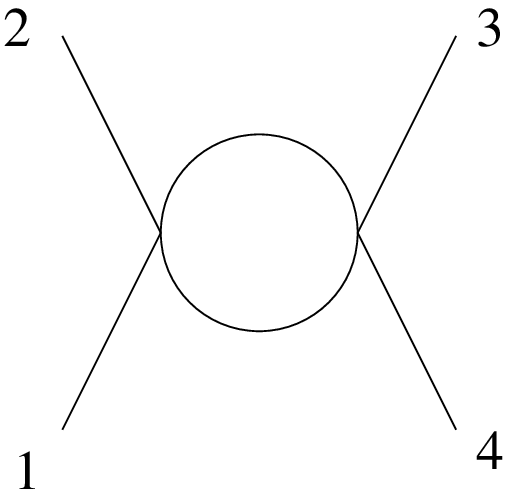}}}
   \:&=\:
      \int\,d^4l\,
       \frac{\delta^{\mbox{\tiny $(4)$}}
        \left(
         \sum_{i=1}^4p^{\mbox{\tiny $(i)$}}
        \right)}{(l-p^{\mbox{\tiny $(1)$}})^2
        (l+p^{\mbox{\tiny $(2)$}})^2}\:=\\
     &=\:
      \int\frac{d^4l_{23}}{l_{23}^2}\frac{d^4l_{41}}{l_{41}^2}
       \delta^{\mbox{\tiny $(4)$}}
        \left(
         P_{12}-l_{41}+l_{23}
        \right)
       \delta^{\mbox{\tiny $(4)$}}
        \left(
         P_{34}+l_{41}-l_{23}
        \right).
 \end{split}
\end{equation}
Using the same change of variables as before and integrating out
all the degrees of freedom that the delta functions allow, one
obtains:
\begin{equation}\eqlabel{eq:Bbl4a}
 \mathcal{I}_2\:=\:
  \int\frac{dz_{23}}{z_{23}}\frac{dz_{41}}{z_{41}}
  \int\langle\alpha,d\alpha\rangle[\tilde{\alpha},d\tilde{\alpha}]
  \frac{s(z)}{\langle\alpha|P_{12}(z)|\tilde{\alpha}]^2},
\end{equation}
where $\alpha$ and $\tilde{\alpha}$ are the spinors related to
$\mu_{41}$. Differently from the scalar box and the scalar triangle,
the scalar bubble does not have a $d\log{\zeta}$ structure.
If we take $\alpha\:=\:\alpha(z)$ and 
$\tilde{\alpha}\:=\:\tilde{\alpha}(z)$ and we trade them for
$z_{12}$ and $z_{34}$, so that the integrands of the box, triangles
and bubbles are functions of the same variables, the above integral
takes the following form:
\begin{equation}\eqlabel{eq:Bbl4b}
 \begin{split}
  \mathcal{I}_2\:&=\:\frac{st}{u^2}
   \int\frac{dz_{23}}{z_{23}}\frac{dz_{41}}{z_{41}}\,
   \frac{dz_{12}\,dz_{34}\,
    a_{12}a_{34}(1+a_{23}z_{23})(1+a_{41}z_{41})}{
    \left[
     -\frac{s}{u}(1+a_{23}z_{23})(1+a_{41}z_{41})
     -\frac{t}{u}(1+a_{12}z_{12})(1+a_{34}z_{34})
    \right]^2}\:=\\
  &=\:\frac{st}{u^2}
     \int\bigwedge_{i=1}^4
     \frac{d\zeta_{\mbox{\tiny $i,i+1$}}}{
      \zeta_{\mbox{\tiny $i,i+1$}}}
      \frac{\zeta_{12}\zeta_{34}}{
      \left[
       -\frac{s}{u}(1-\zeta_{12})(1-\zeta_{34})
       -\frac{t}{u}(1-\zeta_{23})(1-\zeta_{41})
      \right]^2},
 \end{split}
\end{equation}
where the $\zeta_{\mbox{\tiny $i,i+1$}}$ are the ``$d\log$'' 
variables in \eqref{eq:Box4d}. The other bubble integral can 
be obtained from \eqref{eq:Bbl4b} via the label exchange
$2\,\longleftrightarrow\,4$. Notices that the denominator of
the integrand is invariant under such a relabelling and, thus,
it is common to the two scalar bubbles of the one-loop amplitude.
Furthermore, it introduces a new singularity which correspond
to the UV divergence of the bubbles.

\section{Double cuts and the BCFW parametrisation of the one-loop integrand}
\label{app:DCintg}

In Section \ref{subsec:1loopStr} we extensively discuss the one-loop structure with particular reference to the
correspondence between the on-shell forms and the generalised unitarity cuts. In this appendix we show how the
standard representation of the double cuts can be mapped in an ``on-shell-{\it like}'' form. For the sake of 
concreteness, let us focus on the double cut in the $s$-channel of the same four-particle amplitude analysed in 
\ref{subsec:1loopStr}:
\begin{equation}\eqlabel{eq:DCstd}
 \begin{split}
  \Delta_2^{\mbox{\tiny $(s)$}}\mathcal{M}_4^{\mbox{\tiny $(1L)$}}\:&=\:
   \raisebox{-1.2cm}{\scalebox{.32}{\includegraphics{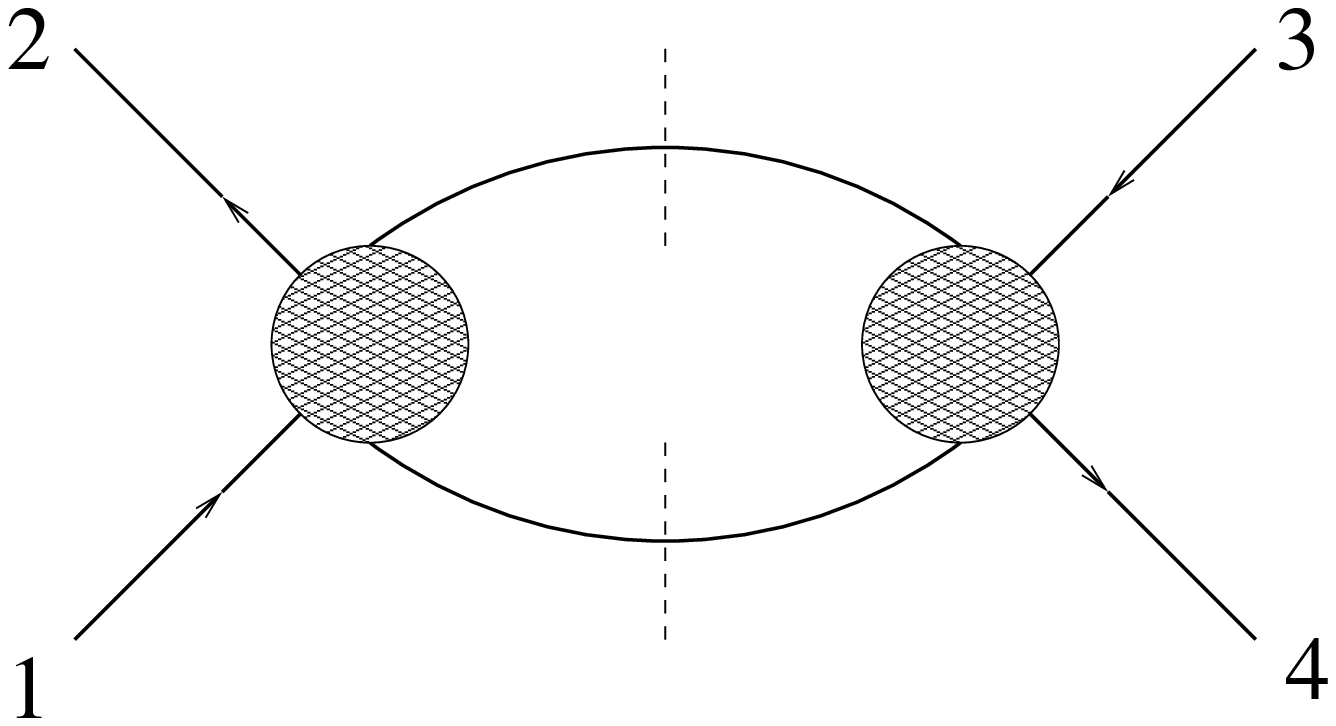}}}\:=\\
  &=\:\sum_{h\,=\,\pm}\int\,d\omega_{23}d\omega_{41}\,
   \mathcal{M}_4^{\mbox{\tiny tree}}\left(-l_{41}^{-h},1,2,l_{23}^{h};\{\tilde{\eta}\}\right)
   \mathcal{M}_4^{\mbox{\tiny tree}}\left(-l_{23}^{-h},3,4,l_{41}^{h};\{\tilde{\eta}\}\right),
 \end{split}
\end{equation}
where 
$d\omega_{i,i+1}\,\equiv\,d^4l_{i,i+1}d^{\mathcal{N}}\,\tilde{\eta}^{\mbox{\tiny $(i,i+1)$}}\,
   \delta(l_{i,i+1}^2)$ is the phase-space of the loop momentum $l_{i,i+1}$, which runs between particle-$i$ and 
-$i+1$ and, as usual, the (super)-momentum $\delta$-functions are contained in the amplitudes 
$\mathcal{M}_4^{\mbox{\tiny tree}}$.

Integrating over the phase-space of $l_{23}$ as well as over both the Grassmann variables 
$\tilde{\eta}^{\mbox{\tiny $(i,i+1)$}}$:
\begin{equation}\eqlabel{eq:DCstd2}
 \begin{split}
  \Delta_2^{\mbox{\tiny $(s)$}}&\mathcal{M}_4^{\mbox{\tiny $(1L)$}}\:=\:
   \sum_{h\,=\,\pm}\int\,d\tau\,\tau\int\langle\mu,d\mu\rangle [\tilde{\mu},d\tilde{\mu}]\,
    \delta\left(P_{12}^2+\tau\langle\mu|P_{12}|\tilde{\mu}]\right)
    \langle\mu|P_{12}|\tilde{\mu}]^{\mathcal{N}}\times\\
  &\times\,
    \delta^{\mbox{\tiny $(2\times2)$}}
   \left(
    \sum_{k=1}^4\lambda^{\mbox{\tiny $(k)$}}\tilde{\lambda}^{\mbox{\tiny $(k)$}}
   \right)
   \delta^{\mbox{\tiny $(2\times\mathcal{N})$}}
   \left(
    \sum_{k=1}^4\lambda^{\mbox{\tiny $(k)$}}\tilde{\eta}^{\mbox{\tiny $(k)$}}
   \right)
   M_4^{\mbox{\tiny tree}}\left(1,2;\,\mu,\tilde{\mu}\right)
   M_4^{\mbox{\tiny tree}}\left(\,\mu,\tilde{\mu};\,3,4\right)
  ,
 \end{split}
\end{equation}
where $l_{41}\,=\,\tau\mu\tilde{\mu}$ due to $\delta(l_{41}^2)$ which implements the cut of this line,
the $\delta$-function in \eqref{eq:DCstd2} is due $\delta(l_{23}^2)$ which puts the other loop-line on-shell,
and $\langle\mu|P_{12}|\tilde{\mu}]^{\mathcal{N}}$ is produced by the integration over the Grassmann variables.

The $\delta$-function in \eqref{eq:DCstd2} can be used to fix $\tau$ to get
\begin{equation}\eqlabel{eq:DCstd3}
 \Delta_2^{\mbox{\tiny $(s)$}}\mathcal{M}_4^{\mbox{\tiny $(1L)$}}\:=\:\mathcal{M}_4^{\mbox{\tiny tree}}
  \int\langle\mu,d\mu\rangle [\tilde{\mu},d\tilde{\mu}]
  \frac{s\langle2,3\rangle\langle4,1\rangle
    \left[
     \langle\mu,1\rangle^{4-\mathcal{N}}\langle3|P_{12}|\tilde{\mu}]^{4-\mathcal{N}}+
     \langle\mu,3\rangle^{4-\mathcal{N}}\langle1|P_{12}|\tilde{\mu}]^{4-\mathcal{N}}
    \right]}{\langle\mu|P_{12}|\tilde{\mu}]^{4-\mathcal{N}}
   \langle\mu,1\rangle\langle4,\mu\rangle\langle2|P_{12}|\tilde{\mu}]\langle3|P_{12}|\tilde{\mu}]}.
\end{equation}
Finally, parametrising the loop spinors $\mu$ and $\tilde{\mu}$ as follows
\begin{equation}\eqlabel{eq:lspar}
 \mu\:=\:\lambda^{\mbox{\tiny $(4)$}}+z_{34}\lambda^{\mbox{\tiny $(3)$}},\quad
 \tilde{\mu}\:=\:\tilde{\lambda}^{\mbox{\tiny $(1)$}}-z_{12}\lambda^{\mbox{\tiny $(2)$}},
\end{equation}
one obtains
\begin{equation}\eqlabel{eq:DCbcfw}
 \begin{split}
  \Delta_2^{\mbox{\tiny $(s)$}}\mathcal{M}_4^{\mbox{\tiny $(1L)$}}\:=\:\mathcal{M}_4^{\mbox{\tiny tree}}
   &\int\frac{dz_{12}}{z_{12}\left(1+\frac{\langle1,3\rangle}{\langle2,3\rangle}z_{12}\right)}
       \frac{dz_{34}}{z_{34}\left(1+\frac{\langle3,1\rangle}{\langle4,1\rangle}z_{34}\right)}\times\\
   &\times\frac{
    \left(-\frac{s}{u}\right)^{4-\mathcal{N}}+
    \left[
     -\frac{t}{u}
     \left(
      1+\frac{\langle1,3\rangle}{\langle2,3\rangle}z_{12}
     \right)
     \left(
      1+\frac{\langle3,1\rangle}{\langle4,1\rangle}z_{34}
     \right)
    \right]^{4-\mathcal{N}}}{
     \left[-\frac{s}{u}-\frac{t}{u}
      \left(
      1+\frac{\langle1,3\rangle}{\langle2,3\rangle}z_{12}
     \right)
     \left(
      1+\frac{\langle3,1\rangle}{\langle4,1\rangle}z_{34}
     \right)
    \right]^{4-\mathcal{N}}},
 \end{split}
\end{equation}
which can be mapped to \eqref{eq:OS2cut} via the simple change of variable 
$a_{i,i+1}z_{i,i+1}\,=\,\zeta_{i,i+1}/(1-\zeta_{i,i+1})$,
$a_{i,i+1}$ being the coefficient of $z_{i,i+1}$ appearing in the measure of \eqref{eq:DCbcfw}.

\section{Some one-loop integrands}\label{app:1lint}

In this appendix, we explicitly write down the on-shell diagrams representing some one-loop integrands. The idea
is to explicitly show how to extract physical information from such a representation. In particular, we
consider the MHV four-particle amplitude with consecutive negative helicity states, the MHV five-particle integrand
$\mathcal{M}_{4}^{\mbox{\tiny $(4)$}}(1^{-},2^{+},3^{-},4^{+},5^{+})$, .. 

\subsection{MHV four-particle amplitude with consecutive negative helicity states}\label{app:1lintM4mmpp}

Let us take consider the BCFW bridge in the $(4,1)$-channel, we have:
\begin{equation}\eqlabel{eq:M4mmppFw}
 \raisebox{-1.5cm}{\scalebox{.23}{\includegraphics{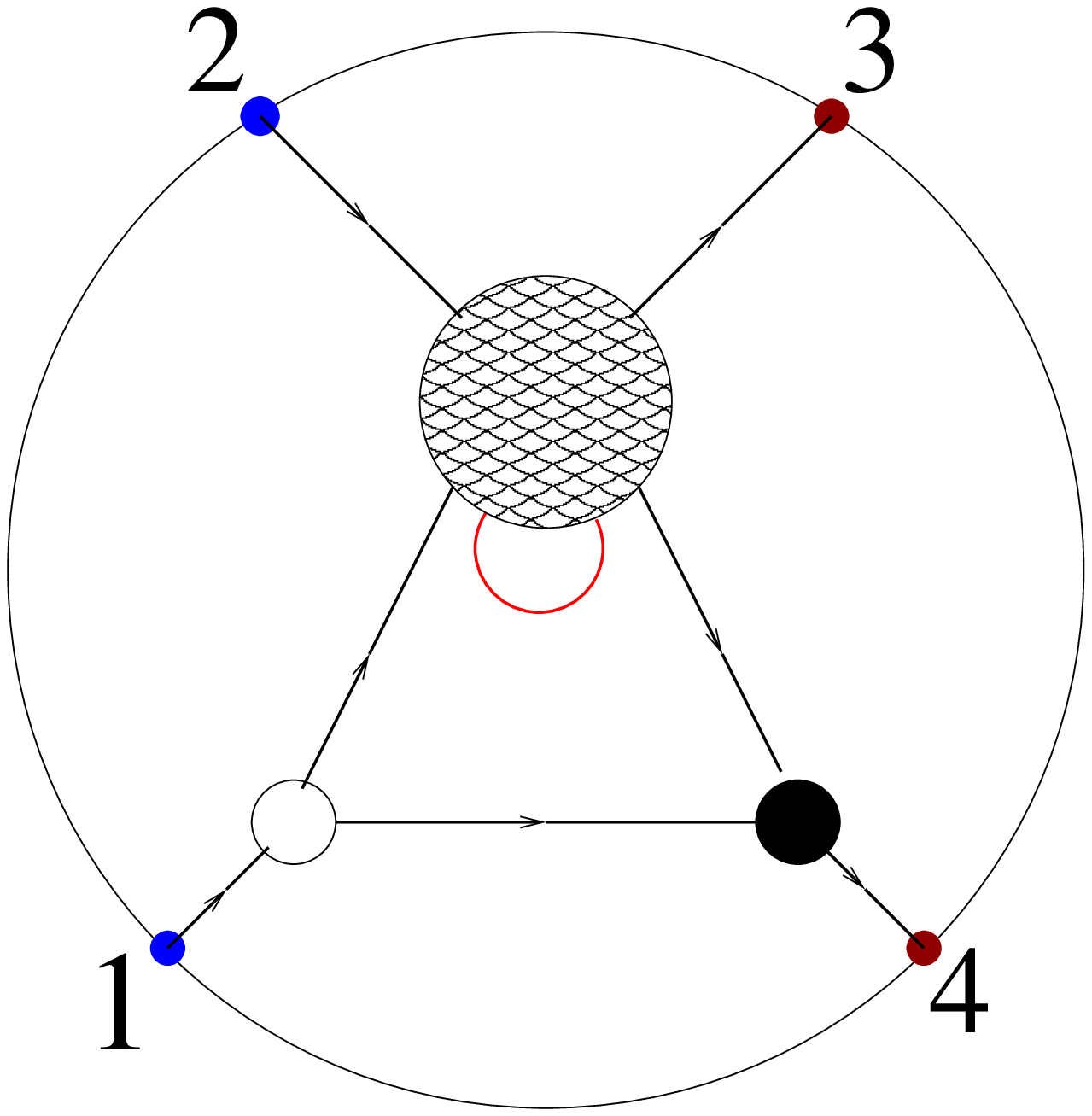}}}\:=\:
 \raisebox{-1.5cm}{\scalebox{.23}{\includegraphics{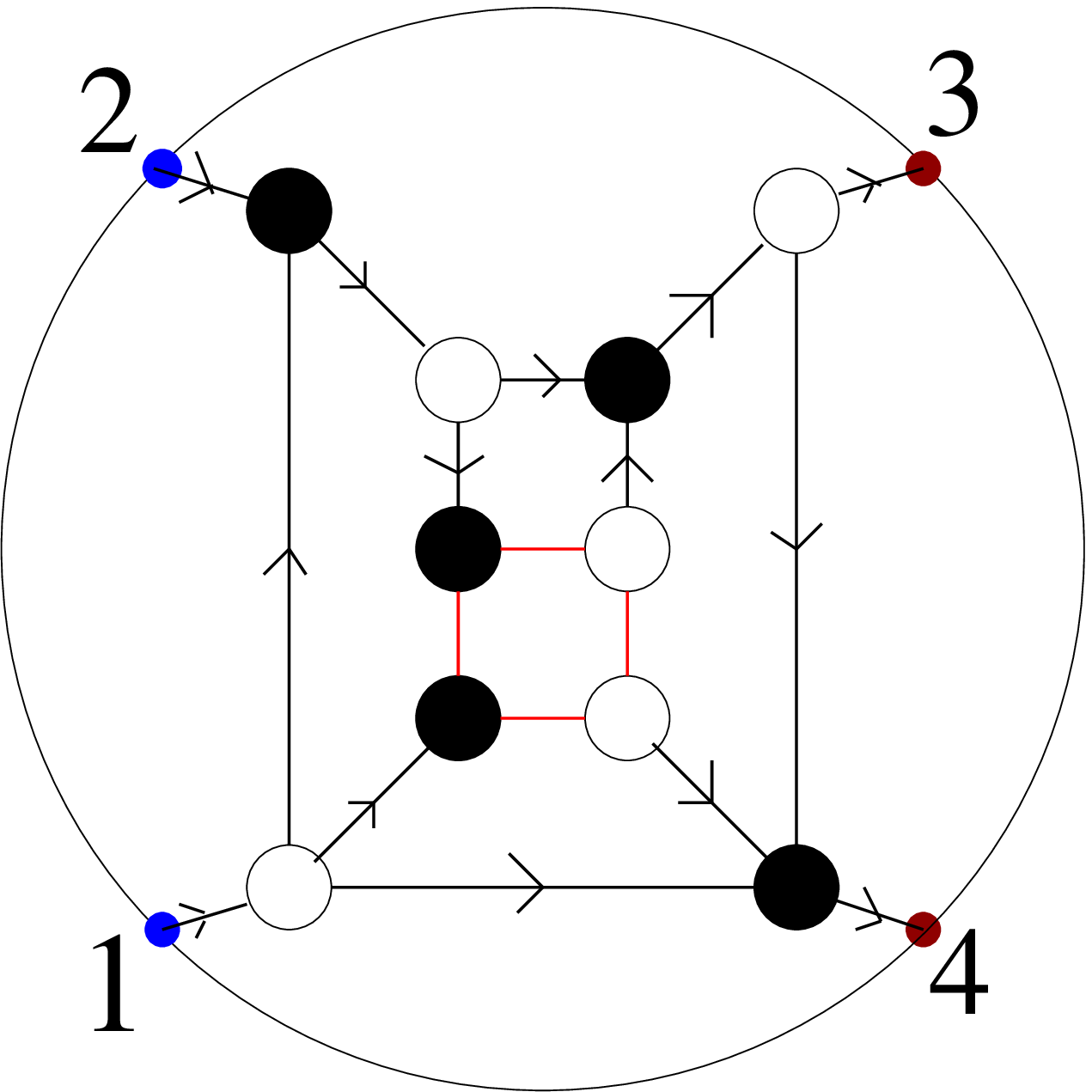}}}\:+\:
 \raisebox{-1.5cm}{\scalebox{.23}{\includegraphics{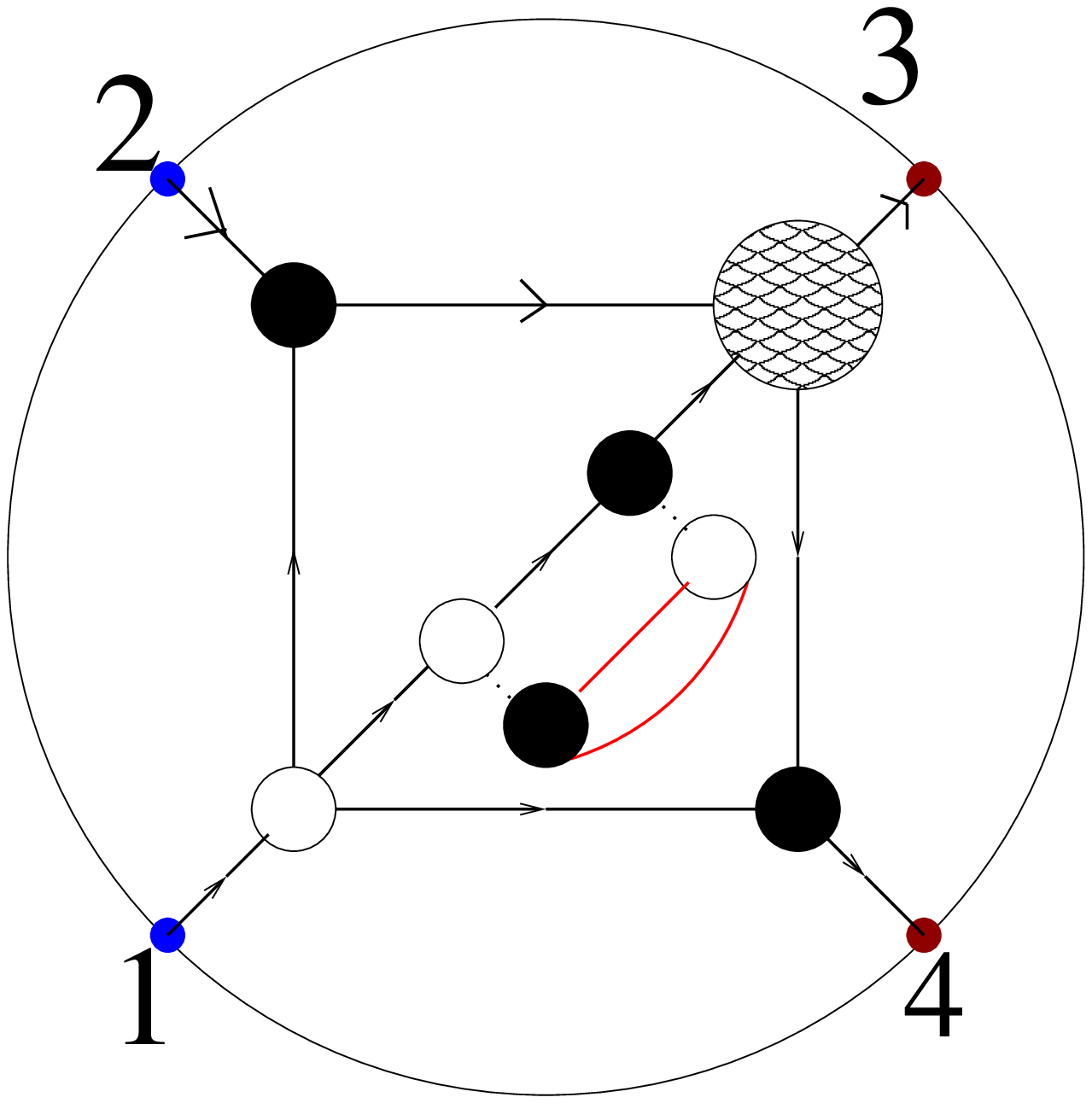}}}\:+\:
 \raisebox{-1.5cm}{\scalebox{.23}{\includegraphics{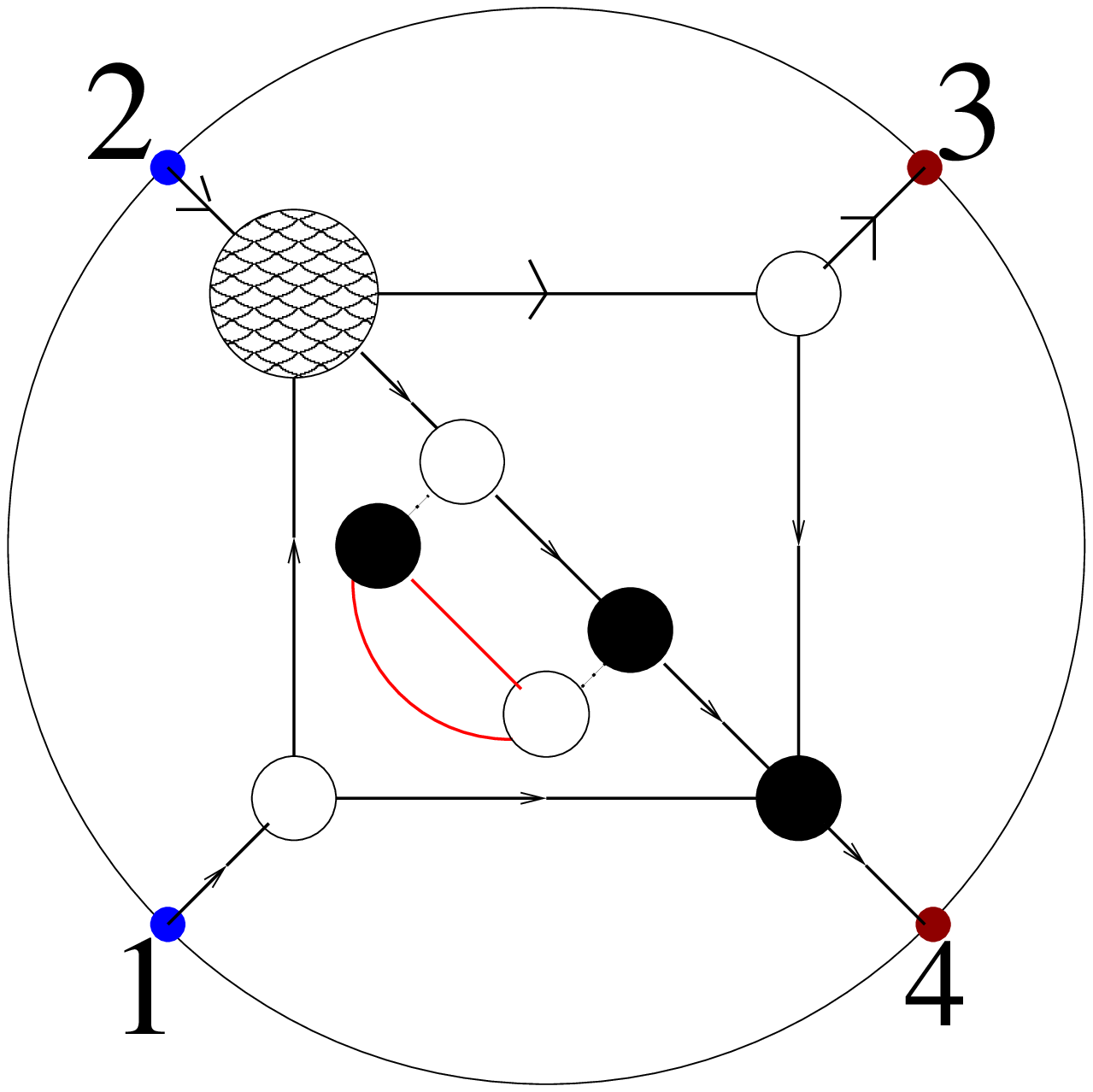}}},
\end{equation}
which has to be thought to be in the quasi-forward regularisation. The first diagram turns out to contain 
all the cut-constructible information. It can be easily checked by integrating over suitable $2$-cycles to
extract the double cuts. The explicit expression for the first term in the right-hand-side of \eqref{eq:M4mmppFw}
can be written as
\begin{equation}\eqlabel{eq:M4mmppCutConstr}
 \raisebox{-1.5cm}{\scalebox{.23}{\includegraphics{1LM4mmppFin.eps}}}\:=\:
 \mathcal{M}_4^{\mbox{\tiny tree}}\,
 \bigwedge_{i\,=\,1}^4\frac{d\zeta_i}{\zeta_i}\,\frac{1+(-\zeta_2)^{4-\mathcal{N}}}{(1-\zeta_2)^{4-\mathcal{N}}},
\end{equation}
where $\zeta_1$, $\zeta_3$ and $\zeta_4$ parametrise the BCFW bridges in the $(1,2)$-, $(3,4)$- and $(4,1)$-channels
respectively, while $\zeta_2$ parametrises the internal red lines\footnote{The lack of a helicity arrow assignment
for the internal red lines has to be understood as a sum over both the multiplets, whose propagation in such lines
turns out to be allowed.}. Notice that explicit expression for the on-shell four-form \eqref{eq:M4mmppCutConstr}
can be straightforwardly obtained on diagrammatic level via mergers and bubble deletions \eqref{BubDel} and 
\eqref{BubDelw}.

It easy to see that the $2$-cycles which allow to extract the double cuts are given by
\begin{itemize}
 \item $t$-channel: 
       $\gamma_2^{\mbox{\tiny $(t)$}}\:=\:\{(\zeta_1,\,\zeta_3)\in\mathbb{C}^2\,|\,\zeta_1\,=\,0\,=\,\zeta_3\}$
       \begin{equation}\eqlabel{eq:M4mmppDct}
        \mathcal{M}_4^{\mbox{\tiny tree}}\,\oint_{\gamma_2^{\mbox{\tiny $(t)$}}}
        \bigwedge_{i\,=\,1}^4\frac{d\zeta_i}{\zeta_i}\,\frac{1+(-\zeta_2)^{4-\mathcal{N}}}{(1-\zeta_2)^{4-\mathcal{N}}}
        \:=\:
        \raisebox{-1.5cm}{\scalebox{.23}{\includegraphics{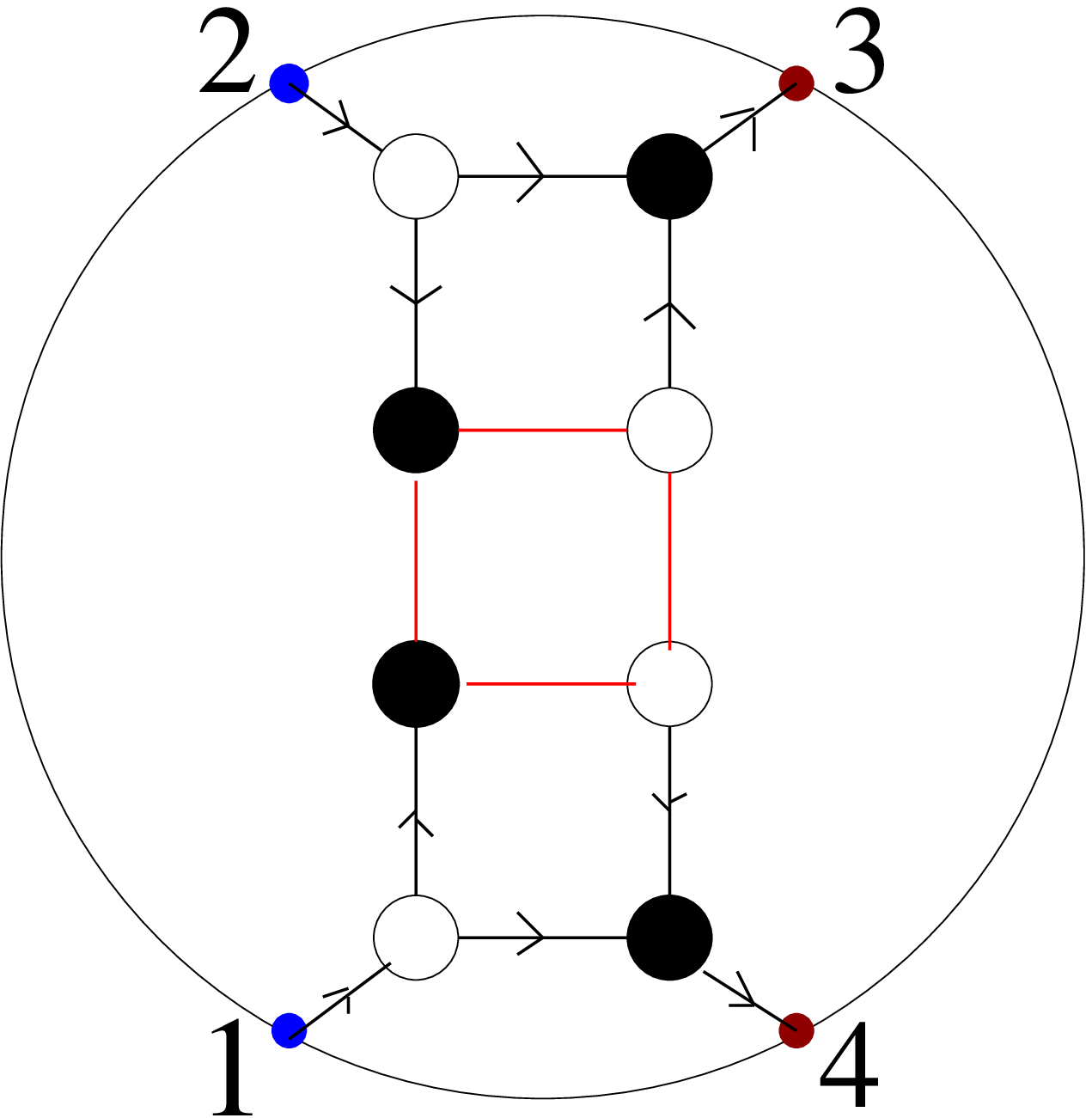}}}
        \:\equiv\:
        \raisebox{-1.5cm}{\scalebox{.23}{\includegraphics{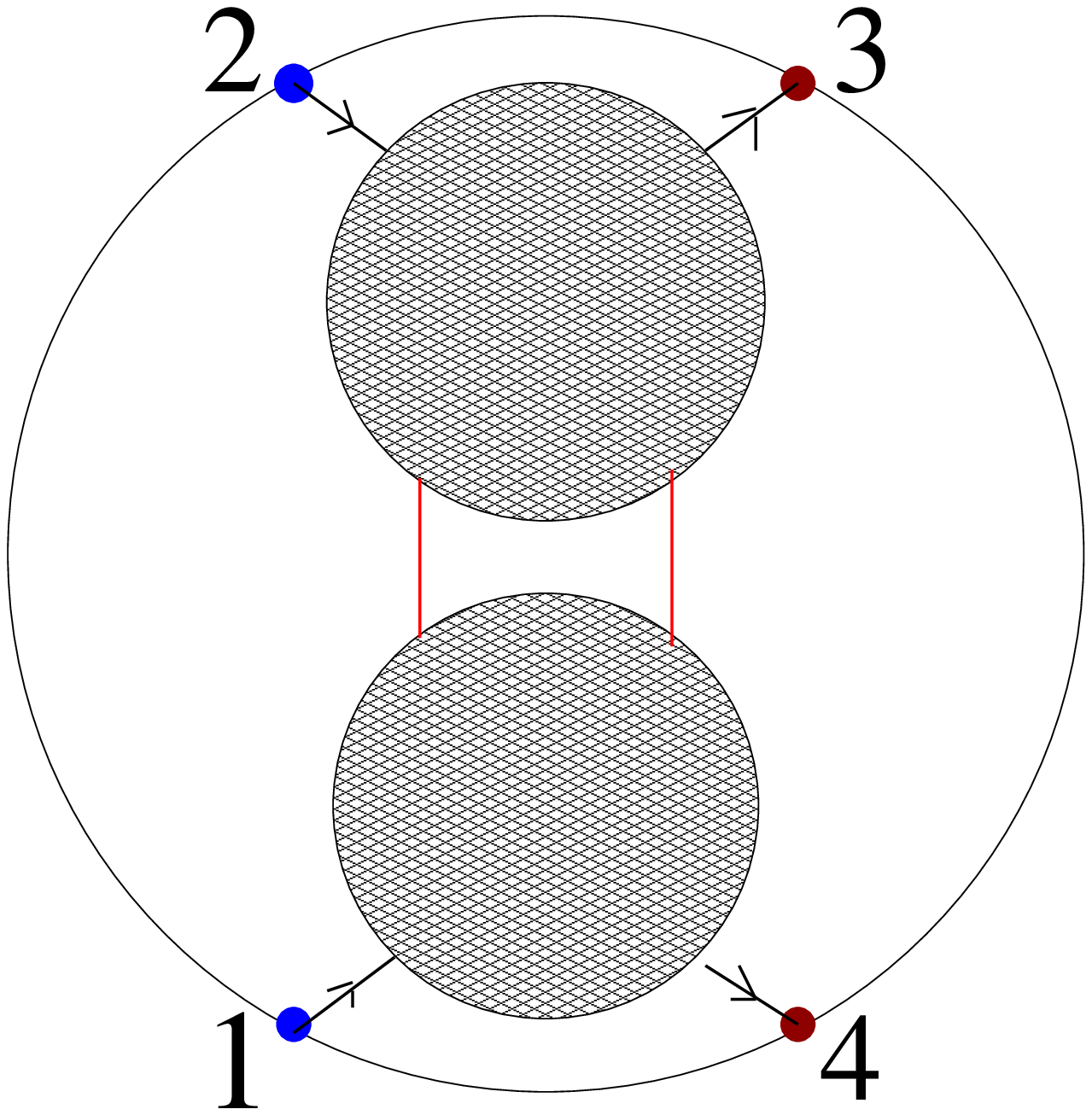}}}
       \end{equation}
 \item $s$-channel: 
       $\gamma_2^{\mbox{\tiny $(s)$}}\:=\:\{(\zeta_2,\,\zeta_4)\in\mathbb{C}^2\,|\,\zeta_2\,=\,\infty,\:\zeta_4\,=\,0\}$
       \begin{equation}\eqlabel{eq:M4mmppDcs}
        \mathcal{M}_4^{\mbox{\tiny tree}}\,\oint_{\gamma_2^{\mbox{\tiny $(t)$}}}
        \bigwedge_{i\,=\,1}^4\frac{d\zeta_i}{\zeta_i}\,\frac{1+(-\zeta_2)^{4-\mathcal{N}}}{(1-\zeta_2)^{4-\mathcal{N}}}
        \:=\:
        \raisebox{-1.5cm}{\scalebox{.23}{\includegraphics{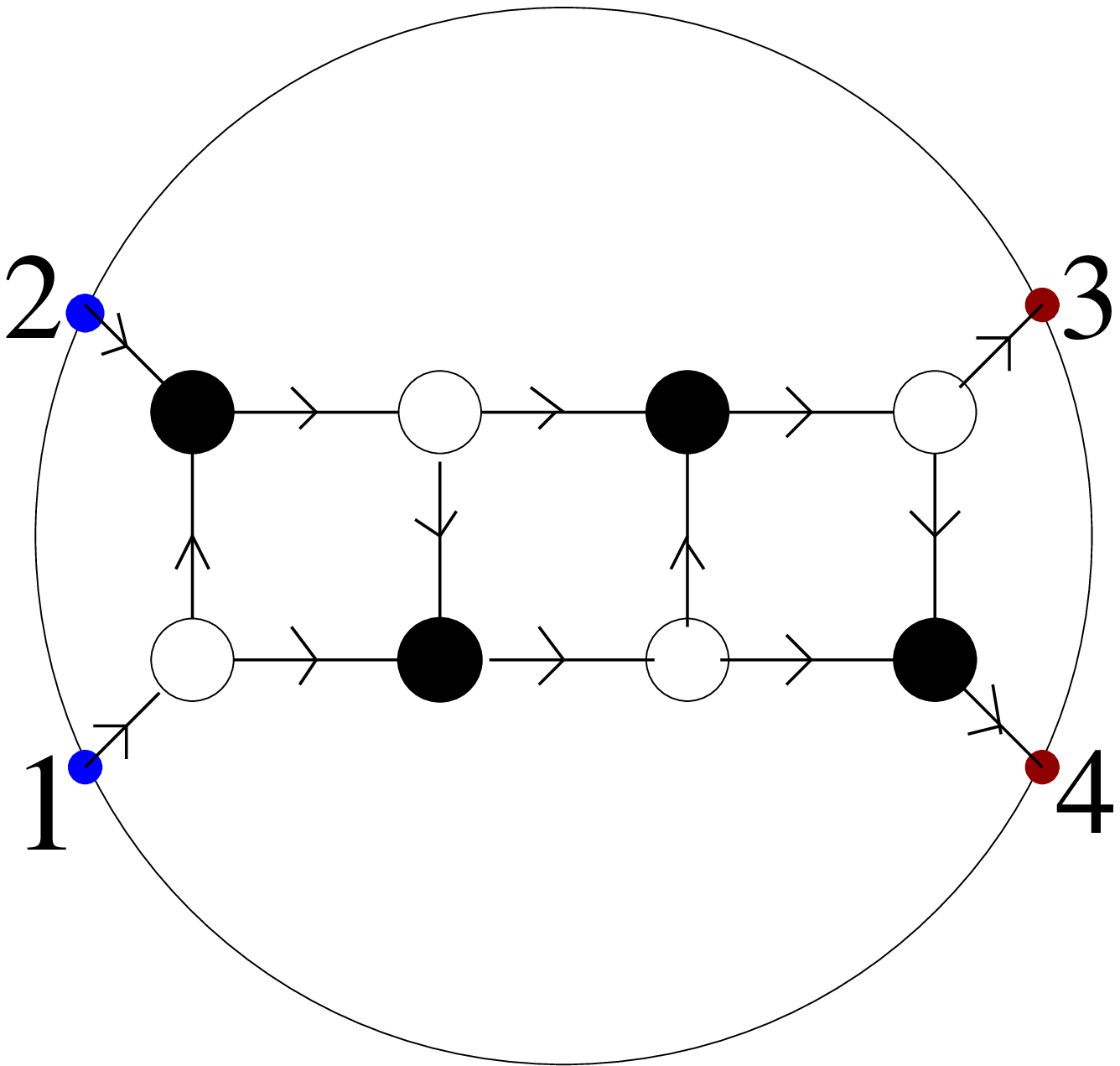}}}
        \:\equiv\:
        \raisebox{-1.5cm}{\scalebox{.23}{\includegraphics{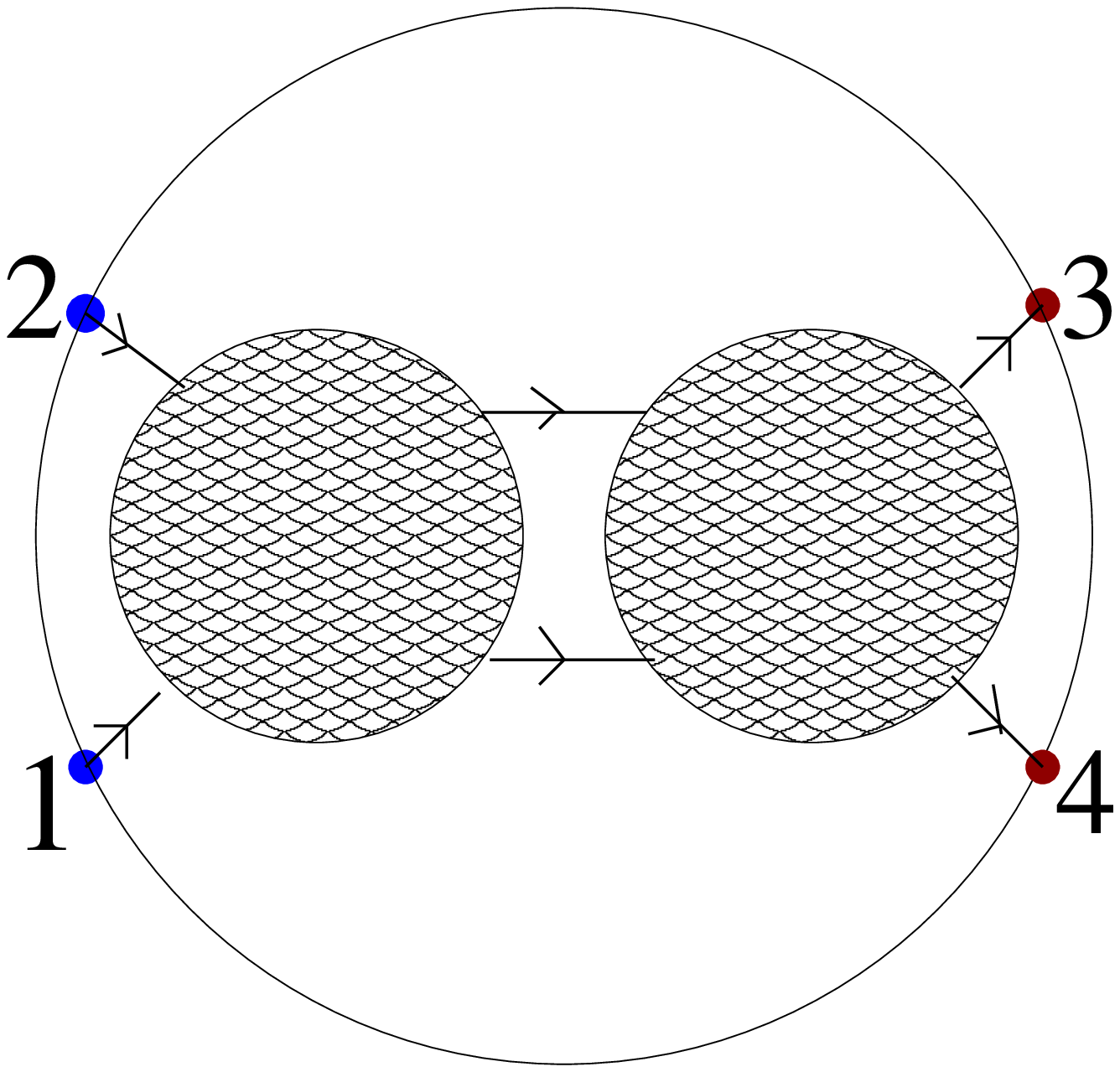}}}
       \end{equation}
\end{itemize}
where the very last diagram in each of \eqref{eq:M4mmppDct} and \eqref{eq:M4mmppDcs} emphasise how the
square sub-diagrams are nothing but tree-level four-particle amplitudes. In the $t$-channel both the multiplets
contribute, while in the $s$-channel just the one which preserves the helicity flows 
$1\,\longrightarrow\,4$ and $2\,\longrightarrow\,3$.

\subsection{Five-particle amplitudes}\label{eq:1lintM5mhv}

For the five particle amplitudes, the recursive relation receives contribution from both factorisation and
forward channels. In particular, there are three classes of terms contributing. For the sake of concreteness, we
focus on the helicity configuration $\mathcal{M}_5^{\mbox{\tiny $(4)$}}(1^{-},2^{+},3^{-},4^{+},5^{+})$, which
shows the following structure:
\begin{equation}\eqlabel{eq:1lM5mpmpp}
 \mathcal{M}_{5}^{\mbox{\tiny $(4)$}}(1^{-},2^{+},3^{-},4^{+},5^{+})\:=\:
 \raisebox{-1.3cm}{\scalebox{.23}{\includegraphics{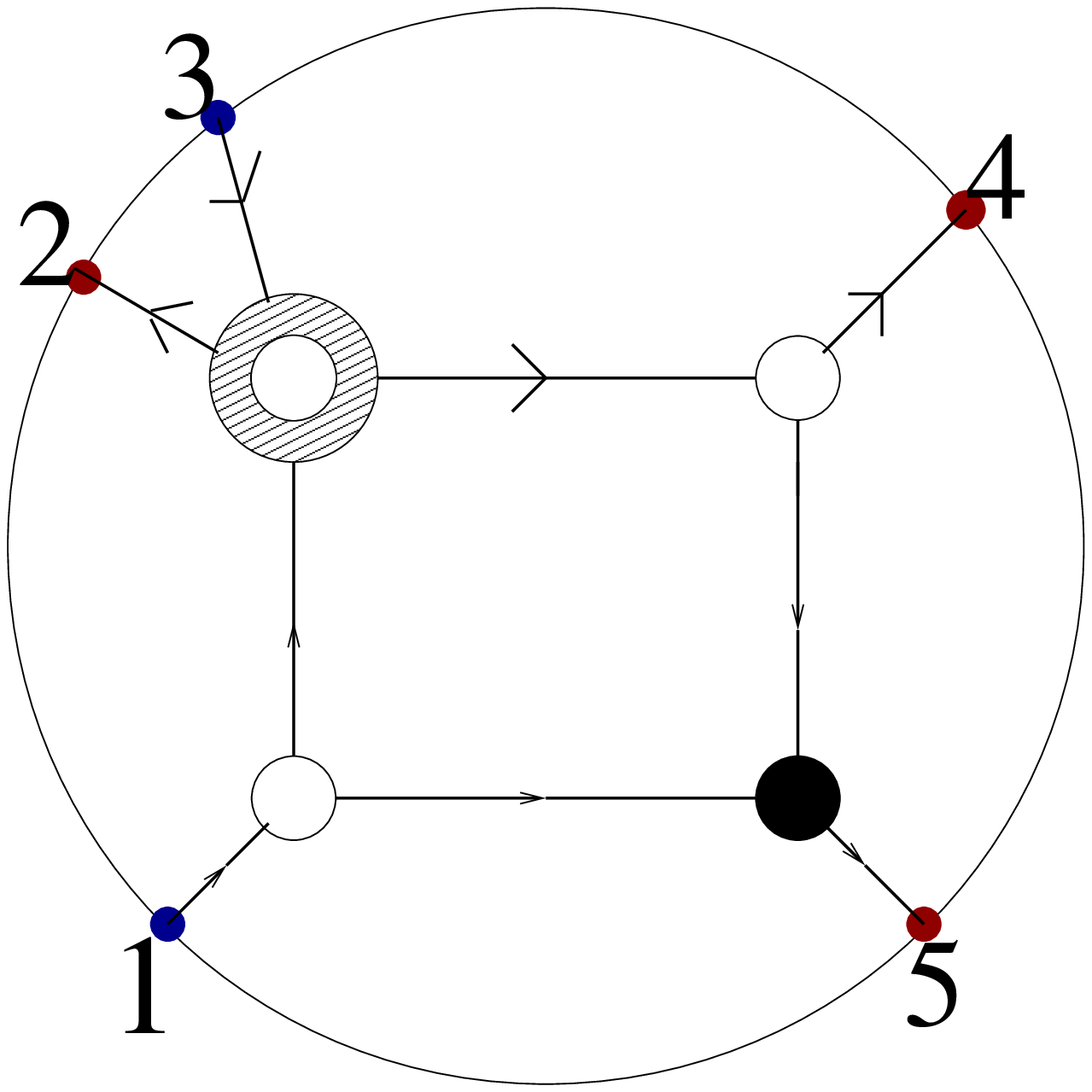}}}\:+\:
 \raisebox{-1.3cm}{\scalebox{.23}{\includegraphics{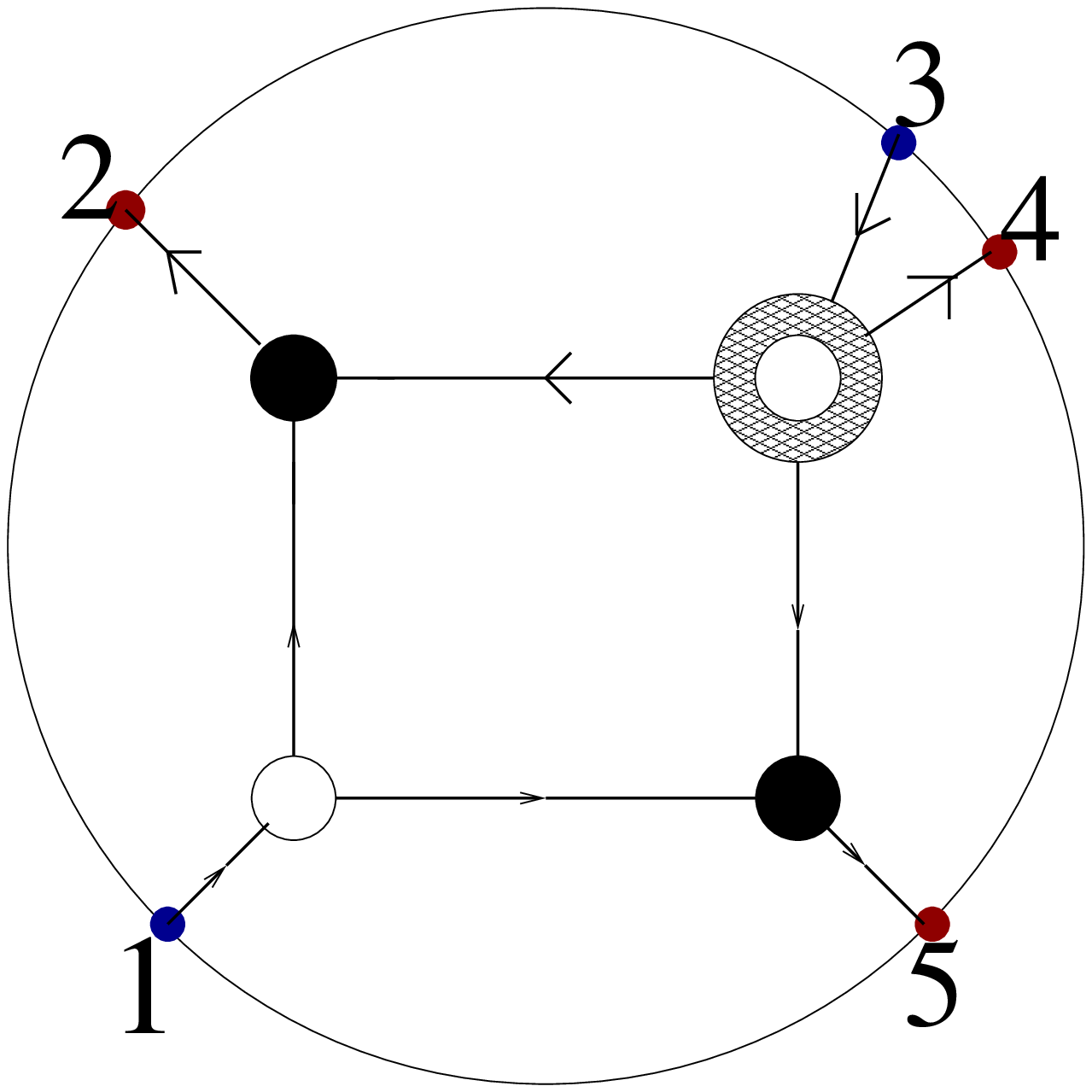}}}\:+\:
 \raisebox{-1.3cm}{\scalebox{.23}{\includegraphics{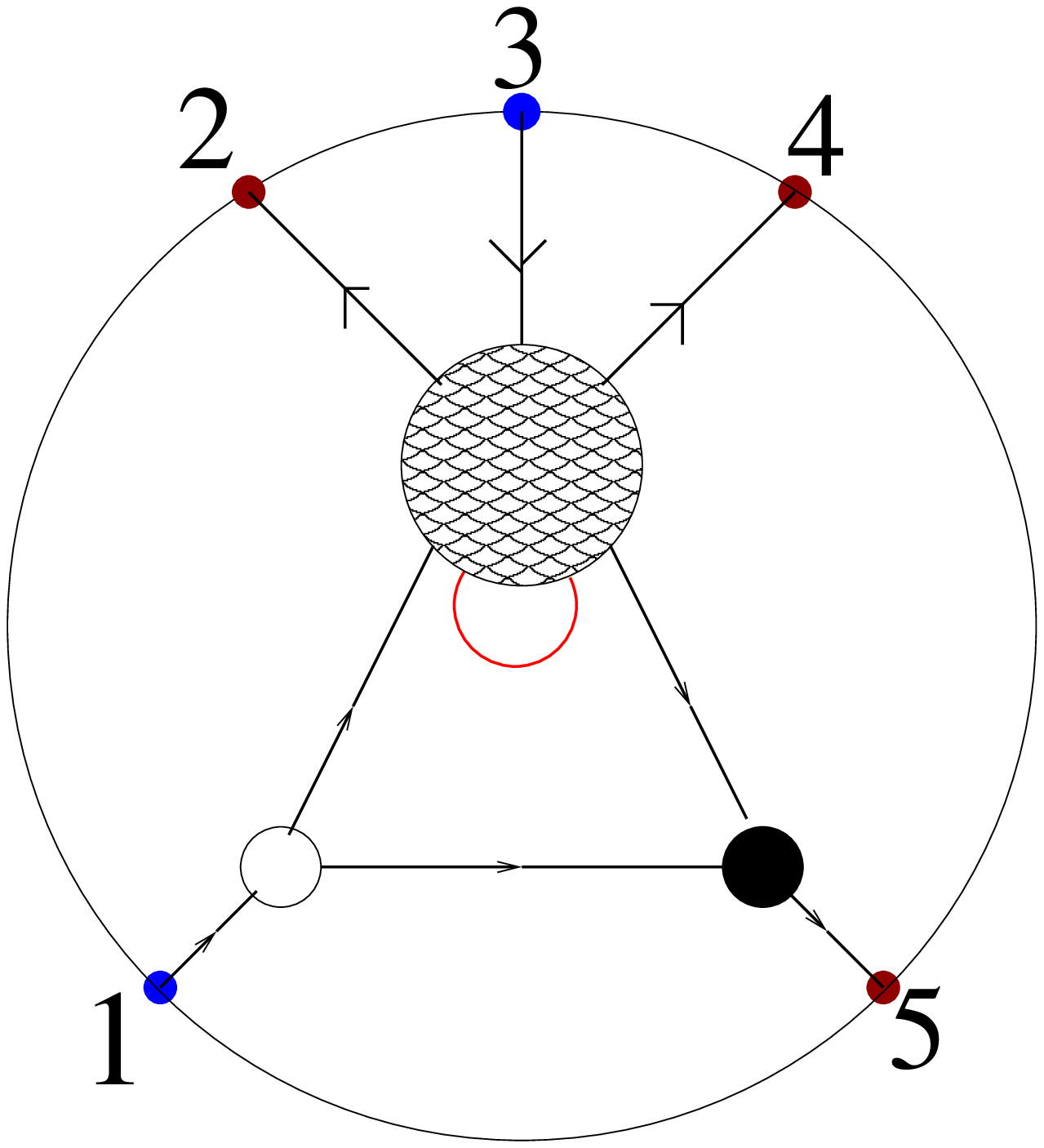}}},
\end{equation}
where the second term contributes just in the case of pure Yang-Mills for which a mass-deformation is introduced
to construct the integrand related to the rational terms. The other two terms encode the cut-constructible 
information for any $\mathcal{N}$. The explicit representation as on-shell diagrams can be written as
\begin{equation}\eqlabel{eq:1lM5mpmpp2}
 \mathcal{M}_{5}^{\mbox{\tiny $(4)$}}\:=\:
 \raisebox{-1.7cm}{\scalebox{.23}{\includegraphics{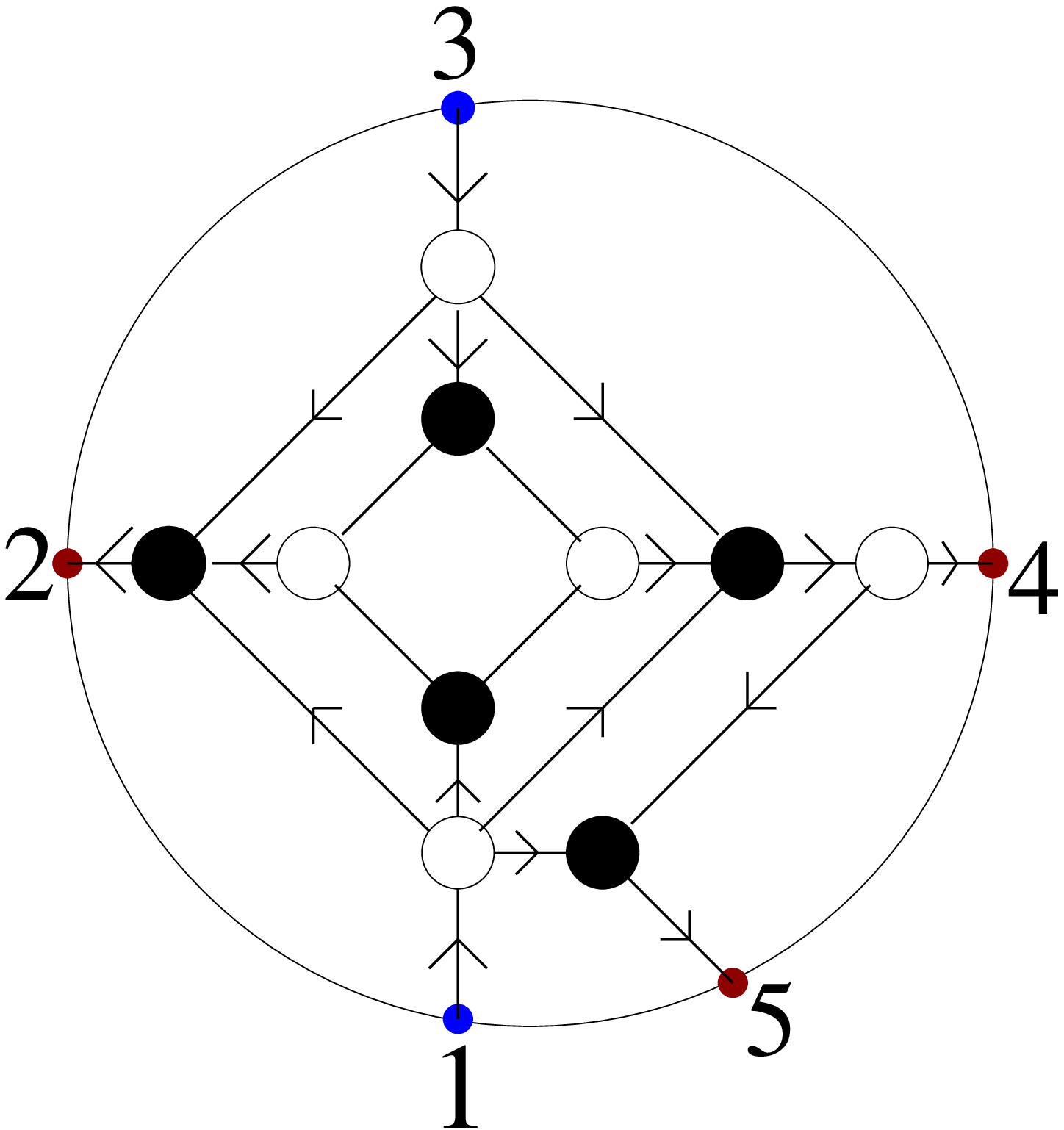}}}\:+\:
 \raisebox{-1.7cm}{\scalebox{.23}{\includegraphics{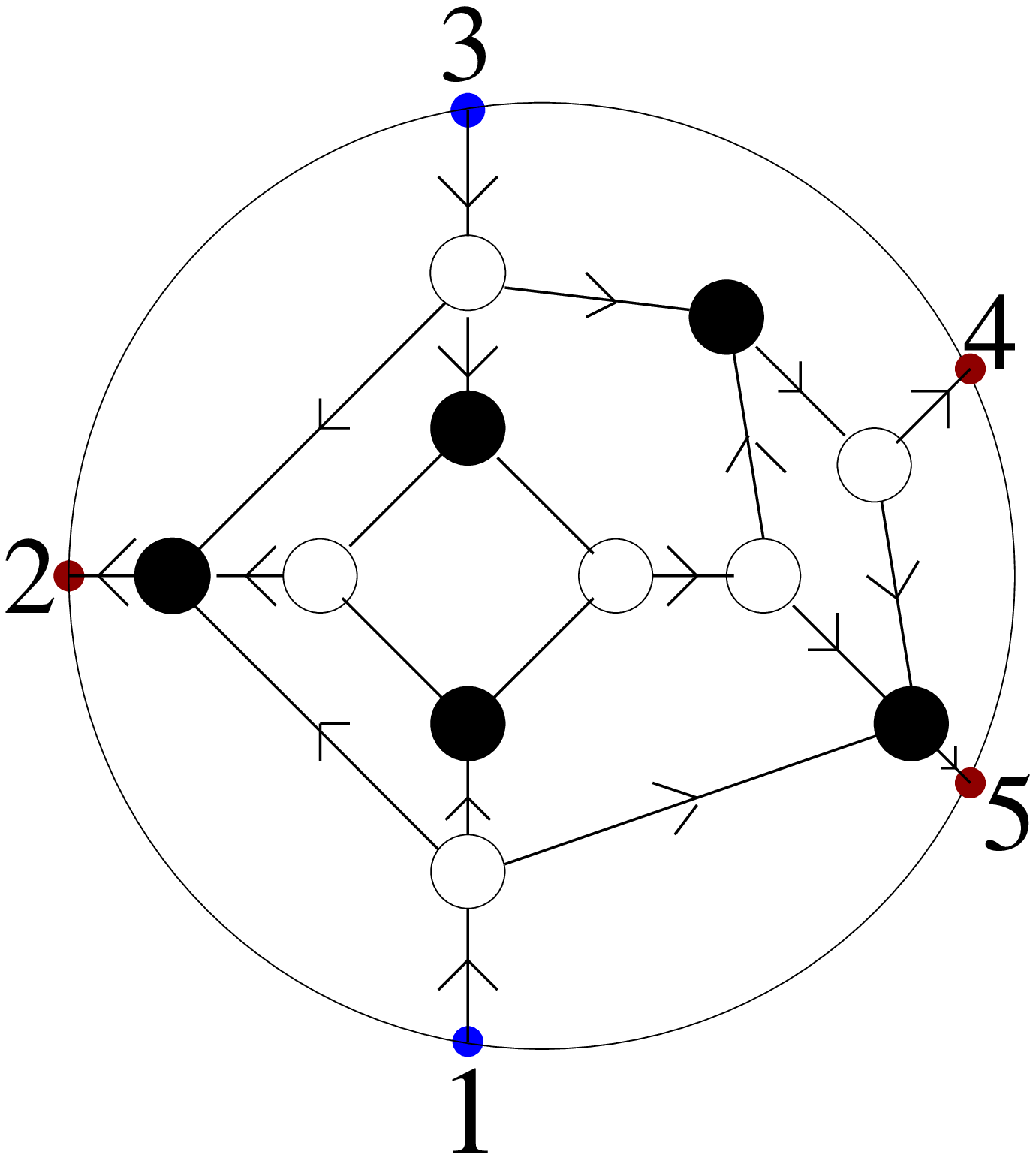}}}\:+\:
 \raisebox{-1.7cm}{\scalebox{.23}{\includegraphics{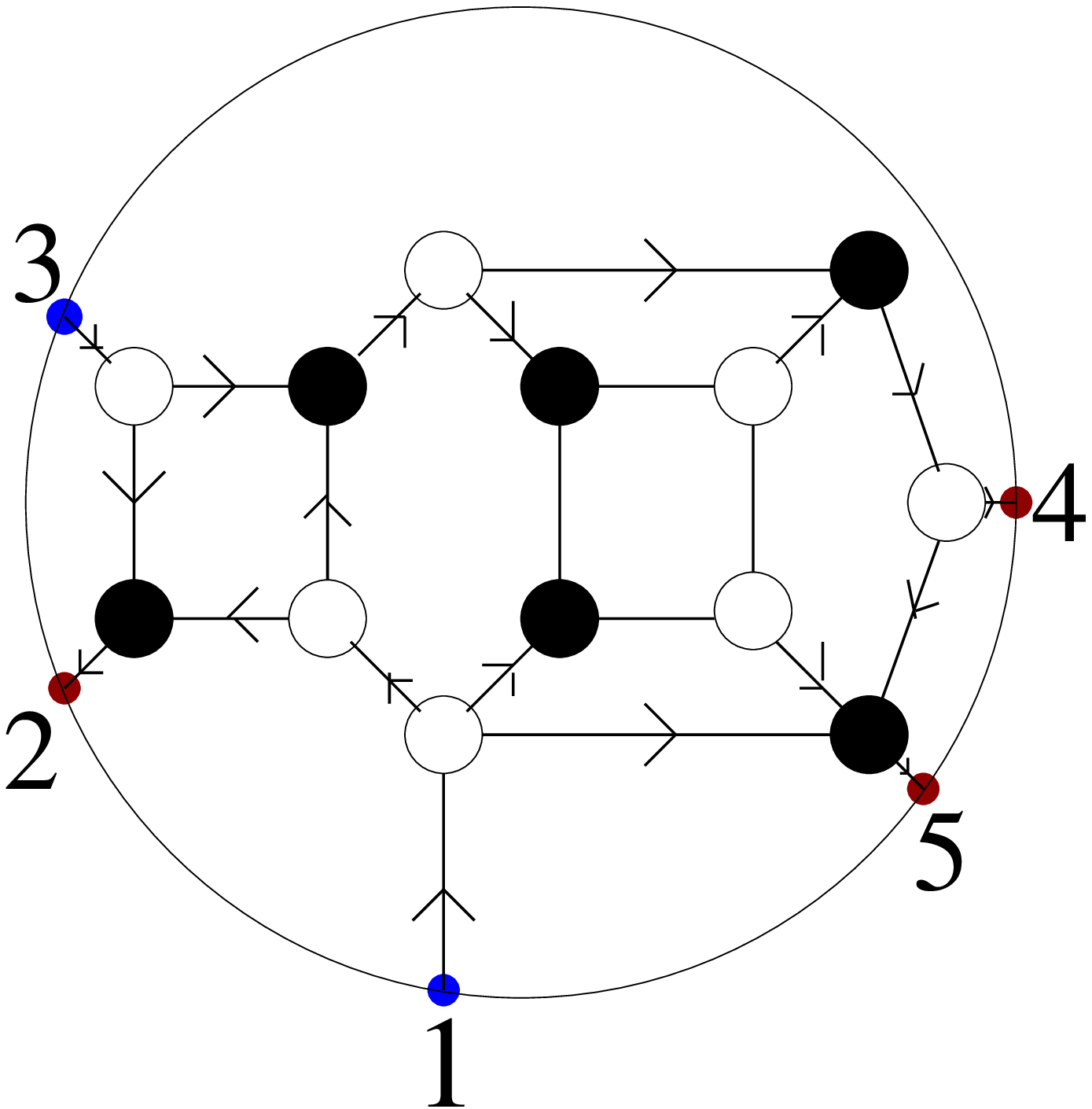}}}
\end{equation}
where the internal lines with no helicity arrow assignment allow for the propagation of both the multiplets,
and the second and third terms come from the forward term in \eqref{eq:1lM5mpmpp}.

It is interesting that, as in the four-particle case, the on-shell processes in \eqref{eq:1lM5mpmpp2} can
be seen as four BCFW bridges applied to a leading singularity. As an example, let us consider the explicit
expression of the second term in \eqref{eq:1lM5mpmpp2}:
\begin{equation}\eqlabel{eq:1lM5mpmppOS2}
 \raisebox{-1.7cm}{\scalebox{.23}{\includegraphics{1LM5mpmpp5.eps}}}\:=\:\mathcal{M}_5^{\mbox{\tiny tree}}
 \bigwedge_{i=4}^2\frac{dz_{i,i+1}}{z_{i,i+1}\left(1+z_{i,i+1}a_{i,i+1}\right)}
 \frac{\left[\mathcal{J}_{a}(z)\right]^{4-\mathcal{N}}+
       \left[\mathcal{J}_{b}(z)\right]^{4-\mathcal{N}}}{
       \left[\mathcal{J}_{a}(z)+\mathcal{J}_{b}(z)\right]^{4-\mathcal{N}}}
\end{equation}
where, as usual, $z_{i,i+1}$ parametrises the BCFW bridge in the $(i,i+1)$-channel, while the $a_{i,i+1}$'s and 
$\mathcal{J}_{a/b}(z)$ are given by
\begin{equation}\eqlabel{eq:1lM5mpmppOS2b}
 \begin{split}
  &a_{12}\:=\:\frac{\langle1,3\rangle}{\langle2,3\rangle},\qquad
   a_{23}\:=\:\frac{\langle1,3\rangle}{\langle1,2\rangle},\qquad
   a_{45}\:=\:\frac{\langle4,1\rangle}{\langle5,1\rangle},\qquad
   a_{51}\:=\:\frac{\langle4,1\rangle}{\langle4,5\rangle},\\
  &\mathcal{J}_{a}(z)\:=\:\frac{\langle5,1\rangle\langle3,2\rangle}{\langle5,2\rangle\langle3,1\rangle}
   \left(1+\frac{\langle4,1\rangle}{\langle5,1\rangle}z_{45}\right)
   \left(1+\frac{\langle3,1\rangle}{\langle3,2\rangle}z_{12}\right),\\
  &\mathcal{J}_{b}(z)\:=\:\frac{\langle1,2\rangle\langle3,5\rangle}{\langle5,2\rangle\langle3,1\rangle}
   \left(1+\frac{\langle1,3\rangle}{\langle1,2\rangle}z_{23}\right)
   \left(1+\frac{\langle3,4\rangle}{\langle3,5\rangle}z_{45}+
         \frac{\langle3,1\rangle}{\langle3,5\rangle}z_{51}\right).
 \end{split}
\end{equation}
The leading singularity encoded in \eqref{eq:1lM5mpmppOS2} can be read off by integrating it over the contour
$T^{4}\,=\,\{z_{i,i+1}\,\in\,\mathbb{C}\,|\,z_{i,i+1}\,=\,0,\,\forall\,i\}$. In the case of the other two
on-shell processes in \eqref{eq:1lM5mpmpp2}, once we integrate over the suitable $T^4$, just one of the two
multiplets which can run in the internal lines with un-fixed decoration contributes to the related leading
singularity.

Further analysing the on-shell process \eqref{eq:1lM5mpmppOS2}, its integration over the following contours,
returns the information contained in three double cuts:
\begin{itemize}
 \item $\gamma_{1}\,=\,\{(z_{51},\,z_{23})\,\in\,\mathbb{C}^2\,|\,z_{23}\,=\,0\,=\,z_{51}\}$
       \begin{equation}\eqlabel{eq:1lM5mpmppDC1}
        \raisebox{-1.7cm}{\scalebox{.23}{\includegraphics{1LM5mpmpp5.eps}}}\:\longrightarrow\:
        \raisebox{-1.7cm}{\scalebox{.23}{\includegraphics{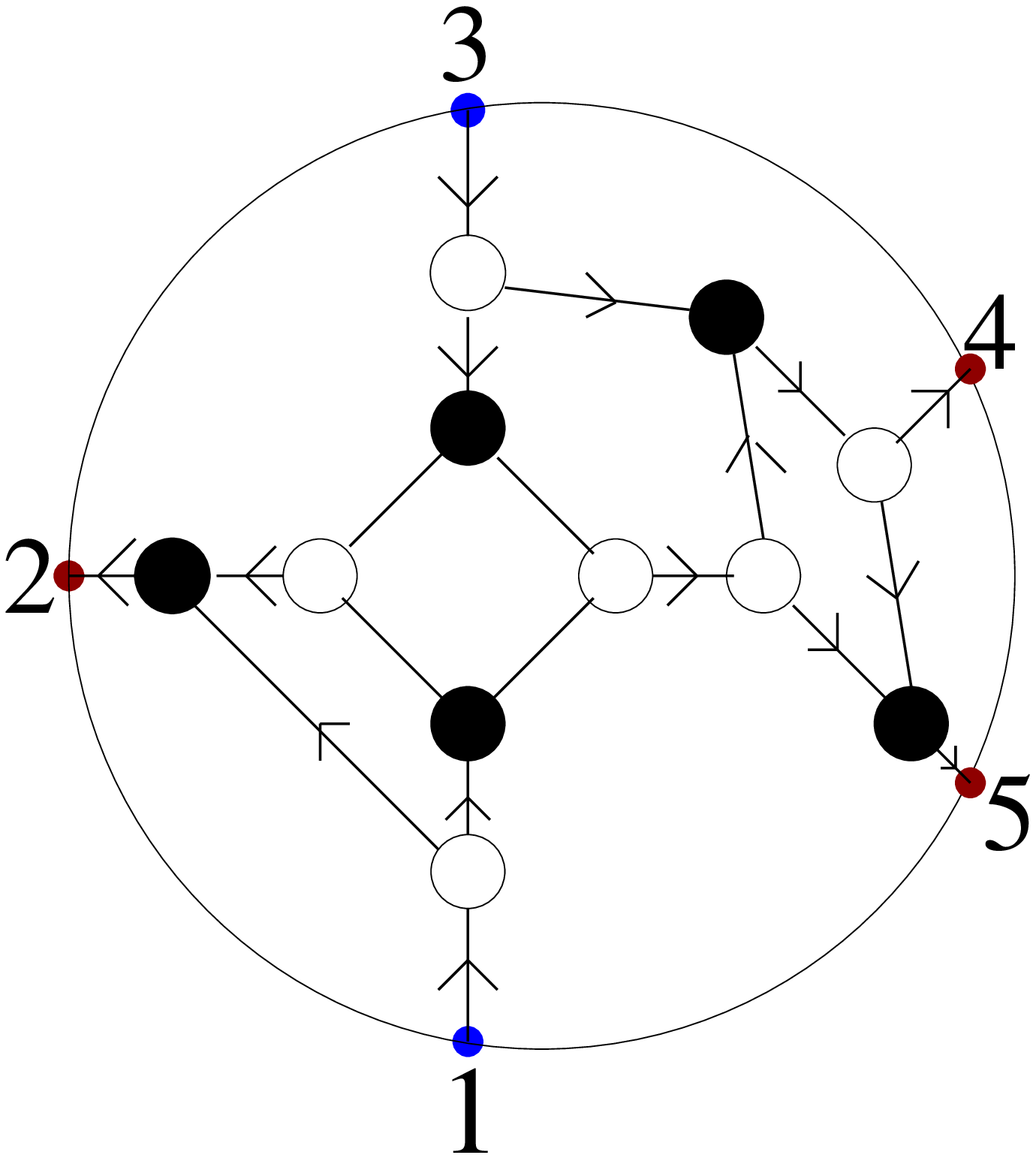}}}\:\equiv\:
        \raisebox{-1.7cm}{\scalebox{.23}{\includegraphics{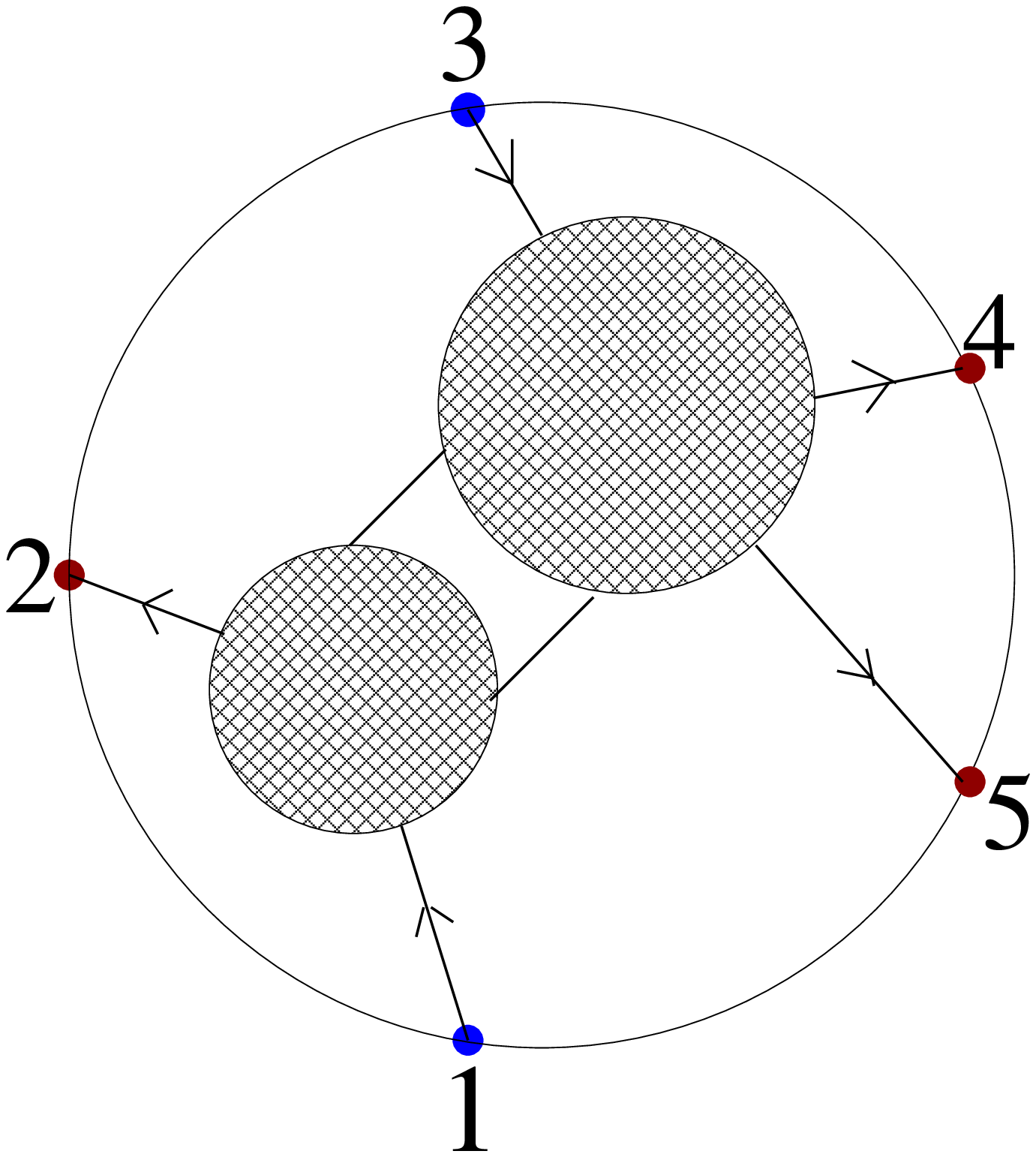}}};
       \end{equation}
 \item $\gamma_{2}\,=\,\{(z_{12},\,z_{45})\,\in\,\mathbb{C}^2\,|\,z_{12}\,=\,0\,=\,z_{45}\}$
        \begin{equation}\eqlabel{eq:1lM5mpmppDC2}
        \raisebox{-1.7cm}{\scalebox{.23}{\includegraphics{1LM5mpmpp5.eps}}}\:\longrightarrow\:
        \raisebox{-1.7cm}{\scalebox{.23}{\includegraphics{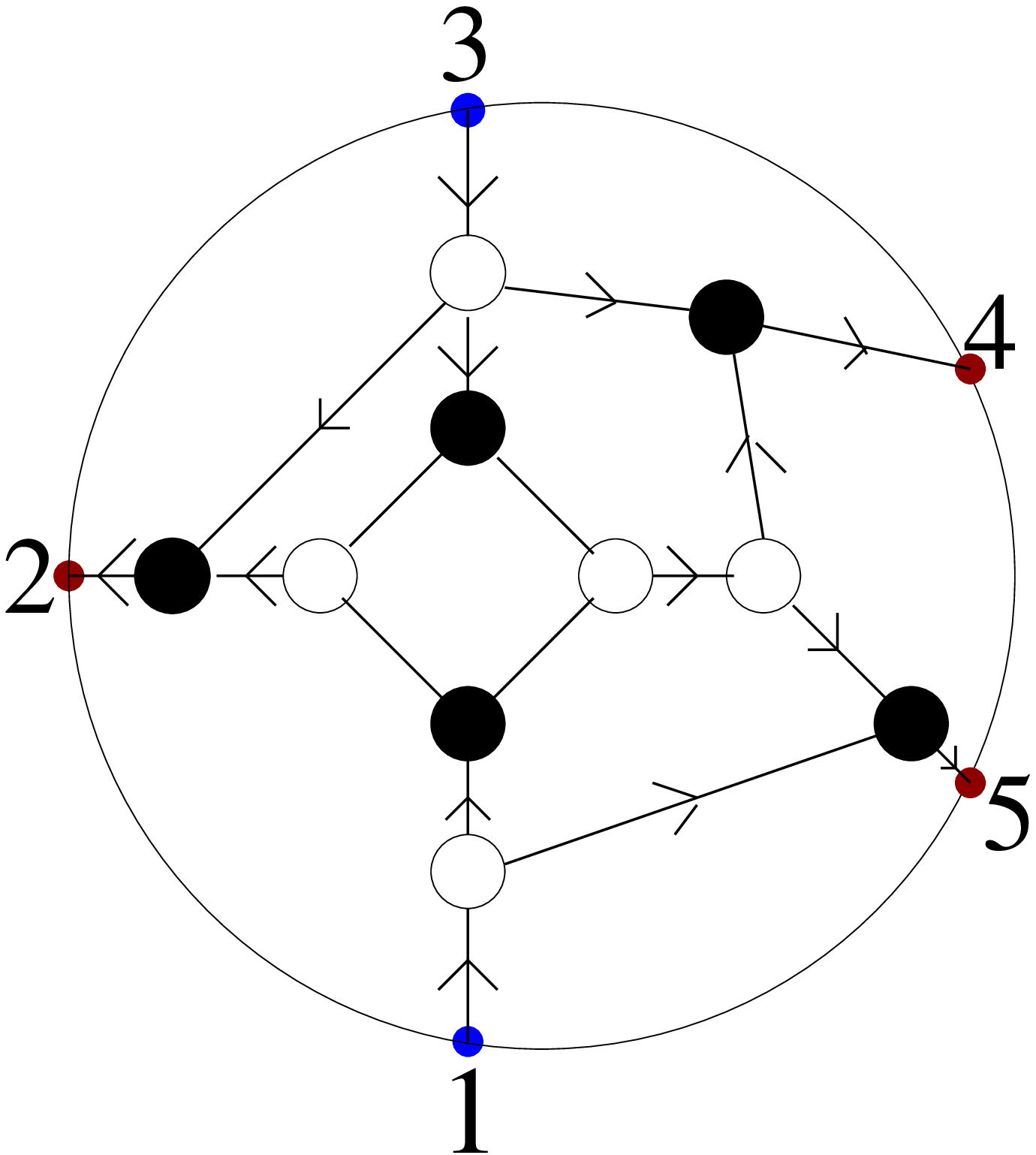}}}\:\equiv\:
        \raisebox{-1.7cm}{\scalebox{.23}{\includegraphics{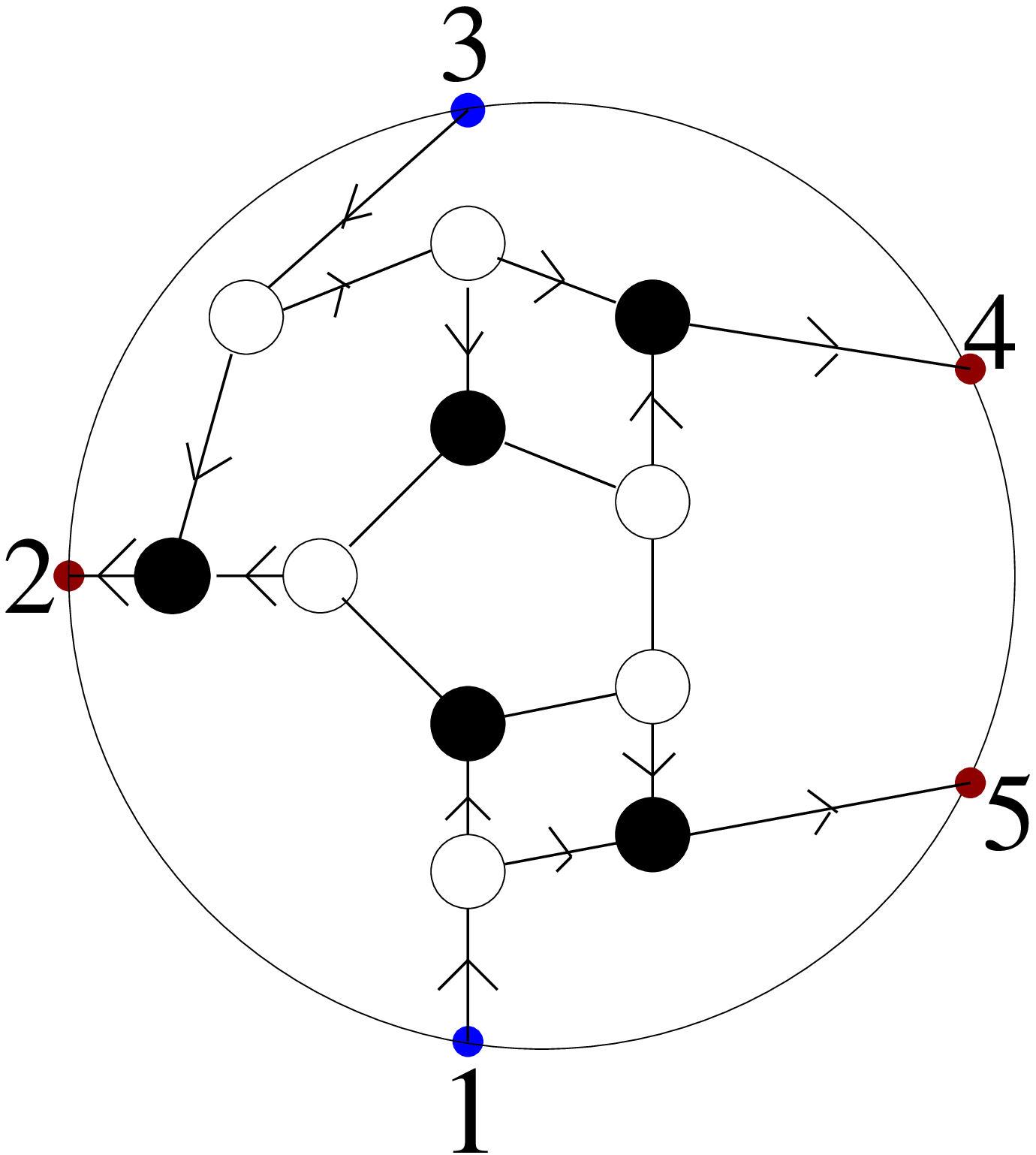}}}\:\equiv\:
        \raisebox{-1.7cm}{\scalebox{.23}{\includegraphics{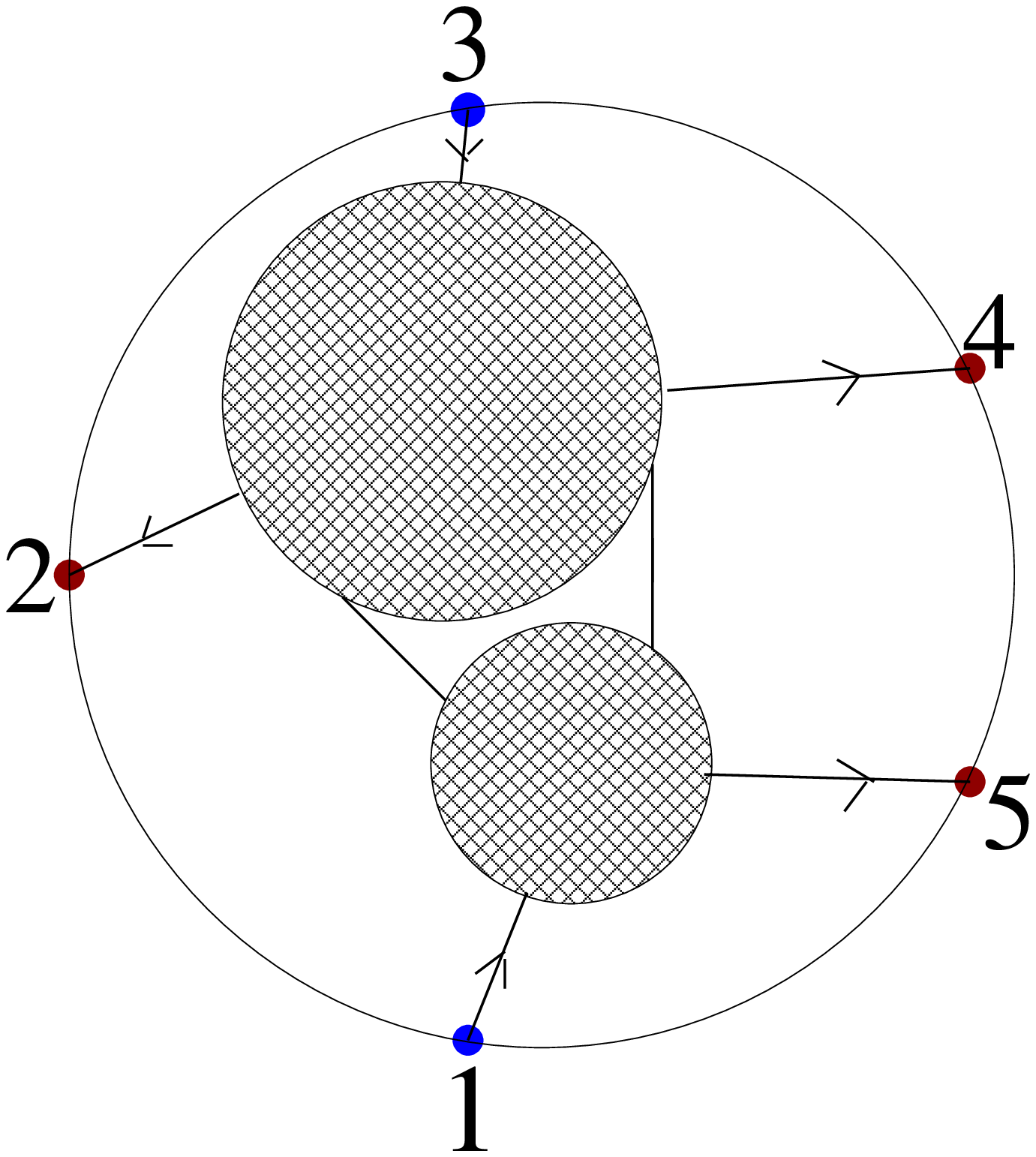}}}
       \end{equation}
       where the third diagram has been obtained via two equivalence operation to make manifest that the
       upper pentagon and box form a tree-level five particle sub-amplitude;
 \item $\gamma_{3}\,=\,\{(z_{23},\,z_{45})\,\in\,\mathbb{C}^2\,|\,z_{23}\,=\,0\,=\,z_{45}\}$
        \begin{equation}\eqlabel{eq:1lM5mpmppDC3}
        \raisebox{-1.7cm}{\scalebox{.23}{\includegraphics{1LM5mpmpp5.eps}}}\:\longrightarrow\:
        \raisebox{-1.7cm}{\scalebox{.23}{\includegraphics{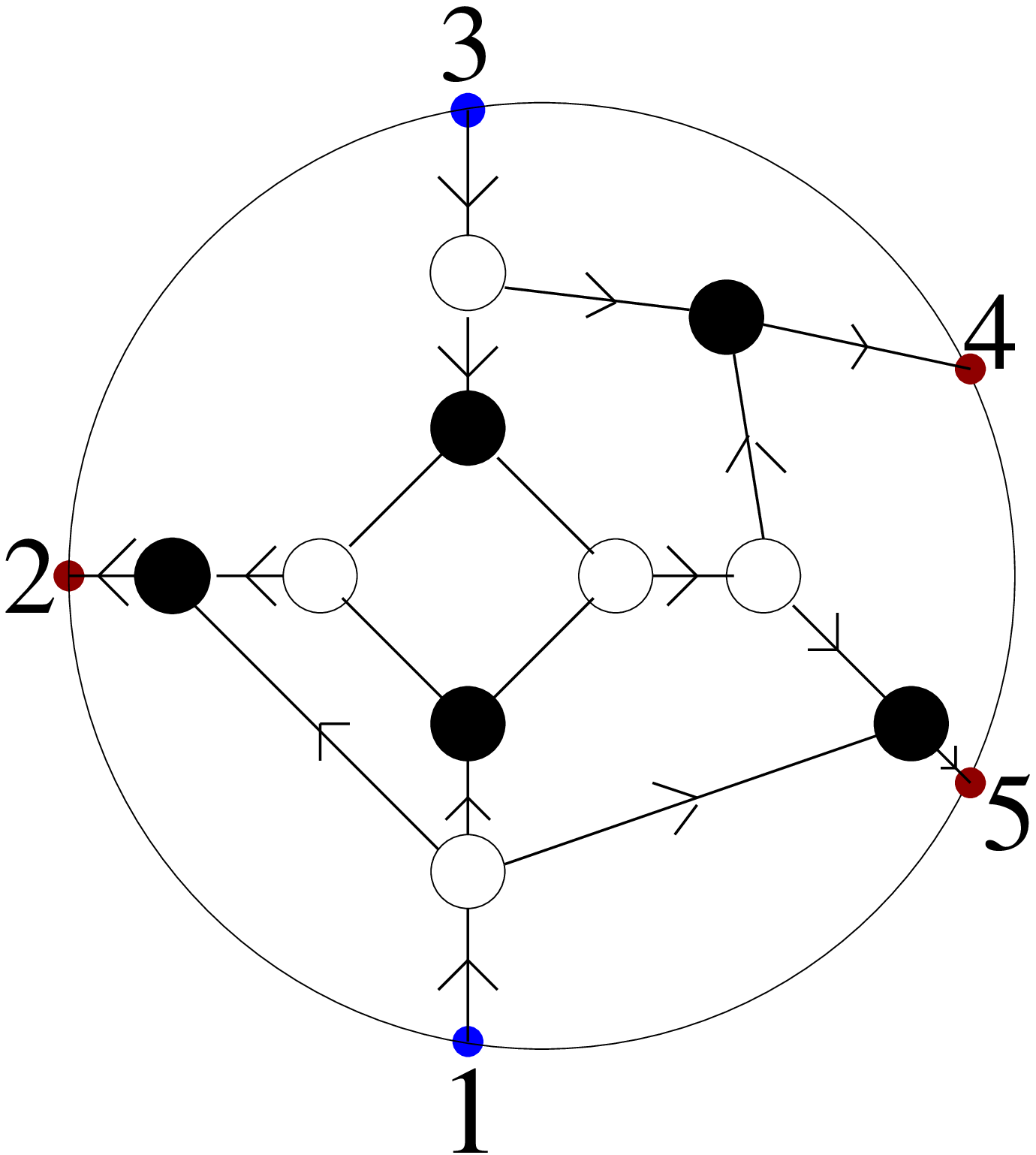}}}\:\equiv\:
        \raisebox{-1.7cm}{\scalebox{.23}{\includegraphics{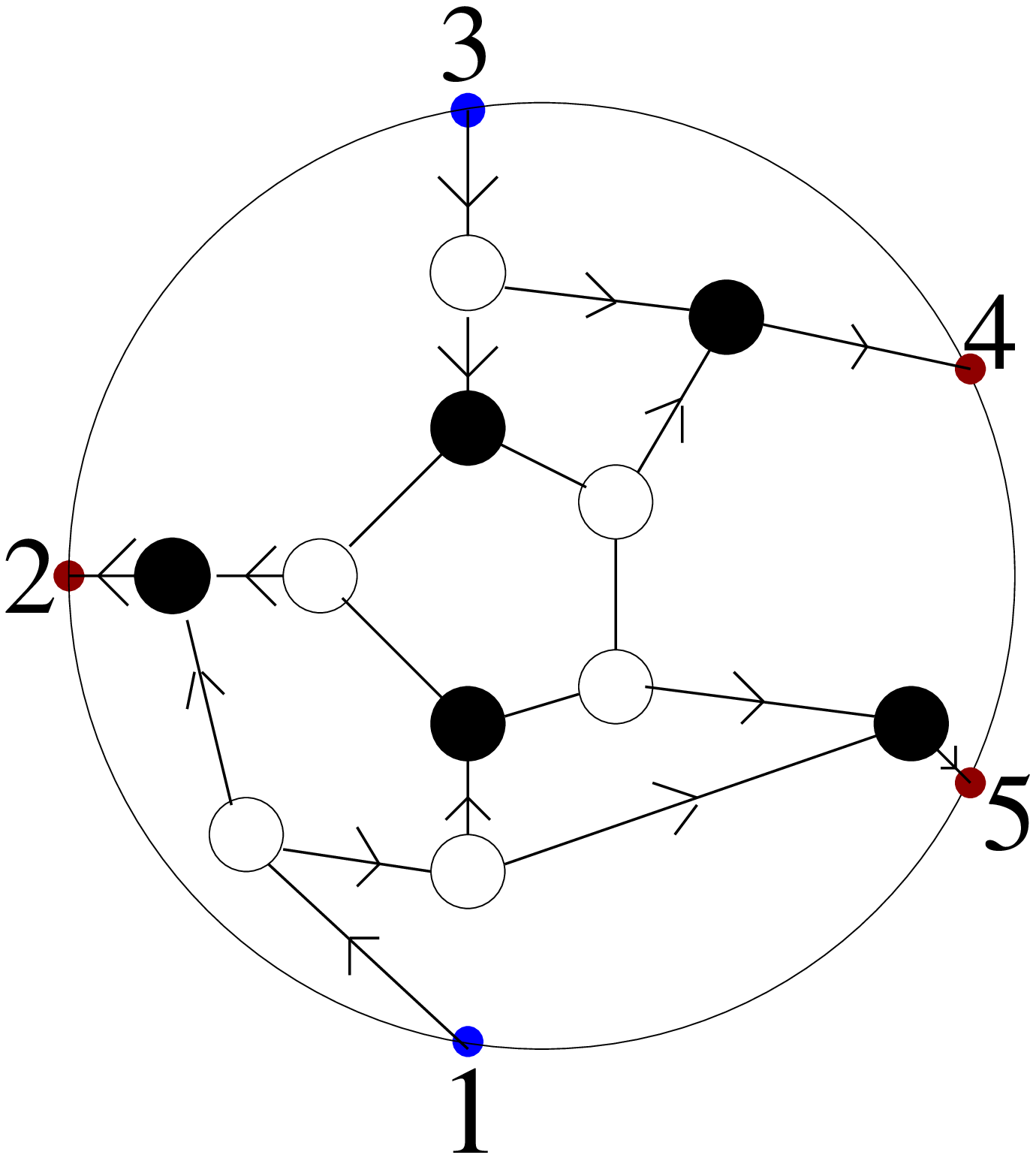}}}\:\equiv\:
        \raisebox{-1.7cm}{\scalebox{.23}{\includegraphics{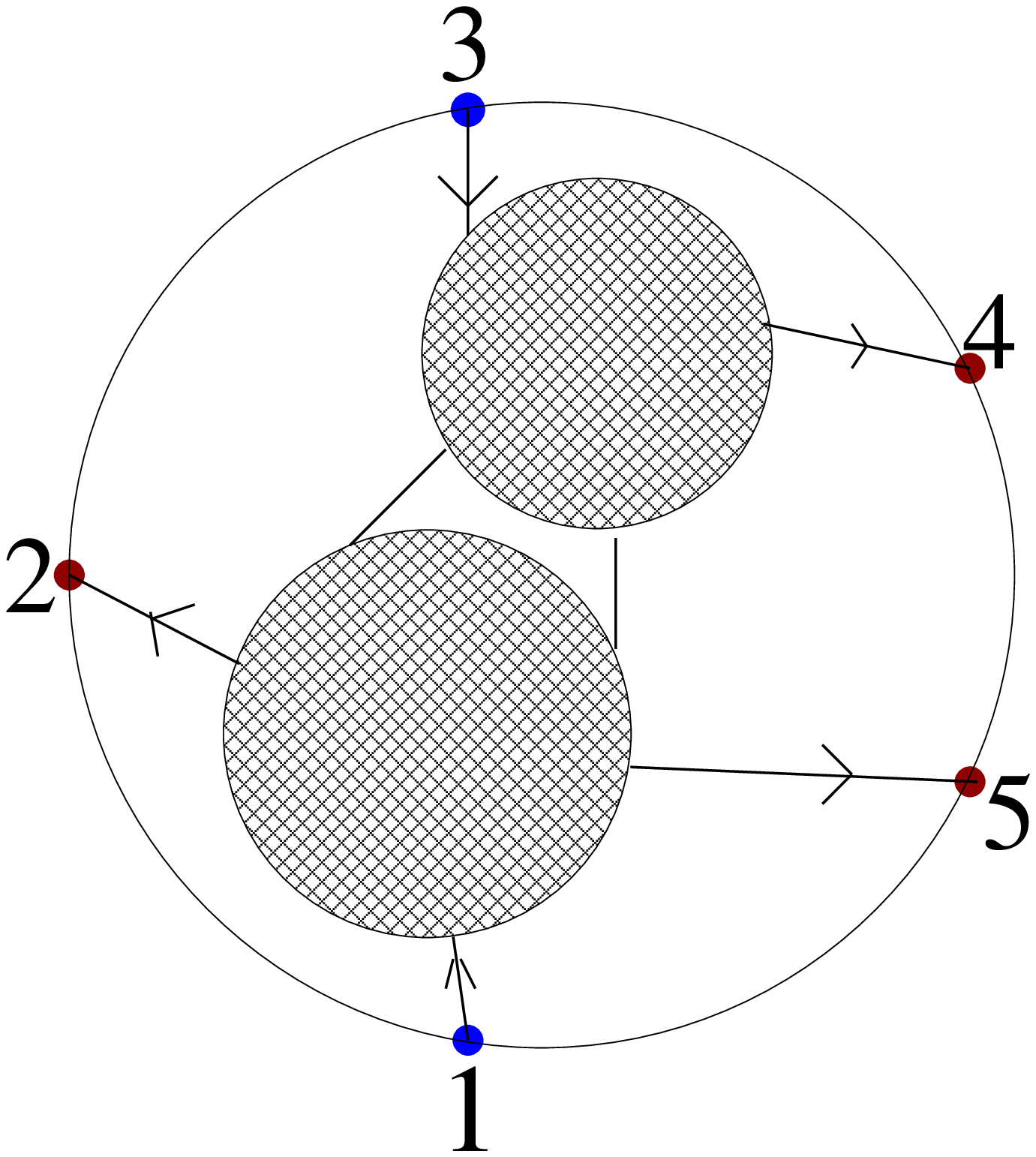}}}
       \end{equation}
\end{itemize}
Thus, the on-shell process \eqref{eq:1lM5mpmppOS2} encodes the information of the double cuts in the $(1,2)$-,
$(3,4)$- and $(5,1)$-channels. It is worth to remark once again that the integration over the contours 
$\gamma_{1}$, $\gamma_{2}$, and $\gamma_{3}$ return an on-shell $2$-form which needs further integration to read
off the coefficients of the more familiar integral basis expansion. However, the on-shell diagrams
\eqref{eq:1lM5mpmpp} beautifully make manifest the full mathematical structure, with the helicity flows
distinguishing between factorisation channels and higher-degree singularities which are signature of the presence
of sub-leading singularities (namely, in the scalar integral basis language, the triangle and bubble structure).

For the sake of completeness, let us decode the information in the other two on-shell processes. The first one
in \eqref{eq:1lM5mpmpp} encodes all the information of the double cut in the $(2,3)$-channel:
\begin{equation}\eqlabel{eq:1lM5mpmppDC4}
 \raisebox{-1.7cm}{\scalebox{.23}{\includegraphics{1LM5mpmpp4.eps}}}\:\longrightarrow\:
 \raisebox{-1.7cm}{\scalebox{.23}{\includegraphics{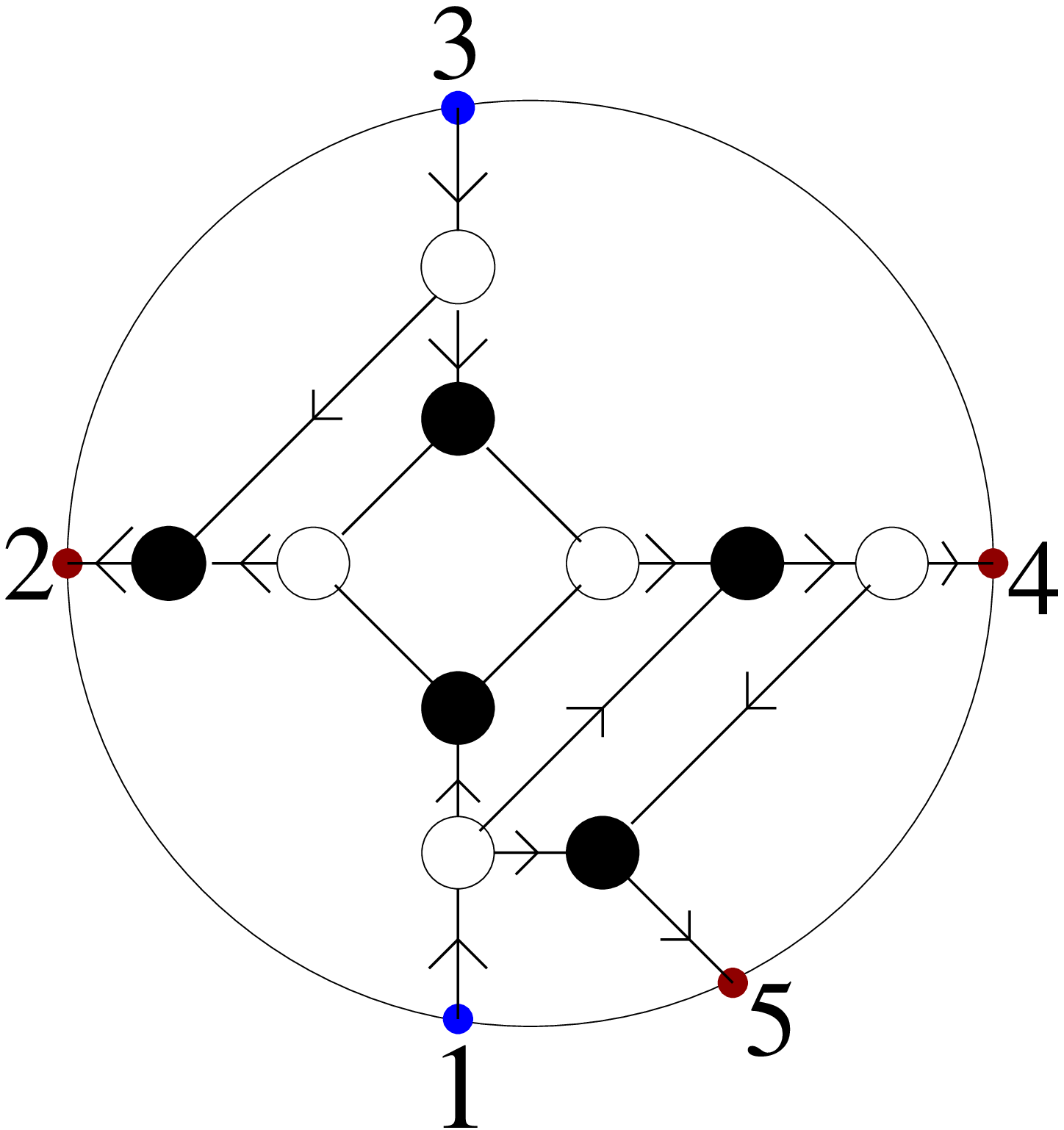}}}\:\equiv\:
 \raisebox{-1.7cm}{\scalebox{.23}{\includegraphics{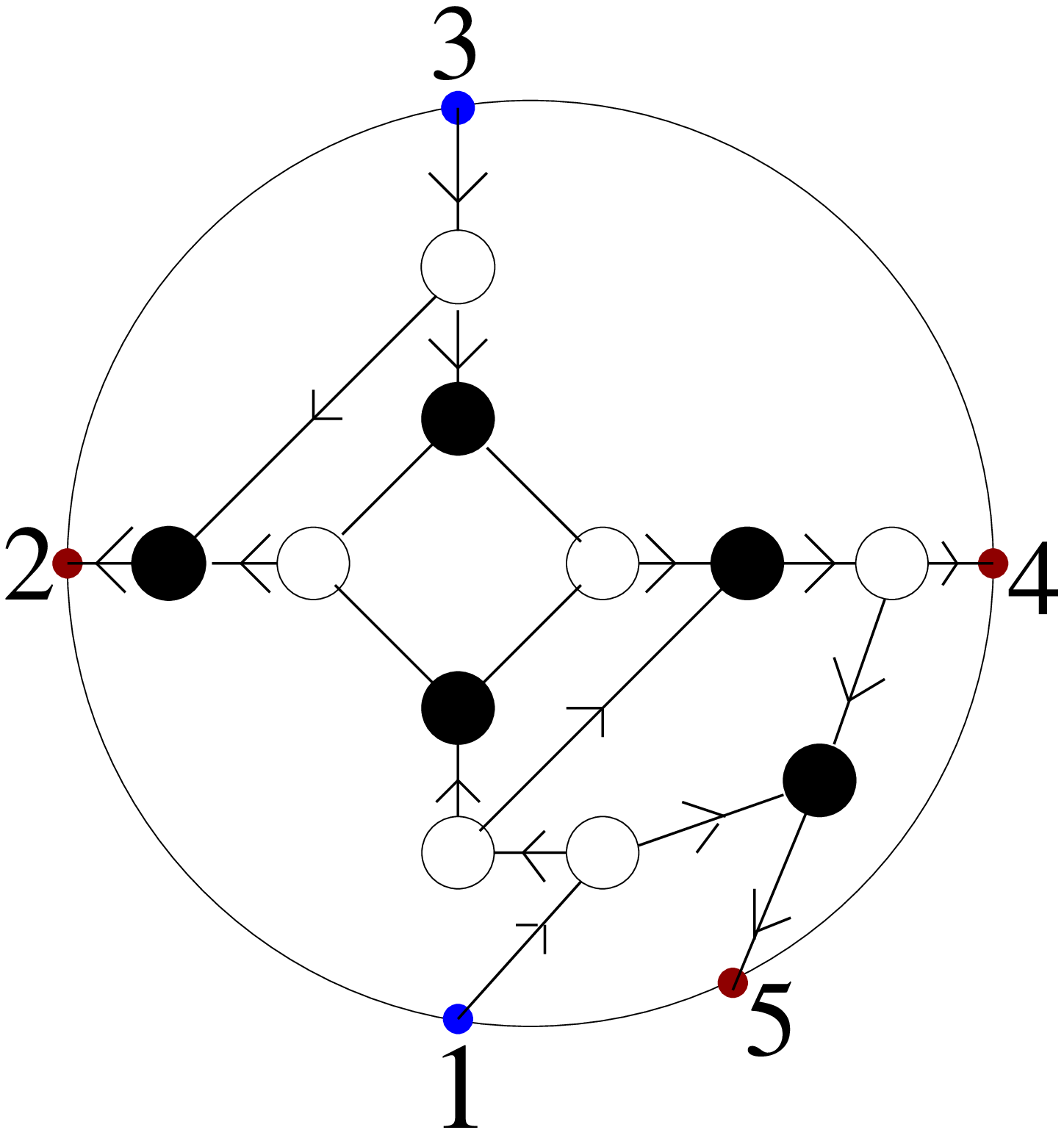}}}\:\equiv\:
 \raisebox{-1.7cm}{\scalebox{.23}{\includegraphics{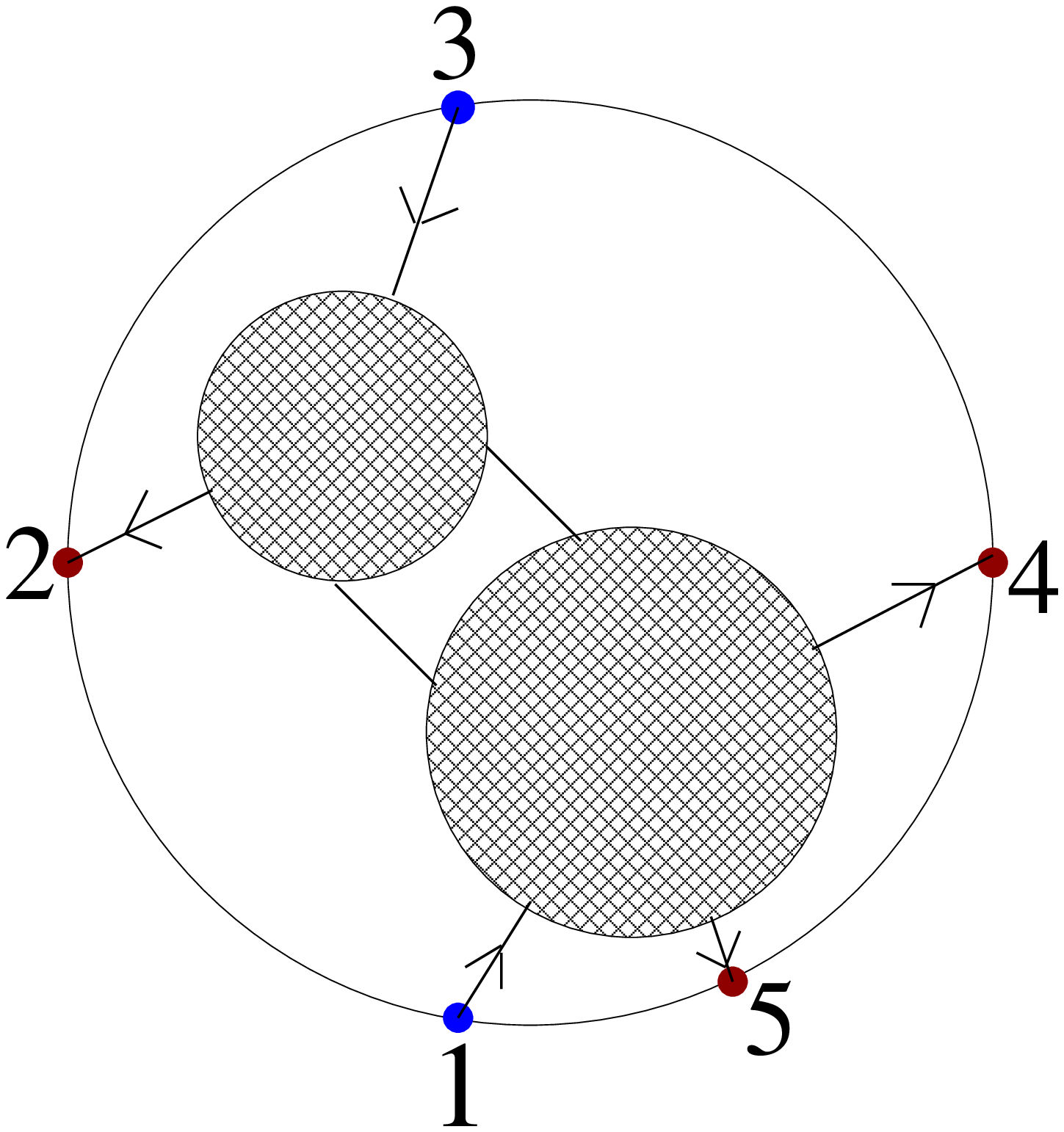}}}
\end{equation}
Notice that when we compute the left-hand-side, the only thing that the BCFW bridge of \eqref{eq:1lM5mpmpp} is
doing is taking the direct product of an on-shell $4$- and $0$-forms localising it in a different region of 
momentum space: the result is a higher point on-shell $4$-form which keeps the very same parametrisation of the 
lower point one. Hence, in \eqref{eq:1lM5mpmppDC4} one is simply integrating over a contour which eliminates
two BCFW bridges of the lower-point amplitude.

In a similar fashion, we can check that the last on-shell process in \eqref{eq:1lM5mpmpp} contains the information
encoded in the remaining double cut:
\begin{equation}\eqlabel{eq:1lM5mpmppDC5}
 \raisebox{-1.7cm}{\scalebox{.23}{\includegraphics{1LM5mpmpp6.eps}}}\:\longrightarrow\:
 \raisebox{-1.7cm}{\scalebox{.23}{\includegraphics{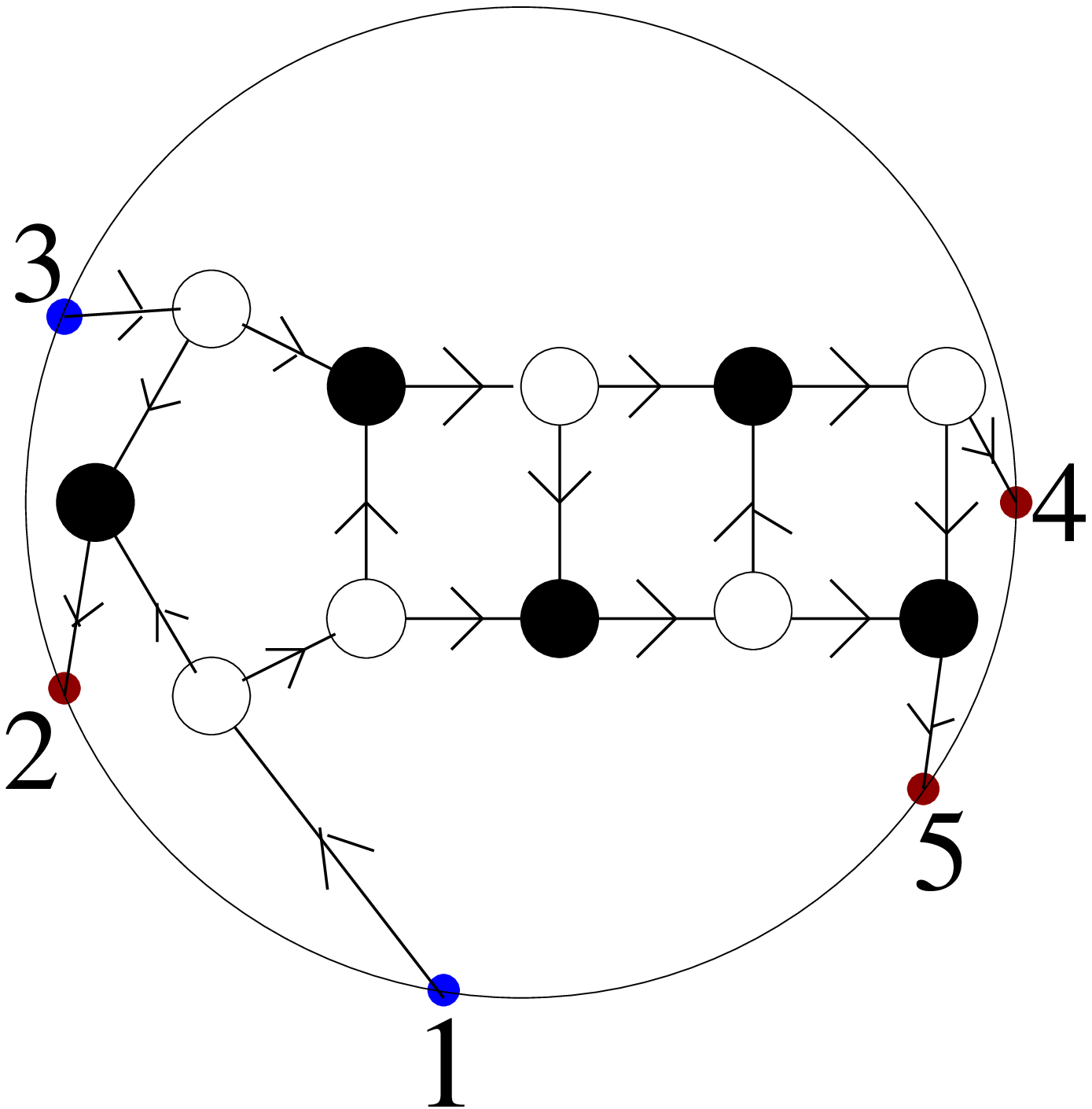}}}\:\equiv\:
 \raisebox{-1.7cm}{\scalebox{.23}{\includegraphics{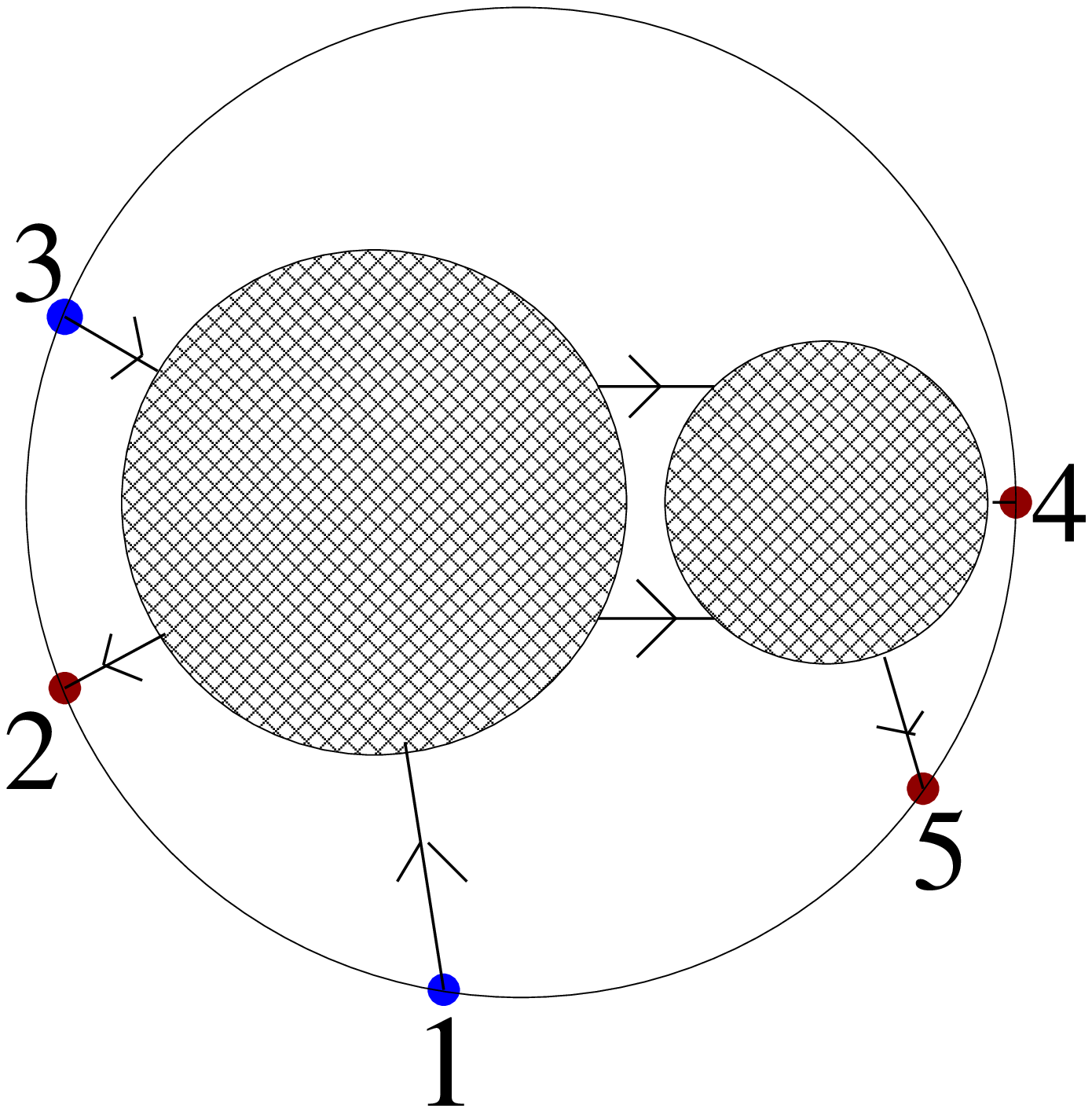}}}
\end{equation}
where one uses equivalence relations as well as the bubble deletion \eqref{BubDelw} on the un-decorated internal
box (integrating over the related higher order singularity).

\bibliographystyle{utphys}
\bibliography{amplitudesrefs}	

\end{document}